%% file: Kuzmin-thesis.tex
\documentclass[11pt,openright,final,smallheadings]{scrbook}

\usepackage[estonian, english]{babel}
\usepackage[utf8]{inputenc}
\usepackage[toc,title,page]{appendix}
\usepackage[T1]{fontenc} 
\usepackage{times}

\usepackage{longtable} 
\usepackage{aas_macros} 
\usepackage{verbatim} 
\usepackage{amsmath} 
\usepackage{amsfonts} 
\usepackage{amssymb}  
\usepackage{amscd} 
\usepackage{graphicx,epsfig} 
\usepackage{subfig} 
\usepackage{float} 
\usepackage{lscape} 
\usepackage{url}
\usepackage{sidecap}
\sidecaptionvpos{figure}{t}

\usepackage[english,estonian]{babel}

\usepackage{accents}

\usepackage{natbib}

\include{newcommands}

\bibpunct{(}{)}{;}{a}{}{,} 

\begin{document}

\frontmatter
\include{headerpages}

\setcounter{tocdepth}{2}
\tableofcontents

\include{introduction0}      
\mainmatter

\include{chapter01}

\include{chapter02} 
\include{chapter03}

\include{chapter04}

\include{chapter05}

\include{chapter06}

\include{chapter07}

\include{chapter08}

\include{chapter09}

\include{chapter10} 
\include{chapter11}

\include{chapter12}

\include{chapter13}

\include{chapter14}

\include{chapter15}

\include{chapter16} 
\include{chapter17}

\include{chapter18}

\include{chapter19}

\include{chapter20}

\include{chapter21}

\include{chapter22}

\part*{Appendix}
\setcounter{figure}{0}

\setcounter{table}{0}

\renewcommand{\thechapter}{\Alph{chapter}.\arabic{chapter}}
\setcounter{chapter}{0}

\include{ch1973}

\include{ch1986}

\include{ch1987}

\backmatter

\include{conclusions}

\addcontentsline{toc}{chapter}{Bibliography}
\bibliographystyle{aa2}
\bibliography{kuzmin}{} 

\end{document}

%% file: newcommands.tex
\newcommand{\bn}{\begin{enumerate}}
\newcommand{\en}{\end{enumerate}}

\newcommand{\bed}{\begin{displaymath}}
\newcommand{\eed}{\end{displaymath}}

\newcommand{\ie}{{\it i.e. }}

\def\Msun{\,{\rm M_{\odot}}}

\def\Mp{\,{\rm M_{\odot} /pc^2}}

\def\Mp3{\,{\rm M_{\odot} /pc^3}}

\newcommand{\be}{\begin{equation}}
\newcommand{\ee}{\end{equation}}
\newcommand{\dd}[1]{{\rm d}#1\,}

\newcommand{\ba}{\begin{array}} 
\newcommand{\ea}{\end{array}}
\newcommand{\pdif}[2]{\frac{\partial #1}{\partial #2}}

\newcommand{\rmd}{\mathrm{d}}

\def\apjl{ApJ}%
\def\apjs{ApJS}%
\def\ao{Appl.~Opt.}%
\def\aap{A\&A}%

%% file: headerpages.tex




\thispagestyle{empty}

\begin{sffamily}
\begin{center}
{DISSERTATIONES ASTRONOMIAE UNIVERSITATIS TARTUENSIS}\\
\textbf{}
\vspace{40mm}

{\LARGE{\textbf{Grigori Kuzmin}}}\\
\vspace{20mm}

{\LARGE{Etudes on the Dynamics of Stellar Systems}}\\[10pt]


\end{center}
\end{sffamily}

\vfill

{\begin{figure*}[h] 
\centering 
\resizebox{0.55\textwidth}{!}{\includegraphics*{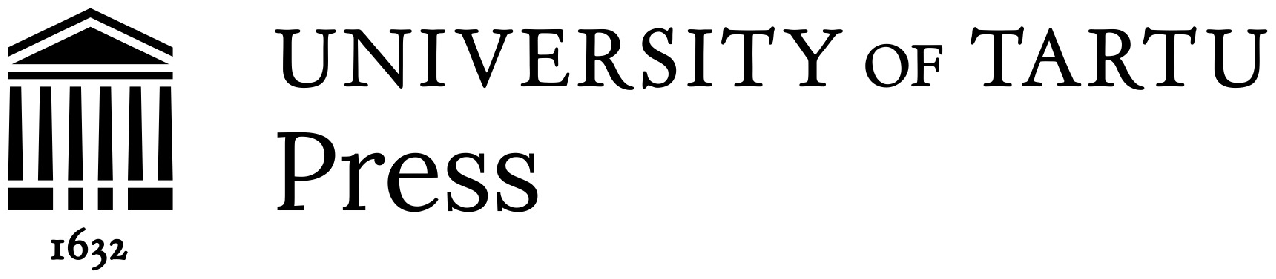}}
\end{figure*} 
}

\newpage
\thispagestyle{empty}
\noindent
This study was carried out at the Institute of Physics and Astronomy,
Estonian Academy of Sciences.

\vspace{5mm}

\noindent
The Dissertation was admitted 
in partial fulfilment of
the requirements for the degree of Doctor of Science in physics and
mathematics,  and allowed for defence by the Council of the
Institute of Physics and Astronomy, 

\vspace{5mm}

\begin{tabbing}

  Opponents: \hspace{0.7cm}

               \= Prof. K. F. Ogorodnikov\\
              \> Leningrad University \\
              \> Russia \\
\\
	       \> Prof. T. A. Agekian\\
              \> Leningrad University \\
              \> Russia \\
\\
	      \> Prof. G. M. Idlis\\
              \> Alma-Ata University \\
              \> Kazakstan \\

\vspace{5mm}\\
            
Leading institute: Sternberg Astronomical Institute, Moscow State
University \\             
\vspace{5mm}\\
Defence:      \> March 6, 1970, University of Tartu, Estonia \\

\end{tabbing}

\vfill
\begin{tabbing}
  Original in Russian --
  I Volume: Chapters 1 -- 10
  \\
  \hskip 33mm  II Volume: Chapters 11 -- 24 
  \\
  \\
ISBN 978-9949-03-839-8
\\
\\
Copyright: Tartu Observatory, 2022 \\
\\
University of Tartu Press 2022\\
www.tyk.ee\\
\end{tabbing}

%% file: introduction0.tex
\chapter{Preface to the English edition}

Grigori Kuzmin was born in Viiburi on 8 April 1917 and lived his first
ten years in Finland. His parents then moved to Tallinn, Estonia,
where Grigori also went to school.  He spoke and wrote Estonian
fluently, although his home language was still Russian. Here, in 1935,
he graduated from the Russian Gymnasium in Tallinn and went on to
study mathematics at the Faculty of Mathematics and Natural Sciences
of the Tartu University. He started as a free student and
matriculated after becoming a citizen of the Republic of Estonia.

From the autumn semester of 1936 he became a frequent visitor to the
Tartu Observatory to take part in a seminar on astrophysics under the
supervision of Ernst \"Opik. Kuzmin graduated cum laude in astronomy in
1940. In 1942, Kuzmin defended his master's thesis ``Bemerkungen zur
Dynamik des kosmischen Staubes'' (Remarks on the Dynamics of Cosmic
Dust).  Kuzmin lectured at the University of Tartu  courses in
stellar dynamics, astrophysics and general and practical astronomy. In
1971 he was awarded a professorship for his pedagogical activities.

{\begin{figure*}[h] 
\centering 
\resizebox{0.50\textwidth}{!}{\includegraphics*{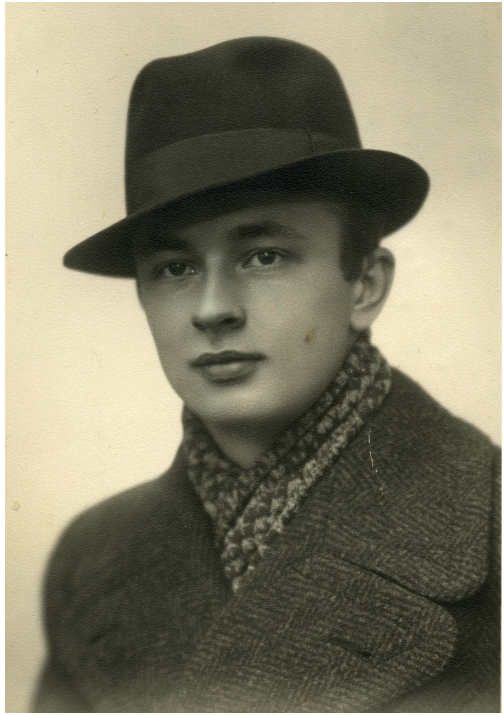}}
\caption{Grigori Kuzmin as student of Tartu University
} 
\end{figure*} 
}

 From 1948 he also worked at the Academy of Sciences, initially at the
 Institute of Physics, Mathematics and Mechanics, later reorganised as
 the Institute of Physics and Astronomy. He obtained his Candidate
 Degree in 1952 and  Doctor of Sciences Degree in 1970.
From 1960 to 1982 he was Head of the Stellar Astronomy Sector.  Kuzmin
retired in 1982 but continued his research activities as a senior
researcher-consultant. 
Kuzmin was editor-in-chief of the Astronomical Calendar (until 1963)
and the Publications of the Tartu Astronomical Observatory.
In 1961 he was elected to  Estonian Academy of
Sciences. He was 1976 - 1979 vice-president and 1979 - 1982 president
of Commission 33 of the  Galactic Structure and Dynamics of the
International Astronomical Union.

After the war  Kuzmin developed the theory of the third integral of
stellar motions, found a new method for calculating models of
galaxy mass distribution, and studied the density of matter in the
Galaxy around the Sun. Professor Pavel Parenago of Moscow University
 worked on similar problems on structure of the Galaxy. He also
developed  a model of the Galaxy and calculated the density of matter
around the Sun. Parenago is considered to be the founder of the modern
school of galaxy research at Moscow University. His results confirmed
Oort's estimate of the density of matter in the Galaxy.
The first serious meeting between the Tartu and Moscow schools took
place at the Session  of Astronomy Council of the Soviet Academy
of Science  in  Tartu,  May 1953. The ``duel'' between Parenago and Kuzmin was
very interesting. Both were dealing with similar problems, but with
very different results. Parenago's authority was very high at the time
and the Moscow establishment did not expect much serious competition
from the province. But the presentations showed that Kuzmin had a much
deeper understanding of the problem and had found a better solution
than Parenago to both the modelling of the structure of the Galaxy and
the problem of matter density. Kuzmin, in his modesty, refrained from
directly criticising Parenago, but his results spoke for
themselves. 
There were two important outcomes of this meeting. Whereas before we
had been treated as mere provincials, from then on Tartu
astronomers were taken as serious players. And   green
light was given for plans to build a new observatory.

In 1964, the new observatory was completed, and  Grigori
Kuzmin continued his work in  T\~oravere.  Kuzmin was a very family-oriented
person. His wife Zoja often visited their children in Tartu. Whenever
Zoja called, Grigori would drive to the station in T\~oravere and pick
her up with parcels.

Kuzmin suffered a heart attack in the early eighties. It happened
after his mother's funeral. The funeral was a lot of work and the
burden fell on Grigori. When the funeral was over and Kuzmin was alone
in his home in Tõravere, he suddenly felt a sharp pain in his
heart. He almost fainted, but luckily he didn't and very carefully
crawled to the phone and managed to call our doctor, Tiiu Kaasik. An
ambulance came immediately and Grigori was taken to Tartu. As we heard
later, the doctors fought for his life for quite a long time. However,
he recovered slowly. After his recovery, he radically changed his way
of life. He gave up smoking completely -- he used to be a passionate
smoker, lighting the next packet of papers from the previous flame. To
exercise his heart muscle, he started to take a brisk walk around the
houses of Tõravere every morning, with a fixed route and time. He
later told us how long it had taken him to complete the route that
morning.

Kuzmin married young, the children came young, and the children
married quickly. So Kuzmin enjoyed the last phase of his life with his
grandchildren. They often visited him. Once, one of the grandchildren
came down with the flu. Kuzmin also fell ill, but his illness took a
serious turn. He had used very strong drugs to treat the heart attack,
which damaged his kidneys, and now his kidneys were failing as a
complication of the flu. Kuzmin was rushed to hospital and on several
occasions his condition was very serious.

We often went to see him. We congratulated him on his 71st
birthday. He was quite upbeat, talked about his scientific problems
and hoped to be out of hospital soon. Indeed, his kidneys had started
to work and he was able to manage without an artificial kidney. One
evening, Zoja Kuzmina was told that Grigori would be home the next
day. A very deep cyclone passed over Estonia during the night and
caused heart problems for many people. Grigori had another heart
attack and the doctors could not save him this time. When Zoja arrived
at the hospital in the morning, Grigori was no more
(22.04.1988). Grigori Kuzmin was buried in the Raadi cemetery not far
from his senior colleague Aksel Kipper.

 
The man to whom we owe perhaps the greatest debt of gratitude for the
continuation and development of Öpik's scientific legacy had passed
away. All of our younger generation of astronomers are directly or
indirectly his students.

Kuzmin taught us how to approach a new problem. At the outset, 
simplifications must be made to obtain an approximate solution to the
problem. Often, an analytical solution can be obtained in this way, so
that the effects of the various factors can be clearly perceived. On
further examination of the problem, it is possible to separate out the
important from the less important factors, which makes it possible to
increase the accuracy of the result considerably without increasing
the computational effort unduly. This approach he apparently 
 learned from Öpik, who  was characterised by a clear
separation of important and unimportant factors.
Kuzmin once told me how he quickly learned to solve new problems. As a
student, he developed the following habit: when a professor started
solving a problem at the blackboard, Kuzmin would try to solve the
problem on his own without looking at the blackboard. In most cases,
the professor finished faster than he did, but not always, even though
the professor already knew the development path and he did not.

{\begin{figure*}[h] 
\centering 
\resizebox{0.80\textwidth}{!}{\includegraphics*{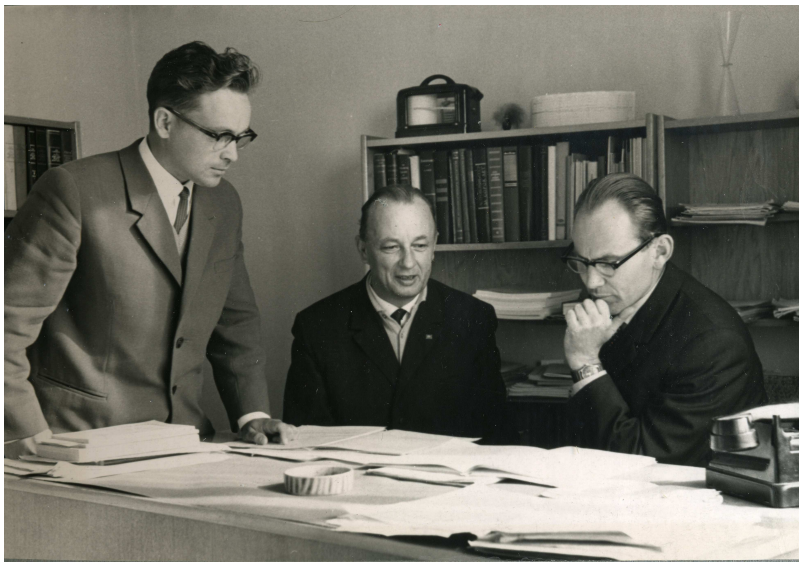}}
\caption{Grigori Kuzmin with his students Sergei Kutuzov and Jaan Einasto
} 
\end{figure*} 
}

Kuzmin  was the main authority in editing and proofreading our
scientific publications. Often his contribution was so serious that it
would be fairer to count him as a co-author. He was, however, very
modest and did not wish to emphasise his authorship. He had a special
knack of finding the right way to put the essence of the problem in a
very compact way. This makes his own work quite difficult to read, as
it is very thought-provoking and requires serious reflection on the
part of the reader.

Kuzmin usually found a solution to new problems
fairly quickly. This was, of course, the result of very intensive
thinking. Once the solution was found, interest waned and writing down
the result was a very laborious task. Then came the time when his
students had a role to play. They helped to organise the initial
typing of the work, because it was difficult for Kuzmin to continue
writing when the text had not been finalised. It was easier to find
imperfections in a cleanly rewritten text than in a rough
manuscript. In this ping-pong style, the final touches were made to
the work, both his own and those of his students.

The preparation of the Kuzmin doctoral thesis is described in Jaan
Einasto memory-book. Here is the story. 

{ In the early seventies, preparations began for the split the 
Institute of Physics and Astronomy into two institutes. In order for the new
institutes to retain their first stage (but staff salaries and the
funding of the institutes as a whole depended on this), both
institutes had to have a sufficient number of doctors of sciences. Such were the
rules of the Soviet bureaucracy, and it was not in our power to change
them. On the astronomers' side, the number of doctors  was quite modest at
that time. On one occasion, when Professor Kipper and I were
discussing our future, and the division of the Institute was under
discussion, Kipper approached me and asked me if I could not formalise
the results of my present work as a doctoral thesis. I was a little
surprised, because I thought it was only natural that Kuzmin should
defend his doctoral thesis first. So I asked for some time to think
about it.

I paced back and forth in my tiny office, wondering what to do. It was
clear that the current method of poking Kuzmin in the back was not
enough, the director had been doing it for years. On the other hand,
it was clear that of all our astronomers, Kuzmin was the most
deserving. I looked through Kuzmin's package of publications to date,
there were quite a number of them, and in any case there should be no
substantive obstacles. The trouble was that the package of papers was
not enough, according to the rules of Soviet bureaucracy the
corresponding typed  manuscript had to be submitted. After deliberating
for a few days, I had an idea: let's rewrite the texts of Kuzmin's
important works as chapters of the dissertation. I also put together
an initial pack of papers. Then I went to Kuzmin himself with the
story. At first Kuzmin didn't want to hear about it, but when we
really started to prepare for typing, Kuzmin picked up the work
package, looked it over, made a few additions and then said OK.

And then the work started on the conveyor belt. First we typed up the
relevant chapter, Kuzmin looked it over and took some time to
think. He was no longer satisfied with the text that had been written 
earlier, but had in the meantime come up with new ideas on how to
develop the problem further or present it better. And so Kuzmin sat up
all night writing a new text, or additions to the chapter. In the
morning, we tried to understand Kuzmin's sodic manuscript as much as
possible and had the addendum typed up in draft form. By lunchtime,
the maestro himself was on the scene, reviewing the manuscript and
making copious additions, sometimes sitting up all night again working
on the text. But time showed that this method was effective, and the
texts of the chapters, with the additions, were assembled into the
final manuscript quite quickly. So, in about half a year, a unique
manuscript was completed, which was so voluminous that we formatted
the dissertation in two volumes. Finding opponents was not difficult
either, Kuzmin was highly respected by all colleagues.

The defence went brilliantly.  As always, we sat at a banquet with our
opponents and guests after the defence. One of the opponents was
Grigory Idlis, who frankly admitted that he had once tried to compete
with Kuzmin and get new results independently of Kuzmin, but soon gave
up, seeing that a race with Kuzmin was a hopeless enterprise. } 

Kuzmin's papers were written in Russian and published in Tartu
Observatory Publications or in Russian-language conference
proceedings. Most Western astronomers do not speak Russian and do not
read observatory publications. In order to present his results, Kuzmin
prepared English-language abstracts of the most important papers and
sent them by post to astronomers in other countries. These received
attention and Kuzmin's authority among astronomers was high. However,
many of his results remained unknown to the English-speaking
readership. Kuzmin's new results were only in the Russian-language
doctoral thesis, which was not published. In order to make his thesis
accessible to other astronomers, it had to be translated into
English. So Peeter Tenjes started translating the chapters of Kuzmin's
thesis and preparing it for publication, including additions to the
doctoral thesis. Recently we had a small conference to celebrate the
200th anniversary of the Old Tartu University Observatory. One of the
main speakers was Tim de Zeeuw, who gave a review of Kuzmin's work. We
discussed the publication of English versions of Kuzmin's papers.

Most chapters of the thesis were translated in full. In some chapters
we used the English summaries written by Kuzmin. We add as appendix
English translations of three papers, written by Kuzmin after the
defence of the thesis. We did not translate review papers, appended to
this dissertation on the present state of stellar system dynamics and
on the development of work on the structure and dynamics of stellar
systems in the Tartu Observatory.  The translation from Russian into
English was made by Peeter Tenjes.  Jaan Einasto, Antti Tamm and Olga
Tihhonova helped to polish the text.  Peeter Einasto prepared scanned
text and figure files.

\vskip 5mm
\hfill Jaan Einasto 



\chapter{Preface}

The author of this thesis began his work on the dynamics of stellar
systems more than 20 years ago, in fact, already in the early 1940's. At
first the spatial and kinematical structure of the Galaxy in
galaxy-vertical direction and the rotation and radial mass
distribution of the Galaxy (and even earlier the M 31 Andromeda
galaxy) were investigated. Gradually these works developed into a more
extensive study of stellar dynamics, covering almost all sections of
it. Its goal was to try to some extent to complete the construction of
classical stellar dynamics, based on the assumption of stationarity or
quasi-stationarity of stellar systems, and thus to give the fullest
possible explanation and description of the observed features of the
spatial and kinematical  structure of regular star systems, and at the
same time to create theoretical background for the study of their
dynamical evolution.

To date, the author has published more than 20 papers on the dynamics
of stellar systems, which served as the basis for this dissertation.
The articles are mostly reproduced in this dissertation without
changes or corrections; only some of them are abridged. However, there
are a number of notes, additions, and comments, sometimes quite
substantial, which give the thesis its own significance. 

In terms of content, the material is divided into three parts. The
first part includes three closely intertwined cycles of  works  on the
determination of the matter density in the vicinity of the Sun, on
the study of the mass distribution in the Galaxy, and on the theory of
the third integral of stellar motion. The second part of the thesis 
includes papers on models of the spatial and kinematical structure
of stellar systems and on some general questions of dynamics of
stellar systems. The third part of the thesis is mainly
devoted to the study of the influence of irregular gravitational forces.

Two review papers are appended to this dissertation -- on the present
state of stellar system dynamics and on the development of work on the
structure and dynamics of stellar systems in the Tartu Observatory. A
bibliography of the author's work is also included.

Each paper, together with supplements, composes one chapter of the 
dissertation. 
The first part of the thesis forms the first volume, the second and
third parts together with the appendices are combined into the second
volume.

\vglue 3mm
\hfill 1969

%% file: chapter01.tex
\part{Gravitational potential and mass distribution of the Galaxy. The
  third integral of stellar motions} 

\chapter[Dynamical density of the Galaxy]{Dynamical density of the
  Galaxy\footnote{\footnotetext ~~Published in Tartu
    Astron. Observatory Publications, vol. 32, pp. 5-43, 1952 with the
    title ``Galactic-equatorial A and K star proper motions
    perpendicular to the galactic plane and Galactic dynamic
    density''} }

\section{Introduction}

When studying the structure and dynamics of the Galaxy, the knowledge
of its dynamical density, \ie the spatial density of gravitating
masses, is very important. Determining the dynamical density of the
Galaxy was the main goal of this work. However, the study of the
kinematics of stars of spectral types A and K, undertaken for this
purpose, was, understandably, of independent interest.

The author's idea for the present work arose from his previous
attempts, in an unpublished paper, to find out the distribution of the
dynamical density of the Galaxy in the vicinity of the Sun (``Notes on
the dynamics of cosmic dust matter'', Manuscript, 1941).

The dynamical density of the Galaxy can be determined by comparing the
spatial distribution of stars with the distribution of their
velocities in the direction perpendicular to the galactic plane. This
method, applied by \citet{Oort:1932}, consists, in a somewhat more
general way, in the following.

Let $R$  is the distance from the Galaxy axis, $\Theta$ is
Galacto-centric longitude and $z$ is the distance from the Galactic
plane (negative to the south of this plane). Let, further,
$v_R, v_\theta$ and $v_z$ be the corresponding velocity components.
Suppose the Galaxy is stationary and has an axial symmetry. Then the
continuity equation for
the Galaxy in the six-dimensional phase space will have  the form \citep{Parenago:1946}:
\be
v_R\frac{\partial\,\Psi}{\partial\,R}+v_z\frac{\partial\,\Psi}{\partial\,z}+
\left(\frac{\partial\,\Phi}{\partial\,R}
  +\frac{v_\theta^2}{R}\right)\,\frac{\partial\,\Psi}{\partial\,v_R} -
\frac{v_Rv_\theta}{R}\frac{\partial\,\Psi}{\partial\,v_\theta}
+\frac{\partial\,\Phi}{\partial\,z}\,\frac{\partial\,\Psi}{\partial\,v_z}
= 0,  
\label{eq1.1}
\ee
where $\Phi$ is the acceleration potential and $\Psi$ is the number 
density of stars in phase space. Suppose that $\Psi$ is an even
function of $v_R$. By integrating the left part of the equation
(\ref{eq1.1}) over all $v_R$ and $v_\theta$ , we obtain
\be
v_z\,\frac{\partial\,f}{\partial\,z}+\frac{\partial\,\Phi}{\partial\,z}\,\frac{\partial\,f}{\partial\,v_z}
=0, 
\label{eq1.2}
\ee
where $f$ is the number of stars in unite volume and unit interval of
$v_z$. General solution of the equation (\ref{eq1.2}) for given $R$
and $\Theta$ is
\be
f = F[v_z^2 - 2(\Phi - \Phi_0)],    
\label{eq1.3}
\ee
where $F$ is an arbitrary function and $\Phi_0$ is the potential in the galactic plane.

Knowing the function $F$ from observations, one can find the spatial
stellar density $D$ as a function of $\Phi - \Phi_0$ 
\be
D=\int_{-\infty}^{\infty} F[v_z^2 - 2(\Phi - \Phi_0)]\dd{v_z},
\label{eq1.4}
\ee
and hence $\Phi - \Phi_0$  as a function of $D$.

If the spatial stellar density is also known
from observations, we are able to calculate $\Phi - \Phi_0$ as a
function of $z$. Dynamical density $\rho$ we get then from the Poisson's
equation (in cylindrical coordinates) 
\be
4\pi\,G\rho=-\frac{\partial^2\Phi}{\partial\,z^2}-\frac{\partial^2\Phi}{\partial\,R^2}
- \frac{1}{R}\frac{\partial\,\Phi}{\partial\,R},
\label{eq1.5}
\ee
where $G$ is the gravitational constant. The first term of the right
side of this equation can be found knowing $\Phi - \Phi_0$; the
remaining terms can be taken from the Galactic rotation data. 

To determine the dynamical density of the Galaxy by this method, it is
thus necessary to know from observations 
$F$ and $D$ functions.  

The spatial stellar density, as a function of $z$, was studied by
several authors, both for all stars and for stars of particular
spectral types. One can, for example, use, as it was done in the present
paper and in the above unpublished paper of the author, the
\citet{Pannekoek:1929} data on the spatial densities A and gK of
stars, derived from the Draper spectral catalogue.

As for the function $F$, we would know it if we knew the $v_z$
distribution near the galactic plane, \ie the function $f(v_z)$ at
$z = 0$, since according to formula (\ref{eq1.3}) in this case
$f = F(v_z^2)$. There are  no direct data on this distribution, but it
is known that the total distribution of $v_z$ for a wider interval of
$z$ values is for many types of stars an approximate normal (Gaussian)
distribution. According to formula (\ref{eq1.3}), this must be normal
if the distribution $v_z$ at $z = 0$ is also normal. In this case, the
distribution $v_z$ is normal at any $z$, since by formula
(\ref{eq1.3}), we then have
\be
f = f_0\,e^{- {v_z^2 -2(\Phi - \Phi_0)  \over  2\sigma_z^2}},
\label{eq1.6}
\ee
where $\sigma_z$ is dispersion of $v_z$, which in our case do not
depend on $z$, and $f_0$ is the value of $f$ at $z=0$ and $v_z=0$. 

From formulae (\ref{eq1.4}) and (\ref{eq1.6}) it follows then
\be
D=D_0\,e^{\frac{\Phi - \Phi_0}{\sigma_z^2}}
\label{eq1.7}
\ee
where $D_0$ is the value of $D$ at $z=0$, and
\be
-\frac{\partial^2\Phi}{\partial\,z^2}=-\sigma_z^2\frac{\partial^2\ln\,D}{\partial\,z^2}. 
\label{eq1.8}
\ee

Starting in the above unpublished paper from \citet{Pannekoek:1929}
data on the $D(z)$ function for A and gK stars and applying formulae
(\ref{eq1.5}) and (\ref{eq1.8}) to both types of stars, the author
obtained, however, although consistent with each other, but unlikely
dependences of the dynamical density $\rho$ on $z$, namely, extremely
fast decreases of $\rho$ with increasing $z$.  Thus a question arose,
is the distribution of $v_z$ normal with sufficient accuracy for
particular types of stars, and whether the formula (\ref{eq1.8}) was
applicable. So the idea emerged to derive the $F$ function directly
from data on the motion of stars near galactic plane, where
$f = F(v_z^2)$.

To solve this problem one can use the data on proper motions of stars
near the galactic equator. Assuming to continue using the Pannekoek
data on the $D(z)$ function for A and gK stars, we have decided to
statistically study the proper motions of galactic-equatorial stars of
these types, which are also the most numerous. The material was the
proper motions along the Galactic latitude calculated from the data of
the General Catalogue by \citet{Boss:1937}, taking into account A
and gK stars up to apparent magnitude 7.0 and within Galactic
latitudes $\pm 3^\circ$.

However, the material chosen in this way, did not allow one to derive
with sufficient certainty the $F$ functions for both types of
stars. The reason for this was the small number of stars used and,
most importantly, the need for large corrections for the observational
errors of proper motions and for the scattering of absolute magnitudes
of stars. Therefore we had to give up the original problem and
restrict ourselves to the narrower task of determining the variance of
$\sigma_z$ for both types of stars. In doing so it became clear that
so far  one need not to specify 
the $F$ functions.
It appears that the $D(z)$ functions for A and gK stars,  obtained by
Pannekoek,  are strongly distorted as 
he did not take
into account the scattering of the absolute magnitudes of these
stars. By using the corrected $D(z)$, one can obtain reasonable
results for the dynamical density without giving up on the assumption
of a normal distribution of $v_z$ for these stars.

The goals of the present paper are therefore reduced to the following:
(1) to deduce from proper motions of galactic-equatorial A and gK
stars the dispersions $\sigma_z$, and (2) applying the assumption of
the normal distribution $v_z$, find on the basis of the obtained
$\sigma_z$ the dynamical density of the Galaxy in the vicinity of the
Sun.

In addition to the dispersions $\sigma_z$, and the dynamical  density $\rho$,
the present paper also obtained, as a side result, data on the
apparent systematic motion of A and gK stars, caused by errors of
precession constants. In addition, on the basis of the result obtained
for the dynamical density, an attempt is made to come to some
conclusions about the structure of the Galaxy as a whole.





\section{Statistical processing and results}

We omit the description of statistical data used and technical details
of the data analysis. 

The processing of statistical data yields for the velocity dispersion of A and gK stars  the
following:
\be
\ba{ll}
{\rm A~ stars:} &\sigma_z=\pm(5.1\pm 0.7)~ {\rm km/s}\\
{\rm gK~ stars} &\sigma_z=\pm(12.2\pm 1.7)~ {\rm km/s} .
\label{eq1.42}
\ea
\ee

The result (\ref{eq1.42}) is the final result of processing our
statistical material. The relatively small values obtained for
$\sigma_z$ were unexpected for the author, since previous studies gave
significantly large values for $\sigma_z$.  However,  recent
studies of  spatial velocities of stars by \citet{Parenago:1950a}
have also obtained, in the case of A stars, rather small values for
$\sigma_z$, although somewhat larger than our values.

\section{Spatial distribution of A and gK stars perpendicular to the
  galactic plane} 

As indicated in the introduction, the final aim of the present work
was to determine the dynamical density of the Galaxy.  For this
purpose, in addition to the dispersions $\sigma_z$, whose values we
derived for A and gK stars from their proper motions, it was necessary
to have data on the spatial distribution of these stars perpendicular
to the galactic plane. We have used here, as already said in the
introduction, the data of \citet{Pannekoek:1929} on the spatial
distribution of A and gK stars on the basis of the Draper
catalogue. It had to be taken into account that these data are
distorted by not taking into account the interstellar absorption and
especially the scattering of the absolute magnitudes of the stars.

To correct the Pannekoek data for interstellar absorption and for the
scattering of absolute magnitudes, it was necessary to solve the
corresponding integral equation. Let $z_\star$ and $r_\star^\prime$ be
the values of $z$ and the projected onto the galactic plane to the
galactic plane of the distance $r^\prime$, found under the assumption
that the absolute magnitude $M$ of stars is equal to some constant
value of $M_\star$, and that there is no absorption. Let $D_\star$ be
the fundamental stellar density $D$, found under the same
assumptions. Then, assuming for simplicity that the ratio
$D(z,r^\prime,l)/D(0,r^\prime,l)$, or briefly $D(z)/D_0$, depends only
on $z$, then, from the stellar statistics integral equation, we find
that the integral equation to determine $D(z)/D_0$, with respect to
$D_\star$, has the form
\be
D_\star(z_\star,r_\star^\prime,l) =
\int_{-\infty}^\infty\frac{D(z)}{D_0}\phi(M^\prime)\dd{M^\prime}, 
\label{eq1.43}
\ee
where
\be
\psi(M^\prime)=\phi[M^\prime -K(z,r^\prime,l)]\left(\frac{r^\prime}{r_\star^\prime}\right)^3\,D_0(r^\prime,l)
\label{eq1.44}
\ee
and (neglecting the distance of the Sun from the galactic  plane)
\be
\frac{z}{z_\star}=\frac{r^\prime}{r_\star^\prime}=10^{-0.2(M^\prime-M_\star)}.
\label{eq1.45}
\ee

In these formulae $\phi$ is the luminosity function,
$M^\prime =m +5 +5\log\pi$ is the absolute magnitude, distorted by
interstellar absorption, and $K$ is the absorption in stellar
magnitudes, and $l$ is the galactic longitude. The function $\psi$, appearing as the kernel in equation
(\ref{eq1.43}), is proportional in the galactic plane to the
distribution function $M^\prime$ at a given $m$ (at a given
$r_\star^\prime$). If multiplied by $D(z)/D_0$, it is proportional to
the distribution function $M^\prime$ at a given $m$ and outside the
galactic plane.

Instead of $D_\star$ and $\psi$ in equation (\ref{eq1.43}) one can
take the averaged $D_\star$ and $\psi$ at a given $z_\star$,
averaging, for instance, at each $z_\star$ over all $l$ and in the
same interval $r_\star^\prime$. If there were no absorption, the
$\psi$ averaged in this way would not depend on
$z_\star$. In reality such a dependence should exist. If, however, we
use only values of order of the equivalent half-thickness of the
absorbing layer, the dependence is weak due to the fact that at any
$r_\star^\prime$ of the order of the equivalent half-thickness of the
absorbing layer, the absorption $K$ is weakly dependent on $z$ and is
furthermore negligible. In order to be able to use the same function
with all $z_\star$, we have confined ourselves to therefore for
averaging $D_\star$ with values of $r_\star^\prime$ smaller than
approximate 160 pc. This upper limit of $r_\star^\prime$ corresponds
to an average value of $r_\star^\prime$ is about 100 pc. Since
Pannekoek had assumed for $M_\star$ a value of $+0.9$ for A stars and
$+0.7$ for gK stars, $r_\star^\prime = 100$ pc, in the galactic plane
corresponds to $m$ values of 5.9 and 5.7. Therefore the averaged
$\psi$ could be assumed to be proportional to the distribution
function $M^\prime$ in the galactic plane at $m=m_0=5.5$, \ie the
$M^\prime$ distribution function we used for the derivation of
characteristic distribution of parallaxes at $m=m_0$.  We only had to
exclude stars of subclass A5, absent in Pannekoek data.

Finding a solution of the integral equation (\ref{eq1.43}) we settled
on the normal $z$-density distribution from theoretical
considerations: 
\be
D=D_0\,e^{-z^2/2\zeta^2},
\label{eq1.46}
\ee
where $\zeta$ is the variance of $z$.  Although $D_\star$ decreases
with $z_\star$ according to a law, very different from Pannekoek's
one, the expression (\ref{eq1.46}) satisfies equation (\ref{eq1.43})
quite well. This can be seen from Table \ref{tab1.7}, which gives the observed
and calculated $D_\star$ values averaged over $l$, $r_\star^\prime$ and
$z_\star$ ($r_\star^\prime$ is less than 160 pc). The unit of $D_\star$
is 1 star per $10^6$ pc$^3$.

\begin{table}
\caption{}
\smallskip
\label{tab1.7}
\centering
\begin{tabular}{|c | c | c | c | c | } 
\hline\hline
  $|z_\star|$    &   \multicolumn{4}{|c|} {$D_\star$} \\
  &\multicolumn{2}{|c|}{ A stars} & \multicolumn{2}{|c|}{ gK stars}\\
\cline{2-5}
 & obs.& calc. & obs. & calc.\\
\hline
  0 -- 30 & 365 & 348 & 215 & 207\\
  30 -- 90& 208&224&162&171\\
  90 -- 150&90&94&128&127\\
  150 -- 210& 40& 36& 95& 94\\
\hline\hline
\end{tabular}
\end{table}

In order to obtain the best agreement with the observed values of
$D_\star$, it would be best, to multiply the calculated $D_\star$
values by a suitable conversion factor, and to chose a suitable value
of $\zeta$. The calculated $D_\star$ values given in the table
correspond to $\zeta= \pm\,96$ pc for A stars and $\zeta=\pm\,205$ pc for
gK stars.

The accuracy of the found values of $\zeta$ depends mainly on the
accuracy of the function $\psi$. The relative error of $\zeta$ is
about the same as the relative error of the parameter $p$, calculated
using $\psi$ and proportional to the mean parallax at $m=m_0$.
Our final result for the vertical spatial density variance is:
\be
\ba{ll}
{\rm A~ stars:} &\zeta=\pm(99\pm 5)~ \mathrm{pc}\\
{\rm gK~ stars} &\zeta=\pm(202\pm 20)~ \mathrm{pc}.
\label{eq1.47}
\ea
\ee
The errors of these results are taken in accordance with the errors of
input parameters 
with some increase due to the inaccuracy of our
determination of $\zeta$ from the Pannekoek data.

If we substitute the density law (\ref{eq1.46}) into formula
(\ref{eq1.8}), we find that $\partial^2\Phi/\partial\,z^2$ is
independent of $z$. In fact $\partial^2\Phi/\partial\,z^2$ must of
course depend on $z$, but since we  are dealing with very small
values of $|z|$ in comparison to the Galactic dimensions, this
dependence is probably not very noticeable. Denoting
$-\partial^2\Phi/\partial\,z^2$ near galactic plane in vicinity of the
Sun by $C^2$, based on formulas (\ref{eq1.8}) and (\ref{eq1.46}) we
have \be C={\sigma_z \over \zeta},
\label{eq1.48}
\ee
Since we have found parameters $\sigma_z$ and $\zeta$, we can compute
the parameter $C$. Using  ({\ref{eq1.42}) and (\ref{eq1.47}) we
  obtain: 
  \be
  \ba{ll}
{\rm A ~stars:} &C=52\pm 7~~ {\rm km/s/kpc},\\
{\rm gK~ stars:} &C=60\pm8~~ {\rm km/s/kpc}.
\label{eq1.49}
\ea
\ee
In error calculation it has been taken into account that errors of $\sigma_z$
and $\zeta$ are only partially independent, since the inaccuracy of $\psi$ causes
either an increase or a decrease of $\sigma_z$ and $\zeta$  at the
same time. The results of (\ref{eq1.49}) are consistent within their
mean errors and on average we can accept: 
\be
C=56\pm 5~ {\rm km/s/kpc}.
\label{eq1.50}
\ee

The constant $C$ relates the characteristic of the spatial
distribution of stars near the galactic plane $\zeta$ with their
kinematical  characteristic  $\sigma_z$. It must have the same
value for stationary stellar subsystems. Knowing this constant and
characteristic $\zeta$ we can for these subsystems calculate
$\sigma_z$ and vice versa. Thus, using the spatial distribution of
long-periodic cepheids at small $|z|$ and $r^\prime< 660$ pc we found,
using the data of \citet{Kukarkin:1949}, $\zeta= 50$ pc, from which
the value follows $\sigma_z=2.8$ km/s. The direct determination of
$\sigma_z$ for these stars was given by \citet{Parenago:1948b}
$\sigma_z=\pm 5.4$ km/s. However, having checked Parenago's
calculations we found an error and obtained
$\sigma_z = \pm(2.9 \pm 1.0)$ km/s, in agreement with the
theoretical result. The stars used by Parengo were taken with weights
corresponding to the mean errors of the conditional
equation for $v_z^2$, while the equilibrium solution gave
$\sigma_z = \pm(3 \pm 3.5)$ km/s.

\section{Dynamical density of the Galaxy. Conclusions on the structure
  of the Galaxy} 

The value of the constant $C$ found in the previous paragraph allows
us to solve the main problem of this paper -- to calculate the
dynamical density $\rho$ of the Galaxy in the vicinity of the
Sun. Since $C^2 = -\partial^2\Phi/\partial\,z^2$ in the Sun's
neighbourhood, the first and the main term in the dynamical  density
formula (\ref{eq1.5}) is thus known. The remaining terms, which are
less significant, we can express through the galactic rotation
constants $A$ and $B$ for the planar subsystems. Since the planar
subsystems rotate practically with a circular velocity, in this case
we can express the Galactic rotation constant
$A=\frac{1}{2}\left(\frac{v_z}{R} -\frac{\dd{v_z}}{\dd{R}}\right)$ and
  $B=-\frac{1}{2}\left(\frac{v_z}{R}+\frac{\dd{v_c}}{\dd{R}}\right)$.
    Given that $v_c^2=-R\,\partial\Psi/\partial\,R$, we find for the sum
    of the second and third terms of formula (\ref{eq1.5}) the
    expression:
    \be
    \frac{\partial^2\Psi}{\partial z^2}
      +\frac{1}{R}\frac{\partial\Psi}{\partial
        R}=-2\frac{\dd{v_c}}{\dd{R}}\frac{v_c}{R}=2(A^2-B^2).
\label{eq1.51}
\ee
The equation for the dynamical density takes the form:
\be
4\pi\,G\rho=C^2-2(A^2-B^2).
\label{eq1.52}
\ee
Taking usually accepted for planar subsystems, $A=+20$ km/s/kpc and
$B=-13$ km/s/kpc, and $C$ according to result (\ref{eq1.50}), we
obtain for dynamical  density 
\be
\rho=(0.34 \pm 0.08)\times 10^{-23} {\rm g/cm^3}
\label{eq1.53}
\ee
or  $0.05\pm 0.01$ solar masses per cubic parsec. The term
$2(A^2-B^2)$ has rather small influence on the result, while the error
of the latter is due mainly to error of $C$. 

The result (\ref{eq1.53}) diverges markedly from the \citet{Oort:1932} result,
obtained with  the same method $\rho = 0.6\times10^{-23}$  g/cm$^3$ or
0.09 solar masses per cubic parsec. This discrepancy is quite 
understandable since the data used by Oort  
for A and gK stars are totally at odds with ours. Note that the result
(\ref{eq1.53}) only slightly exceeds the \citet{Parenago:1945ac}
estimate $\rho = 0.27 \times 10^{-23}$ g/cm$^3$ , found by him from  the
spatial density of stars and their mean mass. 

From the value obtained for the dynamical density, or from the value
of the associated constant $C$, we can obtain some interesting data on
the structure of the Galaxy.  Let us assume, as a rough approximation,
that surfaces of equal density in the Galaxy are rotational ellipsoids
whose ratio of semi-axes is equal to $\epsilon$. In this case
the square of circular velocity, $v_c^2 = -R\,\partial\Psi/\partial R$
at $z=0$ will be calculated by the formula:
\be
v_c^2 = \frac{4\pi G\epsilon}{R}\int_0^R{\rho(a)a^2\dd{a}    \over\sqrt{1-\left(\frac{ae}{R}\right)^2}},
\label{eq1.54}
\ee
where $a$ is the semi-major axis of an ellipsoid of equal density and
$e^2= 1 - \epsilon^2$ is the square of the eccentricity of its
meridian. Judging from the Galactic rotation data by
\citet{Parenago:1948b}, $v_c$ is approximately constant within quite
a wide range $R$. According to (\ref{eq1.54}) this would be the case
if $\rho\,a^2$ is also approximately constant over a wide range $a$.
If $\rho\,a^2 = \mathrm{const}$ we obtain from (\ref{eq1.54}):
\be
v_c^2 = 4\pi\,G\epsilon\frac{\arcsin e}{e}\rho\,R^2.
\label{eq1.55}
\ee
Since $v_c/R=A-B$ and $4\pi\,G\rho=C^2$ (at $v_c= \mathrm{const}$), we get:
\be
\epsilon\frac{\arcsin e}{e}=\left(\frac{A-B}{C}\right)^2.
\label{eq1.56}
\ee
This formula allows us to calculate the ratio of the Galactic 
semi-axes. Using obtained values for $A$, $B$, and $C$, we find that 
\be
\epsilon=0.26 \pm 0.06,
\label{eq1.57}  
\ee
where, in calculating the mean error of the result, we considered the
error of $A-B$   to be of the order of  $\pm\,5$ per cent. The ``dynamical''
ratio of the Galactic axes is therefore 
1/4.   This result indicates that the gravitating  matter
concentration towards Galactic plane is rather moderate and that with
respect to mass distribution the Galaxy is far from being as flattened
as ``optical''.  

The value  $(A - B)^2/C^2$ in formula (\ref{eq1.56}) is related to the
inclination of the acceleration vector to the galactic plane for small
$|z|$.  Indeed, if $|z|$ is small, then 
$-\partial\Psi/\partial\,z=C^2z$ 
 and    $-\partial\Psi/\partial\,R=v_c^2/R=(A-B)^2R$, and we get for
 the tangent of the slope of the acceleration vector:
\be
\frac{\partial\Psi}{\partial\,z}\bigg/\frac{\partial\Psi}{\partial\,R}=
\left(\frac{C}{A-B}\right)^2\frac{z}{R}. 
\label{eq1.58}
\ee
Since  $[(A-B)/C]^2=0.35\pm 0.07$,
it follows that at small $|z|$ the tangent of the inclination vector
of acceleration is approximately 3 times larger than  
the tangent of its inclination, $z/R$, in the case of the spherical
symmetry of the mass distribution. Note that the ratio  $(A-B)/C  =
0.59 \pm 0.06$ is equal with the ratio of the period of a star's
oscillation relative to the galactic plane to the period of 
of the star's orbit around the galactic centre.

Due to the moderate concentration of gravitating  matter
to the Galactic plane the equivalent half-thickness of the Galaxy,
$z_e$, is rather large. In the case of our rough model of the Galaxy
it is calculated by the formula: 
\be
z_e=\epsilon\int_R^\infty\frac{\rho(a)}{\rho(R)}{ \dd{a}\over\sqrt{1-(R/a)^2} }.
\label{eq1.59}
\ee
Assuming that $\rho\,a^2$ is  constant up to  $a=R_0$, where $R_0$ is
the radius of the Galaxy, we find: 
\be
z_e = \epsilon\left(\frac{\pi}{2}-\arcsin\,\frac{R}{R_0}\right)\,R.
\label{eq1.60}
\ee
Since in the vicinity of the Sun the ratio $R_0/R$ is approximately
equal to two, we get
$z_e \approx \frac{\pi}{3}\epsilon\,R=(0.27\pm 0.06)R$.  Thus in the
neighbourhood of the Sun the equivalent half-thickness of the Galaxy
is only about 4 times smaller than the distance to its centre. If we
take the value 7.25 kpc for the latter according to
\citet{Kukarkin:1949}, then we will have for the equivalent
half-thickness of the Galaxy 2.0 kpc. Taking into account the value
obtained for the dynamical density in the neighbourhood of the Sun, we
can calculate the density projected on the galactic plane,
$2\,z_e\rho$. It turns out to be 200 solar masses per square parsec.

In conclusion, we note that our result about the moderate galactic
concentration of gravitational matter agrees with the result by
\citet{Parenago:1949} that a large fraction of all stars in the Galaxy
belong to the subdwarfs forming a ``spherical'' subsystem, and
confirms to a certain extent the idea expressed back in 1943 by
\citet{Eigenson:1943aa} about the dynamical ``sphericity'' of the
Galaxy.
\vglue 3mm
\hfill 1952
 
\section{Appendix added in 1969}

Values of parameters $C$ and $\rho$ obtained in this paper are
underestimated, the density parameter $\rho$ is underestimated by a
factor of two, see Chapter 5 of this Thesis. In addition, according to
new data, parameters $A$ and $B$ are also underestimated. For this
reason the flatness parameter $\epsilon$, and the ratio $z_e/R_\odot$
are overestimated, by a factor about three. The true degree of
``sphericity'' is considerably smaller.

%% file: chapter02.tex
\chapter[On the mass distribution of the Galaxy]{On the mass
  distribution of the Galaxy\footnote{\footnotetext ~~Tartu
    Astron. Observatory Publications, vol. 32, pp. 211-230, 1952.} }  

Knowing the rotation law of a stellar system it is possible to obtain
some data on the mass distribution of that system. Concretely, from
radial acceleration it is possible to calculate ``the surface
density'' of the system, \ie  the mass density projected onto the
equatorial plane of the system \citep[see e.g. ][]{Parenago:1948a}. 

This problem was solved for the Andromeda (M31) and the Triangulum
(M33) nebulae by \citet{Wyse:1942}.  

In their paper Wyse and Mayall started from the ``flat disk'' model,
\ie they assumed that a stellar system has the axial symmetry and is
so flattened, that the radial acceleration in the symmetry plane of
that system is such as if all the system's mass is concentrated in
that plane. As an approximation to the surface density distribution
they proposed  a fifth order polynomial with respect to the distance
from system's centre. 

In present paper we solve the similar problem for the Galaxy. We do
not assume ahead any kind of analytical expression for the surface
density, but find it via numerical integration by using suitable
integral equation. In addition we try to take into account the
``thickness'' of the Galaxy.  
\bigskip

\section{~}
Let us suppose that the isodensity surfaces of Galactic matter are
ellipsoids of revolution with constant axial ratio $\epsilon$. In this
case we may use known formulae for the attraction of infinitely thin
ellipsoidal layer \citep[see e.g.][p.763]{Zhukowski:1950}. Integrating
over the elementary ellipsoidal layers we derive for the radial
acceleration $F_R$ at a distance $R$ from the Galaxy axis 
\begin{equation}
- \frac{F_R}{R}  =  4\pi G \epsilon \int_0^1 \frac{\rho (a) u^2 \dd{u} }{\sqrt{1-e^2u^2}} , 
\label{eq2.1}
\end{equation}
where $G$ is the gravitational constant, $\rho (a)$ -- spatial mass density, 
$$e=\sqrt{1-\epsilon^2} $$
is the eccentricity of the meridional section of an isodensity ellipsoid and
$$u=\frac{a}{a'}.$$
$a$ is the major semiaxis of an isodensity ellipsoid and $a'$ is the
major semiaxis of the ellipsoid confocal to it through the point where
$F_R$ is calculated. Equation (\ref{eq2.1}) enables us to obtain the
expression for the radial acceleration in the equatorial plane of the
Galaxy. In that plane  
$$a'=R$$
and therefore,
\begin{equation}
-F_R R = 4\pi G\epsilon \int_0^R \frac{\rho (a) a^2  \dd{a} }{ \sqrt{R^2-a^2e^2} }. 
\label{eq2.2}
\end{equation}
The left side of the expression is the square of circular velocity.

Equation (\ref{eq2.2}) is an integral equation for $\rho$ when $F_R$ in the galactic plane is known. Having determined $\rho$ we may find the Galactic surface density 
$$\sigma = \int_{-\infty}^{\infty} \rho \,\rmd z ,$$
where $z$ is the distance from the galactic plane. As
$$\rmd z = \frac{\epsilon \,a \rmd a}{\sqrt{a^2-R^2}}, $$
we get
\begin{equation}
\sigma (R) = 2\epsilon \int_R^{\infty}  \frac{\rho (a) \,a \rmd a}{\sqrt{a^2-R^2} }.\label{eq2.3}
\end{equation}

Let us have
$$ \epsilon\rightarrow 0 $$
corresponding to the flat disk model. In this case Eq.~(\ref{eq2.2}) turns to
\begin{equation}
-F_R R = 4\pi G\epsilon \int_0^R \frac{\rho (a) \,a^2 \rmd a}{\sqrt{R^2-a^2} }.
\label{eq2.4}
\end{equation}
If instead of $a$ we take new variable $a^{-1}$ the equation will take
a form similar to the known integral equation for the determination of
spatial densities of globular clusters (Eq.~(\ref{eq2.3}) with
$\epsilon =1$). Using the solution of this equation \citep[see
e.g.][p.224--225]{Parenago:1946} we obtain 
\begin{equation}
4\pi G\epsilon\rho (a)a^2  =  \frac{2}{\pi} \int_0^a \frac{\rmd
  (-F_RR^2)}{\rmd R}  \frac{R \rmd R}{a\sqrt{a^2-R^2}}. \label{eq2.5} 
\end{equation}

Result (Eq.~\ref{eq2.5}) together with Eq.~(\ref{eq2.3}) is exactly
the solution of the integral equation derived by Wyse and Mayall for
the determination of $\sigma$ from $F_R$ for the flat  disk
model. Here $\epsilon\rho$ plays a role of an auxiliary  function,
which may be eliminated when needed. For that we substitute
$\epsilon\rho$ according to Eq.~(\ref{eq2.5}) into Eq.~(\ref{eq2.3}),
where instead of $R$ we designate $R_1$. Changing the order of
integration and integrating over $a$ we find the following expression
for $\sigma (R_1)$ 
\begin{equation}
\sigma (R_1) =  \frac{1}{\pi^2 GR_1^2}  \int_0^{\infty}  \frac{\rmd (-F_RR^2)}{\rmd R} \psi\left( \frac{R}{R_1} \right) \rmd R. \label{eq2.6}
\end{equation}
Here
\begin{equation}
\psi \left( \frac{R}{R_1} \right) =  \left\{
\ba{ll}
\frac{R_1}{R} \left[ {\bf K}\left( \frac{R}{R_1} \right) - {\bf E}\left( \frac{R}{R_1}\right)\right], & \mathrm{for} ~~R<R_1, \\
\noalign{\smallskip}
~~~~~~{\bf K}\left( \frac{R_1}{R}\right) - {\bf E}\left( \frac{R_1}{R}\right), & \mathrm{for }~~R>R_1, \\
\ea
\right.
\label{eq2.7}
\end{equation}
where $\mathbf{ K}$ and $\mathbf{ E}$ are the complete elliptic
integrals of first and second order, respectively. Considering the
amount of calculations for the determination of $\sigma$
Eq.~(\ref{eq2.6}) does not have advantages when compared to
Eqs.~(\ref{eq2.3}) and (\ref{eq2.5}). Having in mind the need to
extrapolate $F_R$ it seems even more convenient to use
Eqs.~(\ref{eq2.3}) and (\ref{eq2.5}) replacing the extrapolation of
$F_R$ with the extrapolation of $\epsilon\rho$. 

Now we must take into account the ``thickness'' of the Galaxy, \ie
that $\epsilon$ is nonzero. Comparing Eq.~(\ref{eq2.4}) with
Eq.~(\ref{eq2.2}) we see that it is possible to use Eq.~(\ref{eq2.4})
and hence its solution in form of Eq.~(\ref{eq2.5}), also in that case
after adding a correction to $-F_RR$ 
$$\Delta (-F_RR) = 4\pi G\epsilon \int_0^R \left[ (R^2-a^2)^{-1/2} - (R^2-a^2e^2)^{-1/2} \right] \rho (a) a^2 \rmd a.$$
In case of sufficiently small $\epsilon$ the subtraction in square
brackets is a very small quantity with an exception of $a$ near to
$R$. Thus we may take $\rho (a) a^2$ out of the integral in form of
$\rho (R)R^2$. The remaining integral equals within a sufficient
precision to $\epsilon$. Therefore, for small $\epsilon$ 
$$\Delta (-F_RR) = 4\pi G\epsilon^2\rho (R)R^2.$$
Because the correction we are dealing with is quite small and it is
sufficient to know only its approximate value, for $\rho$ we may use
Eq.~(\ref{eq2.5}) valid for $\epsilon\rightarrow 0$. Hence we have the
final expression (replacing in Eq.~(\ref{eq2.5}) $a$ with $R'$) 
\begin{equation}
\Delta \left( -F_RR \right)_{R=R'} =  \frac{2}{\pi} \epsilon
\int_0^{R'} \frac{\rmd (-F_R R^2)}{\rmd R} \frac{R \rmd
  R}{R'\sqrt{R'^2-R^2}}. \label{eq2.8} 
\end{equation}

In this way, when making the ``ellipsoidal model'' of the Galaxy in
calculations of the surface density $\sigma$ from the radial
acceleration $F_R$ it is needed to correct $-F_R R$ by the quantity
(\ref{eq2.8}). In this case $\sigma$ is calculated from
Eqs.~(\ref{eq2.5}) and (\ref{eq2.3}) (or from Eq.~\ref{eq2.6}).  

We would like to mention that the assumption of the ellipsoidal
character of the Galaxy restricts the generality of the proposed
method of calculation of $\sigma$ only in case when $\epsilon$ is
assumed to be constant in Eq.~(\ref{eq2.8}), because in general
$\epsilon$ must be assumed to be some function of $R$,  characterising
the dependence of Galactic flatness from distance to the
centre. \footnote{We may start, for example, with an assumption that
  the mass distribution in the Galaxy is a sum of ellipsoidal
  distributions. In this case instead of $\rho$ and $\epsilon$ in
  Eq.~(\ref{eq2.2}) we must use $\rho_i$ and $\epsilon_i$ of the
  individual ellipsoidal distributions and to sum over $i$. If we
  assume $\epsilon_i^2$ to be small quantities, then for $\Delta
  (-F_RR)$ we derive a formula similar to Eq.~(\ref{eq2.8}) with
  $\epsilon$ replaced by  
$$\bar\epsilon (R) = \frac{\sum_i\rho_i (R) \epsilon_i^2 }{\sum_i\rho_i(R)\epsilon_i}.$$ 
}

\section{~}
The data on the rotation of flat Galactic subsystems can be used in
order to determine the radial acceleration $F_R$ in galactic plane,  as
these subsystems rotate with nearly circular velocities, \ie with the
velocities $V$, related to $F_R$ by 
\begin{equation}
V^2= -F_RR. \label{eq2.9}
\end{equation}
For flat subsystems the most complete data include the rotation of
long-period cepheid subsystem, studied by
\citet{Parenago:1948b}. These are the data used in present paper. 

Parenago determined the rotation velocity of the subsystem of long-period cepheids via Camm function
$$f=V'_r \csc (l-l_0) \sec b, $$
where $V'_r$ is the radial velocity corrected from parallactic effect of solar motion;  $l$, $b$ are the galactic coordinates, $l_0$ -- the longitude of galactic centre. Within the precision of stochastic fluctuations the Camm function is related to the angular velocity $\omega (R)$ and to the linear velocity $V(R)$ by the formulae
\begin{equation}
f = R_{\odot} [\omega (R)-\omega (R_{\odot})], ~~~~ V(R)=R\omega (R), \label{eq2.10}
\end{equation}
where $R_{\odot}$ is the solar distance. To determine $V(R)$ Parenago
divides the sample into groups according to $R$. For every group
$\overline{f}$ was calculated -- the mean value of $f$ -- and
thereafter a smooth curve of $\overline{f}(R)$ was drawn and $V(R)$
was calculated. Besides, for $\omega (R_{\odot})$ the value  
$$\omega (R_{\odot})= 33~\mathrm{km/s/kpc} $$
was used, corresponding to the values of Oort constants $A= 20$~km/s/kpc and $B= -13$~km/s/kpc.

The averaging of $f$ was made by Parenago with the weight
$$w = |\sin (l-l_0)~\cos b|,$$
\ie with the weight inversely proportional not to the square, but to
the first power of mean fluctuations (error) of $f$ resulting from
peculiar stellar motions. As a result there is some looseness on the
precision of $\overline{f}$. But it is not essential because of quite
unimportant scatter of weights in groups. In addition, it is important
to mention, that the mean error of $\overline{f}$ was calculated by
Parenago according to formula which is valid only when weights are
inversely proportional to the square of mean error. As a result the
mean error of $\overline{f}$ is underestimated, in some cases even
highly (the mean errors given by Parenago must be multiplied by
$\sqrt{n/\sum w}$, where $n$ is the number of objects in group). This
overestimation of the precision of averaged values of Camm function
leads \citet{Parenago:1947} to make conclusions, which are rather
uncertain on the basis of data at his disposal (the rotation law). But
in the present case  it is not essential, because the mean error of $\overline{f}$ is
small in any case. 

As Parenago used for the solar distance from the galactic centre the
value 8.0~kpc, being slightly larger than the value $R_{\odot}$
accepted today \citep{Kukarkin:1949}, the Parenago's data need a
little correction. The values of $R$ and the values of the rotation
velocity $V$ corresponding to them must be slightly decreased. 

Taking into account that for most of cepheids used by Parenago,  the
distance from the solar radius-vector does not exceed 2~kpc, it is
easy to demonstrate that the decrease of $R$ will be approximately the
same for all (but not too small) $R$ as the decrease of $R_{\odot}$,
\ie equal to 0.75~kpc. The result remains valid also when we take for
the longitude of the Galactic centre not $\rm 325^o$ as it was taken
by Parenago, but according to Kukarkin $\rm 330^o$. At the same time
the values of $\overline{f}$ do not change significantly (not
considering the variations of stochastic fluctuations). 

For the correction of $V$, assuming that $\omega (R_{\odot})$ does not change, we have
\begin{equation}
\Delta V = \omega (R_{\odot})\Delta R + f\Delta\left( \frac{R}{R_{\odot}} \right) , \label{eq2.11}
\end{equation}
where $\Delta R$ and $\Delta\left({R\over R_{\odot}}\right)$ are the corrections for $R$ and ${R\over R_{\odot}}$. Since $\Delta R =  -0.75$~kpc, the first term in this formula equals to $-25$~km/s. The second term is so small that for $|R-R_{\odot}|<  2-3$~kpc it may be neglected.

\begin{figure}[ht]
\centering
\includegraphics[width=110mm]{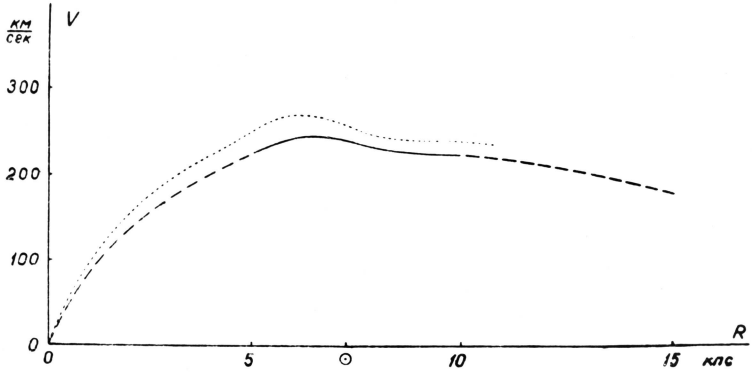}
\caption{Rotation velocity $V$ (in units of km/s) as a function of galactocentric radius $R$ (in kpc) of the Galaxy.}
\label{fig2.1}
\end{figure}

Thus, for $|R-R_{\odot}|< 2-3$~kpc the correction of Parenago's data reduces to decreasing the $R$ by 0.75~kpc  and increasing the $V$ by 25~km/s. The rotational velocity corrected in that way is represented in Fig.~\ref{fig2.1} by continuous line for values of $R$ from 5~kpc to 10~kpc. 

The remaining parts of the rotation curve must be interpolated and
extrapolated, because the range of $R$ in Parenago's data on the
rotation of long-period cepheids is limited. True, there is one value
of $\overline{f}$ for small $R$ being far away beyond the limits of
this interval. But this value of $\overline{f}$ is based on the motion
of three cepheids not belonging to the flat subsystem (they have $|z|
>$ 1.5~kpc). 

Rejecting the value of $\overline{f}$ referred above, evidently not
concerning the flat subsystem, we can interpolate the rotational curve
for small $R$ not in a form of nearly strict line, as it was done by
Parenago, but in a form of significantly curved line, better
corresponding to the expected rotational law for small $R$ both from
theoretical considerations and from the analogy with the rotation law
of the Andromeda nebula \citep[see e.g.][]{Parenago:1948a}. The part
of the rotation curve, interpolated in that way, is represented in
Fig.~\ref{fig2.1} by dashed line. In the same way also the
extrapolated part is represented. The extrapolated part is calculated
from Eqs.~(\ref{eq2.9}) and (\ref{eq2.2}) after the extrapolation of
$\delta = 4\pi G\epsilon\rho a^2$ (see below). 

The rotation velocities plotted in Fig.~\ref{fig2.1} are given also in
Table~ \ref{tab2.1} under the column $V$. In bold are the values of
$V$ resulting from the observational data; interpolated and
extrapolated data of $V$ is given in ordinary form. The table contains
also several other columns, we will describe them below.  

\section{~}

\begin{figure}[ht]
\centering
\includegraphics[width=100mm]{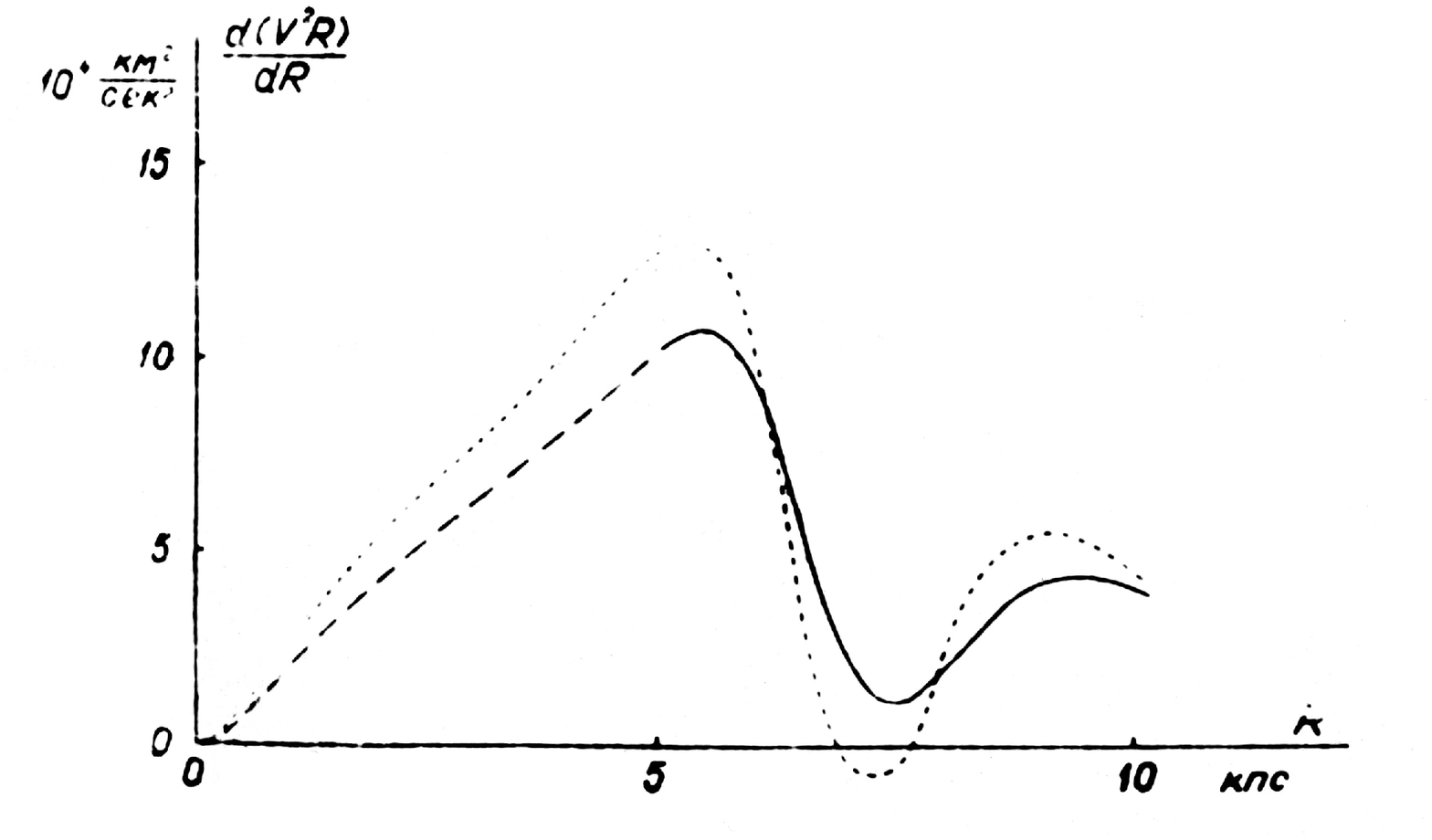}
\caption{Values of $\rmd (V^2R) / \rmd R$ (in units of $10^4 \mathrm{km^2/s^2}$) as a function of galactocentric radius $R$ (in kpc).}
\label{fig2.2}
\end{figure}

As can be seen from the equations of the first section, in order to
determine the surface density of the Galaxy,  one needs to know the
gradient of $-F_RR^2$, \ie 
$$ \frac{\rmd (-F_RR^2)}{\rmd R} =  \frac{\rmd (V^2R)}{\rmd R} = V^2+2RV \frac{\rmd V}{\rmd R}. $$
For $R\le  10$~kpc the values of the gradient are calculated on the
basis the rotation curve given above. These values are given in
Table~\ref{tab2.1} under the column $\rmd (V^2R)/\rmd R$ and in
Fig.~\ref{fig2.2}. The data based on the observational and
interpolated values are given by bold and  continuous lines respectively.

Numerical integration of Eq.~(\ref{eq2.8}) gives us the correction
$\Delta (-F_RR) = \Delta V^2$, transforming the problem of surface
density calculation into the problem of its calculation for the flat
disk model. Besides we supposed $\epsilon = 1/4$ according to our
recent study  \citep{Kuzmin:1952ab}.\footnote{In fact, $\epsilon$ is
  significantly less and equals approximately to 0.1. [Later
  footnote.]} Although the accepted value of $\epsilon$ is quite large
we may use Eq.~(\ref{eq2.8}) because $\epsilon^2$ is always
small. Corrected by $\Delta V^2$ values of $V$ are given in Table~
\ref{tab2.1} under the column $V_1$. The corresponding curve is
plotted in Fig.~\ref{fig2.1} by dotted line, indicating the rotation
curve of the Galaxy in flat disk approximation. 

From the curve $V_1(R)$ we calculated the gradient of $-F_RR^2$,
reduced to the flat disk model. This gradient is given in Table~
\ref{tab2.1} under the column $\rmd (V_1^2R)/\rmd R$ and in
Fig.~\ref{fig2.2} by dotted line. It is seen from Fig.~\ref{fig2.2}
that the reduction to the flat disk model means the amplification of
non-uniform nature of the gradient $-F_RR^2$. This is related to the
fact that for flat mass distribution the circular velocity is more
sensitive to surface density oscillations than in the case of mass
distribution which is extended in $z$-direction. 

It is important to mention that the curves of the gradient of
$-F_RR^2$ in Fig.~\ref{fig2.2} are quite uncertain because even small
uncertainties of the rotation velocities influence the gradient
significantly. But this uncertainty of the gradient curve does not
influence the results of the surface density $\sigma$, because in
calculation of $\sigma$ the values of the gradient of $-F_RR^2$ for
different $R$ are averaged. 

\begin{figure}[ht]
\centering
\includegraphics[width=110mm]{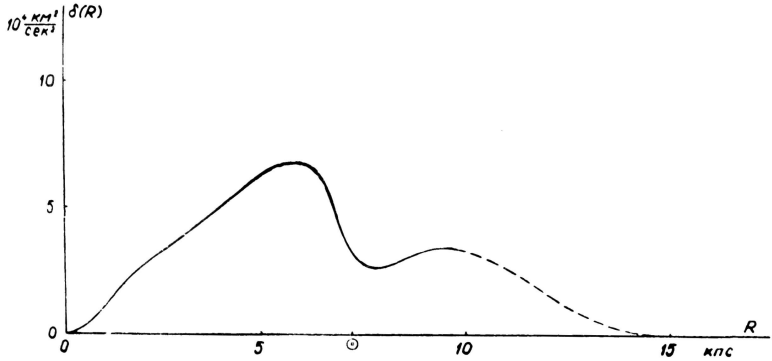}
\caption{Values of $\delta (a)$ (in units of $10^4 \mathrm{km^2/s^2}$) as a function of galactocentric radius $R$ (in kpc).}
\label{fig2.3}
\end{figure}

Using the reduced values of the gradient of $-F_RR^2$ (dotted curve in
Fig.~\ref{fig2.2}), and after performing numerical integration
according to Eq.~(\ref{eq2.5}), we calculated the quantity\footnote{In
  subsequent papers, instead of $\delta (a)$ the function $\mu (a) =
  4\pi\epsilon\rho (a) a^2$ is used, called ``the mass function''. The
  notation $\delta (R)$ (or $\Delta (R)$) will be used to designate
  the mass surface density $\sigma (R)$.  [Later footnote.]}

\begin{equation}
\delta (a) = 4\pi G\epsilon\rho (a)a^2. \label{eq2.12}
\end{equation}
The results are given in Table \ref{tab2.1} under the column
$\delta$. As for all other data in that table the values of $\delta$
are given as a function of $R$, \ie in the galactic plane where
$a=R$. Graphically $\delta$ as a function of $R$ is plotted in
Fig.~\ref{fig2.3}. 

Knowing the function $\delta (R)$ it is possible to find the spatial
matter density $\rho$ in the galactic plane, and to construct the curve $\rho
(R)$. But these results do not carry much weight because of uncertainty of
$\epsilon$ and of the ellipsoidal model of the Galaxy. We would like to mention that
for the solar neighbourhood $\rho\epsilon = 0.9\cdot 10^{-24} \mathrm{g/cm^3}$, 
being quite well in agreement with our previous result \citep{Kuzmin:1952ab} 
$\rho = 3.4\cdot 10^{-24} \mathrm{g/cm^3}$ and $\epsilon =$ 0.26.

The surface density, being most interesting for us, is calculated according to Eq.~(\ref{eq2.3}) from the equation
\begin{equation}
\sigma (R) = \frac{1}{2\pi G} \int_R^{\infty} \frac{\delta (a)}{\sqrt{a^2-R^2}} \frac{\rmd a}{a}. \label{eq2.13}
\end{equation}
To use the equation one needs to extrapolate $\delta (a)$ for $a >
10$~kpc. The extrapolation was done within the assumptions that
$\delta$ is decreasing when $a$ increases, and vanishes when
$a$ tends to some ``effective radius'' of the Galaxy, taken to be 15~kpc.
The true radius of the Galaxy is surely noticeably larger, but for $a >
15$~kpc $\delta$ is probably very small and does not contribute
significantly to the Galactic mass. The extrapolated values of $\delta$
are also given in Table~\ref{tab2.1} and plotted in Fig.~\ref{fig2.3} as the dashed part
of the curve.

\begin{figure}[ht]
\centering
\includegraphics[width=120mm]{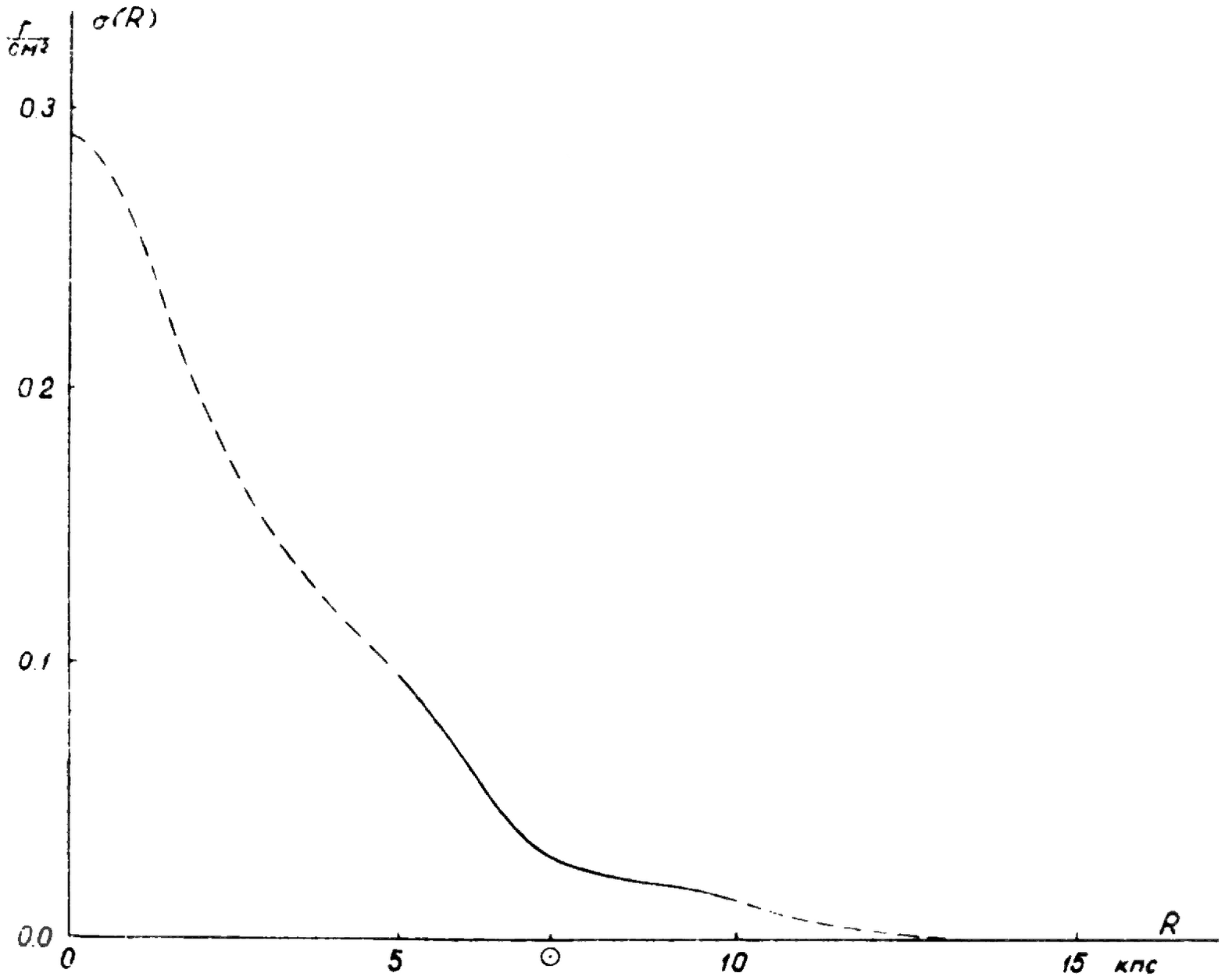}
\caption{Surface densities $\sigma (R)$ (in units of
  $\mathrm{g/cm^2}$) as a function of galactocentric radius $R$ (in
  kpc).}
\label{fig2.4}
\end{figure}

After having performed numerical integration of Eq.~(\ref{eq2.13}) we derive the values of
the surface density of the Galaxy and give them in Table~\ref{tab2.1} in
 $\sigma (R)$ column and in graphical form in
Fig.~\ref{fig2.4}.\footnote{According to contemporary data the distance scale
is by 40~\% ~greater than the one used in paper. Hence the surface density
$\sigma (R)$ and $\rmd\ln\sigma /\rmd R$ decrease proportionally. In addition,
$\sigma (R)$ slightly decreases because of decreasing of $\epsilon$. The
total mass increases. [Later footnote.]} In bold are in Table~ \ref{tab2.1}
more certain values of $\sigma$ being nearly independent of the
extrapolated part of the curve $\delta (R)$. These more certain values of
$\sigma (R)$ are given in Fig.~\ref{fig2.4} by continuous line.

It is seen from Fig.~\ref{fig2.4} that the surface density curve is not completely
smooth. The most certain part of it is highly bended. As a result in
the vicinity of the Sun there is relative deficiency of the density, and
at distances $R$ approximately equal to 5 and 10~kpc, there is relative
excess. This kind of behaviour 
of the density is related to the rotation curve (Fig.~\ref{fig2.1}). The
curve of $\sigma (R)$ is roughly approximated with the exponential law with
the mean gradient of the density logarithm
$${\rmd \log\sigma\over \rmd R} = -0.15~\mathrm{kpc}^{-1}.$$
Approximately similar value of $\rmd \log\sigma /\rmd R$ is in the solar
neighbourhood. This value of $\rmd \log\sigma /\rmd R$ corresponds to
the expected one, because it is slightly greater than the gradient of the
density logarithm for flat subsystems and less than its value for
the intermediate and spherical subsystems of the Galaxy \citep[see e.g.][]{Parenago:1948a}.

\begin{table}
\caption{}
\smallskip
\label{tab2.1}
\centering
\begin{tabular}{|c | c | c | c | c | c | c | c |} 
\hline\hline
$R$    &    $V$    & $V_{\infty}$ & $V_1$ & $\frac{d(V^2R)}{dR}$ & $ \frac{d(V_1^2R)}{
dR}$ & $\delta (R)$ & $\sigma (R)$ \\
\cline{2-7}
kpc & \multicolumn{3}{|c|}{km/s} & \multicolumn{3}{|c|}{$10^4~\mathrm{km^2/s^2}$}
& $\mathrm{ g/cm^2}$\\
\hline
0  &   0 & 508 &   0 & 0.0        &  0.0 & 0.0 & 0.29 \\
1  &  82 & 500 &  94 & 1.8        &  2.3 & 1.0 & 0.26 \\
2  & 136 & 484 & 153 & 4.2        &  5.3 & 2.6 & 0.20 \\
3  & 172 & 464 & 193 & 6.2        &  7.6 & 3.8 & 0.15 \\
4  & 200 & 442 & 223 & 8.0        & 10.0 & 5.0 & 0.12 \\
5  & 224 & 418 & 250 & {\bf 10.1} & 12.6 & 6.3 & {\bf 0.095} \\
6  & 243 & 394 & 270 & {\bf 9.6}  & 10.4 & 6.8 & {\bf 0.063} \\
7  & 243 & 370 & 263 & {\bf 2.2}  & -0.4 & 4.0 & {\bf 0.033} \\
8  & 231 & 349 & 246 & {\bf 2.0}  &  2.2 & 2.7 & {\bf 0.024} \\
9  & 226 & 331 & 242 & {\bf 4.2}  &  5.4 & 3.3 & {\bf 0.020} \\
10 & 224 & 314 & 240 & {\bf 4.1}  &  4.5 & 3.3 & {\bf 0.014} \\
11 & 221 & 298 & ... & ...        & ...  & 2.6 & 0.008 \\
12 & 214 & 284 & ... & ...        & ...  & 1.6 & 0.004 \\
13 & 204 & 272 & ... & ...        & ...  & 0.7 & 0.001 \\
14 & 193 & 261 & ... & ...        & ...  & 0.2 & 0.000 \\
15 & 183 & 251 & ... & ...        & ...  & 0.0 & 0.000 \\
\hline\hline
\end{tabular}
\end{table}

After interpolating the data from Table~\ref{tab2.1} we obtain for the surface density
in the solar neighbourhood ($R=$ $R_{\odot}=$ 7.25~kpc) the value
$$\sigma (R_{\odot}) = 0.030~\mathrm{g/cm^2} = 143 \,{\rm M_{\odot} /pc^2}.$$
This value is less than our previous estimate \citep{Kuzmin:1952ab} of $200
{\,{\rm M_{\odot} /pc^2}}$. Thus the effective half-thickness of the Galaxy in solar
neighbourhood $\frac{1}{2} \sigma (R_{\odot})/\rho (R_{\odot})$ is also
less, namely 1.4~kpc instead of 2.0~kpc.

In Galactic centre $\sigma$ is $\rm 0.29 ~g/cm^2 =$ $\rm 1.4\cdot 10^3
{\,{\rm M_{\odot} /pc^2}}$ and hence exceeds 10 times the surface density in the vicinity of
the Sun. Yet the result is very uncertain. It depends highly on the
interpolation of the rotation curve for small $R$.

Knowing the function $\sigma (R)$ or the function $\delta (R)$,  we may
estimate the total mass of the Galaxy $M$ by using the formulae
\begin{equation}
M = 2\pi\int_0^{\infty} \sigma (R)R\rmd R, ~~~~ GM = \int_0^{\infty} \delta
(a) \rmd a  \label{eq2.14}
\end{equation}
(the second formula results from $4\pi\epsilon\rho (a)a^2\rmd a =$ $\delta
(a) \rmd a/G$ being the mass of the elementary ellipsoidal layer). The
numerical integration give
$$M= 2.0\cdot 10^{44} ~\mathrm{g} = 1.0\cdot 10^{11} \Msun.$$
This result is quite well in agreement with the result by \citet{Safronov:1952}, who
derived $M= 1.1\cdot 10^{11} \Msun$. Recently \citet{Parenago:1952}
obtained for the Galactic mass $1.83\cdot 10^{11} ~\Msun$. This larger
value of the mass was obtained by Parenago because of improbably high
densities at the outer regions of the Galaxy, obtained due to his Galactic
potential formula. 

As we mentioned above, the rotation curve was extrapolated
in the region $R> $ 10~kpc with the help of the
function $\delta (a)$ (Table~\ref{tab2.1} and Fig.~\ref{fig2.1}).
According to Eqs.~(\ref{eq2.2}), (\ref{eq2.9}) and (\ref{eq2.12}) we have
\begin{equation}
V^2 = \int_0^R \frac{\delta (a) \rmd a}{\sqrt{R^2 - a^2e^2}}. \label{eq2.15}
\end{equation}
Calculating according to that formula the rotation velocities also for $R\le
$ 10~kpc we obtain quite good agreement with the initial data. This served
as a check of calculations and correctness of used formulae.

In addition, the function $\delta (a)$ was used to calculate the escape
velocity in the galactic plane. For the square of $V_{\infty}$ we have the
equations 
\begin{equation}
V_{\infty}^2 = 2 \int_R^{\infty} V^2 ~ \frac{\rmd R}{R}, ~~~~ V_{\infty}^2 = 2 
\int_0^{\infty} \delta (a)\chi\left( \frac{R}{a} \right) ~ \frac{\rmd a}{a}. 
\label{eq2.16}
\end{equation}
The expression under the integral in the second equation is the
potential of an elementary ellipsoidal layer in the equatorial plane of
the Galaxy. The function $\chi$ being proportional to that potential has a
form
\begin{equation}
\chi\left( \frac{R}{a}\right) = \left\{
\ba{ll}
\frac{1}{e} \arcsin e & \mathrm{for} ~~R\le a,\\
\noalign{\smallskip}
\frac{1}{e}\arcsin \frac{ae}{R} & \mathrm{for} ~~R\ge a.\\ 
\ea
\right.
\label{eq2.17}
\end{equation}
The results of the calculations are given in Table~\ref{tab2.1}
in the $V_{\infty}$ column. Interpolation of the data from the Table
gives for the solar neighbourhood $V_{\infty} =$ 365~km/s, \ie by 52~per cent
higher value than is the circular velocity (239~km/s). The derived value
is significantly higher than the estimates by \citet{Parenago:1952} and by
\citet{Ogorodnikov:1948}. According to these estimates the escape velocity in
solar neighbourhood exceeds the circular velocity only by 20--30~per cent. But
these estimates are based on the potential expression which can be used only as an
interpolation formula for more or less restricted region of space, and has
some undetermined additive constant. The agreement is significantly better
between our result and the result by \citet{Safronov:1952}, who derived that in
solar neighbourhood the escape velocity exceeds the circular velocity by
45~per cent.

\section{~}

Together with the surface density of the gravitating matter of the Galaxy
it is interesting to determine the surface luminosity density of the Galaxy.

We designate the surface light density by $\sigma^*$, and measure it in
solar luminosities per square parsec.

It is not difficult to find the relation between $\sigma^*$ and Galactic
emission brightness in direction perpendicular to the equatorial plane of
the system. Let us designate that brightness as $I$. Expressing the
brightness in stars of zero-magnitude per square radian we have
\begin{equation}
I = \sigma^* ~10^{-0.4(M_{\odot}-5)} , \label{eq2.18}
\end{equation}
where $M_{\odot}$ is the absolute magnitude of the Sun. Evidently, in
the solar neighbourhood the brightness $I$
equals the doubled brightness of stellar
emission at the galactic pole. Expressing the last quantity in stars of 10-th
magnitude per square degree and designating it by $I'$, we have for
$\sigma^*$ in the solar neighbourhood
\begin{equation}
\sigma^*(R_{\odot}) = 2 \left( \frac{1.8}{\pi} \right)^2 ~10^{0.4(M_{\odot}-
5)} ~I'. \label{eq2.19}
\end{equation}

The emission brightness of stars at the galactic pole can be found from the
distribution of stars according to their magnitudes $m$ in direction of the
galactic pole. It is calculated from the formula
\begin{equation}
I' = \int_{-\infty}^{\infty} A(m) ~10^{-0.4(m-10)} ~\rmd m , \label{eq2.20}
\end{equation}
where $A(m)$ is the number of stars in direction of the galactic pole per
square degree per unit interval of magnitudes. Using the data by
\citet{Seares:1925} we found that $I' =$ 21.8 stars of 10th photographic
magnitudes per square degree. It results that a significant
fraction of that quantity is due to quite bright stars, being hence
probably quite nearby to the galactic plane. This can be seen from Table~\ref{tab2.2}
where the contribution to the general brightness for individual ranges of
$m$ is given. The numbers in parentheses is extrapolated data.

\begin{table}
\caption{}
\smallskip
\label{tab2.2}
\centering
\begin{tabular}
{| c | c | } 
\hline\hline
$m_{pg}$  &  $I'$  \\
\hline
$-\infty$... 3.5 & (5.0) \\
 3.5 ...  7.5    &  5.7  \\
 7.5 ... 11.5    &  6.5  \\
11.5 ... 15.5    &  3.6  \\
15.5 ... 19.5    &  0.9  \\
19.5 ...$\infty$ & (0.1) \\
 \hline
$-\infty ... \infty$ & 21.8 \\
\hline\hline
\end{tabular}
\end{table}

The derived value of $I'$ must be corrected for interstellar light
absorption. It is known \citep[see e.g.][]{Parenago:1948a}. that the
total photographic absorption in direction to the galactic pole is
approximately $0^m.3$. $I'$ must be influenced by the absorption
somewhat less, because a significant fraction of stars contributing to
$I'$ lie within the absorbing layer. Assuming the absorption to be for
example $0^m.2$ we have $I'$ = 26 stars of 10th magnitude per square
degree.

Because the absolute photographic magnitude of the Sun is $+5.3$, the
surface light density in the solar neighbourhood is (Eq.~(\ref{eq2.19}))
$\sigma^*(R_{\odot}) = 22~{\rm L_{\odot} /pc^2}$.

Comparing the result with the value derived for the surface mass density in
the solar neighbourhood, we derive that their ratio is
$$\frac{\sigma (R_{\odot})}{\sigma^*(R_{\odot})} = 6.5. $$
This value is slightly less than the mass-to-luminosity ratio for the
Andromeda and Triangulum nebulae \citep[see e.g.][]{Parenago:1948a}. But it
is still larger than the mass-to-luminosity ratio for regions near to the Sun.
 The last quantity is easy to estimate. For the spatial light
density $\rho^*$ in units of solar luminosities per cubic parsec we have
\begin{equation}\rho^* = D \int_{-\infty}^{\infty} \varphi (M) ~10^{-0.4(M-M_{\odot})} 
~\rmd M , \label{eq2.21}
\end{equation}
where $D$ is the spatial density of stars, $\varphi$ -- the luminosity
function, and $M$ -- the absolute magnitude. Taking $D\varphi (M)$
according to \citet{Parenago:1946}\footnote{As it is known, $\varphi (M)$
for $M>$ +10 is quite uncertain. But this is not essential, because even
the stars with $M>$ +6 practically do not contribute to $\rho^*$.} we
find $\rho^*(R_{\odot}) =$ 0.06 solar photographic luminosities per
$\rm pc^3$.
From other side, according to the author \citep{Kuzmin:1952ab}, the density of
gravitating matter near to the Sun is $\rm 0.05~{\rm M_{\odot} /pc^3}$. Hence,
$${\rho (R_{\odot})\over \rho^*(R_{\odot})} = 0.8.$$

Therefore, in the solar neighbourhood $\sigma /\sigma^*$ is eight times larger
than $\rho /\rho^*$.\footnote{As the mass surface density
$\sigma (R)$ must be taken smaller, the ratio of surface mass and light
densities must decrease up to $\sigma /\sigma^* =$ 4 (at $R=R_{\odot}$).
From the other side, because $\rho$ must be increased nearly two times, the ratio 
$\rho /\rho^*$ increases up to 1.5. The difference between $\sigma
/\sigma^*$ and $\rho /\rho^*$ becomes much less. But it still remains
significant. Correspondingly less is also the difference between
equivalent half-thicknesses. The equivalent half-thickness of gravitating
matter decreases to 0.6~kpc. [Later footnote.]}

This high difference between the ratios of $\sigma /\sigma^*$ and $\rho
/\rho^*$ is related to different concentrations of the gravitating and
luminous matter of the Galaxy. The equivalent half-thickness of the
gravitating matter $\frac{1}{2} \sigma /\rho$ is 1.4~kpc, the equivalent
half-thickness of the luminous matter $\frac{1}{2} \sigma^*/\rho^*$ is only
0.2~kpc. This difference in concentrations of gravitating and luminous
matter was obtained already in our previous paper \citep{Kuzmin:1952ab}. It can
be easily understood because more spheroidal Galactic subsystems have
larger mass-to-light ratios. 

As a conclusion the author thanks A.~S. Sharov for presenting the
result of the paper by Wyse and Mayall, being absent in the library of
Tartu Observatory, and J. Einasto for help in calculations and for
discussions of the paper.
\vglue 3mm
\hfill 1952

%% file: chapter03.tex
\chapter[The third integral of stellar motions]{On the gravitational
  potential of a stationary galaxy and the third integral of stellar
  motions\footnote{\footnotetext ~~Tartu Astron. Observatory, Teated, 
    No.~1, 1954; Report in a common meeting of the Astron. Council of
    Acad. Sci. USSR and the Inst. Phys. and
    Astron. Acad. Sci. Estonian SSR, May 27--29, 1953,
    Tartu. Published also in Notices of the Acad. Sci. Estonian SSR
    {\bf 2}, 368, 1953.}}

One of the basic problems of practical stellar dynamics is the
determination of the gravitational potential of the Galaxy and its
mass distribution. Essential results by Kukarkin, Parenago, etc,
\citep{Parenago:1948a} in analysing the spatial and kinematical
structure of Galactic subsystems enabled to make first steps is
solution of the above mentioned problem. We have in mind first of all
the recent works of P.~P.~Parenago in establishing the problem and
solving it in first approximation \citep{Parenago:1950a,
  Parenago:1952}. The studies by P.~P.~Parenago stimulated growing 
interest to the problems of practical stellar dynamics. Nearly
simultaneously with the recent papers by Parenago also the author of
the present paper published his investigations on the problem of
Galactic mass distribution and gravitational potential
\citep{Kuzmin:1952ab, Kuzmin:1952ac}. Continuing these studies the
author succeeded to derive some new results, in particular on the
problem of the third integral of stellar motion and related problem of
Galactic structure. The aim of the present paper is to give a short
critical review of the present situation with presenting of a series
of new results.

In determining the gravitational potential of the Galaxy the initial
observational data are, firstly, data on the rotation of Galactic
subsystems, and secondly, data on the stellar motion perpendicular to the
galactic plane and the distribution of the spatial density in the same 
direction. 

The data on the rotation of Galactic subsystems can be used to
determine the gravitational potential as a function of the distance
from Galactic axis $R$. In case of flat subsystems the rotational
velocity is nearly equal to the circular velocity enabling to derive
the radial acceleration and hence, the radial gradient of the
potential. However, in order to determine the radial acceleration, it
is recommended and in some cases even necessary to use also data on
the rotation of intermediate and even spherical subsystems. For these
subsystems it is needed to use the statistical equations of motion. We
may use the Jeans equations \citep{Jeans:1922} correcting them for the
effect of velocity ellipsoid's obliquity about the galactic plane. As
we shall demonstrate below the existence of the obliquity beyond the
galactic plane is quite probable.  In addition, the Jeans equations
must be modified by taking into account the Lindblad-Oort relation
\citep{Lindblad:1927, Oort:1928}, connecting the velocity ellipsoid's
axial ratio with the radial gradient of the rotational velocity. (This
relation may be derived from the similar kind of general
considerations as the Jeans equation.\footnote{We have in mind the
  using of hydrodynamic equations of stellar dynamics (``statistical
  equations of motion'') [Later footnote.]})  In this case we have for
galactic plane
\begin{equation}
-R \frac{\partial\Phi}{\partial R} = V^2+q\sigma_R^2 , \label{eq3.1}
\end{equation}
where $\Phi$ is the gravitational potential, $V$ -- the circular velocity,
$\sigma_R$ -- the dispersion of $R$-component of velocities and
\begin{equation}
q = - \frac{\partial \ln (D\sigma_R^2\omega^{-1/2})}{\partial\ln R} + q_1
. \label{eq3.2}
\end{equation}
In the last formula $D$ is the spatial density of stars, $\omega$ -- the
angular rotational velocity and $q_1$ the term, accounting the velocity
ellipsoid's obliquity. This term has a form
\begin{equation}
q_1 = - \left( 1- \frac{\sigma_z^2}{\sigma_R^2} \right) \frac{R}{ R_c},
\label{eq3.3}
\end{equation} 
where $\sigma_z^2$ is the dispersion of velocities $z$-component, $R_c$ is
the distance from the convergent point of velocity ellipsoid's major axis
directions near the galactic plane. [see Appendix A.]

Equation (\ref{eq3.1}) accounts the obliquity of the velocity ellipsoid about the
galactic plane and the radial gradient of the dispersion $\sigma_R$, and is
thus more general than the usually used Oort equation for stellar
motion asymmetry \citep{Oort:1928}. The term
$q\sigma_R^2$,  determining the asymmetric drift,  is
negligibly small for flat subsystems. For intermediate subsystems it is
like a small correction, for spherical subsystems becomes dominating. The
coefficient $q$ is determined mainly by the radial gradient of the spatial
stellar density, because the gradients of $\sigma_R^2$ and of $\omega^{-1/2}$
and the term $q_1$ are probably relatively small. The term $q_1$ is not
easy to determine from observational data; it is needed to limit with
approximate estimations on the basis of theoretical considerations.

The data on the rotation of the subsystem of long-period cepheids, studied
by \citet{Parenago:1948b},  enable quite firmly to estimate the radial
acceleration for distances from $R=$ 5~kpc to $R=$ 10~kpc. For $R>$ 10~kpc
the corresponding data are nearly absent. For $R<$ 5~kpc the radial
gradient is possible to determine only approximately on the basis of the
rotations of planetary nebulae subsystem,  and of intermediate subsystem of
long-period cepheids\footnote{It is known, that part of long-period
cepheids have larger velocity dispersion. On the basis of that and of the
centroid velocity these cepheids belong not to the spherical, but to the
intermediate subsystem as well as the planetary nebulae.},  and on the basis
of the spatial and kinematical structure of spherical subsystems. In last case
the radial acceleration is determined mainly by the second term in
Eq.~(\ref{eq3.1}), enabling to estimate it. It is needed to mention, that this term
was quite firmly determined in the solar neighbourhood, enabling to
determine the circular velocity in the vicinity of the Sun independently
of the Oort constants and of the solar distance from Galactic centre. This
kind of circular velocity determination was done by \citet{Parenago:1950b} by
using Oort formula for asymmetry. The result coincides with the estimate
on the basis of the Oort constants and removes some uncertainty due to
uncertain value of constant $B$.

\begin{figure*}[ht]
\centering
\includegraphics[width=120mm]{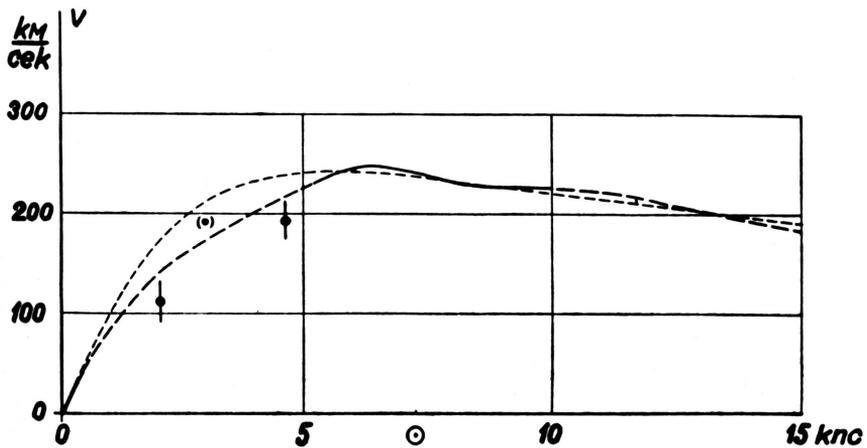}
\caption{Rotation velocities (in units of km/s) as a function of
  galactocentric radius $R$ (in kpc) of the Galaxy.}
\label{fig3.1}
\end{figure*}

The rotation velocity curve of the Galaxy is plotted in Fig.~\ref{fig3.1}. We took
it from our paper \citep{Kuzmin:1952ac}. The continuous part of the curve
corresponds to somewhat smoothed data of the rotation of long-period
cepheid's subsystem at radii from $R=$ 5~kpc to $R=$ 10~kpc. The
extrapolated and interpolated parts of the figure are given by dashed lines.
The three points correspond to the crude circular velocity estimations for
small $R$ (the points with error bars are derived for intermediate
subsystems, a point in parentheses for spherical).

In well-studied region the circular velocity curve has a notable
curvature and a turning point. In that region the velocity curve differs
notably from the smooth Lindblad's rotational law 
\citep{Lindblad:1933}\footnote{The rotational law (\ref{eq3.4}) would be more correct
to call the Eddington-Charlier law. Charlier (C.L. Charlier, Lund Medd.
No. 82, 1917) derived the law (\ref{eq3.4}) by assuming that the velocity distribution
is biaxial Schwarzschild's distribution. But even earlier \citet{Eddington:1915},
proposing the triaxial Schwarzschild's distribution derived more general law,
giving Eq.~(\ref{eq3.4}) for $z=0$. [Later footnote.]}
\begin{equation}
V = \frac{c_3R}{ c_1+c_2R^2} , \label{eq3.4}
\end{equation}
where $c_1$, $c_2$, $c_3$ are constants. True enough, the rotation curve
in Fig.~\ref{fig3.1} may be somewhat distorted due to the errors in long-period
cepheid distances, because of non-perfect extinction corrections. But
nevertheless it seems for us justified to consider the possible reality of
the deviations from Eq.~(\ref{eq3.4}) at the referred interval of $R$. Probably the law
(\ref{eq3.4}) is usable only as an approximate interpolation formula, especially for
flat subsystems. It is evidently not usable as a formula for circular
velocity at large $R$, giving too steep decrease of the radial acceleration
with $R$.

For flat subsystems the referred law is also theoretically not too well
emphasised. It was derived under the assumption that velocity distribution
is precisely Schwarzschild's or generally ellipsoidal. This kind
of assumption can be done when analysing a mathematical model,  but not in
studying the real Galaxy. Without that assumption nothing concrete about the
rotation law can be derived, with the exception that rotation
velocity must practically coincide with the circular one. The velocity
dispersion is small and the velocity distribution is Schwarzschild's one
only when the phase density is proportional to $e^{kf}$, where $k$ is
quite large positive constant and $f$ is a smoothly varying function of the
integrals of motion with only one maximum in velocity space. It is easy
to demonstrate that in this case the resulting velocity distribution is
the Schwarzschild's one by expanding the function $f$ in Taylor's series
around the maximum of the phase density in velocity space. Because $k$ is
large, only 
the first, quadratic with respect to velocities terms, are important. But
although the velocity distribution results to be Schwarzschild independently
of the form of the function $f$, the rotation law depends on the function
$f$, and without knowing it nothing can be said about the rotation law.
True, the function $f$ must be somewhat restricted to avoid too large
radial gradient of spatial stellar density. It gives us that the rotation
velocity nearly coincides with circular velocity. But the restrictions on
the function $f$ depend on the circular velocity law (and not in the
opposite way) and again nothing definite can be said about the rotation
curve. [see Appendix B.]

Hence, the rotation velocity law (\ref{eq3.4}) from one side is not sufficiently
emphasised by the observational data, from other side it is badly emphasised
by the theory. For this reason the analysis of Galactic rotation,
particularly of its flat subsystems, must not be done by determining the
parameters of that law, but by detailed study of  rotation velocities
as a function of $R$, and calculating thereafter the radial acceleration
and the potential. 

One can not agree with \citet{Parenago:1950a, Parenago:1952} accepting in his papers
on the Galactic potential the law (\ref{eq3.4}) for the circular velocity, and
deriving from it the expression for the potential in galactic plane.
For small $R$ the law can be used as an approximate interpolating formula,
but for large $R$ it is completely unacceptable. It
gives for the Galaxy zero total mass, because the corresponding potential
vanishes faster than $R^{-1}$. The zero mass indicates that for some
regions the spatial density is negative. This is just the case. For
large distances from galactic plane Parenago obtained negative densities.

Too fast decrease of the potential for large $R$, giving zero total mass,
results because the radial acceleration according to (\ref{eq3.4}) decreases also
too fast with increasing of $R$. But this influences (decreases) the
potential not only for large $R$,  but for all values of $R$, because the
potential results via integration of the radial acceleration from given
$R$ to $\infty$. This gives highly lowered values for the escape
velocities in Parenago's model (in particular, 300~km/s in the vicinity of
the Sun). But it does not mean that his formula for the Galactic potential
is completely useless. It can be used as more or less approximate
interpolating formula for not too large distances from the Galactic
centre. And there remains unknown a certain additive constant.

Knowing the radial acceleration near the galactic plane it is possible to
derive some data on the mass distribution of the Galaxy. It is possible to
calculate the surface density of the Galaxy, \ie  the density projected
onto galactic plane. The similar problem for the Andromeda and Triangulum
nebulae were solved by \citet{Wyse:1942}, who used a flat disk model.
An attempt to solve the problem for our Galaxy was done in our referred
above paper \citep{Kuzmin:1952ac}, where the problem was discussed in a more
general way, besides it was taken into account a finite ``thickness'' of
the Galaxy. However, the correction for ``thickness'' only slightly influences
the results. Even more less the results must depend from the density
decrease law, when moving away from galactic plane. Assuming the
equidensity surfaces to be spheroids, the surface density $\delta$ is
calculated according to the equation
\begin{equation}
\delta = \frac{1}{ 2\pi} \int \frac{\mu (a) \rmd a}{ a\sqrt{a^2-R^2}} ,
\label{eq3.5}
\end{equation}
where $\mu (a) \rmd a$ is the mass between the equidensity spheroids with
major semiaxis $a$ and $a+ \rmd a$. The function $\mu (a)$, as it was
demonstrated in our paper, is related with the radial acceleration in
the 
galactic plane via equation
\begin{equation}
 - \frac{\partial\Phi}{\partial R} R = G \int_0^R \frac{\mu (a) \rmd a
}{\sqrt{R^2-a^2e^2}}, \label{eq3.6}
\end{equation}
where $G$ is the gravitational constant and $e$ -- the eccentricity of the
meridional sections of equidensity spheroids. For $e=1$ we have a flat
model of the Galaxy. In this case the integral equation (\ref{eq3.6}) reduces to the
Abel's equation, and $\mu (a)$ is soluble by quadratures. To take into
account the ``thickness'' of the Galaxy, it is needed to add a certain
correction to the derived result. An expression for the correction is
given in the referred paper.

\begin{figure*}[ht]
\centering
\includegraphics[width=100mm]{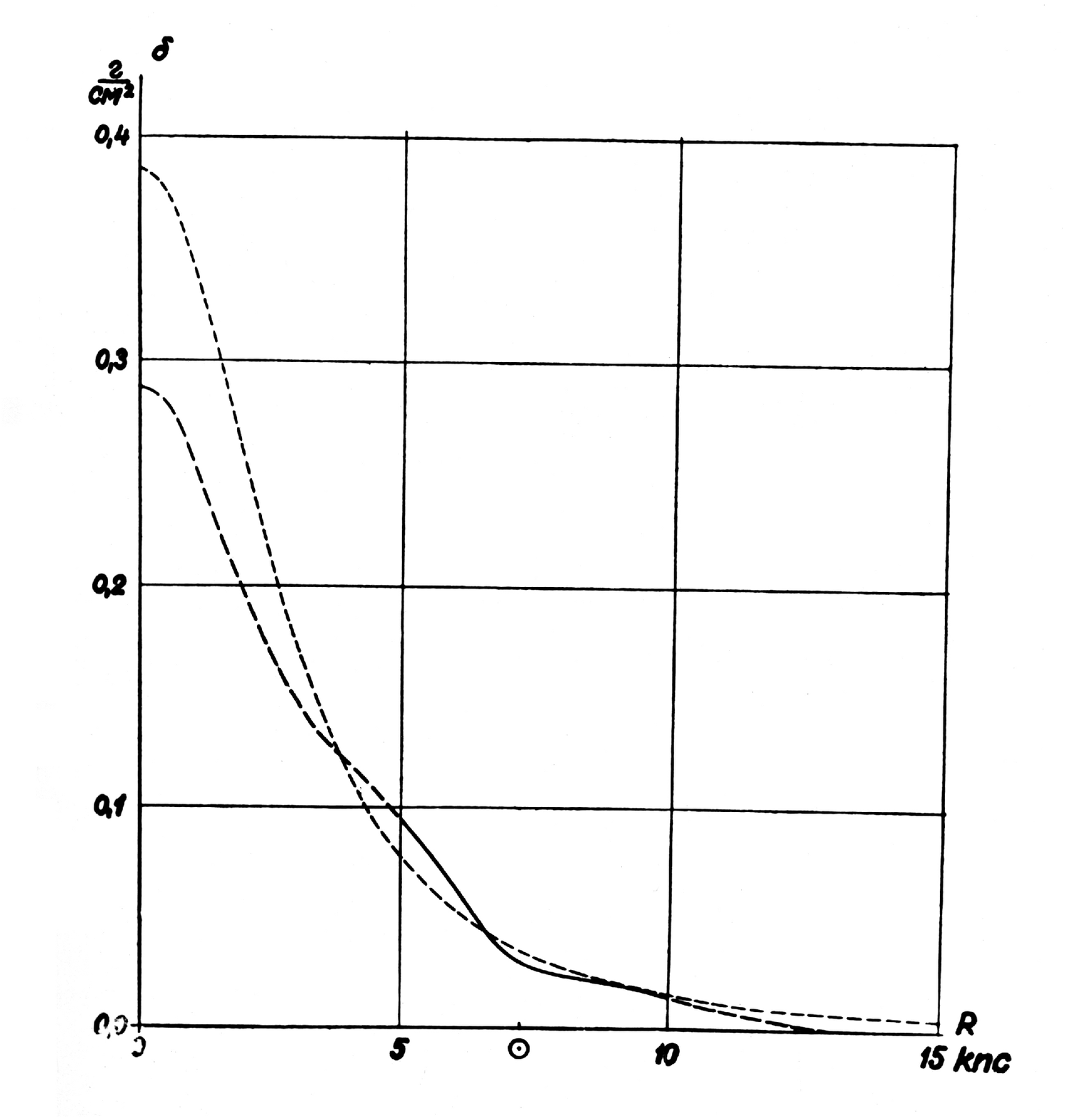}
\caption{Surface densities (in units of g/cm$^2$) as a function of
  galactocentric radius $R$ (in kpc).}
\label{fig3.2}
\end{figure*}

Despite to the fact that data about the radial acceleration near the
galactic plane are far from complete, it is possible to derive quite
firm results for the surface density, especially near to the Sun
\citep{Kuzmin:1952ac}. In Fig.~\ref{fig3.2} the surface density curve
corresponding to the circular velocity curve of Fig.~\ref{fig3.1} is
plotted. Near the Sun the surface density equals to $\rm 140
~M_{\odot}/pc^2$. This is about two times higher than the result
derived by \citet{Oort:1932} from stellar-statistical data, and by
taking into account the interstellar matter\footnote{Because the
  distance scale was overestimated, our surface density must be
  decreased, and the discrepancy with Oort's result decreases. [Later
  footnote.]}. In part the discrepancy may be explained, because Oort
did not take into account the spherical Galactic subsystems. The
surface density curve plotted in Fig.~\ref{fig3.2} can be quite well
approximated by an exponential law with the radial gradient of $\ln\delta
$ equals to $\rm 0.15 ~kpc^{-1}$. The result seems to be acceptable.

It is interesting to note that the details in most firmly established part
of the surface density curve (continuous line) represent the details of
spherical subsystem's density distribution \citep{Kukarkin:1949}. This may serve
as an argument in support to  the considerable role of spherical subsystems
in Galactic mass distribution.\footnote{In a subsequent paper
\citep{Kuzmin:1956b}, we found a more smooth surface density distribution. But some
waves similar to those in Fig.~\ref{fig3.2}, although not so strong, exist. [Later
footnote.]} 

In addition to the surface density the data on the radial acceleration
enable us to estimate the total mass of the Galaxy, and to calculate the
escape velocity at different $R$ in the galactic plane. On the basis of the
discussion above,  evidently the mass and the escape velocity will depend
considerably on the extrapolation of the circular velocity 
for $R>$ 10~kpc. To avoid negative densities, it is reasonable to make the
extrapolation indirectly, by extrapolating the function $\mu (a)$, which
must be non-negative and vanish when $a$ approaches a certain
limit, taken to be the effective radius of the Galaxy. The results, derived by
\citet{Kuzmin:1952ac} for the mass and the escape velocity, are the following:
Galactic mass is $\rm 100\cdot 10^9 ~M_{\odot}$, the escape velocity near
the Sun is 365~km/s. These results agree quite well with the results by
\citet{Safronov:1952},  derived in somewhat different  way. As it was expected, the
escape velocity results to be significantly higher that it was obtained by
Parenago. It exceeds the circular velocity by 125~km/s or 52 per cent.

It is possible to disagree with us, by arguing that large escape
velocity is in contradiction with the data on the asymmetry of stellar
motions, enabling Oort to conclude that the escape velocity exceeds
the circular velocity only by 65~km/s. However, as it was noted
already by \citet{Bottlinger:1932}, the nearly complete absence of
stars with velocities exceeding the circular velocity by more than
65~km/s indicates to the finite dimensions of the Galaxy,  and not to
the escape velocity. We can not agree with these  authors, who take the
difference between the escape and circular velocities to be
65~km/s. That this value is too small, is seen even from the paper by
\citet{Perek:1948}, who used this value in his calculations, and
obtained 
completely unacceptable results for the position of the Sun in the
Galaxy. The Sun lies in his model in a very periphery of the system.

Up to now we analysed the gravitational potential of the Galaxy as a
function of distance from the galactic axis $R$, and the observational
data we used as a function of that distance. Now we study the
potential as a function of the distance from the galactic plane
$z$. Also the data on the kinematics and spatial distribution of stars
in the direction perpendicular to the galactic plane. In
processing  these data it is needed again to use the corresponding
statistical equations by \citet{Jeans:1922}. Yet the using of Jeans
equation is complicated, because we must know the $z$-component
velocity dispersion of stars at various distances from the galactic
plane, but these data are absent at present. The difficulties can be
overcome by using the data on the distribution of velocity
$z$-components near to the galactic plane.  This was done by Oort in
his known paper on the attraction in direction perpendicular to the
galactic plane \citep{Oort:1932}. In fact, Oort used the equation, 
obtained by us in the following form \citep{Kuzmin:1952ab}
\begin{equation}
D = \int_{-\infty}^{\infty} F[v_z^2-2(\Phi -\Phi_0)] \rmd v_z. \label{eq3.7}
\end{equation} 
Here $\Phi $ is the potential, $\Phi_0$ -- the potential in galactic
plane, $v_z$ -- velocity $z$-component, $F(v_z^2)$ -- the distribution
function of velocity $z$-components in galactic plane, and $D$ -- spatial
density of stars. Besides, $D$, $F$, $\Phi$, $\Phi_0$ is assumed to belong
to the same value of $R$. When the function $F$ is known, Eq.~(\ref{eq3.7}) enables
to calculate the spatial distribution as a function of difference $\Phi
-\Phi_0$, enabling further by knowing the spatial density as a function
of $z$ to find the referred difference as a function of $z$. As we shall
see, there is no need to use successive iterations, as it was done by Oort
using in his calculations also Jeans equation.

Equation (\ref{eq3.7}) was derived under the assumption of stationarity and axial
symmetry of the Galaxy. The obliquity of the velocity ellipsoid about the
galactic plane was not taken into account. This kind of obliquity must
cause the explicit dependence of the function $F$ on $z$. But it can be
demonstrated, that for flat subsystems the dependence is not significant.
For intermediate and spherical subsystems the dependence may be
significant. 

For individual subsystems of the Galaxy the velocity $z$-component
distribution may be assumed to be approximately Gaussian. Hence Eq.~(\ref{eq3.7})
takes the  form
\begin{equation}
D = D_0 \exp \left( \frac{\Phi -\Phi_0}{\sigma_z^2}\right) , \label{eq3.8}
\end{equation}
where $D_0$ is the spatial stellar density in galactic plane, and
$\sigma_z$ -- the dispersion of velocity's $z$-component,  being in this
case independent of $z$. Near to the galactic plane the equation turns to
the following
\begin{equation}
D = D_0 \exp \left( - \frac{z^2}{ 2\zeta^2}\right) , \label{eq3.9}
\end{equation}
where $\zeta$ is related with $\sigma_z$ via
\begin{equation}
\frac{\sigma_z}{\zeta} = C , ~~~~ C^2 = -\left[
\frac{\partial^2\Phi}{\partial z^2} \right]_{z=0}. \label{eq3.10}
\end{equation}

Equations (\ref{eq3.8}) and (\ref{eq3.9}) result also from Jeans equation when $\sigma_z$ is
taken to be independent of $z$. They do not take into account the
obliquity of the velocity ellipsoid about the galactic plane,  being
important only for the intermediate and the spherical subsystems.

Galactic potential for most wide range of $z$ was studied by
\citet{Parenago:1952}, who applied Eq.~(\ref{eq3.8}) and used the data
obtained by \citet{Kukarkin:1949} on the spatial density of cepheids
and Mira stars. Unfortunately, because the accepted for $\sigma_z$
values were quite uncertain, and the obliquity of the velocity
ellipsoid about the galactic plane and possible gradient of $\sigma_z$
with $z$ were not taken into account, the results can not be handled
as certain. Significantly more precise results for the potential near
to the galactic plane may be obtained by determining the constant $C$
in the solar neighbourhood. Here the data for several Galactic
subsystems may be used. This kind of analysis was done by Oort in the
referred above paper \citep{Oort:1932}. But the data used by him are
somewhat out of date, and the constant $C$ needs to be revised. A trial
in this direction was done in our referred above paper
\citep{Kuzmin:1952ab}. By processing the data on the proper motions of
A-stars and K-giants near to the galactic equator we derived $C= 56\pm
5$~km/s per kiloparsec. This is less than Oort's result $C= 73$~km/s
per kiloparsec. The value of $C$ was determined also by
\citet{Parenago:1952} from cepheids and Mira stars, and the result was
near to Oort's one. However, his result seems to have smaller
precision than ours.  He used \citet{Kukarkin:1949} data on
short-period cepheids and Mira stars, and extrapolated them for small
$z$. But the extrapolation can be done only very approximately. True
enough, for small $z$ we have the data on the spatial density of
long-period cepheids, but the velocity dispersion $\sigma_z$ for these
stars is very uncertain. Meanwhile, also our value for $C$ in not very
certain, and may be underestimated because of some systematic
errors. But despite of it quite probably $C$ is somewhat less than
Oort's value.

Knowing the potential as a function of $R$ and $z$ it is possible to
calculate the spatial density of matter from the Poisson's equation.
Introducing the Oort constants $A$ and $B$ for the motion in circular
orbit, we have the Poisson's equation for the regions near to the galactic
plane in the form 
\begin{equation}
4\pi G\rho = C^2 - 2(A^2-B^2), \label{eq3.11}
\end{equation}
where $\rho$ is the density. This equation was derived by \citet{Lindblad:1938}.
Later it was independently derived by \citet{Parenago:1952} and by the author
\citep{Kuzmin:1952ab}, and as it seems, first used in real calculations. (Oort in his
referred paper \citep{Oort:1932} also calculated the matter density. But he did
not used the Poisson's equation, but preferred to calculate the density
from Galactic model, consisting of homogeneous spheroids. Later this method
was used by \citet{Safronov:1952}).

In our paper we obtained with help of Eq.~(\ref{eq3.11}) for the density in the
solar neighbourhood a value  $\rm 0.05\pm 0.01 ~M_{\odot}/pc^3$. This result is 
about two times less than Oort's estimate,  and is related with smaller
value of $C$. The derived density only slightly exceeds the density due to
stars. But because our earlier value of $C$ is underestimated, the
real density must be larger, although probably not as large as Oort's
value. 

The mass density in the solar vicinity was determined by \citet{Parenago:1952},
\citet{Safronov:1952}, \citet{Schilt:1950} etc. Schilt obtained quite absurd result,
indicating that the method used by him can not be used. The results by
Parenago and Safronov are similar to Oort's. However, this  coincidence 
does not confirm the correctness of Oort's value. The value of $C$, derived
by Parenago,  is very uncertain, as we mentioned already.   Safronov used
Oort's value of $C$ without determining it again. True enough, Safronov
calculated the density not only from value of $C$,  but also independently
by using galactic rotation, and obtained again a result close  to Oort's
value. It confirms only that the Galactic flatness used by Safronov agrees
with Oort's value of $C$. But the Galactic flatness is a quantity,  needed
himself to be determined from dynamical considerations.

To determine the Galactic flatness,  data on the spatial
density near to the Galactic plane and on the Galactic surface density
are needed,
enabling to determine the equivalent half-thickness of the Galaxy.
According to our results the equivalent
half-thickness of the Galaxy in the solar neighbourhood is 1.4~kpc
\citep{Kuzmin:1952ac}. This result is evidently overestimated,  not only because of
too small value for the spatial density, but also because of slightly high
surface densities. But despite of that it seems probable that the
equivalent half-thickness of the Galaxy in the solar neighbourhood may be as
large as one kiloparsec. If we assume the Galaxy to be spheroidal, this
equivalent half-thickness corresponds to the axial ratio about
1/5. Hence the Galactic flatness is quite moderate. It seems that Galactic flatness
is less than usually accepted, and that the mass of spherical subsystems is
significant. This conclusion agrees with \citet{Parenago:1949} result on large
number of subdwarfs and \citet{Pikelner:1953} results that a significant
amount of the interstellar gas belongs to the spherical subsystem.

Above we discussed the problems of the determination of Galactic
potential and the mass distribution. It remains to present some
results on the possible structure of the Galaxy, related with the
third integral of motion. It is known that one of the problems of the
dynamics of stationary Galaxy is to explain the triaxial velocity
distribution. Usually it is assumed according to \citet{Jeans:1915}
that the phase density of stars is a function of two integrals of
motion -- the energy integral $I_1$ and the angular momentum integral
$I_2$. But in this case the velocity distribution is symmetrical about
the axis, laying in the direction of galactic rotation. In real Galaxy
the velocity distribution is triaxial with the longest axis of the
velocity ellipsoid, directed more or less radially. This difficulty
can be removed if we assume the existence of the third integral of
motion. \citet{Lindblad:1933} proposed the third integral to be the
energy integral in $z$-coordinate\footnote{An additional reference is
  B.  Lindblad, Stockholm Medd. No.~11, 1933. A little earlier the
  form of the integral (\ref{eq3.12}) proposed by
  \citet{Oort:1932}. Therefore, the integral is appropriate to call as
  the Oort-Lindblad integral. [Later footnote.]}
\begin{equation}
I_3 = v_z^2 -2(\Phi -\Phi_0). \label{eq3.12}
\end{equation}
Strictly said, this integral is an integral only then when the potential is a
sum of a function of $R$ and a function of $z$, \ie  when $\partial^2\Phi
/\partial R\partial z =0$. For the Galaxy as a whole the condition
surely is not valid; but when limiting to  the vicinity of galactic
plane and a small range of $R$, the condition may be expected to be
valid with sufficient precision. Thus for stars, moving in nearly circular
orbits, $I_3$ remains nearly constant. It follows now, that for stars of
flat subsystems the phase density can be supposed to be a function of not
only $I_1$ and $I_2$, but also of $I_3$. The velocity distribution for flat
subsystems results to be triaxial.

But the velocity distribution is triaxial not only for flat,  but also for
intermediate and spherical subsystems. For spherical subsystems the
Lindblad's integral can not be used, and it is needed to find the third
integral in more general form. Because the velocity distribution for
individual subsystems is nearly Schwarzschild's one, it is reasonable to
demand that the third integral must permit Schwarzschild's velocity
distribution.\footnote{The third integral being quadratic with respect
to velocities allows the Schwarzschild's ellipsoidal velocity
distribution. But from the quadratic form of the integral does not follow
that the velocity distribution must be necessarily Schwarzschild's one or
even ellipsoidal. We do not agree with Perek (L.Perek, Adv. Astron.
Astroph. {\bf 1}, 165, 1962) classifying our mass distribution model of the
Galaxy, basing on quadratic third integral, to those models
basing on the assumption of ellipsoidal velocity distribution. A model
with ellipsoidal velocity distribution, strictly speaking, is even
impossible, as it was mentioned already by \citet{Eddington:1915}. [Later
footnote.]} For that reason it is needed to search it as a quadratic form in
respect to velocity components. When we assume the third integral to be
independent of galactocentric longitude, it results that the integral
$I_3$ must have the  form 
\begin{equation}
I_3 = (Rv_z-zv_R)^2+z^2v_{\theta}^2+z_0^2(v_z^2-2\Phi^*) , \label{eq3.13}
\end{equation}
where $v_R$, $v_z$ and $v_{\theta}$ are the velocity $R$, $z$ and $\theta$
components respectively, $z_0$ is a constant and $\Phi^*$ is a
function, 
satisfying the conditions
\begin{equation}
\frac{\partial\Phi^*}{\partial R} = \frac{z^2}{z_0^2} \frac{\partial\Phi}{
\partial R} - \frac{Rz}{ z_0^2} \frac{\partial\Phi}{\partial z} , ~~~~~
\frac{\partial\Phi^*}{\partial z} = \left(1+ \frac{R^2}{z_0^2}\right) 
\frac{\partial\Phi}{\partial z} - \frac{Rz}{ z_0^2} \frac{\partial\Phi}{\partial
R} . \label{eq3.14}
\end{equation}
The first two terms in the expression of the integral are the square of the
angular momentum component, directed parallel to the galactic plane. The last
term has a form of the energy integral in $z$-coordinate, but instead of
the potential it contains the function $\Phi^*$.

The function $\Phi^*$ must satisfy simultaneously both equations (\ref{eq3.14}).
Hence $\partial^2\Phi^* /\partial R\partial z$ from both
equations must be the same. It gives us the condition for the existence of
integral (\ref{eq3.13}) 
\begin{equation}
3\left( z \frac{\partial\Phi}{\partial R}-R \frac{\partial\Phi}{\partial z}
\right) - \left( R^2+z_0^2-z^2\right) \frac{\partial^2\Phi}{\partial R\partial
z} + Rz\left( \frac{\partial^2\Phi}{\partial R^2} - \frac{\partial^2\Phi}{
\partial z^2}\right) = 0. \label{eq3.15}
\end{equation}
In its form the condition coincides with the condition,  known from general
dynamics \citep[][Sect. 152]{Whittaker:1904} as the condition for the existence of quadratic in
respect to velocity components integral for planar motions. \citet{Oort:1928} in
his well known paper on the dynamics of the Galaxy also derived the same
condition as the condition of triaxial Schwarzschild's velocity
distribution. He thought that the condition can not be fulfilled, and concluded
that the theory enables only biaxial velocity distribution. The observed
triaxility he explained by accepting that stellar motions satisfy the
condition for 
$\partial^2\Phi 
/\partial R\partial z$. But as we mentioned already, this condition can be
used only for flat subsystems. To explain the triaxility for intermediate
and spherical subsystems it is needed to start from more general condition
(\ref{eq3.15}). If we assume that it is valid within sufficient precision for these
subsystems, the phase density of stars must be a function of $I_1$, $I_2$
$I_3$, and the resulting velocity distribution must be triaxial.

The condition (\ref{eq3.15}) is more general, when compared with $\partial^2\Phi
/\partial R\partial z$, and more easily fulfilled, because it contains an
arbitrary constant $z_0$, which we may choose according to our potential in
the region, where stellar motion take place. Because the triaxial velocity
distribution was observed also for spherical subsystems, it may be supposed
that $z_0$ is approximately constant over quite large regions. It would be
very interesting to determine $z_0$ from observational data. For that
purpose we may use an equation, resulting from the condition (\ref{eq3.15}), when
applied near to the galactic plane, and introducing Oort constants $A$ and $B$
and a quantity $C$. The formula has a form
\begin{equation}
(R^2+z_0^2) \frac{\rmd C^2}{ \rmd R} = -4R~[C^2+B(A-B)]. \label{eq3.16}
\end{equation}
Determination of $z_0$ on the basis of Eq.~(\ref{eq3.16}) is unfortunately quite
complicated because of complications in determining the radial gradient of
$C$. Below $z_0$ will be determined in another way.

In the velocity space the integral (\ref{eq3.13}) is symmetrical about
two orthogonal planes, intersecting along the line
$v_R=v_z=0$.  Because the integral $I_1$ is symmetrical about the same
line, and the integral $I_2$ does not contain $v_R$ and $v_z$, the
referred above planes are symmetry planes of the velocity
distribution. The inclination angle of these planes in respect to the
galactic plane is
\begin{equation}
\tan 2\alpha = \frac{2Rz}{ R^2+z_0^2-z^2}. \label{eq3.17}
\end{equation}
In the galactic plane and on the galactic axis the angle $\alpha$ is
zero or $\rm 90^o$, in general it is different from zero or
$\rm90^o$. Hence outside of the galactic plane the velocity ellipsoid
has some obliquity about the plane. To find the obliquity and related
kinematical effects is quite difficult within contemporary
observational possibilities.  From theoretical arguments the
inclination of the velocity ellipsoid about the galactic plane was
first derived by \citet{Eddington:1915} in his paper on the dynamics
of stellar systems, where he analysed the possibility of
Schwarzschild's triaxial velocity distribution in case of stationary
Galaxy. Later the same conclusion was made by
\citet{Chandrasekhar:1938}. But Chandrasekhar's paper contain some
errors. The method, applicable only in the case of nearly circular
orbits, was used at large distances from the galactic plane, where
orbits highly differ from circular.  Conclusions were made on the
stability of circular orbits, which are meaningless in the present
case.

As we mentioned, the quantity $z_0$ is probably constant over quite large
regions of space. In principle, it is not excluded that it is
approximately constant for the whole Galaxy. In this case the whole
Galactic structure can be treated in a way as $z_0$ being precisely
constant. This assumption may be valid as precisely as is the assumption
of the stationarity or axial symmetry. If we approximate $z_0$ with a
constant,  the condition (\ref{eq3.15}) may be handled as the differential equation
for the potential. The solution of the equation is known from general
dynamics \citep[][Sect. 152]{Whittaker:1904} and has the  form 
\begin{equation}
\Phi = \frac{\varphi (\xi_1)-\varphi (\xi_2) }{ \xi_1^2-\xi_2^2 } ,
\label{eq3.18}
\end{equation} 
where $\varphi$ is an arbitrary function, and $\xi_1^2$ and $\xi_2^2$ are
roots of the equation
\begin{equation}
\frac{R^2}{ \xi^2-z_0^2} + \frac{z^2}{ \xi^2} = 1. \label{eq3.19}
\end{equation}
The same result with slightly different method was derived by \citet{Eddington:1915}. 

\begin{figure*}[ht]
\centering
\includegraphics[width=100mm]{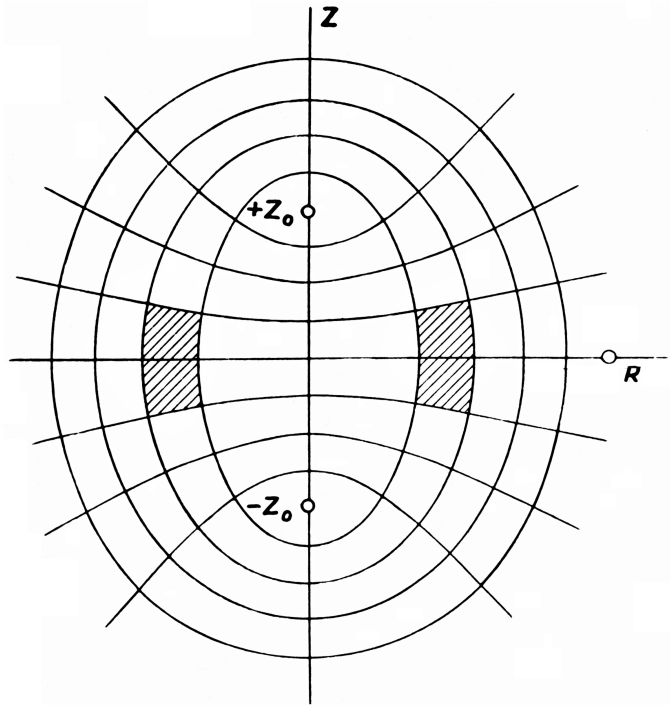}
\caption{Meridional sections of the surfaces of constant $\xi_1$ and $\xi_2$ coordinates.}
\label{fig3.3}
\end{figure*}

As it is seen from Eq.~(\ref{eq3.19}), the surfaces of constant $\xi_1$ and $\xi_2$ are
confocal second order surfaces of revolution. Their common foci lie on the
galactic axis at points $+z_0$ and $-z_0$. A set of meridional sections of
these surfaces is plotted in Fig.~\ref{fig3.3}. It results that tangent planes
of these surfaces are inclined in respect to galactic plane by the same
angle as it was for the symmetry planes of the velocity distribution, \ie 
by angle $\alpha$. Thus the velocity ellipsoid axis must be perpendicular
to these surfaces. Hence they are ``the main velocity surfaces'' according
to \citet{Eddington:1915}. At the same time it may be demonstrated that these
surfaces are enveloping surfaces of stellar orbits. The region where an
orbit lies is bounded by two ellipsoids and two  hyperboloids.  
(In Fig.~\ref{fig3.3} one section of this kind is streaked.)

Can the potential of the Galaxy correspond to the form derived above? Does
it give realistic results? To clarify these questions we choose very
simple expression for $\varphi$
\begin{equation}
\varphi = \Phi^0z_0\sqrt{\xi_0^2+\xi^2} , \label{eq3.20}
\end{equation}
where $\Phi^0$ and $\xi_0$ are constants. The expression for $\varphi$ was
chosen with having in mind that it is even and for large $\xi$ is
proportional to $\xi$. In this case the potential is symmetric about the
galactic plane, and for large distances from Galactic centre inversely
proportional to that distance, while Galactic mass multiplied by the
gravitational constant is $\Phi^0z_0$.

The resulting model of the Galaxy is quite realistic. Deriving
the expressions for the circular velocity and for the quantity $C$ and using
the available data for them we obtained\footnote{Due to changing
distance scale $z_0$ must be larger (the ratio $z_0/R_0\sim 0.5$ must
remain the same). The parameter $\xi_0$ must be reduced approximately twice
due to decreasing of $\epsilon$. [Later footnote.]}
$$\sqrt{\Phi^0} = 425 {\rm ~km/s}, ~~~~~ z_0 = 3.6 {\rm ~kpc}, ~~~~
\xi_0 = 0.6 {\rm ~kpc}. $$
The theoretical circular velocity curve, corresponding to these
values, is 
represented in Fig.~\ref{fig3.1} by dotted line. It does not represent the
observational data in detail,  but general trends are represented quite
well. The quantity $z_0$ results to be approximately twice less than the
solar distance from the Galactic centre. Accordingly the solar position is
plotted in Figs.~\ref{fig3.3} and \ref{fig3.4} (we discuss the later figure below). 

Because $\xi_0$ is significantly less than $z_0$, in first approximation we
may take $\xi_0=0$. In this case the potential in galactic plane is 
\begin{equation}
\Phi = \Phi^0 \frac{1}{\sqrt{1+ \frac{R^2}{ z_0^2}}}. \label{eq3.21}
\end{equation}

The equipotential surfaces in this case are spherical segments with
centres at foci of ``main velocity surfaces'', \ie  on galactic axis at
points $+z_0$ and $-z_0$. For positive $z$ the potential is like all the
Galactic mass in concentrated at southern focus of ``main meridional
surfaces'', for negative $z$ as it is concentrated at northern focus. In
Fig.~\ref{fig3.4} there are represented some meridional sections of the equipotential
surfaces. For $\xi_0=0$ they are given by dashed lines, for $\xi_0 =$
0.6~kpc by continuous lines. As we see, the different value of $x_0$ is 
significant near to the galactic plane.

\begin{figure*}[ht]
\centering
\includegraphics[width=100mm]{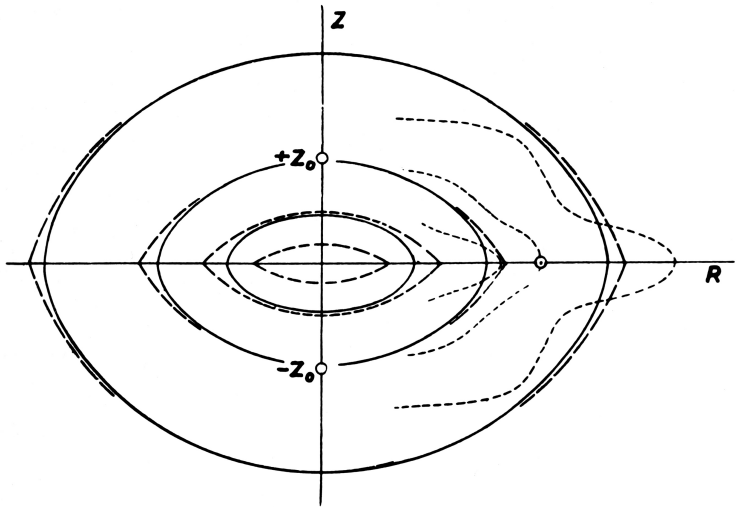}
\caption{Meridional sections of isodensity contours.}
\label{fig3.4}
\end{figure*}

With the exception of nearby to the galactic plane regions, the potential
depends mainly on the surface density of the Galaxy, but as we saw,  the
surface density is determined mainly by the circular velocity curve.
Because the theoretical circular velocity law (Fig.~\ref{fig3.1}) does not differ
highly from the observed one, we may expect that also the general form
of equipotential sections (Fig.~\ref{fig3.4}) is not far from reality. It is
interesting to compare these sections with isodensity contours of the
subsystem of short-period cepheids according to \citet{Kukarkin:1949}. Some of
these contours are given on Fig.~\ref{fig3.4} by short-dashed lines. When the
subsystem of short-period cepheids would not rotate, and would have the
spherical velocity distribution, the isodensity surfaces would coincide with
equipotential surfaces. In fact, as it is seen from the figure,  the
isodensity surfaces are significantly more flattened. The main reason of
the difference is probably the velocity distribution triaxiality. The
rotation of the system is secondary as being too slow.

Using of Poisson's equation demonstrates that the Galactic model,
corresponding to Eqs.~(\ref{eq3.18}) and (\ref{eq3.20}), has everywhere positive density. Hence
it lacks the main shortcoming of the Parenago's model,  which  gives at large
distances from the galactic plane negative densities. For small $\xi_0$ and
for not too large $|z|$ the formulae (\ref{eq3.18}) and (\ref{eq3.20}) give the following
expression for the density
\begin{equation}
4\pi G\rho = \frac{\Phi^0}{ z_0\xi_0} \frac{1}{ \sqrt{1+ \frac{R^2}{ z_0^2}}
\left( 1+ \frac{R^2}{ z_0^2}+ \frac{z^2}{ \xi_0^2}\right)^{3/2}}. \label{eq3.22}
\end{equation}
The isodensity surfaces from this formula are nearly spheroidal. The axial
ratio near to the Galactic centre is $\frac{2}{\sqrt{3}} \frac{\xi_0}{ z_0}$.
As the 
distance from the centre increases,  the axial ratio also increases
very slightly. 
 By taking for $\xi_0$ and $z_0$ the values from above,  the mean
axial ratio is approximately 1/5. For the surface densities in case of
small $\xi_0$ we derive the expression
\begin{equation}
2\pi G\delta = \frac{\Phi^0}{ z_0} \frac{1}{ \left(1+ \frac{R^2}{ z_0^2}
\right)^{3/2} }. \label{eq3.23}
\end{equation}

The surface density curve corresponding to the accepted values of $\Phi^0$
and $z_0$ is plotted on Fig.~\ref{fig3.2} by dashed line. In general
it represents  quite
well the observed curve. Only for large $R$ the surface density decreases
unrealistically slowly (proportional to $R^{-3}$). Too moderate decreasing
of the surface and volume densities at large distances is the main
shortcoming of the discussed model of the Galaxy. The shortcoming is
related, as can be demonstrated, not with a concrete form of the
function $\varphi$,  but with the expression for the potential (\ref{eq3.18}).

Despite of that shortcoming the formula (\ref{eq3.18}) gives in general quite
acceptable model of the Galaxy. We may hope that with some other form of
the function $\varphi$ we shall have even better agreement with the
reality. The main advantage of the Galactic model, basing on the Eq.~(\ref{eq3.18})
for the potential is, that it allows the third single-valued
integral of motion, enables to treat the triaxial velocity distribution
within the theory of the stationary Galaxy, and even now explains a series
of properties concerning the stellar motions. At the same time we do not want
to argue 
that the Galaxy is stationary,  and that all his properties are described
within the present theory. We have in mind the theory of the stationary
Galaxy as an approximate theory, explaining the basic facts and leaving
away details.

As a conclusion we like to say following. We saw that contemporary
observational data enable to make some conclusions about the gravitational
potential of the Galaxy and about its mass distribution. From the
motion of stars around the galactic centre and perpendicular to the
galactic plane, and from dynamical considerations, it is possible to
determine the surface density and the volume density of the Galaxy and its flatness. It
results, that  spherical subsystems of the Galaxy have a considerable
role in galactic dynamics. Further, it results that the Lindblad's
galactic rotational law is not confirmed by observations nor by
theory,\footnote{The rotational law (\ref{eq3.4}) does not have convincing
theoretical emphasis. But it is still quite satisfactory approximation for
the observed rotational law of flat Galactic subsystems in a quite large
range of $R$. To the same degree good approximation is the rotational law,
resulting from the theory of third quadratic integral. With suitable
chosen parameters both laws are similar to each other for a large
interval of $R$. [Later footnote.]}
and the escape velocity results to be significantly higher than it is
assumed usually. The fact of triaxial velocity distribution of stars
enables to make additional conclusions about the possible structure of the
Galaxy. By introducing the third integral of stellar motion we can
explain the triaxial velocity distribution within the theory of the
stationary Galaxy, while the resulting galactic model in general
corresponds to the real observations.
\vglue 3mm
\hfill May 1953

\vglue 5mm

{\bf\Large Appendices added in 1969}
\vglue 5mm
\section{A.  Detailing of Jeans equations}

The first Jeans equation in case of $z=0$ and taking into account
the triaxility of the velocity ellipsoid has the  form
\begin{equation}
\frac{V^2}{ R} - \left[ \frac{1}{ D} \frac{\partial (D\sigma^2_R)}{\partial R}
+ \frac{\sigma^2_R -\sigma^2_{\theta}}{ R} + \frac{\sigma^2_R - \sigma^2_z 
}{ R_c} \right]_{z=0} = - \frac{\partial\Phi}{\partial R}\bigg|_{z=0}
. \label{eq3.A1.1}
\end{equation}
When compared with the original Jeans equation an additional (last)
term in square brackets appears. Taking into consideration the
Lindblad-Oort equation, having the  form
\begin{equation}
1- \frac{\sigma^2_{\theta}}{\sigma^2_R} = - \frac{1}{2} \frac{\rmd \ln\omega}{
\rmd \ln R} , \label{eq3.A1.2}
\end{equation}
we derive Eqs.~(\ref{eq3.1})--(\ref{eq3.3}).

The distance to the point of convergence $R_c$ is related with the
``obliquity'' $\alpha$ of the velocity ellipsoid for $z\ne 0$ 
evidently by
\begin{equation}
\frac{1}{ R_c} = \frac{\partial\alpha}{\partial z}\bigg|_{z=0}. \label{eq3.A1.3}
\end{equation}

\section{B.  Schwarzschild velocity distribution and the circular
velocity law}

Let us assume for the phase density
\begin{equation}
\Psi = \mathrm{const} ~\cdot e^{kf(I_1,I_2,I_3)} , \label{eq3.A2.1}
\end{equation}
where
\begin{equation}
I_1 = v^2_R+v^2_{\theta}+v^2_z-2\Phi , ~~ I_2=Rv_{\theta}, ~~
I_3=v^2_z-2(\Phi -\Phi_0 ) \label{eq3.A2.2}
\end{equation}
are the integral of energy, the integral of momentum and the third
integral in a simplified form. Let us have in velocity space for $z=0$
and for some $R$ a maximum of $f$ at values of $v_{\theta}=V,$ $v_R=v_z
=0$, \ie  for values of integrals
\begin{equation}
I^0_1=V^2-2\Phi_{z=0}, ~~ I^0_2=RV, ~~ I^0_3 =0. \label{eq3.A2.3}
\end{equation}
Expanding $f$ into series in powers of $I_1-I^0_1,$
$I_2-I^0_2,$ $I_3$, we find
\begin{equation}
f = f_0 - f_Rv_R^2 - f_zv_z^2 - f_{\theta}(v_{\theta}-V)^2+... ,
\label{eq3.A2.4}
\end{equation} 
where
$$
f_R=-\left( \frac{\partial f}{\partial I_1}\right)_0, ~~~~~~
f_z = -\left( \frac{\partial f}{\partial I_1}\right)_0 - \left( \frac{\partial
f}{\partial I_3}\right)_0, $$
\begin{equation}
f_{\theta} = - \left( \frac{\partial f}{\partial I_1}\right)_0 - 2
\left( \frac{\partial^2 f}{\partial I_1^2} \right)_0V^2 -2\left(
\frac{\partial^2 f}{\partial I_1\partial I_2} \right)_0 RV - \frac{1}{ 2}
\left( \frac{\partial^2 f}{\partial I^2_2} \right)_0 R^2 , \label{eq3.A2.5}
\end{equation}
(index 0 signifies the value for $I_1=I^0_1,$ $I_2=I^0_2,$ $I_3=0$). The
velocity $V$ is determined by equation
\begin{equation}
2\left( \frac{\partial f}{\partial I_1}\right)_0 V + \left( \frac{\partial f
}{\partial I_2}\right)_0 R = 0. \label{eq3.A2.6}
\end{equation}

For sufficiently large $k$ (flat subsystem) the velocity distribution
is nearly Schwarzschild distribution, because more higher terms
in the expansion of $f$ begin to influence only when phase
density $\Psi$ is nearly zero. Besides, for large $k$ the velocity
$V$ nearly equals with the centroid velocity.

To keep the density and the velocity dispersion for large $k$ from
to strong dependence on $R$, it is needed to demand that the expression
$$\frac{\partial f}{\partial I_1} \frac{\partial I^0_1}{\partial R}+
\frac{\partial f}{\partial I_2} \frac{\partial I^0_2}{\partial R}$$
has a very small value in the vicinity of $I_1=I^0_1(R)$, $I_2=I^0_2(R),$
$I_3=0$ (where the phase density $\Psi$ is still large). But in this
case with a quite high precision
\begin{equation}
\frac{V}{R}= \frac{1}{2} \frac{\rmd I^0_1}{ \rmd I_2^0}, \label{eq3.A2.7}
\end{equation}
and we have
\begin{equation}
V^2=-R \frac{\partial\Phi}{\partial R}\bigg|_{z=0} , \label{eq3.A2.8}
\end{equation}
\ie  $V$ is (nearly) equal to the circular velocity. This
determines the ratio of $\partial f/\partial I_1$ to $\partial
f/\partial I_2$ in the vicinity of $I_1=I_1^0(R)$, $I_2=I^0_2(R)$,
$I_3=0$ and therefore also the ratio $f_{\theta}/f_R$ (it results the
Lindblad-Oort formula for the ratio $\sigma_R /\sigma_{\theta}$). But
the velocity distribution remains nearly Schwarzschild for any
dependence of circular velocity on $R$.

If to take instead of $e^{kf}$ with large $k$ some more steep function of $f$,
the Schwarzschild distribution of velocities replaces with general
ellipsoidal distribution.

%% file: chapter04.tex
\chapter[The third integral and the dynamics of stationary Galaxy]{The
  third integral of stellar motion and the dynamics of  the stationary
  Galaxy\footnote{\footnotetext ~~Tartu  Astron. Observatory
    Publications, vol. 32, pp. 332-368, 1953.} } 
  
It is known from observations that the velocity distribution of stars
is approximately ellipsoidal. The velocity ellipsoid is triaxial with the
long axis directed approximately along the galactic radius, the
intermediate axis is in direction of galactic rotation, and the short axis is
perpendicular to the galactic plane. Triaxial velocity distribution has
been observed for flat as well as for intermediate and spheroidal
subsystems of the Galaxy.

The ellipsoidal velocity distribution was explained within the theory
of a stationary rotating Galaxy \citep[see e.g.][]{Parenago:1946,
  Ogorodnikov:1948}. This theory also explains that the axis of the
velocity ellipsoid laying in direction of galactic rotation is shorter
than the axis parallel to galactic radius. Axial ratio which results
from the theory nearly coincides with the observed one.

However, the theory of a stationary rotating Galaxy meets serious
difficulties while explaining the triaxial shape of the velocity distribution.
Taking into account only two integrals of stellar motion -- the
energy integral and the angular momentum integral -- one obtains that the velocity
distribution resulting from the theory has a form of ellipsoid of revolution
around the axis laying in direction of galactic rotation.

The discrepancy between the theory and the observations can be removed by
introducing a third integral of motion. A form of the third integral
as the energy integral along the $z$-axis was proposed by \citet{Lindblad:1933}.
Unfortunately, this ``integral'' is with sufficient precision 
an integral only for nearly circular orbits. For this reason it can not be used to
explain the triaxial velocity distribution of objects of intermediate and spherical
subsystem. For these subsystems a third integral of some other form needs to be used.

The aim of the present paper is to describe a theory of third
integral of this kind and to apply it to Galactic dynamics. This is the first part of
our study. Here we discuss the foundations of the dynamics of the
stationary Galaxy, the integrals of motion and the third integral. Quite a
considerable part of the paper does not give anything new. However, because
of some lack of clarity in the very foundations of the dynamics of the
stationary Galaxy, it seems suitable to discuss these foundations,  beginning
with the equation of continuity and the Jeans theorem.

\section{~}

The spatial-kinematical  structure of the Galaxy or of some of its
subsystems is given by the density of stars $\Psi$ in six-dimensional
phase space, where coordinates are three orthogonal coordinates of
ordinary space $x,$ $y$, $z$, and three orthogonal velocity components
$u=\dot x$, $v=\dot y$, $w=\dot z$ (points designates time derivatives). For
given $x$, $y$, $z$ the phase density
$\Psi$ is the stellar density in velocity space with coordinates
$u$, $v$, $w$. The stellar density in ordinary space results after the
integration of $\Psi$ over the velocity space.

Generalised velocity vector in six-dimensional phase space is a vector
with components $u$, $v$, $w$, $\dot u$, $\dot v$, $\dot w$.
Divergence of this vector is zero, because in phase space $u$, $v$, $w$
are independent of $x$, $y$, $z$ and the acceleration components $\dot u$,
$\dot v$, $\dot w$ are independent of $u,$ $v,$ $w.$ For this reason the
volume occupied by an element of ``stellar medium'' in phase space remains
constant (Liouville's theorem) and has the properties of
incompressible fluid. Hence, the continuity equation in six-dimensional
space is
\begin{equation}
\frac{D\Psi}{Dt}=0 , \label{eq4.1}
\end{equation}
where $t$ is time and $D/Dt$ is the Stokes operator showing, as it is
known, variation per unit of time at a point moving in phase space
together with a particle (with a star). In long form the Stokes operator is
\begin{equation}
\frac{D}{Dt}= \frac{\partial}{\partial t} + u \frac{\partial}{\partial x} +
v\frac{\partial}{\partial y} + w\frac{\partial}{\partial z} + \dot u\frac{\partial}{
\partial u} + \dot v\frac{\partial}{\partial v} + \dot w\frac{\partial}{
\partial w} , \label{eq4.2}
\end{equation}
and if acceleration is caused by the potential $\Phi$, then
\begin{equation}
\dot u =\frac{\partial\Phi}{\partial x},~~ \dot v=\frac{\partial\Phi}{
\partial y}, ~~ \dot w= \frac{\partial\Phi}{\partial z}. \label{eq4.3}
\end{equation}
In our case $\Phi$ is the gravitational potential of the Galaxy.

In Galactic dynamics the cylindrical coordinates $R$, $\theta$, $z$,
$v_R$, $v_{\theta}$, $v_z$ are usually used, where $R$ is the distance from
galactic axis, $\theta$ is the galactocentric longitude, $z$ is the distance
from galactic plane and $v_R=\dot R$, $v_{\theta}=R\dot\theta$, $v_z=\dot
z$ are the corresponding orthogonal velocity components. In these coordinates
the Stokes operator has the form
\begin{equation}
\frac{D}{Dt} = \frac{\partial}{\partial t} + v_R\frac{\partial}{\partial R} +
v_{\theta} \frac{\partial}{R\partial\theta} + v_z \frac{\partial}{\partial z}
+ \dot v_R \frac{\partial}{\partial v_R} + \dot v_{\theta} \frac{\partial}{
\partial v_{\theta}} + \dot v_z \frac{\partial}{\partial v_z} , \label{eq4.4}
\end{equation}
and
\begin{equation}
\dot v_R = \frac{\partial\Phi}{\partial R} + \frac{v_{\theta}^2}{R} , ~~~~
\dot v_{\theta} = \frac{\partial\Phi}{R\partial\theta} - \frac{v_Rv_{\theta}}{
R}, ~~~~ \dot v_z = \frac{\partial\Phi}{\partial z}. \label{eq4.5}
\end{equation}

The six-dimensional continuity equation (\ref{eq4.1}) or the Liouville's equation,
as it is often called, is the basic equation of Galactic dynamics. This
equation is not fully precise. It takes into account only continuous
motions of stars in phase space in smoothed (mean) gravitational field of
the Galaxy,  and ignores the mixing of stars in phase space due to
encounters. However, within sufficient precision Eq.~(\ref{eq4.1}) may be handled to
be valid, because of smallness of the role of encounters in Galactic 
dynamics.\footnote{Irregularities in Galactic structure make the
irregular gravitational forces to be more powerful than interactions
between individual stars. So the resulting additional term in Liouville's
equation is sufficiently small and must be taken into account only while
studying Galactic evolution. [Later footnote.]}

According to Eq.~(\ref{eq4.1}) the stellar density $\Psi$ in phase space remains
constant in the vicinity of every star, \ie  the points $\Psi =\mathrm{const}$ move
in phase space together with stars. This means that the phase density is
the first integral of motion. But every first integral of motion is in
general a function of some six independent first integrals
of motion
\begin{equation}
\Psi =\Psi (I_1,I_2,...,I_6), \label{eq4.6}
\end{equation}
where
\begin{equation}
I_i=I_i(t,x,y,z,u,v,w), ~~i=1,2,...,6 \label{eq4.7}
\end{equation}
are these six integrals. As equations $I_i=\mathrm{const}$ determine 
six (in general) five-dimensional hyper-surfaces moving in phase space, the
intersection of which gives us the  position of the star, Eq.~(\ref{eq4.6})
signifies that $\Psi$ is constant in phase points moving with stars.
Therefore Eq.~(\ref{eq4.6}) expresses exactly the same as Eq.~(\ref{eq4.1}).

Equation (\ref{eq4.6}) may be obtained also by solving formally Eq.~(\ref{eq4.1}) as a first
order differential equation,  containing partial derivatives of $\Psi$. The
corresponding set of ordinary differential equations is
\begin{equation}
\frac{\rmd x}{u} = \frac{\rmd y}{v} = \frac{\rmd z}{w} = \frac{\rmd u}{ \frac{\partial\Phi}{\partial
x}} = \frac{\rmd v}{ \frac{\partial\Phi}{\partial y}} = \frac{\rmd w}{ \frac{\partial\Phi}{
\partial z}} = \rmd t. \label{eq4.8}
\end{equation}
These equations are just the equations of motion. Their first integrals
are the integrals of motion (Eq.~\ref{eq4.7}), and $\Psi$ must be an arbitrary
function of these integrals. Evidently, $\Psi$ must be single-valued
non-negative function of $I_i$, as only such function can give physical meaning to
the phase density. For the same reason we can use only single-valued functions
of $I_i$.

If the Galactic potential is given as a function of $x$, $y$, $z$ and $t$,
all six integrals $I_i$ can be calculated via integration of Eq.~(\ref{eq4.8}).
Then Eq.~(\ref{eq4.6}) gives us all the sets of possible (and generally
varying in time) spatial-kinematical  structures of Galactic subsystems.
For the Galaxy as a whole the amount of possible configurations is
significantly reduced, because in addition to Eq.~(\ref{eq4.1}) the phase density must
satisfy the Poisson's equation. If $\Psi$ signifies the total phase density,
the mass density $\rho$ in three-dimensional space is calculated as
\begin{equation}
\rho = \iiint\Psi \rmd u \rmd v \rmd w. \label{eq4.9}
\end{equation}
Density $\rho$ is related to potential $\Phi$ via Poisson's equation
\begin{equation}
4\pi G\rho = -\nabla^2\Phi , \label{eq4.10}
\end{equation}
where $G$ is the gravitational constant and
\begin{equation}
\nabla^2= \frac{\partial^2}{\partial x^2} + \frac{\partial^2}{\partial y^2}
+ \frac{\partial^2}{\partial z^2} = \frac{\partial^2}{\partial R^2} +
\frac{\partial}{R^2\partial R} + \frac{\partial^2}{R^2\partial\theta^2} +
\frac{\partial^2}{\partial z^2} \label{eq4.11}
\end{equation}
is Laplace's operator.

Poisson's equation poses restrictions not only on the phase density $\Psi$, but also
on the potential $\Phi$, which can not be given arbitrarily. At first,
the potential can not permit negative 
densities. In addition, the potential must satisfy the boundary conditions at
infinity, related to the finiteness of system dimensions: at
very large distances from the centre of a system the potential must
approach spherical symmetry and be inversely proportional to the
distance from the centre.

\section{~}

The result containing in Eqs.~(\ref{eq4.6}), (\ref{eq4.9}) and (\ref{eq4.10}) is known as the Jeans
theorem \citep{Jeans:1915, Smart:1939}. As we are
interested in the theory of the stationary Galaxy, we will discuss the Jeans
theorem for the potential and the phase density, which are independent of time:
\begin{equation}
\frac{\partial\Phi}{\partial t}=0; ~~~\frac{\partial\Psi}{\partial t} = 0.
\label{eq4.12}
\end{equation}

If the potential is stationary, then all the stars moving through a given point
in phase space must have the same constant phase trajectory in that point. The orbits 
in ordinary and in phase space may be closed or unclosed depending on
the potential as a function of coordinates. In general the orbits are unclosed and
tightly fill with their loops two- or three-dimensional regions in
ordinary space, and two- or more-dimensional hyper-surfaces in phase space.
As the regions of constant phase density move in phase space together
with stars, in case of stationarity the phase density must be constant
along the phase trajectory (Poincar\'e theorem). When the phase trajectory
is unclosed, the phase density must be constant on the whole hyper-surface
filled with the trajectory. But for a stationary potential the phase orbits
are determined by five independent of each other and of time first
integrals of motion, and the hyper-surfaces filled by unclosed orbits 
by four or less integrals of that kind. Therefore,
the first part of the Jeans theorem in case of stationarity has the form
\begin{equation}
\Psi = \Psi (J_1,J_2,... J_n), \label{eq4.13}
\end{equation}
where
\begin{equation}
J_j=J_j(x,y,z,u,v,w) , ~~j=1,2,... n\le 5 \label{eq4.14}
\end{equation}
are the mentioned integrals. If the trajectory of a star is closed and the number of
integrals $J_j$ is five, their values determine five
hyper-surfaces in the phase space, the intersection of which gives the phase
trajectory of a star. If the trajectory in not closed and the number of integrals is
less than five, the intersection of hyper-surfaces give us a hyper-surface
filled with the phase trajectory.

The integrals $J_j$ are not independent of the integrals $I_i$. They are some functions
of these integrals, namely the functions which exclude the dependence of
the resulting integral of time. When potential is given, the integrals
$J_j$ as well as the integrals $I_i$ can be found after the integration of Eq.~(\ref{eq4.8}).
Besides, as $J_j$ are independent of time, the last term in Eq.~(\ref{eq4.8})
must be rejected. The equations will became the differential equations of the
phase trajectory. As a phase trajectory is determined by five
hyper-surfaces, in case of any stationary potential one obtains five
mutually and time-independent integrals. From this fact results that
when a stellar trajectory is not closed, besides the integrals $J_j$ there must
exist additional independent (and independent of time) integrals
\begin{equation}
J_k=J_k(x,y,z,u,v,w), ~~k=n+1,n+2,... 5, \label{eq4.15}
\end{equation}
which determine together with the integrals $J_j$ five hyper-surfaces, giving a
phase trajectory. These integrals are also some functions of the integrals
$I_i$. But they are principally different from the integrals $J_j$ and from the
integrals $I_i$ -- the integrals $J_j$ and $I_i$ are single-valued, but
the integrals $J_k$ are infinitely multiple-valued. Their infinitely 
multiple-valued nature is related to the fact that the phase trajectory is already 
uniquely given by values of $J_j$ as it fills all the hyper-surface
determined by these integrals, and hence other trajectories can not exist there. 
It results now that for the determination of a phase trajectory
it is completely unimportant which values are assigned to $J_k$. Therefore,
although in case of stationarity there always exist five independent and
time-independent integrals, for every unclosed stellar trajectory some of these
integrals are infinitely multiple-valued. These integrals do not
determine discriminating phase trajectories and are not included
in the expression of the phase density. Their inclusion into the phase
density would make it also infinitely
multiple-valued, which does not have physical meaning.\footnote{Following A.
Wintner (The Analytical Foundations of Celestial Mechanics, Princeton,
1941) it is suitable to call the integrals $I_i$ -- non-conservative (in
general) integrals, the integrals $J_j$ -- conservative isolating integrals, and
the integrals $J_k$ -- conservative non-isolating integrals. The integrals
$J_j$ are single-valued or can be reduced into a single-valued form. The
integrals $J_k$ can not be reduced into a single-valued form, they are
infinitely multiple-valued by their nature. The integrals $J_j$ are
like isolating a phase trajectory by determining a hyper-surface where
they are. The integrals $J_k$ do not isolate a trajectory -- they fill
densely all the hyper-surface determined by the integrals $J_j$ (or a part
of it with nonzero measure) and are ergodic. To be more precise, a
trajectory is quasi-ergodic, \ie  it passes not through every point, but
infinitely near to every point of the hyper-surface (or a part of it).
Therefore, in the hyper-surface (or in part of it) are precise not
one but infinite number of trajectories. But any of them are infinitely
nearby to any other, so we have the same trajectory. [Later
footnote.]}

Single-valued and infinitely multiple-valued integrals are well illustrated in
case of general quasi-elastic force field reviewed in detail in the book
by \citet{Pahlen:1937}. When the periods of harmonic oscillations of a particle
along three coordinates are not commensurable, only three independent
integrals are single-valued -- the energy integrals along these three coordinates.
Two integrals relating the phases of oscillations are infinitely
multiple-valued. The trajectory of a particle is unclosed and fills 
rectangular box in ordinary space, but three-dimensional
hyper-surface determined by the energy integrals in phase-space. In the expression of the
phase density there are only the energy integrals, the remaining two
integrals are not included due to their infinitely multiple-valued nature.

The fact that in case of stationarity only single-valued integrals are
included in the expression of the phase density was mentioned several
times in the literature on stellar dynamics. One may refer to the
book by \citet{Pahlen:1937}, also to the paper by \citet{Ogorodnikov:1948},
etc. However there still exists some misunderstanding of  the use of Jeans
theorem. For example \citet{Chandrasekhar:1942} argues that the
physical foundations for ignoring some integrals in the expression of
phase density remain unclear, \citet{Parenago:1946} argues that the fact of 
ignoring some integrals is related to not knowing their precise
values in particular cases. In reality, for unclosed phase trajectories
some integrals are infinitely multiple-valued and do not enter
the expression of phase density by principle.\footnote{A later paper
  about only isolating integrals as arguments of phase density is by
  \citet{Lynden-Bell:1962}. [Later footnote.]}

\section{~}

In order to further clarify the number of integrals in the expression
of the phase density in case of stationarity, we review the number of
single-valued independent integrals $J_j$ for different cases of stationary
potential.

Let us begin with two particular cases of spherically symmetrical
potential: the gravitational potential of a point mass and the potential
of spherically symmetrical quasi-elastic force field. In first case the
potential is proportional to $r^{-1},$ where $r$ is the distance from the
symmetry centre of the potential. In second case the potential is proportional
to $-r^2 + \mathrm{const}$. Both cases belong to these rare examples when
particle's trajectory is closed in both ordinary and phase spaces, and
all five time-independent integrals are single-valued. In ordinary space
the trajectory is an ellipse, lying in the plane which goes through the centre
of symmetry of the potential. In first case the centre of symmetry coincides with one
of the foci of ellipse, in second case with the centre of ellipse. The
scale, the form and the position of ellipse in space are determined by
five parameters -- orbit elements. These parameters or elements are exactly
the values of five integrals.

Let us assume now that potential remains spherically symmetric but
its proportionality to $r^{-1}$ or $-r^2 +\mathrm{const}$ is broken. The trajectory
of a particle in ordinary space remains a planar curve in the plane going
through the centre of symmetry of the potential. But now there appears an
effect called the shift of the line of apsid. The trajectory is unclosed
both in ordinary and phase spaces. In phase space it fills some
two-dimensional hyper-surface, in ordinary space a ring-like region,
the boarders of which are circles with centres in the centre of symmetry of 
the potential. The trajectory is given now by four parameters
determining the inner and outer radii and the position in space of that ring-like
region. Correspondingly the number of single-valued integrals is four. These
integrals are the energy integral and three angular momentum integrals (or
some other integrals as single-valued functions of them). The fifth integral
is infinitely multiple-valued due to the multiple-valued nature of the fifth
parameter -- ``the pericentre's longitude''.

Let us assume next that also spherical symmetry of the potential is broken
but there remains symmetry about some axis. In that case the particle's
trajectory remains unclosed, but in ordinary space it does not lie in a
plane any more. In addition to the shift of the line of apsid there appears a
phenomenon,  that due to the  analogy with celestial mechanics is called ``the shift of the
line of nodes'' and ``secular changes of orbit's inclination and
eccentricity''. In phase space the trajectory fills four-dimensional
hyper-surface, in ordinary space three-dimensional region -- a tube with
symmetry axis coinciding with the symmetry axis of the potential. In general the
trajectory is given only by two integrals -- the energy integral and the integral of 
angular momentum about the potential's symmetry axis. Three remaining
integrals are infinitely multiple-valued.\footnote{Authors suggestion
 that in general case of axisymmetry there exist only two isolating
integrals was somewhat unreasonable. It is only possible to state that
there do not exist other analytical isolating integrals (Poincar\'e's
theory), but nothing about isolating integrals in general. The
same is valid for the case when a stationary potential has no
symmetry whatsoever. And again it is only possible to state that beside the energy
integral there do not exist other analytical isolating integrals. [Later
footnote.]}

Finally we may assume that all the potential's symmetry is broken. In this
case only one time-independent integral can be indicated -- the energy
integral. In phase space particle's trajectory fills all the five-dimensional
hyper-surface determined by the value of energy integral. In ordinary
space it fills three dimensional region, limited by the equipotential
surface where the particle's velocity is zero.

Examples discussed above surely do not exhaust all the classes of
stationary potential allowing some number of independent of each other and
of time integrals. But they evidently demonstrate that the number of these
integrals is in general less than five, and that it 
decreases as the potential becomes more general. Therefore, the
more general is the potential (the less restrictions are posed on the
potential), the smaller is the number of integrals appearing in the
expression of the stationary phase density.

Reduction in the number of integrals in the expressions of the phase
density for more general potentials indicates that for more general
potential the phase density is more restricted; more restricted are also
the resulting velocity distribution and the mass density. But the
potential and the mass density are related via Poisson's equation. For
this reason too general potential or phase density must be come in
contradiction with itself. This is exactly the case. Most general
expressions for the phase density can be derived from five single-valued
integrals of motion. But as we saw this may occur only for very special
class of potentials which not only exclude all the general nature of phase
density (Poisson's equation must remain valid) but also are completely
unrealistic at least for a stellar system as a whole. On the other side
it is possible to assume some very general potential without any symmetry,
where only one single-valued time-independent integral of motion exists --
the energy integral. If the expression for the phase density contains only
the energy integral, the mass density $\rho$ is a function of only
potential $\Phi$, because the energy integral depends on $x,$ $y,$ $z$
only via $\Phi$. But in this case the solution of the Poisson's equation
is equivalent to the problem of finding an equilibrium configuration of
static gravitating fluid. For a finite system the potential must be
spherically symmetric. But the spherically symmetric potential enables not
one, but four independent integrals of motion.

Therefore, the application of the Poisson's equation\footnote{
Similar discussion may be found in \citet{Jeans:1915}. [Later
footnote.]} enables us to conclude that there can not exist a stationary
stellar system, the potential of which permits the existence of five or one
single-valued time-independent integrals of motion.\footnote{To be
precise, that statement is not correct. It is possible to construct a
model of spherical stellar system with finite radius and constant density.
Particular example of such model is homogeneous generalised-polytropic
model. In that case the potential permits all five
conservative isolating integrals of motion. But homogeneous model, surely,
has no relation with real stellar systems. [Later footnote.]} The
remaining cases of two, three or four time-independent single-valued
integrals are generally not in contradiction with the Poisson's equation.

\section{~}

To study all possible forms of stationary potentials and
time-independent single-valued integrals of motion $J_j$, all single-valued solutions 
of the following equation need to be found:
\begin{equation}
\frac{DJ_j}{ Dt}=0, \label{eq4.16}
\end{equation}
while
\begin{equation}
\frac{\partial J_j}{\partial t}=0, ~~~\frac{\partial\Phi}{\partial t}=0.
\label{eq4.17}
\end{equation}
Equation (\ref{eq4.16}) says that integral of motion does not vary at
a point, 
moving in phase space together with a particle.  Eq.~(\ref{eq4.17}) means that
integral $J_j$ and potential $\Phi$ are assumed to be independent of time.
In its long form Eq.~(\ref{eq4.16}) is a partial differential equation of the first
order for $J_j$ and $\Phi$ (c.f. Eqs.~(\ref{eq4.2}) and (\ref{eq4.3})).

We may analyse possible potentials and integrals by choosing a potential
and via Eq.~(\ref{eq4.16}) trying to find the integrals. In this case our problem
reduces to the integration of differential equations for a phase
trajectory. But we may also solve the problem by choosing integrals as
functions of velocity components $u,$ $v,$ $w$,  and trying to find the
potential and the integrals as functions of coordinates $x,$ $y,$ $z$. The
integrals $J_j$ may be given in form of polynomial in powers of $u,$
$v,$ $w$. In this case we try to find the potential and polynomial
coefficients as functions of $x$, $y,$ $z$.

For a stationary stellar system not all solutions of Eq.~(\ref{eq4.16}) are
interesting, only the solutions giving the potential in
sufficiently general form,  and ensuring in this way an enough general
theory. In order to find these potentials,  the second method of solving 
Eq.~(\ref{eq4.16}) has to be applied. In this case for a more general
integral more restrictions result for the potential. Assuming an integral
to be in the form of polynomial in powers of $u,$ $v,$ $z$, the more general
integral will have a higher polynomial order and larger number of coefficients (in
case of some restrictions the number of independent coefficients is
essential). But for larger number of coefficients from solving
Eq.~(\ref{eq4.16}) there will appear also larger number of additional restrictions for the
potential. For this reason, in order to have a potential in most general
form,  we must start not from more general form of integrals but just the
opposite way -- from integrals having the simplest structure: linear or
quadratic in respect to $u,$ $v,$ $w$.

An integral satisfying Eq.~(\ref{eq4.16}) without posing any restrictions on a stationary
potential is the energy integral
\begin{equation}
J_1=u^2+v^2+w^2-2\Phi. \label{eq4.18}
\end{equation}
If the phase density is a function of only the energy integral, the
resulting velocity distribution is spherically symmetric. The centroid
velocity is zero,  and the differential motion of centroids is thus
absent. However as we saw, the potential of a stationary Galaxy must
permit at least one more time-independent single-valued integral of motion.
To have as little restrictions on the potential as possible,  one has to give 
the  form of the integral in a simplest way, \ie linear in
respect to the velocity components
\begin{equation}
J_2=au+bv+cw. \label{eq4.19}
\end{equation}
Together with the energy integral, this integral permits more general
form of velocity distribution,  and in particular the motion of the centroids. The
resulting restrictions on the potential are not large.

To find the restrictions, resulting from the integral (\ref{eq4.19})
and the coefficients $a$, $b$, $c$ as functions of $x$, $y$, $z$, one has to 
substitute Eq.~(\ref{eq4.19}) into Eq.~(\ref{eq4.16}) and to demand that the
resulting equation turns to identity. In this way we have
\begin{equation}
\frac{\partial a}{\partial x} = \frac{\partial b}{\partial y}=
\frac{\partial c}{\partial z} = \frac{\partial a}{\partial y} + \frac{\partial b}{
\partial x} = \frac{\partial a}{\partial z} + \frac{\partial c}{\partial x} =
\frac{\partial b}{\partial z} + \frac{\partial c}{\partial y} = 0; \label{eq4.20}
\end{equation}
\begin{equation}
a\frac{\partial\Phi}{\partial x} + b \frac{\partial\Phi}{\partial y} +
c \frac{\partial\Phi}{\partial z} = 0. \label{eq4.21}
\end{equation}

After differentiating Eq.~(\ref{eq4.20}) with respect to $x$, $y$, or $z$, we have a
set of equations, telling us that all second derivatives of
coefficients $a$, $b$, $c$ are zero. Therefore, $a$, $b$, $c$ are linear
functions of $x$, $y$, $z$. From Eq.~(\ref{eq4.20}) we have also additional
restrictions, which give us after suitable change of coordinates
\begin{equation}
a=ky, ~~b=-kx, ~~c=\mathrm{const}, \label{eq4.22}
\end{equation}
where $k$ is constant. The integral $J_2$ is now in form
\begin{equation}
J_2=k(yu-xv)+cw. \label{eq4.23}
\end{equation}

According to Eq.~(\ref{eq4.22}) the vector lines of the vector $(a,b,c)$ are spirals
with step, equal to $2\pi c/k$ and the axis coinciding with the $z$-axis. It results 
from Eq.~(\ref{eq4.21}) that the potential is constant along the vector lines
of the vector $(a,b,c)$. Therefore, we conclude that the potential,
allowing two independent single-valued time-independent
integrals, has in general a spiral symmetry. However, this class of
symmetry is incompatible with the finite dimensions of the stellar system.
For a finite stellar system only a particular case of spiral symmetry
is possible -- the axial symmetry. Hence in the most general case a
stellar system must have an axially symmetrical potential. In case of axial
symmetry $c=0$ and the integral $J_2$ turns into the angular momentum integral
about $z$-axis, \ie  around symmetry axis of the potential
\begin{equation}
J_2=yu-xv \label{eq4.24}
\end{equation}
($k$ is taken to be unit).

Conclusions about the spiral symmetry of the potential in most general case and
about the axial symmetry of the potential for finite stellar systems are
known as Chandrasekhar's theorem \citep{Chandrasekhar:1939a,
  Chandrasekhar:1942}. Chandrasekhar derived these 
conclusions in a way,  somewhat different from used here. He started not from
demanding the most general form of potential (minimum restrictions), but demanding 
the potential to permit differential centroid motion.
These two demands do not necessarily give same results, as there
exist potentials, allowing differential centroid motion but having not
spiral nor axial symmetry. An example of this kind of potential is the
potential of a general quasi-elastic force field, enabling harmonic
oscillations of a particle along three coordinates with commensurable (but
not equal) periods. However, Chandrasekhar did not discuss the general
problem of the possibility of differential centroid motion, but related it
only with linear terms in the expression of the integral of motion, later being a
general quadratic polynomial in powers of $u$, $v$, $w$. As a result he
derived the same equations as Eqs.~(\ref{eq4.20}) and
(\ref{eq4.21}). In addition to these equations several 
other equations result due to quadratic terms and a free term in the integral,
posing additional restrictions on the potential. These restrictions vanish
only in case where quadratic integral is a function of energy and
linear integrals.

\section{~}

Therefore, the most general potential of a stationary stellar
system has the axial symmetry. By taking for the axis of cylindrical
system of coordinates the symmetry axis of the potential, the condition for
potential axial symmetry can be written in form
\begin{equation}
\frac{\partial\Phi}{\partial\theta}=0.\label{eq4.25}
\end{equation}
Without other restrictions in addition to the stationarity and the axial
symmetry, the only time-independent single-valued integrals are the energy
integral $J_1$ and the angular momentum integral $J_2$, and according to
Eq.~(\ref{eq4.13})
\begin{equation}
\Psi = \Psi (J_1,J_2), \label{eq4.26}
\end{equation}
while in cylindrical coordinates
\begin{equation}
J_1=v_R^2+_{\theta}^2+v_z^2 - 2\Phi; \label{eq4.27}
\end{equation}
\begin{equation}
J_2=Rv_{\theta}. \label{eq4.28}
\end{equation}
From Eq.~(\ref{eq4.26}) it follows that also the phase density must be axially
symmetric
\begin{equation}
\frac{\partial\Psi}{\partial\theta}=0. \label{eq4.29}
\end{equation}
This condition insures the axial symmetry of the mass density, otherwise
the Poisson's equation will not be satisfied.

The result contained in Eq.~(\ref{eq4.26}) indicates the first part of the Jeans
theorem for axisymmetric stationary systems. Often the result is simply called Jeans theorem.

Jeans theorem in its referred narrow meaning remained in fact the basis
for the whole contemporary dynamics of the stationary rotating Galaxy. On
the basis of Eq.~(\ref{eq4.26}) it was possible to explain most essential
characteristics of stellar motion -- ellipsoidal velocity
distribution, galactic rotation, relation between ellipsoidal velocity
distribution and galactic rotation. However, as it was indicated in our
introduction, the theory of a stationary rotating Galaxy, being based only on two
integrals of motion, is in contradiction with observations in one essential
point. According to Eqs.~(\ref{eq4.26})--(\ref{eq4.28}) the velocity distribution must have
axial symmetry about $v_{\theta}$ axis, \ie  the axis in direction
of which according to our formulae occurs the motion of centroids, or in other
words Galactic rotation. In reality this kind of symmetry is not observed.
The velocity distribution is not biaxial but triaxial with the long
axis of the velocity ellipsoid directed approximately along $v_R$ axis
(radially), the intermediate axis along $v_{\theta}$ (in direction of
rotation), and the small axis along $v_z$ (perpendicular to the galactic
plane).

In order to remove the contradiction mentioned above, the theory of stationary stellar systems in context
of velocity distributions is evidently needed to be generalised.  
Hence, in addition to the integrals $J_1$ and $J_2$, an additional independent of them 
time-independent single-valued integral $J_3$ has to be introduced. After the integral is
found, on the basis of Eq.~(\ref{eq4.13}) we have
\begin{equation}
\Psi =\Psi (J_1,J_2,J_3), \label{eq4.30}
\end{equation}
and there will be no difficulties in explaining the triaxial velocity 
distribution. However, if we increase the number of single-valued
independent integrals, additional restrictions for
the potential result. Therefore, the generalisation of the theory in respect to
velocity distribution is at the expense of the generality of the
potential.

It may happen that additional restrictions on the potential due to the
integral $J_3$ will be so strict that it will not possible to agree them
with observational data, or with the restrictions due to Poisson's equation
and with the boundary conditions at infinity. This means that the triaxiality
can not be explained within the stationarity of stellar systems, just as it
is not possible to explain within the stationarity some other effects (vertex
deviation, K-effect etc.), and confirms that the theory of a stationary
Galaxy is valid only as a first, quite crude approximation. But though
the triaxiality of the velocity distribution remains unexplained within the
theory of stationary stellar systems, it may be explained within
``quasi-stationary'' systems. This takes place when it is possible 
to handle the additional integral $J_3$ as ``quasi-integral'',  varying only in large
time-intervals. In this case Eq.~(\ref{eq4.30}) can be used as approximate
formula, and the triaxial velocity distribution is explained within a
``quasi-stationary theory''.

\citet{Lindblad:1933} proposed the third integral as the energy integral in $z$
direction
\begin{equation}
J_3=v_z^2-2[\Phi (R,z)-\Phi_0(R)], \label{eq4.31}
\end{equation}
where $\Phi_0$ is the potential in the galactic plane. This integral is
a precise integral when the stellar motions along $z$-coordinate are
independent of motions along other coordinates, \ie  when
\begin{equation}
\frac{\partial^2\Phi}{\partial R\partial z} = 0. \label{eq4.32}
\end{equation}
The condition (\ref{eq4.32}) can be obtained formally from Eq.~(\ref{eq4.16}) by substituting
$J_3$ according to Eq.~(\ref{eq4.31}), and using the Stokes' operator
in the form of
Eq.~(\ref{eq4.4}). We have then $\partial (\Phi -\Phi_0)/\partial R =0$, giving us
Eq.~(\ref{eq4.32}). General solution of Eq.~(\ref{eq4.32}) as a differential equation is the
potential as the sum of two functions
\begin{equation}
\Phi = \Phi_1(R)+\Phi_2(z). \label{eq4.33}
\end{equation}
Evidently this class of potentials is in contradiction with the finiteness
of the system, and as a result the Lindblad's integral can not be a precise
integral. But for small interval of $R$ values near the galactic plane the
potential may be assumed to have the form of Eq.~(\ref{eq4.33}). For the stars having nearly
circular orbits the Lindblad's integral is a quasi-integral. Hence, the
``quasi-stationary theory'' basing on Lindblad's integral may be used to
explain the triaxial velocity distribution of stars in flat subsystems of
the Galaxy.

However, the triaxial velocity distribution was observed not only for flat but
also for intermediate and spherical subsystems of the Galaxy. For these
subsystems the Lindblad's ``quasi-stationary theory'' can not be used.
So, the integral $J_3$ has to be found in a form, where restrictions on the
potential will be not so strict. As it was mentioned already, in order to
have a potential in most general form, one must use linear and
quadratic integrals. If we demand the existence of other linear integral
than $J_2$, we will have a spherically symmetric potential with three
linear integrals -- the angular momentum integrals. The limitation to
spherical symmetry only is evidently too strict,  and can not be used for
the Galaxy. Hence, we must turn to quadratic integrals. Such
integrals are the energy integral and the Lindblad's integral. But there may
exist also some other quadratic integrals.

\section{~}

Quadratic integrals were studied by several authors:
\citet{Eddington:1915}, \citet{Oort:1928}, \citet{Clark:1937},
\citet{Chandrasekhar:1939a, Chandrasekhar:1942} etc.\footnote{After
  the present paper was sent to print, the author noticed the papers by
  \citet{Camm:1941}, \citet{Fricke:1952}, and \citet{vAlbada:1952} on
  the third quadratic integrals allowing the triaxial velocity
  distribution. [Later footnote.]} All the authors, starting with the
Schwarzschild's ellipsoidal velocity distribution, assumed $\Psi =
e^{-Q}$ or $\Psi = \Psi (Q)$, where $Q$ is general quadratic
polynomial in respect to velocity components. The expression for $\Psi
$ was substituted into the continuity equation (\ref{eq4.1}) and 
the coefficients of $Q$ and the resulting restrictions for
the potential were analysed. As the phase density has the properties of the
integral of motion, in fact the integral $e^{-Q}$ or
$\Psi (Q)$ or simply the integral $Q$ was studied. The assumption that the phase
density is a function of $Q$ only is not necessary, because there are
no reasons to assume the Schwarzschild's or general 
ellipsoidal velocity distribution within high precision.  This assumption leads 
to completely unnecessary limitation of the generality of the theory.

From the authors referred above it was Oort who derived an expression for $Q$ which
may be used as the third integral $J_3$. However Oort supposed
that beside the stationarity and axial symmetry other restrictions for the
potential must not be applied. For this reason he derived $Q$ 
as a function of $J_1$ and $J_2$ only, and the theory remained in the 
frame of Jeans theorem in its narrow meaning. Later Chandrasekhar quite
comprehensibly studied the integral $Q$, but also did not succeed to go
further than Oort in the  theory of the stationary Galaxy.

From the perspective of third integral notable were the analysis by
Eddington and especially by Clark, who related Eddington's results to the
integrals of motion. Clark derived three independent
quadratic integrals: the energy integral and two different from the energy
quadratic integrals.\footnote{We have in mind three integrals,
given by Clark in the proof of his II theorem, and having the most general
form of quadratic integrals studied by him. It should be 
mentioned, that in addition to the integrals of motion, Clark also gives so called
invariant expressions. According of his point of view, the phase density
$\Psi$ must depend also on invariant expressions (despite of the Jeans
theorem). However, these ``invariant expressions'' are nothing but the
integrals of motion in explicit form, and they should not be included into the 
expression of $\Psi$.} Generally speaking they do not permit the axially symmetrical
potential and the differential motion of the centroids. But in particular case the
axial symmetry is possible. In this case one of Clark's integral, different
from the energy integral, turns into the third integral $J_3$ under interest,
and the other to a function of $J_1$, $J_2$ and $J_3$. This
particular case can be used in Galactic dynamics.

Clark's integrals were found in curvilinear ellipsoidal coordinates. Below
we give the derivation of a third integral by using, as it was done also
by Oort in his studies, cylindrical coordinates.

As the third integral is assumed to be quadratic in respect to velocity
components, we write it in form [see Appendix A]
\begin{equation}
J_3= a_{20}v_R^2+2a_{11}v_Rv_z+a_{02}v_z^2+a_{10}v_R+a_{01}v_z+a_{00},
\label{eq4.34}
\end{equation}
where $a_{20}$, $a_{11}$, $a_{02}$ are components independent of velocity,
$a_{10}$ and $a_{01}$ are linear expressions of $v_{\theta}$, $a_{00}$ is
quadratic expression of $v_{\theta}$. We assume that in $a_{10}$,
$a_{01}$, $a_{00}$ we substituted according Eq.~(\ref{eq4.28}) $v_{\theta} =J_2/R$
and temporarily take $J_2$ as a new variable instead of $v_{\theta}$. The
first two coefficients are thus linear functions of $J_2$, and the last one
is quadratic function of $J_2$. We also assume that $J_3$ is independent
of $\theta$, \ie 
\begin{equation}
\frac{\partial J_3}{\partial\theta}=0. \label{eq4.35}
\end{equation}

To find the coefficients of $J_3$ as functions of $R$ and $z$, and
the restrictions for the potential, $J_3$ has to be substituted into
Eq.~(\ref{eq4.16}), and the equation has to equal the identity. As instead
of $v_{\theta}$ we have $J_2$, then by taking into account Eq.~(\ref{eq4.28}) the
Stokes operator in cylindrical coordinates is
\begin{equation}
\frac{D}{Dt}= \frac{\partial}{\partial R}v_R + \frac{\partial}{\partial
z}v_z + \frac{\partial}{\partial v_R}\dot v_R + \frac{\partial}{\partial v_z}
\dot v_z, \label{eq4.36}
\end{equation}
while
\begin{equation}
\dot v_R = \frac{\partial\Phi '}{\partial R}; ~~\dot v_z= \frac{\partial\Phi
'}{\partial z} \label{eq4.37}
\end{equation}
and
\begin{equation}
\Phi '=\Phi - \frac{1}{ 2} \frac{J_2^2}{ R^2}, \label{eq4.38}
\end{equation}
and terms with $\partial /\partial t$ and $\partial /\partial\theta$ are
rejected as not being relevant.

As we see from Eqs.~(\ref{eq4.34}), (\ref{eq4.36}) and (\ref{eq4.37}), the problem of investigating
the integral $J_3$ is similar to studying the quadratic integral in
two-dimensional case. A solution of the problem is known from general
dynamics \citep[Sect. 152]{Whittaker:1904}.

By using Eq.~(\ref{eq4.16}) we have for the coefficients in Eq.~(\ref{eq4.34}) following
equations
\begin{equation}
\frac{\partial a_{20}}{\partial R}=\frac{\partial a_{20}}{\partial z}+
2\frac{\partial a_{11}}{\partial R}=2\frac{\partial a_{11}}{\partial z}+
\frac{\partial a_{02}}{\partial R} = \frac{\partial a_{02}}{\partial z} =
0; \label{eq4.39}
\end{equation}
\begin{equation}
\left. 
\begin{array}{l}
\frac{\partial a_{00}}{\partial R}+2a_{20} \frac{\partial\Phi '}{\partial R}
+ 2a_{11} \frac{\partial\Phi '}{\partial z} = 0, \\
\noalign{\smallskip}
\frac{\partial a_{00}}{\partial z}+2a_{11} \frac{\partial\Phi '}{\partial R}
+ 2a_{02} \frac{\partial\Phi '}{\partial z} = 0;
\end{array}
\right\} \label{eq4.40}
\end{equation}
\begin{equation}
\frac{\partial a_{10}}{\partial R}= \frac{\partial a_{10}}{\partial z}+
\frac{\partial a_{01}}{\partial R}= \frac{\partial a_{01}}{\partial z}=0;
\label{eq4.41}
\end{equation}
\begin{equation}
a_{10} \frac{\partial\Phi '}{\partial R} + a_{01} \frac{\partial\Phi
'}{\partial z}=0. \label{eq4.42}
\end{equation}
To find the expression for $J_3$ this set
of equations has to be solved.

Equations (\ref{eq4.41}) and (\ref{eq4.42}) are similar to
Eqs.~(\ref{eq4.20}) and (\ref{eq4.21}), discussed in 
the analysis of the integral $J_2$. But they differ in a way, that the
coefficients $a_{10}$ and $a_{01}$ as well as ``potential'' $\Phi '$
depend not only on coordinates but also on $J_2$. By substituting in Eq.~(\ref{eq4.42})
$a_{10}$ and $a_{01}$ as linear functions of $J_2$, and $\Phi '$
according to Eq.~(\ref{eq4.38}), and demanding that the equation will equal identity, we
have
\begin{equation}
a_{10}=a_{01}=0. \label{eq4.43}
\end{equation}

Next we analyse Eqs.~(\ref{eq4.39}) and (\ref{eq4.40}). As the coefficients $a_{20}$,
$a_{11}$ and $a_{02}$ are independent of $J_2$, and $\Phi '$ contains $J_2$
only in form of $J_2^2$, the Eq.~(\ref{eq4.40}) is identity only when
\begin{equation}
a_{00}=a_1 +a_2J_2^2, \label{eq4.44}
\end{equation}
where $a_1$ and $a_2$ are some functions of $R$ and $z$. Besides,
Eqs.~(\ref{eq4.40}) can be divided into two pairs of equations
\begin{equation}
\left.
\begin{array}{ll}
\frac{\partial a_1}{\partial R}+2a_{20} \frac{\partial\Phi}{\partial R} +
2a_{11} \frac{\partial\Phi}{\partial z} & = 0,\\
\noalign{\smallskip}
\frac{\partial a_1}{\partial z} + 2a_{11} \frac{\partial\Phi}{\partial R} +
2a_{02} \frac{\partial\Phi}{\partial z} & = 0 
\end{array}
\right\} \label{eq4.45}
\end{equation}
and
\begin{equation}
\left.
\begin{array}{ll}
\frac{\partial a_2}{\partial R}+2 \frac{a_{20}}{ R^3} & = 0, \\
\noalign{\smallskip}
\frac{\partial a_2}{\partial z}+2 \frac{a_{11}}{ R^3} & = 0. 
\end{array}
\right\} \label{eq4.46}
\end{equation}
Differentiating the first equation in (\ref{eq4.46}) with respect to $z$ and the second
with respect to $R$, and eliminating from the derived equations $\partial^2
a_2/\partial R\partial z$, we find
\begin{equation}
\frac{\partial a_{20}}{\partial z}R - \frac{\partial a_{11}}{\partial R} R +
3a_{11} = 0. \label{eq4.47}
\end{equation}
Further, differentiating Eq.~(\ref{eq4.39}) with respect to $R$ or $z$ we find that
all three partial derivatives of $a_{20}$, $a_{11}$, $a_{02}$ with respect to
$R$ and $z$ are zero. Therefore, the coefficients $a_{20}$, $a_{02}$ and
$a_{11}$ are quadratic functions of $R$ and $z$. Substituting them as
quadratic expressions into Eqs.~(\ref{eq4.39}) and (\ref{eq4.47}) we find the restrictions for
these expressions. Results for $a_{20}$, $a_{02}$ and $a_{11}$ can be
presented in forms
\be
\ba{ll}
a_{20} = & b_1+b_2z^2,\\
a_{02} = & b'_1+b_2R^2, \\ 
a_{11} = & -b_2Rz,
\ea
\label{eq4.48}
\ee
where $b_1$, $b'_2$, $b_2$ and zero-point of $z$-coordinate are
arbitrary constants. Substituting these results into Eq.~(\ref{eq4.46}),
integrating the first one over $R$ and the second over $z$,
and demanding that both of them give the same result, we have
\begin{equation}
a_2R^2=b_1+b'_2R^2+b_2z^2, \label{eq4.49}
\end{equation}
where $b'_2$ is again an arbitrary constant.

Now we substitute derived Eqs.~(\ref{eq4.43}), (\ref{eq4.44}),
(\ref{eq4.48}) and (\ref{eq4.49}) into Eq.~(\ref{eq4.34}). 
In addition to other terms there is a term $b'_2J_2^2$. We may neglect it
in the expression of $J_3$. Furthermore, some terms can be combined into terms as
$b_1J_1$, $a_1+2b_1\Phi$ and $(b'_1-b_1)v_z^2$ (taking into account
Eqs.~(\ref{eq4.27}) and (\ref{eq4.28})). First of them may be also neglected. We
designate the remaining terms as $(b'_1-b_1)=b_2z_0^2$, where $z_0$ has the
dimension of length (his geometrical meaning will be clarified later) and
$a_1+2b_1\Phi = -2b_2z_0^2\Phi^*$, where $\Phi^*$ is a function with
dimension of potential. Rejecting the factor $b_2$, common for all terms, we
derive the following expression for $J_3$
\begin{equation}
J_3=(Rv_z-zv_R)^2+z^2v_{\theta}^2+z_0^2(v_z^2-2\Phi^*).\label{eq4.50}
\end{equation}
The function $\Phi^*$ is calculated from equations
\begin{equation}
\left.
\begin{array}{ll}
z_0^2 \frac{\partial\Phi^*}{\partial R} = & z^2 \frac{\partial\Phi}{\partial R}
- Rz \frac{\partial\Phi}{\partial z}, \\
\noalign{\smallskip}
z_0^2 \frac{\partial\Phi^*}{\partial z} = & (R^2+z_0^2) \frac{\partial\Phi
}{\partial z} - Rz \frac{\partial\Phi}{\partial R}. 
\end{array}
\right\} \label{eq4.51}
\end{equation}
These equations follow from Eq.~(\ref{eq4.45}) after substituting $a_{20}$,
$a_{11}$, and $a_{02}$ according to Eq.~(\ref{eq4.48}) and taking into account our
designations.

It can be seen from Eq.~(\ref{eq4.51}), that when $\Phi$ is symmetric about the
plane $z=0$, the function $\Phi^*$ is symmetric about the
same plane. But in this case also the integral $J_3$ is symmetric about
the plane $z=0$ in the sense that $J_3(z,\dot z) = J_3(-z,-\dot
z)$. Therefore it is natural to identify the plane $z=0$, being arbitrary when we derived the
expression for $J_3$, to the galactic plane (the symmetry plane of the Galaxy).

As we see, the integral $J_3$ consists of two parts. The first two terms
give the sum of squares of angular momenta around two orthogonal axis
laying in the plane $z=0$. The last term with the factor $z_0^2$ resembles
the energy integral along $z$-coordinate. For $z_0^2=0$, as it results from
Eq.~(\ref{eq4.51}), the potential has the spherical symmetry. In this case the last
term vanishes and the first two terms are the sum of squares of two
angular momentum integrals. For $z_0^2\rightarrow\infty$ the integral (\ref{eq4.50})
turns into the Lindblad's integral. In this case, according to (\ref{eq4.51}), $\Phi^*$
depends only on $z$, and $\partial^2\Phi /\partial R\partial z$ turns to
zero, just as it has to be for the case of precise Lindblad's integral.

\section{~}

We found an expression for the third integral. Let us discuss now the
resulting restriction on the potential. The restricting condition results
from Eq.~(\ref{eq4.51}). By differentiating the first equation of (\ref{eq4.51}) with respect
to $z$ and the second with respect to $R$ and eliminating $\partial^2\Phi^*
/\partial R\partial z$, we find the condition
\begin{equation}
3\left( z \frac{\partial\Phi}{\partial R}-R \frac{\partial\Phi}{\partial z}
\right) - \left( R^2+z_0^2-z^2\right) \frac{\partial^2\Phi}{\partial
R\partial z} + Rz\left( \frac{\partial^2\Phi}{\partial R^2} -
\frac{\partial^2\Phi}{\partial z^2}\right) = 0. \label{eq4.52}
\end{equation}
This is the restricting condition for the potential resulting from the
existence of the integral (\ref{eq4.50}).

In its form the condition (\ref{eq4.52}) coincides with condition on the 
existence of quadratic integral for two-dimensional problem known from general dynamics
\citep{Whittaker:1904}. The coincidence is not surprising, because as we
saw, the integral $J_3$ and the Stokes operator can be written formally
in a form coinciding with the quadratic integral and the Stokes operator
for two-dimensional problem. Therefore, also the resulting integral (\ref{eq4.50}) is
very similar to the quadratic integral in two-dimensional problem.

The condition (\ref{eq4.52}) was derived also by Oort in the referred above paper
\citep{Oort:1928}. Oort decided that the condition can not be satisfied, and
derived the integral being a function of the energy and the momentum
integrals. As a result the velocity distribution was biaxial. In order to
explain the triaxiality of the velocity distribution, Oort used an
explanation, equivalent with introducing the Lindblad's quasi-integral.

Surely, the condition (\ref{eq4.52}) can not be satisfied precisely for the real
Galaxy. But it is not so strict as the condition (\ref{eq4.32}), corresponding to
Lindblad's integral. This is related to an arbitrary constant $z_0^2$,
contained in Eq.~(\ref{eq4.52}), which may be chosen in a way enabling the
best fulfilment of Eq.~(\ref{eq4.52}) for the Galaxy. When the condition (\ref{eq4.52}) is
satisfied with  sufficient precision, the integral (\ref{eq4.50}) can be handled as a
precise integral, just as $J_1$ and $J_2$ are. In opposite case it may
be used only as quasi-integral, like the Lindblad's one [see Appendix B].
But even as a quasi-integral it can be used for orbits, much more different
from circular ones than it was for the Lindblad's integral. One
only has to choose different $z_0^2$ for different orbits, in a way to ensure
the best satisfaction of (\ref{eq4.52}) for every orbit. In this case $z_0^2$ is a
function of $J_1$, $J_2$, $J_3$ or some approximate function of $R$ (we
discuss orbits still not too different from circular).

In deriving the condition (\ref{eq4.52}) we started with Eq.~(\ref{eq4.51}) determining the
function $\Phi^*$ contained in Eq.~(\ref{eq4.50}). When the condition (\ref{eq4.52}) is valid,
the function $\Phi^*$ may be found in a form, satisfying both equations in
(\ref{eq4.51}). Integrating the second equation in (\ref{eq4.51}) over $z$ and taking into
account that according to the first equation $\Phi^*= \mathrm{const}$ at $z=0$, we
derive
\begin{equation}
z_0^2\Phi^*(R,z)=(R^2+z_0^2)[\Phi (R,z)-\Phi_0(R)]- $$
$$-R\int_0^z \frac{\partial\Phi (R,z)}{\partial R} z \rmd z+ \mathrm{const}. \label{eq4.53}
\end{equation}
We derive another formula for $\Phi^*$ by integrating the first equation
of (\ref{eq4.51}) with respect to $R$, and taking into account that according to the
second equation $\Phi^* -\Phi =\mathrm{const}$ at $R=0$. If Eq.~(\ref{eq4.52}) is
fulfilled both formulae must give the same results to the precision of additive
constant. In this case the integral (\ref{eq4.50}) is the precise integral of
motion. When the condition (\ref{eq4.52}) is not satisfied, the integral (\ref{eq4.50}) can
be used only as a quasi-integral for more or less circular orbits, and the
two equations for $\Phi^*$ will give different results. In this case for
calculation of $\Phi^*$ Eq.~(\ref{eq4.53}) must be used. Equation (\ref{eq4.53}) satisfies
precisely the second equation in (\ref{eq4.51}), and for nearly circular orbits it
satisfies within sufficient precision also the first equation. $z_0^2$ is assumed 
to be chosen in accordance with the condition (\ref{eq4.52}) and
is constant for given orbit.

In order to calculate the function $\Phi^*$, and to apply the integral (\ref{eq4.50})
to real systems, the numerical value of $z_0^2$ has to be known.
Hence there appears the problem of determining the $z_0^2$ from
observational data. To solve it we may use the condition (\ref{eq4.52}) for
$z\rightarrow 0$. Differentiating (\ref{eq4.52}) with respect to $z$ and taking
thereafter $z=0$, we have
\begin{equation}
3 \frac{\partial\Phi}{\partial R}+R \frac{\partial^2\Phi}{\partial R^2}-
4R \frac{\partial^2\Phi}{\partial z^2} -(R^2+z_0^2) \frac{\partial^3\Phi}{
\partial R\partial z^2} = 0. \label{eq4.54}
\end{equation}
Partial derivatives $\partial\Phi /\partial R$ and $\partial^2\Phi /\partial
R^2$ are related to the Oort constants $A$ and $B$ (for
motion in circular orbits) in the  way
\begin{equation}
\frac{1}{ R} \frac{\partial\Phi}{\partial R} = -(A-B)^2; ~~~
\frac{\partial^2\Phi}{\partial R^2}= (A-B)(3A+B). \label{eq4.55}
\end{equation}
Further, according to our designations \citet{Kuzmin:1952ab}
\begin{equation}
\left( \frac{\partial^2\Phi}{\partial z^2}\right)_{z=0} = -C^2.
\label{eq4.56}
\end{equation}
Therefore, Eq.~(\ref{eq4.54}) will have the form
\begin{equation}
(R^2+z_0^2) \frac{dC^2}{ dR} + 4R[C^2+B(A-B)]=0. \label{eq4.57}
\end{equation}
The derived equation can be used to determine $z_0^2$. The values of $A$,
$B$ and $C$ in the vicinity of the Sun are quite well known from the
rotation of flat subsystems of the Galaxy, and from the stellar motions
perpendicular to the galactic plane. Only the radial gradient of $C$
is unknown. Therefore, the problem of the determination of $z_0^2$ reduces
to the determination of that gradient.

Equation (\ref{eq4.54}) leads us to an important consequence on the character of
$z_0^2$. As due to Galactic flattening its equipotential surfaces must
have at $z=0$ higher curvature than the spheres, centres of which coincide
with the centre of the Galaxy, for $z=0$ the expression
$\partial^2\Phi /\partial z^2 < \partial\Phi /R\partial R$ $(<0)$ must be valid. Furthermore,
we may assume that $-\partial\Phi /R\partial R$ or $-\partial^2\Phi
/\partial z^2$ decreases with increasing of $R$, \ie  that $\partial^2\Phi
/\partial R^2 > \partial\Phi /R\partial R$ or $\partial^3\Phi /\partial R\partial
z^2>0$. From Eq.~(\ref{eq4.54}) it results, that when $z_0$ does not depend on $R$
($J_3$ is a precise integral) $z_0^2>0$, \ie  $z_0$ is a real quantity.

\section{~}

Let us discuss now what can we conclude about the velocity distribution
on the basis of the integral (\ref{eq4.50}). By rotating the coordinate axes
$v_R$ and $v_z$ we turn the integral into the sum of squares of
velocities. We derive
\begin{equation}
J_3=z_0^2(\xi_2^2v_1^2+\xi_1^2v_2^2)+z_0^2v_{\theta}^2-2z_0^2\Phi^*,
\label{eq4.58}
\end{equation}
where $\xi_1^2$ and $\xi_2^2$ are roots of the equation
\begin{equation}
z_0^2\xi^4 - (R^2+z_0^2+z^2)\xi^2 +z^2 =0, \label{eq4.59}
\end{equation}
and $v_1$ and $v_2$ are new velocities, replacing after the coordinate
rotation the coordinates $v_R$ and $v_z$. As according to Eq.~(\ref{eq4.59})
$z^2=z_0^2\xi_1^2\xi_2^2$, and the common factor $z_0^2$ in $J_3$ may be
rejected, we have for $J_3$ the following expression
\begin{equation}
J_3 = \xi_2^2v_1^2+\xi_1^2v_2^2 +\xi_1^2\xi_2^2v_{\theta}^2-2\Phi^*.
\label{eq4.60}
\end{equation}

For the rotation angle $\alpha$ of the coordinate axis we get the formula
\begin{equation}
\tan 2\alpha = \frac{2Rz}{ R^2+z_0^2-z^2}. \label{eq4.61}
\end{equation}
The formula gives two values of $\alpha$ between $\rm 0^o$ and $\rm
180^o$. They correspond to the angles between the axes $v_1$ and $v_2$ and
$v_R$ axis, or, in other words, the angles between the axes $v_1$ and $v_2$
and the galactic plane.

Equation (\ref{eq4.59}) can be represented in form
\begin{equation}
\frac{R^2}{ \xi^2-1}+ \frac{z^2}{\xi^2}=z_0^2. \label{eq4.62}
\end{equation}
If we assume $z_0^2$ to be a constant, it results from Eq.~(\ref{eq4.62}) that the
positions of constant $\xi^2$ are the confocal second order surfaces with foci
on $z$ axis at points $z=+z_0$ and $z=-z_0$. The values $\xi^2>1$
correspond to the ellipsoids of revolution, the values $0<\xi^2<1$ to the
two-sheeted hyperboloids of revolution. The turning point $\xi^2=1$
corresponds to $z$ axis, and $\xi^2=0$ to the plane $z=0$. Every point in space 
corresponds only to one ellipsoid and to one hyperboloid.
For that reason two values of $\xi^2$ correspond to every point. These two
values are the roots of the Eq.~(\ref{eq4.59}) $\xi_1^2$ and $\xi_2^2$. The values
$\xi_1$ and $\xi_2$ can be handled as curvilinear coordinates replacing
$R$ and $z$. Let us agree that $\xi_1$ correspond to ellipsoids and $\xi_2$ to
hyperboloids, \ie  $\xi_1^2\ge 1$ and $\xi_2^2\le 1$. Besides we may
assume that $\xi_2>0$ corresponds to the northern part of hyperboloid
($z>0$) and $\xi_2<0$ to the southern part ($z<0$). We assume $\xi_1$ to be
positive (as an analogy with $R$).

The surfaces of constant $\xi^2$ have the property that at every point 
the axes $v_1$ and $v_2$ are perpendicular or tangent to them. 
To proof it let us calculate the  derivative of Eq.~(\ref{eq4.59}) (or
Eq.~\ref{eq4.62}) for $\xi^2 = \mathrm{const}$. From the new equation derived in that way, and
from Eq.~(\ref{eq4.59}) (or Eq.~\ref{eq4.62}) we eliminate $\xi^2$. As a result we have
\begin{equation}
Rz( \rmd R^2 - \rmd z^2 )-(R^2+z_0^2-z^2) \rmd R \rmd z = 0. \label{eq4.63}
\end{equation}
The equation is a differential equation of the surface $\xi^2 = \mathrm{const}$. It
is now easy to find an angle between the galactic plane and the tangent planes of
these surfaces. We obtain the formula coinciding with Eq.~(\ref{eq4.61}). It means that the
axis $v_1$ and $v_2$ are either tangent or perpendicular to the surfaces $\xi^2= \mathrm{const}$. 
Besides, the fact, that for $\xi_2=z=0$ the
expressions of $J_3$ the Eqs.~(\ref{eq4.50}) and (\ref{eq4.58}) must coincide, gives us that the
$v_1$ axis is perpendicular to ellipsoids, and the $v_2$ axis to hyperboloids
(the ellipsoids and hyperboloids intersect orthogonally).

If the phase density depends on $J_1$, $J_2$ and $J_3$, the velocity
distribution must be symmetric about the planes $v_1=0$ and
$v_2=0$, as the integral $J_3$ is symmetric about these planes.
The velocity distribution is in general triaxial in the sense, that the
formally calculated velocity ellipsoid is triaxial. One of
ellipsoid's axis must coincide with the $v_{\theta}$ axis, the remaining
two must be parallel to the axes $v_1$ and $v_2$. In particular case the
distribution may be really ellipsoidal, in general not necessarily.

As the axes $v_1$ and $v_2$ are perpendicular to confocal surfaces of
revolution, determined by Eq.~(\ref{eq4.62}), the axes of velocity ellipsoid are also 
perpendicular to these surfaces. Eddington called this class of surfaces 
the main velocity surfaces. In his paper \citep{Eddington:1915} referred above, he
also derived the main velocity surfaces in form of confocal second order
surfaces. In particular case of axial symmetry they coincide with the
surfaces of revolution (\ref{eq4.62}). In that particular case one of the two
different from the energy integral quadratic integrals by \citet{Clark:1937} turns
into the integral $J_3$ in accordance with Eq.~(\ref{eq4.60}), the other one into a function of
integrals $J_1$, $J_2$ and $J_3$ [see Appendix C].

The derived results for the surfaces $\xi^2= \mathrm{const}$, and for the main velocity
surfaces, correspond to the case when $z_0^2$ is constant, \ie  when $J_3$
is the precise integral of motion. When $J_3$ is a quasi-integral and
$z_0^2$ not precisely a constant, the results surely change. The surfaces
$\xi^2= \mathrm{const}$ and main velocity surfaces no longer coincide, and are no
longer the confocal surfaces of second order. However at distances not too far away 
from the galactic plane, the only place where $J_3$ can actually be used, the
results do not change significantly.

It results from the above considerations concerning the main velocity surfaces and
from Eq.~(\ref{eq4.61}) that outside the galactic axis and the
galactic plane the axis of the velocity ellipsoid are inclined to the
galactic plane. At points, symmetrical to the galactic plane, the axis of
ellipsoid are directed symmetrically. This is related to the
symmetry of the integral $J_3$ about the galactic plane.

Contemporary observational data is not yet able to give hints about the
obliquity of the velocity ellipsoid outside of the galactic plane. But from
theoretical point of view there remain no doubts. First, it was derived by
Eddington in the referred above paper. Later the same was concluded
by \citet{Chandrasekhar:1939b}. However, despite that correct
conclusion, the basis of Chandrasekhar's paper is erroneous. He tried to
generalise the two-dimensional theory of \citet{Lindblad:1927, Lindblad:1936} ellipsoidal
velocity distribution into three dimensions. The referred theory is based on
the study of nearly circular orbits, and its application even in generalised
three-dimensional form must be limited only to regions near to the galactic
plane. Chandrasekhar uses completely formally the method of nearly
circular orbits at large distances from the galactic plane, where orbits
highly differ from the circular ones, and where the method is not
applicable\footnote{In study of highly flattened Galactic subsystems
the method of nearly circular orbits is though suitable for using. Instead of
integrals $J_1$, $J_2$, $J_3$ as arguments of the phase density
the integrals of nearly circular motion should be used. [Later
footnote.]}. Also in general the method of nearly circular orbits, proposed
by Lindblad, does not give nothing new, when compared to results one can acquire directly
from the Jeans theorem and from the integrals $J_1$, $J_2$ and
the Lindblad's quasi-integral. The method only complicates the theory and
narrows its possibilities.

\section{~}

As we saw, the third integral (\ref{eq4.50}) has significant advantages when
compared to the Lindblad's integral, and for that reason it can be used as a
quasi-integral not only for flat Galactic subsystems but also for
subsystems with quite moderate eccentricities. But, in order to explain
the triaxial velocity distribution of spherical subsystems, the integral
$J_3$ can not be used as a quasi-integral, because the orbits of objects in spherical
subsystems differ rather highly from circular. Therefore, if we
try to explain the triaxial velocity distribution of all Galactic subsystems
within the theory of a stationary Galaxy, we have to assume, that the
condition (\ref{eq4.52}) is valid for the whole Galaxy with significant precision,
and that $J_3$ can be handled as a precise integral of motion. For that
reason it seems very interesting to discuss the problem, how precisely
the condition (\ref{eq4.52}) can be valid for the whole Galaxy. Maybe the
condition (\ref{eq4.52}) is valid as precisely as the condition of stationarity or
axial symmetry.

In order to study the validity of the condition (\ref{eq4.52}) for the Galaxy, first
of all Eq.~(\ref{eq4.52}) as a differential equation for $\Phi$ (for $z_0^2= \mathrm{const}$) 
has to be solved. The solution of the equation is known from
general dynamics \citep[see][]{Whittaker:1904}. It can be easily derived after
transforming the equation into canonical form.

For the characteristics of Eq.~(\ref{eq4.52}) we find the equation, coinciding with
Eq.~(\ref{eq4.63}). It results now, that the characteristics are confocal
second order surfaces (\ref{eq4.62}), discussed above. Therefore, in order to transform
Eq.~(\ref{eq4.52}) into canonical
form,  new variables $\xi_1$ and $\xi_2$ are needed.

By taking into account that $\xi_1^2$ and $\xi_2^2$ are the roots of Eq.~(\ref{eq4.59}),
it is easy to derive the relations between $R$ and $z$, and $\xi_1$ and
$\xi_2$
\begin{equation}
\left.
\begin{array}{ll}
R^2= & z_0^2(\xi_1^2-1)(1-\xi_2^2), \\
\noalign{\smallskip}
z^2 = & z_0^2\xi_1^2\xi_2^2. 
\end{array}
\right\}  \label{eq4.64}
\end{equation}
From these equations we find the expressions for $\partial /\partial R$
and $\partial /\partial z$
\begin{equation}
\left.
\begin{array}{ll} 
\frac{z_0^2}{ R} \frac{\partial}{\partial R} = & \frac{\xi_1^2}{
\xi_1^2-\xi_2^2} \frac{1}{\xi_1} \frac{\partial}{\partial\xi_1} - \frac{\xi_2^2}{
\xi_1^2-\xi_2^2} \frac{1}{\xi_2} \frac{\partial}{\partial\xi_2}, \\
\noalign{\smallskip}
\frac{z_0^2}{ z} \frac{\partial}{\partial z} = &
\frac{\xi_1^2-1}{\xi_1^2-\xi_2^2} \frac{1}{\xi_1} \frac{\partial}{\partial\xi_1}
+ \frac{1-\xi_2^2}{\xi_1^2-\xi_2^2} \frac{1}{\xi_2} \frac{\partial}{\partial
\xi_2}. 
\end{array}
\right\} \label{eq4.65}
\end{equation}
By using Eqs.~(\ref{eq4.64}) and (\ref{eq4.65}), we derive from
Eq.~(\ref{eq4.51}) the following quite simple 
equations
\begin{equation}
\frac{\partial\Phi^*}{\partial\xi_1} = \xi_2^2 \frac{\partial\Phi}{\partial
\xi_1}; ~~~ \frac{\partial\Phi^*}{\partial\xi_2} = \xi_1^2 \frac{\partial\Phi
}{\partial\xi_2}. \label{eq4.66}
\end{equation}
Differentiating the first of these equations with respect to $\xi_2$, and the
second with respect to $\xi_1$, and eliminating $\partial^2\Phi^*/
\partial\xi_1\partial\xi_2$, we have
\begin{equation}
\frac{\partial^2}{\partial\xi_1\partial\xi_2} [(\xi_1^2-\xi_2^2)\Phi ]
=0. \label{eq4.67}
\end{equation}
Equation (\ref{eq4.67}) is just Eq.~(\ref{eq4.52}) in canonical form. It can be integrated
immediately giving us the following expression for $\Phi$
\begin{equation}
\Phi = \frac{\varphi_1(\xi_1) - \varphi_2(\xi_2) }{ \xi_1^2-\xi_2^2},
\label{eq4.68}
\end{equation}
where $\varphi_1$ and $\varphi_2$ are arbitrary functions. The derived
result coincides with the Eddington's one \citep{Eddington:1915} in case of
axially symmetrical potential.\footnote{The same expression for
the potential was derived by \citet{Camm:1941}. He
derived also an expression for the density. Analogous expression for the
potential and the density can be found also in a paper by
\citet{Fricke:1952}. But there author chose a non-suitable positions of
foci (not in $z$ 
axis, but in $R$ axis). Also \citet{vAlbada:1952}
positioned foci in $R$ axis, resulting  a wrong conclusion about
non-applicability of the theoretical expression for the potential to real
stellar systems. [Later footnote.]}

By integrating Eq.~(\ref{eq4.66}), and taking into account the expression for $\Phi$,
we find for $\Phi^*$
\begin{equation}
\Phi^* = \frac{\xi_2^2\varphi_1(\xi_1)-\xi_1^2\varphi_2(\xi_2)}{
\xi_1^2-\xi_2^2}. \label{eq4.69}
\end{equation}

In addition to the expressions for $\Phi$ and $\Phi^*$, it is interesting to have a look at the
expressions for the gradients of $\Phi$ with respect to $R$ and $z$. By using Eq.~(\ref{eq4.65}) we find from Eq.~(\ref{eq4.68}) that
\begin{equation}
\left.
\begin{array}{ll}
-z_0^2 \frac{\partial\Phi}{ R\partial R} = & 2 \frac{(\xi_1^2+\xi_2^2)
(\varphi_1-\varphi_2)}{ (\xi_1^2-\xi_2^2)^3} - \frac{\xi_1\varphi_1'+
\xi_2\varphi_2' }{ (\xi_1^2-\xi_2^2)^2} ,\\
\noalign{\smallskip}
-z_0^2 \frac{\partial\Phi}{ z\partial z} = & 2 \frac{(\xi_1^2+\xi_2^2-2)(\varphi_1
-\varphi_2)}{ (\xi_1^2-\xi_2^2)^3} - \frac{(\xi_1-\xi_1^{-1})\varphi_1'
+(\xi_2-\xi_2^{-1})\varphi_2' }{ (\xi_1^2-\xi_2^2)^2} ,
\end{array}
\right\} 
\label{eq4.70}
\end{equation}
where apostrophe signifies the derivative of $\varphi_1$ and $\varphi_2$
with respect to their arguments.

Equation (\ref{eq4.68}) enables also to find an expression for the mass density
$\rho$ on the basis of Poisson's equation (\ref{eq4.10}). For that we use general
expression of Laplace's operator in curvilinear coordinates (see for
example \citet[][Sect. 1]{Subbotin:1948}. In our curvilinear coordinates
$\xi_1$, $\xi_2$ and $\theta$ and axial symmetry the Laplace's operator has
the form
\begin{equation}
\nabla^2 = \frac{1}{ H_1H_2H_{\theta}} \left[ \frac{\partial}{\partial\xi_1}
\left( H_{\theta} \frac{H_2}{ H_1} \frac{\partial}{\partial\xi_1}\right)
+ \frac{\partial}{\partial\xi_2} \left( H_{\theta} \frac{H_1}{ H_2} \frac{\partial
}{\partial\xi_2}\right) \right] , \label{eq4.71}
\end{equation}
where $H_1$, $H_2$, $H_{\theta}$ are Lam\'e coefficients for $\xi_1$,
$\xi_2$ and $\theta$ respectively
$$H_1^2 = \left( \frac{\partial R}{\partial\xi_1}\right)^2 + \left( \frac{\partial
z }{\partial\xi_1}\right)^2 = z_0^2 \frac{\xi_1^2-\xi_2^2}{ \xi_1^2-1},$$
\begin{equation}
H_2^2 = \left( \frac{\partial R}{\partial\xi_2}\right)^2 + \left( \frac{\partial
z }{\partial\xi_2}\right)^2 = z_0^2 \frac{\xi_1^2-\xi_2^2}{ 1-\xi_2^2},
\label{eq4.72}
\end{equation}
$$H_{\theta}^2 = \left( \frac{\partial R\theta}{\partial\theta}\right)^2 =
z_0^2(\xi_1^2-1)(1-\xi_2^2). $$

Substituting Eqs.~(\ref{eq4.68}), (\ref{eq4.71}) and (\ref{eq4.72})
into Eq.~(\ref{eq4.10}), we derive for the mass density $\rho$ the
following expression 
$$
4\pi z_0^2G\rho =2 \frac{(\varphi_1-\varphi_2-\xi_1\varphi_1'+\xi_2\varphi_2')
(2-\xi_1^2-\xi_2^2)}{ (\xi_1^2-\xi_2^2)^3} - $$
\begin{equation}
- \frac{(\xi_1^2-1)\varphi_1'' -(1-\xi_2^2)\varphi_2'' }{ (\xi_1^2-
\xi_2^2)^2 }. \label{eq4.73}
\end{equation}

In order to clarify the applicability of the restriction on the
potential,  resulting from the integral $J_3$, the applications
of formulae for the potential and for the mass density have to be studied. If it is
possible to construct with the help of these formulae a model of the Galaxy,
having physical meaning, and not being in contradiction with observational
data, it will mean that most essential spatio-kinematical  characteristics
of Galactic structure, including the velocity distribution triaxility, can
be explained within the theory of a stationary Galaxy.
Construction of these kind of models along with some other problems,
related with the third integral of stellar motion, will be discussed in
a subsequent paper.
\vglue 3mm
\hfill 1953

\vglue 5mm

{\bf\Large Appendices added in 1969}
\vglue 5mm

\section{A.  Even and odd parts of the quadratic integral}

Using the Liouville's equation, \citet{Idlis:1959} demonstrated that every 
conservative integral consist of two
integrals --- even and odd with respect to velocities. This results
from the fact that motion along a given orbit may be in two opposite
directions. If we fix for the axially symmetric stationary potential
the integral $J_2$, and introduce ``effective potential''
(Eq.~\ref{eq4.38}),  the problem
reduces to planar motion. Hence, the even and odd parts of
the integral (\ref{eq4.34}) with respect of $v_R$ and $v_z$ must be both the
integrals of motion.

As the existence of even and odd in respect to $v_R$ and $v_z$
quadratic integrals put too strict restrictions on the potential, the
odd part of the integral (\ref{eq4.34}) must be rejected, limiting at once with
\begin{equation}
J_3=a_{20}v^2_R+2a_{11}v_Rv_z+a_{02}v^2_z+a_{00}, \label{eq4.A1.1}
\end{equation}
But in this case the condition (\ref{eq4.35})
\begin{equation}
{\partial J_3\over\partial\theta}=0 \label{eq4.A1.2}
\end{equation}
is no more an independent assumption, but a consequence. It can be
derived after substituting $J_3$ to the Liouville's equation. Therefore,
the independence of the integral on $\theta$ results from the need to
minimalize the restrictions on the potential.

The same results come from the paper by \citet{Camm:1941}, where a
particular case when quadratic integral 
depends on $\theta$ (for axially symmetric potential) was studied. The resulting
limitations on the potential are completely unacceptable.

\section{B.  The third quadratic integral as a quasi-integral}

If we handle the third integral as a quasi-integral, not only $z_0$ but
also the zero point of $R$ must depend on orbit, \ie  on the values of
$J_1,$ $J_2$, $J_3$. It was understood by us after the publication of the
first papers on the third integral. This idea was further developed by
\citet{vdHulst:1962}.

Foci, related to the third quadratic quasi-integral in elliptical
coordinates, lie at points
\begin{equation}
R=R_f, ~~~~ z=z_f=\pm z_0 , \label{eq4.A2.1}
\end{equation}
and in general $R_f\ne 0$. Integral has the form
\begin{equation}
J_3 = [(R-R_f)v_z-zv_R]^2+ \frac{R-R_f}{R}z^2v_{\theta} +z^2_f(v^2_z-
2\Phi^* ). \label{eq4.A2.2}
\end{equation}
where
\begin{equation}
z^2_f\Phi^* = [(R-R_f)^2+z^2_f] [\Phi (R,z)-\Phi (R,0)] -
(R-R_f)\int^z_0 \frac{\partial\Phi}{\partial R} z \rmd z. \label{eq4.A2.3}
\end{equation}
The parameters $R_f$ and $z_f$ must be chosen in a way that in the
region of $R,z$, covered by loops of orbit,  the following equation
will be valid as precisely as possible
$$\frac{3J^2_2R_f}{ R^4} +3\left[ z \frac{\partial\Phi}{\partial R} -(R-R_f)
\frac{\partial\Phi}{\partial z} \right] - $$
\begin{equation}
- \left[(R-R_f)^2 +z^2_f-z^2\right] \frac{\partial^2\Phi}{\partial
R\partial z} + (R-R_f) z \left( \frac{\partial^2\Phi}{\partial R^2}
- \frac{\partial^2\Phi}{\partial z^2}\right) = 0. \label{eq4.A2.4}
\end{equation}

\section{C.  Quadratic integrals by Clark}

Let us assume a system of confocal ellipsoidal coordinates $\lambda$,
$\mu$, $\nu$, where $\lambda$ is the major semiaxis of ellipsoids, $\mu$ is the
major semiaxis of hyperboloids of one sheet, and $\nu$ the real semiaxis of
hyperboloids of two sheet. The following relations must be valid
$$\lambda\ge\alpha\ge\mu\ge\beta\ge\nu\ge -\beta ,$$
where $\alpha$ and $\beta$ are constants (the major semiaxis of the
focal ellipse and the real semiaxis of focal hyperboloid).

The integrals by \citet{Clark:1937} (they were contained already in the paper by
\citet{Eddington:1915}) are
\begin{equation}
\left.
\begin{array}{ll}
J_1 & =v^2_{\lambda} + v^2_{\mu} + v^2_{\nu} + \Phi , \\
\noalign{\smallskip}
J^*_2 & =(\mu^2 +\nu^2 )v^2_{\lambda} +(\nu^2 +\lambda^2 )v^2_{\mu} +
(\lambda^2 +\mu^2 )v^2_{\nu} +2\Phi_* ,\\
\noalign{\smallskip}
J^*_3 & =\mu^2\nu^2 v^2_{\lambda} + \nu^2\lambda^2 v^2_{\mu} +
\lambda^2\mu^2 v^2_{\nu} + 2\Phi_{**} , 
\end{array}
\right\} \label{eq4.A3.1}
\end{equation}
where $v_{\lambda}$, $v_{\mu}$, $v_{\nu}$ are the velocity components
along the coordinate lines $\lambda$, $\mu$, $\nu$.

The functions $\Phi_*$ and $\Phi_{**}$ both satisfy three partial differential
equations of the first order. This system of equations gives three
second order differential equations for $\Phi$. These equations
can be solved and we have
\begin{equation}
\left.
\begin{array}{ll}
\Phi = & F_{\lambda}+F_{\mu}+F_{\nu} , \\
\noalign{\smallskip}
\Phi_* = & (\mu^2 +\nu^2 )F_{\lambda}+(\nu^2 +\lambda^2 )F_{\mu}+(\lambda^2
+\mu^2 )F_{\nu} , \\
\noalign{\smallskip}
\Phi_{**} = & \mu^2\nu^2 F_{\lambda} +\nu^2\lambda^2
F_{\mu}+\lambda^2\mu^2 F_{\nu}. 
\end{array}
\right\} \label{eq4.A3.2}
\end{equation}
Here
\begin{equation}
F_{\lambda}={f(\lambda )\over (\lambda^2 -\mu^2 )(\lambda^2 -\nu^2 )},
~F_{\mu} ={f(\mu )\over (\mu^2 -\nu^2 )(\mu^2 -\lambda^2 )},
~F_{\nu} ={f(\nu )\over (\nu^2 -\lambda^2 )(\nu^2 -\mu^2 )}, \label{eq4.A3.3}
\end{equation}
where $f$ is an arbitrary function. Demanding that the functions $\Phi$,
$\Phi_*$, $\Phi_{**}$ have no singularities, next condition must be valid 
for the function $f$ 
$$f(\alpha ) = f(\beta ) = 0.$$

Here we have an example of non-axisymmetric stationary potential
allowing three single-valued integrals of motion. However, the potential
is restricted by quite significant conditions: it is determined by an
arbitrary function of one argument $f$, and by two parameters $\alpha$
and $\beta$. If the argument $f$ is an even function, the potential is
symmetric about three mutually orthogonal planes. As all
integrals are quadratic, they do not permit the differential motion
of centroids.

For
$$\alpha = \beta = z_0 $$
the potential becomes axisymmetric. In this case
$$\lambda = z_0\xi_1 , ~~ \mu = z_0 , ~~ \nu = z_0\xi_2 , $$
and integrals $J^*_2$, $J^*_3$ can be expressed via $J_1,$ $J_2$, $J_3$
$$J_2^* = J^2_2+J_3+z_0^2J_1,$$
$$J^*_3 = J_3.$$
Besides $f(\xi )=\varphi (\xi )(\xi^2 -1)$ and $\Phi_* =\Phi +\Phi^*$,
$\Phi_{**} =\Phi^*$.

The Clark's integrals and many other isolating integrals were analysed
by \citet{Lynden-Bell:1962}.

%% file: chapter05.tex
\chapter[The dynamical parameter $C$ and the matter density]{On the value of
  the dynamical parameter $C$ and the density of matter in the
  vicinity of the Sun \footnote{\footnotetext ~~Published in Tartu
    Astron. Observatory Publications, vol. 33, pp. 3-34, 1955} }

In the paper \citet{Kuzmin:1952ab} published in 1952 we denoted by $C$ the dynamical
parameter determining the change of the gravitational potential of the
Galaxy $\Phi$ in the direction perpendicular to the galactic plane and the
density of matter in this plane $\rho$. Numerical values of $C$ and $\rho$ in the
vicinity of the Sun were derived. In addition, the flattening of the
Galaxy was deduced from dynamical considerations. The obtained value
of $C$ was found to be significantly lower than that obtained from the
earlier study by \citet{Oort:1932}. Accordingly the value of $\rho$ was also
lower. However, recently  \citet{Parenago:1952, Parenago:1954a} has derived values of $C$
and $\rho$ consistent with the Oort values, while \citet{Safronov:1952}
obtained a similar value of $\rho$. The purpose of the present paper is to
clarify the reasons for the difference between our result and the
results of Oort, Parenago, and Safronov, to discuss the accuracy of
these results and the systematic errors in them, and, finally, to
derive the probable values of $C$ and $\rho$.

\section[Method for determining $C$.]{The dynamical parameter $C$ and
  the density of matter in the 
  galactic plane $\rho$. Method for determining $C$}

\subsection{}
The dynamical  parameter $C$ is related to the
Galactic gravitational potential $\Phi$ as follows: 
\be
C^2 = -\left(\frac{\partial^2\,\Phi}{\partial\,z^2}\right)_{z=0},
\label{eq5.1}
\ee
where $z$ is the elevation above the galactic plane. Parameter $C$
thus determines the dependence of $\Phi$ on $z$ near the galactic plane. This
is a necessary complement to the rotational Oort parameters $A$ and $B$
for circular velocity motion. From the known formulas connecting $A$ and
$B$ with circular velocity follows the following expressions for the
relation of $A$ and $B$ with potential:
\be
\left.
\ba{ll}
(A-B)^2 &=-\frac{1}{R}\left(\frac{\partial\,\Phi}{\partial\,R}
\right)_{z=0},\\
(A-B)(3A+B)&=\left(\frac{\partial^2\,\Phi}{\partial\,R^2}
\right)_{z=0},
\ea
\right\}
\label{eq5.2}
\ee
where $R$ is the distance from the galactic
axis. If the potential is symmetric with respect to the galactic axis
and galactic plane, the values of $A$, $B$ and $C$ at a given $R = R_1$
determine completely the behaviour of the potential in the vicinity of
the galactic plane at $R$ close to $R_1$. Decomposing the potential into a
Taylor series, we have
\be
\ba{ll}
\Phi(R,z)&=\Phi_0(R_1)-(A-B)^2\,R_1\,\Delta\,R+\\
&+\frac{1}{2}(A-B)(3A+B)\Delta\,R^2-\frac{1}{2}C^2z^2 + \dots,
\ea
\label{eq5.3}
\ee
where $\Phi_0$ is the value of $\Phi$ at $z = 0$, and $\Delta\,R =R-R_1$.

Using the expansion (\ref{eq5.3}) and assuming that the potential is
stationary, it is not difficult to determine the character of motion
of stars along  orbits, close to circular\footnote{See for instance
  \citet{Parenago:1954}, § 63}. It is known that this motion can be decomposed into a
circular orbit with angular velocity $\omega = A - B$, on which harmonic
oscillations are superimposed on $R, z$ and galactocentric longitude
$\theta$. Thus the cyclical frequency of oscillations along $z$ is equal to $C$,
so the ratio $\omega/C = (A - B)/C$ is equal to the ratio of period of
oscillations along $z$ to the period of orbit around the galactic
center.
If the potential were spherically symmetric, the orbit  would be flat
and both periods would be equal, hence $\omega/C$ would
be equal to unity. In reality the equipotential surfaces are flattened
and $\omega$ is less than $C$. The square of the $\omega/C$ ratio is evidently the
ratio of the radius of curvature of the equipotential surface at $z = 0$
to the distance to the galactic centre (since the said radius is equal
to  the ratio of $\partial\,\Phi/\partial\,R$ to $\partial^2\Phi/\partial\,z^2$  at $z = 0$).

The relationship between the parameter $C$ and the density of matter in
the galactic plane $\rho$ follows from the Poisson equation. Applying this
equation, \citet{Parenago:1952,Parenago:1954a} and the author
\citep{Kuzmin:1952ab} obtained the formula 
\be
4\pi\,G\,\rho=C^2-2(A^2-B^2),
\label{eq5.4}
\ee
where $G$ is the gravitational constant. With the exception of the
outermost peripheral parts of the Galaxy, where $\rho$ is small, $C^2$ is
significantly larger than $A^2$ and $B^2$. Therefore the value of $\rho$ is
determined mainly by the value of $C$ and vice versa.

Besides Eq.~(\ref{eq5.4}) we can write formulas connecting $C$ and
$\omega= A - B$ with $\rho$:
\be
\left.
\ba{ll}
C^2 &= 4\pi\,G\rho\,f,\\
\omega^2=(A-B)^2&=4\pi\,G\rho\,g.
\ea
\right\}
\label{eq5.5}
\ee
Here the multipliers $f$ and $g$ depend on the mass distribution in the
Galaxy. To calculate $f$ and $g$ we decompose the mass distribution into a
sum of inhomogeneous spheroids, the spheroids being approximate
galactic subsystems. The values $f$ and $g$ can then be found as averaged
values of $f$ and $g$ for the individual spheroids. The averaging is
carried out with a weight proportional to the spheroid density at $z =
0$ and a given $R$. The values of $f$ and $g$ for the $i$-th  spheroid are
calculated by the formulas\footnote{These formulas follow from the known
formulas of attraction of the element ellipsoidal layer (see, for
example, \citet{Zhukowski:1950} p. 763ff.)}.

\be
\left.
\ba{ll}
f_i=&\epsilon_i\int_0^R\left(1-\frac{a^2e_i^2}{R^2}
\right)^{-3/2}\frac{\rho_i(a)}{\rho_i(R)}\frac{a^2\dd{a}}{R^2},\\
g_i=&\epsilon_i\int_0^R\left(1-\frac{a^2e_i^2}{R^2}
\right)^{-1/2}\frac{\rho_i(a)}{\rho_i(R)}\frac{a^2\dd{a}}{R^2},
\label{eq5.6}
\ea
\right\}
\ee
where $a$ is the major semiaxis of the isodensity surface, $\rho_i(a)$ is
the corresponding density, $\epsilon_i$ is the ratio of the minor and major axes
of the isodensity surfaces, and $e_i^2= 1 - \epsilon_i^2$. Instead of the sum of
inhomogeneous spheroids, the distribution of masses in the Galaxy can
also be represented by the sum of homogeneous spheroids, as done by
Oort [2] and Safronov [5]. However, if we do not resort to a very
large number of spheroids, the model turns out to be very crude.

From Eq.~(\ref{eq5.6}) we can see that $f_i$ and $g_i$ are the
greater the higher is the mass
concentration of the spheroid towards the centre. Moreover, the effect
of the mass concentration on $g_i$ is greater than on $f_i$. Moreover, $g_i$ is
more sensitive to changes in the flatness of the spheroid, decreasing
as $\epsilon_i$ decreases. If $\epsilon_i$ is very small, then $g_i$
is proportional to $\epsilon_i$, and  $f_i$ is
close to unity. In this case the spheroid approaches the
plane-parallel layer in its properties. Since the Galaxy is rather
strongly flattened, for most spheroids $f_i$ is close to unity and $f$,
appearing in the first Eq.~(\ref{eq5.5}), should therefore also be close to
unity, which agrees with Eq.~(\ref{eq5.4}) where $C^2$ is significantly larger
than $A^2$ and $B^2$. As to the value $g_i$, at those values of $R$ for which the
influence of mass concentration to the centre of the spheroid is
small, $g_i$ is of order $\epsilon_i$. Therefore if $R$ is not very large, then $g$ in the
second Eq.~(\ref{eq5.5}) has an order of mean value of
$\epsilon_i$.  The square of $\omega/C$ has the same order, since 
it is equal to $g/f$. At large $R$ comparable with the Galactic radius, the influence of
the mass concentration towards the centre of the spheroid on $f_i$ and $g_i$,
and hence on $f$ and $g$, becomes large. Therefore, when approaching the
periphery of the Galaxy, $g$, and then $f$, begin to increase rapidly, tending towards infinity. The ratio
$g/f$ approaches  to unity.

\subsection{}
The method employed by \citet{Oort:1932},
\citet{Parenago:1952,Parenago:1954a} and the author \citep{Kuzmin:1952ab}
 to determine $C$, or generally the dependence of $\Phi$ on $z$, is based on
using the relation between the motion and $z$-coordinate distribution of
stars. Starting from from the six-dimensional continuity equation for
a stationary stellar system with axial symmetry, we obtain the
following expression for the joint distribution function of
$z$-components of velocities and $z$-coordinates of stars \citep{Kuzmin:1952ab}:
\be
F = F[v_z^2-2(\Phi(R,z)-\Phi_0(R))],
\label{eq5.7}
\ee
where $v_z$ is the $z$-component of the pecular velocity\footnote{Beside stationarity and axial
symmetry of the stellar system, in deriving Eq.~(\ref{eq5.7}) we have to
assume that the mean value of the velocity component at $R$-coordinate
is zero for every given value of $v_z$.This condition contradicts a
conclusion of the theory of the stationary Galaxy, according to which
the velocity ellipsoid outside the Galactic plane is inclined to the
latter (see \citet{Kuzmin:1953}). However, it can
be shown that for planar subsystems of the Galaxy the inclination of
the velocity ellipsoid does not cause an appreciable error in the
Eq.~(\ref{eq5.7}). In the case of intermediate and especially spherical
subsystems the error of Eq.~(\ref{eq5.7}) can be significant.}.

For the stellar density $D$ according to Eq.~(\ref{eq5.7}) we obtain:
\be
D=\int_{-\infty}^\infty\,F[v_z^2-2(\Phi-\Phi_0)]\dd{v_z}.
\label{eq5.8}
\ee

If we assume that $v_z$ distribution in a given subsystem of the Galaxy
is normal (Gaussian), Eq.~(\ref{eq5.8}) gives the following expression for
the density of such a subsystem:
\be
D=D_0\,e^{\frac{\Phi-\Phi_0}{\sigma_z^2}}.
\label{eq5.9}
\ee
Here $D_0$ is the value of $D$ at $z = 0$ and $\sigma_z$ is the variance of $v_z$, which
in this case is independent of $z$. Knowing $\sigma_z$ and the dependence of $D$
on $z$ for the subsystem with normal distribution $v_z$ we can derive, by
Eq.~(\ref{eq5.9}), the dependence of $\Phi$ on $z$, hence the value of $C$. The
latter  can also be inferred directly from the dependence of $D$ on $z$ in
the vicinity of the galactic plane. For small $|z|$ we can put,
according to (\ref{eq5.3}),
\be
\Phi(R,z)=\Phi_0(R)-\frac{1}{2}C^2z^2,
\label{eq5.10}
\ee
and Eq.~(\ref{eq5.9}) gives
\be
D=D_0\,e^{-\frac{C^2z^2}{2\sigma_z^2}}.
\label{eq5.11}
\ee
Denoting $C^2/{2\sigma_z^2} =1/{2\zeta^2}$ we get
\be
D=D_0\,e^{-\frac{z^2}{2\zeta^2}}
\label{eq5.12}
\ee
and
\be
C=\frac{\sigma_z}{\zeta}.
\label{eq5.13}
\ee

Thus, if the distribution $v_z$ is normal, then for small $|z|$ the
density distribution turns out to be also normal. To determine $C$, we
need to find $\zeta$ 
 from the dependence of $D$ on $z$ at small $|z|$. Eq.~
(\ref{eq5.13}) then gives $C$.

If the $v_z$ distribution of the given subsystem of the Galaxy differs
from the normal one, then to derive the dependence $\Phi$ on $z$ we have to
apply the general Eq.~(\ref{eq5.8}). Calculating by this formula $D$ as a function
of $\Phi$ and knowing $D$ as a function of $z$ from observations, we find the
required dependence of $\Phi$ on $z$. For small $|z|$ the calculation will
give the dependence of $D$ on $z$, comparison of which with the observed
dependence of $D$ on $z$ gives $C$. In practice it is convenient, following
\citet{Oort:1932}, to expand the distribution $v_z$ into a sum of normal
distributions. Then the dependence $D$ on $\Phi$ is obtained as a sum of
distributions (\ref{eq5.9}).

The $v_z$ distribution, if different  from the normal distribution,
changes as a function of $z$;  it depends on  
$z$ and the value of $\sigma_z$.
This change can be seen from Eq.~(\ref{eq5.7}). It is, however, more
convenient to use a expansion of the normal distribution. If a
distribution $v_z$ is represented as a sum of normal distribution,  added with positive
weights, then $\sigma_z$ obviously increases with $|z|$,
since $D$ under a normal distribution $v_z$ with small $\sigma_z$  decreases
with $|z|$ faster than $D$ under a normal distribution $v_z$ with
large $\sigma_z$.
 Otherwise $\sigma_z$ may also decrease with $|z|$. The relation between
the dependence of $D$ and $\sigma_z$ on $z$ is given by the well-known Jeans
equation
\be
\frac{\partial(D\sigma_z^2)}{\partial\,z}=D\frac{\partial\Phi}{\partial\,z}.
\label{eq5.14}
\ee
In the case of normal distribution of $v_z$ this equation is
equivalent  
to Eq.~(\ref{eq5.9}) since in this case $\sigma_z= \mathrm{const}$ and
(\ref{eq5.9}) is a solution of equation (\ref{eq5.14}). 

In the case of a very flat subsystem, Eq.~(\ref{eq5.10}) can be
assumed to be valid in the entire thickness of the
subsystem. Therefore, in the case of a normal $v_z$ distribution, the
density of the very flat subsystem is normal distribution law
(\ref{eq5.12}), with $\zeta$ having the meaning of the variance of
$z$.  For a very flat subsystem
\be
F=F(v_z^2+C^2\,z^2).
\label{eq5.15}
\ee

It follows from the symmetry of this expression with respect to $v_z$
and $Cz$ that the $z$ distribution in the very flat subsystem is the
same as the total distribution of $v_z$ over the entire thickness of
the subsystem.  The ratio of $v_z$ dispersion to $z$ dispersion equals
$C$. The latter result is not difficult to obtain either by
integrating the Jeans equations twice over $z$, or by applying the
virial theorem to the $z$-coordinate motion. Thus, in the case of very
flat subsystems the Eq.~(\ref{eq5.13}) is applicable for any
distribution, unless $\sigma_z$ is taken as the mean of the variance
$v_z$ over the entire thickness of the subsystem, and $\zeta$ is the
variance of $z$.

\section{Studies by Oort, Parenago, Safronov and  author on $C$
and $\rho$}

\subsection{}
The first reliable data on the Galactic
gravitational potential as a function of $z$ in the solar neighbourhood was 
obtained by 
\citet{Oort:1932}. He used the radial velocities of stars at high galactic
latitudes, the spatial velocities of nearby stars, and the data by \citet{vRhijn:1925}
on the dependence of $D$ from $z$ for stars with different
absolute magnitude. Dividing the stars into groups according to their
absolute magnitude Oort presented the distribution of $z$ velocities of
stars in every group as the sum of two or three normal distributions, or
corresponding stellar density distribution as the sum of two or
three distributions of (\ref{eq5.9}). Further, comparing $D(\Phi )$ with $D(z)$ it
was possible to derive $\Phi (z)$. But Oort did not use this possibility
and solved the problem in a more complicated way by calculating
 $\sigma_z$ and $\partial\Phi /\partial z$ as
functions of $z$ with the help of subsequent iterations.
In  calculating $\partial\Phi /\partial z$ he
used the Jeans equation (\ref{eq5.14}). As a result he derived
$\partial\Phi /\partial z$ as a function of $z$ for $0\le |z|\le $ 600~pc.
Oort found the dependence of $\partial\Phi /\partial z$ from $z$ for $0\le |z|\le$
200~pc to be nearly linear, and that in this range
$-\partial^2\Phi /\partial z^2 = $ $\rm 5.62\cdot 10^{-30} ~s^{-2}$. On the
basis of the number of stars used by Oort in velocity distribution
determination (over 500), the precision of van Rhijn data and the
agreement of the results for different absolute magnitudes of stars we
estimate the mean error of $-\partial^2\Phi /\partial z^2$, derived by Oort,
to be 10--15~per cent. Hence, according to Oort in the solar neighbourhood
$$C = 73\pm 4 {\rm ~km/s/kpc}. $$

To determine the matter density in the solar neighbourhood Oort did not use
the Poisson's equation. Instead he used a method, which in its essence
is the
application of Eq.~(\ref{eq5.5}), where $f$ and $g$ are calculated on the basis of
some probable mass distribution of the Galaxy, chosen in a way to obtain a
realistic value for $\omega^2/C^2 = g/f$. He analysed four models of the
Galaxy consisting of homogeneous spheroids, while for two of them he
assumed the presence of a massive local system. Different models gave slightly different
values for the density in the vicinity of the Sun. As the final value Oort
accepted the mean of them, namely $\rm 6.3\cdot 10^{-24} ~g/cm^3$. This
value nearly coincides with the one resulting from Eq.~(\ref{eq5.4}). Taking $C$
according to Oort and $A= $ 20~km/s/kpc and $B=-13$~km/s/kpc, for
example, we have
$$\rho = (6.1\pm 0.8)\cdot 10^{-24} ~\mathrm{g/cm^3} = 0.09\pm 0.01
~\mathrm{M_{\odot}/pc^3} ,$$
where the mean error is determined in fact by the error of $C$, as
$C^2$ significantly exceeds $A^2$ and $B^2$.

Our determination of $C$ was done in 1952, twenty years after
Oort. As the observational data we used proper motions of A and gK stars near the
galactic equator, and the data on the spatial
distribution of these classes of stars. The proper motions were taken
from the General Catalogue by \citet{Boss:1937}, the spatial
distribution from the paper by \citet{Pannekoek:1929}. The
distribution of velocity $z$-component was assumed to be normal for
both classes of stars, and Eqs.~(\ref{eq5.12}) and
(\ref{eq5.13}) were used. We derived the following results.

A-stars: $\sigma_z = 5.1\pm 0.7$~km/s, $\zeta = 99\pm 5$~pc, $C = 52\pm 
7$~km/s/kpc.

gK-stars: $\sigma_z = 12.2\pm 1.7$~km/s, $\zeta = 202\pm 20$~pc, $C = 60
\pm 8$~km/s/kpc.

A significant fraction of the mean error of $\sigma_z$, and nearly all the mean
error of $\zeta$ are caused by the error of mean parallax, prescribed to A and gK
stars of a given visible magnitude. As the error influences
$\sigma_z$ and $\zeta$ in the same manner, when calculating $C$ from
Eq.~(\ref{eq5.13}), this error eliminates. Thus the real mean error of $C$ is less than
the error calculated from formal rules.

On the basis of values of $C$ given above the final value was found:
$$C= 56\pm 5 {\rm ~km/s/kpc},$$
giving according to Eq.~(\ref{eq5.4})
$$\rho = (3.4\pm 0.8)\cdot 10^{-24} {\rm ~g/cm^3} = 0.05\pm 0.01 {\rm
~M_{\odot}/pc^3}. $$
Comparison of our results with Oort's indicates that differences are
significantly larger than it could be expected on the basis of mean errors.
The mean density in the vicinity of the Sun results nearly twice less than
the Oort's value.

In addition to the values of $C$ and $\rho$, we determined also the
flatness of the Galaxy. This was done on the basis of the relation
$\omega^2/C^2 = g/f$. The mass distribution of the Galaxy was represented
by one inhomogeneous spheroid with $\rho (a)a^2/\rho (R)R^2 =1$ and
consequently, according to Eq.~(\ref{eq5.6}) $g=\epsilon\arcsin e /e$ and $f=1$
giving 
\begin{equation}
\frac{\omega^2}{C^2} = \epsilon \frac{\arcsin e}{e}. \label{eq5.16}
\end{equation}
As we may decide on the basis of the rotation law for flat subsystems, the
used effective mass concentration of the spheroid in the centre
approximately corresponds to the real Galactic central mass concentration.
By using the formula above we found that
$$\epsilon = 0.26\pm 0.06.$$
The flatness of the Galaxy is thus quite moderate.

\subsection{}

Nearly simultaneously with our study there appeared the paper by
P.~P.~Parenago on the potential of the Galaxy \citep{Parenago:1952},
where the value of $C$ was derived on the basis of spatial
distribution of Mira stars and of short- and long-period cepheids
collected by \citet{Kukarkin:1949}. By using Eq.~(\ref{eq5.9}) and
ascribing to these classes of stars somewhat mutually related but
otherwise quite arbitrarily chosen values of $\sigma_z$, Parenago
derived $\Phi$ as a function of $z$ for quite long interval of $z$ (up
to $|z|=$ 10~kpc). On that basis he found $C^2/2 = 2550\pm 120$ giving
$C= 71\pm 2$~km/s/kpc. The mean error of the result is highly
underestimated due to ignoring the uncertainties in values of
$\sigma_z$. In a subsequent paper \citet{Parenago:1954a} calculated
$C$ by using for short-period cepheids new absolute luminosity values,
derived by \citet{Pavlovskaya:1954}, and using also the value of
$\sigma_z$, derived for representatives of the spherical subsystems and
ascribed to these stars. As a result he obtained
$$C = 73\pm 14 {\rm ~km/s/kpc}.$$
In this case the mean error better characterises the real precision of
the derived $C$ value, the mean error of $\sigma_z$ was taken into account. 
In addition, the mean error of
absolute luminosities  of short-period cepheids was also taken into account.

As we see, the value of $C$, derived by Parenago, is similar to Oort's one.
But the Parenago's result is very uncertain. Probably the uncertainty is
even larger than it results from the mean error calculated by him. The
main reason of high uncertainty is the fact that the data on the stellar
density of the subsystem of Mira stars and short-period cepheids is absent
or is very uncertain for $|z|<$ 1~kpc, and the data on the density of
long-period cepheids is very uncertain or nearly absent for $|z|>$
0.1~kpc. In that case the relation of the value of $\sigma_z$ for
long-period cepheids to that for Mira stars and short-period cepheids
with the help of Eq.~(\ref{eq5.9}) is extremely uncertain. But $C$ is determined just by 
the value of $\sigma_z$ for the long-period cepheids, because the dependence of
$\Phi$ from $z$ for small $|z|$ is determined only by these stars. It may
be that Parenago's result is no more certain than the
determination of $C$ directly from the observed values of $\sigma_z$ and
$\zeta$ for long-period cepheids. In our paper on the determination of $C$
for long-period cepheids it was found  that $\sigma_z = 2.9\pm 1.0$~km/s and
$\zeta = $ 50~pc, giving $C= 58\pm 20$~km/s/kpc. The mean error of
$C$ is determined by the mean error of $\sigma_z$ (the mean error of
$\zeta$ is relatively small). Because $\sigma_z$ is determined from
proper motions, the possible error for zero-point of period-luminosity
relation of long-period cepheids influences $\sigma_z$ and $\zeta$ in the
same way,  and hence in calculation of $C$ this error eliminates. As we
see, $C$ results to be
more similar to our value. But again the result is very uncertain.

To have the final values for $C$ in the vicinity of the Sun from the data
on the spatial distribution of stars, Parenago used also our
determination of $C$. Besides, he made quite significant corrections to our
value of $\zeta$, because the Pannekoek's data on the spatial distribution
of A and gK stars is distorted by interstellar absorption. As a result of
such correction, instead of $C=56\pm 5$ the value $C=68\pm 7.5$ was
obtained (more correct would be to take $\pm 6$ because the errors of mean
parallax in calculation of $C$ are eliminated, see above). However, we
can not agree with the Parenago's correction, because when we derived
$\zeta$ from Pannekoek's data, the correction for absorption was taken into
account. Pannekoek found the distribution of A and gK stars without taking
into account the absorption and the dispersion of absolute magnitude of
these stars. To derive $\zeta$ from his data, we used the distribution
function of absolute magnitude of A and gK stars brighter than 6.0
magnitude \citep{Opik:1933}, reduced to the galactic equator. This
function is not the distribution of true absolute magnitude $M$, but is
the distribution of absolute magnitude, influenced by the absorption $M' =
m+5+\log\pi$, where $m$ 
is the apparent magnitude and $\pi$ is parallax. While deriving $\zeta$ we
used that part of Pannekoek's data, where the absolute magnitude of stars
is distorted by absorption approximately in the same way as it is for stars
brighter than $6^m.0$ near the galactic equator. Hence the result for
$\zeta$ must be more or less free from absorption.

In addition to the data on the motion and spatial distribution of stars,
Parenago used for the determination of $C$ also the matter density near
the solar neighbourhood. He used the value, resulting directly from the
observations $(6.0\pm 0.5)\cdot 10^{-24} {\rm ~g/cm^3}$, and being the sum of
stellar and diffuse matter densities, with $2.8\cdot 10^{-24} {\rm
~g/cm^3}$ from the stellar matter, and the remaining part from the diffuse
matter, mostly from the interstellar gas. The used value of the density
coincides with the 
value resulting from Oort's $C$ value. As the density of the
interstellar matter is very uncertain, the real error of
Parenago's result may be again significantly higher than the one, estimated by him.

A little before the publication of our paper in 1952 there appeared a
determination of the matter density in the vicinity of the Sun by
\citet{Safronov:1952}. The density was determined with the help of two
methods. In the first one the density was calculated according to
Oort's method, \ie  by using of Eq.~(\ref{eq5.5}). Because $C$ was not
determined anew but Oort's value was used, the derived matter
density nearly coincides with the Oort's result. This result can not
be handled as independent. In the second method the second
equation of (\ref{eq5.5}) was used together with the equation
\begin{equation}
4\pi G\rho ~(f-1) = 2\omega (2 A - \omega ), \label{eq5.17}
\end{equation}
being a consequence of the first equation of (\ref{eq5.5}) and of Eq.~(\ref{eq5.4}). To have
the same value for $\rho$ from both equations one needs to choose the
Galactic mass distribution in a way that $(1 - f)/g$ will be equal to 
$2(2A/\omega - 1)$. The most realistic mass distribution model gave the
value of $\rho$ again near to the Oort's value. This result is an
independent determination. But the precision of the derived value of
$\rho$ is not large, because the data on the Galactic mass distribution
is quite uncertain. The result depends highly on the eccentricity of the
Galaxy. In addition, it is sensitive to the value of $A/\omega $,
determining the Galactic central mass concentration.

\section{Possible systematic errors in determination of $\sigma_z$ by
Oort and the author}

As we saw, our value of $C$ is significantly smaller than the value of $C$
derived by Oort, in fact the difference in values exceed several times
the mean errors of both results. Such a large difference refers that
the results are influenced by rather significant systematic
errors. According to the results, obtained by Parenago and Safronov, it seems
 that the results, obtained by Oort, are more close to reality. However, the
results, obtained by Parenago and Safronov, are not sufficiently certain and
thus, it may be that the result obtained by Oort can be significantly too
high. In order to clarify at least in part the reasons of differences
we review below some possible systematic errors in both determinations
of $C$. Let us begin with the values of $\sigma_z$.

When comparing our results with those of Oort, large systematic
differences are seen. Of course, this comparison is possible only for A and
gK stars, on the basis of which $C$ was determined in our
study. But these stars contribute with a quite significant weight to
Oort's data.

Distribution of $v_z$ for high luminosity stars, in particular for A
and gK stars, was determined by Oort from radial velocities. Radial
velocities for stars between galactic latitudes $\rm \pm 40^o$ and
$\rm \pm 90^o$ were used, also ellipsoidal velocity distribution 
was taken into account. According to the data taken by Oort, the mean
value $|v_z|$ for A and gK stars is $8.8 \pm 0.7$~km/s and $15.3 \pm
1.1$~km/s, respectively. In determination of these values also B6 --
B9 stars were included into A stars, as our A stars correspond to
spectra B8 -- A5 in Harvard classification. Mean errors in these
values were estimated from the numbers of A and gK stars used by Oort
(97 and 112, respectively). To derive dispersions from the mean values
$|v_z|$ one needs to multiply $v_z$ with a coefficient, depending
on the distribution of $v_z$. According to Oort, the distribution of
$v_z$ is quite near the normal distribution and thus it is
justified to take the coefficient equal to $\sqrt{\pi /2}$,
corresponding to precisely normal distribution. In this case we derive
the following values for $\sigma_z$.

\begin{tabular}{lll}
 & A stars & gK stars \\
 Oort  & $\sigma_z = 11.0 \pm 0.9$~km/s, & $\sigma_z = 19.2 \pm 1.4$~km/s \\ 
 Kuzmin & $\sigma_z = 5.1 \pm 0.7$~km/s & $\sigma_z = 12.2 \pm 1.7$~km/s
 \end{tabular}
 
 As we see, the values by Oort are 1.5 -- 2 times larger than our values.
 
\section{Probable values of $C$, $\rho$ and $\epsilon$}

As we have seen, the values of $\sigma_z$ and $\zeta$, following from
the works of Oort and the author, are burdened by various systematic
errors. Unfortunately, not all errors can be accounted for without a
very detailed analysis of the observational data used. Therefore,
based only on what has been outlined above, it is difficult to make a
confident correction to the $C$ values obtained by Oort and the
author. However, some attempt can still be made to deduce the most
probable value of $C$.

We were unable to find significant systematic errors in our values of
$\sigma_z$. The values of $\zeta$, according to the results of the
previous section, should be reduced by 8\% due to of the error of the
used magnitude scale. In addition, $\zeta$ needs to be further reduced
by reduction to $z = 0$ by 2\%.  As a result, $\zeta$ will decrease by
10\%, while $C$, by Eq.~(\ref{eq5.13}), will increase by the same
proportion. The values of $\sigma_z$ at Oort, need to be reduced on
the average by 10\%. On the other hand, the values of $\zeta$,
corresponding to the Oort material, should probably be reduced, though
not by as much. Reducing the Oort value $C$ due to an
$\sigma_z$ error by 10\% , and increasing it due to a systematic error
of $\zeta$ (and a reduction to $z = 0$) by 6\%, we took as the corrected
Oort result, which we present below
together with our corrected value.
\be
\ba{ll}
{\rm Oort~~ (corrected):} &C = 70 \pm 5~~{\rm km/sec/kpc},\nonumber \\
{\rm Kuzmin~~ (corrected):}~& C = 62 \pm 5~~{\rm  km/sec/kpc}.\nonumber
\ea
\ee

The average error of the Oort result is slightly increased due to the
uncertainty of the correction for $\zeta$.The difference between both
determinations is now only $8 \pm 7$, \ie it is within the range of
random errors. The average for both determinations is
\be
C=66\pm 4 ~~{\rm  km/sec/kpc}. \nonumber
\ee

According to \citet{Parenago:1954a} and  \citet{Safronov:1952}, the density of
matter in the vicinity of the Sun is close to the Oort  
uncorrected value. Estimating the error of this result to be $15-20$\%,
 we have
\be
C = 73 \pm 6~~{\rm  km/sec/kpc}. \nonumber
\ee

If we take the weighted average of both $C$ values, we  obtain: 
\be
C = 68 \pm 3~~{\rm  km/sec/kpc}. \nonumber
\ee
and for the density:
\be
\rho= (5.2 \pm 0.5) \times 10^{-24} {\rm g/cm}^3 = 0.077 \pm 0.008\,
M_\odot/pc^3. \nonumber
\ee

As to the Galaxy flatness, using Eq.~(\ref{eq5.16})
and assuming
$\omega=33\pm 2$ km/sec/kpc ($A=20\pm 1$ and $B= - 13\pm 2$), we will
have  an average ratio of half-axes of flatness surfaces
\be
\epsilon= 0.16 \pm 0.03. \nonumber
\ee

The density thus obtained is 1.5 times the value we derived earlier
\citep{Kuzmin:1952ab}. 
The value of $\epsilon$ has responsibly decreased. Instead of the ratio of
half-axis surfaces of isodensity 1:4 we now have the ratio 
1:6. The flattening of the Galaxy is not as small as we previously
believed. However, it is still rather small, so we still have
to assume that the mass of the spherical component of the Galaxy is
quite large.  This is all the more true because the value $\epsilon$ obtained
is the result of averaging for different subsystems of the Galaxy,
where the weight is not the mass of the subsystem, but approximately
its density in the vicinity of the Sun (see section 5.1). If the weight in the
averaging were mass, $\epsilon$ would be much larger.

The above values of $C$, $\rho$ and $\epsilon$ do not, of course, in
any way claim to be definitive values derived from modern
observational data. They can only be used as some guide values until a
more careful determination of $C$ can be made using all the currently
available data on stellar motions and on the spatial distribution
of stars. \footnote{The
real value of $C$ is probably slightly larger than the one derived here. But
on the other hand, $A$ and $B$ must be significantly decreased. Thus
$\rho$ increases up to $\rm 0.09 ~M_{\odot}/pc^3$ and $\epsilon$ decreases
to 0.08. Further determinations of $C$ were made in Tartu by
\citet{Eelsalu:1958},  \citet{Einasto:1964},  \citet{Joeveer:1972tv, 
   Joeveer:1974wa, Joeveer:1975vs, Joeveer:1976te}.  The final
result of the \citet{Eelsalu:1958} analysis is: $C=67 \pm
3$~km/s/kpc.   
The mean value of the  J\~oeveer analysis is $C=70$ ~km/s/kpc, and $\rho_{dyn}=0.09\,
M_\odot/kpc^3$. [Later footnote.]} 

\vglue 3 mm
\hfill 1955

%% file: chapter06.tex
\chapter[The model of the steady galaxy with triaxial distribution of
velocities]{The model of the steady galaxy allowing the triaxial
  distribution of velocities\footnote{\footnotetext
    ~~Astron. Zhurnal, vol. 33, pp. 27--45, 1956 = Tartu
    Astron. Obs. Teated No. 2, 1956.} } 

In order to explain the triaxial velocity distribution within the
theory of the stationary galaxy, in addition to the energy and the
angular momentum integrals, one needs to use a third single-valued
integral. One such integral is suggested in our recent papers
\citep{Kuzmin:1953, Kuzmin:1954}. Our third integral is not related to
the gravitational potential with so strict restrictions as the
$z$-coordinate energy integral proposed by \citet{Lindblad:1933b}, and
explains better the triaxial velocity distribution for significantly
less-flattened Galactic subsystems than the Lindblad's integral. It
may even be that the restrictions on the gravitational potential,
resulting from our third integral, are valid for all Galactic
subsystems, including the spherical ones. In order to clarify this
point, one needs to construct a model for the Galaxy with the referred
restriction being valid, and to compare it with the real Galaxy. In
the paper referred above \citep{Kuzmin:1954} we presented some results
for this kind of analyse, and the resulting model seems quite
realistic. However, at that time we were able to give only preliminary
results. More sophisticated studies of Galactic models, constructed
according to the referred principle, were made later. These results
are summarised in the present paper.

\section{Single-valued integrals of motion and the velocity
distribution} 

Spatio-kinematical  structure of a stellar system, for example
of the Galaxy or of some of its subsystems, is described by the
distribution of stars in the six-dimensional phase space, where the
coordinates are three coordinates of ordinary space and three velocity
components. If we neglect the stellar encounters, the phase density becomes
a function of integrals of motion in the ``smoothed'' gravitational field
of the system. The number of independent integrals of
motion is six. In case of stationarity, the phase density becomes a function
of only time-independent integrals of motion. Although in this case we
have in total five independent of each other and time-independent
integrals of motion, some of them are infinitely multiple-valued, 
and can not be included in the expression of the phase density
\citep{Kuzmin:1953, Pahlen:1947}.

If the gravitational potential of a stellar system, aside from
stationarity, is not restricted further, the only single-valued and
time-independent integral of motion is the energy integral
\begin{equation}
I_1 = v_R^2 +v_{\theta}^2 + v_z^2 - 2\Phi , \label{eq6.1.1}
\end{equation}
where $v_R$, $v_{\theta}$, $v_z$ are the velocity components in cylindrical
coordinates $R$, $\theta$, $z$, and $\Phi$ is the gravitational potential.
As we noted already, the condition for the existence of this integral is
the stationarity of the potential
\begin{equation}
\frac{\partial\Phi}{\partial t} = 0,  \label{eq6.1.2}
\end{equation}
where $t$ is time. Additional restrictions for the potential are needed to enable
 the existence of other single-valued integrals of motion. If we
suppose the axial symmetry of the potential, we get the angular momentum
integral 
\begin{equation}
I_2=R v_{\theta} ,  \label{eq6.1.3}
\end{equation}
where we assume that the symmetry axis of the potential coincides with
the axis of cylindrical coordinates, \ie 
\begin{equation}
\frac{\partial\Phi}{\partial\theta}=0. 
 \label{eq6.1.4}
 \end{equation}
In the theory of the stationary Galaxy these two integrals are
usually taken into account. But if we assume the phase
density to be a function of only these two integrals, we must conclude
that the velocity distribution has axial symmetry about
$v_{\theta}$-axis. This is in contradiction with observations,
which show that the velocity distribution is triaxial.

To remove the above mentioned contradiction between the theory and the
observations, a third single-valued integral of motion needs to be introduced. 
In our earlier papers \citep{Kuzmin:1953, Kuzmin:1954} we proposed the third 
integral to be in a form
\begin{equation}
I_3= (Rv_z-zv_R)^2+z^2v_{\theta}^2+z_0^2(v_z^2-2\Phi^*),  \label{eq6.1.5}
\end{equation}
where $z_0$ is a constant with dimension of length and $\Phi^*$ is the
function satisfying following equations:
\begin{equation}
z_0^2 \frac{\partial\Phi^*}{\partial R} = z^2 \frac{\partial\Phi}{\partial
R} - Rz \frac{\partial\Phi}{\partial z}, ~~~~
z_0^2 \frac{\partial\Phi^*}{\partial
z} = (R^2+z_0^2) \frac{\partial\Phi}{\partial
z} - Rz \frac{\partial\Phi}{\partial R}.  \label{eq6.1.6}
\end{equation}
It is easy to see that the integral $I_3$ satisfies the necessary condition
for every integral of motion $DI_3/Dt = 0$. In addition, 
for $I_3$  to exist, the integrability condition must hold 
for Eq.~(\ref{eq6.1.6}). This condition has the form
\begin{equation}
 3 \left( z \frac{\partial\Phi}{\partial R} - R \frac{\partial\Phi}{\partial z}
\right) - (R^2+z_0^2-z^2) \frac{\partial^2\Phi}{\partial R\partial z} +
Rz\left( \frac{\partial^2\Phi}{\partial R^2} - \frac{\partial^2\Phi}{\partial
z^2} \right) = 0.  \label{eq6.1.7} \end{equation}
This is the restriction for the potential due to the integral $I_3$.

If we assume that the phase density is a function of the three integrals
mentioned above, the resulting velocity distribution is symmetric across two
orthogonal planes through the $v_{\theta}$ axis. The velocity dispersion
ellipsoid is triaxial in agreement with the observations. One of the
ellipsoid's axis coincides with the $v_{\theta}$ axis, the remaining two
have an inclination with respect to the $v_R$ and $v_z$ axes.

The three axes of the velocity ellipsoid determine a triple family of so-called
 main velocity surfaces, \ie the surfaces that are perpendicular to the 
 axes of the velocity ellipsoid at every point. After transforming
the $I_3$ into a sum of squares of velocity components, it is easy to
see \citep{Kuzmin:1953} that the main velocity surfaces are the families of
the second-order confocal surfaces, with the common foci on the symmetry
axis of the system (galactic axis) at points with $z=z_0$ and $z=-z_0$.
The families of the main velocity surfaces consist of ellipsoids of revolution, 
two-sheeted hyperboloids of revolution, and of planes $\theta = \mathrm{const}$.
These intersecting orthogonal surfaces may be handled as the coordinate
surfaces of the curvilinear orthogonal coordinates $\xi_1$, $\xi_2$,
$\theta$. Relation between the coordinates $\xi_1$, $\xi_2$ and $R$,
$z$ is
\begin{equation}
R = z_0\sqrt{(\xi_1^2-1)(1-\xi_2^2)}, ~~~~ z=z_0\xi_1\xi_2.  \label{eq6.1.8}
\end{equation}

\begin{figure*}[ht]
\centering
\includegraphics[width=70mm]{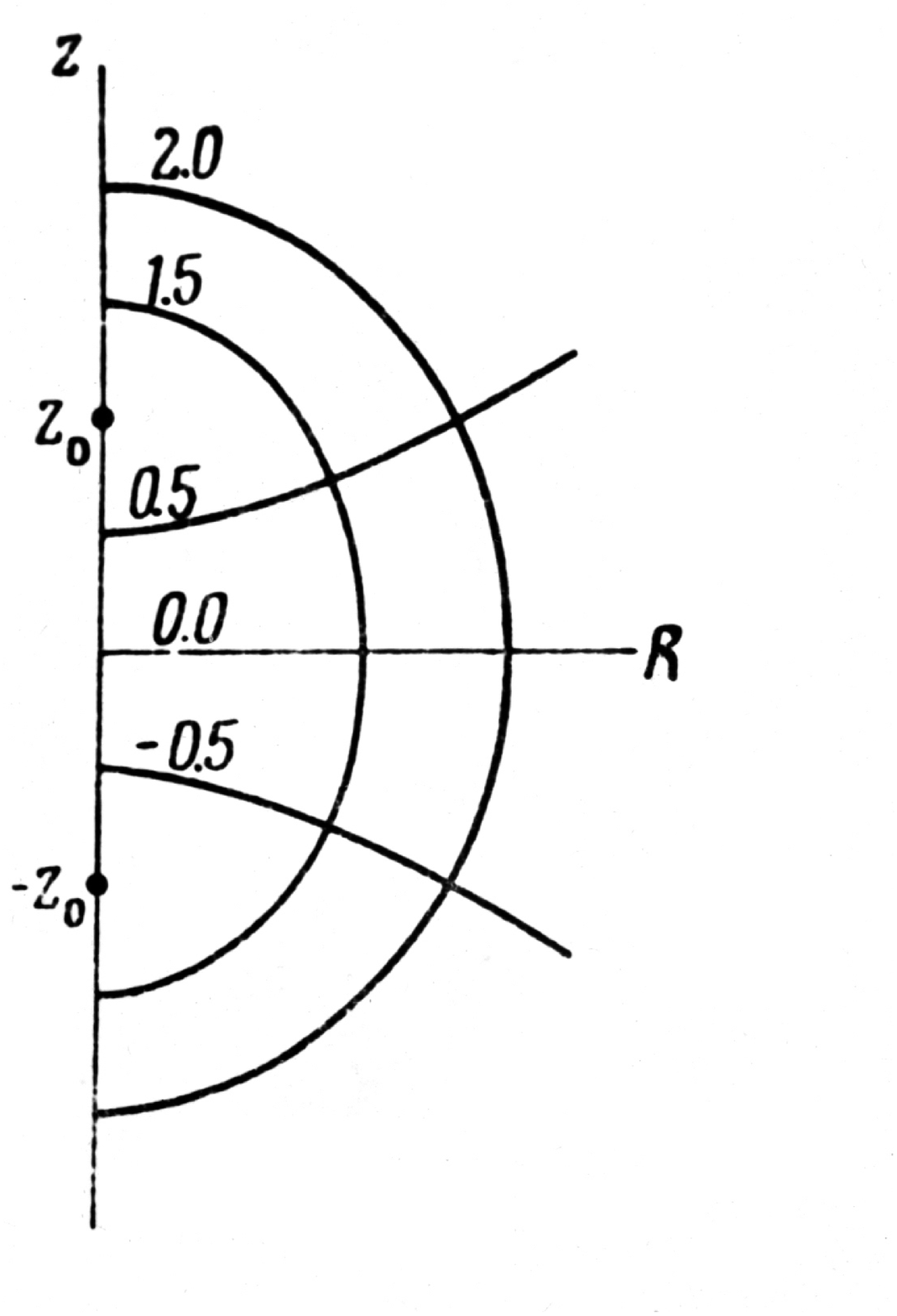}
\caption{Meridional intersections of ellipsoids and hyperboloids.}
\label{fig6.1}
\end{figure*}

Taking $z_0$ as the unit of length, $\xi_1$ equals to the long semi-axis
of the ellipsoids and $|\xi_2|$ to the real axis of hyperboloids.
Hence, for the galactic axis $\xi_1=|z|/z_0$ for $|z|\ge z_0$ and $\xi_2=
z/z_0$ for $|z|\le z_0$. In Fig.~\ref{fig6.1} the meridional intersection of some
ellipsoids and hyperboloids are plotted, while the labels on the curves are
the values of $\xi_1$ and $\xi_2$.

After expressing $I_1$, $I_2$, $I_3$ as functions of coordinates $\xi_1$, $\xi_2$,
$\theta$, and of respective velocity components $v_1$, $v_2$, $v_{\theta}$, 
we obtain \citep{Kuzmin:1953}
\begin{equation}
\begin{array}{ll}
I_1= & v_1^2+v_2^2+v_{\theta}^2 -2\Phi ,\\
\noalign{\smallskip}
I_2= & \sqrt{(\xi_1^2 -1)(1-\xi_2^2)}v_{\theta} ,\\ 
\noalign{\smallskip}
I_3= & \xi_2^2v_1^2+\xi_1^2v_2^2+\xi_1^2\xi_2^2v_{\theta}^2 -2\Phi^* ,\\
\end{array}
\label{eq6.1.9}
\end{equation}
where in the expressions for $I_2$ and $I_3$, we neglected the factors $z_0$
and $z_0^2$ respectively.

\section[Expressions for the potential and the density]{The
  expressions for the potential and the density. The 
conditions for the non-negative density and finite mass}

After introducing the coordinates $\xi_1$ and $\xi_2$, we can rewrite
Eqs.~(\ref{eq6.1.6}) and (\ref{eq6.1.7}) in quite simple form \citep{Kuzmin:1953}
\begin{equation}
\frac{\partial\Phi^*}{\partial\xi_1} = \xi_2^2 \frac{\partial\Phi}{\partial
\xi_1} , ~~~~~ \frac{\partial\Phi^*}{\partial\xi_2} = \xi_1^2 \frac{\partial\Phi
}{\partial\xi_2} \label{eq6.2.1}
\end{equation}
\begin{equation}
\frac{\partial^2}{\partial\xi_1\partial\xi_2} \left[ (\xi_1^2-\xi_2^2)\Phi
\right] =0. \label{eq6.2.2} 
\end{equation}
After solving these equations, we obtain the expressions for the potential
$\Phi$, and for the function $\Phi^*$
\begin{equation}
\Phi = \frac{\varphi_1-\varphi_2}{ \xi_1^2-\xi_2^2} ,\label{eq6.2.3}
\end{equation}
\begin{equation}
\Phi^* = \frac{\xi_2^2\varphi_1-\xi_1^2\varphi_2 }{ \xi_1^2-\xi_2^2},
\label{eq6.2.4} 
\end{equation}
where $\varphi_1$ is an arbitrary function of $\xi_1$, and $\varphi_2$
an arbitrary function of $\xi_2$. Due to $\xi_1\ge 1$ and $\xi_2\le
1$, $\varphi_1$ and $\varphi_2$ may be handled as two different parts of the same
function $\varphi (\xi )$. If this function is continuous and even, the
functions $\Phi$ and $\Phi^*$ are continuous and symmetrical about the
plane $z=0$ (galactic plane).

After applying the Poisson's equation, we can derive an expression for the
mass density. From Eqs.~(\ref{eq6.2.3}) and (\ref{eq6.1.8}) we find \citep{Kuzmin:1953}
\begin{equation}
4\pi Gz_0^2\rho = \frac{2(\varphi_1-\varphi_2-\xi_1\varphi_1'+\xi_2\varphi_2')
(2-\xi_1^2-\xi_2^2) }{ (\xi_1^2-\xi_2^2)^3} -
\frac{(\xi_1^2-1)\varphi_1'' -(1-\xi_2^2)\varphi_2'' }{ (\xi_1^2-\xi_2^2)^2}
, \label{eq6.2.5} 
\end{equation}
where $G$ is the gravitational constant, $\rho$ is the mass density and
apostrophes stand for derivatives.

If the potential is given by Eq.~(\ref{eq6.2.3}) and the density by
Eq.~(\ref{eq6.2.5}), the integral (\ref{eq6.1.9}) exists as a precise integral of motion.
Our problem is to clarify, whether we can construct on the basis of these
formulae a model of the Galaxy, which would correspond at least 
approximately to the reality.

To solve our problem we must first clarify, whether Eqs.~(\ref{eq6.2.3}) and (\ref{eq6.2.5}) have
physical meaning, \ie can we choose the function $\varphi (\xi )$ in a
way that the resulting density will be non-negative everywhere, and the total mass will be
finite. It is easy to see, that this kind of function $\varphi (\xi )$ exists. 

Let us introduce a new arbitrary function $\psi (\xi )$, proportional to
the mass density on the galactic axis
\begin{equation}
\psi (\xi )=4\pi Gz_0^2\rho_{R=0}~ (\xi =z/z_0). \label{eq6.2.6}
\end{equation}
Taking into account, that on the symmetry axis either $\xi_1^2=1$ or $\xi_2^2=1$,
after multiplication of Eq.~(\ref{eq6.2.5}) by $(1-\xi_2^2)^2$ or by
$(\xi_1^2-1)^2$, we have the following equation 
\begin{equation}
(\xi^2-1)^2\psi +2(\varphi -\xi\varphi ')+(\xi^2-1)\varphi '' = 
\mathrm{const}. \label{eq6.2.7}
\end{equation}
Equation (\ref{eq6.2.7}) relates the function $\psi$ to $\varphi$, enabling to
calculate the first function when the second function is known. After
differentiating  Eq.~(\ref{eq6.2.7}), dividing it by $(\xi^2-1)$ and
integrating, we have in addition the following equation
\begin{equation}
(\xi^2-1)\psi + 2\int_0^{\xi}\psi\xi \rmd \xi +\varphi '' = \mathrm{const}.
\label{eq6.2.8} 
\end{equation}
After integrating the last equation twice and demanding that the
function is even, we have
\begin{equation}
\varphi = \xi\int_0^{\xi} (1+\xi^2)\psi \rmd \xi -
(1+\xi^2)\int_0^{\xi}\psi\xi \rmd \xi + c_1\xi^2 +c_2. \label{eq6.2.9}
\end{equation}
The last equation enables to calculate the function $\varphi$ when the
function $\psi$ is given. This equation includes two arbitrary constants
$c_1$ and $c_2$. If we substitute Eq.~(\ref{eq6.2.9}) into Eq.~(\ref{eq6.2.3}), the second
constant factors out, and the first becomes an arbitrary additive term of
the potential.

Using Eqs.~(\ref{eq6.2.7}) and (\ref{eq6.2.8}), we can eliminate from Eq.~(\ref{eq6.2.5}) the
function $\varphi$, and express the density via new arbitrary
function $\psi$. Substituting in Eq.~(\ref{eq6.2.5}) $\varphi -\xi\varphi '$ and
$\varphi ''$ according Eqs.~(\ref{eq6.2.7}) and (\ref{eq6.2.8}), we find
\begin{equation}
4\pi Gz_0^2\rho = g_1^2\psi_1 + 2g_1g_2\psi_{12} +g_2^2\psi_2 , 
\label{eq6.2.10}
\end{equation}
where
\begin{equation}
g_1={\xi_1^2-1\over\xi_1^2-\xi_2^2}, ~~~~
g_2={1-\xi_2^2\over\xi_1^2-\xi_2^2} , \label{eq6.2.11} 
\end{equation}
$\psi_1=\psi (\xi_1),$ $\psi_2=\psi (\xi_2)$ and
\begin{equation}
\psi_{12} = \frac{2\int_{\xi_2}^{\xi_1} \psi\xi \rmd \xi }{ \xi_1^2-\xi_2^2} 
\label{eq6.2.12}
\end{equation}
As the coefficients $g_1$ and $g_2$ are non-negative, we may
conclude from Eq.~(\ref{eq6.2.10}), that the sufficient and necessary condition
of the non-negative density is that the function $\psi$ must be
non-negative, \ie the density on the galactic axis must be non-negative.
From Eq.~(\ref{eq6.2.10}) it is easy to derive the condition for finiteness 
of the system mass. According to Eq.~(\ref{eq6.2.10}) for large $\xi_1$
\begin{equation}
4\pi Gz_0^2\rho = \psi_1 + \left[ 4(1-\xi_2^2)\int_{\xi_2}^{\xi_1}
\psi\xi \rmd \xi + (1-\xi_2^2)\psi_2 \right] \xi_1^{-4}. \label{eq6.2.13} 
\end{equation}
From Eq.~(\ref{eq6.1.8}) results that for large $\xi_1$ the distance $r$ from the
center is also large and equals to $z_0\xi_1$. Therefore,
the mass of a system is finite, when for large
$|\xi |$ the function $\psi$ decreases more rapidly than $|\xi |^{-3}$.
In this case for large $r$ the density decreases with $r$ at least as
$r^{-3}$ and integration gives a finite mass. The mass is easy to
express via $\psi$. For a finite mass for large $r$ the potential is 
$GM/r + \mathrm{const}$, where $M$ is the mass. From Eq.~(\ref{eq6.2.3}) results that
$GM/z_0$ equals to the linear part of $\varphi$ for large $\xi$.
Therefore, by using also Eq.~(\ref{eq6.2.9}) we derive
\begin{equation}
\frac{GM}{ z_0}= \int_0^{\infty} (1+\xi^2)\psi \rmd \xi. \label{eq6.2.14}
\end{equation}

We may conclude, that it is possible to choose a function
$\varphi$ in a way,  that gives us a non-negative density and a finite mass.

\section{Expressions for the density in case of very flattened
stellar systems}

As the density on galactic axis must decrease with increasing
 $z^2$, the function $\psi$ must decrease with $\xi^2$. In this
case the density decreases with $z^2$ not only on galactic axis, but also
outside of it for any $R$ and $z$ (Eq.~\ref{eq6.2.10}). Let us assume that
$\psi$ decreases with $\xi^2$ so rapidly that for small $\xi^2$ in
Eq.~(\ref{eq6.2.10}) we may also neglect the second term when compared with
the third. In this case near the galactic plane we have the
following expression for the density
\begin{equation}
4\pi Gz_0^2\rho = \psi (\xi_2)\xi_1^{-4}. \label{eq6.3.1} 
\end{equation}
According to Eq.~(\ref{eq6.1.8}) we may suppose
\begin{equation}
\xi_1 = \left( 1+ \frac{R^2}{ z_0^2}\right)^{1/2}, ~~
\xi_2= \frac{z}{ z_0}\xi_1^{-1}. \label{eq6.3.2}
\end{equation}
From Eq.~(\ref{eq6.3.1}) it follows that the density in galactic plane varies as
\begin{equation}
\rho_{z=0}=\rho_0 \left( 1+ \frac{R^2}{ z_0^2}\right)^{-2}, \label{eq6.3.3} 
\end{equation}
where $\rho_0$ is the central density.

By denoting the ratio of $z$ axis of isodensity
surface to $R$ axis by $\epsilon$, we have according to Eqs.~(\ref{eq6.3.3})
and (\ref{eq6.2.6}) 
\begin{equation}
\psi =\psi (0)\left( 1+ \frac{\xi^2}{\epsilon^2} \right)^{-2}, \label{eq6.3.4}
\end{equation}
where $\epsilon$ is a function of $\xi$. Substituting (\ref{eq6.3.4})
into (\ref{eq6.2.10}), we realise that the relative error of Eq.~(\ref{eq6.3.1}) near the
galactic plane is of order $\epsilon^2$. This means that Eq.~(\ref{eq6.3.1}) may
be used when
\begin{equation}
\epsilon^2 << 1. \label{eq6.3.5}
\end{equation}
Although the axial ratio of the Galactic isodensity surfaces is not very
small, its square may be handled as a small quantity. For this reason the
present case corresponds just to stellar systems similar to the
Galaxy. 

Hence we may conclude that when the integrals (\ref{eq6.1.9}) exist as
precise integrals, the density distribution near the Galactic plane can
be described with Eq.~(\ref{eq6.3.1}). From this equation it results first, that
in galactic plane the density must vary according to Eq.~(\ref{eq6.3.3}), and
second, the dependence of the density on $z$ must be the same for all
$R$, if the unit for $z$ is taken to be increasing with $R$
proportionally to $(1+R^2/z_0^2)^{1/2}$. The first result seems to be
reliable, but the second seems at first glance to be in conflict
with our understanding of the Galactic structure. It states that the
Galaxy becomes thicker with increasing of $R$. However, the
contradiction is illusive, because the ``thickening'' means that the 
isosurfaces of the relative density $\rho (R,z)/\rho (R,0)$, not of the absolute 
density $\rho (R,z)$, move away from the galactic plane. 
Therefore, there is not impossible that the structure of the
Galaxy can be described with Eq.~(\ref{eq6.3.1}), at least in a first approximation.

\begin{figure*}[ht]
\centering
\includegraphics[width=90mm]{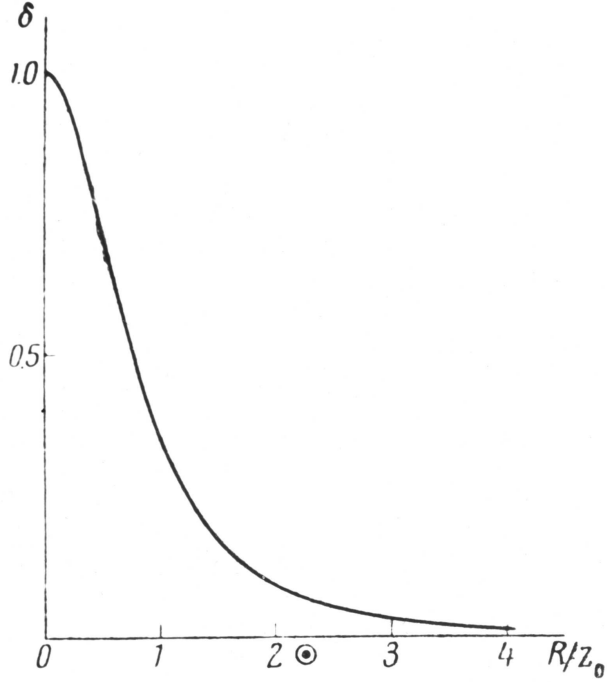}
\caption{Projected density $\delta$ (in units of $\delta$ at $R=0$) as
  a function of galactocentric radius $R$ (in units of $z_0$).}
\label{fig6.2}
\end{figure*}

Equations (\ref{eq6.3.1}) and (\ref{eq6.3.2}) may be used only up to a certain distance from
the galactic plane. When Eq.~(\ref{eq6.3.5}) is valid, the density at larger
distances is too small to give a significant contribution to the density, 
projected onto the galactic plane, \ie to the mass of a column
with unit surface area perpendicular to the galactic plane. Therefore, these
equations can be used to calculate the projected density (``the surface
density''), the equivalent half-thickness and the total mass of our
Galactic model. After integrating the density over $z$, we derive for the
projected density
\begin{equation}
\delta = \frac{\pi}{ 2} z_0\rho_0\bar{\epsilon} \left( 1+ \frac{R^2}{ z_0^2}
\right)^{-3/2}. \label{eq6.3.6}
\end{equation}
The quantity $\bar{\epsilon}$ is calculated according to the formula
\begin{equation}
\bar{\epsilon} = \frac{4}{\pi} \frac{\int_0^{\infty} \psi \rmd \xi }{ \psi
(0)}. \label{eq6.3.7}
\end{equation}
By using Eq.~(\ref{eq6.3.4}) it is easy to demonstrate, that $\bar{\epsilon}$
lies between the minimum and maximum values of $\epsilon$, \ie is some
kind of mean value of $\epsilon$. The calculated dependence of $\delta$
on $R$ is given in Fig.~\ref{fig6.2}, where $\delta$ value at $R=0$ is taken 
as a unit of measure. The same dependence of $\delta$ on $R$ was obtained in
our first paper concerning the third integral \citep{Kuzmin:1953}. At that
time, we did the calculations of the projected density, starting with less general
considerations. In the referred paper we compared the theoretical law for
$\delta (R)$ with the density, derived from the data on Galactic 
rotation \citep{Kuzmin:1952ac}. The results derived from observational data are
very uncertain, but it seems, that the theoretical law represents quite well 
the real dependence of $\delta$ on $R$, or at least may be
used as a first crude approximation. Nevertheless, some doubts arise because
of too slow decrease of $\delta$ with $R$ at large $R$. This is
because at large distances from the Galactic centre our equations give too
slow decrease of $\rho$ with the distance. According to our equations
the Galaxy has extremely undefined boundaries.

Comparing Eq.~(\ref{eq6.3.6}) with (\ref{eq6.3.3}), we find an expression for the
equivalent half-thickness of the Galaxy $z_e$
\begin{equation}
z_e= \frac{\pi}{ 4} z_0\bar{\epsilon} \left( 1+ \frac{R^2}{ z_0^2}
\right)^{1/2}. \label{eq6.3.8} 
\end{equation}
The derived result characterises the ``thickening'' of our Galactic model
with increasing of $R$ referred above. Finally, from Eq.~(\ref{eq6.3.6})
we find an expression for the total mass of the system $M$. After
integration we have
\begin{equation}
M=\pi^2z_0^3\rho_0\bar{\epsilon}. \label{eq6.3.9}
\end{equation}
The same result can be derived on the basis of Eq.~(\ref{eq6.2.14}) with the
condition (\ref{eq6.3.5}).

\section{Expressions for the potential of highly flattened stellar
system} 

Next we shall analyse the expression for the potential, when $\psi$
is a rapidly decreasing function with respect to $\xi^2$. Using Eqs.~(\ref{eq6.2.9}) and (\ref{eq6.2.14})
and taking into account the condition (\ref{eq6.3.5}) and formula (\ref{eq6.3.4}), we
find that for not too small $\xi^2$ the function $\varphi$ may be
expressed as
\begin{equation}
\varphi -\varphi (0)= \frac{GM}{ z_0}(|\xi |-|\bar\xi |), \label{eq6.4.1} 
\end{equation}
where we rejected the quadratic term with an arbitrary factor. $|\bar\xi |$ is 
the value of $|\xi |$ averaged with the weight
$4\psi d\xi$; this quantity is of the order of $\epsilon$ (it results from
(\ref{eq6.3.4})). To find the expression for the 
potential in the regions outside the vicinity of the galactic plane, and for the
potential in the galactic plane, we can use Eq.~(\ref{eq6.4.1}).

\begin{figure*}[ht]
\centering
\includegraphics[width=80mm]{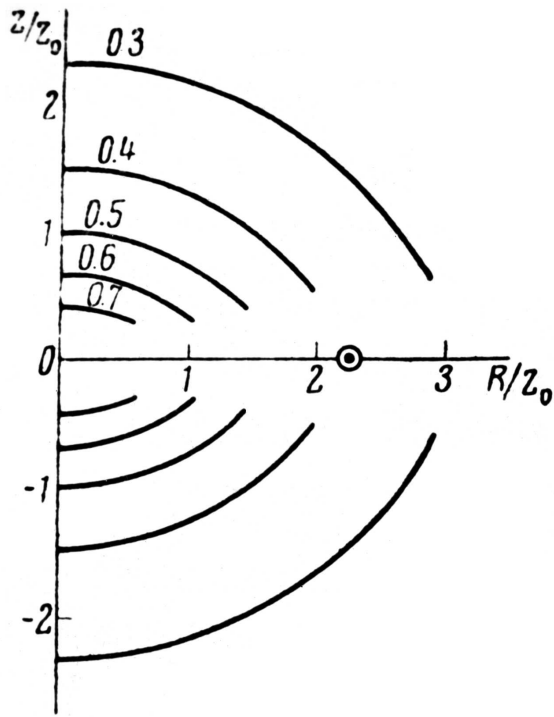}
\caption{Meridional sections of equipotential surfaces. Coordinates
  $R$ and $z$ are in units of $z_0$, Values of potential are in units
  of $GM/z_0$.}
\label{fig6.3}
\end{figure*}

Substituting Eq.~(\ref{eq6.4.1}) into (\ref{eq6.2.3}), we find also that for not very
small $\xi_2^2$, \ie in the regions outside the vicinity of the
galactic plane, 
\begin{equation}
\Phi = \frac{GM}{ z_0(\xi_1+|\xi_2|)}.  \label{eq6.4.2}
\end{equation}
From Eq.~(\ref{eq6.1.8}) it follows, that $z_0(\xi_1+|\xi_2|)$ is the
distance from more distant focus of the main velocity surfaces. Therefore,
the derived result indicates, that when the integrals (\ref{eq6.1.9}) exist as
precise integrals, in the regions outside the vicinity of the  galactic plane the
potential varies as if all the Galactic mass was concentrated in
more distant focus of the main velocity surfaces, \ie for $z>0$ in the 
southern focus,  and for $z<0$ in the northern focus. This is illustrated in
Fig.~\ref{fig6.3}, where according to Eq.~(\ref{eq6.4.2}) meridional sections of some
equipotential surfaces are plotted (the labels on the curves are the
values of $\Phi$ in units of $GM/z_0$).

Extrapolating the above law to the galactic plane, \ie by
substituting in Eq.~(\ref{eq6.4.2}) $\xi_2=0$ and $\xi_1$ according to Eq.~(\ref{eq6.3.2}),
we derive an approximate expression for the potential in the galactic plane
\begin{equation}
\Phi_{z=0} = \frac{GM}{ z_0}\left( 1+ \frac{R^2}{z_0^2} \right)^{-1/2}.
 \label{eq6.4.3}
 \end{equation}
Thus an approximate expression for the escape velocity in the
galactic plane is
\begin{equation}
v_{\infty} = \sqrt{ 2 \frac{GM}{z_0} }\left( 1+ \frac{R^2}{z_0^2}
\right)^{-1/4} ,  \label{eq6.4.4} 
\end{equation}
and the circular velocity is
\begin{equation}
v_c=\sqrt{ \frac{GM}{z_0}}~ \frac{R}{z_0}\left( 1+ \frac{R^2}{z_0^2}
\right)^{-3/4}  \label{eq6.4.5} 
\end{equation}
and the Oort's parameters are
$$A= \frac{3}{4}\sqrt{\frac{GM}{z_0^3}}~ \frac{R^2}{z_0^2} \left( 1+ \frac{R^2}{
z_0^2} \right)^{-7/4}, $$

\begin{equation}
-B= \sqrt{\frac{GM}{z_0^3}} \left( 1+ \frac{1}{4} \frac{R^2}{z_0^2} \right)
\left( 1+ \frac{R^2}{z_0^2} \right)^{-7/4}.  \label{eq6.4.6}
\end{equation}
For the third parameter $C$ ($C^2=-\partial^2\Phi /\partial z^2$ at $z=0$) we derive an
approximate expression from Eq.~(\ref{eq6.3.3}) by assuming $4\pi G\rho =
-\partial^2 \Phi /\partial z^2$, \ie neglecting in Poisson's equation
the terms $\partial^2\Phi /\partial R^2$ and $\partial\Phi /R\partial
R$. Taking into account Eq.~(\ref{eq6.3.9}) we have
\begin{equation}
C= \frac{2}{\sqrt{\pi\bar\epsilon}} \sqrt{ \frac{GM}{z_0^3}} \left( 1+
\frac{R^2}{z_0^2} \right)^{-1}.  \label{eq6.4.7} 
\end{equation}
Relative errors of Eqs.~(\ref{eq6.4.3})--(\ref{eq6.4.7}) are of the
order of $\epsilon$,  and for
more precise calculations corresponding corrections must be taken into
account. To find the expressions for these corrections, more precise equations for $\Phi$ 
and for $\partial^2\Phi /\partial z^2$ in the galactic plane are needed. 
The first formula can be derived by substituting
into Eq.~(\ref{eq6.2.3}) $\varphi_2=\varphi (0)$ and $\varphi_1$ according to
Eq.~(\ref{eq6.4.1}), the second formula by taking into account all terms in Poisson's
equation. As a result we have the following corrections
for ``galactic thickness''
\begin{equation}
\frac{\Delta\Phi}{\Phi}= 2 \frac{\Delta v_{\infty}}{ v_{\infty}}= \frac{\Delta
v_c}{v_c}= \frac{3}{5} \frac{\Delta A}{A}= \frac{4z_0^2+R^2}{4z_0^2-R^2}
~ \frac{\Delta B}{B}= -|\bar\xi |\left( 1+ \frac{R^2}{z_0^2}\right)^{-1/2}
 \label{eq6.4.8} 
 \end{equation}
and
\begin{equation}
\frac{2z_0^2}{2z_0^2-R^2}~ \frac{\Delta C}{C}=- \frac{\pi}{4}\bar\epsilon
\left( 1+ \frac{R^2}{z_0^2}\right)^{-1/2}.  \label{eq6.4.9}
\end{equation}

\begin{figure*}[ht]
\centering
\includegraphics[width=80mm]{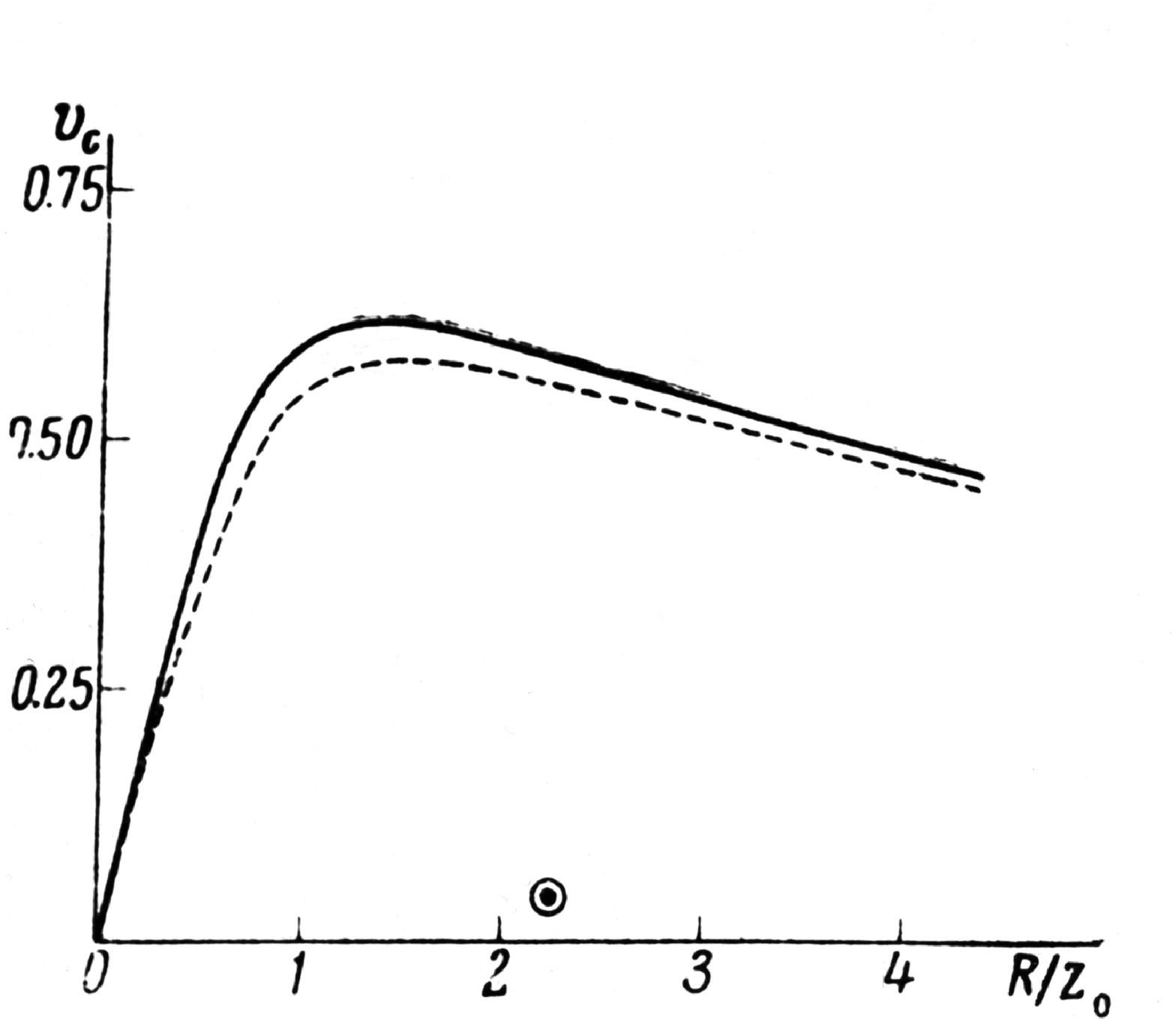}
\caption{Circular velocities of the Galaxy as a function of
  galactocentric radii. Velocities are in units of $\sqrt{GM/z_0}$,
  radii are in units of $z_0$. Continuous line -- circular velocities
  in case of precise integrals Eq.~(\ref{eq6.1.9}), dashed line -- 
  circular velocities in case of finite thickness of the disk.}
\label{fig6.4}
\end{figure*}

When comparing the theory with the observations, the most interesting equation 
from the ones above is Eq.~(\ref{eq6.4.5}), giving the circular velocity in the
Galaxy for the case when the integrals (\ref{eq6.1.9}) are precise. This law is
plotted in Fig.~\ref{fig6.4} by continuous line (the unit of velocity in Fig.~\ref{fig6.4}
is $\sqrt{GM/z_0}$). The circular velocity curve is quite
realistic, resembling the curve obtained from observations.

A comparison of the theoretical circular velocity with the observed
curve were made in our first paper on the third integral \citep{Kuzmin:1954},
where the results presented above were already derived, although on
the basis of less general assumptions. At that time the observed curve was formed 
by  \citet{Parenago:1948b} data on the motion of long-period
cepheids. The theoretical curve coincided in general with
the observed one. Unfortunately, the observed curve was quite uncertain
at that time, so it was difficult to decide, how precisely the
theoretical law corresponds to reality. In addition, as it became clear
later, that the observed curve was considerably influenced by errors on the
zero-point of the period-luminosity relation for cepheids. Now
there is much more data on the circular velocities, and its precision is
significantly higher due to the radio observations by Oort and his
colleagues \citep{vdHulst:1954, Kwee:1954}, which gives us a new
possibility to check our formulae. It turns out that the theoretical law
for circular velocity fits well the radio observations. By choosing
suitably values for $v_c$ at $R=R_{\odot}$, for $\sqrt{GM/z_0}$ and for
$z_0/R_{\odot}$, where $R_{\odot}$ is the distance of the Sun from
the Galactic center, we obtained nearly complete coincidence of the
theoretical law with the radio observations.\footnote{In present
paper we detailed the Galactic model, allowing the third quadratic
integral by demanding the correctness of the model, \ie by demanding
that the density can not be negative anywhere. This gives quite strict
restrictions for the function $\varphi$ (Eq.~(\ref{eq6.2.9}) with $\psi\ge 0$).
Later P.A. Waymann (M.N.R.A.S. {\bf 119}, 34, 1959) and thereafter
G. Hori (Tokyo Contr. No. 31, 1962) approached the problem in a different
way. Not demanding the correctness of the model, they tried using the
potential equation (\ref{eq6.2.1}) to approximate the potential by empirical
Schmidt's model. They succeed at deriving a quite good model. [Later
footnote.]} 
The agreement fails only near the Galactic centre, where the
theory predicts angular circular velocity to be too small. For $R>R_{\odot}$
radio observations do not give the dependence of $v_c$ with
$R$, but it may be expected that also for these regions the theoretical
law will correspond to reality quite well.

Taking into account the available data, including the radio
observations, we re-estimated the parameters in our formulae and
obtained the following values\footnote{Derived values for model
  parameters are not satisfactory at present. $\sqrt{GM/z_0}$, $z_0$
  and $R_{\odot}$ must be increased, $\bar\epsilon$ must be
  lowered. Also Table 1 changes respectively. [Later footnote.]}
$\sqrt{GM\over z_0}=$ 380~km/s; $z_0=$ 3.1~kpc; $\bar\epsilon =$
0.14. For $R_{\odot}$ we used the averaged value $R_{\odot} =
$7.0~kpc, which is close to the value 7.25~kpc derived by
\citet{Kukarkin:1949} from the distribution of short-period cepheids,
that we used in previous papers. True, the luminosity of
short-period cepheids seems to be somewhat lower than it was assumed
till now \citep{Pavlovskaya:1954}, and as a result $R_{\odot}$ needs to be
lowered. From the other side, the radio observations
\citep{vdHulst:1954} give higher values for $R_{\odot}$. For this reason we
decided not to change $R_{\odot}$ significantly.

In accordance with the parameter values given above, we indicate the solar 
position in Figs.~\ref{fig6.2}--\ref{fig6.5}, and give the values of
$v_{\infty}$, $v_c$, $A$, $B$, $C$, $\rho$, $\delta$ and $z_e$ for the
solar neighbourhood and for the centre of the Galaxy in Table \ref{table6.1}. 
The values of $M$ and of $z$-component velocity dispersion $\sigma_z$, 
calculation of which will be described below, are also shown in Table \ref{table6.1}. In
calculations of the values of Table \ref{table6.1}, the corrections 
``for the thickness of the Galaxy'' were take into account where needed, while
assuming $|\bar\epsilon |= $ ${\pi\over 4}\bar\epsilon =$ 0.11
(see below, the model $n=$ 3). The variation of $v_c$ due to the
correction is presented in Fig.~\ref{fig6.4}, where the corrected curve is given by
dashed line. The Table indicates that the parameters are
quite acceptable. (We mention that the values of $A$, $B$, $C$
and $\rho_{z=0}$ from the Table do not satisfy precisely Poisson's
equation. This is because the formulae used for calculations are
precise only up to the first order.)

\begin{table}
\caption{}
\smallskip
\label{table6.1}
\centering
\begin{tabular}{| l | l |  r | r | } 
\hline\hline
      &  &    $R=0$    & $R=R_{\odot}$  \\
\hline
$v_{\infty}$ & (km/s)                & 508     & 334   \\
$v_c$        & (km/s)                 &   0     & 211   \\
$A$          & (km/s/kpc)             &   0     &  18.3 \\
$B$          & (km/s/kpc)             & 109     &  11.8 \\
$C$          & (km/s/kpc)             & 329     &  65   \\
$\rho_{z=0}$ & ${\rm M_{\odot}/pc^3}$ & 2.52    & 0.068 \\
$\delta    $ & ${\rm M_{\odot}/pc^2}$ & 1720    & 114   \\
$z_e$        & (kpc)                  & 0.34    & 0.84  \\
$(\sigma_z)_{z=0}$ & (km/s)           & 70      & 41    \\
\hline
        &   & $M = 104\cdot 10^9 {\rm M_{\odot}}$ &         \\
\hline\hline
\end{tabular}
\end{table}

\section{Concrete models of the Galaxy}

As it was already mentioned, many results of previous chapters were
derived in our first paper on the third integral
\citep{Kuzmin:1954}. But in that paper we derived them on the basis of
less general considerations, namely by studying Galactic models
corresponding to some particular expression of $\varphi$. But as it is
clear now, the results above are independent of the particular form of
$\varphi$ -- they result from the condition (\ref{eq6.3.5}). Therefore,
the results of previous two sections give us the general properties of
all sufficiently flat stellar systems, for which the integrals
(\ref{eq6.1.9}) exist as precise integrals of motion. These properties
quite comprehensibly describe the model of the Galaxy of our interest. 
However, it is recommendable to detail the model by choosing
suitable form of $\varphi$. It will be our aim now.

The function $\varphi$ may be chosen on the basis of data for the
Galactic potential, \ie in the way it was done by \citet{Parenago:1952} in
construction of his model. But because our model
will be probably only a quite crude approximation to the real Galaxy, it is
not practical to devote too much efforts to choosing the function,
especially because data on Galactic potential is very scanty. It seems
to be sufficient to limit ourselves to some analytical and hopefully simple
expression for $\varphi$, with a small number of parameters, which will give 
an acceptable model of the Galaxy.

While searching for a suitable expression for $\varphi$, we tried to generalise
our Galactic model reviewed in our first paper. It results that it is possible to derive 
an acceptable model by taking
\begin{equation}
\varphi ''= \frac{k}{n\xi_0}\left( \frac{\zeta_0}{\zeta}\right)^n , \label{eq6.5.1}
\end{equation}
where $k$, $n$ and $\xi_0$ are constants, while $k>0$,
\begin{equation}
\zeta^2= \frac{\xi^2+\xi_0^2}{1+\xi_0^2} \label{eq6.5.2}
\end{equation}
and $\zeta_0$ is the value of $\zeta$ at $\xi =0$. Taking into
account, that derivative of $\varphi -\xi\varphi '$ equals to
$-\xi\varphi ''$, we have after integration
\begin{equation}
\varphi -\xi\varphi '= \frac{k\xi_0}{n(n-2)}\left( \frac{\zeta_0}{\zeta}
\right)^{n-2} + \mathrm{const}. \label{eq6.5.3}
\end{equation}
By substituting $\varphi '$ found after integration of (\ref{eq6.5.1}), we have
the expression for $\varphi$
\begin{equation}
\varphi = \frac{k}{n\xi_0} \left[ \frac{\xi_0^2}{n-2} \left( \frac{\zeta_0
}{\zeta} \right)^{n-2} + \xi \int_0^{\xi} \left( \frac{\zeta_0}{
\zeta}\right)^n \rmd \xi \right] + \mathrm{const}. \label{eq6.5.4} 
\end{equation}
By substituting $\varphi -\xi\varphi '$ and $\varphi ''$ according to Eqs.
(\ref{eq6.5.1}) and (\ref{eq6.5.3}) into Eq.~(\ref{eq6.2.7}), and by choosing arbitrary constants in a
way that $\psi$ will be finite for $\xi^2=1$, we derive
\begin{equation}
\psi = \frac{2k\zeta_0^{n+2}}{n(n-2)\xi_0^3} \left[ 1- 
\frac{1+ \frac{n}{2}(\zeta^2-1)}{\zeta^n}\right] (\zeta^2-1)^{-2}. 
\label{eq6.5.5} 
\end{equation}

Analyse of Eq.~(\ref{eq6.5.5}) gives us that the function $\psi$ satisfies all the
necessary conditions. First, $\psi$ is non-negative for all $n$, and
decreasing with $\xi^2$ for $n>-2$, giving us non-negative
density everywhere (and for $n>-2$ decreasing with $z^2$). Further, as is seen
from Eq.~(\ref{eq6.5.5}), $\psi$ decreases for all $n$ at large $\xi^2$ not
faster than $\xi^{-4}$. Fulfilment of that condition is necessary,
because otherwise for large $r$ the density on the galactic axis 
decreases faster than $r^{-4}$, and for remaining radii according
to (\ref{eq6.2.13}) as $r^{-4}$. Finally, it can be seen from Eq.~(\ref{eq6.5.5}), that for
$n>1$, $\psi$ decreases at large $\xi^2$ faster than $|\xi |^{-3}$.
This means that for $n>1$ the mass is finite.

For finite mass $GM/z_0$ equals to liner part of $\varphi$
at large $\xi^2$, and we find from Eq.~(\ref{eq6.5.4}) the relation between the mass
and constant $k$
\begin{equation}
\frac{GM}{z_0}= \frac{k}{2n} B\left( \frac{n-1}{2}, \frac{1}{2}\right),
\label{eq6.5.6}
\end{equation}
where $B$ is beta-function. It is easy to see, that every $n$-model
with finite mass ($n>1$) can be derived by summation of models
$n+\Delta n$ ($\Delta n>0$), the masses of which will be distributed
with $\xi_0$ according to
\begin{equation}
dM= \frac{2M}{B\left( \frac{n-1}{2}, \frac{\Delta n}{2}\right)}
(1-u^{-2})^{\frac{\Delta n}{2}-1} u^{-n} \rmd u ~~(u\ge 1), \label{eq6.5.7}
\end{equation}
where $u$ is the ratio of $\xi_0$ to its value for resulting model, and
$M$ is the mass of resulting model. Indeed, substituting in the
expression of $\varphi ''$ for the model $n+\Delta n$ constant $k$
according to (\ref{eq6.5.6}) (where for $M$ will be taken $dM$ according to
(\ref{eq6.5.7})), and integrating from $u=1$ to $u=\infty$, we find $\varphi ''$
for model $n$ (see (\ref{eq6.5.1}) and (\ref{eq6.5.2})).

Comparing Eq.~(\ref{eq6.5.5}) with (\ref{eq6.3.4}), it follows that for sufficiently small
$\xi_0$ the condition (\ref{eq6.3.5}) is valid. In this case, when near to the galactic
plane we may use for the density Eq.~(\ref{eq6.3.1}), where we substitute the
expression of $\psi$ for small $\xi_0^2$ and $\xi^2$, namely
\begin{equation}
\psi =\psi (0)\left( \frac{\zeta_0}{\zeta} \right)^n. \label{eq6.5.8}
\end{equation}
For small $\xi_0^2$, other formulae of last two sections are also valid, 
while for $\bar\epsilon$ in several of them using Eqs.~(\ref{eq6.3.7})
and (\ref{eq6.5.8}) we have
\begin{equation}
\bar\epsilon = \frac{2}{\pi} B\left( \frac{n-1}{2},\frac{1}{2}\right)
\xi_0. \label{eq6.5.9}
\end{equation}
Similarly we may derive an expression for $|\bar\xi |$ for some of these
formulae. We have
\begin{equation}
(n-2) B\left( \frac{n-1}{2},\frac{1}{2}\right) |\bar\xi | = 2\xi_0.
\label{eq6.5.10}
\end{equation}
Equation (\ref{eq6.5.9}) may be used for $n>1$, Eq.~(\ref{eq6.5.10}) for $n>2$. In
both cases $\xi_0$ must be sufficiently small.

Amongst the models resulting from Eq.~(\ref{eq6.5.1}), most interesting are those with
$n=4$ and $n=3$. On the basis of Eq.~(\ref{eq6.5.1})--(\ref{eq6.5.3}),
(\ref{eq6.5.6}) and (\ref{eq6.2.5}) we 
find the density for $n=4$ in the form
\begin{equation}
\rho = \frac{M~\zeta_0^2}{\pi^2~z_0^3~\xi_0^3} \left(
\frac{\zeta_0}{\zeta_1\zeta_2} \right)^4 , \label{eq6.5.11} 
\end{equation}
where $\zeta_1$ and $\zeta_2$ correspond to $\xi =\xi_1$ and $\xi
=\xi_2$. In the centre of the Galaxy $\zeta_1=1$ and $\zeta_2=\zeta_0$,
and the factor in (\ref{eq6.5.11}) is the central density. Expressing
$\zeta_1$ and $\zeta_2$ in (\ref{eq6.5.11}) via $R$, and $z$ (Eqs.~(\ref{eq6.1.8}) and
(\ref{eq6.5.2})), we can conclude that the isodensity surfaces in present case are
similar ellipsoids of revolution with $\epsilon =\zeta_0$. The model with
$n=4$ is thus ellipsoidal.

An ellipsoidal layer does not attract the inner masses, so the
model $n=4$ can be ``cut'' in outer parts without influencing the
existence of the integrals (\ref{eq6.1.9}). The same is valid for models with $n<4$,
as can be derived by summation of $n=4$ models with different
$\epsilon$. It results now that too diffuse boundaries of the Galaxy,
resulting from our theory, is not a serious difficulty.

For the $n=3$ model we derive from Eqs.~(\ref{eq6.5.4}) and (\ref{eq6.5.6})
\begin{equation}
\varphi = \frac{GM}{z_0} \frac{\xi_0}{\zeta_0}\zeta , \label{eq6.5.12}
\end{equation}
and according to (\ref{eq6.2.3}) it follows
\begin{equation}
\Phi = \frac{GM}{\zeta_0\xi_0} \frac{\zeta_0}{\zeta_1+\zeta_2}. \label{eq6.5.13}
\end{equation}
The model corresponds to the model,  reviewed in \citet{Kuzmin:1954}, and is
interesting because of rather simple expression for $\Phi$. The
expression for $\rho$ is somewhat more complex. Equations (\ref{eq6.5.1})--(\ref{eq6.5.3}),
(\ref{eq6.5.6}) and (\ref{eq6.2.5}) for $n=3$ give
\begin{equation}
\rho = \frac{M\zeta_0^2}{4\pi z_0^3\xi_0^3} \left[ \frac{\zeta_0}{\zeta_1
\zeta_2 (\zeta_1+\zeta_2)}\right] [\zeta_1\zeta_2+\zeta_1^2\zeta_2^2 +
(\zeta_1+\zeta_2)^2]. \label{eq6.5.14} 
\end{equation}
[see Appendix A].

The $n=3$ model, as well as all models with $n<4$, may be derived by
summation of $n=4$ models. The mass distribution of $n=4$ with $\xi_0$ is
given by Eq.~(\ref{eq6.5.7}) (with parameters $n=3$ and $\Delta n=1$). From
Eq.~(\ref{eq6.5.7}) results that the scatter of $\xi_0$ (and therefore
also of $\epsilon$) of  added $n=4$ models is significant. For that reason it may be
expected that the model has some similarity with the real Galaxy, 
consisting also of several subsystems with different eccentricity. We
assumed at present that the subsystems can be modelled at least
approximately with $n=4$. For the intermediate and the spherical
subsystems this is probably not very 
far from reality. Unfortunately, the flat subsystems fall outside
of our scheme, because of significantly smaller radial density gradient
when compared with other subsystems (and also too large density
$z$-gradient, inconsistent with the reality). However, the mass
fraction of flat subsystems is not large, and we may neglect them in
the first approximation.

\begin{figure*}[ht]
\centering
\includegraphics[width=100mm]{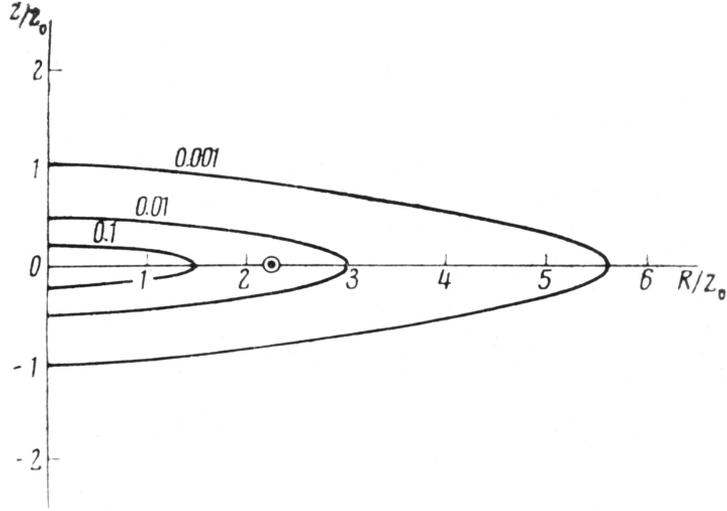}
\caption{Isodensity contours in the meridional plane of the
  Galaxy. Densities are in units of the central density.}
\label{fig6.5}
\end{figure*}

Assuming $\bar\epsilon = 0.14$ and $\xi_0=0.11$ for $n=3$ model, 
we calculated the density at different $R$
and $z$, and constructed the isodensity lines in the meridional plane of the
Galaxy. The result is given in Fig.~\ref{fig6.5}, where the labels on the curves
are densities in units of the central density.  The $n=3$ model 
really resembles the ``smoothed'' Galaxy as we usually assume it.

\section{Velocity $z$-component dispersion and distribution}

Above we described the spatial structure of Galactic models allowing
for the triaxial velocity distribution. It is possible at a certain level 
to characterise the kinematical properties of the models. First of
all, 
it is possible to find the dispersion $v_z$ in galactic plane even without
assuming a concrete expression for $\varphi$. After we choose the
expression for $\varphi$, it is possible to calculate the distribution
function of $v_z$. 

To calculate the dispersion and the distribution function of $v_z$, we start
from six-dimensional continuity equation $D\Psi /Dt=0$, where $\Psi$ is
the phase density. For the stationary stellar system with axial symmetry, the
continuity equation after integration over $v_R$ and $v_z$ is
\begin{equation}
\frac{\partial (Rf\bar v_R)}{R\partial R}+v_z \frac{\partial f}{\partial z}
+ \frac{\partial\Phi}{\partial z} \frac{\partial f}{\partial v_z} =0,
\label{eq6.6.1}
\end{equation} 
where $\bar v_R$ is the averaged $v_R$ for fixed $v_z$, $f$ is the density
in four-dimensional space including the ordinary space and the
$v_z$-space. The first term of the equation is nonzero because of
inclination of the symmetry plane of the velocity distribution with respect to the
galactic plane. Although, for sufficiently flat systems the inclination is
very small. The ratio $\bar{v}_R/v_z$ is of the order of $\frac{1}{2}\sin
2\alpha$, where $\alpha$ is the inclination of the symmetry plane of the
velocity distribution with respect to the galactic plane. It results
now, that $R\bar{v}_R/zv_z$ value is about one or less, due to
$\frac{1}{2}\sin 2\alpha \le z/R$. 
For fixed $v_z$ $f$ is approximately equal to the ordinary density
$\rho$, and $R\bar{v}_R/zv_z$ may be assumed to be slowly varying with
$R^2$ and $z^2$ (essentially not faster than $\rho$ with $R^2$), so the ratio of
the first term in Eq.~(\ref{eq6.6.1}) to the second term is consequently in order of
the ratio $\partial\rho /R\partial R$ to $\partial\rho /z\partial z$,
\ie 
in order of $\epsilon^2$. Hence, when the condition (\ref{eq6.3.5}) is valid, the 
first term in (\ref{eq6.6.1}) is small enough to be neglected. Instead of (\ref{eq6.6.1})
we have\footnote{Equation (\ref{eq6.6.2}) is a basic equation of {\bf
one-dimensional problem of stellar dynamics},  discussed by us already in
Chapters 1, 3, 5 of the D.Sc. Thesis. Solution of one-dimensional problem was discussed also
by Prendergast (A.J. {\bf 59}, 260, 1954) [Later footnote.]}
\begin{equation}
v_z \frac{\partial f}{\partial z}+ \frac{\partial\Phi}{\partial z} \frac{\partial
f}{\partial v_z}=0. \label{eq6.6.2}
\end{equation}

In order to calculate the dispersion of $v_z$ at $z=0$, we multiply
Eq.~(\ref{eq6.6.2}) by $v_z$, and integrate over $v_z$. As a result we have
the Jeans equation
\begin{equation}
\frac{\partial (\rho\sigma_z^2)}{\partial z}= \rho \frac{\partial\Phi}{
\partial z}, \label{eq6.6.3}
\end{equation}
where $\sigma_z$ is the dispersion of $v_z$. By integrating the equation we
derive
\begin{equation}
(\rho\sigma_z^2)_{z=0} = -\int_0^{\infty} \rho \frac{\partial\Phi}{
\partial z} \rmd z. \label{eq6.6.4}
\end{equation}
As a first approximation we neglect the terms 
$\partial^2\Phi /\partial R^2$ and $\partial\Phi /R\partial R$ 
in the Poisson's equation, resulting in
\begin{equation}
\frac{\partial\Phi}{\partial z} = -4\pi G\int_0^z \rho \rmd z, \label{eq6.6.5}
\end{equation}
and Eq.~(\ref{eq6.6.4}) gives us
\begin{equation}
(\rho\sigma_z^2)_{z=0} = \frac{\pi G}{2}\delta^2. \label{eq6.6.6}
\end{equation}
Therefore, according to Eqs.~(\ref{eq6.3.3}), (\ref{eq6.3.6}) and (\ref{eq6.3.9}), we have the following
equation for $\sigma_z$ at $z=0$
\begin{equation}
(\sigma_z)_{z=0} = \sqrt{ \frac{\pi\bar\epsilon}{8} \frac{GM}{z_0}}\left( 
1+ \frac{R^2}{z_0^2}\right)^{-1/2}. \label{eq6.6.7} 
\end{equation}

The relative error of Eq.~(\ref{eq6.6.7}) is of the order of $\epsilon$, and for more
precise calculations is needed to be taken into account, as well as in case
of equations in Sect. 4 the correction for ``galactic thickness''. We derive the
correction by taking into account the neglected terms of Poisson's
equation. Ignoring the dependence of these terms of $z$, and by using for
their calculation the approximate Eq.~(\ref{eq6.4.3}), we obtain the expression for
the correction\footnote{General formula for the correction of
$\sigma_z^2|_{z=0}$ is
$$ \frac{\Delta\sigma_z^2}{\sigma_z^2} = \frac{A^2-B^2}{\pi G\rho_{z=0}}
\frac{|\bar z|}{z_e}.$$
}
\begin{equation}
\frac{z_0^2}{2z_0^2-R} \frac{\Delta\sigma_z}{\sigma_z} = - |\bar\xi |
\left( 1+ \frac{R^2}{z_0^2} \right)^{-1/2}. \label{eq6.6.8}
\end{equation}

The value of $\sigma_z$ at $z=0$ in the vicinity of the Sun and in the
center of the Galaxy is given in Table \ref{table6.1}. The resulting velocity
dispersion is quite large as well as the effective half-thickness of the
Galaxy. However, because the spherical subsystems of the Galaxy have
probably quite large masses, these results may be accepted.

To calculate the distribution function of $v_z$ we use the
solution of Eq.~(\ref{eq6.6.2}), having the form
\begin{equation}
f=f(v_z^2-2\Phi ,R) \label{eq6.6.9}
\end{equation}
and giving us
\begin{equation}
\rho = \int_{-\infty}^{\infty} f(v_z^2-2\Phi ,R) \rmd v_z. \label{eq6.6.10}
\end{equation}
The derived equation is an integral equation for $f$. It has the form of
Abel equation and is soluble by quadratures. When $f$ is known,
also the distribution function of $v_z$ for arbitrary $R$ and $z$ is known.

Let us use the Eq.~(\ref{eq6.6.10}) for the $n=3$ model. By using
Eqs.~(\ref{eq6.3.2}), (\ref{eq6.5.2}), (\ref{eq6.5.13}) and (\ref{eq6.5.14}) or
(\ref{eq6.3.1}) and (\ref{eq6.5.8}) we can conclude, that near to the
galactic plane at a given $R$ the density $\rho$ is approximately
proportional to $(\Phi_0-\Phi )^{-3}$, where $\Phi_0$ is the value of
$\Phi$ at $\xi_2=\xi_0=0$. For such dependence between $\rho$ and
$\Phi$ the solution of Eq.~(\ref{eq6.6.10}) is evidently in the form
\begin{equation}
f=f_0\left( 1+ \frac{v_z^2}{4\sigma_z^2} \right)^{-7/2} ,\label{eq6.6.11}
\end{equation}
where $f_0$ depends on $z$ as $(\Phi_0-\Phi )^{-7/2}$ and $2\sigma_z^2 =
\Phi_0-\Phi$ (coefficient for $\sigma_z^2$ is chosen in a way that the
averaged $v_z^2$ equals to $\sigma_z^2$). The derived equation may be used
only near the galactic plane and for not too large $v_z^2/\sigma_z^2$, in
that case its relative error is of the order of $\epsilon$. Inexactness of
Eq.~(\ref{eq6.6.11}) is related to the inexactness of the relation between $\rho$ and
$\Phi$. However, despite of inexactness of Eq.~(\ref{eq6.6.11}) it enables us to
estimate the general character of $v_z$ distribution for the $n=3$ model.
The resulting distribution of $v_z$ near the galactic plane 
differs significantly from Gaussian, having larger positive
excess and increasing with $z^2$ dispersion. $v_z$ distribution
is just as it may be expected when adding $v_z$ distributions of different
Galactic subsystems.

\section{Generalisation of Lindblad's and Bottlinger's diagrams.
Galactic stellar orbits}

Without detailing more the possible kinematical  structure of the
Galaxy within the present theory, we focus a little on some generalisations,
related to the third integral of motion, and being interesting from a
point of view of Galactic kinematics.

\subsection{Generalisation of Lindblad's diagram}

To have the complete phase-space description of a stellar
system we need to know the phase density $\Psi$. For a stationary system
$\Psi$ is a function of integrals $I_1$, $I_2$ and $I_3$, and of
these integrals only (when the potential is not restricted more than described
in Sect.~1). For that reason within the present theory the
phase-space structure of the Galaxy or of some of its subsystem is
completely known when $\Psi$ is known as a function of $I_1$,
$I_2$, $I_3$. Besides, for the Galaxy as a whole $\Psi$ is restricted in a
way that $\rho$, which is calculated from it, must satisfy the conditions described in
previous Sections. For individual subsystems that restriction is not
necessary. 

$\Psi$ as a function of $I_1$, $I_2$ and $I_3$ can be represented by
isosurfaces in the three-dimensional diagram with $I_1$, $I_2$ and $I_3$ as
coordinate axis. This kind of diagram is a generalisation of the well-known
diagram by \citet{Lindblad:1933b}, and enables to find the velocity distribution
for arbitrary $\xi_1$ and $\xi_2$. Basic properties of the diagram can be
derived when expressing the velocity components $v_1$ and $v_2$ via integrals
$I_1$, $I_2$ and $I_3$. By using Eqs.~(\ref{eq6.1.9}), (\ref{eq6.2.3}) and (\ref{eq6.2.4}) we find
\begin{equation}
(\xi_1^2-1)(\xi_1^2-\xi_2^2)v_1^2=w_1^2, ~~ (1-\xi_2^2)(\xi_1^2-\xi_2^2)
v_2^2=w_2^2, \label{eq6.7.1}
\end{equation}
where $w_1^2$ and $w_2^2$ are the functions $w$ for $\xi =\xi_1$ and $\xi
=\xi_2$ 
\begin{equation}
w^2 =\xi^2(\xi^2-1)I_1 -\xi^2I_2^2 - (\xi^2-1)I_3 + 2(\xi^2-1)\varphi.
\label{eq6.7.2} 
\end{equation}
Equations (\ref{eq6.7.1}) together with the expression for $I_2$ are the equations of
isosurfaces $v_1$, $v_2$ and $v_{\theta}$ in generalised Lindblad's
diagram. Isosurfaces of $v_{\theta}$ are planes, isosurfaces of $v_1$ and
of $v_2$ are parabolic cylinders. For each pair of $\xi_1$ and $\xi_2$ we
have a triple family of these surfaces. The diagram represents the
coordinate surfaces of the velocity space, corresponding to a point in
ordinary space with given $\xi_1$ and $\xi_2$. The region of the
diagram, representing the velocity space of a given point, is limited by two
cylinders
\begin{equation}
w^2(I_1,I_2,I_3,\xi_1)=0, ~~~ w^2(I_1,I_2,I_3,\xi_2)=0, \label{eq6.7.3}
\end{equation}
where $v_1$ and $v_2$ are zero. The parabolic cylinders
(\ref{eq6.7.3}) are analogous to the characteristic parabola of the ordinary Lindblad's
diagram. [see Appendix B.]

\subsection{Elements of stellar orbits and generalisation of
Bottlinger's diagram}

Equations (\ref{eq6.7.3}) may be handled as the equations enabling us to find the
isosurfaces in ordinary space, where $v_1$ and $v_2$ equal to zero for
given $I_1$, $I_2$ and $I_3$. From these equations it follows that the
surfaces where $v_1=0$ are ellipsoids $\xi_1=\mathrm{const}$, the surfaces where $v_2=0$ are
hyperboloids $\xi_2=\mathrm{const}$. Evidently these surfaces are the
enveloping surfaces for stellar orbits, because the velocity component
perpendicular to them is zero. On the basis of Eqs.~(\ref{eq6.2.9}) and (\ref{eq6.7.2}) it is
easy to demonstrate, that for $\psi\ge 0$ and for $I_1$, $I_2$, $I_3$,
allowing for real velocities being less than the escape velocity, there exist two real
solutions for $\xi_1$ and one for $|\xi_2|$. For that reason an orbit has
four enveloping surfaces: two ellipsoids and two sheets of hyperboloid.
Orbit fills ring-like tube limited by these surfaces.

Two values of $\xi_1$, determining the enveloping ellipsoids, and one value of
$|\xi_2|$, determining the enveloping hyperboloid, are suitable to be handled as
the elements of galactic stellar orbits. We designate these elements as
$x'_1$, $x''_1$ and $x_2$, while $x'_1\ge x''_1\ge 1$ and $1\ge x_2\ge 0$.
The first two elements are the maximal and minimal values of $\xi_1$,
corresponding to a given orbit, the third is the maximum value of
$|\xi_2|$. Surely, the orbital elements may be simply $I_1$, $I_2$ and
$I_3$, but it is more convenient to use the elements above, as they give us an idea about
the spatial structure of orbit. Relation between $x'_1$, $x''_2$, $x_2$ and
$I_1$, $I_2$, $I_3$ is given by
\begin{equation}
w^2(I_1,I_2,I_3,x)=0, \label{eq6.7.4}
\end{equation}
while $x'_1$, $x''_1$ and $x_2$ are non-negative roots of the equation.

From Eq.~(\ref{eq6.7.4}) it is easy to derive a generalisation of the diagram,
proposed by \citet{Bottlinger:1931}, and used to determine the orbital elements
from the velocity components. Equation (\ref{eq6.7.4}) is an equation for isosurfaces of
stellar orbits in generalised Lindblad's diagram. When we substitute
$I_1$, $I_2$ and $I_3$ with their expressions, we derive the equation of
isosurfaces of elements for a given point. This is just the equation of
generalised Bottlinger's diagram. On the basis of Eqs.~(\ref{eq6.1.9}), (\ref{eq6.2.3}), (\ref{eq6.2.4})
and (\ref{eq6.7.4}) we have
\begin{equation}
\frac{v_1^2}{x^2-\xi_1^2} + \frac{v_2^2}{x^2-\xi_2^2} + \frac{v_{\theta}^2}{
x^2-1} = -2 \frac{\Phi_{\xi =x} -\Phi}{x^2-\xi^2} , \label{eq6.7.5}
\end{equation}
where $\xi$ in right side is one of the coordinates $\xi_1$, $\xi_2$. From
(\ref{eq6.7.5}) it results, that in velocity space the isosurfaces $x$ are the second
order surfaces.

\subsection{Calculation of orbital elements}

Use of the third integral enables not only to generalise the
Lindblad's and Bottlinger's diagrams, but also to calculate stellar orbits
in general three-dimensional case. For that we need to find the missing
independent integrals of motion. By using Eq.~(\ref{eq6.1.8}) we have
\begin{equation}
\frac{\dd\xi_1}{\dd{t}} = \sqrt{\frac{\xi_1^2-1}{\xi_1^2-\xi_2^2}} \frac{v_1}{z_0} ,
~~ \frac{\dd\xi_2}{\dd{t}} = \sqrt{\frac{1-\xi_2^2}{\xi_1^2-\xi_2^2}} \frac{v_z}{
z_0}, \label{eq6.7.6}
\end{equation}
and together with (\ref{eq6.7.1}) giving
\begin{equation}
\frac{1}{w_1} \frac{\dd\xi_1}{\dd{t}} = \frac{1}{w_2} \frac{\dd\xi_2}{\dd{t}} = \frac{1}{
z_0(\xi_1^2-\xi_2^2)} \label{eq6.7.7}
\end{equation}
or by expressing $\dd{t}=R\dd\theta /v_{\theta}$ via $I_2$ according to (\ref{eq6.1.8})
and (\ref{eq6.1.9})
\begin{equation}
\frac{1}{w_1} \frac{\dd\xi_1}{\dd\theta} = \frac{1}{w_2} \frac{\dd\xi_2}{\dd\theta}
= \frac{(\xi_1^2-1)(1-\xi_2^2)}{I_2(\xi_1^2-\xi_2^2)}. \label{eq6.7.8}
\end{equation}
Integration of Eqs.~(\ref{eq6.7.7}) and (\ref{eq6.7.8}) gives us the integrals we need in forms
\begin{equation}
I_4=U_1-U_2, ~~ I_5=S_1-S_2-\theta , ~~ I_6=T_1-T_2-t , \label{eq6.7.9}
\end{equation}
where $U_1$, $S_1$, and $T_1$ are functions of $I_1$, $I_2$, $I_3$ and
$\xi_1$, and $U_2$, $S_2$ and $T_2$ are functions of $I_1$, $I_2$, $I_3$
and $\xi_2$. Principal values of these infinitely multiple-valued functions
are given by formulae
\begin{equation}
U=-\int_{\xi}^x \frac{\rmd \xi}{w}, ~~ S= -I_2 \int_{\xi}^x \frac{\xi^2 \rmd \xi}{
(\xi^2-1)w}, ~~ T= -z_0 \int_{\xi}^x \frac{\xi^2 \rmd \xi}{w}, \label{eq6.7.10}
\end{equation}
where $\xi =\xi_1$ and $x=x'_1$ or $\xi =\xi_2$ and $x=x_2$. It the
values of $I_1$, $I_2$, $I_3$, $I_4$, $I_5$, $I_6$ are given, we can calculate
according to Eqs.~(\ref{eq6.7.9}) and (\ref{eq6.7.10}) stellar orbit and
position of a star in the orbit at every moment of time.\footnote{As
an example, one orbit for the $n=3$ model was calculated by H. Eelsalu
already in 1953. [Later footnote.]} [see Appendices C,
D.]

Dependence between $\xi_1$ and $\xi_2$ and $\theta$, given by expressions
of integrals $I_4$ and $I_5$, is in general infinitely multiple-valued and
the  same for all values of $I_4$ and $I_5$, when $I_1$, $I_2$ and $I_3$
are given. That means  that the stellar orbit does not depend on the values of 
integrals $I_4$ and $I_5$, and that they are therefore infinitely
multiple-valued. As a result, only $I_1$, $I_2$ and $I_3$ are
single-valued and independent of each other and of time integrals. For
stationary systems the phase density, being single-valued, may be a
function of only these integrals. 

\section{Conclusion}

As a conclusion we like to mention the following. When we suppose the
existence of the third single-valued integral of motion, being independent
of time and of the energy and the angular momentum 
integrals, we apply a certain restriction on the gravitational potential of
a stellar system. This restriction enables to construct a model of the
Galaxy having similarity with the real Galaxy in respect to the potential
and also in respect to phase-space structure. Introduction of the  
third single-valued integral into the theory enables to explain the triaxial 
velocity distribution within the theory of the stationary Galaxy. In
addition, there is a possibility to generalise the Lindblad's and
Bottlinger's diagrams, and to calculate stellar orbits in general
three-dimensional case.

The derived model of the Galaxy surely does not represent the real
Galaxy with high precision. But because the model has a considerable
similarity with the real Galaxy, the third integral of motion may be used
as a quite precise integral for all subsystems of the 
Galaxy. It seems that deviations from the restricting condition on the
Galactic potential due to the third integral are not large.

\vglue 5mm

{\bf\Large Appendices added in 1969}
\vglue 5mm

\section{A.  The mass distribution models}

\subsection{ } 
The mass distribution models discussed in the paper are determined by Eq.~(\ref{eq6.5.1})
\begin{equation}
\varphi ''(\xi ) = \frac{k}{ n\xi_0} \left( \frac{\zeta_0}{\zeta}
\right)^n , \label{eq6.A1.1} 
\end{equation}
where
\begin{equation}
\zeta^2 = \frac{\xi^2 +\xi^2_0 }{ 1+ \xi^2_0} {\rm ~~or~~}
\xi^2 = \frac{\zeta^2 - \zeta^2_0 }{ 1 - \zeta^2_0 } , \label{eq6.A1.2} 
\end{equation}
and
\begin{equation}
\zeta^2_0 = \frac{\xi^2_0 }{1+ \xi^2_0 } , ~~~ 
\xi^2_0 = \frac{\zeta^2_0 }{ 1-\zeta^2_0 }. \label{eq6.A1.3}
\end{equation}

The function $\varphi (\xi )$ and hence the potential can be
expressed via elementary functions when $2n$ is a natural number. In
general for present models $\varphi (\xi )$ is a hyper-geometric
function. But the density can be expressed via elementary
functions independently of $n$.

According to Eq.~(\ref{eq6.2.5}) and by using Eqs.~(\ref{eq6.5.1}) and (\ref{eq6.5.3}) we have the
general formula for the density
$$
\rho = \frac{1}{ 2\pi B( \frac{n-1}{ 2}, \frac{1}{ 2})}~ \frac{M}{
z_0^3\xi_0^3}~ \zeta^{n+2}_0 \left[ \frac{2}{ n-2} \frac{\zeta_2^{-n+2} -
\zeta_1^{-n+2} }{ \zeta^2_1 -\zeta^2_2} (\zeta^2_1 +\zeta^2_2 -2) + \right.$$
\begin{equation}
\left. + (1-\zeta^2_2 )\zeta^{-n}_2 - (\zeta^2_1 -1)\zeta^{-n}_1 
\right] (\zeta^2_1 -\zeta^2_2 )^{-2}. \label{eq6.A1.4} 
\end{equation}
Here we substituted $k$ according to Eq.~(\ref{eq6.5.6}) assuming $n>1$. 
For $n=2$ we have
\begin{equation}
\rho = \frac{1}{ 2\pi}~ \frac{M}{ z_0^3 \xi^2_0}~ \zeta^4_0 
\left[ \frac{\ln\zeta^2_1 -\ln\zeta^2_2 }{\zeta^2_1 - \zeta^2_2} 
(\zeta^2_1 +\zeta^2_2 -2)
+ \frac{1}{\zeta^2_1}+ \frac{1}{\zeta^2_2} -2 \right] (\zeta^2_1
-\zeta^2_2 )^{-2}. \label{eq6.A1.5} 
\end{equation}

The parameter $\zeta_0$ (as well as $\xi_0$) determines the
eccentricity of the model. For $\zeta_0 =0$ the model is flat, for
$\zeta_0 = 1$ spherical. The quantities $\zeta_1$ and $\zeta_2$ are
elliptical coordinates but defined in a different way of $\xi_1$ and
$\xi_2$. The coordinates $R$ and $z$ can be expressed through $\zeta_1$
and $\zeta_2$ according to the formulae
\begin{equation}
R^2 = z_0^2 \frac{(\zeta^2_1 -1)(1-\zeta^2_2 )}{ (1-\zeta^2_0 )^2}, ~~
z^2 = z_0^2 \frac{(\zeta_1^2 -\zeta^2_0)(\zeta^2_2 -\zeta^2_0 )}{ 
(1-\zeta^2_0)^2}. \label{eq6.A1.6} 
\end{equation}

In addition to the $n=4$ and $n=3$ cases discussed in the paper, quite simple
expression for the density results for the limiting case of $n=1$, when
the mass of the model becomes infinitely large. In this case
\begin{equation}
\rho = \rho_0\zeta_0 (1+\zeta_0 )^2 \frac{1+\zeta_1\zeta_2}{
\zeta_1\zeta_2 (\zeta_1 +\zeta_2 )^3}. \label{eq6.A1.7} 
\end{equation}

\begin{figure*}[ht]
\centering
\includegraphics[width=80mm]{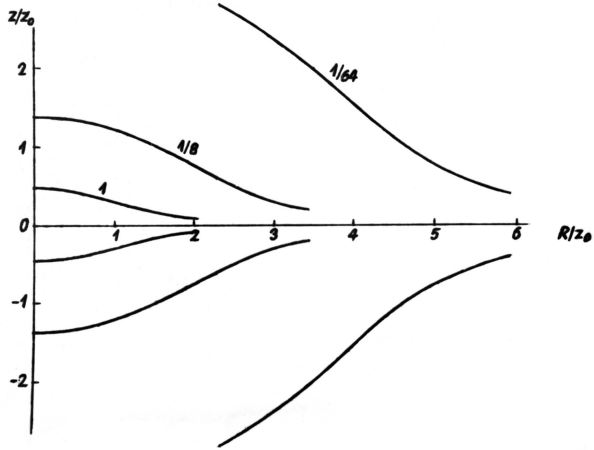}
\caption{Isodensity contours in the meridional plane of the Galaxy for
  the limiting $n=1$ model. The unit for the density is
  $\rho_0\zeta_0$}
\label{fig6.6}
\end{figure*}

For not too large $n$, bending points appear on the isodensity lines in
the meridional plane (when $\zeta_0$ is
small). This is illustrated in Fig.~\ref{fig6.6}, where for the limiting model
$n=1$ three isolines are given. The parameter $\zeta_0$ is taken to be zero
(more precisely a very small quantity). The unit for the density is
$\rho_0\zeta_0$.

For the spherical model ($\zeta_0\rightarrow 1$)
\begin{equation}
\zeta_1 = \zeta = \sqrt{1+r^2/r_0^2}, ~~~ \zeta_2 =1, ~~~ z_0\xi_0 = r_0,
\label{eq6.A1.8} 
\end{equation}
where $r$ is the distance from the model centre. The formula for the
density has the form
\begin{equation}
\rho = \frac{1}{ 2\pi B( \frac{n-1}{ 2}, \frac{1}{ 2})} \frac{M}{ r_0^3} 
\left( \frac{2}{ n-2} - \frac{n}{ n-2} \zeta^{-n+2} + \zeta^{-n} \right)
(\zeta^2 -1)^{-2}. \label{eq6.A1.9} 
\end{equation}
In particular, for $n=4$
\begin{equation}
\rho = \frac{M}{ \pi^2 r_0^3} \zeta ^{-4} = \rho_0\zeta^{-4} , \label{eq6.A1.10}
\end{equation}
and for $n=3$
\begin{equation}
\rho = \frac{M}{ 4\pi r_0^3} \frac{1+2\zeta}{\zeta^3 (1+\zeta )^2} = 
\frac{4}{ 3}\rho_0 \frac{1+2\zeta}{ \zeta^3 (1+\zeta )^2}.\label{eq6.A1.11}
\end{equation}
The spherical model with $n=3$ corresponds to the isochrone model by
M. H\'enon (Ann. d'Ap. {\bf 22}, 126, 1959).

In the limiting case of $n=1$ one has
\begin{equation}
\rho = \rho_0 \frac{4}{\zeta (1+\zeta )^2}. \label{eq6.A1.12} 
\end{equation}

\subsection{ }  
Recently we proposed a new class of models allowing for the
existence of the third quadratic integral. It is determined by the formula
\begin{equation}
\varphi (\xi ) = \frac{\alpha\Phi_0}{\beta +\zeta} \xi^2 , ~~~
\beta = \alpha +1, \label{eq6.A1.13} 
\end{equation}
and
\begin{equation}
\alpha\Phi_0 = \frac{GM\zeta_0}{ z_0\xi_0}. \label{eq6.A1.14} 
\end{equation}

For the potential we have the expression
\begin{equation}
\Phi = \frac{\alpha\Phi_0}{ x}~ \frac{\zeta_0^2 +\beta x+y }{ \beta^2 + \beta
x +y}, \label{eq6.A1.15} 
\end{equation}
where
\begin{equation}
x =\zeta_1 +\zeta_2 , ~~~ y=\zeta_1\zeta_2. \label{eq6.A1.16} 
\end{equation}
At $z=0$ the potential is $\alpha\Phi_0 /(\beta +\zeta_1 )$. This
is the natural generalisation of the potential from the $n=3$ model.

For the density we find after quite complicated calculations on the
basis of Eq.~(\ref{eq6.2.5}) 
$$
\rho =  \frac{1}{ 4\pi G}~ \frac{\alpha\Phi}{ z_0^2\xi_0^2x^3}~
\frac{\zeta_0^2}{\beta^2} \left[ \zeta_0^4 ~ \frac{(1+y)y+x^2 }{ y^3} + \right.
3\zeta_0^2 (\beta^2 -\zeta_0^2 ) \frac{(1+y)(\beta^2 +2\beta x+y) +x^2
}{ y(\beta^2 +\beta x+y)^2} + $$
\begin{equation} 
 + \left. 2(\beta^2 -\zeta_0^2 )^2 \frac{(1+y)(\beta^2 +3\beta x+y)+(\beta^2 +1)x^2
}{ (\beta^2 +\beta x+y)^3} \right]. \label{eq6.A1.17} 
\end{equation}

According to the theorem proved in the paper, the density $\rho$ is
nowhere negative if it is non-negative on the $z$-axis, \ie for
$x=1+y$. From that we derive the condition
\begin{equation}
\left.
\ba{ll}
\zeta_0^2 \le 1- \frac{1}{ 2}\alpha (1+\sqrt{1-4\beta}), & \\
\zeta_0^2 \ge 1- \frac{1}{2}\alpha (1-\sqrt{1-4\beta}). & \\
\ea
\right\}
\label{eq6.A1.18} 
 \end{equation}
When $\beta\ge 1/4$, the density is non-negative independently of
$\zeta_0$. If $\beta < 0$, the density must be
negative somewhere (for all $\zeta_0$).

If $\beta = \zeta_0$, the models coincide with the $n=3$ models. In this
case $\beta = \zeta_0 =0$ and we have a flat model, for $\beta =
\zeta_0 =1$ the isochrone model by H\'enon.

For $\alpha\rightarrow 0$, \ie for $\beta\rightarrow -1$ we have the
model by I.~L. Genkin (Tr. Astrofiz. Inst. {\bf 7}, 16, 1966).

For $\zeta_0 =1$ we have spherical models with
\begin{equation}
\Phi = \Phi_0 \frac{\alpha}{\beta +\zeta} \label{eq6.A1.19}
\end{equation}
and
\begin{equation}
\rho = \rho_0 \frac{\beta (1+2\zeta^2 )+3\zeta }{ 3\alpha\zeta^3 (\beta
+\zeta )^3}. \label{eq6.A1.20}
\end{equation}
For $\alpha\rightarrow$ 0 we have the spherical model by G.~M. Idlis (Astron.
Zh. {\bf 33}, 20, 1956, Izv. Astrofiz. Inst. {\bf 4}, No. 5--6, 1957),
for $\alpha =$ 1 ($\beta =$ 0) --- the model by Schuster, for $\alpha =$
2 ($\beta =$ 2) --- the model by H\'enon.

\begin{figure*}[ht]
\centering
\includegraphics[width=80mm]{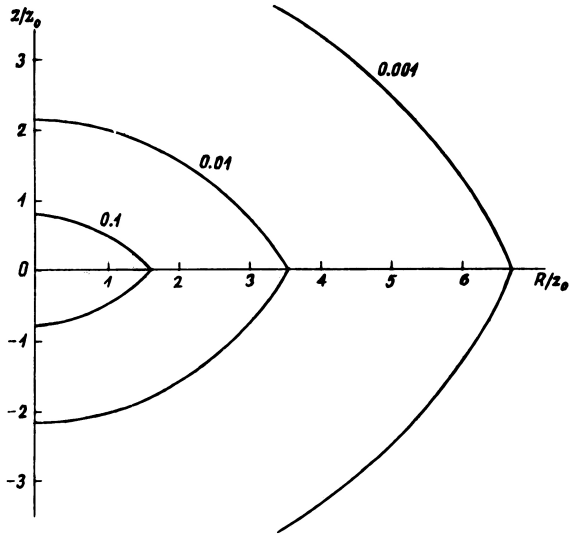}
\caption{Isodensity contours in the meridional plane for a lens-like model.}
\label{fig6.7}
\end{figure*}

If $\zeta_0 =$ 0 the model is flat only for $\beta =$ 0. For
$\beta >$ 0 a lens-like model results. In the simplest case of the
lens-like model $\alpha =$ 1 (quasi-isochrone lens-like model)
\begin{equation}
\Phi = \frac{2\Phi_0}{ x} \frac{x+y}{ 1+x+y}, \label{eq6.A1.21}
\end{equation}
\begin{equation}
\rho = \frac{4}{ 3}\rho_0 \frac{1+2x+y }{ (1+x+y)^2}. \label{eq6.A1.22} 
\end{equation}
In Fig.~\ref{fig6.7} the isodensity lines of this model in the meridional plane
are given.

\section{B.  On the triaxial generalisation of the Lindblad
diagram}

In case of generalisation of the Lindblad diagram into three dimensions
it is comfortable to use the space of $I_1$, $I_2^2$, $I_3$ instead of the space 
of $I_1$, $I_2$, $I_3$. In this space the characteristic
surfaces where the velocity components $v_1$ or $v_2$ turn to zero are
not parabolic cylinders but planes.

The equations of the characteristic surfaces in space of $I_1$,
$I^2_2$, $I_3$ are
\begin{equation}
\frac{w}{\xi^2 -1} = \xi^2I_1 - I_3 - \frac{\xi^2}{\xi^2 -1}I_2^2
+2\varphi (\xi ) = 0, \label{eq6.A2.1} 
\end{equation}
where $\xi$ is $\xi_1$ or $\xi_2$. The characteristic surfaces have
the enveloping surface. It is determined by the equation of the
characteristic surface together with the equation
\begin{equation}
I_1 + \frac{1}{ (\xi^2 -1)^2}I_2^2 +\varphi '(\xi )\xi^{-1} =0. \label{eq6.A2.2}
\end{equation}
The enveloping surface evidently consist of straight lines being
thus a lineated surface.

The part of the enveloping surface corresponding to $\xi\ge$ 1 (\ie 
$\xi =$ $\xi_1$) delineates together with the planes $I^2_2 =$ 0 and $I_3=$
$2\varphi (0)$ a region of physically possible values of $I_1$,
$I_2^2$, $I_3$. The remaining part of the enveloping surface with the
exception of the straight line $\xi =$ 1, \ie 
\begin{equation}
I_1 -I_3 +2\varphi (\xi ) = 0, ~~~ I_2^2 = 0, \label{eq6.A2.3} 
\end{equation}
remains outside of the physical region.

Intersection of the enveloping surface with the plane $I_3 = 2\varphi
(0)$ corresponds to circular orbits. The straight lines forming the
enveloping surface correspond to the orbits $\xi_1 =\mathrm{const}$. In the
$R,z$-plane they are periodic orbits along the $\xi_1$ coordinate
lines. 

 In the velocity space the following ellipse corresponds to the strict line 
 on the enveloping surface along which
 the characteristic plane is tangent to the surface:
\begin{equation}
\frac{v_2^2}{ \xi_1^2 -\xi_2^2}+ \frac{v_{\theta}^2}{\xi_1^2 -1} +
\frac{1}{\xi_1} \frac{\partial\Phi}{\partial\xi_1} = 0 , ~~v_1=0 \label{eq6.A2.4}
\end{equation}
the following hyperboloid corresponds to the strict line $\xi = 1$:
\begin{equation}
-(1-\xi_2^2 )v_1^2+(\xi_1^2 -1) = 2[ \Phi -\Phi^* -\varphi (1)], ~~
v_{\theta} =0 \label{eq6.A2.5} 
\end{equation}
They correspond to the foci of the triple family of
isosurfaces (\ref{eq6.7.5}) of orbit elements (in fact these second-order surfaces
are not confocal). 

\begin{figure*}[ht]
\centering
\includegraphics[width=80mm]{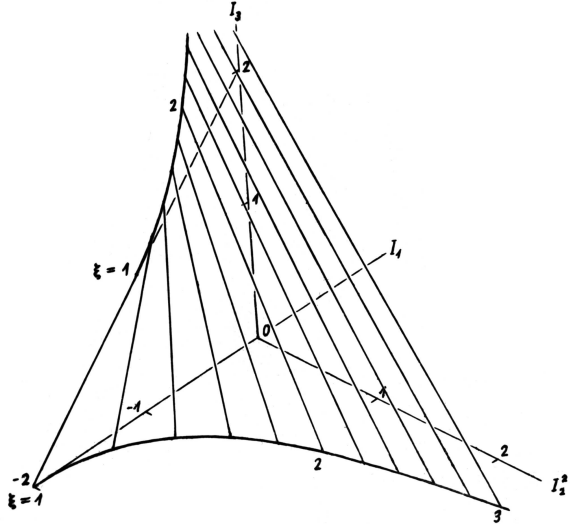}
\caption{Enveloping surfaces for the flat model.}
\label{fig6.8}
\end{figure*}

In order to illustrate the characteristics of the three-dimensional
generalisation of the Lindblad diagrams, the enveloping
surface for the simplest flat model is plotted in Fig.~\ref{fig6.8}. In this case (if we
remove a constant factor)
$$\varphi (\xi )=\xi $$
and the enveloping surface intersects the planes $I_3 =$ 0 and $I_2^2
=$ 0 along the curves with the following parametric equations
$$I_3=0 ~,~~ I_1=(\xi^2 +1)\xi^{-3} ~,~~  I_2^2= (\xi^2 -1)^2
\xi^{-3} ~; $$
$$         I_2^2=0 ~,~~ I_1=\xi^{-1} ~,~~~~~~~~~~~~  I_3=\xi ~.  $$

\section[C. Basic frequencies and periods
for the motion of a star]{C.  Canonical coordinates. Basic frequencies
  and periods  
for the motion of a star. Quasi-Keplerian orbit elements}  

\subsection{ } 
If we take for coordinates
\begin{equation}
q_1=\xi_1 ,~~~ q_2=\xi_2 ,~~~ q_3=\theta. \label{eq6.A3.1}
\end{equation}
the corresponding generalised momenta are
\begin{equation}
p_1=z_0 \frac{w_1}{\xi_1^2 -1} , ~~ p_2=z_0 \frac{w_2}{ 1-\xi_2^2} , ~~
p_3=z_3I_2. \label{eq6.A3.2} 
\end{equation}
These canonical variables can be replaced by the action variables
\begin{equation}
P_i = \frac{1}{ 2\pi}\oint p_i \rmd q_i , ~~ i=1, 2, 3 \label{eq6.A3.3} 
\end{equation}
(where circle integral is taken over one cycle of $q_i$ variation) and
by the corresponding canonically conjugal variables $Q_i$. For
$Q_i$ we have
\begin{equation}
z_0^{-1}Q_i = \frac{1}{ 2} (I_6+t) \frac{\partial I_1}{\partial P_i} - 
\frac{1}{ 2}I_4 \frac{\partial I_3}{\partial P_i} - I_5 \frac{\partial I_2
}{\partial P_i} , \label{eq6.A3.4}
\end{equation}
where the $I_4$, $I_5$, $I_6$ must be replaced by their expressions 
Eqs.~(\ref{eq6.7.9})--(\ref{eq6.7.10})
\begin{equation}
I_4 = U_1-U_2, ~~I_5=S_1-S_2-\theta , ~~I_6=T_1-T_2-t, \label{eq6.A3.5} 
\end{equation}
\begin{equation}
U=\int \frac{\rmd\xi}{ w} , ~~ S=I_2\int \frac{\xi^2 \rmd\xi }{ (\xi^2 -1)\omega},
~~ T=z_0\int \frac{\xi \rmd\xi}{ w}. \label{eq6.A3.6} 
\end{equation}
The partial derivatives $\partial I_1/\partial P_1$ must be
expressed through the reversed partial derivatives $\partial P_1/\partial
I_1$. The variable $P_3=P_{\theta}$ equals simply to
$z_0I_2$ and therefore, $\partial I_2/\partial P_{\theta} = z_0^{-1}$,
$\partial I_2/\partial P_1 =$ $\partial I_2/\partial P_2 = 0$. The
remaining derivatives can be expressed with help of circle integrals
\begin{equation}
U^0=\oint \frac{\rmd\xi}{ w} , ~~~S^0=I_2\oint \frac{\xi \rmd\xi }{ (\xi^2
-1)w}, ~~~ T^0 = z_0\oint \frac{\xi^2 \rmd\xi}{ w}. \label{eq6.A3.7} 
\end{equation}

The variables $P_i$ as functions of $I_1$, $I_2$, $I_3$ are the
(single-valued) integrals of motion. At the same time they are the
adiabatic invariants, \ie if the function $\varphi (\xi )$ determining
the gravitational field slowly varies, while the value of $P_i$ remains constant.

\subsection{ }  
The basic frequencies of the stellar coordinate oscillations
while the star moves are
\begin{equation}
\omega_i = \frac{2\pi}{\tau_i}= \frac{1}{ 2} \frac{\partial I_1}{\partial
P_i}, \label{eq6.A3.8} 
\end{equation}
where $\tau_i$ are the periods. For $\tau_i$ we have the relations
(this can be seen directly from the expressions of the integrals $I_4$,
$I_5$, $I_6$)
\begin{equation}
\tau_1 U_2^0=\tau_2 U_1^0 =\tau_{\theta} \frac{S_1^0U_2^0-S_2^0U_1^0}{
2\pi} = T_1^0U_2^0-T_2^0U_1^0. \label{eq6.A3.9} 
\end{equation}
The formulae for $\tau_1$ and for $\tau_2$ were derived also by H.C. 
van~de~Hulst (B.A.N. {\bf 16}, 235, 1962).

From the analogy with the perturbed Keplerian motions these periods may
be named as anamalistic, draconistic and sideric, respectively.

\subsection{ }  
For the Keplerian ellipse the following expressions for
$P_i$ result 
\begin{equation}
P_1 = I \left( \frac{1}{\sqrt{1-e^2}}-1\right) , ~~P_2=I- |I_2 | , 
~~P_{\theta}=I_2=I\cos i , \label{eq6.A3.10} 
\end{equation}
where $e$ is the orbital eccentricity, $i$ is the orbital inclination,
$I$ is the total kinetic momentum. Now we find
$$\frac{1}{\sqrt{1-e^2}}-1= \frac{P_1}{ |P_{\theta}|+P_2} $$

\begin{equation}
\cos i = \frac{P_1}{ |P_{\theta}|+P_2} \label{eq6.A3.11}
\end{equation}
If we use these formulae for galactic stellar orbits we obtain the
definition of ``quasi-Keplerian'' elements. The definitions were
proposed by D. Lynden-Bell (Observatory {\bf 83}, 23, 1962).

\section[D. Superposition of  flat and lens-like models]{D.  Use of
  elliptic integrals in calculations of 
three-dimensional orbits. A superposition of a flat and lens-like models}

If $w$ is a fourth-order polynomial of $|\xi |$, we can calculate
three-dimensional stellar orbits with the help of elliptic integrals. In
this case the functions $U$, $S$, $T$ are expressed through incomplete
elliptic integrals, the functions $U^0$, $S^0$, $T^0$ and $P$ ($P_1$
and $P_2$) through complete integrals.

This result can be derived for the superposition of flat and
quasi-isochrone lens-like models when
\begin{equation}
\varphi (\xi )=(1-\eta )\varphi_{fl} +\eta\varphi_l , \label{eq6.A4.1}
\end{equation}
where
\begin{equation}
\varphi_{fl} = \Phi_0\cdot |\xi | ,~~~~ \varphi_l =\Phi_0 \frac{2\xi^2}{
1+|\xi |}  \label{eq6.A4.2}
\end{equation}
and $\eta$ is a coefficient characterising the contribution of
lens-like model to the central potential.

For the potential at $z=$ 0 we have
\begin{equation}
\Phi_{z=0} = \Phi_0 \left[ (1-\eta )\zeta^{-1} + \eta \frac{2}{ 1+\zeta}
\right] , \label{eq6.A4.3} 
\end{equation}
where
\begin{equation}
\zeta = \sqrt{1+R^2/z_0^2}. \label{eq6.A4.4} 
\end{equation}

Further, because for flat systems
\begin{equation}
\delta_{fl} = \frac{1}{ 2\pi} \frac{\Phi_0}{ 6z_0}\zeta^{-3} , ~~
|\bar z |_{fl} = 0 , \label{eq6.A4.5}
\end{equation}
and for lens-like systems
\begin{equation}
\delta_l = \frac{1}{\pi} \frac{\Phi_0}{ 6z_0} \frac{1}{\zeta (1+\zeta )^2} ,
~~ |\bar z |_l= \frac{1}{ 2}z_0\zeta , \label{eq6.A4.6} 
\end{equation}
we have for their superposition
\begin{equation}
|\bar z |=z_0 \frac{2\eta\zeta }{ (1-\eta )(1+\zeta )^2 +4\eta\zeta^2}.
\label{eq6.A4.7}
\end{equation}

The expression for $\Phi_{z=0} (R)$ and $|\bar z|(R)$ can be used to
estimate the parameters $\Phi_0$, $z_0$, $\eta$. For the Galaxy
$\eta =$ 0.1 roughly.

%% file: chapter07.tex
\chapter[Some problems concerning the dynamics of the
  Galaxy]{Some problems concerning the dynamics of the
  Galaxy\footnote{\footnotetext ~~Tartu Astronomical Observatory,
    Teated, No.~3, 1956; Report in the First Meeting of the Commission
    on Stellar Astronomy of the 
  Astron. Council of Acad. Sci. USSR, May 1955. Published also in
  Notices of the  
  Acad. Sci. Estonian SSR {\bf 5}, 91 -- 107, 1956. } }

The aim of the present paper is to give a short review of the results
derived by the author on some problems of Galactic dynamics. We review
both theoretical and practical questions of stellar dynamics. Special
emphasise is given to different models of the Galaxy and to using the latest
radio observations of galactic rotation in mass distribution calculations.

\section{The theory of the third integral of the motion of stars}

It is known that the velocity distribution of stars in all subsystems
of the Galaxy is triaxial, \ie  can be represented by triaxial
velocity ellipsoid.\footnote{We have in mind the ellipse calculated
  from the velocity dispersion tensor. The velocity ellipsoid must not
  be confused with the ellipsoidal velocity distribution as it is
  often done. The assumption on ellipsoidal velocity distribution is 
not necessary in the theory of the third integral.} The longest axis
of the velocity ellipsoid is directed approximately along the galactic
radius, the shortest axis -- perpendicular to the galactic
plane. Theoretical explanation of the phenomena meets still certain
difficulties. The theory of stationary stellar systems, being
successful in explaining the basic properties of stellar motions, gave
biaxial velocity distribution. According to Jeans \citep[see
e.g.][]{Parenago:1954} it was usually assumed that the phase density
of a stationary axisymmetric stellar system is a function of two
integrals of motion -- the energy integral and the area integral. In
this case the velocity distribution is biaxial. It is symmetrical
about the axis laying in direction of galactic rotation,
\ie perpendicular to meridional plane. In order to have triaxial
velocity distribution, we must suppose that the phase density is a
function of three integrals of motion. For the stationary system these
integrals are independent of time. In addition, the integrals must be
single-valued, \ie for a given trajectory in the phase space only one
value of the integrals must correspond. The existence of such kind of
integrals is related with the restrictions on the gravitational
potential of the stellar system. For the existence of the energy and
the area integrals it is needed the stationarity and the axial
symmetry of the potential. For the existence of the third
single-valued integral, being different from previous integrals, it is
needed some other additional restriction.

A theory of the third integral was presented in our recent papers
\citep{Kuzmin:1953, Kuzmin:1954, Kuzmin:1956a}. In these studies
concrete expression for the third integral and the differential
equation for the gravitational potential, necessary for the existence
of the integral, were given. The equation is the linear second order
differential equation in partial derivatives of the potential with
respect to coordinates. It is just the additional restriction for the
potential needed for the existence of the third
integral.\footnote{This equation was derived earlier several times as
  the condition allowing the triaxial velocity distribution in
  stationary axisymmetric stellar systems. When preparing the present
  paper for print we heard about the paper by van Albada (Bosscha
  Obs. Contr. No.  1, 1952) where it was derived the expression for
  the third integral being essentially identical with ours. He derived
  also the corresponding differential equation for the potential. But
  van Albada wrongly concludes that the equation has no solutions
  applicable for real stellar systems.}

If the phase density is a function of all three integrals referred above,
the resulting velocity distribution is triaxial. One of the velocity
ellipsoid axis is directed still in direction of galactic rotation, but
there exist another two unequal with each other axis. At points, laying
in the galactic plane, one of them is perpendicular to that plane, the other is
directed along the galactic radius. Outside of the galactic plane these axis
have some angles, different from $\rm 0^o$ or $\rm 90^o$ about the
plane. 
Hence out of the galactic plane the velocity ellipsoid has some obliquity.
The obliquity of the velocity ellipsoid is one of the important
consequences of the theory of the third integral.

The theory of the third integral enables to treat the spatial and kinematical
structure of the Galaxy and of its subsystems more completely than it was
done earlier. Besides, one of the aids for theoretical handling of the
spatial and kinematical structure may be the three-integral generalisation of
the Bottlinger diagram. In addition, there exists a possibility to find the
expressions for remaining two independent integrals, having in this way all
five independent of each other and of time integrals. These remaining two
integrals are generally infinitely multiple-valued, and hence can not
contain in expression of the phase density. But they can be used in
calculations of the stellar orbits in general three-dimensional case. It
is possible to derive also the sixth integral being a function of time and
determining the position of a star in orbit.

\section{Models of the Galaxy resulting from the theory of the third
integral} 

As we mentioned above, the existence of the third integral put
certain restriction on the gravitational potential. The potential must
satisfy a differential equation, leading us to the problem, is the
restriction fulfilled in reality. Because the Galaxy is a steady
system only in first, quite rough approximation, the demand of
stationarity as well as axial symmetry are not satisfied precisely. Hence
the energy integral and the area integral are only approximate integrals.
Evidently, there are no reasons to expect more also from the third
integral. We may use that integral in equal terms with the first two
integrals, when the potential only approximately satisfies the referred
differential equation. To check the possibility of even approximate
correspondence of the theory with reality, we constructed the models of
stellar systems where the condition of the existence of the third integral
is realised \citep{Kuzmin:1954, Kuzmin:1956a}.

The differential equation for the potential $\Phi$ can be solved in
general form. Using the elliptical coordinates $\xi_1$ and $\xi_2$ instead
of the cylindrical coordinates $R$ and $z$, we may write the solution in
following simple form
\begin{equation}
\Phi = \frac{\varphi (\xi_1) - \varphi (\xi_2) }{ \xi_1^2 - \xi_2^2}, 
\label{eq7.1}
\end{equation}
where $\varphi$ is an arbitrary function. The foci of elliptical
coordinates lie on galactic symmetry axis at both sides of galactic plane
at a distance $z_0$ from the plane.\footnote{The coordinates $\xi_1$
and $\xi_2$ are determined in a way, that on galactic axis $z = z_0\xi_2$
for $|z|<z_0$ and $|z| = z_0\xi_1$ for $|z|>z_0$.}

From the Poisson's equation we may derive the expression for the matter
density. It results that when choosing the function $\varphi$ in a suitable
form, we may have the model of stellar system with finite mass and nowhere
negative density. In addition, it is possible to vary the eccentricity of
the model.

In case of sufficiently flattened model, similar to the Galaxy, the
potential in galactic plane ($z=0$) has approximately the form
\begin{equation}
\Phi_{z=0} = \Phi_0 \left( 1+ \frac{R^2}{ z_0^2} \right)^{-1/2} , \label{eq7.2}
\end{equation}
where $\Phi_0$ is the potential in Galactic centre. From Eq.~\ref{eq7.2} the
circular velocity is approximately
\begin{equation}
V = V_0 \frac{R}{z_0} \left( 1+ \frac{R^2}{z_0^2} \right) ^{-3/4},
\label{eq7.3}
\end{equation}
where $V_0 = \sqrt{\Phi_0}$. For the density in galactic plane
$\rho_{z=0}$ and for the projected onto the galactic plane density
$\delta$ we derive
\begin{equation}
\rho_{z=0} = \rho_0 \left( 1+ \frac{R^2}{z_0^2} \right)^{-2}, \label{eq7.4}
\end{equation}
\begin{equation}
\delta = \delta_0 \left( 1+ \frac{R^2}{z_0^2} \right)^{-3/2} , \label{eq7.5}
\end{equation} 
where $\rho_0$ and $\delta_0$ are the values of $\rho_{z=0}$ and $\delta$
for $R=0$.

All these formulae are independent of the concrete form of $\varphi$. It is
needed only that the model is sufficiently flat. But to have a model,
where the density is known at all points in space, we must choose some
concrete form of $\varphi$. In our paper \citep{Kuzmin:1956a} we proposed 
a two-parameter expression for $\varphi$, giving two-parameter family of
models. The parameters determine the eccentricity of the model and the
variation of the density perpendicular to the galactic plane.

The most simple density distribution has the
model, corresponding to inhomogeneous spheroid with the density decreasing
in all directions according to the law similar to Eq.~(\ref{eq7.4}). Another model,
discussed by us in detail, is interesting because of simple forms of the
function $\varphi$ and the potential $\Phi$. In this case the function
$\varphi$ is proportional to $(\xi_0^2 + \xi^2 )^{1/2}$, where $\xi_0$ is
a constant. The isodensity surfaces differ somewhat from ellipsoids of
revolution. Their meridional sections are more curved near their extreme,
compared to the ellipses with the same eccentricity. In addition to these models
we made preliminary calculations also for other models. For some of
them the isodensity surfaces have bending points, and are similar to the
contours of edge-on spiral galaxies. In general the resulting models seem
to be realistic. Therefore we may expect that the condition for the
existence of the third integral is approximately fulfilled for the Galaxy.

The drawback of our models, basing on the third integral, is too slow
decrease of the density at large distances from the centre, as it is evident
from Eqs.~(\ref{eq7.4}) and (\ref{eq7.5}). In addition, the eccentricity of the isodensity
surfaces near the centre is as large as at outer regions. But
according to our understanding the galactic nucleus is quite spheroidal.
Although, changing of the density law at very small distances and at very
large distances from the centre does not influence significantly the
potential at remaining parts of the model, and hence the condition of the
existence of the third integral remains approximately valid. Thus the 
referred drawbacks are not insurmountable for the theory.

\section{Spheroidal models of the Galaxy}

Galactic models, resulting from the theory of the third integral of
stellar motion, can be used in practice for various
approximate calculations. But due to the referred deficiencies of these
models, more useful are other models.

One of the methods of constructing Galactic models is the superposition of
coaxial homogeneous spheroids with different dimensions and eccentricity
used by \citet{Oort:1932, Oort:1941, Oort:1952}. The drawback of the method is that the
resulting model is quite rough with large density jumps. For this reason
more convenient seems to represent the mass distribution of the Galaxy
with the superposition of inhomogeneous spheroids, where the density
decreases smoothly from the centre to the periphery. The properties of
suitably chosen inhomogeneous spheroids may be as simple as are of
homogeneous spheroids. But the number of inhomogeneous spheroids needed for
satisfactory representation of Galactic mass distribution is significantly
smaller. If there is not needed high precision, it is possible to limit
even with one inhomogeneous spheroid.

The inhomogeneous spheroid is characterised by the ratio of short semiaxis
of the isodensity surface to great semiaxis $\epsilon$, and by the density
distribution $\rho (a)$ as a function of great semiaxis of isodensity
surface $a$. Instead of the density it is possible to use the function $\mu
(a)$ or the function $\rho_s(a)$, related with the density in following way
\begin{equation}
\frac{\mu (a)}{4\pi a^2} = \rho_s(a) = \epsilon\rho (a). \label{eq7.6}
\end{equation}
The function $\mu (a)$ is the mass per unit interval of $a$, $\rho_s(a)$ is
the function
calculated from $\mu (a)$ within the assumption of spherical mass
distribution. These functions are suitable to call the mass function and
the spherical density function, respectively.

Knowing the mass function of the spheroidal model, we may calculate the
circular velocity \citep[see][]{Kuzmin:1952ac}
\begin{equation}
V^2 = G \int_0^R \frac{\mu (a)~\rmd a }{ \sqrt{R^2 - a^2e^2}} , \label{eq7.7}
\end{equation}
where $G$ is the gravitational constant and
\begin{equation}
e^2 = 1 - \epsilon^2 \label{eq7.8}
\end{equation}

For the potential in the symmetry plane of the model (in galactic plane)
we have
\begin{equation}
\Phi_{z=0} = G \int_0^{\infty} \mu (a) \chi \left( \frac{a}{R}\right) 
~ \frac{\rmd a}{a} , \label{eq7.9}
\end{equation}
$$ \chi\left( \frac{a}{R} \right) = \left\{ 
\begin{array}{ll}
\frac{1}{e} \arcsin \frac{ae}{R},  &  \mbox{for $a\le R$,}   \\
\noalign{\smallskip}
\frac{1}{e} \arcsin e,                     & \mbox{for $a\ge R$.}
\end{array}  
\right. $$

The mass function allows us to calculate also the projected density
$\delta$ and the total mass $M$ of the model. Calculations are according
to formulae
\begin{equation}
\delta = \frac{1}{2\pi} \int_R^{\infty} \frac{\mu (a) ~\rmd a }{ a\sqrt{a^2 -
R^2}} , \label{eq7.10}
\end{equation}
\begin{equation}
M = \int_0^{\infty} \mu (a)~\rmd a. \label{eq7.11}
\end{equation}

Our spheroidal model, being the most simple from the models based on the
theory of the third integral, belongs also to these class of models,
consisting of one inhomogeneous spheroid. Other models in form of
inhomogeneous spheroids were discussed earlier by \citet{Perek:1948,
  Perek:1951, Perek:1954},  and recently by \citet{Idlis:1954,
  Idlis:1956}. The Idlis model is very interesting 
and is worth to more close discussion.

\section{Idlis model of a stellar system}

In construction of his model Idlis starts from the expression
for potential used by \citet{Parenago:1950a, Parenago:1952} in his
studies of the
potential of the Galaxy. The expressions for the potential and corresponding
expression for the circular velocity are
\begin{equation}
\Phi_{z=o} = \Phi_0 \left( 1+ \frac{R^2}{ R_0^2} \right)^{-1} \label{eq7.12}
\end{equation}
and
\begin{equation}
V = V_0 \frac{R}{R_0} \left( 1+ \frac{R^2}{ R_0^2} \right)^{-1} ,
\label{eq7.13}
\end{equation} 
where $R_0$, $\Phi_0$ and $V_0$ are constants, while $V_0 =
\sqrt{2\Phi_0}$. 

These formulae result from assumptions that for flat subsystems of the
Galaxy the velocity distribution is strictly ellipsoidal. In this case
according to the Oort-Lindblad theory \citep[see e.g.][]{Parenago:1954} the
rotation of flat subsystems is described by Eq.~(\ref{eq7.13}) and $V$ can be
handled as the circular velocity, because the centroid velocity nearly
coincides with the circular velocity. However, it is needed to keep in mind
that in real galaxies the velocity distribution of flat subsystems is not necessarily
strictly ellipsoidal. Hence, Eq.~(\ref{eq7.13}) can be used only as a suitable
interpolating formula.\footnote{This question we discussed quite in
detail in one of our previous papers \citep{Kuzmin:1954}.} The same is valid
also in case of our formula (\ref{eq7.3}), because the condition for the existence
of the third integral may be fulfilled only approximately. 

The law (\ref{eq7.13}) has a drawback -- the corresponding
acceleration along the to galactic radius decreases too rapidly at
large $R$, and it is not possible to find the nonnegative mass
configuration, giving this kind of acceleration. But one may suppose,
following to Idlis, that the law (\ref{eq7.13}) is valid only inside
of the Galaxy. Outside of the Galaxy there will be some other law. In
this case the resulting model of the Galaxy in the form of an 
inhomogeneous spheroid is quite acceptable.

If the spheroid is very flat, the expression for the density results
easily. In this case it is needed to take in Eq.~(\ref{eq7.7}) $e=1$,  and we have
for $\mu (a)$ the Abel integral equation. By solving it and using Eq.~(\ref{eq7.6}), we
derive the following expression for the density at $z = 0$
\begin{equation}
\rho_{z=0} = \rho_0 \left( 1+ \frac{R^2}{ R_0^2}\right)^{-2} 
\left[ \frac{3}{4} + \frac{1}{4}\left( 1-2 \frac{R^2}{R_0^2}\right)
\left( 1+ \frac{R^2}{R_0^2}\right)^{-1/2} \frac{R_0}{R} {\rm arcsh}
\frac{R}{R_0}\right]. \label{eq7.14}
\end{equation}

The expression in square brackets is zero for $R=R^0\simeq
2.7~R_0$, and negative for $R>R^0$. Thus the value $R=R^0$ can be handled
as the model radius, and we assume that the density vanishes outside of that
limit. Because for $R > R^0$ Eq.~(\ref{eq7.12}) is not valid and at $R = R^0$
the potential must be continuous and at infinity it must vanish,
it is needed to add some constant to the potential. Idlis
found that for very flattened model the constant is $0.074~\Phi_0$. This
constant rises significantly the potential at large $R$. Hence at
large $R$ the escape velocity exceeds the circular velocity more than
it was according to Parenago's calculations \citep{Parenago:1950a, Parenago:1952}.

The law (\ref{eq7.14}) gives quite realistic density variation. At small $R$ the
density decreases approximately similar to our model. But at large $R$
the density decrease is faster and vanishes as $R$ approaches to
$R^0$. The finite radius of Idlis model is a significant advantage when
compared to our model.

\begin{figure*}[ht]
\centering
\includegraphics[width=100mm]{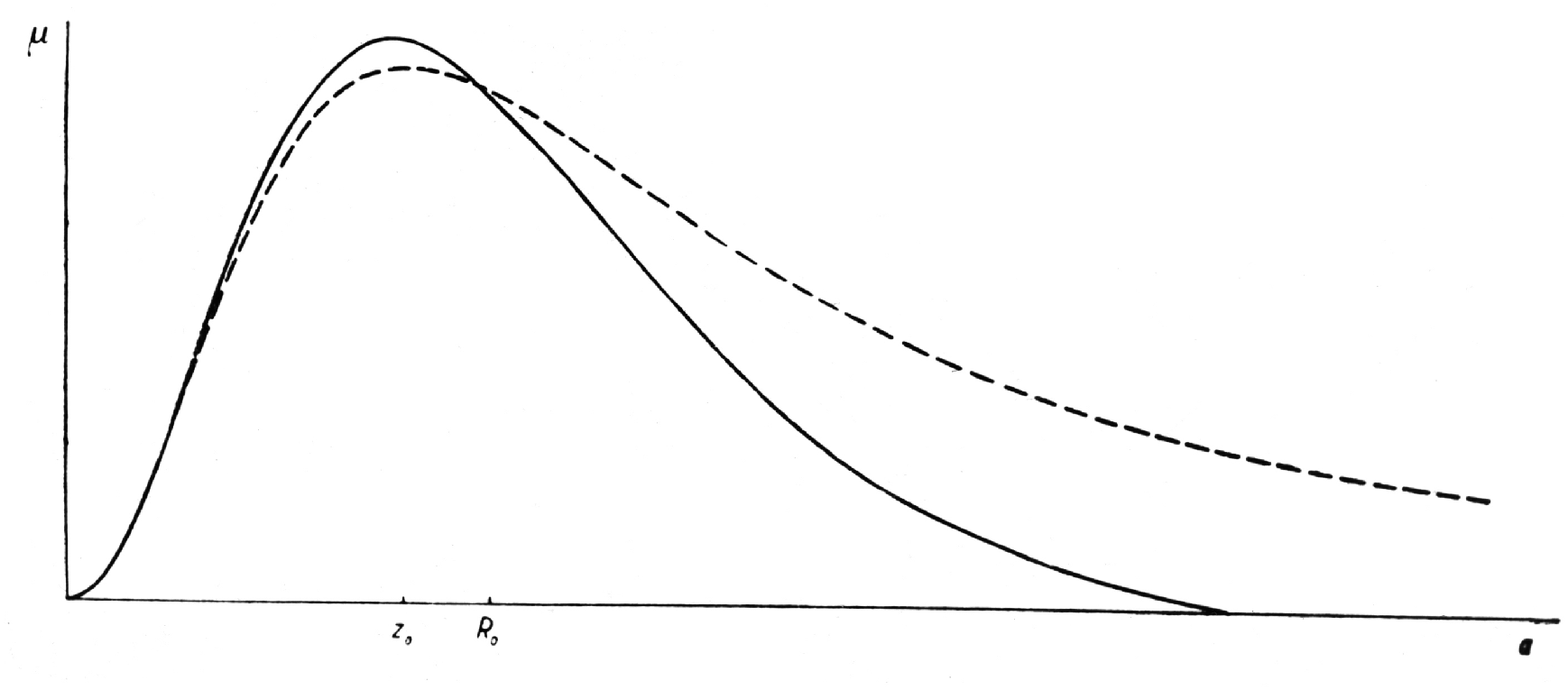}
\caption{}
\label{fig7.1}
\end{figure*}

The difference between our model and Idlis model is plotted in Fig.~\ref{fig7.1}. The 
continuous curve corresponds to the mass function of Idlis model, the
dashed curve to our spheroidal model. The spherical densities at the
centres of both models were taken equal,  and it was chosen $z_0 =
0.8~R_0$. 

With having in mind the practical application of Idlis model,  it would be
recommendable to generalise it for finite $\epsilon$. The generalisation
can be done by using still Eq.~(\ref{eq7.13}) for the circular velocity. In this case
Eq.~(\ref{eq7.7}) can be handled as the integral equation for the mass
function. But reminding that Eq.~(\ref{eq7.13}) is no more than an interpolation
formula, it is more convenient to accept at first the mass function,
corresponding to the density (\ref{eq7.14}), and to calculate the circular velocity
by integration of (\ref{eq7.7}). We did it and it results that the potential in
the symmetry plane and the circular velocity were expressed via elementary
function both inside and outside of the model. Similar calculations
were made also for our spheroidal model.\footnote{These calculations
we made together with J.~Einasto. Derived equations were published only in
1968 \citep{Einasto:1968}. [Later footnote.]} 

\section{Generalised spheroidal model}

Although the Idlis model represents the real Galaxy probably
better than our model, even this model can be only a first quite rough
approximation of reality. To represent the real mass distribution of the
Galaxy more precisely,  it is needed to add several suitably chosen 
inhomogeneous spheroids. These spheroids may be Idlis models with
different axial ratios and radii, for example.

But the mass distribution of the Galaxy may be represented with some
composite model also in other way. We need not to start from some
special models with given in advance density distribution. Contrary, it is
possible to use more general composite model with the mass function
determined from observational data without any ahead fixed model.

This kind of composite model may be constructed also on the basis of
nonhomogeneous spheroids \citep{Kuzmin:1952ac, Kuzmin:1955aa}. The density distribution,
however, is not fixed in advance,  and the number of spheroids is great
(infinite in limiting case).

Every spheroid belonging to the model is characterised by its axial ratio
$\epsilon_i$, and its density function $\rho_i(a)$ or the mass function
$\mu_i(a)$. Summation of $\rho_i$ and $\mu_i$ gives the integrated density
and mass functions
\begin{equation}
\rho (a) = \sum \rho_i(a), ~~~~~ \mu (a) = \sum \mu_i(a), \label{eq7.15}
\end{equation}
characterising the whole model. The integrated density, when we replace the
argument $a$ with the argument $R$,  turns into the density in the symmetry
plane. The integrated mass function characterises the model mass distribution.
Besides, the relation between the integrated mass function and the projected
density of the model, is evidently the same as it was in case of individual
spheroid, and is expressed by Eq.~(\ref{eq7.10}).\footnote{Equation (\ref{eq7.10}) can be
handled as the definition of the mass function for general case.}
Evidently also the total mass is expressed via $\mu (a)$ in the same way,
\ie by Eq.~(\ref{eq7.11}).

In addition to the functions above it is needed to introduce the
flattening functions
\begin{equation}
\epsilon_1(a) = \frac{\sum\epsilon_i\rho_i(a)}{\rho (a)}, ~~~~~
\epsilon_2(a) = \frac{\sum\epsilon_i\mu_i(a)}{\mu (a)} , \label{eq7.16}
\end{equation}
\ie the averaged with weights $\rho_i$ and $\mu_i$ values of
$\epsilon$. These functions, both in its own way, characterise the model.
The function $\epsilon_1(a)$ is related with the mass and density
functions in similar way as it was in case of individual spheroid, \ie by
the equation
\begin{equation}
\frac{\mu (a)}{ 4\pi a^2} = \rho_s(a) = \epsilon_1(a)\rho (a). \label{eq7.17}
\end{equation} 
When all $\rho_i(a)>0$, then evidently $\epsilon_2(a)>\epsilon_1(a)$.

If the mass distribution is sufficiently flattened, then the mass function
is related with the circular velocity and the potential in the symmetry
plane of the Galaxy in a similar way as it was in case of individual
spheroid, \ie by Eqs.~(\ref{eq7.7}) and (\ref{eq7.9}). But in present case it is needed to
use $\epsilon_2(a)$, substituting in these formulae
\begin{equation}
e^2 = 1-\epsilon_2^2(a). \label{eq7.18}
\end{equation}

Using Eq.~(\ref{eq7.7}) as an integral equation we are able to determine the mass
function from the circular velocity as an empirical function. Knowing the
mass function we may calculate the potential in the galactic plane, and
the corresponding escape velocity,  the projected density and the spherical
density. From the later we may move to the density function. But to do it
we need to know the function $\epsilon_1(a)$ with sufficient precision (in
calculation of the mass function we may limit only with approximately
estimated function $\epsilon_2(a)$).

\section{The mass distribution of the Galaxy}

In our earlier paper \citep{Kuzmin:1952ac} we determined the mass
function and the projected density function of the Galaxy from the motion of
the long-period cepheids. Unfortunately these data
cover only quite moderate range of $R$, and are influenced by significant
systematic errors. Recently we made a new determination of the mass
function and the projected density function on the basis of circular
velocities,  derived from radio observations of 21~cm interstellar hydrogen
line made by Dutch astronomers \citep{Kwee:1954}.

The radio observations do not give directly the circular velocity $V$ but
give the differential rotation function $U(x)$
\begin{equation}
U(x) = V(x) - V_{\odot} x, ~~~~ x= \frac{R}{ R_{\odot}}, \label{eq7.19}
\end{equation}
where $R_{\odot}$ and $V_{\odot}$ are the values of $R$ and $V$ in the
vicinity of the Sun. The function $U(x)$ is just the Camm function
multiplied by $x$. The radio observations enable to determine the function
$U(x)$ more or less precisely for the whole range of $x\le 1$. The results
of observations, smoothed in a suitable way,  are given in Table~7.1. For
$x>1$ the radio observations do not allow to determine $U(x)$. Hence by
using of this function we may calculate the mass function only for inner
parts of the Galaxy as regards the Sun. For outer parts the mass
function must be extrapolated.

In calculations of the mass function it is needed to know the circular
velocity near the Sun (we need to move from $U$ to $V$), and the
function $\epsilon_2(a)$ in order to use the integral equation (\ref{eq7.7}).

The results for the mass function only weakly depend from $\epsilon_2(a)$.
For this reason in calculations it was taken $\epsilon_2(a) = \mbox{const} =$ 0.2.
It is somewhat larger than derived by us recently value $\epsilon = 0.16$
\citep{Kuzmin:1955aa}, but the later is the mean value of $\epsilon_1$
and $\epsilon_2$, and 
must be greater than $\epsilon_1$ (see above).\footnote{In fact
$\epsilon$ is twice less and thus $\epsilon_2$
need to be taken $\sim 0.1$. [Later footnote.]}

But the mass function is very sensitive to the circular velocity in the
solar neighbourhood. For determination of $V_{\odot}$ 
the following relation is usually used: 
\begin{equation}
V_{\odot} = (A - B) R_{\odot}, \label{eq7.20}
\end{equation}
where $A$ and $B$ are the Oort constants in the solar neighbourhood. From
quantities in this formula the constant $A$ is known most precisely being
equal to 20~km/s/kpc. The value of $R_{\odot}$ is known slightly worse.
But the radio observations allow to determine 
\begin{equation}
A R_{\odot} = - \frac{1}{2} \left( \frac{dU}{dx} \right)_{x=1}.  \label{eq7.21}
\end{equation}

This value is near to 150~km/s (Table~7.1). Accepting for $A$ the value
referred above, we have $R_{\odot} =$ 7.5~kpc, being quite acceptable
result. Most insufficiently is known the value of $B$. On the basis of
proper motions in GC system we have $B =$ --13~km/s/kpc. But the proper
motions in FK3 system give $B= -7$~km/s/kpc. These two values of $B$ give
for $V_{\odot}$ highly different results, namely $V_{\odot} \simeq $
250~km/s in first case,  and $V_{\odot} \simeq $ 200~km/s in second 
case.\footnote{According to contemporary estimates $R_{\odot} \simeq
$ 10~kpc, $A\simeq$ 15~km/s/kpc and $B \simeq$ --10~km/s/kpc, giving for
the circular velocity $V_{\odot} \simeq $ 250~km/s/kpc. [Later footnote.]}

Hence, using the  Eq.~(\ref{eq7.20}) does not allow to determine the circular velocity
in the vicinity of the Sun with sufficient precision. But its value may be
significantly adjusted during the calculation and extrapolation of the
mass function. 

The extrapolation of the mass function must be done smoothly without jumps.
In addition, there must be taken into account three conditions. First, the
mass function must vanish when $a$ approaches to a certain upper
limit $R_m = R_{\odot} x_m$,  being in fact the Galactic radius. Second, the
mean speed of the mass function decrease when $a$ increases must be in
agreement with the observed radial gradient of stellar density in wide
neighbourhood of the Sun. And third, the potential, corresponding to the
mass function, must be chosen in a way, that the referred Galactic radius
will be in agreement with the real upper limit of stellar velocities in the
vicinity of the Sun. In other words, the stars moving in the vicinity of
the Sun with maximum galactocentric velocities must reach their apogalaxy
just at the boundary of the Galaxy.\footnote{Usually it is wrongly
supposed that the upper limit of velocities corresponds to the escape
velocity,  and it is not taken into account the finite dimensions of the
Galaxy.} 

The mathematical expression of the second condition is
\begin{equation}
\frac{\rmd \log\rho_s }{ \rmd a} = \frac{1}{\mu (a)} \sum \frac{\rmd\log\rho_i }{\rmd a}
\mu_i(a). \label{eq7.22}
\end{equation}

The right side of this expression is the averaged with the weight $\mu_i$
value of the radial gradient of density logarithm for individual
spheroids. Within a certain approximation we may identify the spheroids
with individual subsystems of the Galaxy.

The mathematical expression of the third condition we derive when we use
the energy and the angular momentum integrals for stars, moving in galactic
plane with maximum galactocentric velocities. It has the form
\begin{equation}
2 (\Phi_{\odot} - \Phi_m ) = \left( 1 - \frac{1}{ x_m^2} \right)
\left( V_{\odot} + v_m \right)^2 , \label{eq7.23}
\end{equation}
where $v_m$ is the excess of the upper limit of the solar vicinity's
velocity over the circular velocity, $\Phi_{\odot}$ and $\Phi_m$ are the
values of $\Phi_{z=0}$ in the solar neighbourhood and at the Galactic
radius. 

In calculations we used $v_m =$ 65~km/s from the paper by \citet{Oort:1928}. The
mean radial gradient of the density logarithm was taken 0.2 per kpc. This
value we found on the basis of data by \citet{Parenago:1954}, and considering
that the largest contribution to the mass function results from the
intermediate and the spherical subsystem of the Galaxy as the most massive
ones. For the Galactic radius it was taken the value $x_m =$ 2.5, where
the spatial density is approximately hundred times smaller than in the
vicinity of the Sun (using the density gradient above).

From calculations it results that the mass function can be extrapolated
quite smoothly and in agreement with the accepted values of $v_m$, $x_m$
and the gradient of density logarithm only in the case when the circular
velocity in the solar neighbourhood is near to 250~km/s. Besides, the errors
of the accepted parameters above enables to vary $V_{\odot}$ only within the
limits of $\pm 15$~km/s. Hence the most probable value of the circular
velocity in solar neighbourhood is near to the value of $B$ in GC 
system.\footnote{see previous footnote. [Later footnote.]}

\begin{figure*}[ht]
\centering
\includegraphics[width=100mm]{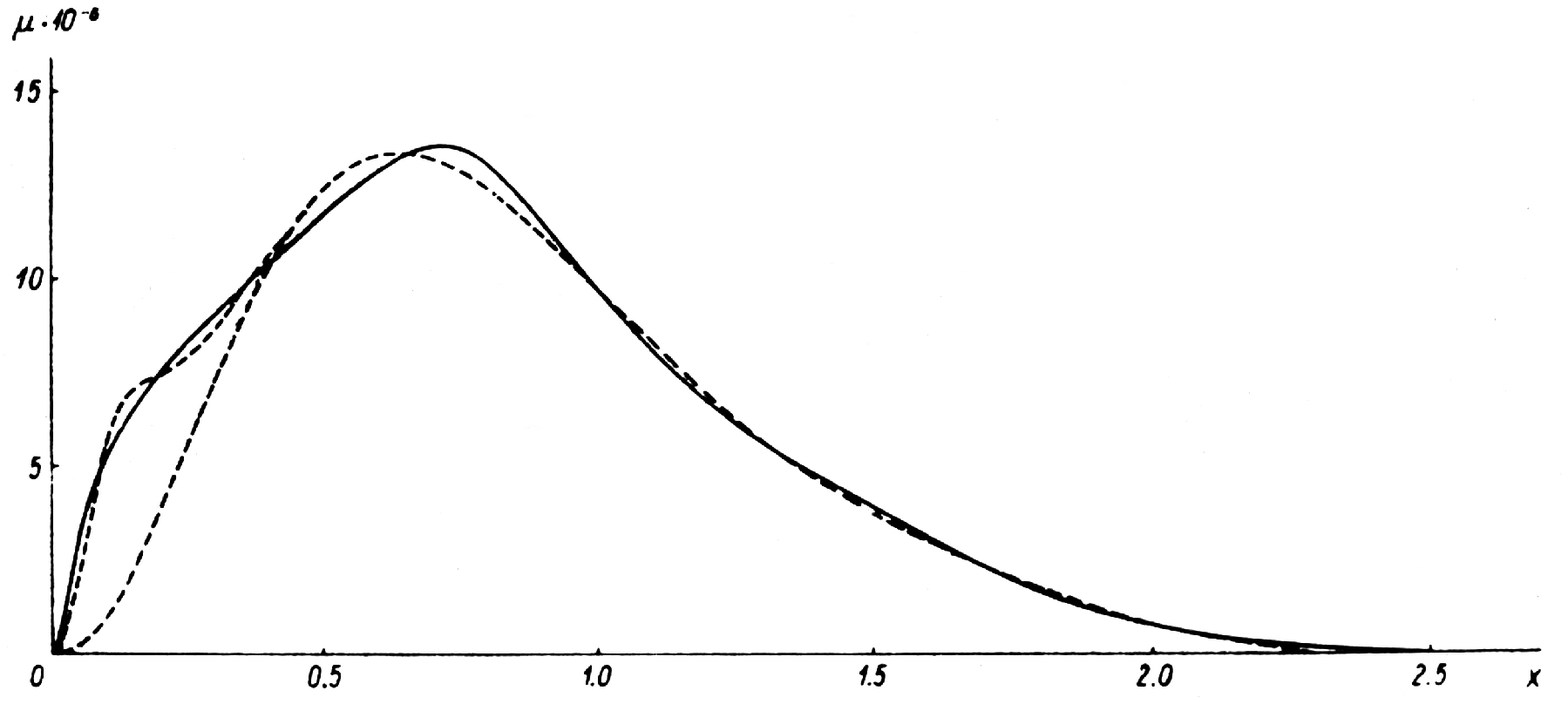}
\caption{}
\label{fig7.2}
\end{figure*}

If we accept $V_{\odot} =$ 250~km/s the resulting mass function is
like to the one, represented by continuous line in Fig.~\ref{fig7.2}
(argument $a$ is replaced by argument $x$).

The corresponding numerical values are given in
Table~7.1.  In the same Table
the values of the circular velocity $V$, the escape velocity in the
galactic plane $V_k = \sqrt{2\Phi_{z=0}}$, the projected density $\delta$
and the spherical
density $\rho_s$ are given. The values of the projected and the spherical
densities at $x=0$ are only approximate estimates. This is because
the function $U(x)$ is uncertain for small $x$,  and the referred densities
are very sensitive to the function $U(x)$ at small $x$. For the total
mass of the Galaxy we derived the following result
$$M = 107\cdot 10^9 {\rm M_{\odot}}.$$

We like to mention that the mass function and the circular velocity
depend only on function $U(x)$ and the value of $V_{\odot}$. The total
mass of the Galaxy, the projected density and the spherical density depend
in addition on the value of $R_{\odot}$,  taken to be 7.5~kpc (see 
above).\footnote{Due to decreasing of $\epsilon$ the functions $\mu
(a)$, $\delta (a)$ and $\rho_s(a)$ decrease somewhat. The increase of
$R_{\odot}$ causes the decrease of the surface density $\delta (a)$
proportionally to $R_{\odot}^{-1}$, and increase of the total mass
proportionally to $R_{\odot}$. The spherical density $\rho_s(a)$ decreases
proportionally to $R_{\odot}^{-2}$, \ie nearly twice. [Later footnote.]}

The derived results are in general similar to the results of our earlier
paper \citep{Kuzmin:1952ac}. But the mass function differs significantly. For
$x>1$ its behaviour is more smooth, and the second maximum, derived by us
earlier, is absent now. For very small $x$ the mass function is much
higher now,  giving higher central density of the Galaxy, and indicating to
the presence of galactic nucleus.

\begin{table}[ht]
\label{table7.1}
\vbox{\halign{
\hfil #\hfil\quad & \quad\hfil #\hfil && \quad\hfil #\hfil\cr
\multispan7 Table 1\hfil\cr
\noalign{\bigskip\hrule\medskip}
$x=R/R_{\odot}$ & $U$ & $V$ & $V_k$ & $\mu\cdot 10^{-6}$ & $\delta\cdot
10^{-3}$ & $\rho_s$ \cr
\noalign{\smallskip}
& \multispan3 \hfil km/s\hfil & $\rm M_{\odot}/pc$ & $\rm M_{\odot}/pc^2$
& $\rm M_{\odot}/pc^3$ \cr
\noalign{\medskip\hrule\medskip}
0.0 &   0 &   0 & 610 &  0.0 & 6.   & 11.    \cr
0.1 & 118 & 143 & 573 &  5.2 & 2.1  &  0.74  \cr
0.2 & 128 & 178 & 542 &  7.5 & 1.39 &  0.26  \cr
0.3 & 124 & 199 & 515 &  9.0 & 1.01 &  0.141 \cr
0.4 & 114 & 214 & 490 & 10.4 & 0.77 &  0.092 \cr
\noalign{\smallskip}
0.5 & 102 & 227 & 467 & 11.7 & 0.61 &  0.066 \cr
0.6 &  88 & 238 & 445 & 12.9 & 0.49 &  0.050 \cr
0.7 &  72 & 247 & 424 & 13.5 & 0.38 &  0.039 \cr
0.8 &  53 & 253 & 404 & 13.0 & 0.28 &  0.029 \cr
0.9 &  28 & 253 & 385 & 11.4 & 0.20 &  0.020 \cr
\noalign{\smallskip}
1.0 &   0 & 250 & 367 &  9.6 & 0.146 & 0.0135 \cr
1.1 &     & 245 & 351 &  8.0 & 0.106 & 0.0093 \cr
1.2 &     & 240 & 336 &  6.7 & 0.078 & 0.0066 \cr
1.3 &     & 234 & 322 &  5.6 & 0.057 & 0.0047 \cr
1.4 &     & 228 & 309 &  4.7 & 0.040 & 0.0034 \cr
\noalign{\smallskip}
1.5 &     & 221 & 298 &  3.8 & 0.028 & 0.0024 \cr
1.6 &     & 215 & 287 &  3.0 & 0.019 & 0.0017 \cr
1.7 &     & 208 & 278 &  2.3 & 0.013 & 0.0011 \cr
1.8 &     & 201 & 269 &  1.6 & 0.008 & 0.0007 \cr
1.9 &     & 195 & 261 &  1.1 & 0.005 & 0.0004 \cr
\noalign{\smallskip}
2.0 &     & 189 & 254 &  0.7 & 0.003 & 0.0002 \cr
2.1 &     & 183 & 247 &  0.4 & 0.001 & 0.0001 \cr
2.2 &     & 178 & 241 &  0.2 & 0.001 & 0.0001 \cr
2.3 &     & 173 & 235 &  0.1 & 0.000 & 0.0000 \cr
2.4 &     & 168 & 230 &  0.0 & 0.000 & 0.0000 \cr
\noalign{\smallskip\hrule}
}}
\end{table}

\section{Application of Idlis model}

Comparison of the mass function, derived from observations (Fig.~\ref{fig7.2})
with the mass function resulting from Idlis model and from our model
(Fig.~\ref{fig7.1}),  indicates only general similarity. It can be seen that
Idlis model represents better the observations. But much better agreement
can be derived by modelling the galactic mass distribution with the
superposition of two Idlis models: one with smaller radius and the other
with larger radius.

Taking the mass function of Idlis model according to Eq.~(\ref{eq7.14})
we found that the best agreement with the empirical mass function
results for
$$m_1 = 1 - m_2 = 0.07, ~~~~ x_1 = 0.46, ~~~~ x_2 = 2.27,$$
where $m_1$ and $m_2$ are the relative masses of smaller and larger models,
respectively, and $x_1$ and $x_2$ their radii. The curve $\mu (a)$ for the
sum of both models is plotted in Fig.~\ref{fig7.2} by dashed line. By dashed line
also the continuation of the larger model for small $a$ is given. It is seen
that the superposition of two Idlis models represent the
observational data quite well.

The smaller model can be identified somewhat provisionally with the
nucleus of the Galaxy. Both the mass and the radius of that model
correspond to the ones, what may be expected for the galactic nucleus.
True enough, the radius of the smaller component seems to be at first glance
too large. But it is needed to have in mind that the smaller model begins
to dominate only when $x<0.2$.

All the calculations referred above, the author plans to present in more
detail in subsequent papers. May be some of the calculations must be made
more precisely. In addition, it is possible that beside the radio
observations it is possible to use some other data on the rotation of the
Galaxy (with the corrections needed). Especially it would be recommendable
to know the function $U$ for $x>1$, where the radio observations do not
allow the determination of $U$.\footnote{Unfortunately this kind of
paper was not published. [Later footnote.]} 

As a conclusion the author expresses his sincere thanks to J.~Einasto
who made the most part of calculations, and made a series of valuable
remarks, in particular on the possibility to improve the value of the
circular velocity in the solar neighbourhood during the process of
calculations and extrapolation of the mass function.
\vglue 3mm
\hfill 1955--1956
\vglue 5mm

{\bf\Large Appendices added in 1969}
\vglue 3mm

\begin{figure*}[h]
\centering
\includegraphics[width=60mm]{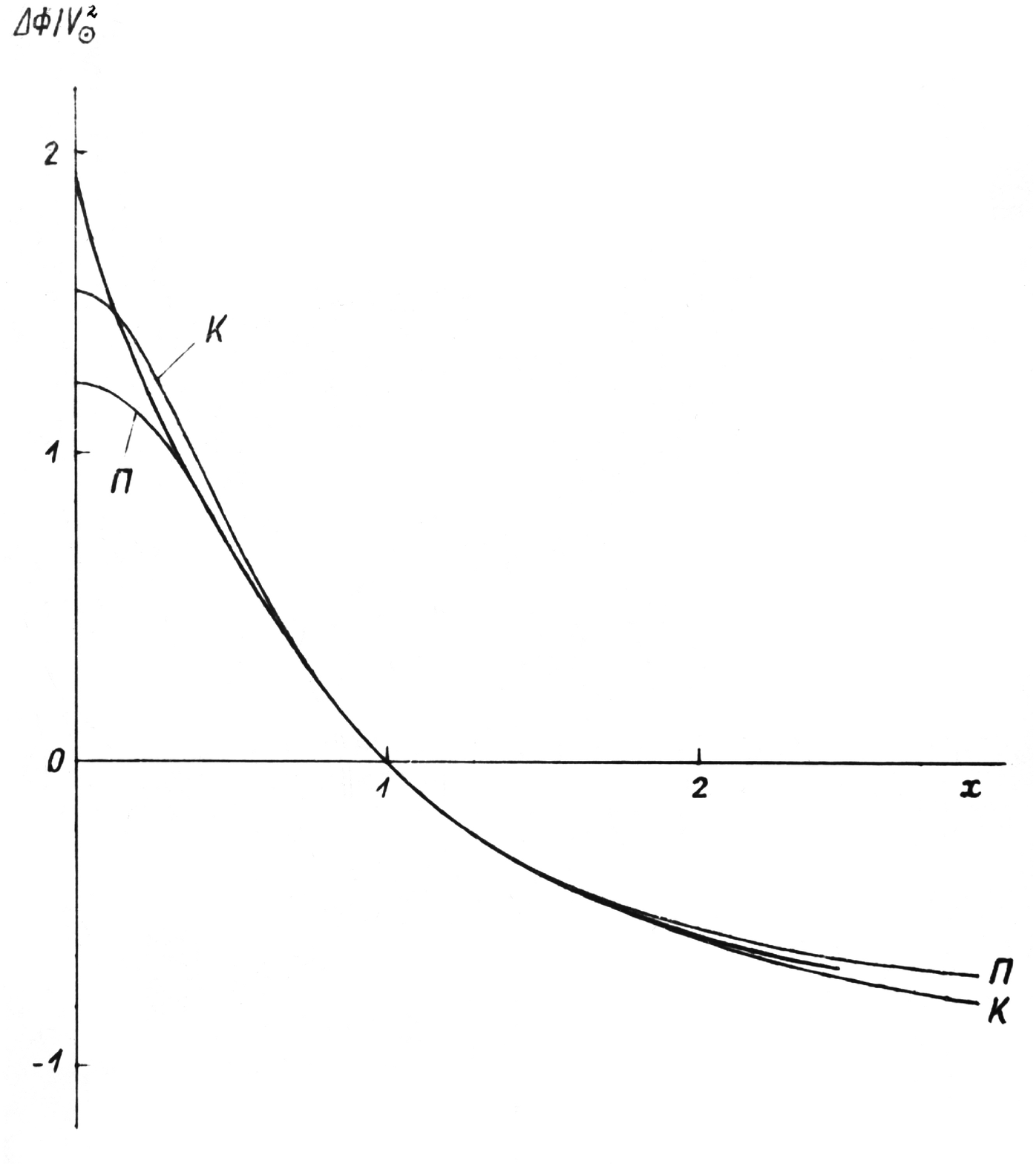}
\includegraphics[width=60mm]{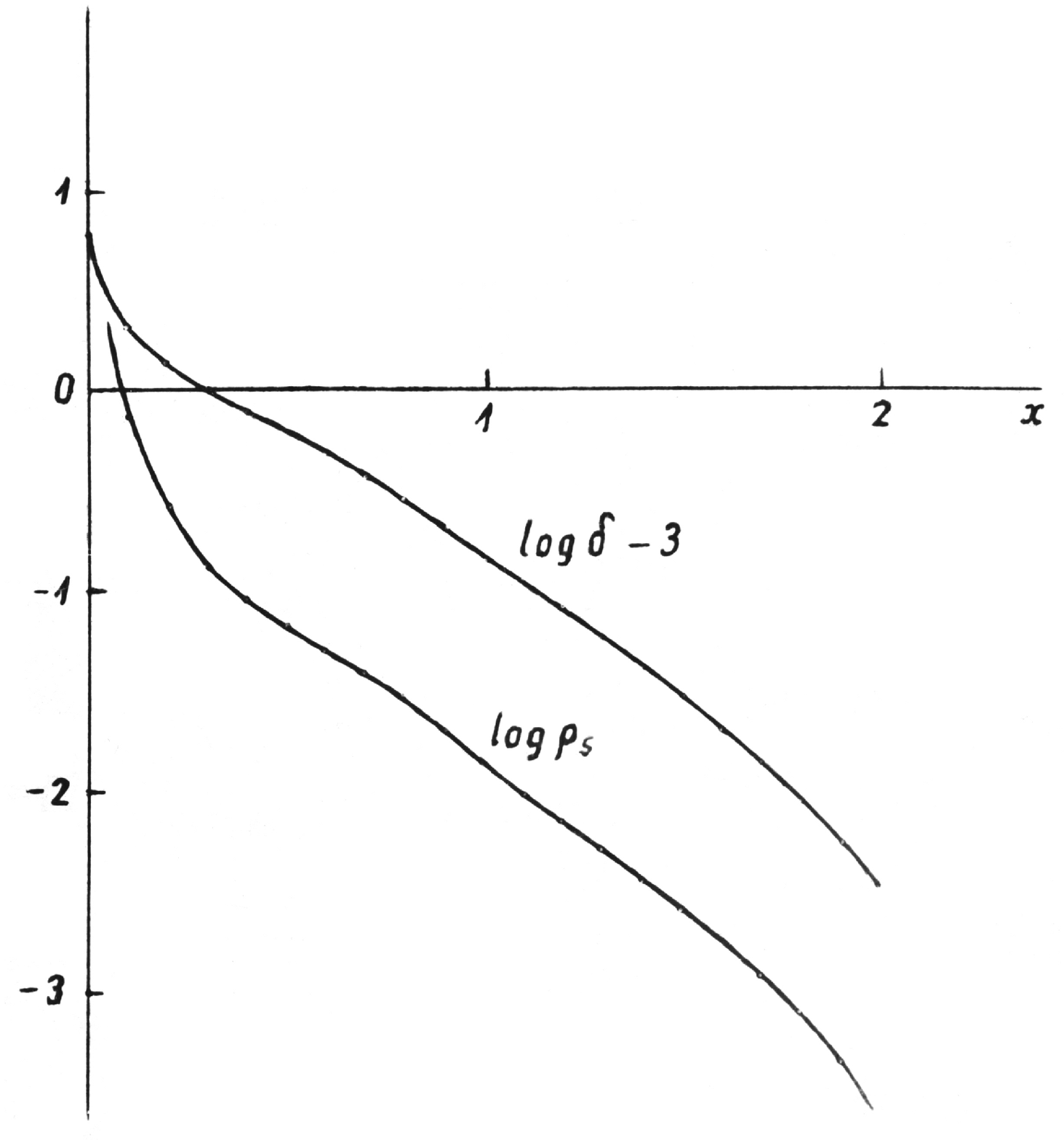}
\caption{}
\label{fig7.A1}
\end{figure*}

\section{A.  Comparison of the Galactic potential in the symmetry plane
  with theoretical laws}

In left panel of the Fig.~\ref{fig7.A1} the curves of the difference $\Delta\Phi = \Phi_{z=0} -
I_{z=0}(R_{\odot})$ -- empirical and theoretical -- are given for the 
law by Parenago-Idlis (P) and for the law resulting from theory of
third quadratic integral (Q). The parameters of theoretical laws are
$V=$ 250~km/s, $A/\omega =$ 0.6 at $R=R_{\odot}$.

\section{B. The density law}

In \citet{Kuzmin:1952ac, Kuzmin:1954} we mentioned that the surface
density of the Galaxy varies according to the exponential law. This is
valid also for the 
spatial density. In right panel of Fig.~\ref{fig7.A1} we give the curves of $\log\delta (x)$ and 
$\log\rho_s (x)$ corresponding to the data from Table~7.1.  As it is seen
for large range of $R$ values the lines are nearly strict.

%% file: chapter08.tex
\chapter[On the theory of the third integral of stellar motion]{On the
  theory of the third integral of stellar
  motion.\footnote{\footnotetext ~~Bull. Abastumani
    Astrophys. Obs. No.~27, 89 -- 92, 1962, Report on the 3rd Meeting of the
    Committee on Stellar Astronomy, October 3--6, 1960, Tbilisi. } } 

In case of the stationary, axisymmetric potential $\Phi$ we have two
isolating (single-valued) conservative integrals of stellar motion  --
the energy and the angular momentum integrals. If we apply a certain
restriction on the potential, there exists also the third integral of
this  type \citep{Kuzmin:1953}. This integral is quadratic in respect
to velocities but is different from the energy integral. 

It is also possible to derive the third integral for general
axisymmetric potential in form of series of powers of coordinates and
velocities.  

By fixing the value of the angular momentum integral $I$, we can
reduce the problem of stellar motion to the two-dimensional one with
the potential 
\be
\tilde{\Phi} =\Phi - \frac{1}{2} \frac{I^2}{R^2}, \label{eq8.1} 
\ee
where $R$ is the distance from the symmetry axis.

To derive the expansion for the third integral we represent $\tilde{\Phi}$ in a form 
\be
\tilde{\Phi} =\tilde{\Phi}_0+\tilde{\Phi}_1 , \label{eq8.2} 
\ee
where $\tilde{\Phi}_0$ is a potential with known third integral. This
potential and the corresponding motion can be called unperturbed. As
every integral of motion must remain constant, we obtain for the
integral $F$ the series 
\be 
F=F_0+F_1+F_2+ \ldots , \label{eq8.3} 
\ee
where
\be
F_n = -\int_0^t \nabla_v F_{n-1} \cdot \nabla_x \tilde{\Phi}_1 \rmd t \label{eq8.4}
\ee
($t$ is time, $\nabla_v$ and $\nabla_x$ are nabla with respect to
velocities and coordinates). Integration is along the unperturbed
orbit. 

For $\tilde{\Phi}$ we use the series
\be
-2\tilde{\Phi} =a_{00}+a_{20} \Delta R^2 + a_{02} z^2 + a_{30} \Delta R^3 + a_{12} \Delta Rz^2 + \ldots , \label{eq8.5}
\ee
where $z$ is the distance from the symmetry plane (the existence of
which is presumed) and $\Delta R$ is counted from the maximum of
$\tilde{\Phi}$. 

By assuming
\be
-2\tilde{\Phi}_0 = a_{00} + a_{20} \Delta R^2 + a_{02} z^2 , \label{eq8.6}
\ee
and correspondingly
\be
F_0 = v_z^2 + a_{02} z^2 , \label{eq8.7}
\ee
we derive the following series for $F$
\be
F= v_z^2 + a_{02} z^2 + \frac{2a_{12}}{4a_{02} - a_{20}} [z v_R v_z - \Delta R  (v_z^2 - a_{02} z^2)] + \ldots . \label{eq8.8} 
\ee
Here $v_R$ and $v_z$ are the velocity components along $R$ and $z$.

The fourth order term in respect to coordinates and velocities was
also derived, but as it is very long, we do not present it in here. As
the previous terms, it is quadratic in respect to
velocities. [q.v. Appendix A.]

Similar series were derived recently by \citet{Contopoulos:1960}.

Is the integral $F$ isolating? Or in other words, does the phase-orbit
lay on a surface of the isoenergetic space or fills all the
isoenergetic space? The answer is given by the known theorem by
\citet{Poincare:1892} on the existence of single-valued
integrals. According to the Poincar\'e's theorem, in general there do
not exist single-valued integrals independent of the energy
integral. The proof of the theorem consists of deriving the series for
the integral by a method similar in some sense to the method
above. Instead of usual coordinates and velocities, the canonical
variables were introduced in a way that coordinates and velocities are
periodic functions of new coordinates. An integral is searched in a
form of double Fourier series, it was found that the integral is a
function of the energy integral, \ie reduces to it. [q.v. Appendix B.] 

Contopoulos in his paper concludes that the Poincar\'e's theorem is
not applicable to the present case. However, the possibility to prove
the theorem depends from the choice of unperturbed potential. If we
take it in ``degenerate'' form as it was done here and by Contopoulos,
we derive an integral being independent of the energy integral, but
the question about its single-valued nature remains open. In order to
prove the theorem, the  unperturbed potential must be taken in more
general form, for example 
\be
\tilde{\Phi}_0 = \varphi_1 (\Delta R) + \varphi_2(z) , \label{eq8.9} 
\ee
where $\varphi_1$ and $\varphi_2$ are different non-quadratic
functions\footnote{In 60-s a valuable series of papers was published
  by G. Contopoulos on the theory of third integral for stellar
  orbits. On the basis of orbit calculations and theoretical
  considerations, he derived that in general cases of axisymmetry and
  plane symmetries for a stationary potential, the third integral is
  isolating in some phase space regions, and non-isolating in other
  regions (Thessaloniki Contr. 10, 12, 13, 17, 18, 22, 29, 31, 38
  (1963--1967)). Ergotic orbits (the third integral is not isolating)
  were found by M. H\'enon and C. Heiles (A.J. {\bf 69}, 73, 1964),
  B. Barbanis (A.J. {\bf 71}, 415, 1965), S. J. Aarseth (Nature {\bf
    212}, 57, 1966). [Later footnote.]}  

Non-isolating character of the third integral must reveal that the
series derived here are not convergent. The series do seem not to
converge, as there are subtractions in denominators of its
coefficients, which may be very small. As a result, arbitrarily
high-order terms may be very large. Probably an explanation is that
the more terms we will take into account, the more complicated the
surface containing the phase-orbit will become. Finally it may result
that the surface (and the orbit on it) will fill all the isoenergetic
space. 

According to the Poincar\'e's theorem, the third integral is not
isolating in general. But in particular cases it may be
isolating. This is the case of quadratic integral. Also, the integral
is isolating for periodic orbits. Two types of such orbits can be
referred to in case of two-dimensional problem. The first type are the
orbits with $z=0$. Motions on these orbits are oscillations along a
segment of $z=0$. The second type are segments of the curve being
symmetrical about $z=0$  and crossing $z=0$ under right angle. The
existence of the orbits of these types is easy to demonstrate but we
do not focus on it here. [q.v. Appendix C.] 

In general, the third integral being non-isolating integral can not be
used as an argument of the phase density. However, instead of the
precise non-isolating integral one may use an approximate
quasi-integral, which is isolating according to its form but slowly
changing its value. Quasi-integral must be introduced in a form
enabling for a phase point to spend as much time as possible near the
surface in isoenergy space, determined by some fixed value of
integral. The more near is the situation to the existence of the
isolating integral, the more precisely the form of the integral is
determined, and the more slowly its value changes. These are the
orbits with small $\Delta R$ and $z$, \ie  nearly circular orbits in
three-dimensional problem. For these orbits the quasi-integral may be
a sum of some first terms of the series derived by us; in fact it is
sufficient to limit with quadratic terms in respect to velocities
terms. 

Variations of the value of quasi-integral are limited by physically
permitted values, and occur probably in extremely complicated way,
having quasi-stochastic character. The corresponding variations of the
phase density can  be treated probably in a similar way as it was done
in analysing the irregular forces, \ie as a diffusion of phase
points\footnote{A similar idea was mentioned also by B. Lindblad
  (Stockholm Medd. 1, 1928). [Later footnote.]}. 

Applying this quasi-integral in form of first terms of the series
(\ref{eq8.8}) gives us the already known obliquity of the velocity
ellipsoid about the galactic plane at points outside of the plane (the
result is known from the theory of quadratic integral
\citep{Kuzmin:1953}) [q.v. Appendix A]. For the $z$-gradient of the
obliquity, the equation for angle $\alpha$ is  
\be
\left( \frac{\partial\alpha}{\partial z} \right)_{z=0} = -  \frac{a_{12}}{4a_{02} - a_{20}} . \label{eq8.10} 
\ee
As $a_{02}$ is significantly larger than $a_{20}$ and is approximately
proportional to the mass density at $z=0$, and $a_{12}$ equals to the
derivative of $a_{02}$ with respect to $R$, Eq.~(\ref{eq8.10}) has the
form 
\be
R \left( \frac{\partial\alpha}{\partial z} \right)_{z=0} \simeq - \frac{1}{4}\left( \pdif{\ln\rho}{\ln R} \right)_{z=0} ,
\label{eq8.11} 
\ee
where $\rho$ is the mass density. In the vicinity of the Sun $\rho$ is
roughly proportional to $R^{-4}$ and hence $R\partial\alpha /\partial
z \sim $ 1. As it was mentioned by us earlier \citep{Kuzmin:1954}, the
$z$-gradient of the obliquity of the velocity ellipsoid influences the
centroid velocity. By taking into account the gradient we derive the
statistical equation of motion in the galactic plane in form 
\be
\bar{v}_\theta^2 + q\sigma_R^2 = v_c^2 , \label{eq8.12} 
\ee
where $\bar{v}_\theta$ is the centroid velocity, $v_c$ is the circular velocity and
\be
q = -\left[ \frac{\partial\ln\rho\sigma_R^2 }{\partial\ln R} + \left( 1- \frac{\sigma_{\theta}^2}{\sigma_R^2} \right) + R \frac{\partial\alpha}{\partial z} \left( 1- \frac{\sigma_z^2 }{\sigma_R^2} \right) \right]_{z=0} . 
\label{eq8.13} 
\ee
$\sigma_R$, $\sigma_\theta$, $\sigma_z$ are the velocity dispersions,
$\rho$ is the mass density or the number density depending on the
definition of the centroid velocity and dispersions. 
\vglue 3mm

\hfill September 1960
\vglue 5mm

{\bf\Large Appendices added in 1969}
\vglue 5mm

\section{A.  The expansion for the integral of motion $F$}

${\bf 1.}$ 
Let us have a general two- or three-dimensional problem of stellar
motion with potential $\Phi$. We designate the orthogonal coordinates
and the corresponding velocities as $x_i$ and $v_i$.  

The condition that the integral $F$ is constant along the stellar trajectory is
\be
v \cdot \nabla_x F + \nabla_x \Phi \cdot \nabla_v F = 0 . \label{eq8.A1.1}
\ee
We try to find the integral $F$ by iterative approximations, beginning
from the known integral $F_0$ for the potential $\Phi_0$. If 
\be
\Phi = \Phi_0 + \sum_{n=1}^{\infty} \Phi_n , \label{eq8.A1.2} 
\ee
where $\Phi_n$ is the correction term of order $n$ for $\Phi_0$, we have for $F$ the series
\be
F = F_0 + \sum_{n=1}^{\infty} F_n , \label{eq8.A1.3} 
\ee
where the successive correction terms $F_n$ are determined by the recurrent formula 
\be
v \cdot \nabla_x F_n + \nabla_x \Phi_0 \cdot \nabla_v F_n = -
\sum_{m=1}^{\infty} \nabla_x \Phi_m \cdot \nabla_v F_{n-m}
.  \label{eq8.A1.4}  
\ee
The left side equals to $\dot F_n$ for the potential $\Phi_0$ and hence
\be
F_n = -\sum_{m=1}^n \int_0^t  \nabla_x \Phi_m \cdot \nabla_v F_{n-m} , \label{eq8.A1.5}
\ee 
where the integration is along orbits with potential $\Phi_0$.  We may
use the differential as well as recurrent integral equations to find
$F_n$. If we combine all the corrections $\Phi_n$ into one general
correction by substituting $\sum\Phi_n$ with $\Phi_1$, we derive for
recurrent integral equation (\ref{eq8.4}). But in fact $\Phi$ is
usually given in form of series and it is useful to arrange the
corrections of $\Phi_0$ according to different orders. In this case
$F_n$ results to be not in a form of infinite power series but in a
form of finite power polynomials. 

Let us assume that $\Phi$ is an even function of $x_2$ (further $x_2 =
z$). For initial zero-order approximation of the potential we take
(two-dimensional problem) 
\be
-2\Phi_0 = a_{00} + a_{20} x_1^2 + a_{02} x_2^2 , \label{eq8.A1.6} 
\ee
corresponding to the zero-order approximation of the integral
\be
F_0 = v_2^2 + a_{02} x_2^2 \label{eq8.A1.7} 
\ee
(another possibility is $F_0=v_1^2+a_{20}x_1^2$).

By expanding $-2\Phi$ into series of powers of $x_1$ and $x_2$ we have for $\Phi_n$
\be
-2\Phi_n = \sum_{l=0}^{[\frac{n+2}{2}]} a_{n-2l+2,2l} ~ x_1^{n-2l+2} ~ x_2^{2l+2} \label{eq8.A1.8} 
\ee
(where [ ] means the integer part).

When using the recurrent integral equation, one must take into account
that in case of the potential $\Phi_0$ the coordinates $x_1$ and $x_2$
of a star oscillate harmonically with frequencies
$\omega_1=\sqrt{a_{20}}$ and $\omega_2=\sqrt{a_{02}}$,
respectively. By expressing $x_i, v_i$ via $t$ we integrate and
thereafter return again to $x_i, v_i$. 

The first two corrections for $F_0$ are the following
\be
\ba{ll}
F_1 = & 2\kappa [(x_1 v_2 - x_2 v_1) v_2 - a_{02} x_1 x_2^2], \\
\noalign{\smallskip}
F_2 = & \lambda (x_1 v_2 - x_2 v_1)^2 - [\kappa a_{12} + \lambda (a_{02} - a_{20})] x_1^2 x_2^2 + (a_{04} -\frac{1}{4}\kappa a_{12}) x_2^4 , 
\ea
\label{eq8.A1.9} 
\ee 
where we designated
\be
\kappa = - \frac{a_{12}}{4a_{02}-a_{20}} , ~~~~ \lambda = - \frac{2a_{22}+3\kappa (2a_{12}-a_{30})}{4(a_{02}-a_{20})} . \label{eq8.A1.10} 
\ee
Recently we found also $F_3$ and $F_4$ but the resulting expressions were very long.
\vglue 3mm

{\bf 2.} 
In our three-dimensional problem for the axisymmetric potential 
\be
x_1 = \Delta R = R - R_0(J) , ~~~x_2 = z , \label{eq8.A1.11} 
\ee
and $\tilde{\Phi}$ means the ``effective potential''
\be
\tilde{\Phi} = \Phi - \frac{1}{2} J^2 R^{-2}. \label{eq8.A1.12}
\ee
$R_0(J)$ is determined from the condition
\be
\frac{\partial\tilde\Phi}{\partial R} \bigg|_{R=R_0(J),z=0} = 0. \label{eq8.A1.13}
\ee
The coefficients $a_{k,2l}$ are determined as the coefficients of the
expansion of $-2\Phi (R_0+\Delta R,z)$ in powers of $\Delta R$ and $z$
and are some functions of $R_0(J)$. $\sqrt{a_{20}}= \omega_R$,
$\sqrt{a_{02}}=\omega_z$ are the harmonic oscillation frequencies for
nearly circular motion. 

For fixed $J$ the integral $F$ is quadratic in velocities up to the
fourth order and has thus within this precision the same properties as
the precise quadratic integral for two-dimensional problem. In
particular, we may relate it to the elliptical system of
coordinates. The directions of coordinate lines determine the
directions of two orthogonal symmetry axis of the integral. The
position of the foci of elliptical coordinates are determined by the
relations 
\be
R_f(J) = R_0(J) - \frac{\kappa}{\lambda}, ~~~z_f(J )= \frac{\pm\sqrt{\lambda -\kappa^2} }{\lambda} . 
\label{eq8.A1.14} 
\ee
\vglue 3mm

{\bf 3.} 
The variable $J$ occurring via $R_0(J)$ in the expression of the
integral can be eliminated by introducing the velocity component
$v_{\theta}=JR^{-1}$. For this we use the relation 
\be
\frac{v_{\theta}^2 - v_c^2}{R} = - \frac{\partial\tilde\Phi}{\partial R} \bigg|_{z=0} , \label{eq8.A1.15}
\ee
where $v_c$ is the circular velocity. In the expansion of the right
side in powers of $\Delta R$ the coefficients 
$a_{k,0}(R_0) = $ $a_{k,0}(R-\Delta R)$ appear. We expand these
coefficients into the Taylor series. By turning the derived expansions
about $\Delta R$ we find the expansion of $\Delta R$ in powers of
$v_{\theta}^2-v_c^2$ or in powers of $\Delta v_{\theta}=$
$v_{\theta}-v_c$. We put this expansion  into the expression of $F$
(before expanding there the coefficients $a_{k,2l}(R_0)=$
$a_{k,2l}(R-\Delta R)$ into  Taylor series). 

The integral $F$ remains an integral of motion after multiplying it by
\be
\exp \left( 2\int_0^{R_0(J)} \kappa \rmd R \right) . \label{eq8.A1.16}
\ee
After multiplication, and designating the new integral as $F$ we find
within the terms of the fourth order with respect to $v_R$, $v_z$,
$\Delta v_{\theta}$, $z$ the expression for $F$ 
$$
F = \exp \left( 2\int_0^R\kappa \rmd R \right) \left\{ (1+ \mu\Delta R^2)v_z^2 - 2 (\kappa +\mu\Delta R) z v_R v_z + \right. $$
\be
+ [\lambda v_R^2 + \kappa R^{-1} v_{\theta}^2 + (a_{02} - a_{20}) \mu\Delta R^2] z^2 - \left. 2 [\Phi (R,z) - \Phi (R,0) - \kappa\int_0^z \frac{\partial\Phi}{\partial R} z \rmd z] \right\} . \label{eq8.A1.17} 
\ee
Here
\be
\mu = \lambda - \kappa^2 - \frac{d\kappa}{dR} = \frac{3a_{20}}{4a_{02} - a_{20}} (\lambda - \kappa R^{-1}) \label{eq8.1.18}
\ee
and
\be
\Delta R = - \frac{2\omega_c}{a_{20}}\Delta v_{\theta} , \label{eq8.A1.19}
\ee
where $\omega_c$ is the angular circular velocity. All the
coefficients in the expression of $F$ are functions of $R$ and not of
$R_0(I)$. We replaced the sum of the terms, that are functions of only
coordinates,  with the finite expression being precise up to the
fourth order terms.  

The derived expression for $F$ corresponds to the precise quadratic integral when
$$ R_f = 0, ~~~ z_f = \pm ~\mathrm{const}.$$
In this case
$$\lambda = \kappa R^{-1} = (R^2 + z_f^2)^{-1}, ~~ \mu = 0.$$
\vglue 3mm

{\bf 4.} 
In the derived approximation the integral $F$ is in fourth power with
respect to velocities. But being quadratic with respect to $v_R$ and
$v_z$, it has two orthogonal symmetry axis in the plane $v_{\theta}=
\mathrm{const}$ (intersecting at point $v_R=$ $v_z=$ 0). For $z=0$ the
symmetry axis of $F$ are directed along the $R$ and $z$ coordinates,
for $z\ne 0$ they are inclined by an angle $\alpha$. The angle
$\alpha$ depends on $v_{\theta}$. For the $z$-gradient of $\alpha$
within the present approximation it results 
\be
\frac{\partial\alpha}{\partial z} \bigg|_{z=0} = \kappa +\mu\Delta R = \kappa + \frac{6\omega_c^2}{4a_{02} -a_{20}} (\kappa -\Delta R) \frac{\Delta v_{\theta}}{v_{\theta}}. 
\label{eq8.A1.20} 
\ee
\vglue 3mm

{\bf 5.} 
The Poisson's equation at $z=0$ has the form
\be
4\pi G\rho_{z=0} = a_{02} + a_{20} - 2\omega_c^2  \label{eq8.A1.21} 
\ee
($G$ is the gravitational constant). For highly flattened stellar
systems, as for example our Galaxy, $a_{02}$ is large when compared
with $a_{20}$ and $\omega_c^2$. Thus approximately 
\be
a_{02} = 4\pi G\rho_{z=0} , \label{eq8.A1.22} 
\ee
and
\be
a_{12} = 4\pi G \frac{\partial\rho}{\partial R} \bigg|_{z=0} , ~~~
2a_{22} = 4\pi G \frac{\partial^2\rho}{\partial R^2} \bigg|_{z=0} . \label{eq8.A1.23} 
\ee

If we also neglect $a_{20}$ in the expressions for $\kappa$ and for
$\lambda$ when compared with $a_{02}$, and $a_{30}$ when compared with
$a_{12}$, we have 
$$
\kappa = - \frac{1}{4} \frac{a_{12}}{a_{02}}, ~~~~ \lambda = - \frac{1}{2} \frac{a_{22}}{a_{02}} + \frac{3}{8} \left( \frac{a_{12}}{a_{02}}  \right)^2 ,$$
and further
\be
\kappa = - \frac{1}{4} \frac{\partial\ln\rho}{\partial R} \bigg|_{z=0} , ~~~~ \lambda = \left[ \frac{1}{ 8} \left( \frac{\partial\ln\rho}{\partial R}\right)^2 - \frac{1}{4} \frac{\partial^2\ln\rho}{\partial R^2} \right]_{z=0} \label{eq8.A1.24}
\ee

Within the same approximation we must neglect the second term in the
expression for the $z$-gradient of $\alpha$. In this case 
\be
\frac{\partial\alpha}{\partial z} \bigg|_{z=0} = \kappa = - \frac{1}{4} \frac{\partial\ln\rho}{\partial R} \bigg|_{z=0} . \label{eq8.A1.25} 
\ee
\vglue 3mm

{\bf 6.} 
Let us have a very flattened stellar system with the exponential density distribution. In this case
\be
\ln\rho = - \frac{R}{\overline{R}} + \mathrm{const} , \label{eq8.A1.26} 
\ee
where $\overline{R}$ is a constant (the density-weighted mean of $R$). Hence
\be
\kappa = \frac{1}{4\overline{R}}, ~~~~ \lambda = \frac{1}{8\overline{R}^2}, \label{eq8.A1.27}
\ee 
\ie in present case $\kappa$ and $\lambda$ are constants.

For $R$ and $z$ of the foci of the elliptical coordinates related to the integral we have
\be
R_f = R_0(J) - 2 \overline{R}, ~~~ z_f = \pm 2 \overline{R} . \label{eq8.A1.28} 
\ee
Therefore, for highly flattened stellar system and for exponential
density distribution $z_f=\pm ~\mathrm{const}$, as in the case for
precise quadratic integral. But in general $R_f$ is nonzero, being
zero only for $R_0=2\overline{R}$. 

Further, in present case
\be
\frac{\partial\alpha}{\partial z} \bigg|_{z=0} = \frac{1}{4\overline{R}}, \label{eq8.A1.29} 
\ee
\ie is also a constant.

The exponential density distribution is quite a good approximation for
the Galaxy, as we see on the basis of our empirical mass distribution
model  of the Galaxy (Chapter 7). If we assume the density $\rho$ to
be proportional to the ``spherical'' density $\rho_s$, in the solar
neighbourhood $\overline{R}$ is roughly $0.25 R_{\odot}$
(corresponding to the density $\sim R^{-4}$). However, for longer
distance interval the exponential density law with somewhat larger
$\overline{R}$ is more suitable 
$$\overline{R} = 0.3 R_{\odot}.$$
This gives us
$$R_f= R_0(J)-0.6R_{\odot}, ~~~z_f=0.6R_{\odot}$$
and
$$R_{\odot} {\partial\alpha\over\partial z}\bigg|_{z=0} \simeq 0.8. $$

\section[B. Poincar\'e's theorem]{B.  Application of the Poincar\'e's
  theorem to the problem of integrals of motion in the dynamics of
  stellar systems.\footnote{\footnotetext ~~The theorem was initially
    proved in connection to three-body problem. Apart from the
    original book by \citet{Poincare:1892}, the theorem is given in
    E. T. Whittaker, Analytical dynamics, Cambridge, 1904 and in
    G. N. Duboshin, Celestial mechanics. Analytical and quantitative
    methods, Moscow, 1964 (in Russian).}} 

Let us assume that the initial potential is (two-dimensional problem)
\be
\Phi_0 = \varphi_1(x_1) + \varphi_2(x_2) \label{eq8.A2.1} 
\ee
and let us introduce canonical variables
\be
p_i = \frac{1}{2\pi} \oint v_i \rmd x_i , ~~~~ q_i = \omega_i(p_i)
\int \frac{\rmd x_i}{v_i} ~~~ (i=1, 2) , \label{eq8.A2.2}  
\ee
where
\be
\omega_i(p_i) = \frac{2\pi}{\oint \frac{\rmd x_i}{v_i}}, \label{eq8.A2.3} 
\ee
and integral is taken along the orbit in the potential $\Phi_0$
(contour integrals are taken over one cycle of variation of $x_i$).  

Hamiltonian $E_0$ (the energy integral) for the potential $\Phi_0$ is a function of $p_i$ alone, and
\be
\frac{\partial E_0}{\partial p_i} = \omega_i(p_i) . \label{eq8.A2.4} 
\ee

For the potential $\Phi_0$
$$\dot p_i = 0, ~~~ \dot q_i=\omega_i(p_i)$$
and hence
$$p_i, ~~ q_i-\omega_i(p_i)t, $$
being constants, are the integrals of motion. Both integrals of motion
$p_i$ are conservative. The remaining two integrals are
non-conservative. Eliminating $t$ from them we have the third
conservative integral  
$$\frac{q_1}{\omega_1(p_1)} - \frac{q_2}{\omega_2(p_2)} .$$
With the exception of periodic orbits this integral is non-isolating.

An orbit in the potential $\Phi_0$ is periodic when the ratio of
frequencies $\omega_2/\omega_1$ is a rational number. If $\varphi_i$
is non-quadratic (arbitrary chosen), the ratio $\omega_2/\omega_1$ is
a continuous function of $p_1, p_2$ and has infinite number of
rational $\omega_2/\omega_2$, and correspondingly infinite number of
$p_1, p_2$, tightly filling all the permitted region (the region of
finite orbits in case of the potential $\Phi_0$). 

Let us assume
\be
\Phi =\Phi_0+\Phi_1 , \label{eq8.A2.5}
\ee
where $\Phi_1$ is an arbitrary correction to $\Phi_0$ (in present case
there is no need to expand $\Phi_1$ into corrections of different
order). Correspondingly the Hamiltonian is 
\be
E=E_0+E_1, ~~~E_1=\Phi_1. \label{eq8.A2.6}
\ee
If $\Phi$ does not reduce to the form of the initial potential
$\Phi_0$, then $E_1$ depends on $q_i$. As $E_1$ is a single-valued
function of $x_i$ and $q_i$ is cyclical coordinate, the function $E_1$
must be a periodic function of $q_i$ with the period of
$2\pi$. Besides, as independently of $E$ the Poisson's brackets $\{
x_i,E \} =v_i$, following must hold 
\be
\{ x_i,E_1 \}  = 0. \label{eq8.A2.7} 
\ee
This is the condition for the non-dependence of $E_1$ on $v_i$.

We try to find the isolating integral $F$ for potential
$\Phi$. Because the isolating integral is a single-valued function (or
can be reduced to), $F$ must be a periodic function of $q_i$. 

Let us assume that integral $F$ exist independently of the correction
$\Phi_1$ and, particularly, independently of the value of the
correction. Further, we assume that $F$ is an {\bf analytical}
function of the value of $\Phi_1$. In other words we assume that while
varying the correction $\Phi_1$, keeping it proportional to himself,
\ie 
$$\Phi = \Phi_0+k\Phi_1 ,$$
the integral $F$ is in form of series
$$F=F_0+\sum_{n=1}^{\infty} k^nF_n .$$

Restriction for $F$ is expressed by the Poisson's brackets
\be
\{ F,E \} =0. \label{eq8.A2.8}
\ee
By substituting temporarily $\Phi$ with $k\Phi_1$ and assuming thereafter $k=1$, we find
\be
F=F_0 +\sum_{n=1}^{\infty} F_n , \label{eq8.A2.9}
\ee
where subsequent corrections $F_n$ are determined by the recurrent formula
\be
\{ F_n,E_0 \} =- \{ F_{n-1},E_1 \} . \label{eq8.A2.10} 
\ee
All $F_n$ must be periodic functions of $q_i$ with period $2\pi$,
otherwise $F$ can not have these properties independent of the value
of $\Phi_1$ (\ie for substituting $\Phi_1\rightarrow k\Phi_1$ it
follows $F_n\rightarrow k^nF_n$). 

As $F_0$ is isolating integral for the potential $\Phi_0$, it does not
depend on $q_i$. Therefore, for the first order correction we have 
\be
\{ F_1,E_0 \} = - \left( \frac{\partial F_0}{\partial p_1} \frac{\partial E_1}{\partial q_1} + \frac{\partial F_0}{\partial p_2} \frac{\partial E_1}{\partial q_2} \right) . \label{eq8.A2.11} 
\ee
The left side in here is $\dot{F}_1$ for the potential $\Phi_1$. Thus,
if $p_1, p_2$ correspond to a periodic orbit in the potential $\Phi_0$
(and these $p_i$ tightly fill all the permitted region of $p_i$) the
condition for $F_1$ to be periodic with $q_i$ is 
\be
\frac{\partial F_0}{\partial p_1} \oint \frac{\partial E_1}{\partial q_1} \rmd t + \frac{\partial F_0}{\partial p_2} \oint \frac{\partial E_1}{\partial q_2} \rmd t = 0 , \label{eq8.A2.12} 
\ee
where the contour integral means the integration over one cycle of
common variation of $x_1, x_2$ in the potential $\Phi_0$. On the other
side, because of the periodicity of $E_1$ in respect to $q_i$ it
results 
\be
\frac{\partial E_0}{\partial p_1} \oint \frac{\partial E_1}{\partial q_1} \rmd t + \frac{\partial E_0}{\partial p_2} \oint \frac{\partial E_1}{\partial q_2} \rmd t =0 . \label{eq8.A2.13} 
\ee
Therefore, when $\partial F_0/\partial p_i$ are not proportional to $\partial E_0/\partial p_i$, then
$$\oint \frac{\partial E_1}{\partial q_i} \rmd t =0 $$
(\ie in expansion of $E_1$ into double Fourier
series\footnote{Contrary to Poincar\'e we did not use here the
  expansion of $E_1$ and $F_1$ to Fourier series.} about $q_i$ all the
terms with the wave numbers, ratio of which equals to
$\omega_2/\omega_1$, must disappear). However, this condition, if not
resulting from general properties of $E$ (as single-valued function of
$x_i$), is not valid in general. Hence, we must assume, that
independently of $p$ the derivative $\partial F_0/\partial p_i$ is
proportional to $\partial E_0/\partial p_i$. But in this case 
\be
F_0 = F_0(E_0), \label{eq8.A2.14} 
\ee
and thereafter we find
$$F_n = \frac{1}{n!} F_0^{(n)} (E_0) E_1^n , $$
\ie 
\be
F=F_0(E) . \label{eq8.A2.15} 
\ee
Altogether, in general, the isolating integral we analysed reduces to the energy integral.

Evidently, these discussions can be generalised into three-dimensional
case. Moreover, we may expand the class of initial potentials
$\Phi$. It is essential to have a possibility to introduce canonical
coordinates $p_i, q_i$ in a way of $q_i$ being cyclic. It is
sufficient, that $\Phi_0$ enables the existence of quadratic integral
being independent of the energy integral (Chapter 6).  

\section[C. On the periodic orbits]{C.  On the periodic orbits in
  two-dimensional problem in case 
  of general symmetric potential} 

The orbits corresponding to a given value of the energy integral $E$
lay in the region restricted by the zero-velocity isocurve of the
potential 
$$\Phi (x_1,x_2) = -E.$$
Let us assume that $\Phi$ is symmetric (and even) with respect to
$x_2$. In this case the zero-velocity curve has the form of an oval
symmetric about $x_1$-axis. Let $P_1$ and $P_2$ be the points of
intersection of the curve with $x_1$-axis, and $P_2$ correspond to
larger $x_1$. Let a point mass be on the zero-velocity curve at point
$P_1$ or point $P_2$, and let it freely ``fall''. It will have a
reciprocal motion along the $x_1$ axis between points $P_1$ and
$P_2$. Evidently the part of $x_1$-axis between $P_1$ and $P_2$ is a
periodic orbit. Now, let a point mass be on the zero-velocity curve
near to the point $P_1$ or $P_2$. In one case the orbit intersects
with $x_1$-axis under a very small angle, in other case under an angle
of nearly $180^0$ (the angle between the directions of the motion and
$x_1$-axis). If we continuously vary the initial coordinate of the
point mass on the zero-velocity curve the angle of intersection with
the $x_1$-axis will also vary continuously. Therefore, there exists
the initial position (at least one) for which the orbit intersects the
$x_1$-axis under the right angle. This gives us the other periodic
orbit, analogous to the first but with curved form in general. 

Apart from the orbits above, there is an infinite number of
``resonance'' periodic orbits (G. Contopoulos, J. D. Hadjidemetriou AJ,
{\bf 73}, 61, 1968, G. Contopoulos, ApJ, {\bf 153}, 83, 1968).

%% file: chapter09.tex
\chapter[The integral equations for mass distribution]{The integral
  equations for mass distribution and some models of the
  galaxies.\footnote{\footnotetext ~~Published in Tartu
    Astron. Observatory Publications, vol. 35, pp. 285-312, 1966. }} 

{\bf Summary}
\bigskip

In a flattened stellar system the circular velocity is mainly
determined by the radial mass distribution. Therefore, if the circular
velocity is known, we can derive the radial mass distribution
without the exact knowledge of the vertical mass distribution.  To
find the radial mass distribution we have to solve an integral
equation, the kernel of which depends on the form of
the spatial mass distribution. If the equidensity surfaces are similar
spheroids with the axes ratio $\epsilon$, the integral equation has the form
\be
V^2 (R) = G \int_0^R \mu (a) \frac{\rmd a}{\sqrt{R^2 - e^2a^2}} , \label{eq9.1.6}
\ee
\citep{Kuzmin:1952ac, Kuzmin:1956b, Burbidge:1959}. There $V(R)$ is the circular
velocity, $R$ the distance from the axis of the system, $a$ the equatorial
radius of an equidensity surface, and $e^2 = 1 - \epsilon^2$. The mass function
$\mu(a)$, \ie  the mass per unit interval of $a$ is to be considered as
the unknown. It is related to the spatial density $\rho(a)$ by formula
\be
\frac{\mu (a)}{4\pi a^2} \equiv \rho_s (a) = \epsilon \rho (a), \label{eq9.1.7}
\ee
and to the projected or surface density $\Delta(R)$ by equation 
\be
\rho_s (a) = - \frac{1}{\pi} \int_a^{\infty} \frac{\rmd \Delta (R)}{\rmd R} \frac{\rmd R}{\sqrt{R^2-a^2}} . \label{eq9.1.9}
\ee

If the non-homogenous spheroid degenerates into a flat non-homogenous
disk, \ie  $\epsilon$ becomes zero, the integral equation assumes the Abelian form 
\be
V^2 (R) = G \int_0^R \mu (a) \frac{\rmd a}{\sqrt{R^2 - a^2}} , \label{eq9.2.1}
\ee
with 
\be
G \mu (a) = \frac{2}{\pi} \int_0^a \frac{\rmd V^2 (R) R}{\rmd R} \, \frac{R \rmd R}{a\sqrt{a^2-R^2}}. \label{eq9.2.2}
\ee
as its solution \citep{Kuzmin:1952ac, Brandt:1960}. The
corresponding equation for the surface density $\Delta(R)$ has the form 
\be
V^2 (R) = - 4GR \int_0^{\infty} \frac{\rmd\Delta (R')}{\rmd R'} \, \psi \left( \frac{R}{R'} \right) \rmd R' , \label{eq9.2.3}
\ee
\citep{Wyse:1942}, and the formula 
\be
\psi (x) = \left\{
\ba{ll}
x \mathbf{D} (x) , &  x < 1, \\
x^{-2} \mathbf{D} (x^{-1}) , & x > 1. 
\ea 
\right.
\label{eq9.2.4}
\ee
can be derived for the solution \citep{Kuzmin:1952ac}. Here $\mathbf{D}$ is the complete elliptic 
integral. If $\epsilon$ is small, we can find $\mu(a)$ by 
a small correction to the solution $\mu_0 (a)$ for $\epsilon = 0$. The solution acquires then the form 
\be
\mu (a) = \mu_0 (a) + \frac{2}{\pi} \epsilon \int_0^a \frac{\rmd \mu_0 (R) R}{\rmd R} 
\frac{R\rmd R}{a \sqrt{a^2-R^2}} .   \label{eq9.2.7} 
\ee
\citep{Kuzmin:1952ac}.

A very general representation of the spatial mass distribution is
possible by the superposition of a set of non-homogenous spheroids of
different flattening. In this case, instead of a single elementary
spheroidal stratum at the given $a$, we have a bundle of such
strata. The mass function $\mu(a)$ represents now the masses of the
superposing bundles of strata. Its relation to the surface density
remains unchanged however. The integral equation for the mass
distribution takes the form 
\be
V^2 (R) = G \int_0^R \mu (a) K(R, a) \frac{\rmd a}{R} , \label{eq9.3.4} 
\ee
with the kernel 
\be
K(R, a) = \frac{1}{\mu (a)} \int_0^{\infty} \vartheta (a, \epsilon) \frac{R}{\sqrt{R^2-e^2a^2}} \rmd\epsilon , \label{eq9.3.5} 
\ee
where $\vartheta(a, \epsilon)$ represents the mass distribution according to
$a$ and $\epsilon$ ($\mu (a) = \int_0^{\infty} \vartheta (a,\epsilon) \rmd\epsilon$). 
If the system is strongly flattened, we have the solution 
\be
\mu (a) = \mu_0 (a) + \frac{2}{\pi} \int_0^a \frac{\rmd\overline{\epsilon}(R) \mu_0 (R) R}{\rmd R} 
\frac{R\rmd R}{a \sqrt{a^2-R^2}}  . \label{eq9.3.8}
\ee
The function $\overline{\epsilon}(a)$ entering into
this solution is the mean $\epsilon$ for the bundle of strata at a
given $a$. In general, $\overline{\epsilon}(a)$ differs from the axes
ratio of the equidensity surfaces.

The mass distribution according to $\epsilon$ is closely related to
the distribution according to the $z$ coordinate. For the mean
$\overline{|z|}_R$ we find the equation 
\be
\frac{1}{2} \Delta (R) \overline{|z|}_R = \int_R^{\infty}  \overline{\epsilon}(a) \rho_s (a) a \rmd a, \label{eq9.4.1}
\ee
which shows that this function is connected with the just mentioned 
function $\overline{\epsilon}(a)$. The formula 
\be
\overline{\epsilon}(R) = - \frac{1}{2\rho_s (R)} \frac{\rmd\Delta (R) \overline{|z|}_R}{R \rmd R} \label{eq9.4.2}  
\ee
enables us to find $\overline{\epsilon}(a)$ providing $\overline{|z|}_R$ is known. Thus,
for the integral equation in question the mean $|z|$ is the most
important characteristic of the $z$-distribution. Equations can be
derived for various other characteristics of this distribution. The
general equation has the form 
$$
\frac{1}{2} (-1)^k \Gamma (k+\kappa ) m_{\kappa} (R) \equiv \int_0^{\infty} \frac{\partial^k \rho (R,z)}{(\partial z^2)^k} 
z^{2(k+\kappa )-1} \rmd z =
$$
\be
= \int_R^{\infty} \frac{\rmd^k}{(\rmd a^2)^k} \left[ \overline{\epsilon^{2\kappa -1}} (a) \rho_s (a) \right] 
( a^2 -R^2)^{k+\kappa -1} a\rmd a . \label{eq9.4.10}
\ee
Here $\kappa$ is an arbitrary number, $k$ is an arbitrary non-negative integer making $k+\kappa$ positive.

If the bundles of elementary spheroidal strata become infinitely thin,
we get ``discrete'' models. They are characterised by two functions $\rho(a)$
and $\epsilon(a)$, the density in the equatorial plane and the axes ratio of
the ``generating'' strata. The density of such a model is given by formulae 
\be
\rho (R,z) = \rho (a) \left[ 1 + \frac{\rmd\ln\epsilon (a)}{\rmd\ln a}
  \left( 1 - \frac{R^2}{a^2} \right) \right]^{-1},
\label{eq9.5.3}
\ee
and 
\be
a^2 = R^2 + z^2 \epsilon^{-2} (a). \label{eq9.5.4}
\ee

The kernel $K$ of the integral equation for the mass distribution has the form 
\be
K(R,a) = \frac{R}{\sqrt{R^2-e^2 (a) a^2}},  \label{eq9.5.5}
\ee
where $e^2 (a) = 1 - \epsilon^2 (a) $.

As a simple example, the discrete model defined by 
\be
\rho (a) = \rho_0 \left( 1 + \frac{a^2}{a_0^2} \right)^{-n} \label{eq9.6.1}
\ee
and 
\be
\epsilon^2 (a) = \epsilon_0^2 \left( 1 + \frac{a^2}{{a'}_0^2} \right)^{-1} \label{eq9.6.2}
\ee
are discussed. For the density we have the expression 
\be
\rho (R,z) = \rho_0 \left( 1 - \frac{z^2}{{a'}_0^2 \epsilon_0^2} \right)^{n-1} \left[ 1+ \frac{R^2}{a_0^2} + 
\left( \frac{1}{a_0^2} - \frac{1}{{a'}_0^2} \right) \frac{z^2}{\epsilon_0^2} \right]^{-n} , \label{eq9.6.4}
\ee
or the expression 
\be
\rho (R,z) = \rho_0 \left( 1 - \frac{z^2}{a_0^2\epsilon_0^2} \right)^{n-1} \left( 1 + \frac{R^2}{a_0^2} \right)^{-n}, \label{eq9.6.6}
\ee
if $a'_0 =a_0$.


The shape of the calculated equidensity surfaces of the discrete models seems
rather unsuitable to describe a galaxy as a whole. Therefore, it is
desirable to consider ``distributed'' models. Two simplified forms of
the distribution function $\vartheta(a, e)$ are suggested: $\mu(a,
\epsilon)\varphi(\epsilon)$  and $\mu(a)\varphi(\epsilon)$, where
$\mu(a,\epsilon)$  is the solution for the mass function in
the case of a single spheroid. The second form is discussed in greater
detail. In this case the kernel $K$ depends only on the ratio $R/a$. As a
further concretisation, the beta-distribution 
\be
\varphi (\epsilon) = \frac{2}{B\left( p, \frac{1}{2} - p \right) } (1-\epsilon^2)^{p-1} \epsilon^{-2p}  \label{eq9.7.7}
\ee
for $\varphi(a)$ has been adopted, which gives the expression 
\be
\rho (R,z) = \frac{2}{B(p,q)} z^{2q-1} \int_r^{\infty} \rho_s (a) (a^2 - r^2 )^{p-1} (a^2 - R^2)^{\frac{1}{2} -p -q} a \rmd a \label{eq9.7.5}
\ee
for the density, and
\be
K\left( \frac{a}{R} \right) = F \left( \frac{1}{2} , p; p+q; \frac{a^2}{R^2} \right)  \label{eq9.7.6}
\ee
for the kernel ($F$ is a hypergeometric function). An interesting case occurs when $p+q=1/2$. 
In this case we have the formulae 
\be
\rho (R,z) = \left( \frac{z}{r} \right)^{-2p} \rho_0 (r) , \label{eq9.7.8}
\ee 
\be
\rho_0 (r) = \frac{2}{B\left( p, \frac{1}{2} - p \right) } \int_r^{\infty} \rho_s (a) \left( \frac{a^2}{r^2} -1 \right)^{p-1} \frac{a\rmd a}{r^2}  \label{eq9.7.9}
\ee
for the density, and the integral equation for the mass distribution reduces to the generalised
Abel equation 
\be
V^2 (R) = G \int_0^R \mu (a) \left( 1- \frac{a^2}{R^2} \right)^{-p} \frac{\rmd a}{R} \label{eq9.7.10}
\ee
with the known solution 
\be
G\mu (a) = \frac{2}{B(p,1-p)} \int_0^a \frac{\rmd V^2(R) R}{\rmd R} \left( \frac{a^2}{R^2} -1 \right)^{p-1} \frac{\rmd R}{a}  . \label{eq9.7.11}
\ee
We call such models conditionally as the ``Abelian'' ones.

As an example, the ``Abelian'' model defined by 
\be
\rho_s(a) = \rho_{s0} \left( 1+ \frac{a^2}{a_0^2} \right)^{-n} \label{eq9.8.1}
\ee
is considered. We have the expression 
\be
\rho (R,z) = \frac{ \sqrt{\pi} \Gamma (n-p)}{\Gamma (n) \Gamma \left( \frac{1}{2} -p\right)} 
\rho_{s0} \left( \frac{z}{a_0} \right)^{-2p} \left( 1+ \frac{r^2}{a_0^2} \right)^{p-n}  \label{eq9.8.4}
\ee
for the density, and 
\be
V^2(R) = 2\pi B \left( \frac{3}{2} , 1-p \right) G \rho_{s0} R^2 
F \left( \frac{3}{2}, n; \frac{5}{2}-p; -\frac{R^2}{a_0^2} \right) \label{eq9.8.6}
\ee
for the circular velocity. If we assume $n + p = 5/2$, the simple law 
\be
V^2(R) = \frac{GM}{a_0}  \frac{R^2}{a_0^2} \left( 1+ \frac{R^2}{a_0^2} \right)^{-3/2} , \label{eq9.8.9}
\ee
known in the theory of the third integral of stellar motion \citep{Kuzmin:1956a}
stands for the circular velocity. 

Although the density distribution in the ``Abelian'' models is by far
not adequately representative, they are of practical interest owing to
the simplicity of solving the Abel equation. As to the value of the
parameter $p$, the linear formula $2p =  1 - \epsilon$ can be suggested. 
\vglue 3mm

\hfill August 1965

%% file: chapter10.tex
\chapter[Solution of integral equations for mass
distribution]{Solution of the integral equations for the mass
  distribution in a spheroidal model\footnote{\footnotetext
    ~~Published in Tartu Astron. Observatory Publications, vol. 35,
    pp. 316-341, 1966. Coauthor S. A. Kutuzov.}} 

\section{The integral equation for the mass function and its solutions}

In determining the radial mass distribution in galaxies from the
rotation velocity of their planar subsystems we often adopt a model in
the form of an inhomogeneous spheroid. In such a model the isosurfaces
of mass density $\rho$ have the shape of similar spheroids, \ie  
\be
\rho=\rho(a), \label{eq10.1. 1}
\ee
where $a$ is equatorial semiaxis of the spheroidal isosurface defined by equation
\be
a^2 = R^2 + \epsilon^{-2} z^2,  \label{eq10.1.2}
\ee
where $R, z$ are cylindrical coordinates, and $\epsilon$ is the ratio
of the polar half-axis of the spheroid to the equatorial one,  assumed
constant. 

The circular velocity $V(R)$ for the spheroidal model is given by formula \citep{Kuzmin:1952ac, Kuzmin:1956b, Burbidge:1959, Kuzmin:1966aa} 
\be
V^2(R) = G\int_0^R \mu(a) \frac{\rmd a}{\sqrt{R^2-e^2a^2}}. \label{eq10.1.3}
\ee
Here $G$ is the gravitational constant,
\be
e^2 = 1-\epsilon^2 \label{eq10.1.4}
\ee
-- the square of the eccentricity of the density isosurface spheroid,  and
\be
\mu(a) = 4\pi\epsilon a^2\rho(a) \label{eq10.1.5}
\ee
is the  mass function which, when multiplied by $\dd{a}$, has the meaning of
the mass,  contained inside the layer between s with $a$ and
$a+\dd{a}$. 
 
Equation (\ref{eq10.1.3}) is an integral equation with respect to the mass
function. If the mass function is found, it is also possible to
determine the surface density $\Delta R$  (the mass of a vertical column
of a unit cross section),  using the formula 
\be  
\Delta R = \frac{1}{2\pi} \int_R^\infty \frac{\mu(a)}{a} \frac{\dd{a}}{\sqrt{a^2-R^2}} . \label{eq10.1.6} 
\ee
The radial mass distribution is thereby determined.

The solution of equation (\ref{eq10.1.3}) depends on $\epsilon$, and so it is possible 
to write
\be
\mu(a) = \mu(a,\epsilon).  \label{eq10.1.7}
\ee
In final analytic form, the solution is known for only two cases, where
$\epsilon=0$ and 1, \ie  for the plane disk model and for the spherical model
(not counting the cylindrical model, $\epsilon\rightarrow\infty$, which is of no practical
interest). For the planar model, the equation turns into an Abel-type
equation, and the solution is 
\be
G\mu(a,0) = \frac{2}{\pi} \int_0^a \frac{\rmd V^2(R) R}{\rmd R} \frac{R}{a} \frac{\rmd R}{\sqrt{a^2-R^2}}  \label{eq10.1.8} 
\ee
For the spherical model, the equation is solved by simple differentiation
\be
G\mu(a,1) = \frac{\rmd V^2(a)a}{\rmd a}.  \label{eq10.1.9}
\ee
Due to considerable flattening in many galaxies $\epsilon$ is usually
small. Therefore, the solution of  $\mu(a,0)$ is a rather good first
approximation. In order to take into account the thickness of the galaxy,
the method of ``thickness correction'' can be applied.
It adds to the solution $\mu(a,0)$ a correction proportional to $\epsilon$,
calculated from the same solution.

Later \citet{Idlis:1961} proposed the interpolation solution
\be
\mu(a,\epsilon) = (1-\epsilon) \mu(a,0) + \epsilon\mu(a,1), \label{eq10.1.10}
\ee
which is simpler and also valid over the whole interval $0\le\epsilon\le 1$. The
solution of (\ref{eq10.1.10}) actually corresponds not to the spheroidal model,
but to a superposition of the planar and spherical models with
proportional parts of  $V^2(R)$ with masses related as $1-\epsilon$ to
$\epsilon$. Nevertheless, this solution is quite close to $\mu(a,\epsilon)$.

But the solution of equation (\ref{eq10.1.3}) can be approached in a different
way. If we represent $\mu(a)$ as a series with powers of $a$, then $V^2(R)$
is also expressed as a power series, and there are simple ratios
($\epsilon$-dependent) between the coefficients of both series. Therefore, if
we represent $V^2(R)$ as a series or approximate it by a power
polynomial, we obtain the required mass function $\mu(a)$ as a series of
polynomials. Such a method for solving equation (\ref{eq10.1.3}) was proposed by
\citet{Burbidge:1959}.

In our previous work \citep{Kuzmin:1966aa}, a generalised integral equation for the mass
function was considered. Here the kernel is obtained by averaging over
$\epsilon$ of the kernel of equation (\ref{eq10.1.3}). The ``specific mass
function'' $\vartheta(a)$,
which defines a density distribution model, is used as a
weight. The generalised mass function $\mu(a)$ --  the integral of the
specific mass function over $\epsilon$ -- does not make as much sense as for the
spherical model. But formula (\ref{eq10.1.6}) remains valid, so the mass function
in any case defines a radial mass distribution.

Special consideration was given to the cases, where the specific mass
function is $\delta$-like with respect to $\epsilon$,  and where its
dependence on $\epsilon$ 
is the same for all $a$. With a $\delta$-like specific mass function, the
integral equation has the same form (\ref{eq10.1.3}), only $e^2$ depends, generally
speaking, on $a$. The corresponding models have been called discrete
models. In the second case the averaging of the kernel in (\ref{eq10.1.3}) does
not depend on its arguments, so you get an equation of the form
\be
V^2(R) = G \int_0^R \mu(a) K\left( \frac{a}{R} \right) \frac{\dd{a}}{R} , \label{eq10.1.11}
\ee
\ie an equation with a kernel, depending only on the $a/R$ ratio, like
the original equation (\ref{eq10.1.3}) (if the multiplier $R^{-1}$ is removed from the
kernel).

The solution of an equation of the form (\ref{eq10.1.11}), in
particular of equation
(\ref{eq10.1.3}), can be written in a form similar to (\ref{eq10.1.8}). To do this, we need
to introduce an auxiliary solving function, or resolvent. The solution
is as follows (if $K(1)\not=0$)

\be
G\mu(a) = \int_0^a \frac{\rmd V^2(R) R}{\rmd R} L\left( \frac{R}{a} \right) \frac{\rmd R}{a} + \frac{1}{K(1)} \frac{\rmd V^2(a) a}{\rmd a} ,  \label{eq10.1.12}
\ee
and the resolvent $L(x)$ has to satisfy the equation
\be
\int_y^1 L(x) K \left( \frac{y}{x} \right) \frac{\rmd x}{x} +
\frac{K(y)}{K(1)} = 1,
\label{eq10.1.13}
\ee
as can be seen by substituting (\ref{eq10.1.12}) into (\ref{eq10.1.11}), or vice versa. The
second terms in expressions (\ref{eq10.1.12}) and (\ref{eq10.1.13}), which appear at
$K(1)\not=\infty$, 
can be omitted if we add to  the resolvent a corresponding $\delta$-form term.

The expression (\ref{eq10.1.12}) can, of course, be applied as a solution of
equation (\ref{eq10.1.11}) only when the resolvent is known or has been found by
solving the integral equation (\ref{eq10.1.13}).

In a previous paper \citep{Kuzmin:1966aa} it was pointed out that there is a rather
extensive class of models (as extensive as the class of spheroidal
models) for which equation (\ref{eq10.1.11}) belongs to the generalised Abel
equation type, and hence has a known resolvent. For these models,  
conventionally called Abelian,
\be
K \left( \frac{a}{R} \right) = \left( 1-\frac{a^2}{R^2} \right)^{-p} ,
~~~ L  \left( \frac{R}{a} \right) = \frac{2}{B(p,1-p)} \left(
  \frac{a^2}{R^2} -1 \right)^{p-1} , \label{eq10.1.14} 
\ee
where the parameter $p$ depends on the flatness of the model. Plane and
spherical models, belonging to the class of spherical
models, are at the same time the limit cases of Abelian models
($p=1/2$ and 0). 

Although Abelian models are of considerable practical interest for
determination of radial mass distribution in galaxies, their vertical
density distribution, however, corresponds poorly to real
galaxies. Therefore, the issue of applying spherical 
models and solving the integral equation (\ref{eq10.1.3}) is not
removed. Consideration of ways to solve equation (\ref{eq10.1.3}) is all the more
justified because this solution can be used to build a more general
model in the way, specified in the previous paper \citep{Kuzmin:1966aa}. The mass
function of such a model is a solution $\mu(a,\epsilon)$ averaged over $\epsilon$
\be
\mu(a) = \int_0^\infty \mu(a,\epsilon) \phi(\epsilon) \rmd\epsilon,
\label{eq10.1.15}
\ee
where $\phi(\epsilon)$ is some weight function.

The formula (\ref{eq10.1.15}) is a generalisation of the interpolation
formula of G. M. Idlis. Obviously, applying formula (\ref{eq10.1.15}) is
equivalent to applying solution (\ref{eq10.1.12}) where the resolvent (including
the $\delta$-term) is obtained from the resolvent $L(x,\epsilon)$ of
equation (\ref{eq10.1.3}) 
by the same averaging over $\epsilon$ as in (\ref{eq10.1.15}).

Below we shall consider two ways of solving the equation (\ref{eq10.1.3}). The first
of these is to apply a expansion $\mu(a,\epsilon)$ to powers of
$\epsilon$ and $1-\epsilon$. The
application of the $\epsilon$ expansion  is a natural generalisation of the
thickness-correction method, which essentially uses the first two
terms of the expansion. The second way is to apply a resolvent in
solving equation (\ref{eq10.1.3}). We will give various expansions and
approximate expressions for the resolvent $L(x,\epsilon)$, and the results of
its computation.

\section{Decomposition of the solution for the flattened mass function}

Let us represent the solution of $\mu(a,\epsilon)$ of the integral
equation (\ref{eq10.1.3}) in the form of 
decomposition
\be
\mu(a,\epsilon) = \sum_{n=0}^{\infty} \mu_n (a,0) \epsilon^n,
\label{eq10.2.1}
\ee
or decomposition
\be
\mu(a,\epsilon) = \sum_{n=0}^{\infty} \mu_n (a,1) (\epsilon-1)^n,
\label{eq10.2.2}
\ee
where
\be
\mu_n(a,\epsilon) = \frac{1}{n!} \frac{\partial^n
  \mu(a,\epsilon)}{\partial\epsilon^n}.
\label{eq10.2.3}
\ee

The zero coefficients of series (\ref{eq10.2.1}) and (\ref{eq10.2.2}) are known. These are
the solutions of $\mu(a,0)$ and $\mu(a,1)$. To  find the other
coefficients, differentiate equation (\ref{eq10.1.3}) $n$ times over $\epsilon$. The circular
velocity is a given function, independent of $\epsilon$, and hence its
derivatives on $\epsilon$ are zero. Using Leibniz's rule for the product,
we obtain
\be
\sum_{i=0}^n \int_0^R \mu_{n-i} (a,\epsilon) \frac{\partial^i}{i! \partial\epsilon^i} \frac{1}{\sqrt{R^2-e^2a^2}} \rmd a = 0.  \label{eq10.2.4}
\ee
Here, the derivative under the integral is expressed in terms of the
Legendre polynomials. Using properties of Legendre polynomials, we can
find that
\be
\frac{1}{i!} \frac{\partial^i}{\partial\epsilon^i} \frac{a}{\sqrt{R^2-e^2a^2}} = (-1)^i \left( \frac{u}{\epsilon} \right)^{i+1} P_i (u),  \label{eq10.2.5}
 \ee
where $P_i(u)$ is a Legendre polynomial of order $i$, and
\be
u = \frac{\epsilon a}{\sqrt{R^2-e^2a^2}} .  \label{eq10.2.6}
\ee

So, we have the equation
\be
\sum_{i=0}^n (-1)^i \int_0^R \mu_{n-i} (a,\epsilon) \left( \frac{u}{\epsilon} \right)^{i+1} P_i(u) \frac{\rmd a}{a} = 0 ,  \label{eq10.2.7}
\ee
or, if we go to integration over $u$,
\be
\sum_{i=0}^n (-1)^i \epsilon^{-i+1}  \int_0^1 \mu_{n-i} (a,\epsilon) \left( \frac{a}{R} \right)^2 u^{i-2}  P_i(u) \rmd u = 0,  \label{eq10.2.7a}
\ee
and
\be
\frac{a}{R} = \frac{u}{\sqrt{\epsilon^2 + e^2u^2}}.  \label{eq10.2.6a}
\ee

The resulting equation is a recurrence integral equation for
successive calculating of functions $\mu_n(a,\epsilon)$. We shall have to apply it
for the cases $\epsilon=0$  and 1. Let us start with the first one.

We can only go directly to the limit $\epsilon\rightarrow 0$ only if
$i=0$ and 1. To find the
limit for all terms as well, we apply the L'Hospital's rule to 
(\ref{eq10.2.7a}), where uncertainties of the form 0 : 0 appear.

Let us apply  the following properties of Legendre polynomials:
\be
\int_0^1 P_i(u) u^{i-2j-2} \rmd u = 0 , ~~~ j=0, 1, ... , k-1 \ge 0 \label{eq10.2.8}
 \ee
and
\be
\left. 
\ba{lr}
(\mathrm{for~odd~} i) &  \int_0^1 P_i(u) u^{-1} \rmd u \\
(\mathrm{for~even~} i) & P_i(0) 
\ea
\right\} = (-1)^k \frac{|i-1|!!}{i!!} ,
\label{eq10.2.9}
\ee
where $k$ is an integer part of $i/2$.

Let us denote 
\be
\mu_{n,j} = \frac{1}{j!} \frac{\partial^j (\mu_n a^2)}{(\partial a^{-2} )^j} a^{-2j-2} ; ~~~ \mu_{n,j}^\ast = \frac{1}{j!} \frac{\partial^j (\mu_n a)}{(\partial a^{-2} )^j} a^{-2j-1} .
\label{eq10.2.10}
\ee
After applying L'Hospital's rule $2k$ times, taking into account (\ref{eq10.2.8}),
the $i$-term of (\ref{eq10.2.7a}) takes the form 
\be
(-1)^i \frac{(2k)!!}{|i-1|!!} \lim_{\epsilon\rightarrow 0} \epsilon^{-i+2k+1} \int_0^1 \mu_{n-i,k} (a,\epsilon) \left( \frac{a}{R}\right)^{2k+2} P_i(u) u^{i-2k-2} \rmd u . \label{eq10.2.11}
\ee
The limit here depends on whether $i$ is even or odd. If $i$ is odd, then,
given (\ref{eq10.2.9}), we get
\be
- (-1)^k \frac{(2k)!!}{(2k+1)!!} \mu_{n-2k-1,k} (R,0).  \label{eq10.2.12}
\ee
If $i$ is even, then we have to go back to
integration over $a$. Then, given (\ref{eq10.2.9}), we find 
\be
(-1)^k \int_0^R \mu_{n-2k,k} (a,0) \left( \frac{a}{R}\right)^{2k}
\frac{\rmd a}{\sqrt{R^2-a^2}},
\label{eq10.2.13}
\ee
or, after some transformation,
\be
(-1)^k \int_0^R \mu^{\ast}_{n-2k,k} (a,0) \frac{\rmd a}{\sqrt{R^2-a^2}}. \label{eq10.2.13a}
\ee
Conversion of (\ref{eq10.2.13}) to (\ref{eq10.2.13a}) is done by  temporally separating the multiplier
$R^{-2k-1}$, by integrating $k$ times over $R^{-2}$ from $R^{-2}$ to $\infty$, integrating by parts $k$ times over $a^{-2}$, and finally, the inverse operation, -- $k$-fold
differentiation over $R^{-2}$ (with temporal substitution of the time
integration by $a/R$).

Thus, we obtain the following recurrent integral equation for finding
functions $\mu_n(a,0)$: 
$$\sum_{k=0}^m (-1)^k \int_0^R \mu^{\ast}_{n-2k,k} (a,0) \frac{\rmd a}{\sqrt{R^2-a^2}}  = $$
\be
= \sum_{k=0}^{n-m-1} (-1)^k \frac{(2k)!!}{(2k+1)!!} \mu_{n-2k-1,k} (R,0),  \label{eq10.2.14}
\ee
where $m$ is an integer part of $n/2$.

The searchable  function appears in equation (\ref{eq10.2.14}) in the term $k=0$ on the
left-hand side. To express it through functions with lower indices, we
have to solve the Abel integral equation. In this case the equation
can be solved immediately with respect to the entire sum
$(-1)^k\mu^\ast_{n-2k,k}$, 
this is what transformation (\ref{eq10.2.13}) - (\ref{eq10.2.13a}) was used for, eliminating
the factor  $(a/R)^{2k}$.

After solving the Abel equation we obtain the expression for the
desired function:
$$ \mu_n(a,0) = -\sum_{k=1}^m (-1)^k \mu^{\ast}_{n-2k,k} (a,0) + $$
\be
+\frac{2}{\pi} \sum_{k=0}^{n-m-1} (-1)^k \frac{(2k)!!}{(2k+1)!!} \int_0^a \frac{\rmd}{\rmd R} \left[ \mu_{n-2k-1,k} (R,0)R\right] \frac{R}{a} \frac{\rmd R}{\sqrt{a^2-R^2}}. \label{eq10.2.15}
\ee
The formula (\ref{eq10.2.15}) allows in principle to compute functions
$\mu_n(a,0)$ 
of any order of $n$, as long as the function  $\mu(a,0)$ is given. But to
find formulas where $\mu_n(a,0)$  is expressed explicitly through
$\mu(a,0)$, 
one needs calculations which become more and more complicated as
one increases $n$. For the first  coefficients of $\mu_n(a,0)$ one
obtains
\be
\ba{ll}
\mu_1 (a,0) = & \frac{2}{\pi} \int_0^a \frac{\rmd\mu(R,0) R}{\rmd R} F_0 \left( \frac{R}{a} \right) 
\frac{\rmd R}{a}, \\
\noalign{\medskip}
\mu_2(a,0) = & - \frac{1}{2} \frac{\rmd\mu(a,0) a}{\rmd a} + \frac{4}{\pi^2} \int_0^a 
\frac{\rmd}{\rmd R} \left[ \frac{\rmd\mu(R,0)R}{\rmd\ln R} \right] F_1 \left( \frac{R}{a} \right) 
\frac{\rmd R}{a} \\
...  & 
\ea  \label{eq10.2.16}
\ee
Here the functions $F_i(x)$ are defined by the following recurrence
formula:
\be
F_{i+1} (x) = \int_x^1 F_i (y) F_0 \left( \frac{x}{y} \right) \frac{\rmd y}{y}, ~~~ F_0(x) = \frac{x}{\sqrt{1-x^2}} ,    \label{eq10.2.17}
\ee
from where we find $F_i(x)$ by successive integration:
\be
F_1(x) = x \boldsymbol{K} (\sqrt{1-x^2}) , ~~ F_2(x) = x\int_x^1 \boldsymbol{K} (\sqrt{1-y^2})  \frac{\rmd y}{\sqrt{y^2-x^2}}, ... , \label{eq10.2.18}
\ee
where $\boldsymbol{K}$  is the complete elliptic integral of the first kind. The
functions $F_i(x)$ up to fourth order are plotted graphically on
Fig.~\ref{Fig10.1}. They are properly normalised, \ie the integral
$F_i(x)$ from $x = 0$ to $x = 1$ is one. Note also that
\be
F_i(x) \rightarrow \frac{1}{2^i} \Pi_{j=1}^i B \left( \frac{j}{2}, \frac{1}{2} \right) (1-x^2)^{\frac{i-1}{2}} , ~~~ \mathrm{if} ~ x\rightarrow 1 ~~~ (i\ge 1) .  \label{eq10.2.19}
\ee

{\begin{figure*}[h] 
\centering 
\resizebox{0.60\textwidth}{!}{\includegraphics*{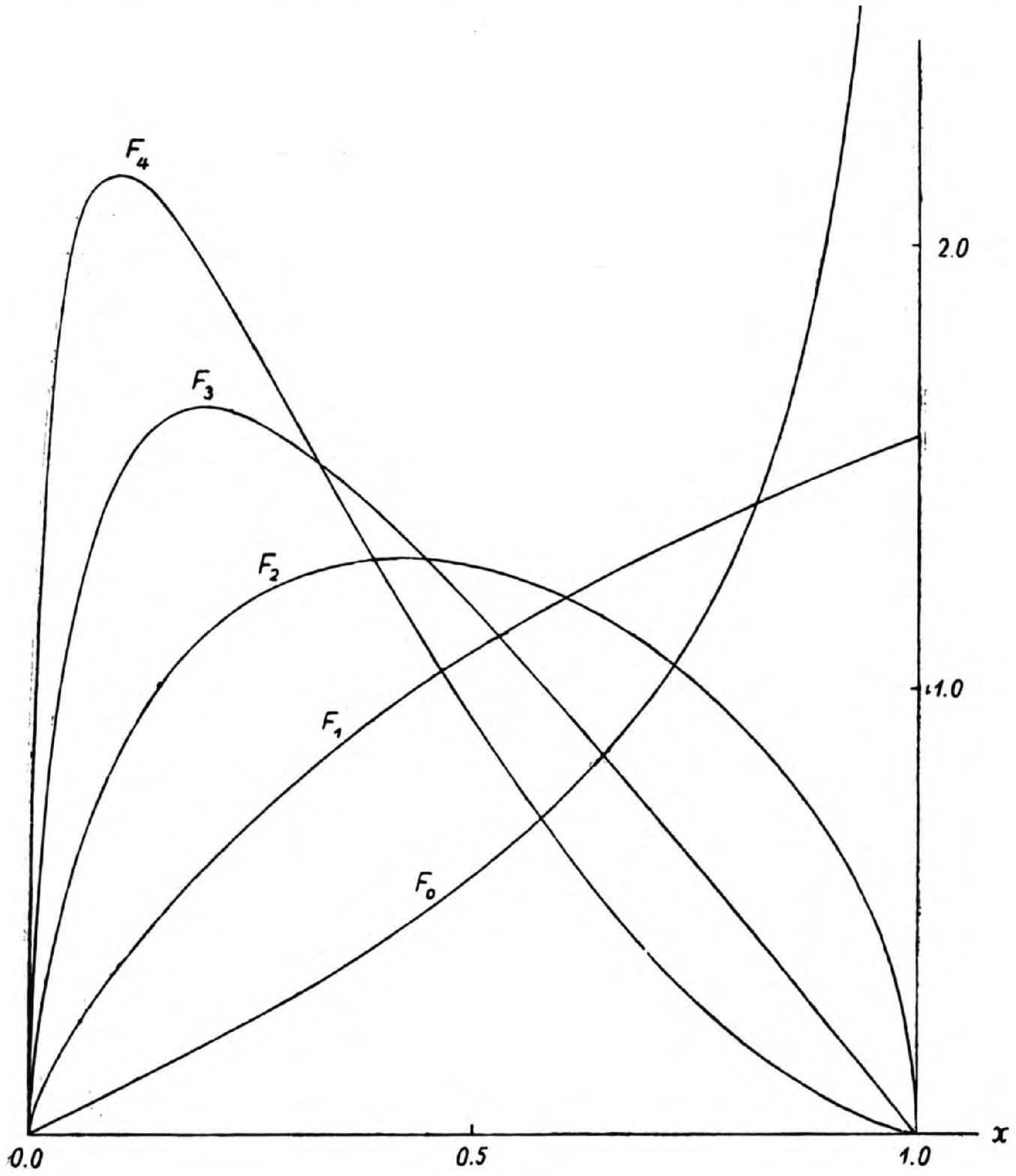}}
\caption{  } 
\label{Fig10.1}
\end{figure*} 
}

We can also derive a decomposition of $F_i(x)$ by powers of
$\sqrt{1-x^2}$.

Obviously, $F_{-1}(x)=\delta(1-x)$. The terms without integrals in
expressions $\mu_n(a,0)$ of even order may therefore be represented by
integrals with the weight function, $F_{-1}(x)$.

In terms, containing functions $F_i(x)$ of first and higher order, one can
integrate by parts and thus lower the order of the derivatives
$\mu(a,0)$. 
Then instead of functions $F_i(x)$ their derivatives over $\ln\,x$
appear. We note that the derivatives of $F_i(x)$  are expressed, like $F_i(x)$ 
iself, by complete elliptic integrals. For instance, the first and
second derivatives are expressed, respectively, through the integrals
$\boldsymbol{D}$ and $\boldsymbol{C}$ (see \ref{eq10.3.21}).

We have expressed $\mu_n(a,0)$ through  $\mu(a,0)$.   But in fact,
not the function $\mu(a,0)$ is given, but the circular
velocity. Therefore, it is of interest to express $\mu_n(a,0)$ explicitly
through the circular velocity, or, what is the same, through
$\mu(a,1)$, 
by virtue of the formula (\ref{eq10.1.9}). For the function $\mu(a,0)$ by formula
(\ref{eq10.1.8}), we have
\be
\mu(a,0) = \frac{2}{\pi} \int_0^a \mu(R,1) F_0 \left(\frac{R}{a}\right) \frac{\rmd R}{a} . \label{eq10.2.20}
\ee
Substituting this expression into formulae (\ref{eq10.2.16}) reduces simply to
replacing $\mu(a,0)$ by  $(2/\pi )\,\mu(a,1)$ and raising the order 
of the function $F_i(x)$ by one:
\be
\left.
\ba{ll}
\mu_1(a,0) = & \frac{4}{\pi^2} \int_0^a \frac{\rmd\mu (R,1) R}{\rmd R}
F_1 \left( \frac{R}{a} \right) \frac{\rmd R}{a} , \\ 
\noalign{\medskip}
\mu_2(a,0) = & -\frac{1}{\pi} \int_0^a \frac{\rmd\mu(R,1)R}{\rmd R}
F_0 \left( \frac{R}{a} \right) \frac{\rmd R}{a} + \frac{8}{\pi^3}
\int_0^a \frac{\rmd}{\rmd R} \left[ \frac{\rmd\mu(R,1) R}{\rmd\ln R}
\right] F_2 \left( \frac{R}{a} \right) \frac{\rmd R}{a} ,\\ 
... & 
\ea
\right\}
\label{eq10.2.16a}
\ee
The function $\mu_i(a,0)$, multiplied by $\epsilon$,  is nothing but a correction term
in the ``galaxy thickness correction''  method \citep{Kuzmin:1952ac}. Involving higher
order terms in series (\ref{eq10.2.1}) should, of course, significantly refine
this method.

Let us now turn to the coefficients of series (\ref{eq10.2.2}). The problem of
finding functions $\mu_n(a,1)$ is incomparably easier than the problem
of finding functions $\mu_n(a,0)$. Substituting $\epsilon=1$ into equation
(\ref{eq10.2.7}) we obtain the recurrence integral equation
\be
\sum_{i=0}^n (-1)^i \int_0^R \mu_{n-i}(a,1) P_i \left(\frac{a}{R}\right) \frac{a^i \rmd a}{R^{i+1}} = 0.  \label{eq10.2.21}
\ee
The function $\mu_n(a,1)$ we are looking for, is in the first term
of the sum. After multiplying by $R$ and differentiating by $R$,
we find (given that $P_i(1)=1$) 
\be
\mu_n(a,1) = \sum_{i=1}^n (-1)^i \left[ \int_0^a \mu_{n-i} (R,1) Q_i \left( \frac{R}{a} \right) \frac{\rmd R}{a} - \mu_{n-i} (a,1)\right] , \label{eq10.2.22}
\ee
here $Q_i(x)$ denote  polynomials
\be
Q_i (x) = x \frac{\rmd}{\rmd x} P_i (x) x^i . \label{eq10.2.23}
\ee

The coefficients we are looking for are:
\be
\left.
\ba{ll}
\mu_1(a,1) = & \int_0^a \mu(R,1) g_1 \left( \frac{R}{a} \right) \frac{\rmd R}{a} + \mu(a,1) , \\
\noalign{\medskip}
\mu_n(a,1) = & \int_0^a \mu(R,1) g_n \left( \frac{R}{a} \right)
\frac{\rmd R}{a} , ~~~ n\ge 2 , 
\ea
\right\}
\label{eq10.2.24}
\ee
where the functions $g_n(x)$ up to the third order are as follows:
\be
\left.
 \ba{ll}
 g_1 = & -2x^2, \\
 g_2 = & -(5-6x^2 + 4\ln x)x^2 , \\
 g_3 = & -(16-30x^2 + 15x^4 + 16\ln x + 4 \ln^2 x) x^2 , \\
... & 
\ea
\right\}
\label{eq10.2.25}
\ee
These are shown in Fig. \ref{Fig10.2}. At $x=1$ the functions $g_n(x)$  are equal
to  $(-1)^n$, except for $g_1(x)$.

{\begin{figure*}[h] 
\centering 
\resizebox{0.60\textwidth}{!}{\includegraphics*{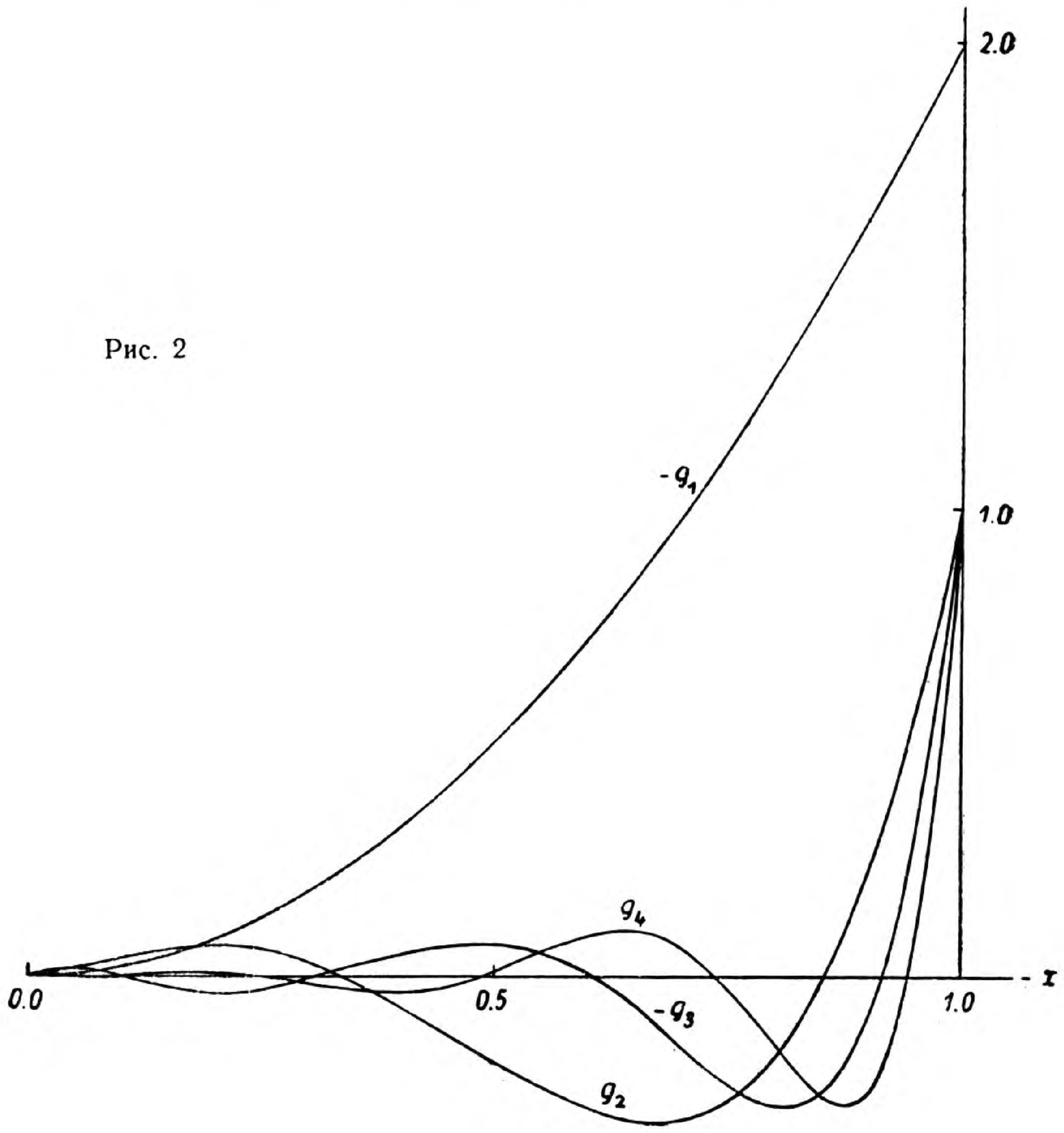}}
\caption{} 
\label{Fig10.2}
\end{figure*} 
}

Thus, the solution of equation (\ref{eq10.1.3}) using series
(\ref{eq10.2.1}) and (\ref{eq10.2.2}) can be considered as found. 

It should be noted that instead of series (\ref{eq10.2.2}) we can use a power
expansion of $e^2$. The coefficients of this series are calculated a
little easier than the coefficients of series (\ref{eq10.2.2}). The recurrent
integral equation, similar to (\ref{eq10.2.21}), has instead of
Legendre polynomials  simply constants.

\section{Resolvents  of the integral equation for the mass function}

Let us now solve the equation (\ref{eq10.1.3}) using the resolvent, \ie  using
the formula (\ref{eq10.1.12}). If the resolvent is found, then we have the
solution of equation (\ref{eq10.1.3}) in a finite form.

For equation (\ref{eq10.1.3})
\be
K \left( \frac{a}{R} \right) = \frac{R}{\sqrt{R^2-e^2a^2}}. \label{eq10.3.1}
\ee
The solution of (\ref{eq10.1.12}) is written in the form
\be
\mu(a,\epsilon) = \int_0^a \mu(R,1) L \left( \frac{R}{a} , \epsilon
\right) \frac{\rmd R}{a} + \epsilon\mu(a,1),
\label{eq10.3.2} 
\ee
and the equation for the resolvent (\ref{eq10.1.13}) turns out to be
\be
\int_0^1 L(x,\epsilon) \frac{\rmd x}{\sqrt{x^2- e^2y^2}} = 1 -
\frac{\epsilon}{\sqrt{1-e^2y^2}}.
\label{eq10.3.3} 
\ee
We have to solve this equation.

First of all we should note that we can easily find
an expression for moments of the desired function.

Let
\be
M_\nu = \int_0^1 L(x) x^\nu \rmd x + \frac{1}{K(1)}  \label{eq10.3.4}
\ee
is  the moment of order $\nu$  (it can also be non-integer) of the resolvent
$L(x)$  together with the delta-like term corresponding to the second
term, in (\ref{eq10.1.12}) or (\ref{eq10.3.2}). Let us differentiate the equation  (\ref{eq10.1.13})
over $y$ and then multiply by $y^{1+\nu}$ to integrate from 0 to 1. Then, after some transformation, we find
\be
\frac{1}{M_\nu} = K(1) - \int_0^1 K' (y) y^{1+\nu} \rmd y . \label{eq10.3.5}
\ee
In our particular case
\be
\frac{1}{M_\nu (\epsilon)} = \frac{1}{\epsilon} - e^2 \int_0^1
\frac{y^{2+\nu} \rmd y}{ (1-e^2y^2)^{3/2}}.
\label{eq10.3.6} 
\ee
The integral here converges if $\nu$ is greater than $-3$. If $\nu$
is greater than $-1$, we can integrate by parts:
\be
\frac{1}{M_\nu (\epsilon)} = (1+\nu) \int_0^1 \frac{y^{\nu} \rmd y}{ \sqrt{1-e^2y^2}} .\label{eq10.3.6a}
\ee
Obviously, the moments $M_\nu(\epsilon)$ must have a simple relation to the
coefficients $A_n(e)$ in the method of \citet{Burbidge:1959}. Indeed,
\be
A_\nu (e) = \frac{e^{1+\nu}}{(1+\nu) M_\nu (\epsilon)} = \frac{1}{2} B \left( \frac{1+\nu}{2}, \frac{1}{2}; e^2 \right) . \label{eq10.3.7}
\ee
Here the function on the right is an incomplete beta function.

The expression for moments allows us to investigate the
 resolvent for small $x$. Assume
\be
L(x) = \beta x^{1+\alpha} ~~~ \mathrm{if} ~~ x\rightarrow 0. \label{eq10.3.8}
\ee
Then
\be
M_\nu = \frac{\beta}{(2+\nu) + \alpha} ~~~ \mathrm{if} ~~ 2+\nu \rightarrow -\alpha . \label{eq10.3.9}
\ee
Therefore $\alpha$ and $\beta$ can be found from the equations
\be
\frac{1}{M_{-2-\alpha}} = 0,  \label{eq10.3.10}
 \ee
\be
\frac{1}{\beta} = \frac{\rmd M_\nu^{-1}}{\rmd\nu}\bigg|_{\nu=-2-\alpha} .  \label{eq10.3.11}
\ee

Equation (\ref{eq10.3.10}) is conveniently solved using a series expansion. By
decomposing under the integral in (\ref{eq10.3.6}) into a power series
of $2+\nu$,
 we find after integration
\be
\frac{1}{M_\nu (\epsilon )} = \epsilon + e^2 (2+\nu) \sum_{i=0}^\infty (-1)^i f_{i+1} (\epsilon) (2+\nu)^i ,  \label{eq10.3.12}
\ee
where the coefficients of the series are determined by the recurrence
formula 
%
\be
f_{i+1} (\epsilon) = \frac{1}{e} \int_0^e f_i (\epsilon) \rmd e; ~~~ f_0(\epsilon) = \frac{1}{\epsilon} .  \label{eq10.3.13}
\ee
The coefficients are close to one. By taking out the factor
$[1+(2+\nu)]^{-1}$, 
we obtain a transformed series whose sum is determined
essentially by the first term  $f_1=e^{-1}\arcsin\,e$. This makes it easy
to solve equation (\ref{eq10.3.10}) by successive approximations. For the same
reason it is easy to calculate the derivative of $M_\nu^{-1}$ over $\nu$.

If we keep only the term $f$ in the above transformed series, we obtain
the following approximations: 
\be
\alpha(\epsilon) \simeq \frac{\epsilon}{\epsilon + e\arcsin e} , ~~~
\beta(\epsilon) \simeq \frac{e\arcsin e}{ (\epsilon + e\arcsin e)^2
}. \label{eq10.3.14} 
\ee
The first gives $a$ with an error of about $0.01$.  The error of the
second formula is three times larger.

The values of the functions $\alpha(\epsilon)$ and $\beta(\epsilon)$
for different $\epsilon$ are as follows:
\begin{tabbing}
$\epsilon$~~~~~ =~~~~ \= ~0.0~~~ \= ~0.1~~~ \= ~0.2~~~ \= ~0.3~~~ \= ~0.4~~~ \= ~0.5~~~ \= ~0.6~~~ \= ~0.7~~~ \= ~0.8~~~ \= ~0.9~~~ \= ~1.0~~~\\
$a(\epsilon)$ = \> .000 \> .065 \> .133 \> .206 \> .284 \> .368 \> .460 \> .564 \> .688 \> .824 \> 1.000\\
$\beta(\epsilon)$ = \> .637 \> .618 \> .594 \> .564 \> .528 \> .484\>  .430 \> .362 \> .276 \> .162 \> .000 
\end{tabbing}

Let us now proceed to the solution of equation (\ref{eq10.3.3}). Consider
decompositions similar to those of (\ref{eq10.2.1}) and (\ref{eq10.2.2}) for the solution of
the original equation (\ref{eq10.1.3}), viz.
\be
L(x,\epsilon) = \sum_{n=0}^\infty L_n (x,0) \epsilon^n   \label{eq10.3.15}
 \ee
 and
\be
L(x,\epsilon) = \sum_{n=0}^\infty L_n (x,1) (\epsilon -1)^n,
\label{eq10.3.16}
\ee
where
\be
L_n (x,\epsilon) = \frac{1}{n!} \frac{\partial^n L(x,\epsilon)}{\partial\epsilon^n} . \label{eq10.3.17}
\ee
In order to find coefficients $L_n(x, 0)$ and $L_n(x, 1)$, it is appropriate
to use as auxiliary functions the coefficients $\mu_n(a,0)$ and $\mu_n(a, 1)$,
for which we derived recurrence formulas in the previous section.

It follows from formula (\ref{eq10.3.2}) that if we put
\be
\mu(a,1) = \delta (a-a_0),  \label{eq10.3.18}
\ee
where $a_0$ is an arbitrarily fixed value of $a$, then for $a > a_0$
\be
L_n \left( \frac{a_0}{a}, \epsilon \right) = \mu_n (a,\epsilon) a. \label{eq10.3.19}
\ee
Using the results of the previous paragraph, we find the following
expressions for functions $L_n(x, 0)$ up to order three:
\be
\left.
\ba{ll}
L_0(x,0) = & \frac{2}{\pi} F_0 (x), \\
\noalign{\smallskip}
L_1 (x,0) = & - \frac{4}{\pi^2} \frac{\rmd F_1 (x)}{\rmd\ln x}, \\
\noalign{\smallskip}
L_2 (x,0) = & \frac{1}{\pi} \frac{\rmd F_0 (x)}{\rmd\ln x} +
\frac{8}{\pi^3} \frac{\rmd^2 F_2 (x)}{(\rmd\ln x)^2}, \\ 
\noalign{\smallskip}
L_3 (x,0) = & - \frac{4}{3\pi^2} \left[ \frac{\rmd F_1 (x)}{\rmd\ln x} + 2 \frac{\rmd^2 F_1 (x)}{(\rmd\ln x)^2} \right] - \frac{16}{\pi^4} \frac{\rmd^3 F_3 (x)}{(\rmd\ln x)^3}. \\ 
... & 
\ea
\right\}
\label{eq10.3.20}
\ee
Here
\be
\left.
\ba{ll}
F_0(x) = & \frac{x}{(1-x^2)^{1/2} }, ~~ \frac{\rmd F_0(x)}{\rmd\ln x} = \frac{x}{(1-x^2)^{3/2} }, \\
\noalign{\smallskip}
\frac{\rmd^2 F_0(x)}{(\rmd\ln x)^2} = & \frac{3x}{(1-x^2)^{5/2} } - \frac{2x}{(1-x^2)^{3/2} }, \\
\noalign{\smallskip}
\frac{\rmd F_1(x)}{\rmd\ln x} = & x \boldsymbol{D} (\sqrt{1-x^2}), ~~
\frac{\rmd^2 F_1(x)}{(\rmd\ln x)^2} = x \boldsymbol{C} (\sqrt{1-x^2}).
\\ 
... & 
\ea \right\}
\label{eq10.3.21}
\ee

For functions $L_n(x, 1)$ we have
\be
\left.
\ba{ll} 
L_0(x,1) = & 0, \\
l_n(x,1) = & g_n(x), ~~~ n\ge 1
\ea
\right\}
\label{eq10.3.22}
\ee

In addition to the above expansion we can also use the expansion
\be
L(x,\epsilon) = L(1,\epsilon) \sum_{n=0}^\infty l_n (\epsilon) (x^2
-1)^n,
\label{eq10.3.23}
\ee
where
\be
L(1,\epsilon)l_n(\epsilon)  = \frac{1}{n!} \frac{\partial^n
  L(x,\epsilon)}{(\partial x^2)^n}\bigg|_{x=1} , \label{eq10.3.24} 
 \ee
and
\be
L(1,\epsilon) = \frac{e^2}{\epsilon},  \label{eq10.3.25}
\ee
as is obtained from equation (\ref{eq10.3.3}) at $y\rightarrow  1$.

Let us differentiate equation (\ref{eq10.3.3}) $n + 1$ times by $y^2$,
changing  first the
 integration variable to $x/y$ (or $y/x)$. Turning to
the limit  $y = 1$, we obtain
\be
\frac{1}{2} \sum_{i=0}^n (n-i)! l_{n-i} (\epsilon) \left( \frac{\partial}{\partial y^2}\right)^i \frac{y^{-2(n-i)-2}}{\sqrt{1-e^2y^2}}\bigg|_{y=1} = \frac{\epsilon}{e^2} \left( \frac{\partial}{\partial y^2} \right)^{n+1} \frac{1}{\sqrt{1-e^2y^2}}\bigg|_{y=1} .  \label{eq10.3.26}
\ee
By differentiating with Leibniz's rule, we obtain a recurrence formula for the coefficient
decomposition (\ref{eq10.3.23}):
\be
\sum_{i=0}^n l_{n-i} (\epsilon) q_{n,i} (\epsilon) =
\frac{(2n+1)!!}{2^n} \left( \frac{e}{\epsilon} \right)^{2n}
, \label{eq10.3.27} 
\ee
where $q_{n,i}(\epsilon)$ are the following polynomials:
\be
q_{n,i} (\epsilon) = i! \, \sum_{j=0}^i (-1)^{i-j}
\frac{(n-j)!}{(i-j)!} \frac{|2j-1|!!}{(2j)!!} \left(
  \frac{e}{\epsilon} \right)^{2j} . \label{eq10.3.28} 
\ee
Using the recurrence formula, we find
\be
\ba{ll}
l_0(\epsilon) = & 1, ~~~ l_1 (\epsilon) = \frac{1}{\epsilon^2} , ~~~
l_2 (\epsilon) = \frac{5}{4} \frac{e^2}{\epsilon^4} , \\ 
\noalign{\smallskip}
l_3 (\epsilon) = & \frac{(2+35e^2) e^2}{24\epsilon^6} , \\
... & 
\ea \label{eq10.3.29}
\ee
We could similarly obtain a power expansion of $L(x, \epsilon)$ simply from the
difference $1 - x$. But the proof becomes more complicated. Equation
(\ref{eq10.3.3}) then has to be differentiated in terms of $y$ instead of $y^2$. 
Instead of monomials corresponding to product $(1 -e^2y^2)^{-1/2}$ by $y^2$
for $y = 1$, we obtain Legendre polynomials.

Let us further consider the behaviour of the resolvent in the
neighbourhood of its peak at $\epsilon=0$,  $x= 1$.

Denote by
\be
r=\sqrt{1-x^2 + \epsilon^2}, ~~~~ \tan\vartheta = u = \frac{\sqrt{1-x^2}}{\epsilon} , ~~~ v= \frac{\sqrt{1-y^2}}{\epsilon} .  \label{eq10.3.30}
\ee
It follows from equation (\ref{eq10.3.3}) that
\be
L(x,\epsilon) \rightarrow \frac{1}{\epsilon}\psi(u) =
\frac{1}{r}\chi(\vartheta) ~~~ \mathrm{if} ~ r\rightarrow
0,  \label{eq10.3.31} 
\ee
where the function  $\psi(u)$ satisfies the equation
\be
\int_0^v \frac{\psi (u) u \rmd u}{\sqrt{1+v^2-u^2}} = 1- \frac{1}{\sqrt{1+v^2}}.  \label{eq10.3.32}
\ee
The solution of this integral equation can be given by the series
$$ \psi(u) = 1- u^2 + \frac{5}{4} u^4 - \frac{37}{24} u^6 + \frac{353}{192} u^8 - ...  $$
\be
= \frac{2}{\pi} u^{-1} \left[ 1- \left( \frac{2}{\pi} - \frac{1}{2} \right) u^{-2} + \left( \frac{12}{\pi^2} - \frac{5}{\pi} + \frac{3}{8} \right) u^{-4} - ... \right] .\label{eq10.3.33}
\ee
They follow from series (\ref{eq10.3.15}) and (\ref{eq10.3.23}),  applied to the vicinity of
the peak. From series (\ref{eq10.3.33}) we find corresponding series for
function  $\chi(\vartheta)=\psi(u)\,\sec\vartheta$: 
$$  \chi(\vartheta) = 1- \frac{1}{2} \sin^2\vartheta + \frac{1}{8} \sin^4\vartheta + \frac{1}{48}\sin^6\vartheta  - \frac{1}{384} \sin^8\vartheta - ... $$
\be
= \frac{2}{\pi} \left[ 1 + \left( 1-  \frac{2}{\pi} \right) \cos^2\vartheta  + \left(  \frac{3}{2} - \frac{8}{\pi} + \frac{12}{\pi^2} \right) \cos^4\vartheta  + ... \right] . \label{eq10.3.34}
\ee
The values of function $\chi(\vartheta)$ obtained with series (\ref{eq10.3.34}) are as
follows:
\begin{tabbing}
$\sin^2\vartheta = $  \= 0.0~~~~\= 0.1~~~~\= 0.2~~~~\= 0.3~~~~\= 0.4~~~~\= 0.5~~~~\= 0.6~~~~\= 0.7~~~~\= 0.8~~~~\= 0.9~~~~\= 1.0\\
$\chi(\vartheta) = $   \> 1.00  \> .951 \> .905 \> .862 \> .821 \> .784 \> .749 \> .717 \> .687 \> .661 \> .637 
\end{tabbing}

Expressions (\ref{eq10.3.31}) are the first terms of series
\be
L(x, \epsilon) = \frac{1}{\epsilon} \sum_{i=0}^\infty \psi_i (u)
\epsilon^{2i} = \frac{1}{r} \sum_{i=0}^\infty \chi_i (\vartheta )
r^{2i} . \label{eq10.3.35} 
\ee
Here $\psi_0(u)\equiv \psi(u)$,
$\chi_0(\vartheta)\equiv\chi(\vartheta)$. For the remaining
coefficients, we can obtain decompositions similar to (\ref{eq10.3.33}) and
(\ref{eq10.3.34}). However, only for $\chi_1(\vartheta)$ do the series still
appear to be converging well enough.

\section{Calculation of the resolvent; approximate formulas}

Above we got series that can be used to calculate the resolvent. But
it is possible to solve equation (\ref{eq10.3.3}) numerically as well -- by
successive approximations. To do this we need to increase the ``solving
power'' of the kernel of the equation. Let us apply the Abel method: 
we multiply the equation (\ref{eq10.3.3}) by $y/\sqrt{y^2 - y_i^2}$, integrate
over $y$ from $y_1$ to  unity, change the order of integration, and
differentiate over $y_1$. Then we obtain a new integral equation
(omitting the index at $y$)
\be
\int_y^1 L(x,\epsilon) \frac{\epsilon x\rmd x}{\sqrt{x^2-y^2} (x^2
  -e^2 y^2)} =  e^2 \frac{\sqrt{1-y^2}}{1-e^2y^2}.   \label{eq10.4.1} 
\ee
Although the kernel of the equation is not transformed into a delta function, as
in the case of the Abel equation, it has become much "steeper" than
the original equation.

The solution of equation (\ref{eq10.4.1}) can be found in the following way.

First we find the first approximation. For it we can take an average
value $L(x, \epsilon)$ derived from the equation, formed with the kernel as a
weight. For the corresponding value of $x$ we can take some average of
$x$. It is convenient to take a quadratic mean. These averages are
expressed as follows
\be
\overline{L} (y,\epsilon) = e^2 \lambda (y,\epsilon) \frac{y}{\sqrt{1-e^2y^2}},  \label{eq10.4.2}
 \ee
\be
\overline{x^2} (y,\epsilon) = e^2 y^2 +\epsilon \lambda(y,\epsilon) y \sqrt{1-e^2y^2},  \label{eq10.4.3}
 \ee
where
$$
\lambda(y,\epsilon) = \frac{z}{\arcsin z}, ~~ z=\sqrt{\frac{1-y^2}{1-e^2y^2}}. 
$$
(Function $\lambda(y,\epsilon)$ varies within a comparatively narrow range, from
$2/\pi$ at $y=0$ to unity at $y= 1$).

Thus, as a first approximation we can assume the parametric relation
\be
L(x,\epsilon) = \overline{L} (y,\epsilon) , ~~~ x^2 = \overline{x^2} (y,\epsilon) .  \label{eq10.4.4}
\ee
This is exact for $\epsilon=0$, $\epsilon=1$, $x=0$  and $x =
1$, whereas  for $\epsilon=1$ and $x=1$
the first derivatives of $\epsilon$ and $x$ are also exact, respectively.

Given the first approximation, substitute it into equation (\ref{eq10.4.1}) and
take the resulting difference between the right and left sides of the
equation as a new given function in the right part of the equation. We
now solve the equation with respect to the first approximation using
the same approach. Then we can search for the correction to the
correction, and so on.

Such calculations were done by Herbert Niilisk, under one of the
authors of this paper (Kuzmin), as early as 1955. However, the work
was not completed at that time. 
Only a few years ago the authors of this paper returned to the
question of solving the integral equation of mass distribution, in
particular to the question of the resolvent $L(x, \epsilon)$. The resolvent was
recalculated using the series considered in the previous
section. These series and formulas for the limit behavior of the
resolvent allowed us to compute its values in rather wide overlapping
neighbourhoods $\epsilon = 0$, $\epsilon = 1$, $x = 0$, and $x =
1$. For the remaining small 
region the values of the resolvent could be quite confidently
interpolated using numerically derived results.

{\begin{figure*}[h] 
\centering 
\resizebox{0.60\textwidth}{!}{\includegraphics*{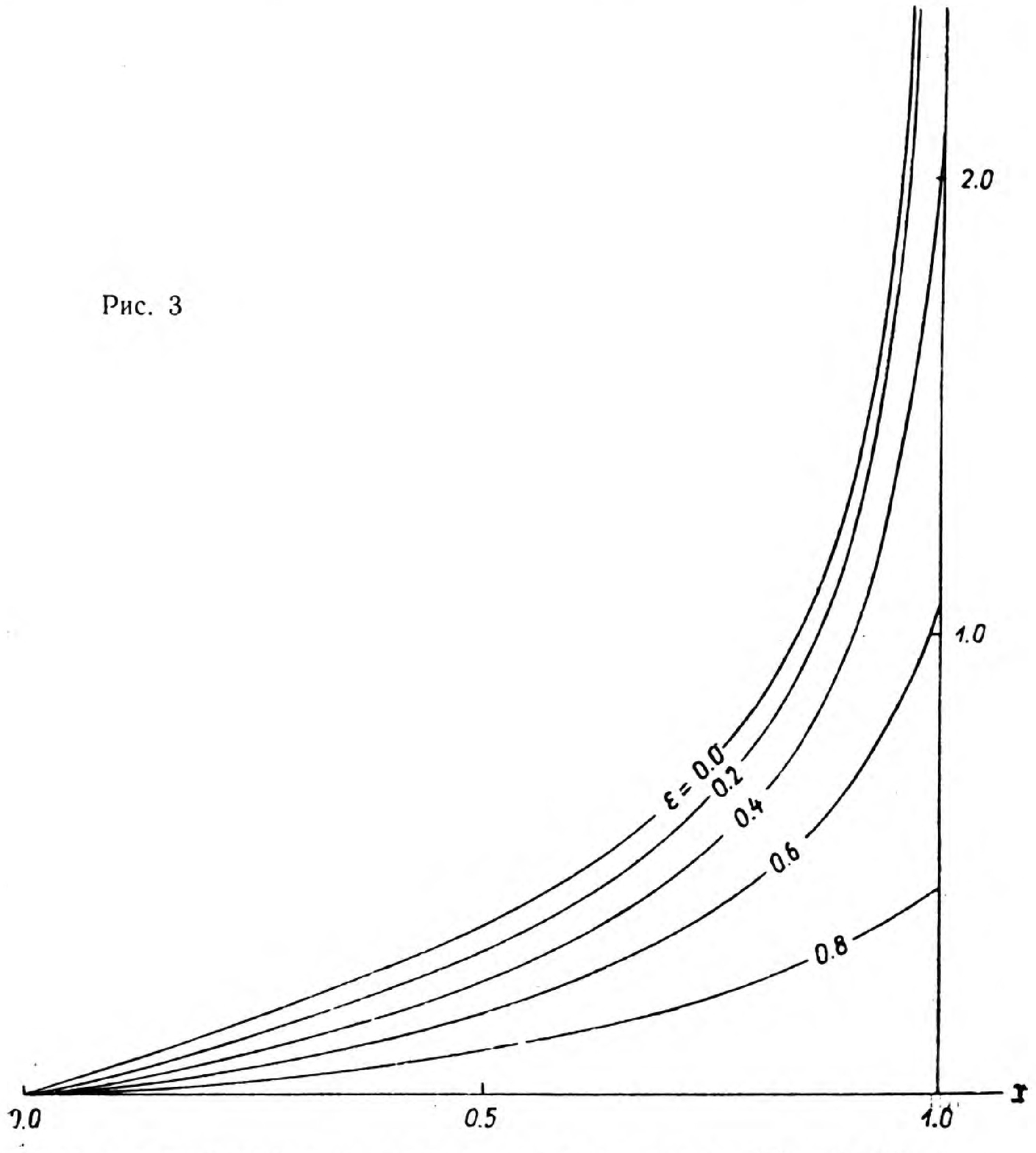}}
\caption{} 
\label{Fig10.3}
\end{figure*} 
}

The results of the calculations are  illustrated by a graph (Fig. \ref{Fig10.3}).

The use of the resolvent for the solution of the integral mass
distribution equation has the disadvantage that if we need to change
the value of $\epsilon$, the equation has to be solved anew by calculating the
integral in the solution (\ref{eq10.3.2}) as a function of $a$. This does not
appear in the interpolation method of G. M. Idlis, where it is
sufficient to compute $\mu(a, 0)$ and $\mu(a, 1)$, and then we can find
$\mu(a, \epsilon)$ by interpolating on any $\epsilon$ according to formula (\ref{eq10.1.10}). The
Idlis method corresponds to a resolvent in the form
\be
L(x,\epsilon) = L(x,0) (1-\epsilon),  \label{eq10.4.5}
\ee
\ie  instead of an exact resolvent, a linear interpolation
between $L(x, 0)$ and $L(x, 1)=0$. The resolvent in this form is
very different from the true resolvent. Idlis's method, however, can
be greatly improved by adding to the expression (\ref{eq10.4.5}) a
further quadratic term.

We shall require that for $\epsilon = 1$ we have not only the equality of the
approximate and true resolvent, but also their differentiate  over  $\epsilon$. Then for
the approximate resolvent we will have
$$ L(x,\epsilon) = L(x,0) (1-\epsilon) - [ L(x,0) + L_1 (x,1)]  \epsilon(1-\epsilon) =  $$
\be
= L(x,0) (1-\epsilon)^2 - L_1 (x,1) \epsilon (1-\epsilon)  \label{eq10.4.6}
\ee
$ (L_1(x,\epsilon) = \partial L / \partial\epsilon)$ or by inserting expressons for $L(x,0)$ and $L_1(x,1)$,
\be
L(x,\epsilon) = \frac{2}{\pi} \frac{x}{\sqrt{1-x^2 }} (1-\epsilon)^2 +
2x^2 \epsilon (1-\epsilon ). \label{eq10.4.6a} 
\ee
The corresponding solution to the integral mass distribution equation
of mass distribution has the form
\be
\mu(a,\epsilon) = (1-\epsilon)^2 \mu(a,0) + 2\epsilon (1-\epsilon) \int_0^a \mu(R,1) \frac{R^2\rmd R}{a^3} + \epsilon\mu (a,1) . \label{eq10.4.7}
\ee
Using formula (\ref{eq10.4.7}) instead of formula (\ref{eq10.1.10}) does not complicate the
calculations, since the integral appearing here is simple to
calculate. The correction is very significant. However, the deviations
of the resolvent (\ref{eq10.4.6}) from the true one are in general as large as
the deviations of the resolvent (\ref{eq10.4.5}). But large deviations take place
in this case only in a rather narrow interval of $x$ values near
$x = 1$ 
(starting from about $x= 0.8$). The mass function $\mu(a, \epsilon)$ is only
weakly affected if $\mu(a,1)$ changes smoothly enough.

On the influence of inaccuracy of the resolvents (\ref{eq10.4.5}) and (\ref{eq10.4.6}) on
the solution for the mass function can be seen by comparing their
moments with the corresponding moments of the exact resolvent
(including the delta term). If the function $\mu(a, 1)$ changes smoothly,
it can be approximated by a quadratic or cubic polynomial. Therefore
it is sufficient to consider the moments of the lowest orders.

The moments of the exact resolvent are calculated by the formula
(\ref{eq10.3.6}). For moments of integer order we find

\be 
\left.
\ba{ll}
M_{-2} (\epsilon) =  \frac{1}{\epsilon} , &    M_{-1} (\epsilon) = 1,  \\
\noalign{\smallskip}
M_0 (\epsilon) = \frac{e}{\arcsin e} ,    &     M_{1} (\epsilon) = \frac{1}{2} (1+\epsilon) ,   \\
\noalign{\smallskip} 
M_2 (\epsilon)  = \frac{2}{3} \frac{e^3 }{\arcsin e - e\epsilon}  &  M_{3} (\epsilon) = \frac{3}{4} \frac{(1+\epsilon)^2}{2+\epsilon} .
\ea \right\} \label{eq10.4.8}
\ee
As for the approximated resolvents (\ref{eq10.4.5}) and (\ref{eq10.4.6})
the first moments are expressed by the linear formula
\be
M_\nu (\epsilon) = M_\nu (0) (1-\epsilon) + \epsilon ,  \label{eq10.4.9}
\ee
and the moments of the second are expressed by the quadratic formula
\be
M_\nu (\epsilon) = M_\nu (0) (1-\epsilon)^2 + \frac{2}{3+\nu} \epsilon (1-\epsilon) + \epsilon . \label{eq10.4.10}
\ee
Here
\be
M_\nu (0) = \frac{1}{\pi} B \left( \frac{2+\nu}{2}, \frac{1}{2} \right) , \label{eq10.4.11}
\ee
and the term $\epsilon$ appears as a consequence of the delta term of the
resolvent.

In all three cases $M_{-1} = 1$. The moments of $M_1(\epsilon)$ are the same. The
moments of the other orders are different.

The difference between the moments, as well as between the resolvents
themselves, is zero at $\epsilon=0$  and 1. It reaches a maximum near the
middle of the interval. Assuming $\epsilon = 0.5$, we obtain by formula (\ref{eq10.4.8})

$M_0 = 0.827, ~~M_1 = 0.750, ~~M_2 = 0.705, ~~M_3 = 0.675$.

\noindent The formulas (\ref{eq10.4.9}) and (\ref{eq10.4.10}) give

$M_0 = 0.818, ~~M_1 =0.750, ~~M_2= 0.712, ~~M_3= 0.689$

\noindent and

$M_0 = 0.826, ~~M_1 =0.750, ~~M_2 = 0.706, ~~M_3= 0.677$.

As we see, the moments of the resolvent (\ref{eq10.4.6}) differ from the exact
resolvent moments very little, an order of magnitude smaller than the
moments of the resolvent (\ref{eq10.4.5}).

The formula (\ref{eq10.4.7}) seems to be accurate enough for practical
applications. However, it can still be considerably refined. For this
purpose let's add a correction term to the linear resolvent (\ref{eq10.4.5}) (in
the first line of formula (\ref{eq10.4.6})) with some factor depending on $\epsilon$ in
order to achieve the coincidence of one more moment of the approximate
and exact resolvent. Let's assume a coincidence of the angular
momentum $M_3$. Then the above factor will take a very simple form,
$3/(2+\epsilon)$. Instead of formula (\ref{eq10.4.6a}) we obtain

\be
L(x,\epsilon) = \frac{4}{\pi} \frac{x}{\sqrt{1-x^2}} \frac{(1-\epsilon)^2}{2+\epsilon} + 6x^2 \frac{\epsilon (1-\epsilon)}{2+\epsilon} . \label{eq10.4.12}
\ee
Accordingly formula (\ref{eq10.4.7}) is replaced by the following:

\be
\mu(a,\epsilon) = \frac{2(1-\epsilon)^2}{2+\epsilon} \mu(a,0) +
\frac{6\epsilon (1-\epsilon)}{2+\epsilon} \int_0^a \mu(R,1)
\frac{R^2\rmd R}{a^3} + \epsilon\mu(a,1) .  \label{eq10.4.13}
\ee
The expression (\ref{eq10.4.10}) for the momentum changes in a similar way. Now
the approximate and exact resolvents  have the same moments $M_{-1}$,
$M_1$,  $M_3$, 
and, in addition, a moment of some non-integer negative order between
$\nu=-1$ and $-2$. The calculation shows that the $M_0$ and $M_2$ moments differ
from the corresponding moments of the exact resolvent by less than
0.0005. The moment $M_4$ is also identical with the same
accuracy. Therefore, for practical applications the accuracy of
formula (\ref{eq10.4.13}) is more than sufficient.

\vskip 5mm
\hfill November 1965

%% file: chapter11.tex
\part{Spatio-kinematical  structure of self-gravitating stationary
  stellar systems and some general problems of dynamics of stellar 
  systems}

\chapter[Models of stationary stellar systems with axial
symmetry]{Models of stationary self-gravitating stellar systems with
  axial symmetry\footnote{\footnotetext ~~Bull. Abastumani
    Astrophys. Obs. No.~27, 82 -- 88, 1962. Report on the 3rd Meeting of the
    Committee on Stellar Astronomy, October 3 -- 6, 1960,
    Tbilisi. Coauthor S. A. Kutuzov.}} 

\section{Models of stationary self-gravitating stellar systems with axial
symmetry}

For the majority of stellar systems, the assumptions of stationarity and 
axial symmetry are good approximations. For this reason the
construction of models of stationary self-gravitating stellar systems is an
important task of stellar dynamics.

Most of the models constructed by now concerned spherical systems, the
more general case of axisymmetric systems remained little studied.

In the present report we attempt to construct models of this
more general class of systems.

In case of the stationary stellar system the phase density depends on
phase (\ie  on coordinates and velocities) only through the conservative
isolating (single-valued) integrals of motion. If the system is
self-gravitating, the mass density in coordinate space, which is the integral of
the phase density over the velocity space, satisfies together with the
gravitational potential the Poisson's equation. In order to construct a 
model for the stationary self-gravitating system, we must thus find the phase
density as a function of phase and the gravitational potential as a
function of coordinates so that both conditions are valid.

In case of stationarity and axial symmetry there exist in general two
conservative isolating integrals of motion -- the energy integral and the
angular momentum (area) integral.

If the potential has a certain additional restriction, there exists
the third integral, which is quadratic in respect to velocities
\citep{Kuzmin:1953}. If three integrals exist,  
the velocity distribution can be triaxial.

When constructing the spherical models, the starting point is usually the
phase density as a function of integrals, and the aim is to find the
potential. However, already \citet{Eddington:1916} proposed to use an
alternative approach, where one starts from potential or from the
mass density, and the aim is to find the phase density as a function
of integrals and thereafter of phase. Eddington's method seems to be
more appropriate and we use it to construct models for general case of
axial symmetry.

In order to have a model, which resembles more or less the real Galaxy, it is desirable
to begin with the potential allowing for the existence of the third integral.
However, finding a solution for the phase density as a function of three
integrals has certain difficulties. For this reason in present paper we
limit ourselves with the solutions, that depend only on the energy integral and the area
integral, and produce thus biaxial velocity distribution.

If the phase density is a function of only  integrals of energy and  area, then the
mass density is given by
\begin{equation}
\rho = \frac{2\pi}{R} \iint_{v^2\ge 0} \Psi (E,I) \rmd E  \rmd I, \label{eq11.1}
\end{equation}
where
\begin{equation}
E=\Phi - \frac{1}{2}(v^2+w^2), ~~~ I=Rw. \label{eq11.2} 
\end{equation}
Here $\rho$ is the mass density in coordinate space, $\Psi$ is the phase
density, $\Phi$ is the gravitational potential, $R$ is the distance from
the symmetry axis, $v$ is the velocity in the meridional plane, $w$ is the
velocity perpendicular to this plane, $E$ is the negative energy integral
and $I$ is the area integral.

In (\ref{eq11.1}) the mass density $\rho$ is a function of $\Phi$ and $R$
\begin{equation}
\rho = \rho (\Phi ,R) . \label{eq11.3}
\end{equation}
In order to find the phase density, the mass density must be in form of
(\ref{eq11.3}). In this case, the phase density results from (\ref{eq11.1}) as a solution
of an integral equation. Only non-negative solutions have physical meaning. 
The solution can not be unique, because the anti-symmetrical part
of $\Psi$ in respect to $I$ does not contribute to the mass density and
may be chosen arbitrarily with the only limitation  that the total phase density
must be non-negative. The symmetrical part of the phase density can be derived
uniquely. It must be non-negative.

Solution for the symmetrical part of the phase density can be derived
in the 
form of series. By assuming the mass density $\rho$ for small $\Phi$ to be
proportional to $\sim\Phi^{\kappa}$, we have the following series for the mass
density and for the symmetrical part of the phase
density\footnote{It is possible to demonstrate that if $k$ is a
natural number, then $\kappa$ is also a natural number. [Later footnote.]
[Appendix A] }
\begin{equation}
\rho = \Phi^{\kappa} \sum_{k=0}^{\infty} p_k(\Phi R^2) \Phi^k, ~~~ \Psi = E^{\kappa -3/2}\sum_{k=0}^{\infty} P_k(I^2) E^k, \label{eq11.4} 
\end{equation}
where $p_k$ and $P_k$ are polynomials of the order $k$ with respect to their
arguments. The coefficients of polynomial $P_k$ may be calculated from the
coefficients of polynomial $p_k$ according to formulas derived by \citet{Fricke:1952}.
 In addition, the series for $\rho$ with respect to
positive and negative powers of $R$, and the corresponding series for the
symmetrical part of $\Phi$ with respect to positive and negative powers of
$I$,  can be used. 

In case where $\rho$ steeply increases with $\Phi$ and reaches infinity as
$\Delta\Phi^{-\nu}$ when approaching to a certain curve on plane
$\Phi ,R$, where $\Delta\Phi$ is the distance from the curve
along $\Phi$, we may also use the series
\begin{equation}
\rho =\Delta\Phi^{-\nu} \sum_{k\ge 0} h_k(R)\Delta\Phi^k , ~~~ \Psi =\Delta E^{-\nu-3/2} \sum_{k\ge 0} H_k(I)\Delta E^k.
\label{eq11.5} 
\end{equation}
Here $\Delta E$ is the distance from the curve along $E$ on the plane
$E,I$ where the phase density becomes infinite. It is possible to
find the curve by knowing the curve of infinite density on plane $\Phi ,R$. The
functions $H_k$ can be found from the functions $h_k$. Using of series (\ref{eq11.5})
gives us a solution, which is non-symmetrical with respect to $I$, and can be
used as a solution for the total phase density. Also the
symmetrical part of the solution can be derived if needed.  

In addition to the phase density, it is also interesting to obtain the
solutions for the velocity dispersions and for the projections of the
phase density into the coordinate axis and into the planes in velocity
space. It is suitable to call these functions ``the functions for
partial description of the model'' in order to differentiate them from
the phase density, which gives the complete description of the
model. It is easy to derive the projection of the phase density into
an axis of meridional plane and the corresponding velocity dispersion
by quadratures from the partial description functions. We obtain
\begin{equation}
f(U,R)= \frac{1}{\pi} \int_0^U \frac{\partial\rho}{\partial\Phi} \frac{\rmd\Phi
}{\sqrt{2(U-\Phi )}}, \label{eq11.6}
\end{equation}
where
\begin{equation}
U =\Phi - \frac{1}{2}u^2
\end{equation}
and
\begin{equation}
\sigma_u^2= \frac{1}{\rho}\int_0^{\Phi} \rho \rmd\Phi. \label{eq11.7} 
\end{equation}
In (\ref{eq11.6}) $u$ is one of the meridional velocity components. If we choose
$u$ in $z$-direction, Eqs.~(\ref{eq11.6}) and (\ref{eq11.7}) give us the distribution
function and the dispersion of $z$-component of the velocities, depending
 only marginally on the triaxiality of the velocity
distribution for flat models. 

In the present paper we used the potential and the mass density from
the family of models with $n=3$ by \citep{Kuzmin:1956a}.
These models allow for the existence of a third integral, the property we
were not able to use until now. Besides, this family of models has another
remarkable property: if suitably choosing the units, one can express
$\rho (\Phi ,R)$ and thus also the symmetrical part of the phase
density with the same functions for the whole family. The function
$\rho (\Phi ,R)$ may be expressed in a finite form [Appendix B]
\begin{equation}
\rho = \Phi^4 ~ \frac{2-R^2\Phi^2-\Phi\sqrt{1-R^2\Phi^2} }{
(\sqrt{1-R^2\Phi^2}-\Phi )^3} \label{eq11.8} 
\end{equation}

According to (\ref{eq11.8}), $\rho$ increases with $\Phi$ from zero for $\Phi =0$ to
infinity when the denominator vanishes. For powers $\kappa$ and $\nu$,
which characterise the increase of $\rho$ for small $\Phi$ and $\Delta\Phi$,
we have
\begin{equation}
\kappa = 4, ~~~~~ \nu = 3. 
\end{equation}

From the series (\ref{eq11.4}) one can conclude that the symmetrical part of the
phase density in present case is non-negative, \ie  has a
physical meaning [q.v. Appendix B].

To derive the solution we used all the series referred above. They enabled
us to calculate the symmetrical part of the phase density for small $E$, for
small and large $I$ and for small $\Delta E$. For intermediate values of $E,I$
the series converge badly or are completely useless. But for
these regions, it was possible to derive the functions we need quite
reliably via interpolation. Thereafter we moved from the symmetrical part
of phase density to the total phase density. For small $\Delta E$ the
solution was already known, for remaining regions of $E,I$, it was possible to derive the solution with a certain amount of arbitrarity by assuming
the phase density to be a monotonically increasing function of $I$. [q.v.
Appendix C.]

\begin{figure*}[h]
\centering
\resizebox{.5\columnwidth}{!}{\includegraphics{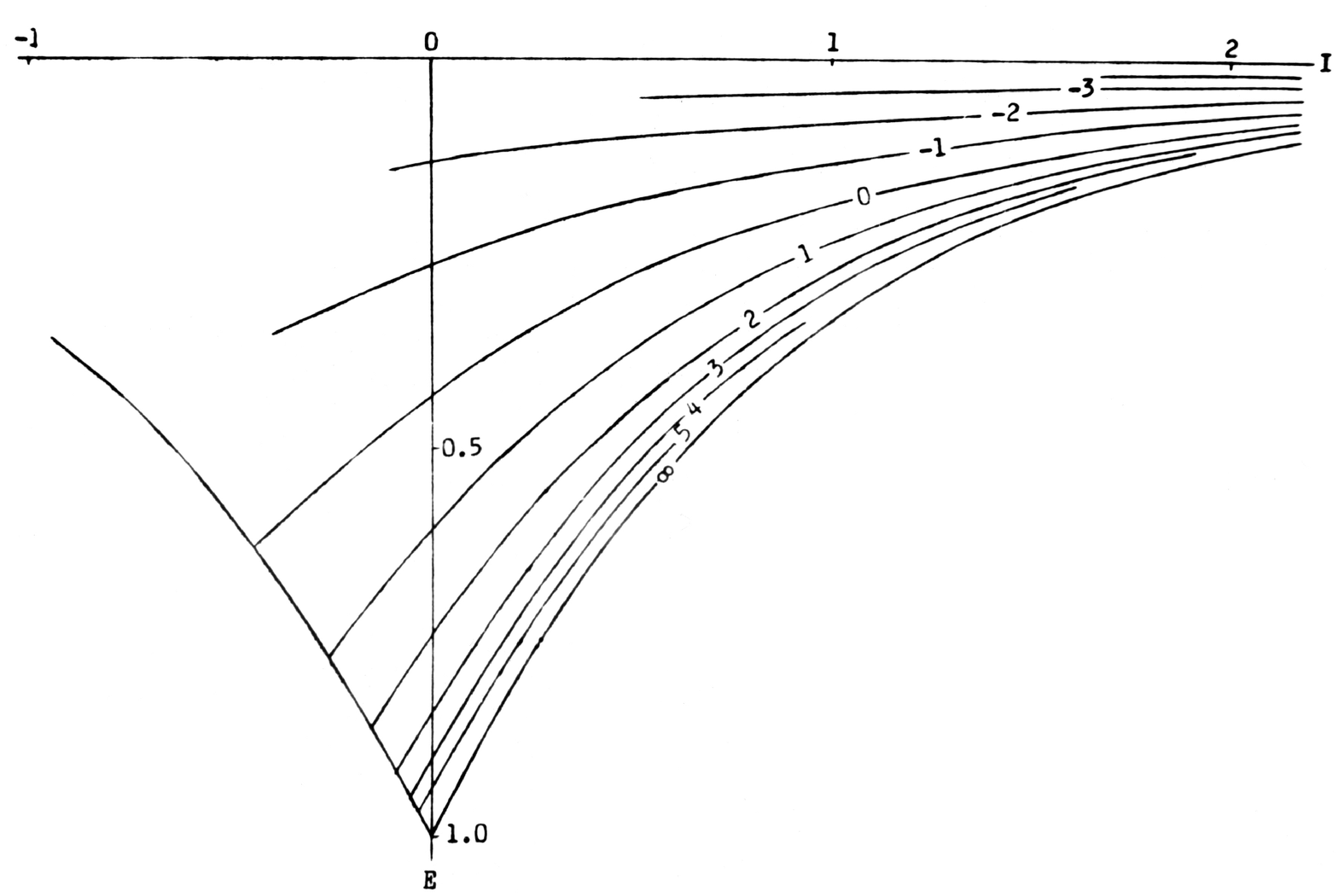}}
\caption{}
\label{fig11.1}
\end{figure*}

The results are given in Fig.~(\ref{fig11.1}) in form of logarithm of the phase density.
The diagram is the Lindblad diagram for the present family of models. With
the help of the diagram it is possible to calculate the velocity distribution for all
models of the family at every point of the coordinate space. For this we must
use the expressions for the integrals of motion and for
the potential as the function of coordinates for a concrete model of interest. 
[q.v. Appendix D.]

From the family of models a model suitable for our Galaxy has been chosen.
To choose the model parameters we used the data on 
the rotation of the Galaxy and the data on the gradient of the
acceleration in $z$-direction in the vicinity of the Sun. This data was
compared with theoretical predictions of the $n=3$ model.

For the resulting model the circular velocity in the vicinity of the Sun
is 190~km/s,\footnote{According to the recent data the circular
velocity in the vicinity of the Sun is 250~km/s. Therefore, the flatness
of the model is not sufficient and the velocity dispersion when compared
with the circular velocity is too large. [Later footnote.]}
the ratio $A/\omega = 0.64$, the dispersion of the meridional
component of the velocity is 23~km/s. Approximate calculation of the
total $z$-component velocity dispersion, when taking into account
all Galactic subsystems, gives the similar value. To not have too diminished
dispersion in $R$-direction due to the biaxial 
velocity distribution, we made our model slightly thicker. For this
reason the meridional dispersion was increased up to 31~km/s, the
mean value of real dispersions in $R$ and $z$ directions.

\begin{figure*}[th]
\centering
\resizebox{.5\columnwidth}{!}{\includegraphics{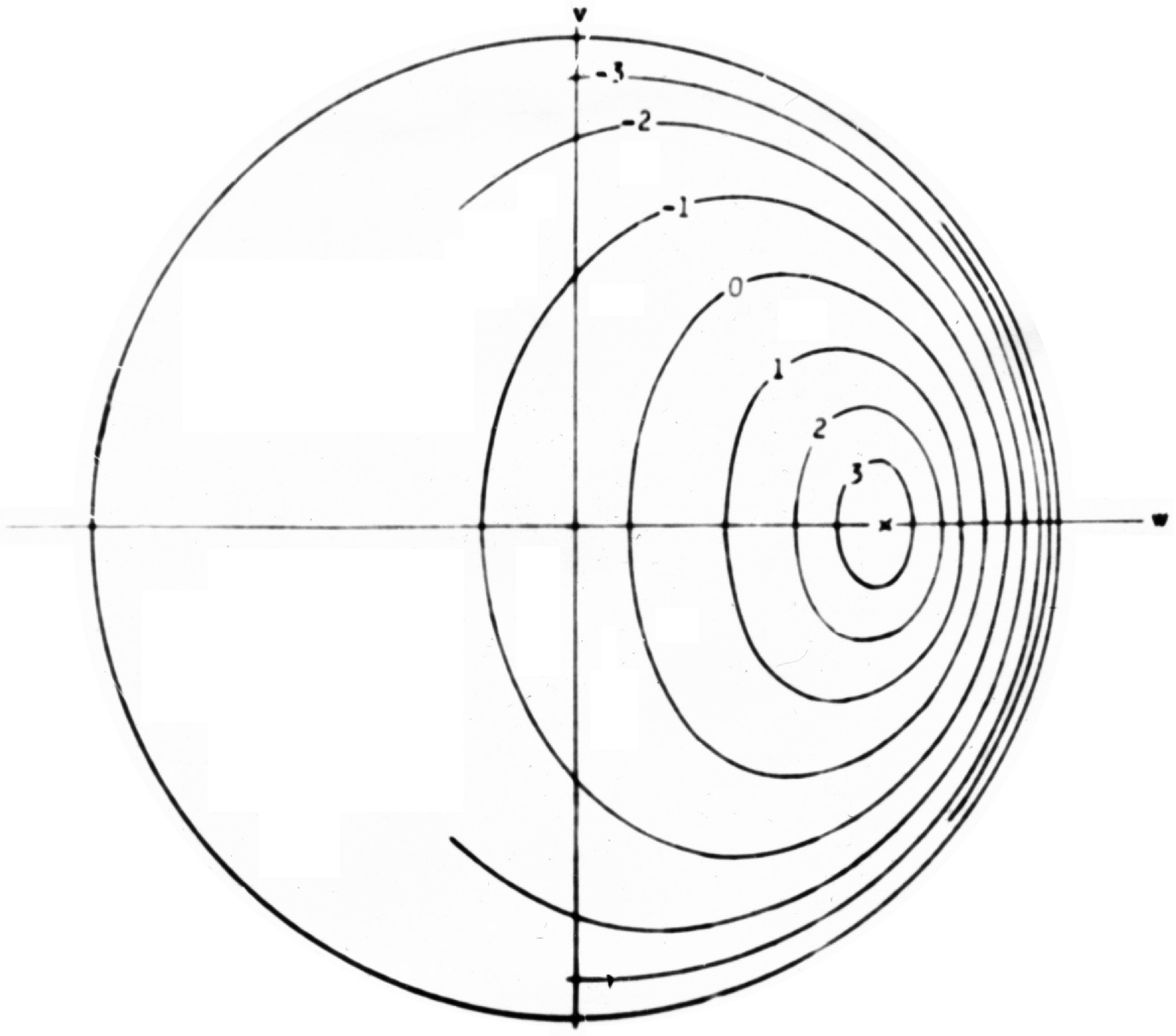}}
\caption{}
\label{fig11.2}
\end{figure*}

In Fig.~(\ref{fig11.2}) we give the isocurves of the logarithm of the phase density
in the plane $u=0$ in velocity space in the solar neighbourhood for our
thickened model. The circular velocity is marked by a cross. As can be
seen, the theoretical velocity distribution is remarkably similar to
the observed total distribution for all Galactic subsystems.
Some evident asymmetry is superposed to the general ellipsoidality. In
addition, there is a ``cutoff'' in the side of large velocities. The
limit, where the phase density becomes very small, 
is smaller than the escape velocity, and corresponds roughly to the
Oort limit (circular velocity +65~km/s). It is interesting to mention that
despite the fact that the model mass density decreases with radius
quite slowly, the cut velocity distribution is clearly seen.

\begin{figure*}[th]
\centering
\resizebox{.5\columnwidth}{!}{\includegraphics{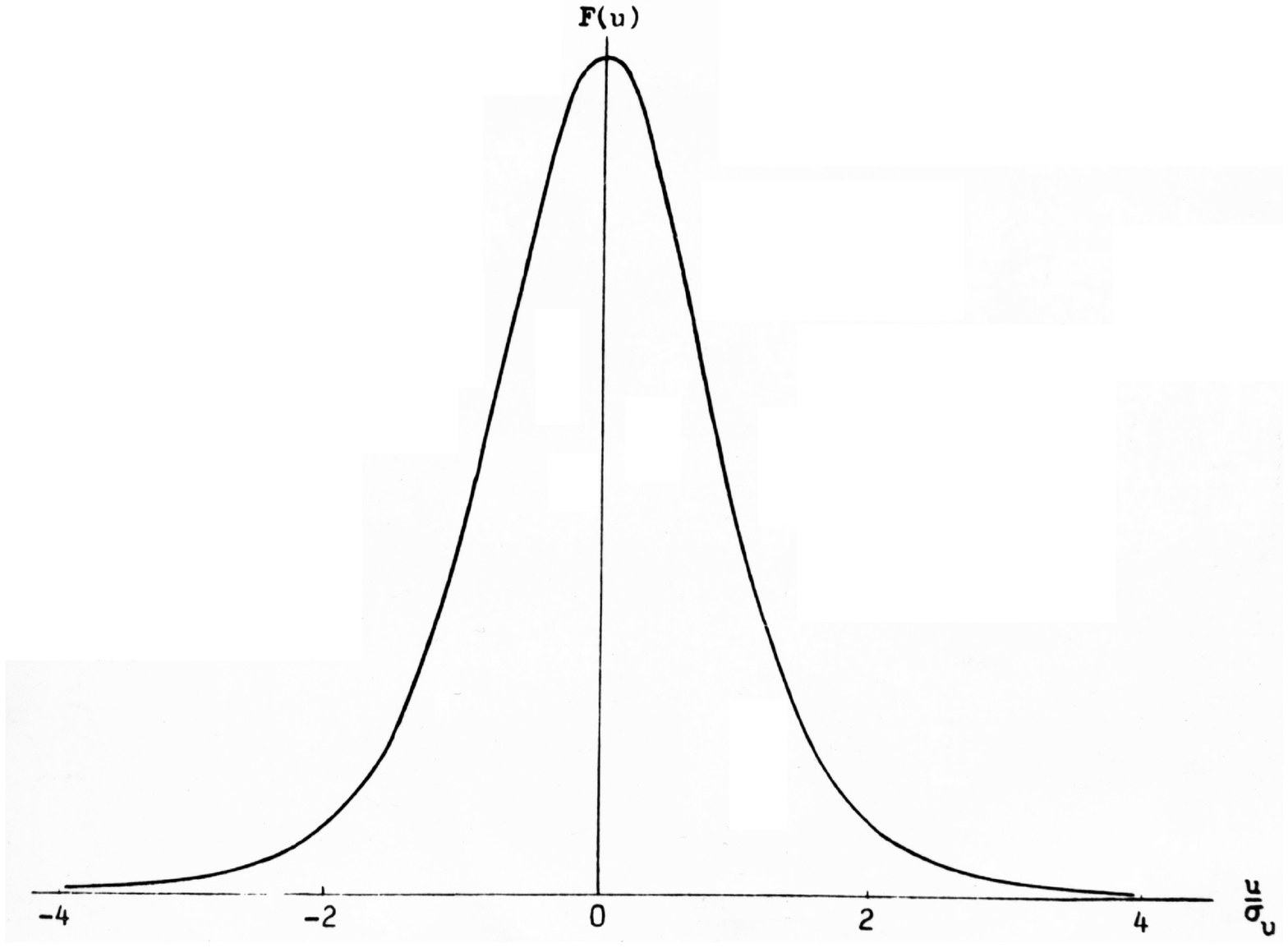}}
\caption{}
\label{fig11.3}
\end{figure*}

In Fig.~\ref{fig11.3} the distribution curve for the meridional velocity component in
the vicinity of the Sun is plotted. The distribution differs 
significantly from gaussian and corresponds roughly to the observed total
distribution of the velocity $R$ and $z$ components.

However, despite the general similarity
between the theoretical and the observed velocity distributions, there
exist also differences. The asymmetry, cutoffs and excess of the
theoretical distribution are too small. In other words, the flat and the
spherical components of the Galaxy are represented too little in
the theoretical model, and the intermediate component too much. Better
agreement may be expected for the model with $n<3$. The further project is
therefore to study these classes of models. And it would be
desirable to have a solution for the phase density as a function of
three integrals. [q.v. Appendix G]

We also analysed the limiting case of the model with $n=3$, when the model
approaches the spherical one and $\Psi$ is a function of $E$ alone. The
resulting spherical model is identical to the isochrone model by \citet{Henon:1959,Henon:1960a}
 [Appendix E]. In the central parts of the model the velocity
distribution is nearly Maxwellian. It is possible that this model
corresponds more to real spherical stellar systems than the well-known
Schuster density distribution model.

\vglue 3mm 
\hfill September 1960

\vglue 5mm

{\bf\Large Appendices added in 1969}
\vglue 5mm

\section{A.  The series for the density and the phase density}

$\mathrm{1.^o}$~  Substituting the series (\ref{eq11.4}) for the density and the phase
density 
\begin{equation}
\rho = \Phi^{\kappa} \sum^{\infty}_{k=0} p_k(\Phi R^2) \Phi^k ,
~~~~~ \Psi = E^{\kappa -3/2} \sum^{\infty}_{k=0} P_k(I^2) E^k  \label{eq11.A1.1} 
\end{equation}
into (\ref{eq11.1}) we obtain that the polynomial coefficients
\begin{equation}
p_k(\Phi R^2) = \sum^k_{l=0} a_{kl} (\Phi R^2)^l,
~~~~~ P_k(I^2) = \sum^k_{l=0} A_{kl} I^{2l}  \label{eq11.A1.2} 
\end{equation}
are related by
\begin{equation}
a_{kl} = 4\sqrt{2}~\pi ~ \frac{2^l}{2l+1} B (\kappa +k - \frac{1}{2} , l +\frac{3}{2}) A_{kl}. \label{eq11.A1.3} 
\end{equation}

In addition to these expansions we have the series in powers of $R^2$ and $I^2$
\begin{equation}
\rho = \sum^{\infty}_{k=0} q_k(\Phi ) R^{2k}, ~~~~~ \Psi = \sum^{\infty}_{k=0} Q_k(E) I^{2k}, \label{eq11.A1.4} 
\end{equation}
where the functions $q_k(\Phi )$ are given via $Q_k(E)$
\begin{equation}
q_k(\Phi ) = 4\sqrt{2}~\pi ~ \frac{2^k }{ 2k+1} \int^{\Phi}_0 Q_k(E) (\Phi -E)^{k+1/2} \rmd E. \label{eq11.A1.5} 
\end{equation}
By solving this generalised Abel equation we find
\begin{equation}
Q_k(E) = \frac{1}{2\pi^2 |2k-1|!!} \int^E_0 q_k^{(k+2)}(\Phi ) (\Phi - E)^{-1/2} \rmd\Phi. \label{eq11.A1.6} 
\end{equation}

Further, we have the expansions in powers of $R^{-1}$ and $I^{-2}$
\begin{equation}
\rho = R^{-\kappa} \sum^{\infty}_{k=0} s_k(\Theta )R^{-k}, ~~~ \Theta = \Phi^2 R^2,
\end{equation}
\begin{equation}
\Psi = I^{-\kappa +3/2} \sum^{\infty}_{k=0} S_k(J) I^{-2k}, ~~~ J = 2 E I^2.   \label{eq11.A1.7} 
\end{equation}
At the same time it can be demonstrated that
\begin{equation}
\tilde{s}_k(\Theta ) = s_k(\Theta )\Theta^{-\delta} = \frac{2\pi}{(l+\frac{1}{2})l} \int^{\Theta}_0 S^{(l)}(J) ~J^{-l-2\delta -1} ~(\Theta - J)^{l+1/2} ~\rmd J, \label{eq11.A1.8} 
\end{equation}
where
\begin{equation}
l+\delta = \frac{\kappa +k }{2} - 1, ~~~ \delta = 0, \frac{1}{2}.  \label{eq11.A1.9}
\end{equation} 
By solving the generalised Abel equation we find
\begin{equation}
S^{(l)}(J) = \frac{1}{\pi^2} J^{l+2\delta +1} \int^J_0 \tilde{s}_k^{(l+2)}(\Theta ) ~(J-\Theta )^{-1/2} ~\rmd\Theta . \label{eq11.A1.10} 
\end{equation}
Therefore, $S(J)$ can be derived with $l$-times integration (can be done in general form analytically).
\vglue 3mm

$\mathrm{2.^o}$~ In the series (\ref{eq11.5})
\begin{equation}
\Delta R = \Phi_0 (R) -\Phi , ~~~ \Delta E = E(I) - E, \label{eq11.A1.11} 
\end{equation}
and it is assumed that the density and the phase density will be
infinite at the curve $\Phi = \Phi_0 (R)$ on the $\Phi ,R$-plane and
at the curve $E = E_0(I)$ on the $E,I$-plane (Lindblad diagrams),
respectively. The curve $E = E_0(I)$ is an enveloping curve for parabolas
\begin{equation}
v^2 \equiv 2(\Phi -E) - I^2 R^{-2} = 0  \label{eq11.A1.12} 
\end{equation}
corresponding to $\Phi = \Phi_0 (R).$ Thus the parametric
equations of the curve are
\begin{equation}
E_0 = \Phi_0 (R) + {1\over 2} \Phi_0 '(R) R, ~~~ I^2_0 = - \Phi_0 '(R) R^3. \label{eq11.A1.13} 
\end{equation}

The derivation of the series (\ref{eq11.5})
\begin{equation}
\rho = \Delta\Phi^{-\nu} \sum h_k(R) \Delta\Phi^k , ~~~~~ \Psi \sim \Delta E^{-\nu -3/2} \sum_{k=0}^{k<\nu} H_k(I) \Delta E^k ,
\label{eq11.A1.14} 
\end{equation} 
is quite long. We derived the relation
\begin{equation}
h_k(R) = 4\sqrt{2}~\pi \sum^k_{i=0} \sum^{2i}_{j=0} B(\nu -k, i+\frac{3}{2}) ~g_{ij}(R) \frac{H^{(j)}_{k-i}(I_0(R)) }{ j!} , \label{eq11.A1.15} 
\end{equation}
enabling to find step by step $H_k(I).$ In this expression $g_{ij}$ are
given via derivatives of $E_0(I_0).$ In particular we have
\begin{equation}
g_{00}^2 = [R^2 E''_0(I_0(R)) + 1]^{-1} = \frac{1}{4} \frac{R\Phi_0 ''(R)}{ \Phi_0 '(R)} + \frac{3}{4} . \label{eq11.A1.16} 
\end{equation}

\section[B.  Expression for the density]{B.  Expression for the
  density. Specification of the series for the density and the phase
  density} 

$\mathrm{1.^o}$~ The expressions for the dimensionless potential and the
density used by us are (Chapter 6, the $n = 3$ model)
\begin{equation}
\Phi = \frac{1}{ \zeta_1+\zeta_2} , \label{eq11.A2.1} 
\end{equation}
\begin{equation}
\rho = \frac{(\zeta_1 +\zeta_2 )^2 + \zeta_1\zeta_2 + \zeta_1^2\zeta_2^2 }{ (\zeta_1 +\zeta_2 )^2 \zeta_1^3 \zeta_2^3} \label{eq11.A2.2}, 
\end{equation}
where dimensionless $R$ and $z$ are related with $\zeta_1$ and
$\zeta_2$ according to formulas
\begin{equation}
R^2 = (\zeta_1^2 -1)(1-\zeta_2^2 ), ~~~~~ z^2 = (\zeta_1^2 -\zeta_0^2 )(\zeta_2^2 -\zeta_0^2 ), \label{eq11.A2.3} 
\end{equation}
where $\zeta_0$ is a parameter characterising flatness of the
model. Expressing $\zeta_1$ and $\zeta_2$ via $\Phi$ and $R$, we
derive a formula (\ref{eq11.8})
\begin{equation}
\rho = \Phi^4 ~ \frac{2- R^2\Phi^2 -\Phi\sqrt{1-\Phi^2 R^2}}{ (\sqrt{1-\Phi^2 R^2} - \Phi )^3} . \label{eq11.A2.4} 
\end{equation}

$\mathrm{2.^o}$~ In the series for the density in powers of $\Phi$ for 
the polynomial $p_k(\Phi R^2)$, we find coefficients $a_{kl}$ after some
calculations 
\begin{equation}
a_{kl} = \frac{(k-l)(k+l+5)+4(l+1) }{2lB({k-l+1\over 2}, l)},
\label{eq11.A2.5} 
\end{equation}
Now we can calculate the polynomial $P_k(I^2)$ coefficients $A_{kl}$ for
the series of the phase density in powers of $E.$ 

Further, evidently
\begin{equation}
q_0(\Phi ) = \frac{\Phi^4}{ (1-\Phi )^3} (2-\Phi ). \label{eq11.A2.6} 
\end{equation}
This function gives us the density as a function of the potential on the
symmetry axis of the model. In general for the functions $q_k(\Phi )$ in
the series for the density in powers of $R^2$ we have the
expressions of following type
\begin{equation}
q_k(\Phi ) = \frac{\Phi^{4+2k} }{ (1-\Phi )^{3+k}} r_k(\Phi ), \label{eq11.A2.7}
\end{equation}
where $r_k(\Phi )$ is a polynomial of power $k+1$.

For the function $s_k(\Theta )$ in the series for the density in
powers of $R^{-1}$ we find $(\Theta = \Phi^2 R^2)$
\begin{equation}
s_k(\Theta ) = \frac{\Theta^{2+k/2} }{ (1-\Theta )^{3/2 +k/2}} (1+k)(2+ \frac{k}{2}-\Theta ). \label{eq11.2.8} 
\end{equation}

From the functions $q_k(\Phi )$ and $s_k(\Theta )$ we can move to the
functions $Q_k(E)$ and $S_k(J)$ in the series for the phase density
in powers of $I^2$ and $I^{-2}$. We derive the
expressions consisting of elementary functions but they are
quite long.

The function $Q_0(E)$ being the phase density in case of $I=0$
determines the spherically symmetric velocity distribution on the
symmetry axis of the model.
\smallskip

$\mathrm{3.^o}$~ For the function $\Phi_0 (R)$ we have the expression
\begin{equation}
\Phi_0 (R) = \frac{1}{\sqrt{1+R^2}}. \label{eq11.A2.9} 
\end{equation}
This is the potential on the symmetry axis for the model $\zeta_0
\rightarrow 0,$ \ie  for the flat model. 

For the curve $E=E_0(I)$ on the Lindblad $E,I$ diagram we derive the
parametric equations
\begin{equation}
E_0= \frac{2+R^2 }{ 2(1+R^2)^{3/2}}, ~~~~~ I_0^2 = \frac{R^4}{ 
(1+R^2)^{3/2}}. \label{eq11.A2.10} 
\end{equation}
This curve is the enveloping curve for the characteristic parabolas on
the Lindblad diagram for the flat model. The curve consist of two
parts $I\ge 0$ and $I\le 0$, intersecting at point $E=1,$ $I=0.$ In
the expansion of the phase density in powers of $\Delta E$ we count
$\Delta E$ from one branch of the enveloping curve, let say from the
branch $I\ge 0$ (if the stellar system rotates in direction of
positive $I$). In order to obtain a solution also for negative $I$,
the formally $R^2<0$ needs to be assumed, enabling the enveloping
curve to continue beyond the intersection point $E=1,$ $I=0$.

Knowing the equation for the enveloping curve we can find a function
$g_{ij}(R)$ in (\ref{eq11.A1.15}). In particular
\begin{equation}
g_{00} = \frac{1}{2} \sqrt{\frac{4+R^2 }{ 1+R^2}}. \label{eq11.A2.11} 
\end{equation}
This function has the meaning of the axial ratio of velocity ellipsoid for
nearly circular orbits in case of flat model.

Because $\nu =3$ and in the expansion of the phase density in powers of
$\Delta E$ we can obtain only three terms, only three functions $h_k(R)$
are interesting in the expansion of the density in powers of $\Delta\Phi$.
In particular, we find 
\begin{equation}
h_0(R) = \frac{1}{ (1+R^2)^5}, \label{eq11.A2.12} 
\end{equation}
giving immediately
\begin{equation}
H_0(I_0(R)) = \frac{105}{32\sqrt{2}\pi} \frac{1}{ (4+R^2)^{1/2} 
(1+R^2)^{9/2}}. \label{eq11.A2.13}
\end{equation}
Expressions for $H_1(I)$ and $H_2(I)$ are longer.

\section{C.  The total phase density}

 The expansions for the phase density discussed above, with the exception 
of the one in powers of $\Delta E$, give us only the even part of the phase
density with respect to $I$. The odd part remains quite arbitrary. One of the
possibilities to calculate the total phase density is to interpolate
the coefficients $A_{kl}.$ For the even part of the phase density we have the
expansion
\begin{equation}
\Psi_{\mathrm{even}} = E^{5/2} \sum^{\infty}_{k=0} \sum^{\infty}_{l=0} 
A_{k+l,l} E^k (EI^2)^l  \label{eq11.A3.1} 
\end{equation}
(in this formula $\kappa = 4$). In order to have
odd terms being in ``equal rights'' with even terms in this expansion, it is natural to
suppose the expression for the total phase density
\begin{equation}
\Psi = E^{5/2} \sum^{\infty}_{k=0} \sum^{\infty}_{m=0}  A_{k+m/2, m/2} E^k (E I^2)^{m/2} . \label{eq11.A3.2} 
\end{equation}
As we have the general expression for $A_{kl}$, which is valid also for 
fractional number indices, the coefficients necessary for calculations
result in simple way.

 \section{D.  Projections of the phase density. Velocity dispersion}

Simple expression can be derived for the projection of the phase
density in the meridional plane. Designating the projection as $\varphi$
we have after some calculations
\begin{equation}
\varphi = \varphi (V,R) = \frac{1}{2\pi} \frac{\partial\rho}{\partial\Phi}
\bigg|_{\Phi =V} , \label{eq11.A4.1} 
\end{equation}
where 
\begin{equation}
V = \Phi - \frac{v^2}{2}, \label{eq11.A4.2} 
\end{equation}
and $v$ is the total meridional velocity.

For the projection of the phase density $f$ onto some axis in the
meridional plane we derive the equation (\ref{eq11.6}),
\begin{equation}
f = f(U,R) = \frac{1}{\pi} \int^U_0 \frac{\partial\rho}{\partial\Phi}  \frac{\rmd \Phi}{\sqrt{2(U-\Phi )}} , \label{eq11.A4.3} 
\end{equation}
where
\begin{equation}
U=\Phi - \frac{1}{2}u^2 \label{eq11.A4.4}
\end{equation}
and $u$ is the orthogonal component of $v$.

For the dispersion of $u$ we find
\begin{equation}
\sigma^2_u = \frac{1}{\rho} \int^{\Phi}_{0} \rho \rmd\Phi. \label{eq11.A4.5} 
\end{equation}
This formula can be derived also from the second Jeans
hydrodynamic equation.

\begin{figure*}[t]
\centering
\resizebox{.6\columnwidth}{!}{\includegraphics{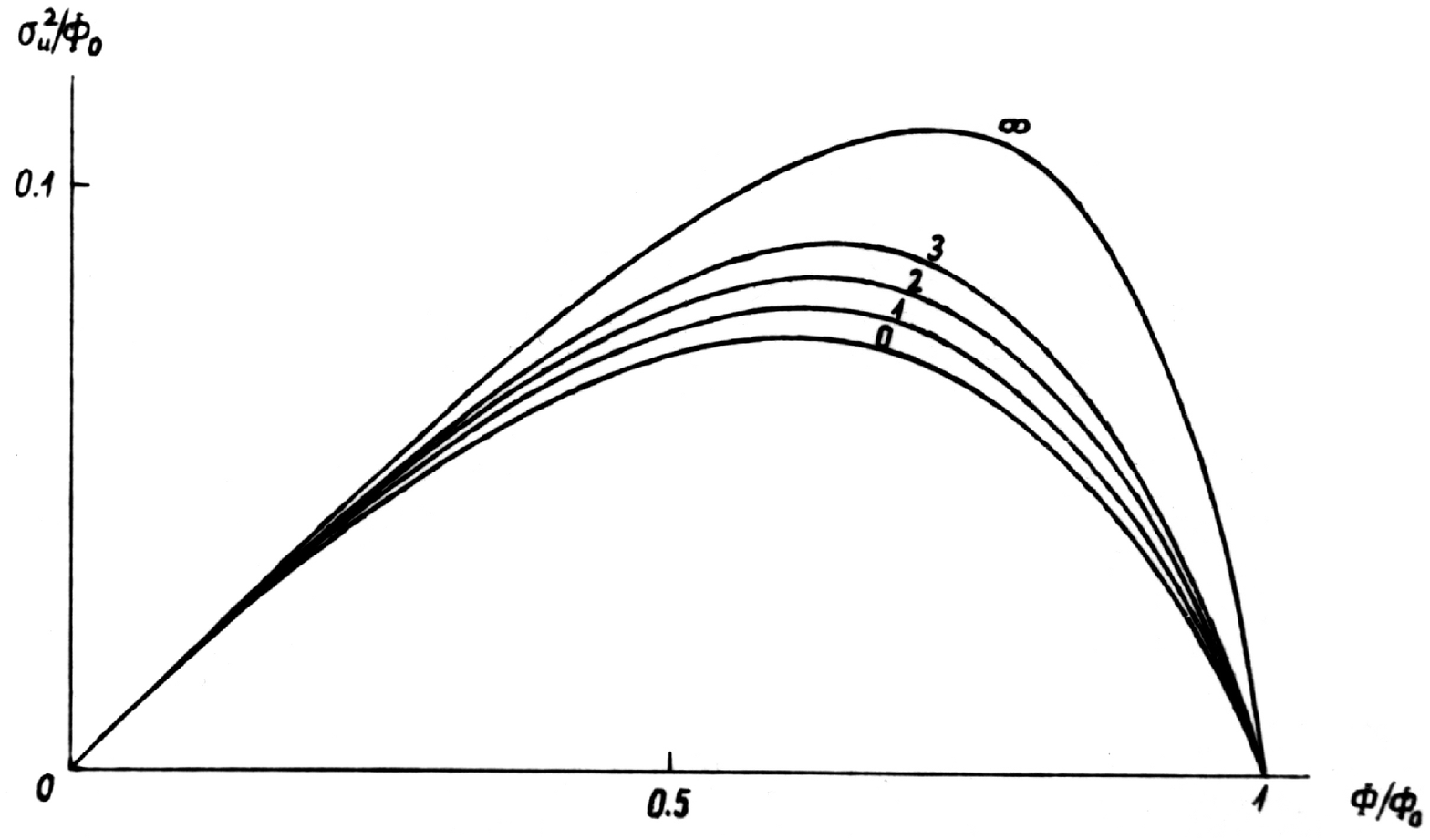}}
\caption{}
\label{fig11.D1}
\end{figure*}

Fig.~\ref{fig11.D1} shows the dispersion $\sigma_u$ versus
potential for our models. For all $R$ the unit of $\sigma_u^2$ and $\Phi$
is $\Phi_0 (R).$

\section{E.  Spherical model}

In case of $\zeta_0\rightarrow 1$ our model becomes spherical. As it
was mentioned in Appendix A for Chapter 6, the result is the isochrone
model by \citet{Henon:1959, Henon:1960a} with the dimensionless potential and density
\begin{equation}
\Phi = \frac{1}{1+\zeta}, \label{eq11.A5.1}
\end{equation}
\begin{equation}
\rho = \frac{1+2\zeta }{ \zeta^3 (1+\zeta )^2}, \label{eq11.A5.2} 
\end{equation}
where
\begin{equation}
\zeta = \sqrt{1+r^2}, \label{eq11.A5.3} 
\end{equation}
and $r$ is dimensionless distance from the centre of the model.

For the phase density of the model we have
\begin{equation}
\Psi = Q_0(E), ~~~ E\le \frac{1}{2} . \label{eq11.A5.4} 
\end{equation}
The expression that we find for $Q_0(E)$ coincides with the 
\citet{Henon:1960a}, q.v. \"U.-I. K. Veltmann,
Tartu Astr. Obs. Publ. {\bf 35}, 27, 1966 result.

\section{F.  The model by Lynden-Bell}

D. Lynden-Bell (M.N.R.A.S. {\bf 123}, 447, 1962) proposed the
potential 
\begin{equation}
\Phi = [(1+r^2)^2 - \eta R^2]^{-1/4}, \label{eq11.A6.1} 
\end{equation}
where $\eta$ is the parameter characterising the eccentricity of the
model. For $\eta = 0$ the model is spherical and coincides with the
Schuster one.

By expressing the density via $\Phi$ and $R$, we derive a very simple
equation  
\begin{equation}
\rho = (3-\eta )\Phi^5 + \frac{5}{2} \eta (4-\eta )\Phi^9 R^2. \label{eq11.A6.2}
\end{equation}
From this expression we can find the solution for the even part of the phase
density in form of two terms proportional to $E^{7/2}$ and $E^{13/2}I^2.$

Unfortunately, the model by Lynden-Bell can not be very flat. If $\eta >
\sim 0.8$ the model becomes ring-like, if $\eta > 3$ the model looses
physical meaning.

\section[G. Phase density as a function of three integrals of
  motion]{G.  Possibilities to construct the model of flat stellar
  system with the phase density as a function of three integrals of
  motion} 

The potential taken by us as the basis of mass distribution allows for
the existence of the third quadratic integral of motion. For that
model we can try to find the phase density as a function of three
integrals $I_1,$ $I_2,$ $I_3.$ In this case we have a model with
triaxial velocity distribution in accordance with observations.

In case of 
\begin{equation}
\Psi = \Psi (I_1,I_2,I_3), \label{eq11.A7.1} 
\end{equation}
we have the integral equation
\begin{equation}
\rho (R,z) = 4\iiint \Psi (I_1,I_2,I_3)  \frac{\partial (v_1,v_2,v_{\theta})}{\partial (I_1,I_2,I_3)}
\rmd I_1\rmd I_2 \rmd I_3 , \label{eq11.A7.2} 
\end{equation}
where $v_1,$ $v_2,$ $v_{\theta}$ are the velocity components in ellipsoidal
coordinates $\xi_1,$ $\xi_2,$ $\theta$ (arbitrary
orthogonal velocity components may be used). The factor 4 appears because
for given 
$I_1,$ $I_2,$ $I_3$ correspond four points in velocity space.

If three integrals $I_1,$ $I_2,$ $I_3$ have the forms discussed in Chapter
6 the Jacobian is
\begin{equation}
\frac{\partial (v_1,v_2,v_{\theta})}{\partial (I_1,I_2,I_3)} = \frac{1}{4w_1w_2} , \label{eq11.A7.3} 
\end{equation}
where
\begin{equation}
w^2 = \xi^2 (\xi^2 -1)I_1 - \xi^2 I_2^2 - (\xi^2 -1)I_3 +  2(\xi^2 -1)\varphi (\xi ) \label{eq11.A7.4} 
\end{equation}
(the notations $w$ and $\varphi$ have no relation with the same
notations used above). Integrals $I_1,$ $I_2,$ $I_3$ may be chosen
also in some other form. In this case the Jacobian also changes.

The integral equation does not enable to determine the phase density (the even
part of it) uniquely. For this reason we must restrict our
phase density in a way that a function we are looking for is a
function of only two integrals. This can be done by assuming
\begin{equation}
\Psi (I_1,I_2,I_3) = F(I_1,I_2,I_3) \psi (I_1,I_2), \label{eq11.A7.5} 
\end{equation}
where $F(I_1,I_2,I_3)$ is a given function and $\psi (I_1,I_2)$ is
unknown function. The integral equation now has the form
\begin{equation}
\rho (R,z) = \iint K(I_1,I_2;R,z) \psi (I_1,I_2) \rmd I_1 \rmd I_2, 
\label{eq11.A7.6}
\end{equation}
with the kernel
\begin{equation}
K(I_1,I_2;R,z) = 4\int \frac{\partial (v_1,v_2,v_{\theta})}{\partial (I_1,I_2,I_3)} F(I_1,I_2,I_3) \rmd I_3. \label{eq11.A7.7} 
\end{equation}
With respect to the even part of the function $\psi (I_1,I_2)$ this equation
gives a unique solution and determines the even part of 
phase density.

In the simplest Jeans approximation
\begin{equation}
F(I_1,I_2,I_3) = \mathrm{const} = 1 \label{eq11.A7.8} 
\end{equation}
and the kernel has the form
\begin{equation}
K(I_1,I_2;R,z) = \left\{
\ba{ll}
\pi z_0/R, & \mathrm{for} R^2(I_1+2\Phi )\ge z_0^2I_2^2,\\
\noalign{\smallskip}
0, & \mathrm{for} R^2(I_1+2\Phi ) <  z_0^2I_2^2.\\
\ea 
\right.
\label{eq11.A7.9} 
\end{equation}
The integral equation reduces to (\ref{eq11.1}) ($I_1 = -2E,$ $I=z_0I_2$).

One possibility to solve non-Jeans approximation is to introduce instead of
the energy integral $I_1$ an integral
\begin{equation}
I'_1 = \frac{I_1 }{ 1+pI_3} \label{eq11.A7.10} 
\end{equation}
and to take
\begin{equation}
F(I'_1,I_2,I_3) = \frac{1}{1+pI_3} \label{eq11.A7.11} 
\end{equation}
or in other words to take the phase density in form
\begin{equation}
\Psi (I_1,I_2,I_3) = \frac{1}{1+pI_3} \psi \left( \frac{I_1}{1+pI_3},I_2 \right). 
\label{eq11.A7.12} 
\end{equation}
Here $p$ is the parameter characterising the elongation of the
velocity distribution along $\xi_1$-coordinate. For different $I_2$ it may
be different.

When compared with Eq~\ref{eq11.A7.3} an additional factor $(1+pI_3)$ appears
now in Jacobian, it reduces with $F$ and after integration we find
\begin{equation}
K(I'_1,I_2;R,z) = \left\{
\ba{ll}
\frac{\pi z_0}{R} \frac{1}{\sqrt{(1-pI'_1\xi^2_1 )(1-pI'_1\xi^2_2 )}}, & 
\mathrm{for} I^0_3(I'_1,I_2;\xi_1 )\ge I^0_3(I'_1,I_2;\xi_2 ),\\
\noalign{\smallskip}
0, & \mathrm{for} I^0_3(I'_1,I_2;\xi_1 )<   I^0_3(I'_1,I_2;\xi_2 ),\\
\ea
\right. 
\label{eq11.A7.13} 
\end{equation}
where
\begin{equation}
I^0_3(I'_1,I_2;\xi ) = \frac{\xi^2 (\xi^2 -1)I'_1 - \xi^2 I_2^2 + 2(\xi^2 -1) \varphi (\xi ) }{ (\xi^2 -1)(1-pI'_1\xi^2 )} \label{eq11.A7.14} 
\end{equation}
is the value of $I_3$ in the case where $w$ vanishes.

In Jeans approximation, the curve discriminating the
regions in $I_1,I_2$ diagram where $k>0$ from the regions where $k\equiv 0$ is parabola. In
non-Jeans case this curve in $I'_1,I_2$ diagram is the third-order curve or
has even more complicated form if $p$ depends on $I_2.$

The value of the parameter $p$ can be estimated by looking at the axial ratios
of nearly ellipsoidal isosurfaces of the phase density in the velocity
space near to its maximum. In case of sufficiently flat model for $z=0$ we
have the following expression for the ratio of $R$ and $z$ semi-axis of
ellipsoidal isosurfaces $k_z$
\begin{equation}
k_z^2 -1 = p \frac{2+R^2 }{\sqrt{1+R^2}} 
\end{equation}
(when using of dimensionless $R$ and dimensionless potential). As
near to the Sun $k_z \simeq$ 3.5, a realistic estimate for (dimensionless)
$p$ is $p\simeq$ 1 or a bit more. Surely, it will be better to assume
$p$ to be a function of $I_2$, giving that in the formula above $p$
will be a function of $R.$

\section{H.  A model of very flat self-gravitating stellar system}

The problem of the construction of the model of very flat
self-gravitating axisymmetric and stationary stellar system is soluble
quite simply by quadratures. The problem reduces into
one-dimensional problem of stellar systems (cf. Chapters 1, 3, 6).

In this case we may use the integrals of nearly-circular motion
\citep{Kuzmin:1961aa}. 
\begin{equation}
I_1 = v_R^2 +k^2_R(R) v^2_{\theta}, ~~~ I_2 = R, \label{eq11.A8.1} 
\end{equation}
where $k_R$ is axial ratio of the velocity ellipsoid
\begin{equation}
k_R^{-2} = \frac{1}{4} \frac{R\Phi ''_0(R) }{ \Phi '_0(R)} + \frac{3}{4}, \label{eq11.A8.2} 
\end{equation}
where $\Phi_0 (R)$ is the potential at $z=0$ and may be arbitrarily
given. For the third integral we may take the Oort-Lindblad integral
\begin{equation}
I_3 = v^2_z - 2[\Phi (R,z) - \Phi_0 (R)]. \label{eq11.A8.3} 
\end{equation}

Substituting 
\begin{equation}
\Psi = \Psi (I_1,I_3,R) \label{eq11.A8.4} 
\end{equation}
into the equation for the density we have
\begin{equation}
\rho (R,z) = \frac{\pi}{k_R} \iint \frac{\Psi (I_1,I_3,R)}{\sqrt{I_3+2(\Phi - \Phi_0 )}} \rmd I_1 \rmd I_2. \label{eq11.A8.5} 
\end{equation}
The density satisfies the Poisson's equation having in
present case the form
\begin{equation}
\frac{\partial^2\Phi}{\partial z^2} = -4\pi G\rho (R,z). \label{eq11.A8.6} 
\end{equation}

With the Poisson's equation $\rho (R,z)$ and $\Phi (R,z)-\Phi_0 (R)$ are
mutually determinable and one of them can be given arbitrarily. This is
the first solution of our problem. By knowing both functions we can
find $\rho$ as a function of $\Phi -\Phi_0$ and $R$, and thus we can begin to
solve the integral equation for the phase density. some restrictions must be imposed 
in order to have a unique
solution for the phase density. As an
example, we can introduce an integral $I'_1$
\begin{equation}
I'_1 = I_1 +k_z^2(R)I_3, \label{eq11.A8.7} 
\end{equation}
where the function $k_z(R)$ is an arbitrary function, and to suppose
\begin{equation}
\Psi = \Psi (I'_1). \label{eq11.A8.8} 
\end{equation}
In this case the velocity distribution is ellipsoidal with the axial
ratios of velocity ellipsoid $1 : k_R^{-1} : k_z^{-1}.$

The integral equation will have the form
\begin{equation}
\rho (R,\Phi -\Phi_0 ) = \frac{2\pi}{ k_Rk_z}  \int^{\infty}_{2k^2_z(\Phi -\Phi_0 )} \Psi (I'_1)\sqrt{I'_1 + 2k_z^2 
(\Phi -\Phi_0 )} \rmd I'_1. \label{eq11.A8.9} 
\end{equation}
This equation is a special case of generalised Abel equation and is
soluble by quadratures, so the model construction is completed.

Instead of the phase density we may find the projection of phase density
onto the $v_z$ axis
\begin{equation}
f(I_3,R) = \frac{\pi}{k_R} \int \Psi (I_1,I_3,R) \rmd I_1, \label{eq11.A8.10} 
\end{equation}
without any assumptions about the dependence of $\Psi$ on $I_1$ and $I_3$.
The integral equation for $f$
\begin{equation}
\rho (\Phi -\Phi_0,R)=\int^{\infty}_{2(\Phi -\Phi_0 )}
\frac{f(I_3,R) \rmd I_3 }{\sqrt{I_3+2(\Phi -\Phi_0 )}} \label{eq11.A8.11} 
\end{equation}
is again the Abel equation and is soluble by quadratures.

Another possibility to solve the problem is not to give $\Phi -\Phi_0$
or $\rho$ but to give $\Psi (I_1,I_3,R)$ or $f(I_3,R)$. In this case we
derive by quadratures $\rho (R,\Phi -\Phi_0 )$ and we solve
the Poisson equation in form
\begin{equation}
\frac{\partial^2\Phi}{\partial z^2} = 4\pi G\rho (\Phi -\Phi_0 ,R).
\label{eq11.A8.12} 
\end{equation}
The result is soluble by quadratures, and the problem is solved.

The construction of the self-gravitating model of very flat stellar system
was studied by \citet{Vandervoort:1967}, but his
equations were very complicated.

%% file: chapter12.tex
\chapter[On the virial theorem and its applications]{On the virial
  theorem and its applications.\footnote{\footnotetext ~~Tartu
    Astron. Observatory Publications, vol. 34, pp. 10--17, 1963.} }

\section{The virial theorem for a stationary stellar system}

The following quantity is called virial
\be
W = \sum \boldsymbol{r} \boldsymbol{f} , \label{eq12.1} 
\ee
where $\boldsymbol{r}$ is a position vector for a point mass and $\boldsymbol{f}$ is a
force vector acting on a point mass. Summation is over all points
belonging to the system.

Usually in the virial expression (\ref{eq12.1}), $\boldsymbol{r}\boldsymbol{f}$ is understood
as the scalar multiplication of vectors $\boldsymbol{r}$ and $\boldsymbol{f}$ (the Clausius
virial). But a natural generalisation is to interpret $\boldsymbol{r}\boldsymbol{f}$ as
a dyad, thus attributing the virial $W$ the tensor character.

In a stellar system $\boldsymbol{f}$ is the gravitational force acting on a star. If
$\boldsymbol{g}$ is the acceleration of a star, then
\be
W=\sum m \boldsymbol{r} \boldsymbol{g} , \label{eq12.2} 
\ee
where $m$ is stellar mass. By approximating a stellar system with the
continuous mass distribution, $W$ has the form
\be
W = \int \rho \boldsymbol{r} \boldsymbol{g} \rmd V , \label{eq12.3} 
\ee
where $\rho$ is the density and $\rmd V$ is volume element.

From the identity
$$(\boldsymbol{r}\boldsymbol{r})'' = 2(\boldsymbol{v}\boldsymbol{v} +\boldsymbol{r}\boldsymbol{g}_s) ,$$
where $\boldsymbol{v}$ is the velocity vector and $s$ is the symmetric part of
the tensor, results the tensor virial theorem for a stationary stellar
system\footnote{The 
method we used to derive Eq.~(\ref{eq11.4}) assumes finiteness of
$\overline{\boldsymbol{r}\boldsymbol{r}}$. However, Eq.~(\ref{eq11.4}) is valid also for infinite
$\overline{\boldsymbol{r} \boldsymbol{r}}$ (as an example, for spherical Schuster or
H\'enon models). For stationary systems we have the hydrodynamic
equations of stellar dynamics
$$\nabla ( \rho\overline{\boldsymbol{v}\boldsymbol{v}} ) = \rho\boldsymbol{g} , $$
where the averaging is done over volume element. After multiplication of the
equation by $\boldsymbol{r}$ and integration over the space occupied by the
stellar system we derive Eq.~(\ref{eq11.4}) ($W=W_s$, otherwise due to the
force momenta the system can not be stationary). [Later
footnote]}
\be
M\overline{\boldsymbol{v}\boldsymbol{v}} + W_s = 0 , \label{eq12.4} 
\ee
where $M$ is the mass of the system and bar designates the averaging.

By taking the trace of Eq.~(\ref{eq11.4}) we obtain the usual virial theorem for
stationary stellar systems. It is known, that the virial, according to its usual
meaning, \ie  the trace of the tensor $W$ for a self-gravitating 
systems\footnote{When having not a self-gravitating stellar system but some
kind of subsystem, then while using the virial theorem we must take into
account all the forces, \ie  the virial of our subsystem must include the
self-gravitation as well as the gravitation resulting from other subsystems.
[Later footnote]}
equals to its potential energy.

The virial theorem for stationary stellar systems may be applied also for
systems which are stationary in rotating coordinates, and for
non-self-gravitating systems (stationary perturbing field). In such cases
the virial tensor consists of parts resulting from 1) self-gravitation, 2)
tidal forces, 3) centrifugal forces, 4) Coriolis forces.

If the system has a rigid body rotation, the Coriolis' part of the virial
vanishes, $\overline{\boldsymbol{v}\boldsymbol{v}}$ is the velocity dispersion
tensor. 

\section{The virial of the inhomogeneous ellipsoid}

For inhomogeneous ellipsoid the isodensity surfaces are similar 
coaxial ellipsoids. Their semi-axis (we assume them to be directed along 
the coordinate axis $x_1$, $x_2$, $x_3$) are
$$a_1u, ~a_2u, ~a_3u $$
respectively, where $a_1$, $a_2$, $a_3$ are constants and $u$ depends on
the dimensions of isosurfaces. The density $\rho$ of inhomogeneous
ellipsoid is a function of $u$.

It is known, that the elementary ellipsoidal layer (between $u$ and $u + \rmd u$)
does not attract an inner mass point (Newton's theorem). Thus the
potential energy of the gravitational interaction between the elementary
ellipsoidal layer and an inner mass point is independent of the position
of the latter.

Similar theorem may be proved also for the virial: the virial of the
elementary ellipsoidal layer resulting from an inner mass point is
independent of the position of the point.

In order to prove this, let us draw a chord through a given point P until it
intersects with the ellipsoidal layer. Let the length and the direction of
the chord be vector $\boldsymbol{h}$. The space angle $\rmd\Omega$ around the chord cuts
out two masses from the layer. The total virial resulting from these two
masses is
\be
-G m \rho\cdot \frac{\boldsymbol{h}\boldsymbol{h} }{ h^2} \cdot h \rmd h \cdot \rmd\Omega , \label{eq12.5}
\ee
where $G$ is the gravitational constant, $m$ -- the mass of the point mass at
P, $\rmd h$ -- the length of the part of chord in ellipsoidal layer (it is the
same at both sides of the chord).

From the properties of elementary ellipsoidal layer results, that
the quantity $h \rmd h$ is independent of the position of P and depends only
on the direction of the chord. The same is valid also for $\boldsymbol{h}\boldsymbol{h}/h^2$. 

The virial of an ellipsoidal layer results after integration of Eq.~\ref{eq11.5}
over all directions of a hemisphere. The result is independent of the
position of point P, which was to be proved.

From this theorem can be concluded, that the virial of an elementary
ellipsoidal layer caused by inner masses is independent of their
distribution. Therefore, we can calculate the virial by assuming the inner
mass distribution to be homogeneous. It forms homogeneous ellipsoid.

The component of the gravitational attraction along the coordinate $x_i$
inside and on the surface of the homogeneous ellipsoid is
$$\pi G\alpha_i\rho x_i ,$$
where $\rho$ is the density of the homogeneous ellipsoid, $\alpha_i$ is a
coefficient depending on the axial ratios of the ellipsoid. The tables of
coefficients $\alpha_i$ were published by \citet{Mineur:1939AnAp}.

It is easy to find the virial of an ellipsoidal layer resulting from the inner mass
$M(u)$ by using the above equations for the components of acceleration. 
The outer mass does not contribute to the virial of the layer.

After the integration over all layers, we derive for the diagonal components of
tensor $W$ that
\be
W_{ii} = -\pi G\alpha_ia_i^2 \int_0^{\infty} M(u)\rho (u) u \rmd u .
\label{eq12.6}
\ee
For present choice of coordinates the non-diagonal elements equal to zero.

By taking into account the inhomogeneous distribution of the inner
mass we have\footnote{For a stellar system which is stationary in non-moving
coordinates from Eqs.~(\ref{eq11.6}) and (\ref{eq11.4}) we have
$$ \overline{v_i^2} = \frac{\alpha_i}{4} \frac{a_i^2}{a_1a_2a_3} G\left(
\frac{\overline{M(u)}}{u}\right) , $$
where averaging is taken with weights proportional to the mass of
elementary ellipsoidal layer. This formula is valid also for a subsystem,
if a subsystem and the system as a whole can be approximated by concentric
coaxial and similar inhomogeneous spheroids. In this case $M(u)$ means the
total inner mass. The averaging is done with weights proportional to the mass
of elementary ellipsoidal layer of the subsystem. [Later footnote]}
\be
M(u) = 4\pi a_1a_2a_3 \int_0^u \rho (u) u^2 \rmd u . \label{eq12.7} 
\ee
For $u\rightarrow\infty$ the mass $M(u)$ approaches the total mass $M$ of
the system.

Similar results were obtained by \citet{Wijk:1949}. He expected the results
to be only approximate but in fact they are precise.

If we introduce the ``van Wijk' mean''
\be
\bar{\rho} = \frac{3 \int_0^{\infty} \rho (u)u \rmd u \int_0^u \rho (u)u^2 \rmd u}{\int_0^{\infty}\rho (u)u^4 \rmd u } \label{eq12.8} 
\ee
(in designations being independent of the unit of $u$) we may write
Eq.~(\ref{eq11.6}) in form
\be
W_{ii} = -\pi G\alpha_i\overline{x_i^2}~\bar{\rho} M , \label{eq12.9} 
\ee
where $\overline{x_i^2}$ is the mean of $x_i^2$. The equation is
proportional to $a_i^2$.

\section{The ellipsoidal stellar system having a rigid body rotation} 

Let us approximate a stellar system with inhomogeneous ellipsoid. We
also assume that our system rotates about the axis $x_3$ like a rigid
body with an angular velocity $\omega$ and is stationary in a system of
coordinates rotating with the same velocity.

The gravitational virial is expressed by Eq.~(\ref{eq12.9}).

The components of the centrifugal acceleration are
$$\omega^2x_1, ~\omega^2x_2, ~0 .$$
Hence for the centrifugal virial we have
\be
W_{11}'=\omega^2\overline{x_1^2}M, ~~W_{22}'=\omega^2\overline{x_2^2}M, ~~W_{33}' = 0. \label{eq12.10} 
\ee
In total we have
\be
\ba{ll}
W_{11} = & -(\pi G\alpha_1\bar{\rho} -\omega^2) \overline{x_1^2}M, \\ 
\noalign{\smallskip}
W_{22} = & -(\pi G\alpha_2\bar{\rho} -\omega^2 )\overline{x_2^2}M, \\
\noalign{\smallskip}
W_{33} = & - \pi ~G~\alpha_3~\bar{\rho} ~\overline{x_3^2}~M . 
\ea
\label{eq12.11} 
\ee

If we assume a spherical velocity distribution
\be
\overline{v_1^2}=\overline{v_2^2} = \overline{v_3^2} = \sigma^2
\label{eq12.12} 
\ee
from the virial theorem we have
\be
(\pi G\alpha_1\bar{\rho} -\omega^2)\overline{x_1^2} = (\pi G\alpha_2\bar{\rho} -\omega^2)\overline{x_2^2} = \pi G\alpha_3\bar{\rho} ~\overline{x_3^2} = \sigma^2 \label{eq12.13} 
\ee

These equations are analogous to those determining the equilibrium of
homogeneous MacLaurin and Jacobi ellipsoids. But in the present case the
ellipsoids are inhomogeneous and the density of the usual homogeneous
ellipsoids of these types is replaced by van Wijk's mean density.

By using the results on the equilibrium of the homogeneous ellipsoids, we
have the following condition for the stationary stellar system having a
rigid body rotation and approximated with inhomogeneous ellipsoids
\be
\frac{\pi G\bar\rho}{\omega^2}\ge 2.22 .\label{eq12.14} 
\ee
This condition is satisfied by ellipsoids of revolution -- ``inhomogeneous
MacLaurin ellipsoids''. Equality in Eq.~(\ref{eq12.14}) corresponds to the ellipsoid
of revolution having the ratio of the polar axis to the equatorial one 
\be
\epsilon = 0.368 . \label{eq12.15} 
\ee

For
\be
\frac{\pi G\bar{\rho}}{\omega^2} \ge 2.67 \label{eq12.16}
\ee
triaxial ellipsoids are possible -- ``inhomogeneous Jacobi ellipsoids''.

The model of inhomogeneous ellipsoid with rigid body rotation can be
used to study the elliptical galaxies, galactic nuclei and probably
barred galaxies.

The idea of using the equilibrium ellipsoids in stellar dynamics was
formulated by \citet{Ogorodnikov:1958}.

\section{Ellipsoidal star cluster in a galaxy}

Let us assume a star cluster moving on a circular orbit in a symmetry
plane of an axially symmetric galaxy. We approximate the cluster with
inhomogeneous ellipsoid with the axes $x_1$ and $x_3$ always directed
along $R$ and $z$ coordinates of cylindrical coordinate system related
to a galaxy. We assume that the cluster is stationary and has a rigid
body rotation in these coordinates.

The components of tidal and centrifugal acceleration together are
$$\kappa_1x_1,~~~ 0,~~~ \kappa_3x_3 . $$
Here
\be
\kappa_1= \frac{\partial F_R}{\partial R} - \frac{F_R}{R} , ~~ \kappa_3 = \frac{\partial F_z}{\partial z} , \label{eq12.17} 
\ee
where $F_R$ and $F_z$ are the components of gravitational acceleration in
a galaxy along $R$ and $z$ respectively. Hence the components of the
virial due to tidal and centrifugal forces are
\be
W'_{11}=\kappa_1\overline{x_1^2}M, ~~W'_{22} =0, ~~W'_{33}=\kappa_3\overline{x_3^2}M . \label{eq12.18} 
\ee
By adding them to the components of gravitational virial, we have
\be
\ba{ll}
W_{11} = & -(\pi G\alpha_1\bar{\rho} -\kappa_1) \overline{x_1^2}M, \\ 
\noalign{\smallskip}
W_{22} = & -\pi G\alpha_2\bar{\rho} ~\overline{x_2^2}M, \\
\noalign{\smallskip}
W_{33} = & -(\pi G\alpha_3\bar{\rho} -\kappa_3)\overline{x_3^2}M. 
\ea
\label{eq12.19}
\ee

Under the assumption of spherical velocity distribution we find from the
virial theorem
\be
(\pi G\alpha_1\bar{\rho} -\kappa_1)\overline{x_1^2} =  \pi G\alpha_2\bar{\rho} ~\overline{x_2^2} = (\pi G\alpha_3\bar{\rho} -\kappa_3 ) \overline{x_3^2} = \sigma^2  \label{eq12.20} 
\ee

Similar results were derived by \citet{Mineur:1939AnAp} when he approximated a
cluster with homogeneous ellipsoid. He calculated the tables for the
solutions of the equations. Mineur' ellipsoids were generalised by \citet{Wijk:1949} for the case of inhomogeneous ellipsoids by introducing the mean 
$\bar{\rho}$.

On the basis of Mineur' results the condition for the existence of a
stationary cluster approximated by inhomogeneous ellipsoid is
\be
\frac{\pi G\bar{\rho}}{\kappa_1} \ge \chi \left( -\frac{\kappa_3}{\kappa_1}
\right) . \label{eq12.21} 
\ee
The function $\chi$ slowly rises with the argument, beginning from $\chi
=$ 3.56 for $\kappa_3/\kappa_1 = $ 0. Equality in Eq.~(\ref{eq12.21}) corresponds to
the axial ratio
\be
\frac{a_2}{a_1} = 0.51 . \label{eq12.22} 
\ee
The ratio $a_3/a_2$ decreases when $-\kappa_3/\kappa_1$ increases beginning
from one for $\kappa_3/\kappa_1 = 0$.

The parameters $\kappa_1$ and $\kappa_3$ are related to Oort-Kuzmin
parameters $A$, $B$, $C$ by
\be
\kappa_1 = 4A(A-B), ~~~\kappa_3 = -C^2. \label{eq12.23} 
\ee
In our Galaxy in the vicinity of the Sun $\sqrt{\kappa_1}$ is about 
50~km/s/kpc, $-\kappa_3/\kappa_1 \sim$ 2.0 (in Mineur' tables it
corresponds to $\chi =$ 4.2). Now we find the condition (\ref{eq12.21}) for the solar 
neighborhood\footnote{These results have to be revised by taking
into account new observational data. For the lowest value of $\bar{\rho}$ in the
vicinity of the Sun we find $3.5\cdot 10^{-23} {\rm g/cm^3}=$ $0.5 {\rm
M_{\odot}/pc^3}$ and the corresponding axial ratios of inhomogeneous
ellipsoid are~ 1 : 0.51 : 0.37.}
$$\bar{\rho}\ge 5\cdot 10^{-23} {\rm g/cm^3} = 0.8 {\rm M_{\odot}/pc^3} .$$
Equality corresponds to ellipsoid with axial ratios
$$a_1: a_2: a_3 = 1: 0.51: 0.40 .$$

This condition for $\bar{\rho}$ is probably a better stability condition for a
star cluster when compared with the well-known Bok's condition \citep{Bok:1934}. 
\vglue 3mm
\hfill March 1963

%% file: chapter13.tex
\chapter[On the existence of symmetry planes in stellar systems]{On
  the existence of symmetry planes in stellar
  systems.\footnote{\footnotetext ~~Tartu  Astron. Observatory
    Publications, vol. 34, 18 -- 37, 1963.}}

\section{Theorem on the existence of a symmetry plane}

According to the well-known theorem by \citet{Lichtenstein:1933} the rotating equilibrium configuration of inhomogeneous gravitating fluid has a symmetry plane perpendicular to the rotational axis. Within certain conditions the similar theorem can be used in case of stellar systems.

If we neglect the irregular forces, the phase density $\Psi$ becomes a function of integrals of regular motion of stars $I(\boldsymbol{r}, \boldsymbol{v}, t)$ ($\boldsymbol{r}$ -- position vector, $\boldsymbol{v}$ -- velocity vector, $t$ -- time). Let the gravitational potential of a system $\Phi$ belong to the family of potentials allowing the existence of the integrals of stellar motion in a form 
\be
I=I(\Phi ,\boldsymbol{r}, \boldsymbol{v}) \label{eq13.1}
\ee
where the dependence of $I$ on his arguments is the same for the whole family (we reject the argument $t$ as being unimportant). We assume that $\Psi$ is a function of these integrals. In this case the density of stars in the coordinate space, being integral of $\Psi$ over the velocity space,
is
\be
\rho = \rho (\Phi ,\boldsymbol{r}). \label{eq13.2} 
\ee

Let us suppose
\be
\frac{\partial\rho}{\partial\Phi}\ge 0 \label{eq13.3}
\ee
and that in a suitably chosen system of coordinates $\rho$ has a form
\be
\rho = \rho (\Phi ,x,y), \label{eq13.4} 
\ee
or
\be
\rho = \rho (\Phi ,x,y,z^2), \label{eq13.5} 
\ee
while
\be
\frac{\partial\rho}{\partial z^2} < 0. \label{eq13.6} 
\ee

For every isodensity surface, let us draw in $z$-direction a family of chords through the region where $\rho$ exceeds the value of $\rho$ on the isosurface. Let at the centre of chords $z=$ $z_m(\rho ,x,y)$ and $z_m$ have an upper and lower bound.

Chords can be treated as elementary, close to each other, homogeneous gravitating columns. In this case $\rho$ consists of their densities and $\Phi$ -- their potentials. The potential of a column is symmetric about the plane $z=z_m$ and decreases when moving away along $z$.

From Eqs.~(\ref{eq13.4}) and (\ref{eq13.3}) results that $\Phi$ is the same at both ends of every column. From $z_m$ being unbound and from the properties of column potential results that all $z_m$ must be identical (otherwise the equality of potentials at both ends of a column is not possible at least for columns with $z_m$ at upper and lower bounds of $z_m$). The equality of all $z_m$ means that the configuration has a symmetry plane at $z=$ $z_m= \mathrm{const}$.

If Eq.~(\ref{eq13.5}) is valid, the potential at both ends of columns is equal only when $z_m= 0$. If $z_m > 0$, from Eqs.~(\ref{eq13.3}) and (\ref{eq13.6}) results that at the ``upper'' end of the column the potential is larger than at the ``lower'' end. If $z_m< 0$ the situation is just the opposite; as the upper edge of $z_m$ is non-positive, and the lower edge is non-negative, $z_m = 0$ must hold. Therefore, we have a symmetry plane at $z= 0$.

\section{The case when the isosurfaces of the density and of the potential coincide}

Let us assume that the isosurfaces of $\rho$ and $\Phi$ coincide, \ie
\be
\rho = \rho (\Phi), \label{eq13.7}
\ee
and that Eq.~(\ref{eq13.3}) is valid. In that case from the theorem in previous section for any $z$-direction (if $z_m$ are bounded), we have a symmetry plane. This is possible only when the configuration is spherically symmetric. 

The conclusion about spherical symmetry is tightly related to our assumption about $z_m$ being unbound in any $z$-direction. Without this assumption other configurations may result from Eq.~(\ref{eq13.7}). But according to their properties they are infinite.

Examples of this kind of infinite configurations are the cylindrically symmetric and the plane parallel configurations. In both cases  the isosurfaces of $\rho$ and $\Phi$ coincide, being coaxial circular cylinders or parallel planes.

It is easy to demonstrate the existence of a symmetry plane for plane parallel configurations. One doesn't even need to use the theorem from previous section with the condition (\ref{eq13.3}). The existence of a symmetry plane follows from Poisson's equation
\be
\Phi ''(z)+4\pi G\rho (\Phi ) = 0 \label{eq13.8} 
\ee
($G$ is the gravitational constant). The solution of this equation is symmetrical about $(z-z_m)$, where $z_m$ is an arbitrarily given value of $z$ for which $\Phi '(z) =0$. 

\section{An application of the theorem to the existence of a symmetry plane}

$\mathrm{1.^o}$ ~Let us assume a stationary self-gravitating stellar system. From stationarity follows the existence of the integral of type (\ref{eq13.1}) in form of the energy integral
\be
I=v^2 -2\Phi . \label{eq13.9} 
\ee
Let us assume that $\Psi =\Psi (I)$. In this case
\be
\rho = 4\pi \int_0^{\infty} \Psi (I) v^2 \rmd v = \rho (\Phi ), \label{eq13.10}
\ee
and we find that
\be
\rho '(\Phi ) = 4\pi\int_0^{\infty} \Psi \rmd v \ge 0 .\label{eq13.11}
\ee
Now we may use the results from previous section. If the system is finite in the sense that $z_m$ is bounded for all $z$, we have the spherically symmetrical configuration.

For spherical symmetry there exist also other integrals of motion which may serve as arguments of the stationary phase density (the angular momentum integral). The fact that omission of other integrals leads us to the case where such integrals exist, proves that for any finite stationary self-gravitating stellar system there must exist at least two integrals. This fact was mentioned earlier by us \citep{Kuzmin:1953} and even before by \citet{Jeans:1919}.
\vglue 3mm

$\mathrm{2.^o}$ ~Let us have a self-gravitating system which is stationary in rotating coordinates. In this case the ordinary energy integral is replaced by Jacobi integral
\be
I = v^2-2\Phi -\omega^2R^2 , \label{eq13.12} 
\ee
where $v$ is the velocity in rotating coordinates, $\omega$ is the angular velocity and $R$ -- the distance from rotation axis. Let us assume again $\Psi =\Psi (I)$, then
\be
\rho = \rho (\Phi + \frac{1}{ 2}\omega^2R^2) .\label{eq13.13} 
\ee
Equation (\ref{eq13.3}) is still valid. From the theorem presented in Sect.~1 we conclude that such kind of finite system must have a symmetry plane perpendicular to the rotation axis. The existence of other symmetry planes does not follow from the theorem.
\vglue 3mm

$\mathrm{3.^o}$ ~Let us assume that a stationary system is not self-gravitating, but disturbed by tidal forces from other system. We limit ourselves to the quadratic terms of the perturbed potential (we assume it to be stationary as well). For a suitably chosen system of coordinates we have the integral 
\be
I=v^2-2\Phi -\sum_i \kappa_i x_i^2. \label{eq13.14}
\ee
Here the last term, where $\kappa_i$ are constants and $x_i$ are orthogonal coordinates, signifies the sum of centrifugal and tidal potentials. For $\Psi =\Psi (I)$ we have
\be
\rho = \rho (\Phi + \frac{1}{ 2}\sum_i\kappa_ix_i^2) \label{eq13.15}
\ee
and condition (\ref{eq13.3}). From the theorem in Sect.~1 follows that for $\kappa_i \le 0$ the configuration has the symmetry plane perpendicular to the axis $x_i$.

As an example, let us have a stationary star cluster moving on circular orbit in a stationary axisymmetric galaxy. In this case $\kappa_1 >0$, $\kappa_2 =0$, $\kappa_3<0$ (q.v. Chapter 12). Therefore, the cluster has the symmetry planes $x_2=\mathrm{const}$ and $x_3=0$. The existence of the third plane does not follow from the theorem.
\vglue 3mm

$\mathrm{4.^o}$ ~Let us have a stationary self-gravitating system with rotational symmetry. In this case we have two integrals of type (\ref{eq13.1}) -- the energy integral and the momentum integral
\be
I_1=v^2-2\Phi , ~~~I_2=Rv_{\theta} , \label{eq13.16} 
\ee
where $R$ is the distance from the symmetry axis, $v_{\theta}$ -- the velocity along $\theta$-coordinate in cylindrical coordinates. We assume $\Psi = \Psi (I_1,I_2)$. Then
\be
\rho = 2\pi \int_{-\infty}^{\infty} \int_{v_{\theta}}^{\infty} \Psi (I_1,I_2) v \rmd v \rmd v_{\theta} = \rho (\Phi ,R). \label{eq13.17}
\ee
We find that
\be
\frac{\partial\rho}{\partial\Phi} = 2\pi \int_{-\infty}^{\infty} \Psi_{v=v_{\theta}} \rmd v_{\theta} \ge 0. \label{eq13.18} 
\ee
From the theorem in Sect.~1 follows the existence of the symmetry plane perpendicular to the symmetry axis (we assume also that configuration is finite in the sense used here).

\section{The case of the third integral}

Let us assume that in case of rotational symmetry in addition to the integrals (\ref{eq13.16}) there exist one more argument in $\Psi$ -- the third integral in form of \citep{Kuzmin:1953} 
\be
I_3=(Rv_z-zv_R)^2+z^2v_{\theta}^2+z_0^2(v_z^2-2\Phi^*), \label{eq13.19}
\ee
where $v_R$, $v_{\theta}$, $v_z$ are the velocity components in cylindrical coordinates $R$, $\theta$, $z$, and $z_0$ is a constant. The integral exists if the potential of a system has Eddington-Camm form \citep{Eddington:1915, Camm:1941}
\be
\Phi = {\varphi (\xi_1) -\varphi (\xi_2) }{ \xi_1^2-\xi_2^2}, \label{eq13.20} 
\ee
and the function $\Phi^*$ has the form
\be
\Phi^*= \frac{\xi_2^2\varphi (\xi_1)-\xi_1^2\varphi (\xi_2) }{\xi_1^2-\xi_2^2}. \label{eq13.21} 
\ee
In these expressions $\varphi$ is an arbitrary continuous function with continuous first and second derivatives, $\xi_1$ and $\xi_2$ are elliptical coordinates related to $R$ and $z$ via 
\be
R=z_0\sqrt{(\xi_1^2-1)(1-\xi_2^2)},~~~ z=z_0\xi_1\xi_2  \label{eq13.22}
\ee
and $\xi_1\ge 1$, $|\xi_2|\le 1$. Isosurfaces of $\xi_1$ and $\xi_2$ are confocal ellipsoids and two-sheeted hyperboloids of revolution with the foci at $R=0$ and $z=\pm z_0$.

The existence of a symmetry plane can be proved with the aid of the Poisson's equation for $R=0$.

If $R=0$ then $z=\pm z_0\xi_1$ for $|z|\ge z_0$ and $z=z_0\xi_2$ for $|z| \le z_0$. Because $\varphi$ may have an arbitrary additive constant, we may take $\varphi (1)=0$. In this case for $R=0$
\be
\Phi_0(z)=\Phi_0^*(z)= \frac{z_0^2}{ z^2-z_0^2} \varphi\left( \frac{z}{ z_0}  \right) . \label{eq13.23} 
\ee
For $|z|\ge z_0$ the function $\varphi$ and therefore also $\Phi_0(z)$ must be even.

For $R=0$ the integrals have forms
\be
\ba{ll}
I_1 = & v_R^2+v_z^2-2\Phi_0(z), \\
\noalign{\smallskip} 
I_2 = & 0, \\
\noalign{\smallskip}
I_3 = & z^2v_R^2+z_0^2v_z^2 -2z_0^2\Phi_0(z) . 
\ea
\label{eq13.24}
\ee
Hence, for $\rho$ in case of $R=0$ we derive
\be
\rho_0=2\pi \int_{-\infty}^{\infty} \int_0^{\infty} \Psi (I_1,0,I_3) v_R \rmd v_R \rmd v_z = \rho_0(\Phi_0,z^2). \label{eq13.25}
\ee
For derivatives $\partial\Phi /\partial R$ and $\partial^2\Phi /\partial R^2$ in case of $R=0$ we find
\be
\left( \frac{\partial^2\Phi}{\partial R^2}\right)_0 = \left( \frac{\partial\Phi}{ R\partial R}\right)_0 = \frac{z}{ z^2-z_0^2}\Phi '_0(z) - \frac{2z_0^2 }{ (z^2-z_0^2)^2}  [\Phi_0(z)-\Phi_0(z_0)]. \label{eq13.26}
\ee
The Poisson's equation for $R=0$ results now
\be
\Phi ''_0(z) + \frac{2z}{ z^2-z_0^2}\Phi '_0(z) - \frac{4z_0^2 }{(z^2-z_0^2)^2} [\Phi_0(z)-\Phi_0(z_0)] + 4\pi G\rho_0(\Phi_0,z^2)=0. \label{eq13.27} 
\ee
As the functions $\Phi_0(z)$ and $\Phi_0'(z)$ are continuous, the solution of Eq.~(\ref{eq13.27}) may be uniquely continued through the singular points $z=\pm z_0$. If $|z|\ge z_0$ only the even solution of Eq.~(\ref{eq13.27}) is usable. By continuing the solution into the region $|z|<z_0$ we have the even solution there as well. Thus, the function $\varphi (\xi )$ is symmetric and has the symmetry plane at $z=0$.
\vglue 3mm

{\hfill March 1963}

%% file: chapter14.tex
\chapter[On the theory of integrals of stellar motion]{On the theory
  of integrals of stellar motion\footnote{\footnotetext ~~Published in
    Tartu Astron. Observatory Publications, vol. 34, pp. 457-481,
    1964}} 

{\bf Summary}
\bigskip

The integrals of motion are classified according to the degree of
completeness with which they determine the gravitational potential.

Integrals of the first kind are ``determining'':  if given in a concrete
form, they determine completely the corresponding gravitational
field. Example: the integral of energy.

Integrals of the second kind are ``cylindrical'': their equisurfaces in
the velocity space are parallel cylinders, non-circular in general. In
the coordinate space they are characterised by the vector lines which
determine the direction of the cylinders. If the characteristic vector
lines have orthogonal surfaces, the integral of the second kind is
``surface determining'': it determines completely the behaviour of the
potential on these surfaces. But the potential has an arbitrary term
whose equisurfaces coincide with the orthogonal surfaces. Example: the
two-dimensional energy integral. Example of a "``degenerate'' integral of
the second kind: the integral of total angular momentum.

Integrals of the third kind are ``plane'': their equisurfaces in the
velocity space are parallel planes. In coordinate space they are
characterised by vector lines which are orthogonal to these
planes. The integrals of such kind are ``line-determining'': they
determine completely the behaviour of the potential on the
characterising vector lines. The potential has an arbitrary term whose
equilines coincide with characterising vector lines. Example: the
one-dimensional energy integral. Example of a degenerate integral of
this kind: the integral of angular momentum.

The integrals of the
third kind imply minimal restrictions on the gravitational potential.

Various forms of the dependence of the integral on the velocities are considered.

In the case of the integrals of the third kind (``plane integrals'') the
characterising vector lines are equidistant: the mutual distances
between the neighbouring lines do not vary along the lines. If the
integral is conservative, it is reduced to the form 
\be
I = v_z^2 - 2 U(z) \label{eq14.2.17}
\ee
(one-dimensional energy integral) or to the form 
\be
I = R^2 v_\theta^2 - 2 U (\theta ) . \label{eq14.2.19}
\ee
Here $v_\theta$ and $v_z$ are velocities in cylindrical coordinates
and $U$ are functions of coordinates. The corresponding forms of the
potential are  
\be
\Phi = I(z) + V(x,y,t) \label{eq14.2.18}
\ee
and 
\be
\Phi = \frac{1}{R^2} U(\theta ) + V(R,z,t). \label{eq14.2.20}
\ee
Functions $V(x,y,t)$ and $V(R,z,t)$ are arbitrary and are not
determined by the integrals. The degenerate plane conservative
integrals are represented by the forms 
\be
I = z_z  \label{eq14.2.21}
\ee
and 
\be
I = R v_\theta + k v_z, \label{eq14.2.23}
\ee
to which the potential 
\be
\Phi = V(x,y,t) \label{eq14.2.22}
\ee
and the potential with helical symmetry 
\be
\Phi = V(R,z-k\theta , t) \label{eq14.2.24}
\ee
correspond, $k$ is a constant.

In real stellar systems, only the integral (\ref{eq14.2.23}) with $k=0$ can be
correct. Thus, the conservative integral which imposes minimal
restrictions on the potential and which can correspond to a real
stellar system is reduced to the integral of angular momentum and the
restrictions to the axial symmetry \citep[][theorem]{Idlis:1961}.

From among the integrals of the second kind only the ``circular
cylindrical'' conservative integrals are considered. They are reduced
to the form 
\be
I = v_x^2 + v_y^2 - 2 U(x,y), \label{eq14.3.16} 
\ee
(two-dimensional energy integral) and to the form
\be
I = (\boldsymbol{r} \times \boldsymbol{v} )^2 - 2 U \left(
  \frac{\boldsymbol{r}}{r} \right) \label{eq14.3.18}  
\ee
with corresponding potentials 
\be
\Phi = U(x,y) + V(z,t) \label{eq14.3.17} 
\ee
and 
\be
\Phi = \frac{1}{r^2} U\left( \frac{\boldsymbol{r}}{r} \right) + V(r,t). \label{eq14.3.19} 
\ee
If the integrals are degenerate, we have the forms 
\be
I = v_x^2 + v_y^2 , \label{eq14.3.20} 
\ee
\be
\Phi = V(z,t), \label{eq14.3.21}
\ee
\be
I = ( \boldsymbol{r} \times \boldsymbol{v} )^2 , \label{eq14.3.22}
\ee
\be
\Phi = V(r,t). \label{eq14.3.23}
\ee
Only the integral of total angular momentum (\ref{eq14.3.22}) can be correct for a real
stellar system.

From among the integrals of the first kind the ``spherical''
non-conservative integral is considered. This form can be reduced to
\citet{Chandrasekhar:1942} spherical quadratic integral. The same is true
with the integral whose dependence on velocities is not given in
advance, but which has the potential as an explicite argument \citep[][form]{Idlis:1961}.

All the forms of integrals considered turn out to be the integrals, 
quadratic (or linear) in the velocities. The more general
``ellipsoidal'' integral can also be reduced to quadratic one, as shown
by \citet{Garwick:1943}.

Chandrasekhar's spherical non-conservative integral represents no
other than the Jacobi integral in coordinates and time,  transformed by
\citet{Schurer:1943} transformation. Consequently, this integral, by itself,
has nothing to do with the spiral structure of galaxies,  contrary to
Chandrasekhar's well known theory.

The ellipsoidal non-conservative quadratic integral, obtained by
Chandrasekhar, is a combination of the spherical integral and the
angular momentum integral. A more general form of the non-conservative
ellipsoidal integral has been found recently by \citet{Genkin:1962}. These
integrals can be obtained from the integral of energy, the angular
momentum integral, and the third conservative integral by Schurer's
transformation.

Quadratic non-conservative integrals have an essential deficiency as
pointed out by \citet{Kurth:1949}. Assuming the phase density to be a function
of these integrals, we come to a contradiction with the Poisson
equation, except the case when the space density is independent of
coordinates, which cannot correspond to a real stellar system. The
only way to avoid this difficulty is to assume that the mass of the
stellar system varies with time (for example by dissipation of stars
from the system due to the irregular gravitational forces). In this
way we can obtain a model of a non-stationary stellar system without
any inner contradiction. The variation in time of such a model should
be, however, of a very special kind due to the special character of
the integrals used, which is by far not necessary in the dynamics of
non-stationary stellar systems.
\medskip

\hfill May 1964

%% file: chapter15.tex
\chapter[Hydrodynamic models of spherical stellar
systems]{Hydrodynamic models of spherical stellar
  systems.\footnote{\footnotetext ~~Tartu Astron. Observatory
    Publications, vol. 36, pp. 3-37, 1968. Coauthor
    \"{U}.-I.~K. Veltmann.}} 

{\bf Summary}
\bigskip

A steady stellar system with complete spherical symmetry is described
hydrodynamically by three functions: the space density, $\rho(r)$, the
radial velocity dispersion, $\sigma_r(r)$, and the transversal velocity
dispersion, $\sigma_t(r)$, $r$ being the distance from the centre of the
stellar system. They are mutually connected by the hydrodynamic
equation 
\be
\frac{\rmd \rho\sigma_r^2}{\rmd r} + \frac{2}{r} \rho (\sigma_r^2 -
\sigma_t^2 ) = \rho \frac{\rmd\Phi}{\rmd r} . \label{eq15.1.1} 
\ee
In addition, we have the Poisson equation 
\be
\frac{\rmd^2 \Phi}{\rmd r^2}  + \frac{2}{r} \frac{\rmd\Phi}{\rmd r} = -4\pi G\rho \label{eq15.1.2}
\ee
for the gravitational potential $\Phi(r)$.

It is possible to obtain a model of a steady spherical stellar system
``purely hydrodynamical'', without any specific assumption about the
phase density. For this purpose we need two additional relations
between the hydrodynamic functions, the potential $\Phi$ and $r$. These
relations enable us to eliminate the velocity dispersions and to solve
at first the hydrodynamic equation, and thereafter the Poisson
equation. A similar method in the phase-density description is to fix
the phase density as function of the integrals of motion. This enables
us to deduce the function $\rho(\Phi, r)$ and then to solve the Poisson
equation.

In the manner described we can obtain hydrodynamically the polytropic
models, for instance, if we admit the relations 
\be
\sigma_r^2 (r) = (1+\lambda r^2) \sigma_t^2 = [\Phi (r) - \Phi (r_e) ] / (n+1) . \label{eq15.2.6} 
\ee
Here $r_e$ is the radius of the model, $n$ the index of polytrope and $\lambda$ a constant. 
If $\lambda= 0$, we obtain usual polytropic models, otherwise the generalised ones,
for which the dispersions $\sigma_r^2$ and $\sigma_t^2$ are not equal.

More general polytropic models (generalised polytropic models of the
second kind) are hydrodynamically defined by the only relation
\be
n\sigma_r^2 + (1+\lambda r^2) \sigma_t^2 = \Phi (r) - \Phi (r_e) . \label{eq15.2.8}
\ee
In this case the second relation remains arbitrary.

The Schuster-Eddington polytropic model, $n=5$, $\lambda= 0$, described by
simple formulae 
\be
\rho (r) = \rho (0) [1 + (r/r_0)^2 ]^{-5/2}  \label{eq15.3.5}
\ee
and 
\be
\Phi (r) = \Phi (0) [1+(r/r_0)^2 ]^{-1/2} \label{eq15.3.6}
\ee
for the density and the potential, is
one of the most important polytropic model. There exists another very
interesting polytropic Schuster model, $n=3$, $\lambda= r_0^2$, with radially
prolate velocity distribution ($r_0$ is the characteristic distance in
the Schuster law, assumed later to be unity). We call these two models
as ``standard Schuster models'' I and II respectively.

It is possible to use for obtaining of hydrodynamic models instead
of ``deductive'' method, described above, the method, in which the order
of the solution of the equations is inverse. Instead of one additional
relation we fix the density function $\rho(r)$. Then we solve the Poisson
equation and thereafter the hydrodynamic equation using the remaining
additional relation. This ``inductive'' method has been systematically
worked out at Tartu for the phase-density description. It is by far
superior in practice, because this  is the only function, which is known
from observations. Moreover, we have no difficulty in obtaining
various models with finite mass, but with an infinite radius. It is
known that irregular gravitational forces tend to increase the radius
of stellar system as much as possible.

To establish the necessary additional relation we can depart from the relation 
\be
n(r) \sigma_r^2 + (1+\lambda r^2)\sigma_r^2 = \Phi (r) - \Phi (r_e) , \label{eq15.4.1} 
\ee
in which we replace in the general case the constant
index of polytrope $n$ by a variable effective index $n(r)$. Eliminating
$\sigma_r^2(r)$ from the hydrodynamic equation (\ref{eq15.1.1}) we find the expression
$$
\sigma_r^2 (r) = \int_0^r \exp\left\{ -2 \int_s^r \left[ 1+
    \frac{n(u)}{1+\lambda u^2} \right] \frac{\rmd u}{u} \right\} \cdot 
$$
\be
\cdot \frac{\rho (s)}{\rho (r)} \frac{1+\lambda s^2}{s^2} \phi^{\prime} (s) \rmd s \label{eq15.4.6}
\ee
(or a similar expression in which we integrate from $r$ to $r_e$) as its solution. Here 
\be
\phi (r) = [\Phi (r) - \Phi (r_e) ] (1+\lambda r^2)^{-1} r^2.
\ee

It should be noticed that we can attribute any value of $\lambda$ for a given
model, if we choose appropriate function $n(r)$. However, for a certain
value of $\lambda$ the expression for $n(r)$ may become very simple. For
instance, if we choose for a generalised polytropic model its ``own''
value of $\lambda$,  the function $n(r)$ becomes a constant.

We adopt the expression 
\be
\frac{n(r)}{1+\lambda r^2} = n-m \frac{\mu r^2}{1+\mu r^2} =\frac{n+(n-m) \mu r^2}{1+\mu r^2} \label{eq15.5.1}
\ee
for the function $n(r)$, where $\lambda$, $\mu$, $m$,
$n$ are parameters. The parameter $n$ has the meaning of $n(0)$. For $\lambda= 0$
and infinite models $n - m=n(\infty)$. The additional relation now takes the form 
\be
[n+(n-m)\mu r^2 ] \sigma_r^2 + (1+\mu r^2) \sigma_t^2 = 
[\Phi (r) - \Phi (r_e) ] (1+\mu r^2) /(1+\lambda r^2)  . \label{eq15.5.2}
\ee
It defines a wide class of hydrodynamic models, which
contains the generalised polytropic models as special cases $\lambda= m = 0$
and $\lambda=\mu$,  $m = n$. The solution for $\sigma_r^2(r)$ takes the form
\be
\sigma_r^2 (r) = \int_0^r \left( \frac{s^2}{r^2} \right)^{n+1} \left( \frac{1+\mu s^2}{1+\mu r^2}\right)^{-m} 
\frac{\rho (s)}{\rho (r)} \frac{1+\lambda s^2}{s^2} \phi '(s) \rmd s. \label{eq15.5.3}
\ee
Evidently, the condition $\lambda > -r_e^{-2}$ must be fulfilled, 
except  the case $\lambda=\mu$.

To  concretise the density $\rho(r)$ we adopt in this paper the Schuster
law, which is a good approximation to the observational data for many
spherical systems (from  globular clusters, open clusters to
clusters of galaxies). The additional relation and the solution for
$\sigma_r^2 (r)$ take the form 
\be
[n+(n-m)\mu r^2 ] \sigma_r^2 + (1+\mu r^2) \sigma_t^2 = 
\Phi (r) (1+\mu r^2) /(1+\lambda r^2)  \label{eq15.6.3}
\ee
and 
\be
\sigma_r^2 (r) = \Phi (r) \int_0^r \left( \frac{s^2}{r^2} \right)^{n+1} \left( \frac{1+\mu s^2}{1+\mu r^2}\right)^{-m} 
\left( \frac{1+s^2}{1+r^2}\right)^{-3} \frac{2+s^2-\lambda s^4}{(1+s^2)(1+\lambda s^2)} \frac{\rmd s}{s} , \label{eq15.6.6}
\ee
where $\Phi(r)$ is to be inserted according to 
\be
\Phi (r) = \Phi (0) (1+r^2)^{-1/2} . \label{eq15.6.2}
\ee

This family of the Schuster models includes both standard
models. Besides of the series of generalised polytropic models, they
belong to the series of models $\sigma_r^2/\Phi = \mathrm{const}$. For the latter the
velocity dispersion is expressed by the simple formulae 
\be
\sigma_r^2 / \Phi = 1 / (n+1) \label{eq15.6.9}
\ee
and
\be
\sigma_t^2 / \Phi = [1+\frac{1}{2} (n-3) r^2 ] / (n+1)(1+r^2) . \label{eq15.6.11}
\ee

For the velocity dispersion at small $r$ the expansions 
\be
\frac{\sigma_r^2}{\Phi} = \frac{1}{n+1} \left[ 1+ \frac{m \mu -(n+1) \lambda - \frac{1}{2} (n-5)}{n+2} r^2 + \dots \right]    \label{eq15.7.1}
\ee
and 
\be
\frac{\sigma_t^2}{\Phi} = \frac{1}{n+1} \left[ 1+ \frac{2m \mu -2(n+1) \lambda + \frac{1}{2} n(n-5)}{n+2} r^2 + \dots \right] \label{eq15.7.2}
\ee
can be derived. If we try to obtain the asymptotic expansion, the
distinct division of our family of models into two subfamilies, $\lambda=0$
and $\lambda \not=0$, ($\lambda > 0$, in general) becomes evident. For $\lambda=0$ the
asymptotic expansions are given by 
\be
\frac{\sigma_r^2}{\Phi} \sim \frac{1}{2(n-m-2)} \left( 1- \frac{m \mu^{-1} - n +m + 5}{n-m-3} r^{-2} + \dots \right) \label{eq15.7.5}
\ee
and 
\be
\frac{\sigma_t^2}{\Phi} \sim \frac{1}{2(n-m-2)} \left[ n-m-4 + \frac{3m \mu^{-1} - (n -m)(n-m-5)}{n-m-3} r^{-2} + \dots \right] , \label{eq15.7.6}
\ee
for $\lambda\not=0$ by
\be
\frac{\sigma_r^2}{\Phi} \sim \frac{1}{2(n-m+2)} \left[ 1 + \frac{m \mu^{-1} - 2(n-m+2)\lambda^{-1} -m+n+1}{n-m+3} 
r^{-2} + \dots \right] \label{eq15.7.12}
\ee
and 
$$
\frac{\sigma_t^2}{\Phi} \sim \frac{1}{2(n-m+2)} \left[ m-n - \right.
$$
\be
\left. - \frac{3m \mu^{-1} - 6(m-n+2)\lambda^{-1} - (m-n-1)(m-n)}{n-m+3} 
r^{-2} + \dots \right]  .  \label{eq15.7.13}
\ee
In the latter case an additional condition at $r\rightarrow\infty$
must be fulfilled. As consequence of this 
condition the subfamily $\lambda\not= 0$ becomes three-parametric,  like the
subfamily $\lambda= 0$. From the asymptotic expansions the necessary
conditions for $\sigma_r^2(r)$ being non-negative can be derived.

The integral, involving in formula  (\ref{eq15.6.6}) for $\sigma_r^2$, 
can be expressed in terms of the  
Lauricella generalised hypergeometric function of the
kind $D$. Its integral representation has the form 
\be
B(a,c-a) F(a;b_i ; c; z_i) = \int_0^1 \Pi_i (1-z_i t)^{-b_i} t^{a-1}
(1-t)^{c-a-1} \rmd t ,
\label{eq15.8.1}
\ee
where  $a$, $b_i$, $c$ are the parameters and $z_i$ the arguments. In
section 8 of the original paper some 
properties of this function are listed: special values, reduction of
arguments, transformations, recurrent formulae, derivatives, power
expansions (square brackets are used in formulae to distinguish one of
the parameters or one of the arguments). 
\smallskip

Sections 9 and 10 of the original paper were omitted in the Thesis and
they were not summarised also in here.  
\smallskip

{\begin{figure*}[h] 
\centering 
\resizebox{0.48\textwidth}{!}{\includegraphics*{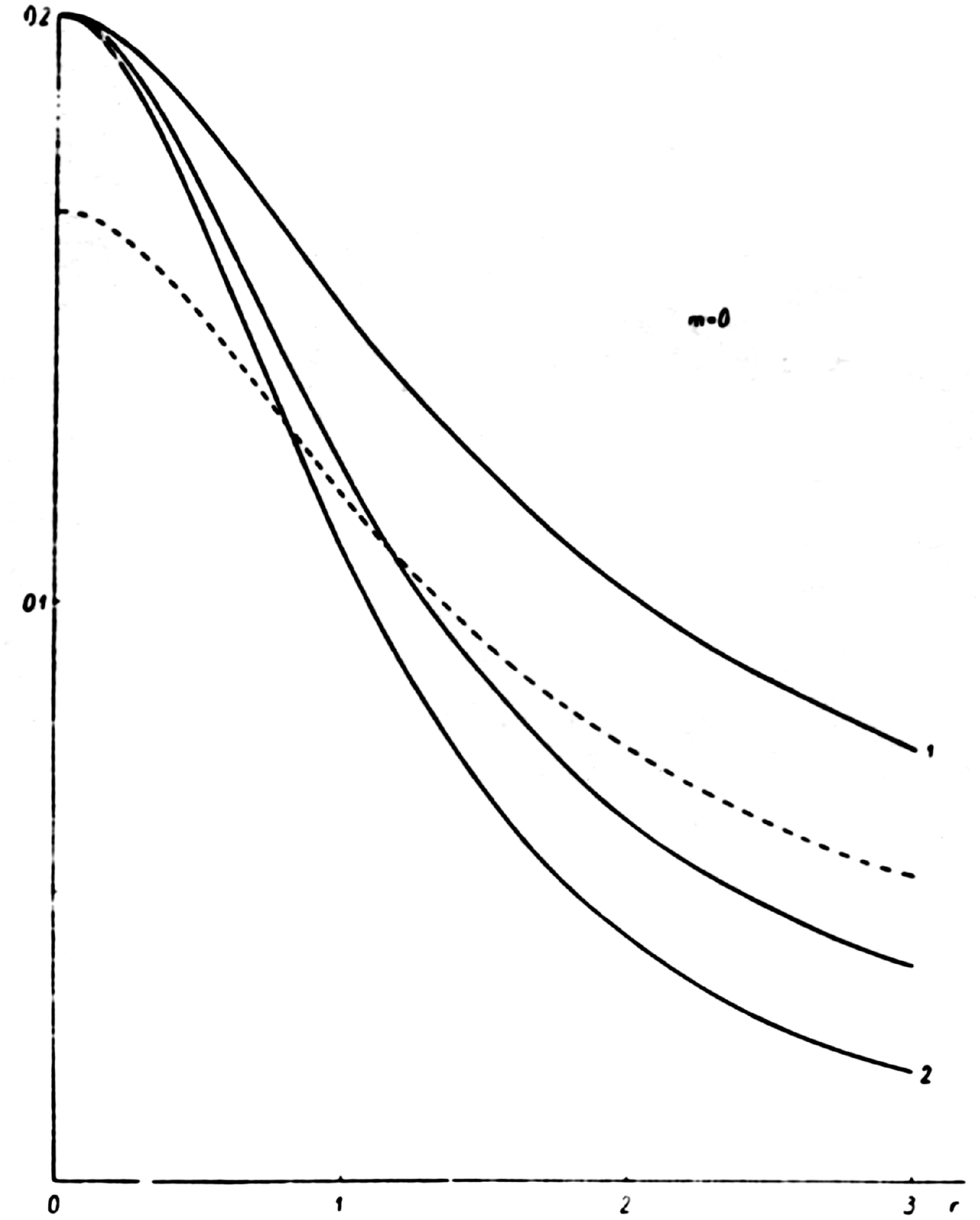}}
\resizebox{0.48\textwidth}{!}{\includegraphics*{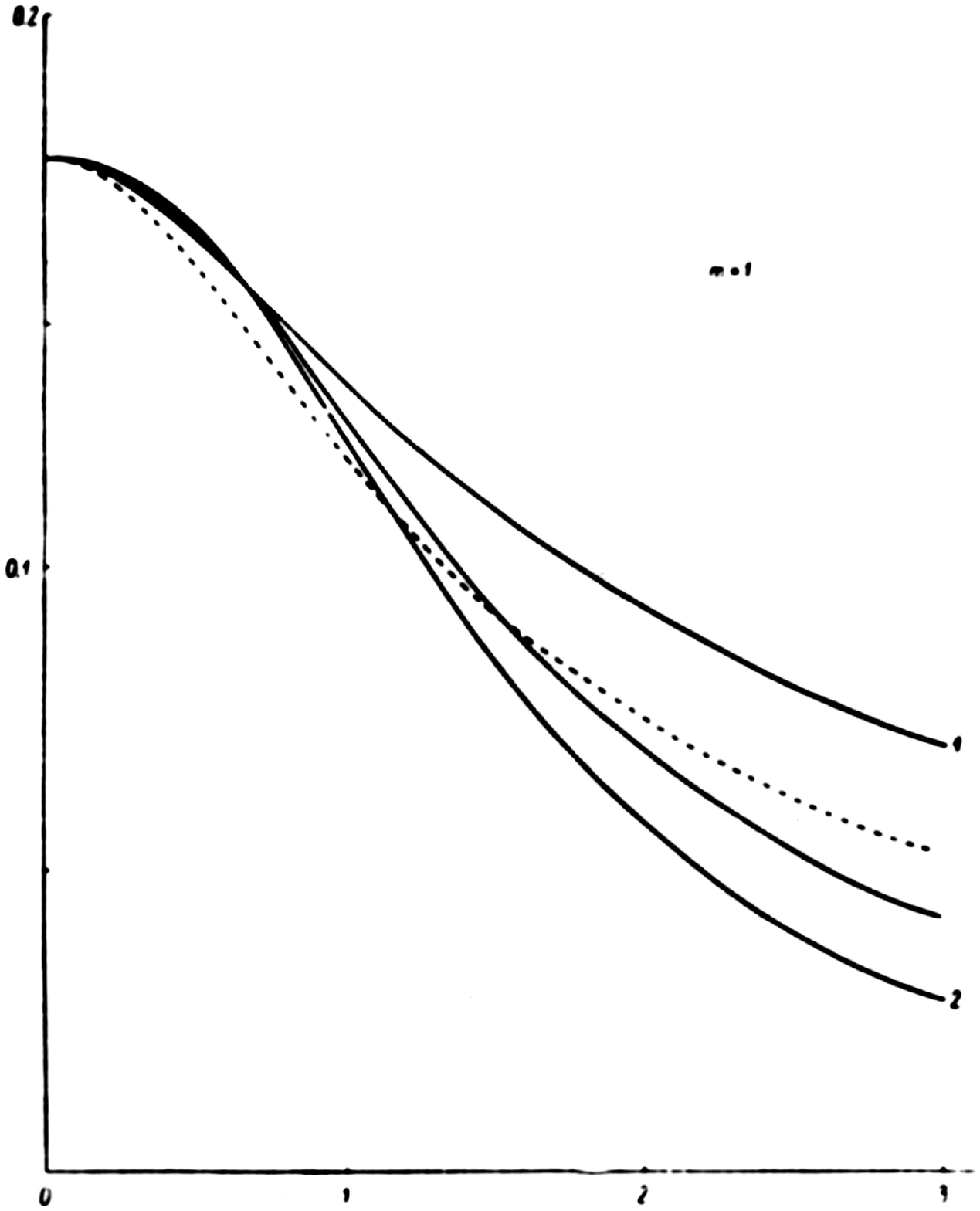}}\\
\resizebox{0.48\textwidth}{!}{\includegraphics*{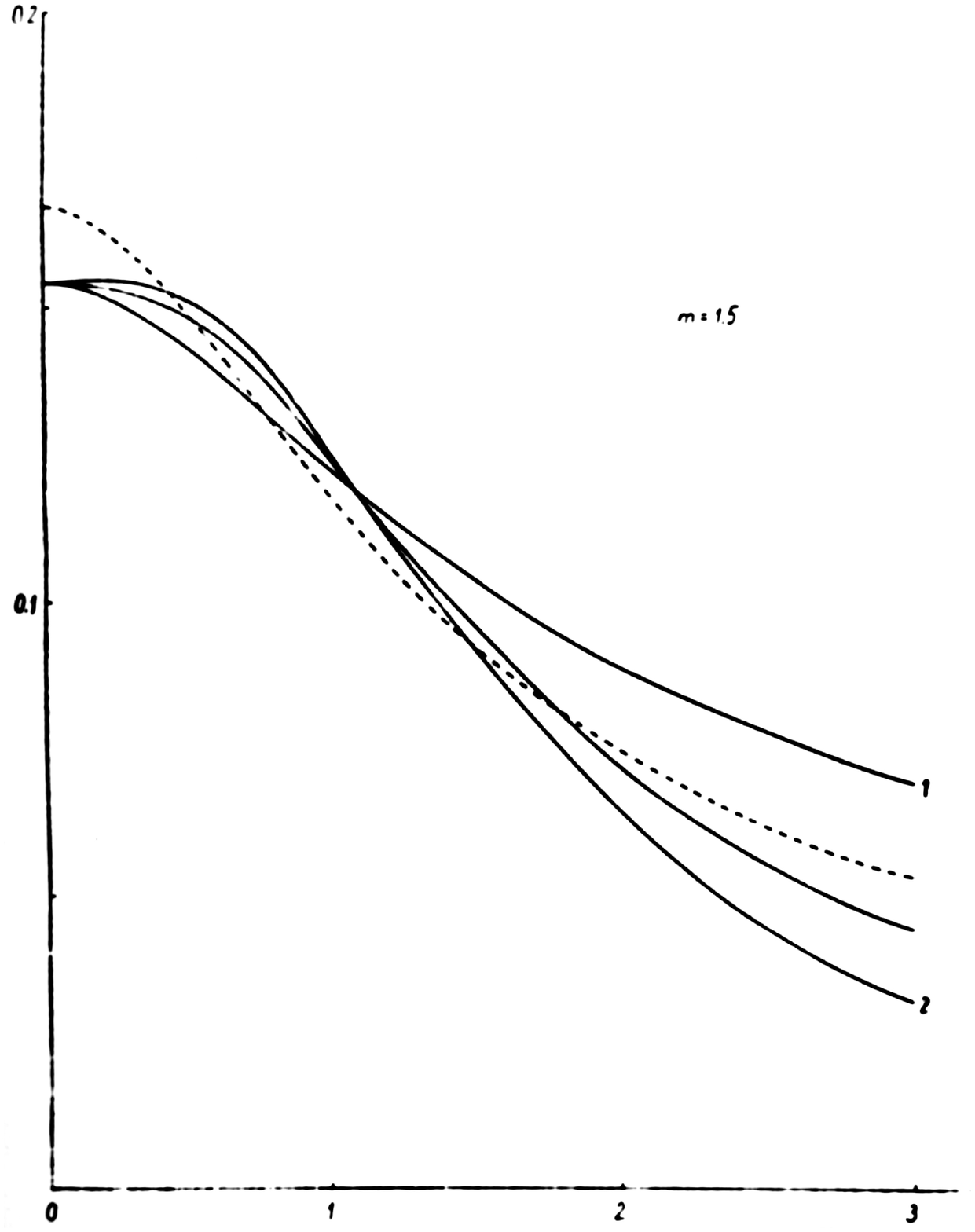}}
\resizebox{0.48\textwidth}{!}{\includegraphics*{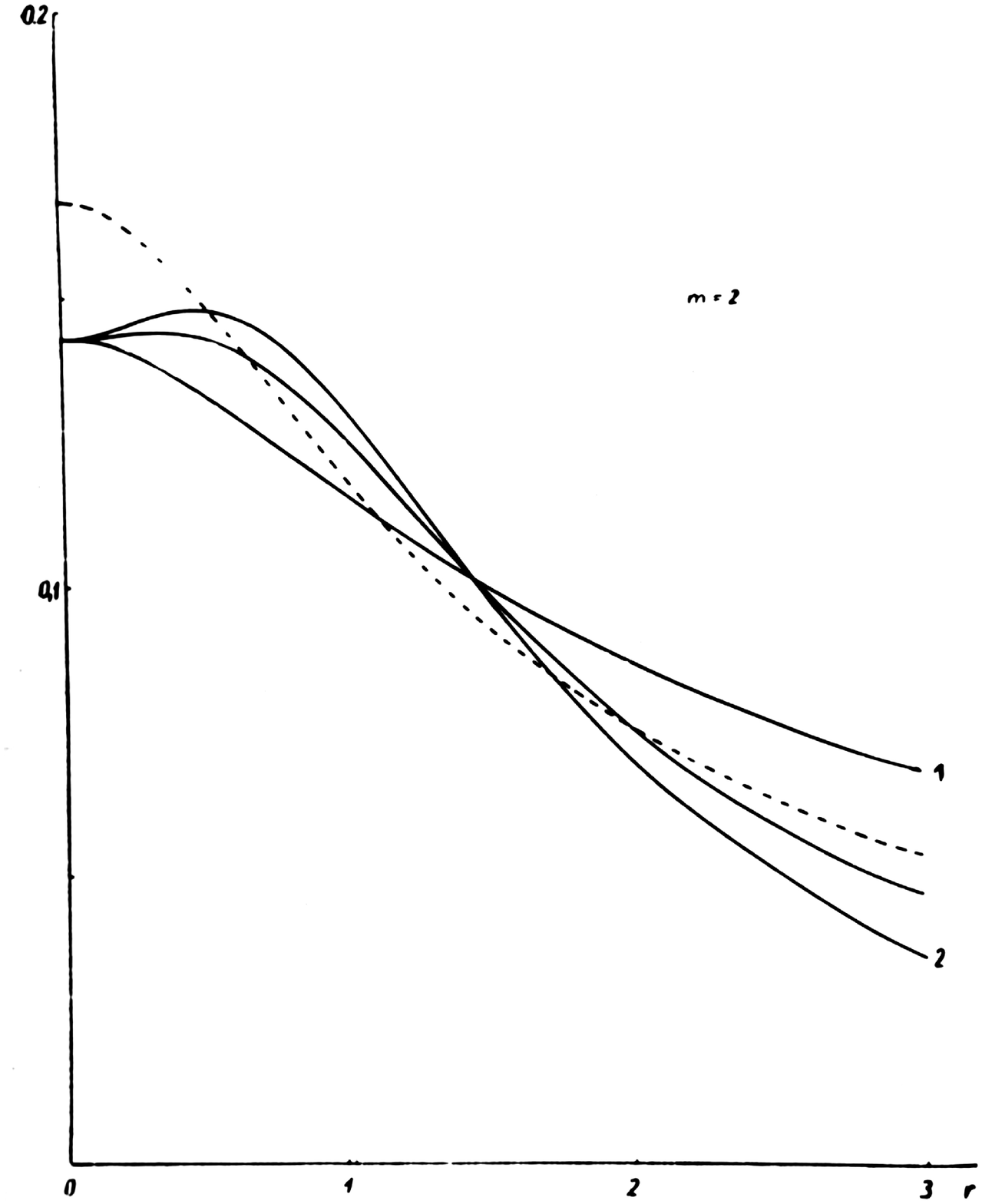}}\\
\caption{Dispersions $\sigma_r^2(r)$ and $\sigma_t^2(r)$ are shown by
  solid lines, marked by 1 and 2 respectively.  For comparison the mean
  dispersion $\sigma^2(r)$ and the dispersion for the
  Schuster-Eddington model $\sigma_r^2(r)$ are plotted by solid unmarked and dashed
  curves respectively. Dispersions are given in units of the potential in the
  centre $\Phi(0)$.  Data are given for model parameter $m=0,~1,~1.5,~2$.}
  \label{Fig15.2}
\end{figure*} 
}

The models $\lambda= 0$, $\mu=1$  have been studied in more detail. The
expression for $\sigma_r^2$ reduces in this case to the formula 
\be
\frac{\sigma_r^2}{\Phi} = \frac{1}{n+1} F(1, m+3; n+2; y) -
\frac{y}{2(n+2)} F(1, m+4; n+3; y) ,  \label{eq15.11.1} 
\ee
which after some transformations takes the from 
\be
\frac{\sigma_r^2}{\Phi} = \frac{1}{2(n-m-2)} \left[ 1+
  \frac{n-2m-5}{n+1} F(r) \right],
\label{eq15.11.6}
\ee
where
\be
F(r) = F(1, n-m-2; n+2; -r^2 ) . \label{eq15.11.7}
\ee

In the first stage of this work the models $\lambda= 0$, $m = 0$ were
considered. The values of the parameter $n$ were chosen mainly in the
interval $4 \le n < 5$, in which the velocity distribution is
prolate. 

Later, however, we found the possibilities to obtain essentially
better models, in which the two properties, expected from the theory
of the irregular gravitational forces,  are taken into account: the
prolateness of the velocity distribution in the outer regions of the
model,  and the approach to the isothermal state in the inner regions.

The value $n(\infty) = n -m = 4$, which corresponds to the extreme
prolateness of the velocity distribution in the outer region of the
model, has been adopted. The various values $m \ge 0$ were taken to obtain
various degrees of increase of the effective index of polytrope toward
the centre. 

In Fig. \ref{Fig15.2}  calculated velocity dispersions $\sigma_r^2(r)$,
$\sigma_t^2(r)$  and the mean dispersion 
\be
\sigma^2 (r) = \frac{1}{3} \sigma_r^2 (r) + \frac{2}{3}\sigma_r^2 (r) \label{eq15.12.2}
\ee
are given. For comparison the velocity dispersion $\sigma_r^2(r)$ in the
Schuster-Eddington model is plotted by dashed curves.

As seen, the graphs take the shape of a ``bird head'' for larger $m$: the transversal velocity dispersion becomes larger than the radial dispersion at small $r$. On the other hand, the ``isothermal plateau'', where the mean dispersion is approximately constant, arises.

Apparently, the models with $m$ between 1 and 2 ($n$ between 5 and 6) are
the most suitable, from among of the models considered, to describe
hydrodynamically the spherical stellar systems, in which the irregular
gravitational forces are acting. 

\medskip
\hfill October 1967

%% file: chapter16.tex
\chapter[Lindblad diagram and isochronic models]{Lindblad diagram and
  isochronic models.\footnote{\footnotetext ~~Tartu
    Astron. Observatory Publications, vol. 36, pp. 470-500,
    1968. Coauthor \"{U}.-I.~K. Veltmann.}} 

{\bf Summary}
\bigskip

We use in this paper, like in some previous ones, the Lindblad diagram to
visualise the structure of a stellar system. The integrals of motion,
usually two in number, entering into the phase density as arguments,
serve as the coordinates of the Lindblad diagram. The so-called
characteristic lines bound the regions, everyone of which represents
velocity space at given space coordinates. The totality of these more
or less overlapping regions restricted by the envelope of the
characteristic lines represents the phase space. Hence, if isolines of
the phase density are drawn on the diagram, one gets a picture of the
entire spatial-kinematical structure of a stellar system.

If the phase density depends on three integrals, the diagram becomes
three-dimensional \citep{Kuzmin:1956a}. In the case of more than three
integrals of motion the regions representing the velocity spaces do
not longer overlap and are occupying separate three-dimensional
hypersurfaces in the space of the integrals of motion.

Characteristic lines (surfaces) together with their envelope form the
frame of the Lindblad diagram, which is fixed,  provided the
gravitational potential of the system is given. But the isolines of
the phase density may vary, since the solution for the phase density,
in spite of the mass distribution being the same, is not unique.

In a previous paper \citep{Kuzmin:1962aa} the two-dimensional Lindblad
diagram was used to illustrate the structure of a family of
self-gravitating phase models found for steady axially symmetric
stellar systems. The spherical limiting case of these models is
identical with the isochronic model with spherical velocity
distribution, found independently from these models by \citet{Henon:1959, Henon:1960}. 
The Lindblad diagram for spherical stellar systems with
non-spherical velocity distribution was discussed in two other papers \citep{Veltmann:1965, Veltmann:1966a}. The Schuster law for the space density was used
there.

In the present paper the discussion of the Lindblad diagram for
spherical stellar system is continued. The aim of the paper is to
study the kinematical structure of the  isochronic models with a
non-spherical velocity distribution. For comparison, the Schuster
models are also treated.

We assume for a spherical steady stellar system the complete spherical
symmetry. In this case the phase density depends on two arguments --
the energy integral, $I_1$, and the angular momentum integral, $I_2$. In their
expressions 
\be
I_1 = v_r^2 + v_t^2 - 2\Phi (r) \label{eq16.2.1} 
\ee
and 
\be
I_2 = r v_t, \label{eq16.2.2} 
\ee
$v_r$ and $v_t$ denote the radial and
transversal velocity components, $\Phi(r)$ is the gravitational potential and
$r$ is the distance from the centre of the system. The potential is related
to the total mass density $\rho(r)$ of the system by the Poisson equation
\be
\Phi ''(r) + 2\Phi ' (r) /r + 4\pi G\rho (r) =0, \label{eq16.2.3} 
\ee
or by its solution 
\be
\Phi (r) = 4\pi G \left[ r^{-1} \int_0^r \rho (r) r^2 \rmd r + \int_r^{\infty} \rho (r) r \rmd r \right] . \label{eq16.2.4}
\ee

We consider $-I_1$ and $I_2^2$ as the coordinates on the Lindblad
diagram. The overlapping regions $D(r)$, which represent the velocity
spaces for given values of $r$, are restricted by the characteristic
straight lines 
\be
I_1 - r^{-2} I_2^2 + 2\Phi (r) =0 \label{eq16.2.8} 
\ee
and by the $I_1$ axis. A net of parallel straight isolines 
\be
v_r^2 = I_1 -r^{-2} I_2^2 + 2\Phi (r) , ~~~~ v_t^2 = r^{-2} I_2^2 , \label{eq16.2.6}
\ee
$v_r^2=\mathrm{const}$, $v_t^2=\mathrm{const}$ can be drawn for every such region.

Because the integration of the phase density over the velocity space
gives the space density, we have for the phase density $\Psi$  the integral equation 
\be
\rho (r) = \pi r^{-2} \iint_{v_r^2 \ge 0} \Psi (-I_1 , I_2^2 ) v_r^{-1} \rmd I_1 \rmd I_2^2 . \label{eq16.2.7} 
\ee
In the latter equation the densities need not
necessarily be the total mass densities unlike $\rho(r)$ in the Poisson equation.

The envelope of the characteristic lines and $I_1$ axis together bound
the area $D$ of physically possible values of the integrals of
motion. The area $D$ represents the entire phase space. The equations
\be
\ba{rl}
-I_1 = & 2\Phi (r) + \Phi ' (r) r , \\
\noalign{\smallskip}
I_2^2 = & - \Phi ' (r) r^3 ,
\ea
\label{eq16.2.9} 
\ee
are the parametric equations of the envelope. It is a regular
descending concave curve, which is tangent to the $I_1$ axis at its
highest point, $-I_1=2\Phi_0$, $I_2^2=0$, $\Phi_0$ being the potential
at the centre of the system. The envelope behaves here like the parabola 
which depends on the central density $\rho_0$. At
great values of $I_2^2$ the envelope tends to the $I_2^2$ axis and can
be approximated by the hyperbola 
\be
3(2\Phi_0 + I_1 )^3 = 16\pi G \rho_0 I_2^2 . \label{eq16.2.11} 
\ee
The latter depends on $M$, the total mass.

Dimensionless coordinates $\chi$ and $\xi$ on the Lindblad
diagram are introduced according to formula 
\be
\chi = -I_1 /2\Phi_0, ~~~~ \xi = I_2^2 / 2\Phi_0 r_0^2 .  \label{eq16.3.1} 
\ee
The coordinate $\chi$
means the negative energy per unit mass, measured in units $\Phi_0$, and the
coordinate $\xi$ a half of the squared angular momentum per unit mass,
measured in units $\Phi_0 r_0^2$, where $r_0$ is a certain characteristic
distance. Further, the dimensionless potential $\Phi$, squared velocity
components $v_r^2$, $v_t^2$ (unit $\Phi_0$), density (unit $\rho_0$) and phase density
(unit $\Phi_0^{-1/2}\rho_0$) are introduced. The transitions 
\be
\left.
\ba{ll}
r_0, ~~ \Phi_0, ~~ \rho_0 \rightarrow 1, ~~~~ -I_1 \rightarrow 2x, ~~ I_2^2 \rightarrow 2\xi , & \\
\noalign{\medskip}
4\pi G \rightarrow 4\pi G \rho_0 r_0^2 / \Phi_0 = \rho_0 r_0^2 / \int_0^{\infty} \rho (r) r \rmd r & 
\ea
\right\} 
\label{eq16.3.4} 
\ee
transform the formulae into dimensionless ones. In particular, the
equation for the characteristic lines becomes 
\be
\left\{ 
\ba{ll}
\chi = & \Phi (r) + \frac{1}{2} \Phi ' (r) r , \\
\noalign{\smallskip}
\xi = & -\frac{1}{2} \Phi ' (r) .
\ea
\right.
\label{eq16.3.6}  
\ee
The coordinates of the highest point of the envelope of the characteristic lines are
now $\chi=1$, $\xi=0$. The equations of the limiting parabola and hyperbola
becomes 
\be
(1-x)^2 = k_1\xi  ~~~~ \mathrm{when} ~ \xi\rightarrow 0, \label{eq16.3.7}
\ee
and 
\be
x^{-1} = k_2 \xi   ~~~ \mathrm{when} ~ \xi\rightarrow 0, \label{eq16.3.8} 
\ee
with constants $k_1$and $k_2$ according to 
\be
k_1 = 8\pi G \rho_0 r_0^2 /3\Psi_0 = 2\rho_0 r_0^2 /3 \int_0^\infty \rho (r) r \rmd r \label{eq16.3.9} 
\ee
and
\be
k_2 = (2\Phi_p r_0 / GM)^2 = \left[ 2r_0 \int_0^\infty \rho (r) r \rmd r \bigg/ \int_0^\infty \rho (r) r^2 \rmd r \right]^2 . \label{eq16.3.10}
\ee

Thereafter, envelopes of the characteristic lines for the Schuster
model and the isochronic model are 
discussed. The expression of the dimensionless potential and the density has the form 
\be
\Phi (r) = \zeta^{-1} , ~~~~~ \rho (r) = \zeta^{-5} \label{eq16.4. 2} 
\ee
for the Schuster model, and 
\be
\Phi (r) = 2 (\zeta + 1)^{-1} \label{eq16.5. 1} 
\ee
and 
\be
\rho (r) = \frac{4}{3} \zeta^{-3} (\zeta + 1)^{-2} (2\zeta +1) \label{eq16.5.2} 
\ee
for the isochronic one, with $\zeta$ expressed by 
\be
\zeta^2 = 1+\kappa r^2 . \label{eq16.4.1}
\ee
In the latter formula $\kappa$ is a constant, which depends on the choice of the characteristic
distance $r_0$. The envelope for the Schuster model is a part of the fourth degree algebraic curve 
\be
4 x y^3 - 27 x^2 + 18 xy + y^2 + 4 =0 \label{eq16.4.7} 
\ee
($y = x - \kappa\xi$). In the case of
the isochronic model the envelope turns out to be a part of the hyperbola 
\be
\kappa x \xi = (1- x)^2 . \label{eq16.5.4}
\ee
We have $k_1=2\kappa$, $k_2=4\kappa$ for the Schuster
model , and $k_1=k_2=2\kappa$ for the isochronic one. We take $\kappa=1$
in the first case and $\kappa=2$ in the second case. The constants are now equal and
therefore the curvatures of the envelopes at their highest points are
the same. 

The integral equation (\ref{eq16.2.7}) 
for the phase density in dimensionless variables takes the form 
\be
\rho (r) = 2\sqrt{2} \pi r^{-2} \iint_{D(r)} \Psi (x,\xi ) [ \Phi (r) -x -r^{-2}\xi ]^{-1/2} \rmd x \rmd\xi . \label{eq16.6.1}
\ee
In previous papers \citep{Veltmann:1964, Veltmann:1965, Veltmann:1966a, Veltmann:1966b} the
generalised polytropic models have been studied. For such models the
phase density has the form $\Psi_0\kappa^\alpha\psi(\xi)$.
Here the constant $\Psi_0$ is chosen according to 
\be
\Psi_0^{-1} = 4\sqrt{2} \pi B(\alpha + 1, 3/3) , \label{eq16.6.3}
\ee
which makes  $\psi(0)=1$. The exponent $\alpha$ defines the index of polytrope, $n=\alpha+3/2$.
The integral equation (\ref{eq16.6.1}) takes now the form of the generalised Abel equation 
\be
g(\eta ) = n \int_0^\eta (\eta -\xi )^{n-1} \psi (\xi ) \rmd\xi  \label{eq16.6.5}
\ee
or 
\be
f(\eta ) = n \int_0^1 (1-s)^{n-1} \psi (\eta s) \rmd s \label{eq16.6.6}
\ee
with given functions defined
by 
\be
g(\eta ) = \eta^n f(\eta ) , \label{eq16.6.7}
\ee
\be
f(\eta ) = \rho (r) [ \Phi (r)]^{-n} , \label{eq16.6.8}
\ee
\be
\eta (r) = \Phi (r) r^2 . \label{eq16.6.9}
\ee
As the solution we have the expression
\be
\psi (\xi ) = \frac{1}{\Gamma (n+1) \Gamma (l-n)} \frac{\rmd^l}{\rmd\xi^l} 
\int_0^\xi (\xi - \eta )^{l-n-1} g(\eta ) \rmd\eta , \label{eq16.6.10}
\ee
which reduces to 
\be
\psi (\xi ) = \frac{1}{n!} g^{(n)} (\xi ) , \label{eq16.6.11}
\ee
if $n$ is a natural number.

The solutions of the integral equation (\ref{eq16.6.5}) for the Schuster and
isochronic models are considered according to the
papers  \citet{Veltmann:1964, Veltmann:1965, Veltmann:1966a, Veltmann:1966b}. For the isochronic model the function $\psi(\xi)$ can be
expressed in terms of the Appel  hypergeometric functions, according
to formulae 
\be
\psi (\xi ) = \frac{2}{3} \psi_2 (\xi ) + \frac{1}{3} \psi_3 (\xi ) ,\label{eq16.8.6}
\ee
where
\be
\psi_i (\xi ) = F(n+1; 2-n, i; 1; -\frac{1}{2}\xi , -\xi ) \label{eq16.8.7} 
\ee
or
\be  
\psi_i (\xi ) = f_i (\xi) h_i (\xi) , \label{eq16.8.8}
\ee
\be
f_i (\xi) = \vartheta^{-i} \left( \frac{\vartheta+1}{2}  \right)^{n-2},   \label{eq16.8.9}
\ee
\be
\vartheta =1+\xi , \label{eq16.8.10}
\ee
\be
h_i (\xi) = F(-n; 2-n,i;1; z_1, z_2 ) ,    \label{eq16.8.11}
\ee
\be
z_1 = \xi / (2+\xi) , ~~~~ z_2 = \xi / (1+\xi) . \label{eq16.8.12}
\ee
If $n$ is a natural number,  the solution reduces 
to the elementary functions for both class of models. For example, we have formulae 
\be
\psi(\xi) = \frac{5}{4} \vartheta^{-7} - \frac{5}{12} \vartheta^{-5} + \frac{1}{6} ~~~~~~(n=4) \label{eq16.8.13} 
\ee
and 
\be
\Psi(\xi) = \frac{5}{3} \vartheta^{-6} - \frac{2}{3} \vartheta^{-4}  ~~~~~~~~~~ (n=3) \label{eq16.8.14}
\ee
for isochronic models with $n=3$ and $n=4$.

If $n$ is small enough, the velocity distribution in a generalised
polytropic models is radially prolated,  and the isolines of the phase
density on the Lindblad diagram are raising. If $n=4$ for the Schuster
model and $n=3$ for the isochronic model, then  the radial prolateness
becomes extremely large at  $r\rightarrow\infty$,  \ie the ratio of the velocity
dispersions  $\sigma_r^2/\sigma_t^2$ tends to infinity. In this limiting case both
models have a slightly negative phase density at large $\xi$. 

The extreme radial prolateness of the velocity distribution at the
periphery of a spherical stellar system is expected from the
considerations of the theory of irregular gravitational forces. On the
other hand, theory predicts for the central parts of the stellar system
a velocity distribution close to the Maxwellian one. From this
standpoint the generalised polytropic models have a serious
deficiency. It is impossible to fit both conditions simultaneously. If
the velocity distribution in such a model is extremely prolate at
large $r$, it differs appreciably from Maxwellian distribution at small
$r$, since the increase of the phase density with increasing $\chi$ is too
slow. Moreover, for both investigated classes of models the condition
of the non-negativeness of the phase density is slightly infringed in
this case.

To obtain better models the stellar systems are considered as the
superpositions of the polytropic models. We suppose that the phase
density follows the formulae 
\be
\Psi(x,\xi) = \sum_n \sum_m a_{mn} \Psi_{0mn} x^{n-3/2} \psi_{mn} (\xi) , \label{eq16.10.1}
\ee
\be
\psi_{mn} (\xi) = (1+p_{mn} \xi)^{-m-n-1} , \label{eq16.10.2} 
\ee
where $m$ and $n$ are integers and $p_{mn}$ non-negative
coefficients. If all $p_{mn}$ are positive 
for the smallest $n$, the velocity distribution is extremely prolate
at $r\rightarrow\infty$. To find the coefficients $a_{mn}$
we have to represent the space densities of formal subsystems according to formula 
\be
\left( 
\ba{c}
m+n \\
m
\ea
\right) \rho_{mn} = \Phi^n \sum_{l=0}^m \left(
\ba{c}
n+l-1\\
l
\ea
\right) (1+p_{mn} \eta )^{l-m-1}   \label{eq16.10.5} 
\ee
(for $\eta$ see (\ref{eq16.6.9})).

The examples are examined in the case of the Schuster model and of the
isochronic model. For the Schuster model we take $p_{mn}=p$, $m=0$.
Three subsystems are introduced according to formulae 
\be
\Psi(x,\xi) = \frac{32\sqrt{2}}{7\pi^2} \Theta^{-5/2} u^{5/2} \left[
  u+p\left( \frac{7}{10} - \frac{4}{3} u^2\right) \right]
, \label{eq16.11.6} 
\ee
\be
u = x /\Theta , ~~~~~ \Theta =1+p\xi .\label{eq16.11.7}
\ee
If  $p=0$, the model reduces to the classical Schuster-Eddington
model. For $p>0$ the radial prolateness of the velocity distribution
appears. It increases with increasing of the parameter $p$,  being always
extremely great at $r\rightarrow\infty$. The inequalites 
\be
\Psi(x,\xi) \ge 0, ~~~~ \mathrm{if} ~ p \le 30/19 \label{eq16.11.9} 
\ee
\be
\partial\Psi(x,\xi) /\partial x \ge 0, ~~~~ \mathrm{if} ~p \le 14/17   \label{eq16.11.10}
\ee
\be 
\partial\Psi(x,\xi) /\partial\xi \le 0, ~~~~ \mathrm{if} ~p \le 36/35 \label{eq16.11.11}
 \ee
determine the values of the parameter $p$, for which the phase density is a
non-negative function, increasing with $\chi$ and decreasing with $\xi$. The
new generalised Schuster models are essentially better than  generalised
polytropic Schuster models, but the velocity distribution in the
central parts of the models is still far from the Maxwellian one.

For the isochronic model two examples are studied. At first we suppose
$p=1$ and decompose the stellar system into four subsystems according to 
\be
\rho(r) = \frac{1}{3} (1-q)\rho_{03} + \frac{5}{9} q \rho_{23} - \frac{7}{36} (3-q) \rho_{04} +
\frac{5}{12} (3-q) \rho_{24} ,\label{eq16.12.4}
\ee
where $q$ is a parameter and 
\be
\ba{ll}
\rho_{03} = \Phi^3\vartheta^{-1} , ~~~~ & \rho_{23} = \frac{1}{10} \Phi^3 (\vartheta^{-3} + 3\vartheta^{-2} + 6\vartheta^{-1} ),\\
\rho_{04} = \Phi^4\vartheta^{-1} , ~~~~ & \rho_{24} = \frac{1}{15}
\Phi^4 (\vartheta^{-4} + 34\vartheta^{-2} + 10\vartheta^{-1} ) 
\ea
\ee
The phase density is expressed by formula 
\be
\Psi(x,\xi) = \frac{2\sqrt{2}}{45\pi^2} x^{3/2} \vartheta^{-7} \{ 5\vartheta [ 5q+3(1-q) \vartheta^2] + 
2(3-q)x (15-7\vartheta^2) \} . \label{eq16.12.5}
\ee
If $q=3$, the model
reduces to the generalised polytrope isochronic model $n=3$. The
phase density is a non-negative function increasing with $x$ and
decreasing with $\xi$, if the conditions 
\be
\Psi(x,\xi) \ge 0, ~~~\partial\Psi(x,\xi) /\partial x \ge 0, ~~~~ \mathrm{if} ~ -3/2 \le q \le 1\label{eq16.12.7}
\ee
and
\be
\partial\psi(x,\xi) /\partial\xi \le 0, ~~~~ \mathrm{if} ~ -2/3 \le q \le 1 \label{eq16.12.8}
\ee
are fulfilled. But as in the previous case, the velocity distribution in the central parts of
the model differs strongly from the Maxwellian law.

We obtain much better isochronic models supposing as for the Schuster
model $p_{mn}=p$, $m=0$. The number of formal subsystems is now,
however, infinite. For the phase density we found formula 
$$
\Psi(x,\xi) = \frac{\sqrt{2}}{3\pi^2} \Theta^{-5/2} u^{3/2} \left\{ \frac{4}{5} (1-p) u \left[ F_{22} \left( \frac{1}{2} u \right) + F_{13} 
\left( \frac{1}{2} u\right) \right] + \right.
$$
\be
\left.  + p \left[ F_{12} \left( \frac{1}{2} u\right) + F_{03} \left( \frac{1}{2} u \right) \right] \right\}  \label{eq16.13.8}
\ee
with 4 Gauss hypergeometric functions 
\be
F_{ij} \left( \frac{1}{2} u\right) = F \left( i+1, i+j+1; i+j-\frac{1}{2}; \frac{1}{2} u\right) .   \label{eq16.13.9} 
\ee
This formula can be transformed to the form 
$$
\Psi(x,\xi) =  \frac{\sqrt{2}}{3\pi^2} \Theta^{-5/2} u^{3/2} \left[ (1-p) F_{21} \left( \frac{1}{2} u \right) + \right.
$$
\be
\left. + pF_{12} \left( \frac{1}{2} u \right) + (2p-1) F_{03} \left( \frac{1}{2} u \right) \right]   \label{eq16.13.11} 
\ee
with 3 Gauss hypergeometric functions. In formulae 
$$
\Psi(x,\xi) = \frac{\sqrt{2}}{12\pi^2} \Theta^{-5/2} (2-u)^{9/2} \cdot
$$
$$
\cdot \left[ ( a_0 + a_1 u + a_2 u^2 + a_3 u^3 + a_4 u^4 ) \sqrt{u(2-u)} + \right. 
$$
\be
\left. + 2(b_0 + b_1 u + b_3 u^2 ) \arcsin \sqrt{u/2} \right]  \label{eq16.14.1} 
\ee
and 
\be
\ba{ll}
a_0 = 27 - 39 p, & b_0 = -27 + 39p = -a_0, \\
a_1 = -33 + 139p, & b_1 = 42 - 24 p, \\
a_2 = 80 - 178p, & b_2 = 12 -24 p. \\
a_3 = -30 + 62 p, & \\
a_4 = 4-8p, & 
\ea
\label{eq16.14.2} 
\ee
these hypergeometric functions are expressed by elementary functions.

If  $p=0$, the new model turns into the isochronic model with the
spherical velocity distribution. If $p>0$, we have a prolate velocity
distribution, which is extremely prolate 
at great $r$. The phase density is a non-negative function, increasing
with $\chi$ and decreasing with $\xi$, if the parameter $p$ obeys the conditions
\be
\Psi(x,\xi) , \ge 0, ~~~~ \mathrm{if} ~~ p \le (32+9\pi ) / (16 + 3\pi ) \simeq 2.371 ,  \label{eq16.15.1}
\ee
\be
\partial\Psi /\partial x , \ge 0, ~~~~ \mathrm{if} ~~ p \le (1184+375\pi ) / (720 + 225\pi ) \simeq 1.655 ,  \label{eq16.15.4} 
\ee
and 
\be
\partial\Psi /\partial\xi , \le 0, ~~~~ \mathrm{if} ~~ p \le (832+255\pi ) / (480 + 135\pi ) \simeq 1.806 . \label{eq16.15.5}
\ee
%

Tables and Figures are not included in the present Summary and they can be seen from the original paper (in Russian).
The Table 2 contains the values of the phase density on
the $\chi$ axis for $p=0$  and $p=1$. If we take the mean values of both
columns with weights $1-p$  and $p$  we obtain  $\Psi(\chi,0)$ for any value of
the parameter $p$. 

On Fig. 3 the curves of natural logarithms $\Psi(\chi,0)$ are drawn
for $p=0$ and  $p=1$. As seen, they differ only slightly from straight lines
in a wide range of $\chi$. Consequently, if $p$ do not essentially exceed
unity, the velocity distribution in the central parts of the model is
close to the Maxwellian distribution, which is smoothly truncated at
$\chi\rightarrow 0$. 

When the phase density on the $\chi$ axis has been calculated, it can be
easily found for any other point by formula 
\be
\Psi(x,\xi) = \Psi(u,0) \Theta^{-5/2} , ~~~~ u=x/\Theta , ~~~ \Theta = 1+p\xi .  \label{eq16.16.1}
\ee
On Fig. 4 the Lindblad diagram with the  $\ln(\Psi/\Psi_0)$  isolines is presented for the
case $p=1$. If $p<1$, the inclination of isolines decreases and they
become more straight. If  $p>1$,  the inclination and the curvature of
isolines increases and the diagram becomes like the Lindblad diagram
of the generalised polytropic isochronic models $n=3$ (Fig. 2),
without negative region at right, however.

The new isochronic models seem to be a good approximation to the
structure of isolated spherical stellar systems, in which the
influence of the irregular gravitational forces has been effective. 

\medskip
\hfill October 1967

%% file: chapter17.tex
\part{Time evolution of the structure of stellar systems. The effect
  of irregular forces} 

\chapter[The effect of stellar encounters and evolution of star
clusters]{The effect of stellar encounters and the evolution of star
  clusters\footnote{\footnotetext ~~Tartu Astron. Observatory
    Publications, vol. 33, 75 - 102, 1957.}} 

Gravitational interactions between the stars during their
encounters are quite essential in the evolution of star clusters. As a
result of this kind of interactions, a cluster disrupts little by
little by loosing stars (and the energy). The mass and the dimensions
of a cluster decrease, the density increases.

The importance of the role of the encounters between the cluster stars
was first emphasised by \citet{Ambartsumian:1938}, who calculated the
approximate disruption (dissipation) rate of a cluster. Similar
results were derived a little later by \citet{Spitzer:1940}, who also
analysed the variations of cluster dimensions and density. Later, the
encounter effect in star clusters was studied by \citet{Chandrasekhar:1942,
Chandrasekhar:1943a, Chandrasekhar:1943b, Chandrasekhar:1943c} and
thereafter by a number of authors \citep{Gurevich:1950,
  Skabitskii:1950, Minin:1952, Woolley:1954, Hoerner:1957}. As a 
result of these studies (true enough, being not in all cases free of
errors and insufficient arguments), the evolution of clusters outlined
by Ambartsumian and Spitzer was specified and detailed at a certain
lever.

Having in mind the need to analyse the cluster evolution in more detail, 
the purpose of the present paper is to derive the equations allowing to
describe the evolution of clusters more precisely than it was done before.
In this paper we analyse the equations enabling to follow the variations
of the phase density and the gravitational potential of the cluster. We also study
the formulas necessary to calculate the encounter function. We treat the
clusters as isolated, spherical, almost steady systems of 
big number of gravitating particles of constant mass. This approximations
are quite common in stellar dynamics when studying star clusters, and the
properties of these kind of systems were studied in detail \citep{Eddington:1913, Eddington:1915a, Eddington:1916, Jeans:1916, Bruggencate:1927, Shiveshwarkar:1936, Kurth:1949, Kurth:1949a, Kurth:1950, Kurth:1955, Camm:1952, Woolley:1954a, Woolley:1956, Woolley:1956a}. The cluster model that we
use corresponds probably quite well to globular clusters. In case of open
clusters the similarity is worse and in some case even completely absent
(very sparse clusters can not be handled even approximately as isolated).

\section{The phase density, the potential and the encounter function}

When treating star clusters as systems consisting of big number of
stars, we use the phase density $\psi$ for the description of their structure. 
The phase density is the density of particles in six-dimensional
phase space, where the coordinates are three rectangular coordinates of
ordinary space and three corresponding velocity components. Due to the
discrete nature of medium in phase space, the phase density is the smoothed
density with the properties of probability density. The phase density can
be defined as the number density of particles, the mass density or in some
other way. It is important to distinguish between the total phase
density and the partial phase density, latter corresponding to the unit interval
of some particle's characteristic, e.g. to unit interval of mass.

Integration of the total phase density as the phase mass density over the
velocity space gives us the ordinary mass density in three-dimensional
physical space $\rho$, and enables to derive the gravitational
potential of the system $\Phi$. Potential $\Phi$, calculated in this
way,  is
smoothed or regular potential, corresponding to continuous mass
distribution in physical space. A randomly varying
irregular potential is added to the regular potential, resulting from real discreteness of mass
distribution. The existence of that potential is revealed in gravitational
interactions during stellar encounters in physical space.

According to our assumption the number of stars is big, so the real
gravitational potential of the system is highly smoothed and differs only
slightly from regular potential. Hence the motion of stars is governed by
the regular potential, and the role of encounters is very small. Encounters may
be revealed only after sufficiently long time, much longer than
the ``revolution time''. This conclusion is confirmed by the calculations
of the relaxation time and the mean free path of particles \citep{Chandrasekhar:1942}. 

To clarify the ways of system evolution due to encounters, we must follow
the time evolution of the phase density $\psi$. Evolution of the phase
density is caused by the motion of stars in phase space. This motion consist
of smooth regular motion, governed by the regular potential, and of more
or less impulsive motions related to stellar encounters in physical space
and governed by irregular potential. Correspondingly the evolution of
the phase density is caused by non-stationarity of regular motion and by
non-compensation of impulsive displacements. We designate the variation of the phase
density per unit time resulting from the encounters as
$\chi$. This is called the encounter function. 

At the initial stage of evolution of the system the variation of the phase
density,  resulting from non-stationarity of the regular gravitational
potential, can be very large. But as it was mentioned already by
\citet{Eddington:1916, Eddington:1921}, after quite short time -- of
the order of few revolution times
-- the system evolves due to the ``mixing'' process into a
nearly-stationary state. Although the ``mixing'' process does not decrease
the speed of phase density variation, it looses the correlation of phase
density variations between different points of velocity space. As a result
these variations will be like stochastic fluctuations, and the mean
variation of the phase density in arbitrarily small volume element of the
velocity space will tend to zero. The encounter effect is also
supporting the establishment of stationarity,  
smoothing the inhomogeneities in
particle distribution in velocity space.

Hence, if leaving aside the initial short stage of the system evolution,
when non-stationarity may be high,\footnote{This stage of evolution in
case of spherical systems was comprehensibly studied by \citet{Kurth:1951}. His
results concern the very beginning of the evolution, when the results of
the mixing process have not yet appeared. Kurth used his theory to analyse
contemporary globular clusters. It is difficult to agree with this kind of
application, because this initial stage of evolution is evidently finished
for these clusters long ago.} we can assume that the system is in nearly
stationary state, and the deviations are described by slow action of the
encounter effect, which changes step by step the structure of the system.

Having in mind approximately spherical form of most clusters, we may
suppose that our system also has the spherical symmetry.

In case of the spherical symmetry, the system density $\rho$ and thus the
potential $\Phi$ are functions of only the distance from system centre $r$
and time $t$
\be
\rho = \rho (r,t), ~~~~~\Phi = \Phi (r,t). \label{eq17.1.1} 
\ee
We may assume the velocity distribution to be symmetrical about the radial
direction. In this case $\psi$ and $\chi$ are functions of $v_t$, $v_r$,
$r$, $t$ only, while $v_t$ and $v_r$ are the tangential and radial
components of the particle velocity $\boldsymbol{v}$.

If the spherical system is precisely stationary, then the phase density is a
function of two integrals of motion -- the energy integral and the total
angular momentum integral. Thus instead of arguments $v_t$ and $v_r$ it is
suitable to use other arguments, supposing
\be
\psi = \psi (p,q,r,t), ~~~~~\chi = \chi (p,q,r,t), \label{eq17.1.2}
\ee
where $p$ and $q$ are related to velocity in following
way\footnote{Designations $p$ and $q$  here do not have any
relation with canonical variables. [Later footnote.]}
\be
2p = 2\Phi - v^2, ~~~~~2q = r^2v_t^2, \label{eq17.1.3}
\ee
or
\be
v^2 = 2(\Phi - p), ~~~~ v_t^2 = 2q/r^2, ~~~~ v_r^2 = 2(\Phi - p -
q/r^2). \label{eq17.1.4} 
\ee
The arguments $p$ and $q$ are negative energy and half of the square
of angular momentum per unit mass respectively. In spherically symmetric system
$q$ remains constant during the regular motion, and is thus an integral of
motion. In a stationary system also $p$ would be an integral of motion. In
this case $\psi$ would be a function of $p$ and $q$ only. In reality,
however, $p$ is not constant, and the speed of its variation for a regularly
moving point is described by the equation $\dd\Phi /\dd{t} - \boldsymbol{v}~\nabla\Phi =
\partial\Phi /\partial t.$ But as the system is nearly stationary,
$p$ varies very slowly and $\psi$ is a function of mainly $p$ and $q$.

The region of permitted values of $p$ and $q$ for given $r$ and $t$ is
limited by the condition $v_r^2 \ge 0$. Two values of $v_r$ correspond
to every possible set of $p$ and $q$ 
inside that region, being equal in size but
with the opposite sign, therefore the functions $\psi$ and $\chi$ are
two-valued. One branch of them corresponds to moving away from the centre
 ($v_r >0$), the other approaching to the centre
($v_r<0$). For $v_r=0$ both branches coincide.

For the particles belonging to the system, the region of permitted values
of $p$ and $q$ is limited also by the condition $p>0$, because the values
$p\le 0$ correspond to velocities equal or larger than the escape
velocity (assuming $\Phi =0$ for $r=\infty$). As a consequence of
encounters, the particles may obtain all possible values of $p$ and $q$, and
fill all the referred region. This means that the system radius is
infinite (when the system is isolated). But in this case some part of the
particles will have $p\le 0$. These particles do not belong to the system
any more and escape freely, as encounters act slowly. Hence their phase
density is nearly zero. Therefore, we may suppose $\psi >0$ for $p>0$, and
$\psi =0$ for $p\le 0$. $\psi$ must be smooth bounded function,
as encounters smooth the irregularities of particle distribution. The
encounter function may have for $p>0$ both positive and negative values.
For $p\le 0$ it is positive up to some maximum value obtainable by a
particle during encounter. This part of the encounter function describes
the escape of particles from the system -- the dissipation.

\section{The equation for the phase density}

In case of regular motion the volume of the phase space remains
constant (Liouville's theorem). Hence regular motion does not change the
phase density at comoving point. At comoving point the phase density
varies only because of stellar encounters. The variation of the phase
density is thus described by the equation 
\be
\frac{D\psi}{Dt} = \chi, \label{eq17.2.1}
\ee
where $D/Dt$ is the Stokes derivative for regular motion.

Equation (\ref{eq17.2.1}) is the well known kinetic Boltzmann equation. In stellar
dynamics this equation is usually used by assuming $\chi =0$, \ie 
neglecting the encounter effect. In the present case, it is essential 
to take it into account. In general form Eq.~(\ref{eq17.2.1}) was discussed
in stellar dynamics by \citet{Charlier:1917}, and later by
\citet{Chandrasekhar:1943a, Chandrasekhar:1943b, Chandrasekhar:1943c}
and \citet{Skabitskii:1950}. Mainly the
right side of the equation was analysed, but not the application of the equation as a
whole. Trying to simplify the equation, Skabitski as a first
approximation substituted 
the left side of Eq.~(\ref{eq17.2.1}) with $\partial\psi /\partial t$. 
However, this substitution is unjustified, because in this case the
effect of encounters is reduced to the variation of the velocity
distribution of particles only. In reality the variation of the spatial distribution
is as important.\footnote{Assuming $\partial\psi
/\partial t = \chi$ Skabitski derived $\partial\rho /\partial t = 0$,
expressing simply the conservation of the number and the
mass of stars when dissipation is absent. He didn't derive dissipation
because of the used forms of $\psi$ and $\chi$. A  similar error was made
by \citet{Minin:1952}. He in-explicitly assumed that the escape of particles
from a given volume element of the system influences only the density in
that volume element. In fact the influence spreads over all the
space available for escaped particles.} But despite of that Eq.~(\ref{eq17.2.1}) for
the spherical systems will have quite simple form.

Due to the spherical symmetry we may put $\psi$ and $\chi$
according to Eq.~(\ref{eq17.1.2}) into Eq.~(\ref{eq17.2.1}). Taking into account that in case of
regular motion $p$ varies with the speed $\partial\Phi /\partial t$, $r$
with the speed $v_r$ and $q$ remains constant we have the equation
\be
\frac{\partial\psi}{\partial t} + \frac{\partial\psi}{\partial p} \frac{\partial
\Phi}{\partial t} + \frac{\partial\psi}{\partial r} v_r = \chi .
\label{eq17.2.2}
\ee 
Because at non-moving point of the phase space $p$ varies also with the
speed $\partial\Phi /\partial t$ (see Eq.~(\ref{eq17.1.3})), and $q$ and $r$ remain
constant, the sum of first two terms on left side of the equation is the
speed of phase space variation at non-moving point of the phase space. Hence
the equation may be written in form
\be
\left( \frac{\partial\psi}{\partial t} \right) + \frac{\partial\psi}{
\partial r} v_r = \chi , \label{eq17.2.2a}
\ee
where $(\partial /\partial t)$ is the time derivative at non-moving point
of the phase space. Equation (\ref{eq17.2.2a}) differs from Skabitski's first 
approximation equation by including the second term on the  left side. This
term takes into account the redistribution of particles in the physical
space. 

Let us take now into account the approximate stationarity of the system.

Equation (\ref{eq17.2.2}) concerns both branches of the phase density and the
encounter function ($v_r>0$ and $v_r<0$). Designating the half-sums of
both branches of $\psi$ and $\chi$ as $\Psi$ and $X$ and their
half-differences as $\Delta\psi$ and $\Delta\chi$, we may replace
Eq.~(\ref{eq17.2.2a}) with the pair of equations
\be
\left( \frac{\partial\Psi}{\partial t} \right) + \frac{\partial\Delta\psi }{
\partial r} |v_r| = X, ~~~~~ \left( \frac{\partial\Delta\psi}{\partial t}
\right) + \frac{\partial\Psi}{\partial r} |v_r| = \Delta\chi . \label{eq17.2.3}
\ee
In collisionless system the stationarity would be strict, $\Psi$ would
not depend explicitly on $r$, and $\Delta\psi$ would be zero. In reality, the
system is not strictly stationary. But because it is close to stationarity
and the encounter effect is very small, also $(\partial\psi /\partial t)$
and $\chi$ are very small and moreover $(\partial\Delta\psi /\partial t)$
and $\Delta\chi$ are very small. Therefore, on the basis of Eq.~(\ref{eq17.2.3}) we
may suppose
\be
\Psi = \Psi (p,q,t) . \label{eq17.2.4}
\ee
Now we may eliminate from the first equation of (\ref{eq17.2.3}) the function
$\Delta\psi$ and the argument $r$. To do this we first average the terms of
the equation with the weight $\rmd r/|v_r|$ over all possible values of $r$ for
given $p$, $q$, $t$. After averaging the second term on left side of the
equation will vanish, because at bounds of $r$ we have $\Delta\psi =0$. we replace 
the quantities $\partial\Phi /\partial t$ and $X$ in the first term (in long form) and on the 
right side of the equation with their
averaged values $\overline{\partial\Phi /\partial t}$ and $\overline{X}$, which 
are some functions of only $p$, $q$, $t$. Hence we have [q.v. Appendix A]
\be
\frac{\partial\Psi}{\partial t} + \frac{\partial\Psi}{\partial p}
\frac{\overline{\partial\Phi} }{\partial t} = \overline{X} , \label{eq17.2.5}
\ee
or in other form (we use it further)
\be
\left( \frac{\partial\Psi}{\partial t}\right) - \frac{\partial\Psi}{\partial
p} \left( \frac{\partial\Phi}{\partial t} - \frac{\overline{\partial\Phi} }{
\partial t} \right) = \overline{X} . \label{eq17.2.5a} 
\ee
In these equations
\be
\tau \frac{\overline{\partial\Phi} }{\partial t} = \int_{r_1}^{r_2} 
\frac{\partial\Phi}{\partial t} \frac{\rmd r}{|v_r|} , ~~~~~
\tau \overline{X} = \int_{r_1}^{r_2} X \frac{\rmd r}{ |v_r|} , \label{eq17.2.6}
\ee
while
\be
\tau = \int_{r_1}^{r_2} \frac{\rmd r}{|v_r|}. \label{eq17.2.7}
\ee
Here $v_r$ was substituted according to Eq.~(\ref{eq17.1.4}), $r_1$ and $r_2$ are the
roots of the equation $v_r = 0$ for given $p$, $q$, $t$.

As $\rmd r/v_r = \rmd t$ and $p$ vary very slowly in case of regular motion
($q$ remains constant), we may handle the averaging, used in deriving
Eq.~(\ref{eq17.2.5}), as  averaging over time at regularly moving point.
In this case the averaging spreads over the time interval when $r$ changes
from $r_1$ to $r_2$. The referred time-interval (it is the half-period of
particle oscillation along $r$) evidently equals to $\tau$. The averaging
can be interpreted in another way, namely as the averaging over the phase
space for given $p$, $q$, $t$. This results because the phase space volume
element is $2\pi v_t~\rmd v_t \rmd v_r\cdot 4\pi r^2 \rmd r$ or
according to (\ref{eq17.1.4}) $8\pi \rmd p \rmd q\cdot \rmd r/|v_r|$. 

We can use Eq.~(\ref{eq17.2.5}) for real calculations to
study the variation of the phase density due to encounters. According
to this equation at a point moving in $p$,$q$ plane parallel to the axis
$p$ with the velocity $\rmd p/\rmd t = \overline{\partial\Phi /\partial t}$, the
variation speed of $\Psi$ equals to $\overline{X}$. If $\Psi$ is given at
a certain initial moment of time, and if we are able to calculate
$\overline{\partial\Phi /\partial t}$ and $\overline{X}$ from known
$\Psi$, then the equation enables to find $\Psi$ at any moment of time.

Equation (\ref{eq17.2.5}) can be supplemented by the equation allowing to find
$\Delta\psi$. Comparing Eq.~(\ref{eq17.2.5a}) with the first equation of (\ref{eq17.2.3}) we
have 
\be
\frac{\partial\Delta\psi}{\partial r} |v_r| = X - \overline{X} - 
\frac{\partial\Psi}{\partial p} \left( \frac{\partial\Phi}{\partial t} -
\frac{\overline{\partial\Phi}}{\partial t} \right) \label{eq17.2.8}
\ee
Because $X$ and $\partial\Phi /\partial t$ are very small, also
$\Delta\psi$ is very small compared to $\Psi$. Hence $\Psi$ and $X$
nearly coincide with $\psi$ and $\chi$.

\section{The equation for the potential. Initial conditions}

The function $\Psi (p,q,t)$ describes the spatio-kinematical 
structure of our system of particles only when we know its potential $\Phi
(r,t)$, because $p$ is related to the velocity $\boldsymbol{v}$ via potential.
Therefore, to study the evolution of the system, we must find not only the
evolution of $\Psi$, but also the evolution of $\Phi$. In addition, as we
saw, the evolution of $\Psi$ can not be determined without knowing the
evolution of $\Phi$. We must keep the potential in mind also when
establishing the initial conditions for our problem. Giving the function
$\Psi$ at some initial moment we must also give the corresponding function
$\Phi$.

It is possible to find the relation between $\Psi$ and $\Phi$, and the equation for
$\partial\Phi /\partial t$, starting with Poisson's
equation. Having in mind the spherical symmetry, the Poisson's equation is
\be
\frac{\partial^2}{\partial r^2} (r\Phi ) + 4\pi Gr\rho (r,t) = 0,
\label{eq17.3.1} 
\ee
where $G$ is the gravitational constant. If under the phase
density we assume the phase mass density, then the mass density in physical space
$\rho$ is
\be
\rho = \int_{(V)} \psi \rmd V, \label{eq17.3.2} 
\ee
where $V$ is the volume element in velocity space and integration is over
all that space. Because $\rmd V =$ $2\pi v_t~\rmd v_t \rmd v_r =$ $2\pi \rmd p \rmd q /r^2
|v_r|$ (see Eq.~(\ref{eq17.1.4})) and the sum of both branches of $\psi$ is $2\Psi$,
Eq.~(\ref{eq17.3.2}) has now the form
\be
\rho (r,t) = P(\Phi (r,t),r,t) = \frac{4\pi}{r^2} \iint \Psi
\frac{\rmd p \rmd q}{|v_r|} , \label{eq17.3.3} 
\ee
where integration is over all $p$ and $q$ satisfying the condition $v_r^2
\ge 0$. The argument $t$ is contained in the function $P$ via $\Psi$, the
arguments $\Phi$ and $r$ via $v_r$ (see Eq.~(\ref{eq17.1.4})). Substituting
Eq.~(\ref{eq17.3.3}) into Eq.~(\ref{eq17.3.1}) we derive the equation
\be
\frac{\partial^2}{\partial r^2} (r\Phi ) + 4\pi GrP(\Phi ,r,t) = 0. 
\label{eq17.3.4} 
\ee
Equations (\ref{eq17.3.3}) and (\ref{eq17.3.4}) just relate the functions $\Psi$ and $\Phi$. We
can use these equations if the initial conditions for our problem are
given. 

As the argument $t$ in Eqs.~(\ref{eq17.3.3}) and (\ref{eq17.3.4}) is not essential and
$\Psi$ is independent of $r$, they coincide with equations describing the
stationary spherical stellar system. To have a physically meaningful model,
the function $\Psi$ must be non-negative and go to zero for $p\le 0$.
The potential $\Phi$ must satisfy the boundary conditions $r\Phi = 0$ for
$r=0$ and $r \partial (r\Phi )/\partial r =0$ for $r=\infty$. The first of
them results from the demand to have zero mass at the centre of the system, the
second -- from the demand of the finite total mass of the system and from the
condition $\Phi =0$ for $r=\infty$. We may start the model construction
by giving the function $\Psi (p,q)$ and taking into account the
considerations above. Then after finding the function $P(\Phi ,r)$ from
Eq.~(\ref{eq17.3.3}), we derive $\Phi (r)$ from Eq.~(\ref{eq17.3.4}). But we may also give first
the function $\Phi (r)$ and the function $\rho (r)$ related to it. We find the
function $\Psi$ with the help of Eq.~(\ref{eq17.3.3}). The relation between
$\Psi$ and $\Phi$ is not one-valued. Because the functions $\Phi$ and
$\rho$ determine the function $P(\Phi ,r)$ only along the curve $\Phi
=\Phi (r)$, then $\Phi$ evidently does not determine $\Psi$ in a single way.
But it may result that also $\Psi$ does not determine $\Phi$ in one-valued
form. This occurs when the referred boundary conditions for $\Phi (r)$
happen to be insufficient for one-valued determination.

The models of stationary spherical stellar systems were studied by
different authors. As an example, already \citet{Jeans:1916} and \citet{Eddington:1916}
studied the models with spherical velocity distribution and with the density
distribution similar to the density distribution of polytropic gaseous
sphere. Afterwards these kind of models were studied by \citet{Camm:1952} who
generalised them for the case of ellipsoidal velocity distribution. Also
other models were analysed (see for example \citet{Woolley:1954a,
  Woolley:1956}. Usually in the model 
construction the function $\Psi (p,q)$ was given and the potential
$\Phi$ was found as the solution of Eq.~(\ref{eq17.3.4}). But already Eddington
mentioned that it is possible to start from the function $\rho (r)$ (or
$\Phi (r)$). He analysed the case when $\Psi$ is independent of $q$, \ie 
when the velocity distribution is spherical. In this case $P$ is
independent of $r$ and Eq.~(\ref{eq17.3.3}) turns into Abel integral equation (after
integration over $q$).

By giving the initial conditions for our problem, it is suitable to
use the models, where the encounters cause smooth variations without
changing the model structure in a radical way. In such model $\Psi$
must be a smooth bounded function, vanishing only for $p\le 0$. Hence
the radius of the model must be infinite. The mass of the model must
be surely finite.  From the models studied up to now, only one
satisfies these conditions.  This is the well known model with
$\Psi\sim p^{7/2}$, and the Schuster density distribution thoroughly
studied already by \citet{Eddington:1916} and recently by \citet{Skabitskii:1950}
and \citet{Minin:1952}. But even this model is not completely suitable for
us. If we use in calculations of an encounter function the approximate
formulas given further in Section 5, we must suppose that
$\partial\Psi /\partial p$ remains finite when $p$ decreases to
zero. This condition is in contradiction with the law $\Psi\sim
p^{7/2}$, and is moreover incompatible with the spherical velocity
distribution. On the basis of Eq.~(\ref{eq17.3.3}), on can demonstrate that
within this condition the system mass is finite ($\Phi\sim r^{-1}$,
$\rho r^3\rightarrow 0$ for $r\rightarrow\infty$) only when $\sqrt{q}
~\partial\Psi /\partial p\rightarrow 0$ for $p\rightarrow 0$ and
$q\rightarrow\infty$.\footnote{It means that for small $p$ the
function $\Psi$ decreases with $q$, \ie  in outer regions of spherical
stellar systems the velocity distribution becomes radially
elongated. [Later footnote.]} Hence, to give the initial conditions we
can not use the developed models but need to construct new ones.

Now we discuss the equation for $\partial\Phi /\partial t$. It was obtained
from Poisson's equation. Differentiating (\ref{eq17.3.1}) with respect to $t$ we have
\be
\frac{\partial^2}{\partial r^2} \left( r \frac{\partial\Phi}{\partial t}
\right) + 4\pi Gr \frac{\partial\rho}{\partial t} = 0 . \label{eq17.3.5}
\ee
We must substitute into this equation
\be
\frac{\partial\rho}{\partial t} = \int_{(V)} \left( \frac{\partial\psi}{
\partial t} \right) \rmd V , \label{eq17.3.6}
\ee
where $(\partial /\partial t)$ again means differentiating with respect to
$t$ in non-moving point of the phase space. Going to variables $p$, $q$,
adding both branches of $\Psi$ and substituting $(\partial\Psi /\partial
t)$ according to (\ref{eq17.2.5a}) we find
\be
\frac{\partial^2}{\partial r^2} \left( r \frac{\partial\Phi}{\partial t}
\right) + 4\pi Gr \left[ F\left( \frac{\partial\Phi}{\partial t} - 
\frac{\overline{\overline{\partial\Phi}} }{\partial t} \right) + H \right]
= 0, \label{eq17.3.7}
\ee
where
\be
\ba{ll}
F &=  \frac{4\pi}{r^2} \iint \frac{\partial\Psi}{\partial p} \frac{\rmd p \rmd q }{
|v_r|} , \\
\noalign{\smallskip}
F \frac{\overline{\overline{\partial\Phi}} }{\partial t} &=  \frac{4\pi}{ r^2}
\iint \frac{\partial\Psi}{\partial p} \frac{\overline{\partial\Phi}}{
\partial t} \frac{\rmd p \rmd q }{ |v_r|} ,  \\ 
\noalign{\smallskip}
H &=  \frac{4\pi}{ r^2} \iint \overline{X} \frac{\rmd p \rmd q }{ |v_r|} . 
\ea
\label{eq17.3.8}
\ee
The same result can be derived when differentiating Eq.~(\ref{eq17.3.4}) with respect
to $t$. 

The function $\overline{\overline{\partial\Phi /\partial t}}$ in Eq.~(\ref{eq17.3.7})
is twice averaged $\partial\Phi /\partial t$. But double averaging turns to
single averaging. Substituting $\overline{\partial\Phi /\partial t}$
according to Eq.~(\ref{eq17.2.6}) into the second formula of (\ref{eq17.3.8}), and changing the
order of integration, we may write
\be
rF \frac{\overline{\overline{\partial\Phi}} }{ \partial t} =
\int_0^{\infty} r' \frac{\partial\Phi ' }{\partial t} f(r,r') \rmd r' ,
\label{eq17.3.9}
\ee 
where
\be
f = \frac{4\pi}{ r r'} \iint \frac{\partial\Psi}{\partial p} \tau^{-1} 
\frac{\rmd p \rmd q }{ |v_r v_r'|} , \label{eq17.3.10}
\ee
while apostrophe in $\partial\Phi /\partial t$ and $v_r$ means that the
argument $r$ is replaced with the argument $r'$, and integration in
Eq.~(\ref{eq17.3.10}) is over all $p$ and $q$ within the conditions $v_r^2\ge 0$,
$v_r'^2\ge 0$. Equations (\ref{eq17.3.9}) and (\ref{eq17.3.10}) were written in form, where the
first of them includes the term $r' \partial\Phi '/\partial t$ in a
similar way as (\ref{eq17.3.7}). But surely the term $\partial\Phi '/\partial t$ is
averaged, not $r'\partial\Phi '/\partial t$.

If the functions $\Psi$, $\Phi$, and $X$ are known at a given moment of
time, then we may find $F$ and $H$ as functions of $r$, and $f$ as
function of $r$ and $r'$. Equation (\ref{eq17.3.7}) can be solved as the equation for
$r\partial\Phi /\partial t$. To solve it we use the boundary conditions
$r\partial\Phi /\partial t = 0$ for $r=0$ and $\partial (r\partial\Phi
/\partial t)/\partial r = 0$ for $r=\infty$.

Equation (\ref{eq17.3.7}) together with (\ref{eq17.2.5}) form a set of equations. Knowing the
functions $\Psi$, $\Phi$ and $X$ at initial moment of time we may find
according to Eq.~(\ref{eq17.3.7}) $\partial\Phi /\partial t$. Calculating thereafter
$\overline{\partial\Phi /\partial t}$ we will find $\partial\Psi /\partial
t$ from Eq.~(\ref{eq17.2.5}). Thus the functions $\Psi$ and $\Phi$ are known for the
subsequent moment of time. Calculating also the encounter function $X$ we
can find again $\partial\Phi /\partial t$ and thereafter $\partial\Psi
/\partial t$. In that way, solving step by step Eqs.~(\ref{eq17.2.5}) and (\ref{eq17.3.7}) we
may determine $\Psi$ and $\Phi$ at any moment of time.

\section{Calculation of the encounter function}

Above we derived the equations allowing to follow the evolution of a
spherical system of gravitating particles caused by the encounter effect.
Now we begin to deal with calculations of the encounter function. From
technical viewpoint this is perhaps the most difficult part of our
problem. 

As our system consists of big number of particles, the most
important encounters are the ones having small impact parameter compared to 
the orbit's dimensions. For the same reason the encounter time is short when
compared with revolution time. Hence the encounter effect can be assumed to
be similar to the elastic collision of particles (with the difference that
during the encounters we have a completely different distribution of deflection
angles). And besides, the encounters can be handled within sufficient
precision as the encounters between two bodies. True, the encounters with
impact parameter, exceeding the mean distance between particles, seem to
resemble multiple collisions. But due to the weakness of
interactions in case of distant encounters these interactions add together in
a way, which is similar to a series of two-body 
encounters.\footnote{Quite often it was assumed  following \citet{Charlier:1917} and
others, that encounters with impact parameter exceeding the
mean distance between stars have little effect, and they can be neglected.
In fact the contribution of these encounters to the encounter
function is quite significant. In statistical physics, when discussing the
interaction between charged particles, these encounters were taken into
account by using two-body encounter model. This was done for example in
a paper by \citet{Landau:1937}. Recently the need to take into account distant
encounters was mentioned again by \citet{Cohen:1950}.}

If treating the encounter of particles as two-body elastic collisions,
the velocity vector of the centre of mass and the values of
particle velocities in respect to each other must remain unchanged
during the encounter. If $\boldsymbol{v}_1$ and $\boldsymbol{v}_2$ are the velocities of
particles with masses $m_1$ and $m_2$, then these velocities must change
is a way that the centre of mass in velocity space remains in place, and the
vector of relative velocity of particles $\boldsymbol{w}$ keeps its value but
rotates by some angle $\varphi$ -- the deflection of the motion direction.
The value and the plane of the angle depend on the impact parameter $D$
and on the impact azimuthal angle $\vartheta$. The behaviour of the particle
velocities with respect to the centre of mass $\boldsymbol{u}_1$ and $\boldsymbol{u}_2$ are
similar to $\boldsymbol{w}$. During the encounter the particles relocate their
positions in velocity space, remaining in diametrically opposite points on
the spheres described around the centre of mass by radii
\be
u_1 = \frac{m_2}{ m_1 + m_2} w , ~~~~~
u_2 = \frac{m_1 }{ m_1 + m_2} w. \label{eq17.4.1} 
\ee
If we take the positions of particles on the referred spheres before the
encounter to be the poles of spherical coordinates, then after the
encounter the spherical coordinates of their positions are $\varphi$,
$\vartheta$. 

During the encounter the particle velocities $\boldsymbol{v}_1$ and $\boldsymbol{v}_2$
turn into some new velocities $\boldsymbol{v}'_1$ and $\boldsymbol{v}'_2$,, which can be
calculated from $\varphi$ and $\vartheta$. On the other side, during some
particular encounter the velocities $\boldsymbol{v}'_1$ and $\boldsymbol{v}'_2$ may turn
again into 
$\boldsymbol{v}_1$ and $\boldsymbol{v}_2$. Let us designate the phase number densities of
particles with masses $m_1$ and $m_2$ as $\psi_1$ and $\psi_2$ for $\boldsymbol{v}_1$ and $\boldsymbol{v}_2$. For $\boldsymbol{v}'_1$ and $\boldsymbol{v}'_2$ the phase number
densities are $\psi_1'$ and $\psi_2'$. Further, let us designate the
function of relative deflection frequency per unit of solid angle as
$\Omega$, and the effective interaction radius of particles in physical
space as $D_0$. If $\boldsymbol{v}_1$ and $\boldsymbol{v}_2$ lie within the volumes $\rmd V_1$
and $\rmd V_2$ of the velocity space respectively, and the deflection of the
motion direction lies within the solid angle $\rmd\omega$ (=$\sin\varphi
\rmd\varphi \rmd\vartheta$), then the encounters remove from the volume $\rmd V_1$
per unit time the number of particles with masses $m_1$, equal to 
$\psi_1 \rmd V_1\cdot \psi_2 \rmd V_2\cdot \pi D_0^2\omega\cdot \Omega\cdot
\rmd\omega$. On the other side, if $\boldsymbol{v}'_1$ lies within the solid angle,
taken with respect to the centre of mass, and $\boldsymbol v'_2$ within the volume
$\rmd V'_2$, then the encounters add to the volume $\rmd V_1$ per unit of time the
number of particles equal to $\psi_1 \rmd\omega\cdot \psi_2 \rmd V'_2\cdot \pi
D_0^2\omega\cdot \Omega\cdot \rmd V_1$. Besides, if we suppose that in both
cases the center of mass lies within the same volume, then $\rmd V'_2 = \rmd V_2$.
Designating the encounter function of particles $m_1$ with respect to
particles $m_2$ as $\chi_{1,2}$, we derive for $\chi_{1,2}$ after the
integration the  well 
known Boltzmann expression\footnote{In stellar dynamics the
expression for the encounter function was derived first by \citet{Charlier:1917}.}
\be
\chi_{1,2} = \iint (\psi_1'\psi_2' - \psi_1\psi_2)\cdot \pi D_0^2
\omega\cdot \Omega \rmd\omega \rmd V_2 . \label{eq17.4.2}
\ee
Integration is over all the velocity $\boldsymbol{v}_2$ space and over all
deflections. 

The encounter function $\chi_{1,2}$ can be calculated also from the flux
of the particles with mass $m_1$ per unit surface perpendicular to the
flux. If we designate the vector of this flux $\boldsymbol{i}_{1,2}$ then from the
continuity equation in velocity space we have
\be
\chi_{1,2} = -\nabla\boldsymbol{i}_{1,2} , \label{eq17.4.3} 
\ee
where $\nabla$ must be taken in velocity space. Let us assume that
particle displacement along the great circles of the spheres $u= \mathrm{const}$ and
let the number of particles with deflection angles greater or equal to
$\varphi$ be $2\pi\sin\varphi\cdot J$. Taking again $\boldsymbol{v}'_1$ in solid
angle $\rmd\omega$ with respect to the centre of mass and $\boldsymbol{v}'_2$ in volume
$\rmd V'_2$, we find that the flux of particles through the perpendicular
surface element $\rmd S$ in vicinity of $\boldsymbol{v}_1$ equals to $\psi_1' \rmd\omega
\cdot \psi_2 \rmd V'_2\cdot \pi D_0^2\omega\cdot u_1J\cdot \rmd S$, while again we
can suppose $\rmd V'_2 = \rmd V_2$. Now we find the expression for $\boldsymbol{i}_{1,2}$

\be
\boldsymbol{i}_{1,2} = \iint\psi_1'\psi_2'\cdot \pi D_0^2\omega\cdot
u_1\boldsymbol{J} \rmd\omega \rmd V_2, \label{eq17.4.4} 
\ee
where the vector $\boldsymbol{J}$ must be directed perpendicular to the vector
$\boldsymbol{w} = \boldsymbol{v}_2 - \boldsymbol{v}_1$, with the azimuthal angle perpendicular to
the azimuthal angle of $\boldsymbol{v}'_1$. Besides, according to the definition
of $J$ we have

\be
J = \frac{1}{\sin\varphi} \int_{\varphi}^{\pi} \Omega \sin\varphi 
\rmd\varphi . \label{eq17.4.5}
\ee

Surely, it is easier to calculate $\chi_{1,2}$ from Eq.~(\ref{eq17.4.2}), but in
next Section, where we derive approximate formula for $\chi_{1,2}$, it is more
convenient to use Eqs.~(\ref{eq17.4.3}) and (\ref{eq17.4.4}).

As $\psi_1$ and $\psi_2$ can be handled as functions of $p$, $q$,
$r$, $t$ Eq.~(\ref{eq17.4.2}) or Eqs.~(\ref{eq17.4.3}) and (\ref{eq17.4.4}) must give $\chi_{1,2}$ as a
function of the same arguments. Besides, as $\psi_1$ and $\psi_2$
were two-valued functions of $p$ and $q$, the function $\chi_{1,2}$ is
also two-valued. But the difference between two branches of $\psi_1$ and
$\psi_2$ ($v_r>0$ and $v_r<0$) is not large. Hence, when we substitute the
functions $\psi_1$ and $\psi_2$ with half-sums of two branches in both of
them, we derive within sufficient precision the half-sum of both branches
of $\chi_{1,2}$. Just in this sense we need to know the encounter function
when using the equations of previous Sections. But Eq.~(\ref{eq17.4.2}) and
Eqs.~(\ref{eq17.4.3}) and (\ref{eq17.4.4}) do not give directly the total encounter function
$\chi$. To have $\chi$ as the variation of particle $m_1$ number phase
density, we must integrate $\chi_{1,2}$ over all values of $m_2$. If we are
interested in $\chi$ as the variation of the total mass phase density (in
this sense the encounter function was used in Sect.~3), we must multiply
the previous encounter function by $m_1$ and integrate over $m_1$.

In order to use Eqs.~(\ref{eq17.4.2}) and (\ref{eq17.4.4}), it is necessary to have some concrete
expressions for $\Omega$ and $J$. They can be derived if the deflection
angle $\varphi$ as a function of impact parameter $D$ is known. Functions
$\Omega$ and $J$ are related to $D$ and $\varphi$ in the following way
\be
\Omega = - \frac{1}{\pi\sin\varphi} \frac{D}{ D_0}~ \frac{\rmd D}{\rmd\varphi} , 
~~~~~ J = \frac{1}{2\pi\sin\varphi} \frac{D^2}{D_0^2} . \label{eq17.4.6}
\ee
When $D$ is not large compared to orbit dimension, the
relation between $D$ and $\varphi$ can be taken as in two-body
problem. For larger $D$ our formulas are not valid. Although, as the
role of very distant encounters is relatively small, we may approximately
use our formulas for all encounters with $D$ up to orbital dimensions. We take $D_0$ for the
orbital dimension. The precise value of $D_0$ is
not important, because the encounter effect depends only very weakly on
$D_0$ (logarithmically). It is natural to assume that $\varphi\rightarrow 0$
for $D\rightarrow D_0$. In this case, $D$ can be assumed as a function of
$\varphi$ in form of the following simple interpolation
formula 
\be
D^2 = D_0^2 \frac{1+\cos\varphi}{2} \frac{\lambda -1}{ \lambda -
\cos\varphi} , \label{eq17.4.7} 
\ee
where
\be
\lambda -1 = \frac{2G^2 (m_1+m_2)^2 }{ D_0^2 w^4}. \label{eq17.4.8} 
\ee
Equation (\ref{eq17.4.7}) differs from the analogous formula for two-body problem only
because instead of unity there is $\lambda$ in the denominator of its second term. 

By inserting (\ref{eq17.4.7}) into (\ref{eq17.4.6}) we derive the expressions for functions $\Omega$
and $J$. The expression for $J$ is straightforward, the expression for $\Omega$ is
\be
\Omega = \frac{1}{4\pi} \frac{\lambda^2 -1 }{ (\lambda - \cos\varphi )^2} .
\label{eq17.4.9} 
\ee
In case of great number of gravitating particles the right side of (\ref{eq17.4.8})
is in general very small, and thus $\lambda$ is close to one. Hence
Eq.~(\ref{eq17.4.9}) gives very large values of $\Omega$ for small $\varphi$, \ie 
large relative frequency of small deflections. This is the well known
property of gravitational encounters, being highly different from elastic
collision of spherical particles, for which, as it is known, the deflection
distribution is uniform. Equation (\ref{eq17.4.9}) gives uniform distribution when
$\lambda$ is large, meaning that $m_1+m_2$ must be large or $w$ must be
small. In this case the deflection angle distribution will be really close to uniform.

\section{Approximate formula for the encounter function}

It is quite complicated and time consuming to 
compute the encounter function on the basis of formulas above. The computations can be
significantly simplified by taking into account that in most
cases displacement of particles in velocity space due to encounters is very small. 
This allows us to calculate approximately $\chi_{1,2}$ by expanding
$\psi_1$ and $\psi_2$ in Taylor series and keeping only the first terms. 

Let us use Eqs.~(\ref{eq17.4.3}) and (\ref{eq17.4.4}) to calculate
$\chi_{1,2}$. Keeping in expansion of $\psi_1'\cdot\psi_2'$ 
only linear terms and rejecting the
terms which disappear in integration over $\vartheta$, we derive Eq.~(\ref{eq17.4.4})
in the following form
\be
\boldsymbol{i}_{1,2} = \iint (u_2\psi_1\nabla_J\psi_2 -
u_1\psi_2\nabla_J\psi_1 ) \sin\varphi \cdot \pi D_0^2w \cdot u_1 J \rmd\omega
\rmd V_2, \label{eq17.5.1} 
\ee
where vector $\nabla_J$ is the component of $\nabla$ along $\boldsymbol{J}$.
Equation (\ref{eq17.5.1}) is similar to the equation derived by \citet{Landau:1937} and
used by \citet{Skabitskii:1950}.

Let us integrate (\ref{eq17.5.1}) over the deflection angles. With the help of
Eqs.~(\ref{eq17.4.1}), (\ref{eq17.4.6})--(\ref{eq17.4.8}) we find
\be
\boldsymbol{i}_{1,2} = \pi G^2m_2 \int (m_1\psi_1\nabla_{\perp}\psi_2 - 
m_2\psi_2\nabla_{\perp}\psi_1 ) L w^{-1} \rmd V_2 , \label{eq17.5.2} 
\ee
where
\be
L = \frac{1}{\lambda -1} \int J\sin\varphi \rmd\omega = \frac{1}{\lambda -1}
\int \Omega (1-\cos\varphi ) \rmd\omega = \frac{\lambda +1}{2} \ln \frac{\lambda +
1 }{ \lambda -1} -1  \label{eq17.5.3} 
\ee
and vector $\nabla_{\perp}$ is the component of $\nabla$ perpendicular to
$\boldsymbol{w} = \boldsymbol{v}_2 - \boldsymbol{v}_1$. As $w^2\nabla_{\perp} = (w^2 - \boldsymbol{w}
\boldsymbol{w} )\nabla$, from (\ref{eq17.5.2}) results the following equation for $\boldsymbol{i}_{1,2}$ 
\be
\boldsymbol{i}_{1,2} = \pi G^2 m_2 (m_1\boldsymbol{b} \psi_1 - m_2 {\bf B} \nabla\psi_1
). \label{eq17.5.4} 
\ee
Here
\be
\ba{ll}
{\bf B} = & \int (w^2 -\boldsymbol{w}\boldsymbol{w})w^{-3}L\psi_2 \rmd V_2, \\
\noalign{\smallskip}
\boldsymbol{b} = & \nabla {\bf B} = 2\int \boldsymbol{w} w^{-3}L\psi_2 \rmd V_2,
\ea
\label{eq17.5.5}
\ee
while the expression for $\boldsymbol{b}$ was derived under the assumption that
$L$ is independent of the direction of $\boldsymbol{w}$.

As the velocity distribution is symmetrical about the radial
direction, vectors $\boldsymbol{b}$ and $\nabla\psi_1$ in Eq.~(\ref{eq17.5.4}), and also two
main axes of tensor $\bf B$ lie on the plane parallel to $\boldsymbol{v}$ and
$\boldsymbol{v}_t$. Vector $\boldsymbol{i}_{1,2}$ lies on the same plane. If in addition
the velocity distribution is spherically symmetric, vectors $\boldsymbol{b}$
and $\nabla\psi_1$ are parallel to $\boldsymbol{v}$. In this case also
one of the main axis of tensor $\bf B$  
lies in the same direction, while the other two main
axes perpendicular to $\boldsymbol{v}$ are equal to each other. Vector
$\boldsymbol{i}_{1,2}$ is also parallel to $\boldsymbol{v}$. 

Using the relation between $\bf B$ and $\boldsymbol{b}$ we may rewrite (\ref{eq17.5.4}) in
form 
\be
\boldsymbol{i}_{1,2} = \pi G^2 m_2 [(m_1+m_2)\boldsymbol{b}\psi_1 - m_2\nabla {\bf
B}\psi_1] . \label{eq17.5.4a} 
\ee
Equation (\ref{eq17.5.4a}) along with (\ref{eq17.4.3}) can be seen as the diffusion equation
of particles in velocity space [q.v. Appendix B]. It is similar to
\citet{Chandrasekhar:1943a, Chandrasekhar:1943b, Chandrasekhar:1943c}
diffusion equation. It is easy to see that the 
first term in (\ref{eq17.5.4a}), which is similar to the corresponding term in
Chandrasekhar's equation, accounts for the systematic variation of $\boldsymbol{v}_1$
or the dynamical friction. Indeed, the encounters of particles $m_1$ and
$m_2$ cause the systematic variation of $\boldsymbol{v}_1$, equal to
$-\boldsymbol{u}_1(1-\cos\varphi )$ per one encounter,  or  
$- \psi_2 \rmd V_2 \cdot \pi D_0^2w \cdot \Omega \rmd\omega \cdot \boldsymbol{u}_1
(1-\cos\varphi )$ in volume $\rmd V_2$, solid angle $\rmd\omega$ and unit
time. We find after integration that systematic variation of
$\boldsymbol{v}_1$ per unit time is equal to $\pi G^2m_2
(m_1+m_2)\boldsymbol{b}$ (see (\ref{eq17.4.1}), 
(\ref{eq17.5.3}) and (\ref{eq17.5.5})). Assuming the spherical symmetry of $\psi_2$, and that $L$
is independent of $w$, our expression for dynamical friction coincides
with Chandrasekhar's expression, derived under the same assumptions. As it
is seen from (\ref{eq17.5.5}), the method used to calculate the
$\boldsymbol{b}$ in case of $\rmd L/\rmd w =0$ 
is similar to the method for calculating the gravitation acceleration, where
instead of density we have $\psi_2$, and instead of the gravitational
constant we have $-2L$. Hence in case of $\rmd L/\rmd w =0$ and spherical
symmetry of $\psi_2$, the dynamical friction depends only on $\psi_2$ for
$v_2\le v_1$, just as it was derived by Chandrasekhar.

The second term in (\ref{eq17.5.4a}) accounts (approximately) for the
fluctuating part of $\boldsymbol{v}_1$ variation. It is not difficult
to see that $2\pi G^2m_2^2{\bf B}$ is (within sufficient precision)
the tensor of second moments of $\boldsymbol{v}_1$ variation per unit
time. Meanwhile, we mention that the fluctuating term must be
proportional to $\nabla {\bf B}\psi_1$ and not ${\bf B}\nabla\psi_1$,
as it was taken by Chandrasekhar. Invalidity of Chandrasekhar's
expression is easy to show in case of $m_1\rightarrow 0$. If
$m_1\rightarrow 0$ and if among particles $m_1$ and $m_2$ there is
statistical equilibrium, then $\nabla\psi_1\rightarrow 0$. Just this
results from Eq.~(\ref{eq17.5.4a}) with
$\nabla {\bf B} = \boldsymbol{b}$ (or from Eq.~(\ref{eq17.5.4})). But
if we replace the second term in (\ref{eq17.5.4a}) with the one
proportional to ${\bf B}\nabla\psi_1$, we obtain the wrong conclusion
that $\nabla\psi_1$ is finite (because $\bf B$ is
finite).\footnote{The diffusion equation was treated correctly in
  paper by \citet{Cohen:1950}. However, we can not agree with their
  opinion that equation, similar to (\ref{eq17.4.2}), can not be used in
  case of distant encounters. In reality, the diffusion equation
  results as a consequence of that formula (or of equivalent formulas
  (\ref{eq17.4.3}) and (\ref{eq17.4.4}). Their remark that Landau's
  equations \citep{Landau:1937} do not allow to account the dynamical
  friction is also misleading.}

Inserting (\ref{eq17.5.4}) (or (\ref{eq17.5.4a})) into
(\ref{eq17.4.3}), and accounting for the relation between $\bf B$ and
$\boldsymbol{b}$, we derive for the encounter function the following
expression
\be
\chi_{1,2} = \pi G^2m_2 [ m_1a\psi_1 + (m_2-m_1)\boldsymbol{b}\nabla\psi_1 +
m_2{\bf B}\nabla\nabla\psi_1 ] , \label{eq17.5.6} 
\ee
where
\be
a= -\nabla\boldsymbol{b} = 2\int \frac{\rmd L}{\rmd w} w^{-2}\psi_2 \rmd V_2 \label{eq17.5.7} 
\ee
(or $a= 8\pi L\psi_2$ in case of $\rmd L/\rmd w =0$). Equation (\ref{eq17.5.6}) together with
(\ref{eq17.5.5}) and (\ref{eq17.5.7}) can be derived also from Eq.~(\ref{eq17.4.2}). But the calculations
are more complicated, as in expansion of $\psi_1'\psi_2'$ the second order terms must be used 
(as it was done by \citet{Landau:1937}).

Above we used $p$ and $q$ as arguments of the phase density, and hence it
is recommendable to move also in Eq.~(\ref{eq17.5.6}) to these variables (q.v.
(\ref{eq17.1.4})). Considering that $\nabla = - \boldsymbol{v}\partial
/\partial p + r^2\boldsymbol{v}_t\partial /\partial q$ we find 
$$\chi_{1,2} = \pi G^2m_2 \left[ m_1a\psi_1 + (m_2-m_1) \left( a_1
\frac{\partial\psi_1 }{\partial p} + a_2 \frac{\partial\psi_1 }{\partial q}
\right) + \right. $$
\be
 \left. + m_2 \left( a'_1 \frac{\partial\psi_1}{\partial p} + a'_2
\frac{\partial\psi_1 }{\partial q} + a_{11} \frac{\partial^2\psi_1}{
\partial p^2} + 2a_{12} \frac{\partial^2\psi_1}{\partial p\partial q}
+ a_{22} \frac{\partial^2\psi_1 }{\partial q^2 } \right) \right] .
\label{eq17.5.8}
\ee 
Here, $a_1 = -\boldsymbol{b}\boldsymbol{v}$, $a_2 = r^2\boldsymbol{b}\boldsymbol{v}_t$, $a'_1 = - {\bf B}
\nabla\boldsymbol{v} =$ $ - {\rm Sp}{\bf B}$, $a'_2 = r^2{\bf B}\nabla\boldsymbol{v}_t = $
$r^2{\rm Sp_t}{\bf B}$, $a_{11} = {\bf B}\boldsymbol{v}\boldsymbol{v}$, $a_{12} = -r^2{\bf
B}\boldsymbol{v}\boldsymbol{v}_t$, $a_{22} = r^4{\bf B}\boldsymbol{v}_t\boldsymbol{v}_t$. If the function
$\psi_2$ is known, the coefficients $a$, $a_1$, $a_2$ etc. can be
calculated as functions of $p$ and $q$. Then, knowing $\psi_1$ we derive
from (\ref{eq17.5.8}) the function $\chi_{1,2}$ as a function of the same arguments.

As was mentioned in Sect.~1, the encounter function for $p\le 0$
describes the escape of particles from system. Yet, because $\psi_1=0$
for $p\le 0$, Eq.~(\ref{eq17.5.8}) gives $\chi_{1,2} =0$ for $p<0$. This result is
caused by the approximate nature of formulas in present Section. But they
still enable, although approximately, to describe the escape of
particles. Indeed, if the phase density gradient is finite 
on the inner surface of the escape velocity sphere, then we have a finite flux of
particles, that have the escape velocity. We may identify this flux with the
flux of particles escaping the system, similarly as it was done by
\citet{Chandrasekhar:1943a, Chandrasekhar:1943b, Chandrasekhar:1943c}
in his calculations of cluster dissipation  
speed.\footnote{As it was mentioned above, we can not agree with the
form of fluctuating term in Chandrasekhar's equation. Thus his
calculations need revision. Meanwhile, the
calculations of the escape speed of stars with different masses from the
cluster on the basis of Ambartsumian-Spitzer theory, done in his book
\citep{Chandrasekhar:1942}, are also disputable. In these calculations,
the uniform energy distribution over all range of masses is assumed. 
This assumption is unjustified, because the kinetic energy of stars is
restricted by the escape velocity.}

The same result for the escape speed of a particle can be derived from
Eq.~(\ref{eq17.5.6}). According to that formula, the behaviour of the encounter function on
the surface of escape velocity sphere is like the delta-function. Supposing
$\psi_1$ to be expressed as a function of $p$ and $q$ and using Eq.~(\ref{eq17.5.8}),
we derive $\chi_{1,2}$ for $p=0$ in the following form
\be
\chi_{1,2}^{(0)} = \pi G^2m_2^2\left( a_{11}
\frac{\partial\psi_1}{\partial p} \right)_0 \delta (p), \label{eq17.5.9} 
\ee
where index 0 under the parentheses designates the limit when $p$
vanishes. The particles, that correspond to the distribution (\ref{eq17.5.9}),
can be handled just as escaping particles. True, in reality the behaviour of
the encounter function at $p=0$ is not precisely like the delta-function,
but has some, although not large, ``width'', with about half of them
having $p>0$. But these particles form very rarefied media without further
influence on the system dynamics. We can suppose that they leave the system.

The condition $(\partial\psi /\partial p)_0>0$, that was necessary for dissipation according to
equations of previous Section, may be not fulfilled for
real systems, even if the system dissipates. However, if we use these formulas in
description of the encounter effect, we shall find that the
necessary condition for starting the dissipation is immediately present.
Hence it is meaningful to include it among the initial conditions (it was
mentioned in Sect.~3).

With this we finish the discussion about the equations and formulas
for our problem. As we saw, the basic equations are
Eq.~(\ref{eq17.2.5}) for the phase density and Eq.~(\ref{eq17.3.7})
for the potential. Simultaneous solution of these equations allows to
follow the evolution of the system, caused by the encounter effect. It
is quite complicated to solve these equations, especially because of
complications in calculations of the encounter function according to
Eq.~(\ref{eq17.4.2}). But as we saw, the encounter function can be
simplified somewhat by using the approximate formula (\ref{eq17.5.8}).

In the present paper we avoided drawing any conclusions about the possible
evolution of star clusters. But in subsequent papers we hope to use the
theory presented above for this purpose.\footnote{Later the theory
developed here was generalised and applied in case of highly flattened
Galactic subsystems (Chapter 22). Studies on the dynamical evolution of
spherical stellar systems were not continued. We like to refer to the
following papers on the evolution of spherical systems under the influence
of irregular gravitational forces: \citet{Henon:1961tz}; 
\citet{Michie:1963ua, Michie:1961wt}; and 
\citet{Agekyan:1963,Agekyan:1964to}. 
[Later footnote.]}
\vglue 3mm

{\hfill 1957}

\vglue 5mm

{\bf\Large Appendices added in 1969}
\vglue 5mm

\section[A. Adiabatic invariants and
integrals of motion]{A. Using the adiabatic invariants and
integrals of motion in case of variable potential}

For slowly varying gravitational potential we have adiabatic invariant
\be
P_r = \frac{1}{2\pi} \oint v_r \rmd r. \label{eq17.A1.1} 
\ee
The invariant $P_r$ is constant only in average. If a star moves, we have for the
time derivative
\be
\dot{P}_r = \frac{\partial P_i}{\partial E} \dot{E} + \frac{\partial P_i}{
\partial t} = - \frac{1}{\omega_r} \left( \frac{\partial\Phi}{\partial t} - 
\frac{\overline{\partial\Phi} }{\partial t} \right) ,\label{eq17.A1.2} 
\ee
where $E$ is the energy integral (in original paper and in Chapter 17
$E=-p$) and $\omega_r$ is the frequency of $r$ oscillations of the star.

If we use in $\Psi$ the argument $P_r$ instead of $E$, \ie 
\be
\Psi = \Psi (P_r,I,t) , \label{eq17.A1.3}
\ee
where $I$ is the integral of angular momentum (in original paper and in
Chapter 17 it was designated as $I^2=2q$), we have instead of Eqs.~(\ref{eq17.2.5}) and
(\ref{eq17.2.8})
\be
\frac{\partial\Psi}{\partial t}= \overline{X}, ~~~~
\frac{\partial\Delta\psi}{\partial r} |v_r| = X - \overline{X} - 
\frac{\partial\Psi}{\partial E} \frac{\partial\Phi}{\partial t}. \label{eq17.A1.4}
\ee

Instead of $P_r$ we can use also the integral of motion in slightly
variable gravitational field
\be
K_r = P_r - \int \dot{P}_r \rmd t = P_r - \int \left(
\frac{\partial\Phi}{\partial t} - \frac{\overline{\partial\Phi}}{\partial
t} \right) \frac{\rmd r}{ v_r} . \label{eq17.A1.5} 
\ee
If
\be
\Psi = \Psi (K_r,I,t) , \label{eq17.A1.6} 
\ee
we derive equations
\be
\frac{\partial\Psi}{\partial t} = \overline{X} , ~~~~
\frac{\partial\Delta\psi}{\partial r} |v_r| = X - \overline{X} .
\label{eq17.A1.7} 
\ee
However, these simplifications are only formal and do not simplify the
solution of equations because we must find the functions $P_r(E,I,t)$ or 
$K_r(E,I,r,t).$

\section{B. The encounter function}

$\mathrm{1.^o}$ ~The formula for the encounter function
\be
X_{1,2} = -\nabla_v \boldsymbol{i}_{1,2} = - \nabla_v \cdot (\boldsymbol{a} \Psi_1 ) +
\nabla_v \cdot\nabla_v ({\bf A}\Psi_1 ), \label{eq17.A2.1} 
\ee
where
\be
{\bf A} = \pi G^2 m_2^2 {\bf B}, ~~~~ \boldsymbol{a} = \pi G^2 m_2 (m_1+m_2) 
\boldsymbol{b}, \label{eq17.A2.2} 
\ee
\be
\ba{ll}
{\bf B} = & \int (w^2 - \boldsymbol{w}\boldsymbol{w} )w^{-3}L\Psi_2 \rmd V, \\
\noalign{\smallskip}
\boldsymbol{b} = & \nabla_v {\bf B} = 2\int \boldsymbol{w} w^{-3}L\Psi_2 \rmd V 
\ea
\label{eq17.A2.3}
\ee
is the Fokker-Planck equation for stellar encounters.
\vglue 3mm

$\mathrm{2.^o}$ ~In the case of equal stellar masses
\be
X_{1,2} = X, ~~ \Psi_1 = \Psi_2 = \Psi , ~~ m_1=m_2=m . \label{eq17.A2.4} 
\ee

Assuming the spherical velocity distribution
\be
\Psi = \Psi (p), ~~~~ p = -E = \Phi -v^2/2 \label{eq17.A2.5} 
\ee
the encounter function has the spherical symmetry in velocity space.

Assuming for simplicity
\be
L = L_0 = \mathrm{const} , \label{eq17.A2.6} 
\ee
the Fokker-Planck equation gives us in case of spherical velocity
distribution 
\be
X = 8\pi^2 G^2m^2L_0 \left( f_0\Psi -f_1\frac{d\Psi}{dp}+f_2 \frac{d^2\Psi}{
dp^2}\right) , \label{eq17.A2.7} 
\ee
where we find following expressions for functions $f_1,$ $f_2$ and $f_3$
\be
\ba{ll}
f_0 = & \Psi \\
\noalign{\smallskip}
f_1 = & \int^p_0 \Psi \rmd p + (\Phi -p)^{-1/2} \int^{\Phi}_p (\Phi
-p)^{1/2} \Psi \rmd p \\
\noalign{\smallskip}
f_2 = & \frac{2}{3}(\Phi -p) \left[ \int^p_0\Psi \rmd p +(\Phi -p)^{-3/2}
\int^{\Phi}_p (\Phi -p)^{3/2} \Psi \rmd p \right] . 
\ea 
\label{eq17.A2.8} 
\ee
For the most simple polytropic model of a spherical star system, the
phase density $\Psi$ is proportional to $p^s,$ $s=n-3/2,$ where $n$ is
the polytrope index. In this case the integrals in expressions for $f_1$
and $f_2$ will be the elementary and incomplete B-functions, and if
$2s$ is a natural number, the incomplete B-function reduces into
elementary function. 

\begin{figure}[ht]
\centering
\includegraphics[width=70mm]{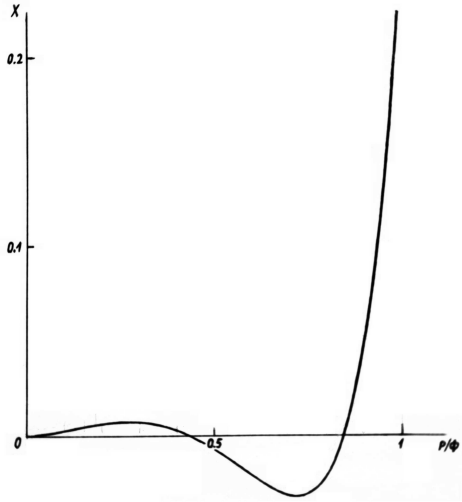}
\caption{The encounter function for the
Schuster-Eddington model. Unit for $X$ is
$\Psi^2_{p=\Phi}$.}
\label{fig17.1}
\end{figure}

In Fig.~(\ref{fig17.1}) the encounter function for the
Schuster-Eddington model is given, \ie  for $s=7/2.$ Unit for $X$ is
$\Psi^2_{p=\Phi}$ (without a factor before parentheses in
Eq.~\ref{eq17.A2.7}). The encounter function for Schuster-Eddington model was
derived also by \citet{Skabitskii:1950}.

\vglue 3mm

$\mathrm{3.^o}$ ~Although in case of spherical velocity distribution the
encounter function has the spherical symmetry, the averaged encounter
function 
\be
\overline{X} = \oint X v_r^{-1} \rmd r \Bigg/ \oint v_r^{-1} \rmd r, \label{eq17.A2.9}
\ee
occurring in our basic expression (\ref{eq17.2.5}) doesn't have spherical symmetry.

\begin{figure}[ht]
\centering
\includegraphics[width=80mm]{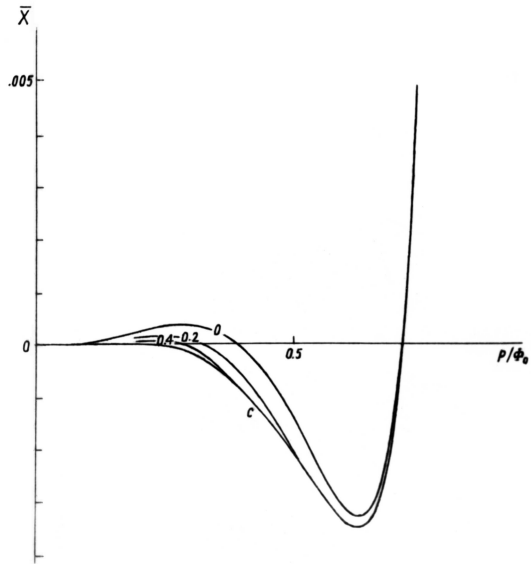}
\caption{$\overline{X}$ for Schuster-Eddington model for
different angular momentum integrals $q$.}
\label{fig17.2}
\end{figure}

In order to illustrate it, we calculated $\overline{X}$ for the
Schuster-Eddington model. Figure \ref{fig17.2} presents $\overline{X}$ for
different angular momentum integrals $q$. The units for $p,$ $q$ and
$\overline{X}$ are $\Phi_0$, $\Phi_0 r_0^2$ and $\Psi^2_{p=\Phi_0}$
respectively ($\Phi_0$ is potential at the centre of the model and
$r_0$ is characteristic length in Schuster law). Line c corresponds to
the circular velocity.

The lack of spherical symmetry in the averaged encounter function means
that although the irregular forces do not explicitly destroy the
spherical symmetry of the velocity distribution (just the opposite
-- they tend to remove non-sphericity if it will occur), the velocity
distribution becomes non-spherical. This results due to the spatial
redistribution of stars caused by simultaneous action of regular and
irregular forces.

As can be seen from figure, the derivative $\partial\Psi /\partial q$ for
small $p$ is negative, \ie  the velocity distribution in the
periphery of the spherical stellar system becomes radially elongated
(the term $\frac{\overline{\partial\Phi}}{\partial
t} \frac{\partial\Psi}{\partial p}$ in Eq.~(\ref{eq17.2.5}) does not influence the
result; it causes only slight shift in $p$ of the curves of $\Psi$).
\vglue 3mm

$\mathrm{4.^o}$ ~Derivation of the encounter function in case of spherical
velocity distribution is quite simple. It is possible to use
not only the Fokker-Planck equations but also more precise
formulas (Boltzmann formula in fact), which take into account large
velocity variations in nearby encounters. This kind of calculations were
made by \citet{Woolley:1954} and  \citet{Agekyan:1959wt}.
They were done also by us (unpublished).
\vglue 3mm

$\mathrm{5.^o}$ ~In case of non-spherical velocity distribution, the calculations
of the encounter function are highly complicated. Although the
use of Fokker-Planck equation simplifies the problem, it
remains complicated, as we need to calculate the vector $\boldsymbol{b}$ and the
tensor {\bf B}. For this reason it is recommended to simplify the
expression for the encounter function. One of the possibilities is to use,
following \citet{Chandrasekhar:1943a,
  Chandrasekhar:1943b}\footnote{\citet{Chandrasekhar:1943} in
  Fokker-Planck equations (despite of the relation $\boldsymbol{b} = 
\nabla {\bf B}$)}
\be
{\bf A} = A, ~~~ \boldsymbol{a} = -\beta \boldsymbol{v}, \label{eq17.A2.10} 
\ee
where $A$ and $\beta$ are scalars independent of velocity.

The expression for the encounter function will have form
\be
X = \beta\nabla_v (\boldsymbol{v} \Psi ) + A\nabla^2_v \Psi . \label{eq17.A2.11}
\ee
For the encounters between stars of equal masses we have the relation
\be
A=\beta\sigma^2 , \label{eq17.A2.12} 
\ee
where $\sigma^2$ is the mean of the components of velocity
dispersion. This relation expresses that $\sigma^2$ remains constant in
encounters (kinetic energy conservation).

While moving to the variables $p,$ $q$ we find
\be
\ba{ll}
X = & \beta \left( 3\Psi - [3\sigma^2 +2(\Phi -p)]
\frac{\partial\Psi}{\partial p} + 2(r^2\sigma^2
+q) \frac{\partial\Psi}{\partial q} + \right. \\
& \left. +2\sigma^2 (\Phi -p) \frac{\partial^2\Psi}{\partial p^2} +
2r^2\sigma^2 q \frac{\partial^2\Psi}{\partial q^2} \right) , 
\ea
\label{eq17.A2.13} 
\ee 
where $\beta$, $\sigma$, $\Phi$ are functions of $r$. In order to
compare this formula to the more precise formula, we calculated $X$ for
the Schuster-Eddington model (spherical velocity distribution,
$\sigma^2 =\Phi /6$). The resulting curve is similar to the one
represented in Fig.~(\ref{fig17.1}) with the exception that minimum and maximum of
the curve are shifted to small $p$.

%% file: chapter18.tex
\chapter[Dynamics of stellar systems with the encounter effect]{Dynamics of stellar systems with allowance for the encounter effect.\footnote{\footnotetext ~~Tartu Astron. Observatory Teated No. 6, 1, 1963; Report on the 2nd Meeting of the Committee on Stellar Astronomy, November 1957.} } 

Due to gravitational interactions in stellar encounters, the velocity distribution of stars changes, and at the same time there occurs the spatial redistribution of stars. Up to recent time it was assumed that encounters are important only in star clusters. In the studies on encounters only encounters of the cluster stars with the field stars were taken into account. However, in 1938--40 \citet{Ambartsumian:1938} and \citet{Spitzer:1940} demonstrated that it is essential to take into consideration also encounters between the cluster member stars themselves. Due to encounters clusters continuously little by little loose their members and change their dimensions. If a cluster is sufficiently dense, the encounters between its members are very important, and as a result the cluster dimensions decrease. The gravitational interaction between stars tends to turn the velocity distribution of stars into a spherical one. But the spatial redistribution of stars, related to the encounters, tends in general to oppose the establishment of the spherical velocity distribution. As it was demonstrated recently by \citet{Woolley:1956} and \citet{Hoerner:1957}, in outer parts of clusters the velocity distribution appears to be radially elongated.\footnote{Similar conclusion was made independently by us \citep{Kuzmin:1957aa} but not expressed sufficiently clearly (Chapter 17). [Later footnote.]} 

In very large stellar systems, for example in our Galaxy, the stellar encounters have nearly no influence on the dynamics. For this reason, it was assumed, that the encounters can be neglected in the Galaxy. But in 1951--52 \citet{Spitzer:1951, Spitzer:1953} from one side and \citet{Gurevich:1954} from other side independently demonstrated that the encounter effect may play important role in the Galaxy. These kind of encounters are the encounters of individual stars with large stellar or diffuse matter aggregates, and the encounters between the aggregates themselves. As a result the energy of peculiar velocities increases on account of rotational energy of the Galaxy (angular momentum is conserved). Just as in case of the star clusters, the encounters do not produce the spherical velocity distribution. The velocity distribution remains triaxial. The encounters explain why the velocity dispersion perpendicular to the galactic plane is smaller than the velocity dispersion in radial direction.  

A stellar system is described by the phase density and the gravitational potential related to it. To have a complete description of the evolution, caused by stellar encounters, one needs to follow the variation of the phase density. This analysis has  not been done yet, or has been done quite incompletely, without taking into account the redistribution of stars in coordinate space. For this reason our aim is to derive the equations, enabling to solve the problem with sufficient accuracy.

Thanks to the mixing effect, a stellar system after some ``revolution time'' evolves into the nearly stationary state, and subsequent variation of the system is caused by encounters. As the influence of encounters during the revolution time is not large, the system can be handled as nearly stationary. In stationary stellar system the phase density is a function of single-valued integrals of motion only. In case of star clusters (those, which can be handled as isolated) we may use the spherically symmetrical model of stellar system, where the phase density is a function of the energy integral and the total angular momentum integral. In case of galaxies, the axisymmetrical models have to be used, where the phase density is a function of the energy integral, the angular momentum integral, and some kind of third integral or quasi-integral. If we take into account stellar encounters, the phase density becomes also an explicit function of coordinates (although the dependence is slight) and time. 

Let us divide the phase density into two parts: the main part $\Psi$, being a function of only integrals of motion of a stationary system and of time, and the small part $\Delta\Psi$, being in addition a function of coordinates. We also assume that the value of $\Delta\Psi$, averaged along the osculating orbit, is zero.\footnote{Under the osculating orbit we mean the orbit in stationary gravitational field, coinciding with a given variable field at a certain moment of time. [Later footnote.]} Starting from known kinetic Boltzmann equation, and assuming the stellar encounter effects to be sufficiently slow, we have the following equation for $\Psi$ and $\Delta\Psi$ \footnote{Initially in Egs.~(\ref{eq18.1}) and (\ref{eq18.2}) instead of $\partial I_i/\partial t$ there was $\dot{I}_i$. But because $I_i$ are the integrals in stationary gravitational field, we have $\dot{I}_i = \partial I_i/\partial t$. As it was done in Appendices of Chapter 17, we can introduce the adiabatic invariants $P_i$ or integrals $K_i$ for slowly varying gravitational field instead of $I_i$ as arguments of $\Psi$. Because $\overline{\dot{P}_i} = 0$ (but $\dot{P}_i \ne 0$) and $\dot{K}_i = 0$, Egs.~(\ref{eq18.1}) and (\ref{eq18.2}) will simplify. But the simplification is interesting only when the expressions for $P_i$ or $K_i$ are known (Chapters 19 and 22). [Later footnote.]}
\be
\frac{\partial\Psi}{\partial t} + \sum \frac{\partial\Psi}{\partial I_i} ~ \frac{\overline{\partial I_i}}{\partial t} = \overline{X}, \label{eq18.1} 
\ee

\be
\frac{\partial\Delta\Psi}{\partial s}~v + \sum \frac{\partial\Psi}{\partial I_i} \left( \frac{\partial I_i}{\partial t} - \frac{\overline{\partial I_i} }{\partial t} \right) = X - \overline{X} , \label{eq18.2}
\ee
where $t$ is time, $I_i$ -- integrals of motion, $s$ and $v$ -- the path length and the velocity on the osculating orbit and $X$ is the encounter function, \ie  the rate of the direct phase density variation due to encounters. The bar denotes the time averaging along the osculating orbit. It can be replaced by averaging over the region of space filled by  loops of orbit. The weight in averaging is in this case the volume of the space corresponding to the unit of volume in space of  integrals.\footnote{In case of stationarity the density of stars on orbit is proportional to $v^{-1}$. Let us replace the orbit with infinitely thin bunch of orbits, determined by the volume element in space of integrals. In case of stationarity the phase density is constant along the phase trajectory, and hence the density in ordinary space is proportional to the volume element filled by stars in velocity space. Therefore the averaging along the orbit with the weight $\rmd s/v$ is identical to spatial averaging with the weight, proportional to the referred volume. [Later footnote.]}  

To solve Egs.~(\ref{eq18.1}) and (\ref{eq18.2}) we must know the derivatives $\partial I_i/\partial t$ and the encounter function $X$. The integrals $I_i$ are the functions of time via gravitational potential. From the Poisson's equation we derive the following equation for the evolution of the potential $\Phi$
\be
\nabla^2 \frac{\partial\Phi}{\partial t} + 4\pi G \frac{\partial\rho}{\partial t} = 0, \label{eq18.3}
\ee
where $\rho$ is the mass density, $G$ -- the gravitational constant. If the phase density is the phase mass density, then $\rho$ in the integral of $\Psi$ over the velocity space. Expressing $\partial\Psi /\partial t$ at its non-moving point, we find for the derivative $\partial\rho /\partial t$
\be
\frac{\partial\rho}{\partial t} = \int \left[ \overline{X} + \sum \frac{\partial\Psi}{\partial I_i} \left( \frac{\partial I_i}{\partial t} - \frac{\overline{\partial I_i} }{\partial t} \right) \right] ~\rmd V, \label{eq18.4}
\ee
where $dV$ is the volume element in the velocity space.

The equations above were initially derived by us for the spherical stellar systems \citep{Kuzmin:1957aa}. But the equations remain principally the same also for axisymmetric systems. For spherical systems from derivatives $\partial I_i/\partial t$ only the derivative of the energy integral remains, being equal to $-2\partial\Phi /\partial t$.\footnote{Here we have in mind the energy integral according to Lindblad $I_1 = v^2 -2\Phi$ (and not $E = \frac{1}{2} v^2 - \Phi$). [Later footnote.]} The derivative of the angular momentum integral is zero. For axisymmetric systems we have the derivatives from the energy integral and from the third integral.

An essential problem in solving the equations is the calculation of the encounter function. As the displacements of stars or of some other objects in the velocity space, caused by single encounter, are very small in general, we may apply the diffusion equation studied first by \citet{Landau:1937}. Let us have objects with masses $m_1$ and $m_2$, and phase densities $\Psi_1$ and $\Psi_2$. In this case according to the diffusion equation the encounter function for objects with mass $m_1$ is in form
\be
X_{1,2} = - \pi G^2 \nabla ( m_1 m_2 \nabla {\bf B}\cdot \Psi_1 - m_2^2 {\bf B} \nabla\Psi_1), \label{eq18.5}
\ee
where {\bf B} is the tensor, depending on the velocity of objects with mass $m_1$, and can be calculated with the aid of the phase density $\Psi_2$. The operator $\nabla$ should be applied in the velocity space.

If the axisymmetric system is sufficiently flattened, and if the stellar orbits are nearly circular, then in our equations we may use the integrals of nearly circular motion. This simplifies our problem. By using the integrals of nearly circular motion, it is quite easy to calculate the variations of the velocity dispersions and the systematic radial shift of stars. The increase of the velocity dispersions, caused by encounters, is not necessarily related to the radial contraction of the system, as it was assumed by \citet{Idlis:1955}. The contraction occurs only in the inner parts of the system, in the outer parts the system expands.

Because of complications in calculations, related to the equations above, we have not jet applied them in practice. The results of solving the equations will be very interesting. In particular, it would be interesting to clarify, what is the velocity distribution in the Galaxy due to encounters. Analysing the encounters of stars with massive nebulae, Spitzer and Schwarzschild derived in their paper the velocity distribution with negative excess, being in disagreement with observations. But the calculations were done within too simplified assumptions. Maybe by carrying more detailed calculations the velocity distribution will be completely different. Good agreement between the theoretical and observed axial ratio of velocity ellipsoid seems to confirm that encounters are quite important in the Galaxy.
\vglue 5mm
{\hfill 1957}

%% file: chapter19.tex
\chapter[On the variation of the dispersion of stellar velocities]{On
  the variation of the dispersion of stellar
  velocities.\footnote{\footnotetext ~~Tartu Astron. Observatory
    Publications vol. 33, pp. 351 -- 368, 1961.}} 

In papers by Prof. T. Rootsm\"ae, in particular in his posthumous
paper, published in the same volume of Publications,\footnote{We had in
  mind the paper: \citet{Rootsmae:1961}
  [Later footnote.]} the kinematical  criteria of stellar ages
is used. When establishing these criteria, it is essential to know
what are the kinematical  characteristics of just-born stars in different
epoch, and how these characteristics evolve in time. Both questions
are not sufficiently well elaborated yet. In the present paper we focus on
the second question. We study the evolution of the velocity
dispersions, caused by the variation of regular gravitation field of
the Galaxy, and by the irregular gravitational forces. In addition, we
discuss the systematic motions of the stars, caused by the same
reasons. 

\section{~}

Let us focus at first on the stationary regular potential, where we neglect the irregular forces.

Let $R$, $\theta$, $z$ be cylindrical coordinates and $v_R$,
$v_{\theta}$, $v_z$ -- the corresponding velocity components. Further,
let $\Phi$ be the regular gravitational potential, symmetrical about
the axis $R=0$ and the plane $z=0$.   

We limit ourselves with the approximation of nearly circular orbits, \ie when the Galaxy is highly flattened.

The theory of nearly circular orbits is well known. Following
\citet{Chandrasekhar:1942} we define  equations of motion 
\be
\ddot{R} = \frac{\partial\Phi}{\partial R} + \frac{I^2}{R^3}, \label{eq19.1.1}
\ee
\be
\ddot{z} = \frac{\partial\Phi}{\partial z}, \label{eq19.1.2}
\ee
where dots designate time derivatives, and
\be
I=Rv_{\theta} \label{eq19.1.3}
\ee
is the angular momentum integral.

Let $\omega (R)$ be the angular velocity for the motion on circular
orbit with radius $R$ (laying in plane $z=0$). The velocity $\omega$
is related to the potential according to the equation 
\be
\omega^2 = - \frac{1}{R} \left( \frac{\partial\Phi}{\partial R}\right)_{z=0} . \label{eq19.1.4}
\ee

Let us define the ``mean radius'' of the orbit $\overline{R}$ as the
radius of the circular orbit  motion,  which corresponds to the
same $I$ as in case of given orbit. Evidently for $\overline{R}$ the
following equation is valid  
\be
\overline{R}^2 \omega (\overline{R}) = I. \label{eq19.1.5}
\ee

As the orbit is assumed to be nearly circular, then
\be
\Delta R = R -\overline{R} \label{eq19.1.6}
\ee
and $z$ are small. Hence Eqs.~(\ref{eq19.1.1}) and (\ref{eq19.1.2}) will have forms
\be
(\Delta R)\ddot {} = -\omega_R^2(\overline{R})\Delta R, \label{eq19.1.7} 
\ee
\be
\ddot{z} = -\omega_z^2 (\overline{R}) z, \label{eq19.1.8}
\ee
where
\be
\omega_R^2 = 4\omega^2 + R \frac{\partial\omega^2}{\partial R} =  - \left( \frac{3}{R} \frac{\partial\Phi}{\partial R} + \frac{\partial^2\Phi}{\partial R^2} \right)_{z=0}, \label{eq19.1.9} 
\ee
\be
\omega_z^2 = - \left( \frac{\partial^2\Phi}{\partial z^2} \right)_{z=0} . \label{eq19.1.10} 
\ee

From Eqs.~(\ref{eq19.1.7}) and (\ref{eq19.1.8}) it results, that
$\Delta R$ and $z$ oscillate harmonically about $\Delta R=0$ and
$z=0$, while $\omega_R$ and $\omega_z$ are corresponding frequencies
(we assume $\omega_R$ and $\omega_z$ to be real, corresponding to
stable circular orbits). 

From Eqs.~(\ref{eq19.1.7}) and (\ref{eq19.1.8}) we find the energy integrals
\be
E_R = \frac{1}{2} ( v_R^2 + \omega_R^2\Delta R^2 ), \label{eq19.1.11}
\ee
\be
E_z = \frac{1}{2} (v_z^2 +\omega_z^2 z^2). \label{eq19.1.12} 
\ee
These are the energies of $R$ and $z$ oscillations, and they supplement
the energy of motion in circular orbit with radius $\overline{R}$ [see
Appendix A]. 

The integral (\ref{eq19.1.11}) can be modified by introducing the $\theta$-velocity
\be
v'_{\theta} = v_{\theta} - v_c , \label{eq19.1.13} 
\ee
where $v_c = R\omega (R)$ is the circular velocity. It is quite easy to derive the relation
\be
v'_{\theta} = - \frac{1}{2} \frac{\omega_R^2}{\omega} \Delta R . \label{eq19.1.14}
\ee 
This transforms the integral (\ref{eq19.1.11}) into the form found by \citet{Lindblad:1927}
\be
E_R = \frac{1}{2} (v_R^2 + k^2 v'_{\theta}{}^2 ), \label{eq19.1.15} 
\ee
where 
\be
k = 2 \frac{\omega}{\omega_R} . \label{eq19.1.16} 
\ee

In Eqs.~(\ref{eq19.1.11}), (\ref{eq19.1.12}) and (\ref{eq19.1.15}) the
argument of $\omega_R$ and $\omega_z$ is $\overline{R}$. But within
the precision used in here it is possible to replace $\overline{R}$
with $R$. 

After forming the time averaged quantities, we have
\be
\overline{v_R^2} = k^2 \overline{v'_{\theta}{}^2} = \omega_R^2 \overline{\Delta R^2} = E_R, \label{eq19.1.17}
\ee
\be
\overline{v_z^2} = \omega_z^2 \overline{z^2} = E_z . \label{eq19.1.18} 
\ee
In case of stationarity these time averages equal to spatial
averages. After averaging over $E_R$ and $E_z$ we find 
\be
\sigma_R^2 = k^2\sigma_{\theta}^2 = \omega_R^2\eta^2 = \overline{E}_R, \label{eq19.1.19} 
\ee
\be
\sigma_z^2 = \omega_z^2\zeta^2 = \overline{E}_z , \label{eq19.1.20} 
\ee
where $\sigma_R$, $\sigma_{\theta}$, $\sigma_z$, $\eta$ and $\zeta$
are the dispersions of $v_R$, $v_{\theta}$, $v_z$, $\Delta R$, and $z$,
respectively. The centroid moves along the circular orbit with the
circular velocity. 

We derive the same results when studying the phase density. According
to the Jeans theorem we may suppose the phase density to be a function
of integrals of motion $E_R$, $E_z$, and $\overline{R}$, whereby
$\overline{R}$ may be replaced within sufficient precision with
$R$. The velocity distribution is elliptical about $v_R$ and
$v_{\theta}$. We stress, that this ellipticity results from smallness
of $\Delta R$, and is independent from the form of the potential. The
form of the potential determines only the axial ratio $k$ of the
``velocity ellipsoid''' to be a function of $R$. 

The result, according to which the ratio of $R$ and $\theta$ axis of
the velocity ellipse equals to $k = 2\omega /\omega_R$, is well known
as the Lindblad formula \citep{Lindblad:1927}. 

\section{~}

Let us assume now, that the gravitational potential slowly varies in time and hence $\omega$, $\omega_R$, $\omega_z$ vary, and $\overline{R}$ varies according to (\ref{eq19.1.5}).

If we neglect the variations of $\omega_R$ and $\omega_z$, we derive
previous results from Eqs.~(\ref{eq19.1.7}) and (\ref{eq19.1.8}), only
the expressions (\ref{eq19.1.11}) and (\ref{eq19.1.5}) will slightly
change having now the forms 
\be
E_R = \frac{1}{2} (v'_R{}^2 + \omega_R^2 \Delta R^2) = \frac{1}{2} (v'_R{}^2 + k^2v'_{\theta}{}^2 ), \label{eq19.2.1} 
\ee
\be
v'_R = v_R - \overline{v}_R , \label{eq19.2.2}
\ee
where $\overline{v}_R = d\overline{R}/dt$. The velocity
$\overline{v}_R$ may be identified with $R$-velocity of centroid
($\theta$-velocity of centroid is $v_c$). 

Let us take into account the variation of $\omega_R$ and
$\omega_z$. As a result of their variation $E_R$ and $E_z$ will
vary. Differentiating $E_R$ and $E_z$ with respect to time and taking
Eqs.~(\ref{eq19.1.7}) and (\ref{eq19.1.8}), we find 
\be
\dot{E}_R = \frac{1}{2} \frac{\rmd\omega_R^2}{\rmd t} (\Delta R)^2 , \label{eq19.2.3}
\ee
\be
\dot{E}_z = \frac{1}{2} \frac{\rmd\omega_z^2}{\rmd t} z^2 . \label{eq19.2.4} 
\ee
Averaging these results by taking into account (\ref{eq19.1.17}) and
(\ref{eq19.1.18}),  and averaging again over $E_R$ and $E_z$,  gives us 
\be
\frac{\rmd\overline{E}_R}{\rmd t} = \frac{1}{\omega_R} \frac{\rmd\omega_R}{\rmd t} \overline{E}_R , \label{eq19.2.5} 
\ee
\be
\frac{\rmd\overline{E}_z}{\rmd t} = \frac{1}{\omega_z} \frac{\rmd\omega_z}{\rmd t} \overline{E}_z . \label{eq19.2.6}
\ee

Equations (\ref{eq19.2.5}) and (\ref{eq19.2.6}) enable to find the
variation of dispersions $v_R$, $\sigma_{\theta}$, $\sigma_z$, $\eta$,
$\zeta$. Using Eqs.~(\ref{eq19.1.19}) and (\ref{eq19.1.20}) and also
the relation (\ref{eq19.1.16}) we find 
\be
\frac{1}{\sigma_R} \frac{\rmd\sigma_R}{\rmd t} = - \frac{1}{\eta} \frac{\rmd\eta}{\rmd t} = \frac{1}{2\omega_R} \frac{\rmd\omega_R}{\rmd t}, \label{eq19.2.7} 
\ee
\be
\frac{1}{\sigma_{\theta}} \frac{\rmd\sigma_{\theta}}{\rmd t} = \frac{3}{2\omega_R} \frac{\rmd\omega_R}{\rmd t} - \frac{1}{\omega} \frac{\rmd\omega}{\rmd t} , \label{eq19.2.8} 
\ee
\be
\frac{1}{\sigma_z} \frac{\rmd\sigma_z}{\rmd t} = - \frac{1}{\zeta} \frac{\rmd\zeta}{\rmd t} = \frac{1}{2\omega_z} \frac{\rmd\omega_z}{\rmd t} . \label{eq19.2.9}
\ee

These results are valid in the vicinity of orbit, having the mean radius
$\overline{R}$, varying together with $\omega$. Because
$\rmd\overline{R}/\rmd t = \overline{v}_R$, then evidently 
\be
\frac{\dd}{\dd{t}} = \frac{\partial}{\partial t} + \overline{v}_R \frac{\partial}{\partial R} . \label{eq19.2.10}
\ee

Let us find now the expression for $\overline{v}_R$, and also for the
centroid $z$-velocity $\overline{v}_z$. The velocity $\overline{v}_R$
results easily from Eq.~(\ref{eq19.1.5}} (where
$I=\mathrm{const}$). It gives 
\be
\overline{v}_R = - \frac{1}{2\omega} \frac{\rmd\omega}{\rmd t} R . \label{eq19.2.11} 
\ee
It is easy to derive $\overline{v}_z$. Because of proportionality with
variation of $E_z$, the distribution of $z$ remains self-similar. Now,
having in mind the variation of $\zeta$, we conclude that 
\be
\overline{v}_z = - \frac{1}{2\omega_z} \frac{\rmd\omega_z}{\rmd t} z . \label{eq19.2.12} 
\ee
The distribution of $\Delta R$ also remains self-similar. Variation of
$\Delta R$ distribution does not influence $\overline{v}_R$  (within
precision used in here). And the velocity distribution remains
self-similar.  

Equations (\ref{eq19.2.11}) and {\ref{eq19.2.12}} allow to find the
expression for three terms in divergence of the centroid velocity
vector. They have forms 
\be
\frac{\partial\overline{v}_R}{\partial R} = - \frac{2}{\omega_R} \frac{\rmd\omega_R}{\rmd t} + \frac{3}{2\omega} \frac{\rmd\omega}{\rmd t} , \label{eq19.2.13}
\ee
\be
\frac{\overline{v}_R}{R} = - \frac{1}{2\omega} \frac{d\omega}{dt} , \label{eq19.2.14} 
\ee
\be
\frac{\partial\overline{v}_z}{dz} = - \frac{1}{2\omega_z} \frac{\rmd\omega_z}{\rmd t} . \label{eq19.2.15} 
\ee
The second and the third of these equations result directly, the first
one requires some calculations using Eqs.~(\ref{eq19.1.5}) and
(\ref{eq19.2.11}) [see Appendix A]. 

Comparing the results (\ref{eq19.2.13}) and (\ref{eq19.2.14}) with
(\ref{eq19.2.7}) and (\ref{eq19.2.8}), and (\ref{eq19.2.15}) with
(\ref{eq19.2.11}), it can be seen that 
\be
\frac{1}{\sigma_R} \frac{\rmd\sigma_R}{\rmd t} + \frac{1}{\sigma_{\theta}} \frac{\rmd\sigma_{\theta}}{\rmd t} = - \frac{\partial\overline{v}_R}{\partial R} - \frac{\overline{v}_R}{R}, \label{eq19.2.16} 
\ee
\be
\frac{1}{\sigma_z} \frac{\rmd\sigma_z}{\rmd t} = - \frac{\partial\overline{v}_z}{\partial z} . \label{eq19.2.17} 
\ee
Derived relations agree with the \citet{Lindblad:1950} ``adiabatic''
theory,  where the multiplication of volume of the velocity ellipsoid
and the volume element of the ``stellar medium'' (in ordinary space)
remains constant. The Lindblad theory is an integral expression of the
Liuoville theorem on the conservation of the phase volume. 

The results of the present Section have some similarity with the
results of the Chandrasekhar theory of non-stationary stellar systems
\citep{Chandrasekhar:1942}. The Chandrasekhar's theory is more general
in the sense that there are no restrictions on the flatness of the
system or the subsystem, and on the evolution speed of the
potential. But in case of very flat subsystems and slowly varying
potential our theory is more general than the Chandrasekhar's theory.  

\section{~}

Let us discuss now the effect of irregular forces. Irregular forces
influence the structure of Galactic subsystems directly and
indirectly. In latter case the influence is via the gravitational
potential.\footnote{This double nature of action of irregular forces
  was missed by \citet{Idlis:1957ab}, leading him to some wrong
  results.} These variations of the spatio-kinematical characteristics
of flattened Galactic subsystems, discussed above, may serve as the
result of indirect influence of irregular forces. Now we discuss their
direct action.

We handle the irregular forces as impulses abruptly changing the
stellar velocities. We assume that the action of irregular forces is
quite slow, similar to the variations of the potential discussed
above. In this case the structure of the Galaxy at every moment of
time is quasi-stationary. 

On the basis of Eqs.~(\ref{eq19.1.15}) and (\ref{eq19.1.12}) we have
for $\overline{E}_R$ and  $\overline{E}_z$ 
\be
\overline{E}_R = \frac{1}{2} (\sigma_R^2 + k^2\sigma_{\theta}^2), \label{eq19.3.1} 
\ee
\be
\overline{R_z} = \frac{1}{2} (\sigma_z^2 + \omega_z^2\zeta^2) . \label{eq19.3.2} 
\ee
Now we derive the variations of $\overline{E}_R$ and $\overline{E}_z$ due to irregular forces
\be
\frac{\rmd\overline{E}_R}{\rmd t} = \frac{1}{2} \left( \frac{\delta\sigma_R^2}{\delta t} + k^2 \frac{\delta\sigma_{\theta}^2}{\delta t} \right) , \label{eq19.3.3} 
\ee
\be
\frac{\rmd\overline{E}_z}{\rmd t} = \frac{1}{2} \frac{\delta\sigma_z^2}{ \delta t} , \label{eq19.3.4} 
\ee
where $\delta /\delta t$ designates variation per unit time without
the spatial redistribution of stars. 

Equations (\ref{eq19.3.3}) and (\ref{eq19.3.4}) can be represented also in form
\be
\frac{\rmd\overline{E}_R}{\rmd t} = \left( \frac{1}{\sigma_R} \frac{\delta\sigma_R}{\delta t} + \frac{1}{\sigma_{\theta}} \frac{\delta\sigma_{\theta}}{\delta t} \right) \overline{E}_R , \label{eq19.3.5} 
\ee
\be
\frac{\rmd\overline{E}_z}{\rmd t} = \frac{1}{\sigma_z} \frac{\delta\sigma_z}{ \delta t} \overline{E}_z , \label{eq19.3.6} 
\ee
where we substituted $\sigma_R/\sigma_{\theta}$ instead of $k$ (Eq.~\ref{eq19.1.19}). 

As the discussed variations result only from direct action of
irregular forces, we may neglect the variations of $\omega_R$ and
$\omega_z$, and also $k=2\omega /\omega_R$. In this case, having in mind
Eqs.~(\ref{eq19.1.19}) and (\ref{eq19.1.20}), we derive 
\be
\frac{1}{\sigma_R} \frac{\rmd\sigma_R}{\rmd t} = \frac{1}{\sigma_{\theta}} \frac{\rmd\sigma_{\theta}}{\rmd t} = \frac{1}{\eta} \frac{\rmd\eta}{\rmd t} = \frac{1}{2} \left( \frac{1}{\sigma_R} \frac{\delta\sigma_R}{\delta t} + \frac{1}{\sigma_{\theta}} \frac{\delta\sigma_{\theta}}{\delta t} \right) , \label{eq19.3.7} 
\ee
\be
\frac{1}{\sigma_z} \frac{\rmd\sigma_z}{\rmd t} = \frac{1}{\zeta} \frac{\rmd\zeta}{\rmd t} = \frac{1}{2\sigma_z} \frac{\delta\sigma_z}{\delta t} . \label{eq19.3.8} 
\ee
Due to the redistribution of stars in space, the real variations of
$\sigma_R$, $\sigma_{\theta}$ and $\sigma_z$ differ from their
variation in absence of redistribution. From Eq.~(\ref{eq19.3.7})
follows that the redistribution of stars keeps the area of the
velocity ellipse constant. It also restores the axial ratio of
ellipse,  
if the irregular forces change this ratio. In addition, the
redistribution of stars restores the ratio $\sigma_z/\zeta$, causing
the $\sigma_z$ variation to decrease twice due to the corresponding
variation of $\zeta$ (Eq.~\ref{eq19.3.8}). 

The redistribution of stars in space causes some of the potential
energy and circular motion energy turn to the peculiar motion energy
and vice versa. 

It is not difficult to obtain the expressions for the variation of
kinetic energy of peculiar $R$ and $\theta$ motion, caused by the
redistribution of stars. Let us designate 
\be
K = \frac{1}{2} (\sigma_R^2 + \sigma_{\theta}^2) \label{eq19.3.9}
\ee
as the specific kinetic energy of peculiar $R$ and $\theta$ motion. On
the basis of Eq.~(\ref{eq19.1.19}) we have 
\be
K = \frac{k^2+1}{2k^2} \overline{E}_R. \label{eq19.3.10}
\ee
Calculating derivatives $d/dt$ of (\ref{eq19.3.10}) and $\delta
/\delta t$ of (\ref{eq19.3.9}), and using Eq.~(\ref{eq19.3.3}), we find 
\be
\left( \frac{\rmd}{\rmd t} - \frac{\delta}{\delta t} \right) K = - \frac{k^2-1}{4k^2} \left( \frac{\delta\sigma_R^2}{\delta t} - k^2 \frac{\delta\sigma_{\theta}^2 }{ \delta t} \right) , \label{eq19.3.11} 
\ee
or from Eqs.~(\ref{eq19.1.19}) and (\ref{eq19.3.10})
\be
\left( \frac{\rmd}{\rmd t} - \frac{\delta}{\delta t} \right) K = - \frac{k^2-1}{k^2+1} \left( \frac{1}{\sigma_R} \frac{\delta\sigma_R}{\delta t} - \frac{1}{\sigma_{\theta}} \frac{\delta\sigma_{\theta}}{\delta t} \right) K . \label{eq19.3.12} 
\ee
The expression in parenthesis on the right side of
Eq.~(\ref{eq19.3.1}) is the relative variation of the ratio
$\sigma_R/\sigma_{\theta}$, if the redistribution of stars is
absent. When the irregular forces tend to ``round'' the velocity
ellipse, \ie  when $\sigma_R/ \sigma_{\theta}$ approaches one, the
spatial redistribution of stars causes the increase of the kinetic
energy of peculiar $R$ and $\theta$ motion on the account of the
circular motion energy. If the irregular forces tend to ``elongate''
the velocity ellipse, the opposite process occurs. In both cases 
$k\ne 1$, being valid for $\partial\omega /\partial R \ne 0$, \ie  in
case of differential rotation. 

Now we derive the expressions for $R$ and $z$ components of the
systematic motions,  caused by irregular forces. If irregular forces
change the $\theta$ velocity of the centroid, then the spatial
redistribution of stars restores its equilibrium value to be equal to
$v_c$ for flat subsystems. As a result, the systematic motion of stars
appears along $R$. From the expression of the angular momentum
integral we have 
\be
\frac{\overline{\rmd I}}{\rmd t} = R \frac{\delta\overline{v}_{\theta}}{ \delta t} = R^2 \frac{\delta\omega}{\delta t} . \label{eq19.3.13} 
\ee
As $\Delta R$ is small, we may replace $\overline{\rmd I/\rmd t}$ by
$\rmd I/\rmd t$, handling $I$ as a function of $R$, and thereafter
take $\overline{v}_R = \rmd\overline{R}/\rmd t$. Using
Eqs.~(\ref{eq19.1.5}) and (\ref{eq19.1.9}) we find 
\be
\overline{v}_R = \frac{k}{\omega_R} \frac{\delta\omega}{\delta t} R . \label{eq19.3.14}
\ee 
On the other side, if we take $\overline{v}_z$ to be approximately
proportional to $z$, then using the expression for $\rmd\zeta /\rmd t$
we derive 
\be
\overline{v}_z = \frac{1}{2\sigma_z} \frac{\delta\sigma_z}{\delta t} z .\label{eq19.3.15} 
\ee

The spatial redistribution of stars along $R$ coordinate, related to
the variation of $\sigma_R$ and $\sigma_z$, must cause some additional
systematic motions along the galactic radius. But within the precision
used in here the value of that motion is neglectfully  small.  

\section{~}

The irregular forces, caused by individual stars, are very weak in the
Galaxy, and can not play a significant role. But, as it was indicated by
\citet{Spitzer:1953} and \citet{Gurevich:1954}, the role of irregular
forces caused by the clouds of diffuse matter and of stars may be
essential. They  found that these forces may significantly increase
the velocity dispersion of stars. They also mentioned that the action
of irregular forces alone can explain why $\sigma_z$ is less than
$\sigma_R$. 

Using the theory of irregular gravitational forces, presented by the
author earlier \citep{Kuzmin:1957aa} and having in mind that stellar
masses are small compared with cloud masses,  we may derive the
following formulas for the variation of velocity dispersion of stars
along $i$-th coordinate $\sigma_i$, and for the centroid velocity along
the same coordinate $\overline{v}_i$ [see Appendix C]. 
\be
\frac{\delta\sigma_i^2}{\delta t} = \overline{ [ w^2-w_i^2+2w_i(v_i- \overline{v}_i) ] \tau^{-1} } , \label{eq19.4.1} 
\ee
\be
\frac{\delta\overline{v}_i}{\delta t} = \overline{ w_i\tau^{-1}}, \label{eq19.4.2} 
\ee
where $w$ is the cloud velocity with respect to star, $w_i$ -- $i$-component of this velocity, and
\be
\tau^{-1} = 2\pi G^2 m^2 n L(x) w^{-3} , \label{eq19.4.3} 
\ee
while 
\be
x^2 = \frac{r}{Gm} w^2. \label{eq19.4.4} 
\ee
In last formulas $G$ is the gravitational constant, $m$ -- cloud
mass, $n$ -- spatial density of clouds, $r$ -- effective action radius
of irregular forces, $L(x)$ -- function, which increases starting from
small $x$ and up to $x\sim 1$ proportionally to $\sim x^4$, thereafter
with smaller speed, and finally for large $x$ proportionally to
$\sim\ln x$. The quantity $\tau$ characterises the time of action of
irregular forces as a function of $w$. 

Equations (\ref{eq19.4.1}) and (\ref{eq19.4.2}) would be significantly
simpler, if it would be  possible to assume that $\tau$ is independent of
$w$, \ie  $L(x)$ is proportional to $x^3$. Within a certain
approximation this  can be done. The clouds form quite flattened
subsystem of the Galaxy. The action radius of irregular forces is of
the order of thickness of that subsystem. As the clouds give  notable
contribution to the Galactic density only near the symmetry plane of
the Galaxy, the mean $w^2$ for highly flattened subsystem of stars
probably does not exceed $Gm/r$ more than 1 -- 2 orders of
magnitude. Hence $x$ is in the average an order of unity. For these values
of $x$ $L(x)$ increases still quite rapidly. 

Supposing $\tau$ to be independent of $w$, we find the following formulas
\be
\tau \frac{\delta\sigma_R^2}{\delta t} = \kappa_R^2 + 3 (\sigma^2 - \sigma_R^2) , \label{eq19.4.5} 
\ee
\be
\tau \frac{\delta\sigma_{\theta}^2}{\delta t} = \kappa_{\theta}^2 + 3
(\sigma^2 -\sigma_{\theta}^2 ) , \label{eq19.4.6} 
\ee
\be
\tau \frac{\delta\sigma_z^2}{\delta t} = \kappa_z^2 + 3 (\sigma^2 - \sigma_z^2 ) , \label{eq19.4.7} 
\ee
and 
\be
\tau \frac{\overline{\delta v_{\theta}}}{\delta t} = \Delta \overline{v}_{\theta} . \label{eq19.4.8} 
\ee
Here $\kappa_R^2$, $\kappa_{\theta}^2$, $\kappa_z^2$ are the mean
squares of the cloud velocities with respect to the centroid of stars, 
and are perpendicular to the axes $v_R$, $v_{\theta}$, $v_z$,
respectively, 
\be
\sigma^2 = \frac{1}{3} (\sigma_R^2 + \sigma_{\theta}^2 + \sigma_z^2 ) \label{eq19.4.9}
\ee 
is the square mean velocity dispersion and
$\Delta\overline{v}_{\theta}$ denotes the velocity difference for
clouds and stars. The quantity $\tau$ must be taken in these formulas
to be a function of $\sigma$. But for small $\sigma$, corresponding to
flat subsystems of the Galaxy, this dependence is weak. For large
$\sigma$ the quantity $\tau$ sharply increases, and our formulas are
inexact. 

From Eqs.~(\ref{eq19.4.5})--(\ref{eq19.4.7}) and (\ref{eq19.4.9}) it follows that
\be
\tau \frac{\delta\sigma^2}{\delta t} = \kappa^2 , \label{eq19.4.10} 
\ee
where
\be
\kappa^2 = \frac{1}{3} (\kappa_R^2 + \kappa_{\theta}^2 +\kappa_z^2 ) . \label{eq19.4.11} 
\ee
$3\kappa^2$ is the doubled specific kinetic energy given to stars by the clouds during the time $\tau$.

Equations (\ref{eq19.4.5}) -- (\ref{eq19.4.7}) indicate that there is a
tendency to uniform the kinetic energy along the three velocity
components. In case of small $\sigma$ this may be opposed by the
inequality between $\kappa_R$, $\kappa_{\theta}$, $\kappa_z$. But
probably the difference between these quantities for flat subsystems
is not large. 

As the system of clouds is rotating with the circular velocity (as all
very flat subsystems), then with respect to these subsystems
$\Delta\overline{v}_{\theta} =0$ and thus $\kappa_R^2 =
\sigma_{\theta}^2 + \sigma_z^2$, $\kappa_{\theta}^2 = \sigma_R^2 +
\sigma_z^2$, $\kappa_z^2 = \sigma_R^2 + \sigma_{\theta}^2$, where
$\sigma_R$, $\sigma_{\theta}$, $\sigma_z$ are cloud velocity
dispersions at present. If for clouds $\sigma_R >
\sigma_{\theta},\sigma_z$ as it was for stars, then $\kappa_R <
\kappa_{\theta},\kappa_z$. But on the other side,
Eqs.~(\ref{eq19.4.5}) -- (\ref{eq19.4.7}) do not take into account
distant encounters between stars and clouds very correctly. Due to the
flatness of the cloud system and its differential rotation, in case of
distant encounters, the irregular forces act mainly only in $R$
direction. Hence $\kappa_R^2$ must be taken somewhat larger when
compared with $\sigma_{\theta}^2 + \sigma_z^2$ for clouds. Probably we
do not make a large error when accepting for flat subsystems 
\be
\kappa_R = \kappa_{\theta} = \kappa_z = \kappa . \label{eq19.4.12}
\ee

For relative variation of dispersions we derive from
Eqs.~(\ref{eq19.4.5})--(\ref{eq19.4.7}) the following formulas
($\sigma_R/\sigma_{\theta} = k$) 
\be
\tau\left( \frac{1}{\sigma_R} \frac{\delta\sigma_R}{\delta t} +  \frac{1}{\sigma_{\theta}} \frac{\delta\sigma_{\theta}}{\delta t} \right) = \frac{(k^2+1)(\kappa^2+3\sigma^2) }{ 2\sigma_R^2} -3 , \label{eq19.4.13} 
\ee
\be
\tau\left( \frac{1}{\sigma_R} \frac{\delta\sigma_R}{\delta t} - \frac{1}{\sigma_{\theta}} \frac{\delta\sigma_{\theta}}{\delta t} \right) = - \frac{(k^2-1)(\kappa^2+3\sigma^2)}{2\sigma_R^2}, \label{eq19.4.14} 
\ee
\be
\frac{\tau}{\sigma_z} \frac{\delta\sigma_z}{\delta t} = \frac{\kappa^2+3\sigma^2}{2\sigma_z^2} - \frac{3}{2} . \label{eq19.4.15} 
\ee
These formulas give concrete expressions for the right sides of
Eqs.~(\ref{eq19.3.7}), (\ref{eq19.3.8}) and
(\ref{eq19.3.12}). Equation (\ref{eq19.4.13}) gives the relative
variation of the area of velocity ellipse during the time $\tau$. This
variation remains invariant in case of spatial redistribution of
stars, 
related to the action of irregular forces. Equation (\ref{eq19.4.14})
gives the relative variation of the axial ratio of the velocity
ellipsoid during the time $\tau$ in case of the absence of
redistribution of stars (redistribution restores this ratio). The
irregular forces tend to round the ellipse. Therefore, according to
Eq.~(\ref{eq19.3.1}), the redistribution of stars causes the
transformation of some energy of circular motion into the energy of
peculiar $R$ and $\theta$ motion. 

\section{~}

As with respect to flat subsystems $\Delta\overline{v}_{\theta} = 0$,
from Eq.~(\ref{eq19.4.8}) it follows that for them 
$\delta\overline{v}_{\theta}/\delta t = R\delta\omega /\delta t =
0$. Thus the action of irregular forces does not cause systematic
radial motions in very flat subsystems (Eq.~\ref{eq19.3.14}). Less
flattened subsystems rotate more slowly. With respect to them
$\Delta\overline{v}_{\theta} >0$, and consequently for them
$\delta\overline{v}_{\theta} > 0$. These subsystems must expand
radially. The cloud system, because of the conservation of the
angular momentum, must have $\delta\overline{v}_{\theta} <0$, \ie
must contract. As much as the stars of very flattened subsystems are
related to cloud system, these radial motions must be given over to
these subsystems. Because $\tau^{-1}$ is small compared to $\omega_R$,
the radial motion speed may be only small part of
$\Delta\overline{v}_{\theta}$.  

Although some moderate radial motions may exist in individual
subsystems of the Galaxy, in the Galaxy as a whole the radial mass
transformation is probably nearly absent, at least at present
evolution stage. Hence the values of $\omega$ and $\omega_R$, which
depend mainly on the radial mass distribution, vary very little. If
$\omega$ varies very little, its variation can not cause significant
radial motion (because of Eq.~(\ref{eq19.2.11})). 

Because $\omega_R$ varies very little, we can suppose that the
variation of $\sigma_R$ and $\sigma_{\theta}$ is caused by the direct
action of irregular forces. Hence we may suppose that the total
variation of $\sigma_R$ and $\sigma_{\theta}$ is described by
Eqs.~(\ref{eq19.3.7}) and (\ref{eq19.4.13}). 

But it is not justified to neglect the variation of $\omega_z$. If the
irregular forces vary $\sigma_z$ and thus also $\zeta$, then the
dynamical density of the Galaxy in the vicinity of the symmetry plane
$\rho$ must vary, and as a result $\omega_z^2\simeq 4\pi G\rho$ must
vary. 

Accounting for the variation of $\omega_z$, we derive on the basis of
Eqs.~(\ref{eq19.2.9}) and (\ref{eq19.3.8}) the following expressions
for the total relative variation of $\sigma_z$ and $\zeta$ 
\be
\frac{1}{\sigma_z} \frac{\rmd\sigma_z}{\rmd t} = \frac{1}{2} \left( \frac{1}{\sigma_z} \frac{\delta\sigma_z}{\delta t} + \frac{1}{\omega_z} \frac{\rmd\omega_z }{\rmd t} \right) , \label{eq19.5.1} 
\ee
\be
\frac{1}{\zeta} \frac{\rmd\zeta}{\rmd t} = \frac{1}{2} \left( \frac{1}{\sigma_z}  \frac{\delta\sigma_z}{\delta t} - \frac{1}{\omega_z} \frac{\rmd\omega_z}{\rmd t} \right) . \label{eq19.5.2} 
\ee

Let us designate
\be
\frac{1}{\omega_z} \frac{\rmd\omega_z}{\rmd t} = - \frac{\alpha}{\tau} . \label{eq19.5.3}
\ee
In this case according to Eqs.~(\ref{eq19.3.7}), (\ref{eq19.5.1}), (\ref{eq19.5.2}), (\ref{eq19.4.13}), (\ref{eq19.4.15}) we have the following results [see Appendix D]
\be
\frac{\tau}{\sigma_R} \frac{\rmd\sigma_R}{\rmd t} = \frac{\tau}{\sigma_{\theta}} \frac{\rmd\sigma_{\theta}}{\rmd t} = \frac{(k^2+1)(\kappa^2+3\sigma^2)}{4\sigma_R^2} - \frac{3}{2} , \label{eq19.5.4} 
\ee
\be
\frac{\tau}{\sigma_z} \frac{\rmd\sigma_z}{\rmd t} = \frac{\kappa^2+3\sigma^2}{4\sigma_z^2} - \frac{3}{4} - \frac{\alpha}{2} , \label{eq19.5.5} 
\ee
\be
\frac{\tau}{\zeta} \frac{\rmd\zeta}{\rmd t} = \frac{\kappa^2+3\sigma^2}{4\sigma_z^2} - \frac{3}{4} + \frac{\alpha}{2} . \label{eq19.5.6} 
\ee

As the density in every subsystem at $z=0$ is proportional to
$\zeta^{-1}$ and $\omega_z^2\simeq 4\pi G\rho$, then by using
Eq.~(\ref{eq19.5.6}) we find an additional relation [see Appendix D] 
\be
\left( \frac{\overline{\kappa^2+3\sigma^2}}{3\sigma_z^2} \right) = 1+ 2\alpha , \label{eq19.5.7} 
\ee
while the averaging is done with a weight, which is proportional to the
subsystem density at $z=0$. In Eq.~(\ref{eq19.5.7}) we did not account
for the contribution of the cloud subsystem to the density. If it is
large and varies in a different way, compared to the stellar subsystem
in average, then the coefficient in $\alpha$ has to be slightly
changed. 

On the basis of Eqs.~(\ref{eq19.5.4}) and (\ref{eq19.5.5}) we may
conclude that the relative variations of $\sigma_R$ and $\sigma_z$ try
to establish the equality between their variations. Let us suppose
that this kind of equilibrium is established within some
approximation, and the ratio $\sigma_R /\sigma_z$ varies only
little. Then from Eqs.~(\ref{eq19.5.4}) and (\ref{eq19.5.5}) we derive
the relation  
\be 
(\kappa^2+3\sigma^2) \left( \frac{k^2+1}{\sigma_R^2} - \frac{1}{\sigma_z^2} \right) = 3 - 2\alpha  . \label{eq19.5.8} 
\ee

In solar neighbourhood $\sigma_R/\sigma_z$ is nearly the same for all
subsystems. From Eq.~(\ref{eq19.5.8}) we find that this is the case
when [see Appendix D] 
\be
\frac{\sigma_R^2}{\sigma_z^2} = 1 + k^2 \label{eq19.5.9}
\ee
and 
\be
\alpha = \frac{3}{2} . \label{eq19.5.10} 
\ee

With help of Eq.~(\ref{eq19.5.9}) we find
\be
\frac{3\sigma^2}{\sigma_z^2} = k^2 + k^{-2} + 3 . \label{eq19.5.11} 
\ee
Thus Eqs.~(\ref{eq19.5.4})--(\ref{eq19.5.7}) will have forms
\be
\frac{\tau}{\sigma_R} \frac{\rmd\sigma_R}{\rmd t} = \frac{\tau}{\sigma_{\theta}} \frac{\rmd\sigma_{\theta}}{\rmd t} = \frac{\tau}{\sigma_z} \frac{\rmd\sigma_z}{ \rmd t} = \frac{\kappa^2}{4\sigma_z^2} + \frac{k^2 + k^{-2} -3 }{4} , \label{eq19.5.12}
\ee
\be
\frac{\tau}{\zeta} \frac{\rmd \zeta}{\rmd t} = \frac{\kappa^2}{4\sigma_z^2} + \frac{k^2+ k^{-2} +3}{4} , \label{eq19.5.13} 
\ee
and
\be
\kappa^2\overline{\sigma_z^{-2}} = 9 - (k^2+k^{-2}) . \label{eq19.5.14} 
\ee

Relation (\ref{eq19.5.9}) is quite interesting. It seems to agree well
with the observed axial ratio of velocity ellipsoid. In solar
neighbourhood $\sigma_{\theta}/ \sigma_R = 0.63$ and thus $k^2 =
2.5$. Therefore Eq.~(\ref{eq19.5.9}) gives $\sigma_z/\sigma_R =$ 0.53,
which coincides precisely with the observed value
\citep{Einasto:1961}. 

The second terms in Eqs.~(\ref{eq19.5.12}) and (\ref{eq19.5.13}) for
$k^2 = 2.5$ are near to zero and equal to 3/2, respectively. Thus the
velocity dispersion and $\zeta$ increase in time for any value. And if
the dispersion is large, its increase stops, but the increase of
$\zeta$ continues.  

For $k^2 = 2.5$ Eq.~(\ref{eq19.5.14}) gives $\kappa^2
\overline{\sigma_z^{-2}} \simeq 6$ or
$\kappa^2\overline{\sigma_R^2}\simeq 1.7$. We may conclude now that
the peculiar velocities of the cloud must be quite significant,
although not very large, because $\overline{\sigma_z^{-2}}$ and
$\overline{\sigma_R^{-2}}$ can have quite significant values [see
Appendix D]. 

The derived results have some similarity with the results by
\citet{Spitzer:1953} and \citet{Gurevich:1954}. The presented theory
is quite rough yet. But it might be that it represents the influence
of irregular forces on the kinematics and dynamics of the Galaxy
somewhat better than earlier theories. 
\vglue 5mm
{\hfill November 1961}

\vglue 5mm
{\bf\Large Appendices added in 1969}
\vglue 5mm

\section[A.  Integrals in slowly varying
  gravitational field]{A.  Integrals of nearly circular motion in slowly varying
  gravitational field. Vertex deviation}  

For a stationary system we have the integrals of nearly circular
motion (\ref{eq19.1.11}), (\ref{eq19.1.12}) and (\ref{eq19.1.5}) 
\be
E_R= \frac{1}{2} (v_R^2 +\omega^2_R x^2), ~~ E_z = \frac{1}{2} (v^2_z + \omega^2_z z^2), ~~ I = \omega{\overline{R}}^2 , \label{eq19.A1.1} 
\ee
where
\be
x = \Delta R = R-\overline{R}. \label{eq19.A1.2} 
\ee

If the gravitational field slowly varies in time we have the adiabatic invariants 
\be
P_R = \frac{1}{ 2\pi} \oint v_R \rmd R = \omega^{-1}_RE_R, ~~ P_z = \frac{1}{2\pi} \oint v_z \rmd z = \omega^{-1}_z E_z, ~~ P_{\theta}=I . \label{eq19.A1.3}
\ee 
During the motion of a star, $P_R$ and $P_z$ oscillate, being
invariants only on average. We derive the integrals of motion
according to formulas 
\be
K_R=P_R-\int \dot{P}_R \rmd t, ~~~~ K_z=P_z-\int \dot{P}_z \rmd t, \label{eq19.A1.4} 
\ee
giving
\be
\ba{ll}
K_R = & \frac{1}{2} \omega^{-1}_R ( \dot{x}^2+ \omega^2_R x^2 + \frac{\dot{\omega}_R}{\omega_R} x \dot{x} ), \\\noalign{\smallskip}
K_z =  & \frac{1}{2} \omega^{-1}_z  ( \dot{z}^2+ \omega^2_z z^2 + \frac{\dot{\omega}_z}{\omega_z} z \dot{z} ).  
\ea
\label{eq19.A1.5} 
\ee
Now we introduce new variables
\be
v'_R = v_R-\overline{v}_R, ~ v'_{\theta} = v_{\theta}-\overline{v}_{\theta}, ~  v'_z = v_z-\overline{v}_z ~~~~(\overline{v}_{\theta }= v_c). \label{eq19.A1.6} 
\ee
Within sufficient precision we may assume
\be
\dot{x} = v'_R + \frac{\partial\overline{v}_R }{\partial R} x . \label{eq19.A1.7} 
\ee
We find the derivative $\partial\overline{v}_R /\partial R$ from relations
\be
\frac{\partial\overline{v}_R}{\partial R} = \frac{\rmd}{\rmd t} \left( \frac{\rmd\overline{R}}{\rmd I} \right) \Bigg/ \frac{\rmd\overline{R}} {\rmd I} \label{eq19.A1.8} 
\ee
and
\be
\frac{\rmd\overline{R}}{\rmd t} = \overline{v}_R = - \frac{1}{2} \frac{\dot{\omega}}{\omega} . \label{eq19.A1.9} 
\ee
Because
\be
\frac{\rmd I}{\rmd\overline{R}} = \frac{\omega_R}{k}\overline{R} , \label{eq19.A1.10} 
\ee
we have
\be
\frac{\partial\overline{v}_R}{\partial R} = - \frac{1}{2} \frac{\dot{\omega}_R}{\omega_R} + \frac{3}{2} \frac{\dot{k}}{k} . \label{eq19.A1.11} 
\ee
Further, we may substitute
\be
\dot{z} = v'_z + \frac{\partial\bar v_z}{\partial z} z, \label{eq19.A1.12} 
\ee
and
\be
\frac{\partial\overline{v}_z}{\partial z} = - \frac{1}{ 2} \frac{\dot{\omega}_z }{\omega_z} . \label{eq19.A1.13} 
\ee
Finally
\be
\omega_R x = -kv'_{\theta} . \label{eq19.A1.14} 
\ee
Therefore we have (within the precision sufficient in here)
\be
\ba{ll}
K_R = & \frac{1}{2} \omega^{-1}_R (v'^2_R +k^2v'^2_{\theta} - \frac{3\dot{k}}{\omega_R} v'_Rv'_{\theta} ), \\
\noalign{\smallskip}
K_z  = & \frac{1}{2} \omega^{-1}_z (v'^2_z + \omega^2_z z^2), \\
\noalign{\smallskip}
I = & \omega R^2 . 
\ea
\label{eq19.A1.15} 
\ee
In the expression for $I$ we substituted $\overline{R}$ with $R$. This
can be done also for $\overline{R}$ as argument of $\omega$,
$\omega_R$, $\omega_z$. 

According to the Jeans theorem, the integrals $K_R,$ $K_z$, $I$ can be
used as the arguments of the phase density  
\be
\Psi = \Psi (K_R,K_z,I). \label{eq19.A1.16} 
\ee
Velocity distribution, projected onto the $v_z=0$ plane, is elliptical
with a centre at $v_{\theta}=$ $\overline{v}_{\theta}$ $(=v_c)$, $v_R
=$ $\overline{v}_R.$ The axes of the velocity ellipse are inclined
with respect to $R$ and $\theta$ by a small angle 
\be
\gamma = - \frac{3}{2} \frac{\dot{k}}{ (k^2-1)\omega_R}, \label{eq19.A1.17} 
\ee
\ie  there is ``vertex deviation'' for $\dot{k}\ne 0.$

Our theory of slightly non-stationary flat stellar system (subsystem)
is consistent with \citet{Chandrasekhar:1942} theory if we take 
\be
\frac{\dot{\omega}_R}{\omega_R} = \frac{\dot{\omega}}{\omega} = 2 \frac{\dot{\varphi}_1}{\varphi_1}, ~~~~~ \frac{\dot{\omega}_z}{\omega_z}= 2 \frac{\dot{\varphi}_2 }{\varphi_2}, \label{eq19.A1.18} 
\ee
where $\varphi_1$ and $\varphi_2$ are the functions introduced by Chandrasekhar. However, as in this case $\dot{k} =0$, the vertex deviation is absent.

\section[B.  Vertex deviation from  irregular gravitational forces]{B.
  The vertex deviation resulting from the irregular gravitational
  forces}  

Because of irregular forces the uniform distribution of stellar
$R$-oscillation (and also of $z$-oscillation) phases is destroyed. 

By taking into account that
\be
v_R = \sqrt{eE_R}\cos\theta_R , ~~~~ kv'_{\theta} = - \sqrt{2E_R}\sin\theta_R \label{eq19.A2.1} 
\ee
\be
\dot{\theta}_R = \omega_R , \label{eq19.A2.2} 
\ee
where $\theta_R$ is the phase of $R$-oscillations, we find
\be
\ba{ll}
\frac{\rmd\sigma^2_R}{\rmd t} = & 2k\omega\overline{v_Rv'_{\theta}}, \\
\noalign{\smallskip}
k^2 \frac{\rmd\sigma_{\theta}^2}{\rmd t} = & -2k\omega_R\overline{v_Rv'_{\theta}}. 
\ea 
\label{eq19.A2.3}
\ee
If the action of irregular forces would stop, the variations of
$\sigma^2_R$ and $\sigma^2_{\theta}$ would be like damped oscillations
because of mixing due to $R$-gradient of $\omega_R.$ Yet the irregular
forces support the nonuniform distribution according to the phase of
$R$-oscillations. Adding to the variations of $\sigma^2_R$ and
$\sigma^2_{\theta}$, resulting from nonuniform distribution in
$\theta_R$, the variations resulting from direct action of irregular
forces we have 
\be
\ba{ll}
\frac{\rmd\sigma^2_R}{\rmd t} = & 2k\omega_R \overline{v_Rv'_{\theta}} + \frac{\delta\sigma^2_R }{\delta t} , \\
\noalign{\smallskip}
k^2 \frac{\rmd\sigma^2_{\theta}}{\rmd t} = & -2k\omega_R \overline{v_Rv'_{\theta}} + k^2 \frac{\delta\sigma^2_{\theta} }{\delta t} . 
\ea
\label{eq19.A2.4} 
\ee
Because of quasi-stationarity the left parts of these equations are equal. Hence
\be
2k \omega_R\overline{v_Rv'_{\theta}} = - \frac{1}{2}\left( \frac{\delta\sigma^2_R}{\delta t} -k^2 \frac{\delta\sigma^2_{\theta} }{\delta t} \right) . \label{eq19.A2.5} 
\ee
But
\be
\overline{v_Rv'_{\theta}} = -\gamma (k^2-1)\sigma^2_{\theta} \label{eq19.A2.6}
\ee
(for small angles $\gamma$). Consequently
\be
\gamma = \frac{k }{4(k^2-1)\omega_R\sigma^2_R} \left( \frac{\delta\sigma^2_R}{\delta t} - k^2 \frac{\delta\sigma^2_{\theta} }{\delta t} \right) \label{eq19.A2.7} 
\ee
Further according to Eq.~(\ref{eq19.A4.1}) we have
\be
\gamma = - \frac{1}{2\tau\omega_R} \frac{\sigma_*^2 + p\sigma^2 }{\sigma_R\sigma_{\theta}} . \label{eq19.A2.8} 
\ee

\section[C. Variations of the mean velocity and velocity
dispersion]{C. Variations of the mean velocity and velocity dispersion
  due to the direct action of irregular forces} 

Evidently
\be
\ba{ll}
\frac{\delta\overline{v}_i }{\delta t} = & \overline{a}_i , \\
\noalign{\smallskip}
\frac{\delta\sigma^2_i }{\delta t} = & 2 [ \overline{(v_i-\overline{v}_i )a_i} + \overline{A}_{ii} ], 
\ea 
\label{eq19.A3.1} 
\ee
where $a_i$ is the $i$-th component of mean velocity variation vector,
due to irregular forces (dynamical friction), and $A_{ii}$ is the
$ii$-th component of the dispersion tensor of the velocity
variation. Averaging is done over stars in unit volume. 

In \citet{Kuzmin:1957aa} (Chapter 17) we gave the expressions for
vector $\boldsymbol{b}$ and tensor {\bf B}, related to $\boldsymbol{a}$
and {\bf A} according to relations 
\be
\boldsymbol{a} = \pi Gm^2(1+\mu ) \boldsymbol{b} , ~~~ {\bf A} = \pi Gm^2{\bf B},  \label{eq19.A3.2} 
\ee
where $\mu$ is the ratio of the masses of ordinary star and disturbing
star. By using the expressions for $\boldsymbol{b}$ and {\bf B} and
assuming $\mu =0$, we derive Eqs.~(\ref{eq19.4.1})--(\ref{eq19.4.3})
and thereafter within the simplification we have done we have
Eqs.~(\ref{eq19.4.5})--(\ref{eq19.4.8}). 

\section[D. Variation of the velocity dispersion]{D. Variation of the
  velocity dispersion. The ratio of dispersions}  

In following papers (Chapters 20 and 22) we decided to replace in
Eqs.~(\ref{eq19.4.5})--(\ref{eq19.4.7}) the coefficient 3 with $2p$
while $p\simeq 1.$ In this case by taking $\kappa_i = 2\sigma^2_*,$ we
have 
\be
\tau \frac{\delta\sigma^2_i}{\delta t} = 2 \left[ \sigma^2_* + p(\sigma^2 -
\sigma^2_i ) \right] . \label{eq19.A4.1} 
\ee
Further, assuming
\be
\frac{\dot{\omega}_z}{\omega} = - \frac{\alpha}{\tau} , \label{eq19.A4.2} 
\ee
we find for the variation of the velocity dispersions
\be
\ba{ll}
\frac{\tau}{\sigma^2_R} \frac{\rmd\sigma^2_R}{\rmd t} = & \frac{\tau}{\sigma^2_{\theta}} \frac{\rmd\sigma^2_{\theta}}{\rmd t} = \frac{(k^2+1)(\sigma^2_* +p\sigma^2 ) }{\sigma^2_R} -2p, \\
\noalign{\medskip}
\frac{\tau}{\sigma^2_z} \frac{\rmd\sigma^2_z}{\rmd t} = & \frac{\sigma^2_* +p\sigma^2}{\sigma^2_z} -p -\alpha . 
\ea
\label{eq19.A4.3} 
\ee

From these equations we have
\be
\ba{ll}
\tau \frac{\rmd\sigma^2_R /\sigma^2_z }{\rmd t} = & \frac{\sigma^2_R}{\sigma^2_z} \left[ \left(\sigma^2_* + p\sigma^2 \right) \left( \frac{k^2+1}{\sigma^2_R} - \frac{1}{\sigma^2_z} \right) +\alpha -p \right] ,\\
\noalign{\smallskip}
\tau \frac{\rmd\sigma^2}{\rmd t} = & \lambda\sigma^2_* + (\lambda -2)p\sigma^2 -
\frac{1}{3} (\alpha -p) \sigma^2_z , 
\ea
\label{eq19.A4.4} 
\ee
where
\be
\lambda = 1+ \frac{k^2+k^{-2}}{3}. \label{eq19.A4.5} 
\ee
Supposing $\alpha$, $p$, $k$, $\sigma_*$ to be constants, we may
eliminate $t$ from these equations. This gives us the differential
equation of the first order, and the solution gives us the dependence
of $\sigma^2 / \sigma^2_z$ on $\sigma^2 /\sigma^2_*$. For the initial
condition we may demand that $\sigma_R^2 /\sigma_z^2$ must be finite
for $\sigma^2 /\sigma^2_*  =0$. 

For 
\be
\alpha = p \label{eq19.A4.6} 
\ee
we derive the solution
\be
\frac{\sigma^2_R}{\sigma^2_z} = k^2 +1. \label{eq19.A4.7} 
\ee
For general case we find
\be
\frac{\sigma^2_R}{\sigma^2_z} = k^2 + 1 + \frac{(\alpha -p)p}{\lambda (1+p)} \frac{\sigma^2}{\sigma^2_*} + ... \label{eq19.A4.8} 
\ee

In the vicinity of the Sun, the solution with $\alpha =p$ is suitable. But this is a good approximation probably also for a quite general case.

After introducing $\sigma^2_*$ and $p$ Eq.~(\ref{eq19.5.7}) will have a form
\be
\left( \frac{\overline{\sigma^2_* + p\sigma^2} }{ \sigma^2_z}\right) = p + 3\alpha . \label{eq19.A4.9} 
\ee
Accepting the solution with $\alpha =p$ we find
\be
\sigma^2_* \overline{\sigma^{-2}} = p \frac{4-\lambda}{\lambda} .  \label{eq19.A4.10} 
\ee
$\lambda\simeq 2$ in the solar neighbourhood, but as a more crude
approximation it can be valid everywhere (in Hooke's field $\lambda
=5/3$, in Newtonian field 29/12). Therefore, if $p\simeq 1$ we have
$\sigma^2_*\simeq 1/\overline{\sigma^{-2}}$. This is quite acceptable
estimate for $\sigma^2_*$.

%% file: chapter20.tex
\chapter[On the dynamics of the non-stationary Galaxy]{On the dynamics
  of the non-stationary Galaxy\footnote{\footnotetext ~~Tartu
    Astron. Observatory Teated No.~6, 19, 1963; Report on the Meeting
    of the Working Group on Stellar Kinematics and Dynamics of the
    Astron. Council of Academy of Sciences USSR, January 1962.} } 

\section{~}

The theory of non-stationary stellar systems by
\citet{Chandrasekhar:1942} has became highly known and achieved high
authority within the recent decades. Chandrasekhar's theory is focused
on the variation of spatio-kinematical  structure of a stellar system
due to the variation of the gravitation potential. The theory starts
with postulating the existence of precisely ellipsoidal velocity
distribution in the system. This assumption makes the results of the
theory quite specific. 

For very flat non-stationary subsystems of the Galaxy it is fairly easy to develop a more general and flexible theory than the Chandrasekhar's theory. In this theory the gravitational potential $\Phi$ is assumed to be symmetrical with respect to an axis and a plane. But in addition, unlike in Chandrasekhar's theory, one needs to suppose that $\Phi$ is varying quit slowly with time.

In this case, the usual theory of nearly circular orbit can be applied with some modifications to the motion of stars in the subsystem. Circular orbits are replaced by quasi-circular orbits, harmonic oscillations about the circular orbit by quasi-harmonic oscillations. The angular velocity $\omega$ of motion on the quasi-circular orbit, and the frequencies $\omega_R$ and $\omega_z$ of quasi-harmonic oscillations along the cylindrical coordinates $R$ and $z$ are connected to the properties of the potential $\Phi$ in usual manner. However, $\omega$, $\omega_R$, $\omega_z$ are functions of not only $R$ but also of time $t$.

Time variations of the radius of quasi-circular orbit and of the amplitudes $a_R$ and $a_z$ of the oscillations are 
\be
\frac{\dot{R}}{R} = - \frac{\dot{\omega}}{2\omega} , ~~~~ \frac{\dot{a}_R}{a_R} = - \frac{\dot{\omega}_R}{2\omega_R} , ~~~~ \frac{\dot{a}_z}{a_z} = - \frac{\dot{\omega}_z}{2\omega_z} . \label{eq20.1} 
\ee

As in ordinary theory of nearly circular orbits, the isocurves of $a_R$ in the velocity space are ellipses. The axial ratio of ellipse is the same as in usual theory
\be
k = \frac{2\omega}{\omega} . \label{eq20.2} 
\ee
But unlike in the ordinary theory, the line of the vertices of the ellipses has the deviation from the $R$ direction. The deviation angle $\gamma$ is 
\be
\gamma = - \frac{3\dot{k} }{ 2(k^2-1)\omega_R } . \label{eq20.3} 
\ee

In a way similar to the usual theory, we may define the state of quasi-stationary motion as the state, where the distribution of stars with galactocentric longitude on quasi-circular orbit $\theta$ and with the phases $\theta_R$ and $\theta_z$ of $R$ and $z$ oscillations is uniform. Then, as in case of usual theory, the velocity distribution is elliptical, while the velocity ellipses describe the equidensity lines in the velocity space. This result can be derived also from Jeans theorem by using the integrals of motion.

For the centroid velocities we have
\be
\bar{v}_R = - \frac{\dot{\omega}}{2\omega} R, ~~~~ \bar{v}_{\theta} = R\omega, ~~~~ \bar{v}_z = - \frac{\dot{\omega}_z}{2\omega_z} z . \label{eq20.4} 
\ee
It can be seen, that the K-effect turns out to be anisotropic in
general. This means that the expansion or the contraction in $R$,
$\theta$ and $z$ directions occurs with different velocities. The
K-effect anisotropy distorts the differential rotation by deviating
the ``kinematical  center'' of the Galaxy from $R$ direction by angle 
\be
\gamma ' = - \frac{2\dot{k} }{ (k^2-1)\omega_R} . \label{eq20.5} 
\ee

The velocity dispersions $\sigma_R$, $\sigma_{\theta}$, $\sigma_z$ and the dispersion of $z$ coordinate are now functions of $R$ and $t$. Besides, as in usual theory, following relations hold
\be
\sigma_R = k\sigma_{\theta} , ~~~~~ \sigma_z = \omega_z\zeta . \label{eq20.6}
\ee

For time variations of the dispersions we have the expressions
\be
\frac{\dot{\sigma}_R}{\sigma_R} = \frac{\dot{\sigma}_{\theta}}{\sigma_{\theta}} + \frac{\dot{k} }{k} = \frac{\dot{\omega}_R}{2\omega_R} , ~~~~~ \frac{\dot{\sigma}_z}{\sigma_z} = - \frac{\dot{\zeta}}{\zeta} = \frac{\dot{\omega}}{2\omega_z} . \label{eq20.7} 
\ee
Here we have in mind the variation with $R$ moving with centroid $R$-velocity, \ie 
\be
\frac{\rmd}{\rmd t} = \frac{\partial}{\partial t} + v_R \frac{\partial}{\partial R}. \label{eq20.8}
\ee

The developed theory is evidently significantly more general than the
Chandrasekhar's theory. In Chandrasekhar's theory the relative
variations of $\omega$, $\omega_R$ and $\omega_z$, and also of the
velocity dispersions $\sigma_R$ and $\sigma_z$ are only functions of
time and not of $R$. 

Further, in Chandrasekhar's theory the rotational velocity $\bar{v}_{\theta}$ of a stellar system and the ratio of dispersions $\sigma_R/\sigma_{\theta}$ are fixed functions of $R$. In our theory this dependence is determined by $\omega$ and $\omega_R$ as functions of $R$, which can be given arbitrarily. And finally, Chandrasekhar's theory does not give the deviation of vertices.

All these highly specific results of Chandrasekhar's theory are related to the postulate about strictly ellipsoidal distribution of velocities. In our theory as well as in usual theory of nearly circular orbits, the ellipsoidal velocity distribution (or to be more precise, the elliptical distribution) is assumed only as some kind of approximation.

\section{~ }

Although the theory presented above describes the deviation of
vertices, it does not explain the value of observed deviation. First,
the observed deviation of vertices is too large; second, it predicts
the deviation of the ``kinematical  centre'' to be of the same order
as the deviation of vertices. In reality, no deviation of
``kinematical  centre'' is detected. 

We must assume that the condition of quasi-stationarity is not precisely valid. If the distribution of stars with $\theta$, $\theta_R$, $\theta_z$ is not completely uniform, there will be oscillations of the spatial density, the centroid velocity vector and the velocity dispersion tensor. These oscillations, if they are not damped too quickly, can explain according to \citet{Lindblad:1955} the observed deviation of vertices.

Slowly these oscillations will be damped, because other stars oscillating with respect to circular orbits of different radii are moving in the neighbourhood of a given point, and $\omega$, $\omega_R$, $\omega_z$ are functions of $R$. This process of approaching to a stationary state is usually called ``mixing''.

Let us have for the distribution of stars in $\theta$, $\theta_R$, $\theta_z$ for $t=0$ a Fourier expansion term $\exp (i\Theta )$, where
$$\Theta = l\theta + m\theta_R + n\theta_z $$
($l$, $m$, $n$ are natural numbers). In this case the corresponding expansion term of the phase density at a given point of the velocity space oscillates according to the formula
\be
e^{i\Omega t} \cdot e^{i\frac{\rmd\Omega}{\rmd R}\Delta R t} , \label{eq20.9} 
\ee
where
\be
\Omega = l\omega + m\omega_R + n\omega_z , \label{eq20.10} 
\ee
and $\Delta R$ is the distance along $R$ from the circular orbit. It is proportional to the difference between $v_{\theta}$ and circular velocity.

The oscillations of the phase density are mixed in the course of time owing to the second factor in Eq.~(\ref{eq20.9}). The amplitude of the oscillations averaged over arbitrarily small volume of velocity space tends to zero for $ t\rightarrow\infty $.

It is suitable to take the characteristic damping time $T$ of the  oscillations of the space density and of the ``velocity body'' according to the formula
\be
T^{-1} = \bigg| \frac{\rmd\Omega}{\rmd R}\bigg| ~ \frac{\sigma_R}{\omega_R}
\label{eq20.11}
\ee
($\sigma_R/\omega_R$ equals to mean square of $\Delta R$).

If the $\Omega$ gradient is very small for a set of $l$, $m$, $n$, the phase space oscillations mix very slowly. In the system of coordinates rotating with velocity
\be
\omega^* = \frac{\Omega}{l} , \label{eq20.12} 
\ee
the flat subsystem is almost stationary, although it can be non-stationary in the non-rotating coordinates and not have axial symmetry. Apart from the usual three integrals of nearly circular motion, that form the arguments of the phase density, there exists fourth integral [see Appendix A].

To explain the vertex deviation as well as other possible deviations from the ``normal'' state of the first and second moments of velocities in the flat subsystem, $\Omega$ gradient must be assumed to be small for $0 < |m| + |n| \le 2$. Indeed, as it was pointed out by \citet{Lindblad:1955}, in the Galaxy 
$$2 \omega - \omega_R$$
varies very slowly with $R$. In the system of coordinates rotating with velocity 
$$\omega^* = \omega - \frac{1}{2} \omega_R , $$
the flat subsystem of the Galaxy is almost stationary. In the solar neighbourhood $\omega$ exceeds $\omega^*$ about 3 times [see Appendices
B and C].

\section{~ }

In Section I we treated the variation of spatio-kinematical
characteristics of the flat subsystem of the Galaxy due to the
variation of the gravitational potential. In that case the action of
the irregular gravitational forces leads to the variation of the
spatial distribution of stars, and consequently to the variation of
the potential.  The discussed variation of the spatio-kinematical  characteristics is the indirect effect of irregular forces. Now we shall discuss the direct action of irregular forces. 

The action of the irregular forces can be treated within a certain approximation as the action of momentary impulses, abruptly changing the position of a star. If the impulses a not large or are not frequent, the subsystem can be treated as quasi-stationary at every moment. The spatial redistribution of stars has sufficient time to re-establish the equilibrium values of the velocity centroid and the dispersion ratios $\sigma_R/\sigma_{\theta} = k$ and $\sigma_z/\zeta = \omega_z$. In this case the variations of the dispersions due to irregular forces are described by
\be
\frac{\dot{\sigma}_R}{\sigma_R} = \frac{\dot{\sigma}_{\theta}}{\sigma_{\theta}} = \frac{1}{2} \left( \frac{\dot{\sigma}_R}{\sigma_R} + \frac{\dot{\sigma}_{\theta}}{\sigma_{\theta}} \right)_0 ~~~~~ \frac{\dot{\sigma}_z}{\sigma_z} = \frac{\dot{\zeta}}{\zeta} = \frac{1}{2}  \left( \frac{\dot{\sigma}_z}{\sigma_z} \right)_0 . \label{eq20.13} 
\ee
The subindex zero means here the direct variation of dispersions, \ie  the variations ``before'' the spatial redistribution of stars.

In case of spatial redistribution of stars, while restoring the quasi-stationary state, some of the nearly circular motion energy turns into the energy of peculiar motion parallel to the galactic plane, and half of $z$-motion energy variation turns into the potential energy variation in the same direction. 

We may add to Eq.~(\ref{eq20.13}) the formula for the vertex deviation, caused by irregular forces
\be
\gamma \frac{k^2-1}{k} \omega_R = \frac{1}{2} \left( \frac{\dot{\sigma}_R }{\sigma_R} - \frac{\dot{\sigma}_{\theta} }{\sigma_{\theta}} \right)_0 .
\label{eq20.14} 
\ee

According to \citet{Spitzer:1953} and \citet{Gurevich:1954}, the source of irregular forces may be expected to be the massive clouds of stars and diffuse matter.

Taking into account the big mass of these clouds, the following approximate formula can be derived
\be
\left( \frac{\rmd\sigma_R^2}{\rmd t} \right)_0 = \frac{2}{\tau} \left[ \sigma_*^2 + p (\sigma^2 - \sigma_R^2 ) \right] \label{eq20.15} 
\ee
and two other similar formulas, where $\tau$ is the time, characterising the rate of irregular forces, $\sigma^*$ is the quantity characterising mainly the dispersion of the peculiar velocities of clouds, $p$ is a parameter being approximately equal to one,\footnote{In our paper in Tartu Astron.Obs. Publ., 33, 351, 1961 we chose $p=3/2$, but it would be better to adopt a somewhat smaller value.} and $\sigma$ is the square mean velocity dispersion of stars along three coordinate axes. Relation (\ref{eq20.15}) shows that irregular forces tend to round the velocity ellipsoid. But in practice this does not occur because of the spatial redistribution of stars.

Using Eq.~(\ref{eq20.15}) we have
\be
\ba{ll}
\frac{\dot{\sigma}_R}{\sigma_R} = \frac{\dot{\sigma}_{\theta} }{\sigma_{\theta}} = & \frac{1}{\tau} \left( \frac{k^2+1}{2} ~ \frac{\sigma_*^2 + p\sigma^2 }{\sigma_R^2} - p \right) , \\
\noalign{\medskip}
\frac{\dot{\sigma}_z}{\sigma_z} = \frac{\dot{\zeta}}{\zeta} = & \frac{1}{2\tau} \left( \frac{\sigma_*^2 + p\sigma^2 }{\sigma_z^2} - p\right) 
\ea 
\label{eq20.16}
\ee
and
\be
\gamma = - \frac{1}{2\tau\omega_R} \frac{\sigma_*^2 + p\sigma^2 }{\sigma_R \sigma_{\theta}} . \label{eq20.17} 
\ee

To obtain the total variation of dispersions $\sigma_R$, $\sigma_{\theta}$, $\sigma_z$ and $\zeta$, these results must be added to  those in Sect.~I. $\omega$ and $\omega_R$ vary very slowly due to irregular forces, and the corresponding terms in variations of $\sigma_R$ and $\sigma_{\theta}$ can be neglected. The same is valid for the vertex deviation. But the variation of $\omega_z$ must be taken into account, because this quantity must vary significantly due to the spatial redistribution of stars. Hence we must add to the relative variations of $\sigma_z$ and $\zeta$ in Eq.~(\ref{eq20.13}) their variation according to Eq.~(\ref{eq20.7}).

Within some approximation it can be assumed that the ratio $\sigma_R/\sigma_z$ is near to its equilibrium value, \ie  to the value where its time derivative is zero. It is also known that for different subsystems of the Galaxy the ratio $\sigma_R/\sigma_z$ (as well as $\sigma_R /\sigma_{\theta}$) is nearly the same. Assuming the ratio $\sigma_R /\sigma_z$ to be near its equilibrium value and to be the same for all flat subsystems we find
\be
\tau \frac{\dot{\omega}_z}{\omega_z}= - p \label{eq20.18} 
\ee
and 
\be
\frac{\sigma_R^2}{\sigma_z^2} = 1+k^2 . \label{eq20.19} 
\ee

Equation (\ref{eq20.19}) may be written in form
\be
\frac{1}{\sigma_R^2} + \frac{1}{\sigma_{\theta}^2} = \frac{1}{\sigma_z^2} 
\label{eq20.19'} 
\ee

This relation agrees well with observations.

Therefore, it seems that the observed ratio $\sigma_R/\sigma_z$ can be explained by the theory of irregular forces. The irregular forces have probably a considerable role in dynamics of the Galaxy.

The vertex deviation, however, cannot be explained by the action of the irregular forces.
\vglue 5mm
{\hfill January 1962}

\vglue 8mm

{\bf\Large Appendices added in 1969}
\vglue 8mm

\section{A.  The fourth isolating integral for nearly circular motion} 

Non-conservative integrals for nearly circular motion are
\be
J_{\theta} = \theta -\omega t, ~~ J_R = \theta_R -\omega_R t, ~~ J_z = \theta_z -\omega_z t. \label{eq20.A1.1} 
\ee
From these we can combine an integral
\be
J = \Theta - \Omega t, \label{eq20.A1.2} 
\ee
where
\be
\Theta = l\theta +m\theta_R +n\theta_z , ~~~~ \Omega = l\omega +m\omega_R +n\omega_z . \label{eq20.A1.3} 
\ee
Let us assume
\be
\rmd\Omega / \rmd R =0 \label{eq20.A1.4} 
\ee
and let us introduce the rotating system of coordinates, where galactocentric longitude is
\be
\theta '=\theta -\omega^* t, \label{eq20.A1.5} 
\ee
\be
\omega^* = \Omega /l . \label{eq20.A1.6} 
\ee
In these coordinates we have an isolating conservative integral of nearly circular motion
\be
J = l\theta ' +m\theta_R +n\theta_z . \label{eq20.A1.7} 
\ee

In the Galaxy $\Omega$ is nearly constant for $l=2$, $m=-1$, $n=0$, \ie  for
\be
\Omega = 2\omega -\omega_R , \label{eq20.A1.8} 
\ee
and
\be
\omega^* = \omega - \frac{1}{2}\omega_R . \label{eq20.A1.9} 
\ee
In this case
\be
J = 2\theta ' -\theta_R . \label{eq20.A1.10} 
\ee

\section[B.  Model of the Galaxy allowing the fourth integral of
circular motion]{B.  A model of the Galaxy allowing the fourth
  integral of nearly circular motion}  

As
\be
\omega^2_R = 4\omega^2 + R \frac{\rmd\omega^2}{\rmd R} \label{eq20.A2.1} 
\ee
and assuming $\omega - \frac{1}{2}\omega_R = \omega^* = \mathrm{const}$ we have the differential equation
\be
\frac{1}{2} x \frac{\rmd y}{\rmd x} + \frac{2y-1}{y} =0 , \label{eq20.A2.2} 
\ee
where
\be
y = \frac{\omega}{\omega^*}, ~~~ x = \frac{R}{R^*} , \label{eq20.A2.3} 
\ee
and $R^*$ is the scale parameter. The solution of this equation is 
\be
e^{2(y-1)}(2y-1) x^8 = 1 , \label{eq20.A2.4} 
\ee
where the constant of integration is chosen in a way that for $x=1$ $y=1.$

\begin{figure*}[ht]
\centering
\includegraphics[width=80mm]{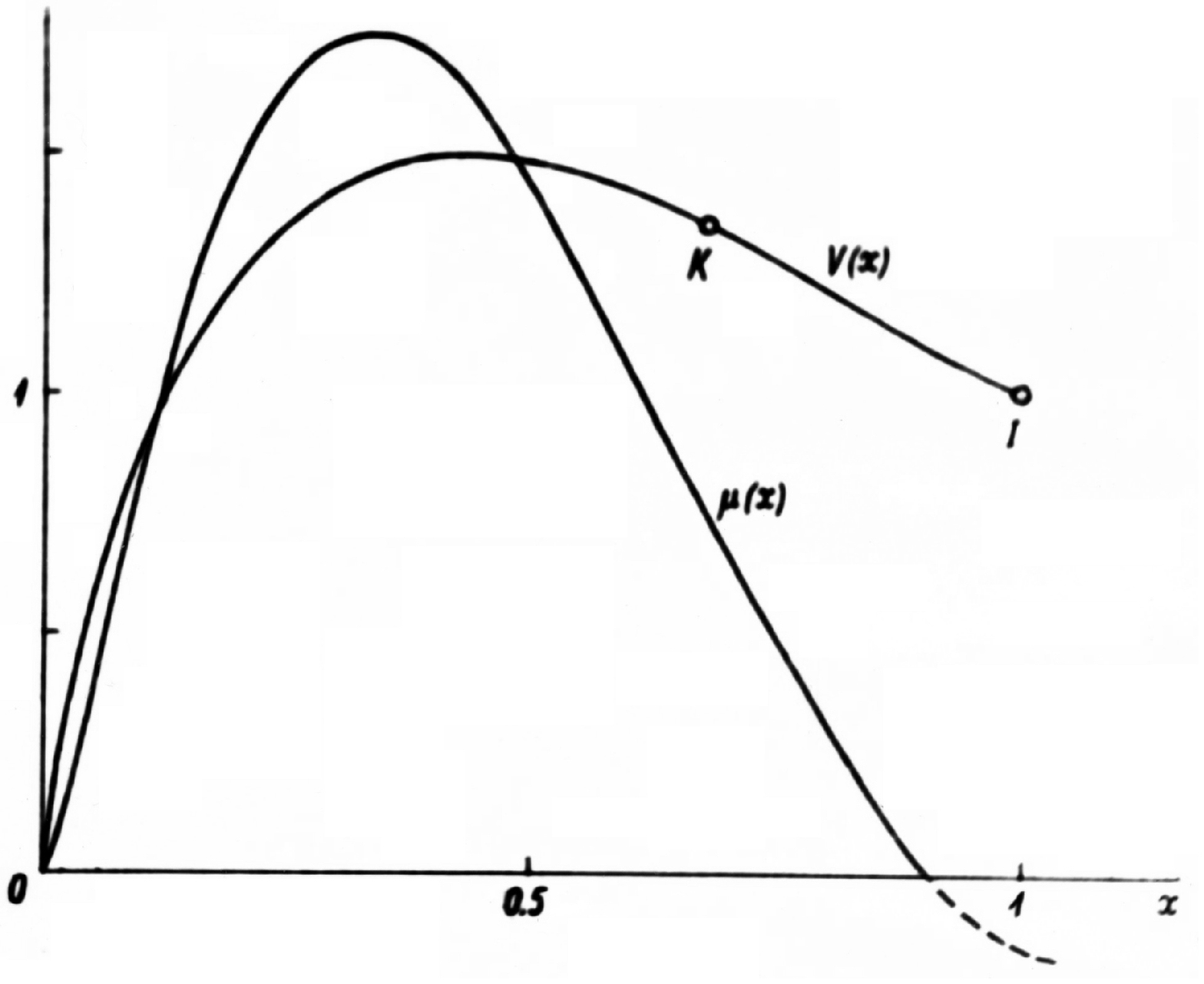}
\caption{}
\label{fig20.A1}
\end{figure*}

The circular velocity $V$ in units of $R^*\omega^*$ equals to $xy$. The curve $V(x)$ is plotted in Fig.~(\ref{fig20.A1}). The point $K$ for $y=2$ corresponds to the Keplerian nearly circular orbits, the point $I$ for $x=y=1$ corresponds to the circular orbits at the limit of stability. In the same Figure the mass function $\mu (x)_{\epsilon =0}$ in units of $G^{-1}(R^*\omega^* )^2$ is given. The model is quite acceptable with the exception of outermost regions, where the mass function vanishes too rapidly (the limit of the model).

\section[C.  Non-axisymmetric highly flattened subsystem]{C.
  Non-axisymmetric highly flattened subsystem, stationary in rotating
  coordinates}  

The phase density of this kind of subsystem is
\be
\Psi = \Psi (E_R,E_z,J,R) , \label{eq20.A3.1} 
\ee
where
\be
E_R = \frac{1}{2} (v^2_R +kv'^2_{\theta} ), ~~~ E_z = \frac{1}{2} (v^2_z +\omega^2_z z^2) \label{eq20.A3.2} 
\ee
(Chapter 19) and
\be
J = 2\theta ' -\theta_R . \label{eq20.A3.3} 
\ee
As in $J$ there is the term $2\theta '$, the subsystem is symmetric in
a sense that its properties are similar in diametrically opposite
points with respect to the galactic centre. 

\begin{figure*}[ht]
\centering
\includegraphics[width=80mm]{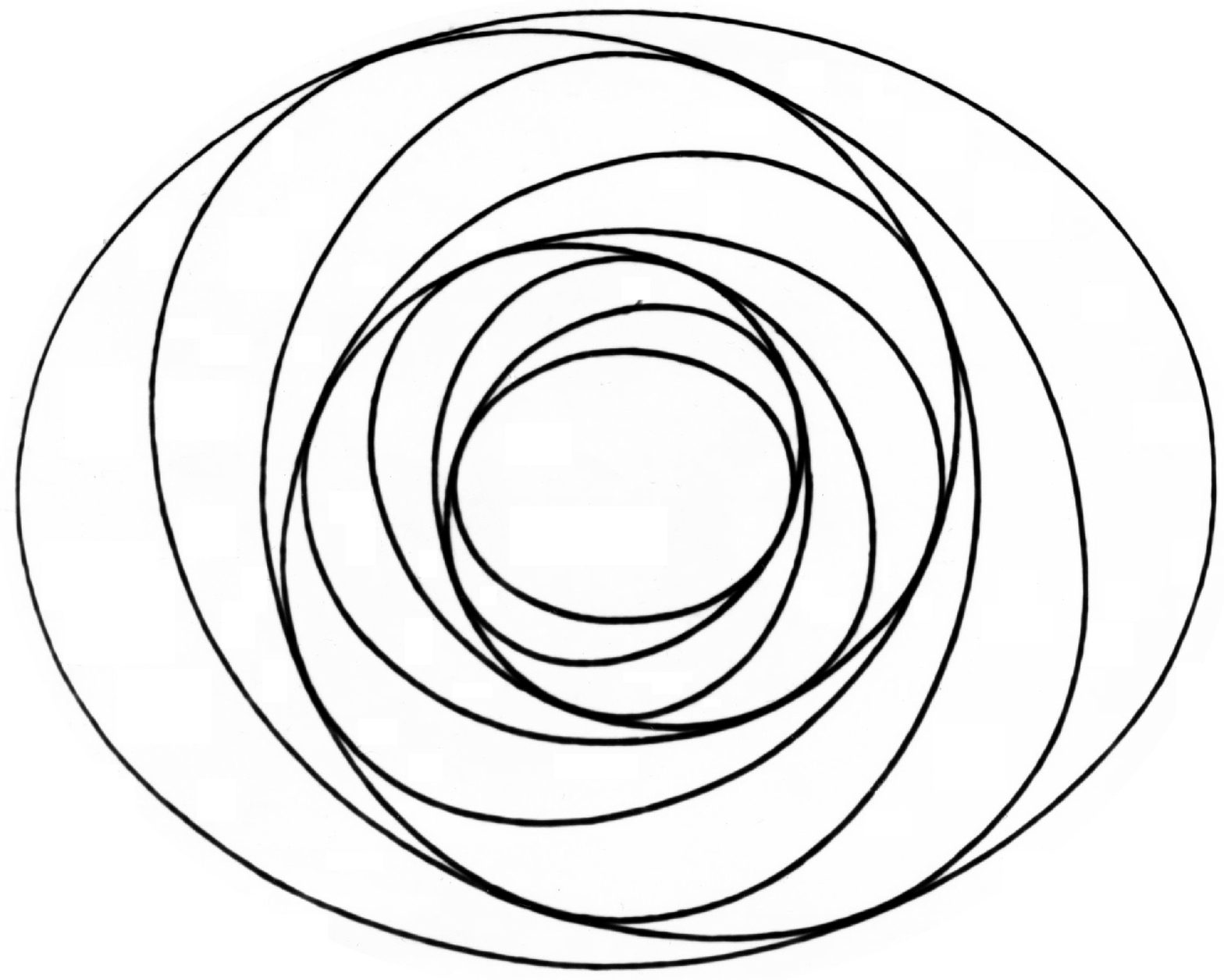}
\caption{}
\label{fig20.A2}
\end{figure*}

In particular, the subsystem may have two-armed spiral structure. It
is not difficult to construct a model for this kind of subsystem. In
coordinates where the subsystem is stationary, the nearly circular
orbits are closed and have oval form with a centre at the centre of
the Galaxy. Let us assume for simplicity that there are no velocity
dispersions, \ie  orbits are not intersecting. In this case it is not
difficult to choose a configuration of orbits with given (or depending
on dimensions) form giving us a spiral structure. An example is
demonstrated in Fig.~(\ref{fig20.A2}). 

In a stationary system of coordinates the spiral arms rotate with the
angular velocity $\omega^*$, being smaller than the rotation velocity
$\omega$ of the subsystem. Hence, they are similar to the density
waves spreading in subsystem in the opposite direction to the
rotation. 

The total gravitational field is axisymmetric. Therefore, our
subsystem must have very small mass. Otherwise, its gravity would
destroy the axial symmetry of the general gravitational field. 

Wave theory of spiral structure taking into account the gravitation of
waves is developed recently by Lin and others.

%% file: chapter21.tex
\chapter[Hydrodynamics of stellar systems]{Hydrodynamics of stellar systems.\footnote{\footnotetext ~~Trudy Astrophys. Inst. Acad. Sci.  Kazachstan SSR, vol. 5, 70, 1965 = Tartu Astrophysical Observatory, Teated No. 15a, 1965. Report on the Meeting ``Kinematics and dynamics of stellar systems and interstellar medium'' October 10 -- 16, 1963, Alma-Ata.}} 

\section{~}

From theoretical point of view the most complete statistical description of a stellar system is the phase density. But in practice this function is difficult to determine. It is easier to determine and to operate with the spatial density, the centroid velocity and the velocity dispersion tensor. This is why the hydrodynamic equations play an important role in the stellar dynamics when studying the spatio-kinematical structure and the gravitational field of stellar systems. Besides, they can be used to certain extent for studying the different forms of oscillations and evolutionary phenomena in stellar systems.

In stellar dynamics the hydrodynamic equations were introduced by \citet{Jeans:1922}, who first used them to study the Galaxy. Later \citet{Lindblad:1927} derived theoretical relation between the velocity dispersions along $R$ and $\theta$ coordinates having, as it was understood later, the hydrodynamic interpretation. \citet{Oort:1928} confirmed Lindblad's result and analysed the Jeans equations in case of Schwarzschild velocity distribution. The Jeans equations in case of triaxial velocity ellipsoid were studied by the author \citep{Kuzmin:1954, Kuzmin:1956a, Kuzmin:1962ab}.

\citet{Lindblad:1927a} first used the hydrodynamic equations to study the variations in stellar systems. Amongst his further papers on this topic, special attention deserves a paper, where he derived the ``adiabatic'' theorem of the hydrodynamics of stellar systems \citep{Lindblad:1950}.\footnote{The following papers are also very important in the  hydrodynamics of stellar systems: \citet{Lindblad:1936}; \citet{Courtez:1950}. [Later footnote]} 

Amongst recent studies on the hydrodynamics of stellar systems we mention interesting papers by \citet{Davydov:1955}, \citet{Ogorodnikov:1957}, and \citet{Agekyan:1963} on the hydrodynamics of spherical stellar systems.

\section{~}

Most convenient method to derive the hydrodynamic equations is to use the transfer equation [q.v. Appendix A]. Transferred quantity may be the stellar weight, the velocity component or the dyad of the residual velocity components, giving us the following equations, respectively
\be
\frac{\rmd\rho}{\rmd t} + \rho \frac{\partial v_l}{\partial x_l} = 0 ,\label{eq21.1}
\ee
\be
\frac{\rmd v_m}{\rmd t} + \frac{1}{\rho} \frac{\partial\rho w_{lm}}{\partial x_l} = \frac{\partial\Phi}{\partial x_m} + \frac{\rmd v_m}{\rmd t_0} , \label{eq21.2} 
\ee
\be
\frac{\rmd w_{mn}}{\rmd t} + w_{lm} \frac{\partial v_n}{\partial x_l} + w_{ln} \frac{\partial v_m}{\partial x_l} + \frac{1}{\rho} \frac{\partial\rho w_{lmn}}{\partial x_l} = \frac{\rmd w_{mn}}{\rmd t_0} . \label{eq21.3}
\ee
Here $t$ is time, $x$ -- rectangular coordinate, $\rho$ -- spatial weight density, $v$ -- centroid velocity component, $\rho w$ -- component of tensor of peculiar velocity momenta, $\Phi$ -- gravitational potential. Time-derivative designates the variation speed at the point comoving with centroid, the time-derivative with subindex zero stands for the variation, due to the irregular gravitational forces. As usual, the summation is performed over repeating indices (the index $l$). 

This form of equations is independent of the weight ascribed to stars. But surely the meaning of the equations depends on weight. We expect that the weight does not change in time. Otherwise additional terms appear in the equations. 

The first equation, known as the continuity equation, says that the weight density is inversely proportional to the volume element of the centroid continuum. In other words, the amount of the weight in the element of centroid continuum remains constant.\footnote{In case of the dissipation of stars from the system, one corresponding term must be added  in Eq.~(\ref{eq21.1}). [Later footnote.]} Just this, and not the continuity of the medium, is the meaning of the equation.

The second equation is the statistical equation for stellar motion, or the local centroid velocity equation. We may call it as the macro-motion equation. The  second term with opposite sign gives the centroid acceleration, caused by nonuniform transfer of macro-motion, terms on the right side are the gravitational acceleration and the acceleration due to irregular forces (the dynamical friction).

It is suitable to call the third equation the micro-motion equation, because the velocity dispersion tensor $w_{mn}$ is the most important characteristic of residual motion, or micro-motion. Contrary to the previous equations, this equation is much less known is stellar dynamics, and remained nearly unused. The second and the third terms take into account the mutual dependence of micro- and macro-motions. The fourth term characterises the non-uniformity of the micro-motion transfer, the right side of the equation reflects the effect of irregular forces on the micro-motions. 

\section{~}

In hydrodynamics or in gas dynamics the equation of macro-motion corresponds to the generalised Euler equation. If $\rho$ is the mass density then $\rho w_{mn}$ is the stress tensor. In hydro- or gas dynamics the stress consist of isotropic pressure and viscous stress, being opposite to the deformation of the element of medium. In gas dynamics, the viscous nature of the stress anisotropy results from the equation of micro-motion (\ref{eq21.3}), when the effect of irregular forces is sufficiently intense. 

The trace of the tensor $w_{mn}$ is proportional to the temperature, and as a result the trace of the micro-motion equation gives the temperature variation. The last term on the left side of the equation accounts for the heat conductivity and can be expressed via temperature gradient. In gas dynamics, this property of the heat conduction results from the equation, which is higher by an order than Eq.~(\ref{eq21.3}), and again one needs to suppose that the intensity of irregular forces is sufficiently high.

For the same condition, it is possible to express the right sides of the equations of micro- and macro-motion (if they are nonzero, which is the case for individual gases and their composition). Hence, if the external forces are given, the equations of gas dynamics can be reduced to the closed set of equations, where the number on unknown functions is equal to the number of equations.

In the hydrodynamics of stellar systems we have a highly different situation. Here we can not suppose the high intensity of irregular forces. Most important here is the regular gravitational field, while the irregular gravitational forces are relatively weak. Thus the hydrodynamic equations of stellar dynamics do not form a closed set of equations. The medium, consisting of stars,  can not be handled simply as viscous heat-conducting gaseous medium. The behaviour of stellar medium has special characteristics, although some properties resembling viscosity and heat conductivity may appear due to irregular forces.

\section{~}

In the first approximation, a stellar system can be assumed to be stationary and axisymmetric, and the irregular forces can be neglected. In addition, it can be assumed that the centroid motion reduces to the rotation about the symmetry axis, \ie  in cylindrical coordinates $v_R = v_z =0$. In this case, the equation for the density reduces to identity, and the equations of macro-motion are
\be
- \frac{v_{\theta}^2}{R} + \frac{1}{\rho} \left( \frac{\partial\rho w_{RR} }{\partial R} + \frac{\partial\rho w_{Rz}}{\partial z} + \rho \frac{w_{RR} - w_{\theta\theta} }{R} \right) = \frac{\partial\Phi}{\partial R} ,
\label{eq21.4} 
\ee
\be
\frac{1}{\rho} \left( \frac{\partial\rho w_{zz} }{\partial z} + \frac{\partial\rho w_{Rz} }{\partial R} + \rho \frac{w_{Rz} }{R} \right) =  \frac{\partial\Phi}{\partial z} . \label{eq21.5} 
\ee
The third equation, resulting from the equation of macro-motion reduces to identity if we assume that two axis of the velocity ellipsoid lie in the meridional plane, \ie  $w_{R\theta} = w_{z\theta} = 0$.

Equations (\ref{eq21.4}) and (\ref{eq21.5}) are Jeans equations \citep{Jeans:1922}, generalised for the case of triaxial velocity ellipsoid. They relate the spatio-kinematical  characteristics of a stationary axisymmetric system or its subsystem to the gravitational field, and also the characteristics of subsystems to each other, as all the subsystems lie in the same gravitational field.

Within the assumptions above the equations of micro-motion have the forms
\be
w_{RR} \left( \frac{\partial v_{\theta}}{\partial R} + \frac{v_{\theta} }{R} \right) - 2w_{\theta\theta} \frac{v_{\theta} }{R} + w_{Rz} \frac{\partial v_{\theta} }{\partial z} + \frac{1}{\rho} \frac{\partial\rho w_{RR\theta} }{\partial R} + \frac{2w_{RR\theta} - w_{\theta\theta\theta} }{R} + \frac{1}{\rho} \frac{\partial\rho w_{Rz\theta} }{\partial z} = 0, \label{eq21.6} 
\ee
\be
w_{zz} \frac{\partial v_{\theta} }{\partial z} + w_{Rz} \left( \frac{\partial v_{\theta} }{\partial R} + \frac{v_{\theta} }{R} \right) + \frac{1}{\rho} \frac{\partial\rho w_{Rz\theta} }{\partial R} + \frac{2w_{Rz\theta} }{R} + \frac{1}{\rho} \frac{\partial\rho w_{zz\theta} }{\partial z} = 0. \label{eq21.7} 
\ee 
Remaining four equations reduce to identities, if we assume that the third order moments of the peculiar velocities, odd in respect to $R$ and $z$, are equal to zero $w_{RRR} =$ $w_{Rzz} =$ $w_{R\theta\theta} =$ $w_{z\theta\theta} =$ $w_{zRR} =$ $w_{zzz} = 0$.

If the system or the subsystem is sufficiently flattened, we may neglect all terms in parentheses in Eq.~(\ref{eq21.4}), the second and third terms in Eq.~(\ref{eq21.5}), and all terms except the first two in Eq.~(\ref{eq21.6}). In this case Eq.~(\ref{eq21.4}) reduces to the equality between the centroid velocity and the circular velocity, Eq.~(\ref{eq21.5}) becomes Jeans second equation \citep{Jeans:1922}, and Eq.~(\ref{eq21.6}) becomes Lindblad's relation between the velocity dispersions in $R$ and $\theta$ directions \citep{Lindblad:1927}.\footnote{Lindblad relation was derived for hydrodynamic equations by Courtez (q.v. first footnote), by \citet{Davydov:1955} and by us (unpublished report at Astron. Inst. Sternberg, Moscow, 1950). [Later footnote.]} 

Denoting the angular circular velocity and the frequencies of the angular oscillation of a star on the circular orbit in the $R$ and $z$ directions $\omega$, $\omega_R$, $\omega_z$ respectively, we have for highly flattened subsystem 
\be
\frac{v_{\theta}}{R} = \omega, ~~~~~ \frac{\partial v_{\theta}}{\partial R} + \frac{v_\theta}{R} = \frac{1}{k} \omega_R , \label{eq21.8} 
\ee
\be
\frac{\sigma_R}{\sigma_{\theta}} = \frac{2\omega}{\omega_R} = k , ~~~~~ \frac{\sigma_z}{\zeta} = \omega_z . \label{eq21.9} 
\ee
Here $\sigma_R$, $\sigma_{\theta}$, $\sigma_z$ are the dispersions of the $R$, $\theta$, $z$ velocities respectively, and the quantity $\zeta$ characterises the thickness of the system. If we neglect the variation of $\sigma_z$ with $z$, then $\zeta^{-2} = -\partial^2\ln\rho /\partial z^2 = \mathrm{const}.$

\section{~}

When applying the hydrodynamic equations of stellar systems to study their evolution, very serious problem is that the system of equations is not closed.

The problem simplifies significantly in case of flat rotating systems, when the transfer terms of micro-motions, containing the third order moments of peculiar velocities, can be dropped. If we can also neglect the irregular forces, the system of equations becomes closed, and the equations of micro-motion reduce to the equations, describing the mutual transformation of macro- and micro-motion. It is suitable to call these equations of micro-motion adiabatic [q.v. Appendix D].

From these equations it follows that the velocity ellipsoid is deformed, and rotates just as the centroid continuum element, but in the opposite direction. In particular, if the centroid continuum element expands, the velocity ellipsoid contracts respectively and vice versa. This gives us the Lindblad's \citep{Lindblad:1950} adiabatic theorem, according to which the volume of the velocity ellipsoid varies in time proportionally to the spatial density of the  stars. Lindblad derived his theorem on the basis of quite complicated analysis, but it results easily from the equations of micro-motion.

\section{~}

Adiabatic equations of micro-motion together with remaining hydrodynamic equations allow to study the oscillations of the flattened system or subsystem [q.v. Appendix E].

It is not difficult to solve the equations, supposing that the density of the system and the gravitational field have axial symmetry and stationarity. In this case the oscillations are the oscillations of the stellar motion.

Solution for the velocity dispersion tensor can be written in form
$$w_{mn} = c_{mn} + u_{m1} u_{n1} + u_{m2} u_{n2} , $$
while
\be
c_{RR} = k^2 c_{\theta\theta} = a_0,~~~~~ \mathrm {and~ remaining} ~ c_{mn}=0, \label{eq21.10}
\ee
$$u_{R1} + iku_{\theta 1} = a_1 e^{-i\omega_R t} , ~~~ u_{z1}= 0,$$
$$u_{R2} + iku_{\theta 2} = a_2 e^{-i(\omega_R\pm\omega_z) t} , ~~~ u_{z2} = b , $$
where $a_0$, $a_1$, $a_2$, $b$ are arbitrary constants.

For the centroid velocity it results
\be
v_R + ikv'_{\theta} = \alpha e^{-i\omega_R t} \pm i\omega_z z \frac{a_2}{b} e^{-i(\omega_R\pm\omega_z) t} , ~~~ v_z = 0, \label{eq21.11}
\ee
where $v'_{\theta}$ is the centroid velocity along $\theta$ counted from the circular velocity, and $\alpha$ is an arbitrary constant.

On the basis of that solution, the velocity ellipsoid is elongated when rotating with angular frequency $\omega_R$, and is oblique when rotating with angular frequency $\omega_R+\omega_z$ or $\omega_R-\omega_z$. Centroid describes elliptical epicycles with the same frequencies, while the amplitude of the second frequency is proportional to $z$ and the phase is  shifted by $\rm 90^o$, when compared to the phase of ellipsoid's obliquity. 

All these oscillations must be gradually damped due to the phase differences appearing between neighbouring points of space. It is difficult to follow the damping process on the basis of hydrodynamic equations. It is only possible to conclude that for more flattened systems the damping is slower.

The vertex deviation, observed for flat subsystems of the Galaxy, can  serve as a reference to the presence of the discussed oscillations in these subsystems.

\section{~}

Hydrodynamic equations of stellar systems can be used to study the secular variations in flattened stellar systems or subsystems. But in this case, the irregular forces have to be accounted for.

As the variations are slow, the structure of the system is near to the quasi-stationary state (neglecting the oscillations discusses above). Thus the centroid rotates with nearly circular velocity, the vertex deviation is small and the Lindblad's relation for the velocity dispersions in $R$ and $z$ directions is valid.  The relation between the velocity dispersion in $z$ direction and the thickness of the system also remains valid.

From the equations of micro-motion results the following expression for the secular variation of velocity dispersions [q.v. Appendix E].
\be
\frac{\dot{\sigma}_R}{\sigma_R} + \frac{\dot{\sigma}_{\theta}}{\sigma_{\theta}} = - \left( \frac{\partial v_R}{\partial R} + \frac{v_R}{R} \right) + \left( \frac{\dot{\sigma}_R}{\sigma_R} + \frac{\dot{\sigma}_{\theta}}{\sigma_{\theta}} \right)_0 , \label{eq21.12} 
\ee
\be
\frac{\dot{\sigma}_z}{\sigma_z} = - \frac{\partial v_z}{\partial z} + \left( \frac{\dot{\sigma}_z}{\sigma_z} \right)_0 , \label{eq21.13} 
\ee
where subindex zero still means the variation caused by the irregular forces.

The conservation of angular momentum enables to derive from Eq.~(\ref{eq21.8})\footnote{See Chapter 19, Appendix A. [Later footnote.]}
\be
2 \frac{v_R}{R} = - \frac{\dot{\omega} }{\omega} , ~~~~ \frac{\partial v_R }{\partial R} + \frac{v_R}{R} = \frac{\dot{k}}{k} - \frac{\dot{\omega}_R }{\omega_R} . \label{eq21.14} 
\ee
Using also Eq.~(\ref{eq21.9}) we find
\be 
\frac{\partial v_z}{\partial z} = \frac{\dot{\zeta} }{\zeta} = \frac{\dot{\sigma}_z }{\sigma_z} - \frac{\dot{\omega}_z}{\omega_z} , \label{eq21.15} 
\ee
\be
\frac{\dot{\sigma}_R}{\sigma_R} - \frac{\dot{\sigma}_{\theta} }{\sigma_{\theta}} = \frac{\dot{k}}{k} = \frac{1}{2} \left( \frac{\partial v_R }{\partial R} - \frac{v_R}{R} \right) . \label{eq21.16} 
\ee
Equation (\ref{eq21.16}) together with Eq.~(\ref{eq21.12}) gives the individual variations of $\sigma_R$ and $\sigma_{\theta}$.

On the basis of Eq.~(\ref{eq21.16}) and the equations of micro-motion we derive the following expression for the secular part of vertex deviation
\be
\gamma = \frac{ \frac{3}{4} \left( \frac{\partial v_R}{\partial R} - \frac{v_R}{R} \right) - \frac{1}{2} \left( \frac{\dot{\sigma}_R}{\sigma_R} - \frac{\dot{\sigma}_\theta}{\sigma_\theta} \right)_0 }{ \frac{\partial v_\theta }{\partial R} - \frac{v_\theta}{R} } .
\label{eq21.17}
\ee 
The deviation is surely not large and has no relation with the observed one in the Galaxy.

Part of the vertex deviation, caused by the irregular forces, may be handled as resulting from the viscosity of the stellar medium. The following formula is approximately valid\footnote{Equation (\ref{eq21.18}) results from expression for the variation of dispersions, derived in Chapters 19 and 22 (if we take $p=1$). In order to have the same meaning for $\tau$ as it was in Chapters 19, 20, 22, $s^2 = 2(\sigma_*^2 +\sigma^2)$ must be taken. [Later footnote.]}

\be
2 \left( \frac{\dot{\sigma}_R}{\sigma_R} - \frac{\dot{\sigma}_{\theta} }{\sigma_{\theta}} \right)_0 = \frac{s^2}{\tau} \left( \frac{1}{\sigma_R^2} - \frac{1}{\sigma_{\theta}^2} \right) , \label{eq21.18} 
\ee
where $s$ is the square mean velocity of the disturbing bodies with respect to the stars, and $\tau$ is the time characterising the rate of the action of irregular forces. Using this formula we find the expression for the kinematical  viscosity coefficient
\be
\nu = \frac{1}{4} ~ \frac{s^2}{\omega_R^2} ~ \frac{1}{\tau} . \label{eq21.19} 
\ee
In gas dynamics the viscosity coefficient increases with $\tau$, but in dynamics of stellar systems it is just the opposite [q.v. Appendix F].

The oscillations and secular variations in flattened subsystems, discussed above, are closely related to the properties of nearly circular motion of stars. Starting with the theory of nearly circular orbits, \citet{Lindblad:1954} analysed the oscillations with frequencies $\omega_R$ already ten years ago. Later we found the expression for the damping time of different oscillations and studied the secular variations. We also derived the theoretical formula for the ratio of velocity dispersions in $R$ and $z$ directions \citep{Kuzmin:1961aa, Kuzmin:1963ae}. It can be seen from present paper that the hydrodynamic equations can be successfully used for solving this kind of problems.
\vglue 5mm
{\hfill September 1963}

\vglue 5mm

{\bf\Large Appendices added in 1969}
\vglue 5mm

\section[A.  Derivation of hydrodynamic equations of the dynamics of
stellar system]{A.  On the derivation of hydrodynamic equations of
  the dynamics of stellar systems}  

Usually the derivation of the hydrodynamic equations is started from the Liouville-Boltzmann equations
\be
\frac{\partial\Psi}{\partial t}+v_l \frac{\partial\Psi}{\partial x_l} + \frac{\partial\Phi}{\partial x_l} \frac{\partial\Psi}{\partial v_l} = \left( \frac{\rmd\Psi}{\rmd t} \right)_0 . \label{eq21.A1.1} 
\ee
Multiplication of the equation by 1, $v_m$, $v_mv_n$, etc. and integration over the velocity space gives the hydrodynamic equations of the 0, 1, 2,  etc. order. 

Yet it is more comfortable to use the transfer equation without the direct use of Liouville-Boltzmann equation. The transfer equation for a quantity $q$ (with the conservation of the ``weight'' of a particle) is
\be
\frac{\partial\rho\bar{q} }{\partial t} + \frac{\partial\rho\overline{qv_l} }{\partial x_l} = \rho\dot{\bar{q}} , \label{eq21.A1.2} 
\ee
where the averaging is done over the volume element. The right side takes into account the variation of $q$ both under the regular and irregular forces. Hydrodynamic equations of the 0, 1, 2, etc. order result from substituting $q=1,$ $v_m$, $v_mv_n$, etc.

Evidently the transfer equation can be reduced to the form
\be
\frac{\rmd\rho\bar{q} }{\rmd t} + \frac{\partial\rho q(v_l-u_l) }{\partial x_l} = \rho\dot{\bar{q}} , \label{eq21.A1.3} 
\ee
where
$$ \frac{\rmd}{\rmd t} = \frac{\partial}{\partial t}+u_l \frac{\partial}{\partial x_l} , $$
and $u_l$ is arbitrarily given vector field. For $q$ we may take $v_m-u_m$, $(v_m-u_m)(v_n-u_n)$, ... (instead of $v_m$, $v_mv_n$, ...) and $u_l$ can be set identical to $\bar v_l$ (in the paper $\bar v_l$ was used simply as $v_l$). This simplifies substantially the derivation of the necessary equations.

\section[B.  Hydrodynamic equations for  stationary axisymmetric
stellar system]{B.  Hydrodynamic equations of the order of $-1$ for
  the stationary axisymmetric stellar system} 

Equations of the order of $-1$ for the stationary and axially symmetrical system can be derived from the Liouville's equations in cylindrical coordinates $R$, $\theta$, $z$. By multiplying the equation by $v_R$ or by $v_z$ and integrating over the velocity space we have the ordinary Jeans equations (generalised, if taking into account the obliquity of the velocity ellipsoid for $z\ne 0$). But we can multiply the Liouville's equation by $v^{-1}_R$ or $v^{-1}_z$ and integrate. In particular, instead of the second Jeans equation
\be
\frac{\partial\rho\sigma^2_z}{\partial z} = \rho \frac{\partial\Phi}{\partial z} \label{eq21.A2.1} 
\ee
($\sigma^2_z$ is the dispersion of $v_z$) we have
\be
\sigma^2_z \frac{\partial\rho}{\partial z} = \kappa\rho \frac{\partial\Phi}{\partial z} , \label{eq21.A2.2} 
\ee
where
\be
\kappa = \sigma^2_z \int \frac{\partial\Psi}{v_z\partial v_z} \rmd V  \label{eq21.A2.3} 
\ee
(integration is over the velocity space).

The equation of the order of $-1$ has the following advantage compared to the Jeans equations. It does not contain the gradient of the velocity dispersion, which is difficult to obtain from observations.

If the distribution of $v_z$ is Gaussian, then $\kappa =1$ and $\sigma^2_z = \mathrm{const}.$ The subtract $\kappa -1$ measures the deviation of the $v_z$ distribution from the Gaussian distribution, and is similar to the excess of the distribution.

For the $z$-gradient of $\sigma^2_z$ we find
\be
\frac{\partial\sigma^2_z}{\partial z} = -(\kappa -1) \frac{\partial\Phi}{\partial z} . \label{eq21.A2.4} 
\ee
For $z=0$ we have
\be
\ba{ll}
\sigma^2_z \frac{\partial^2\ln\rho}{\partial z^2}\bigg|_{z=0} &=  \kappa \frac{\partial^2\Phi}{\partial z^2}\bigg|_{z=0} , \\
\noalign{\medskip}
\frac{\partial^2\sigma^2_z}{\partial z^2}\bigg|_{z=0} &=  -(\kappa -1) \frac{\partial^2\Phi}{\partial z^2}\bigg|_{z=0} . 
\ea
\label{eq21.A2.5}
\ee

Equations of the order $-1$ were proposed by us more than 20 years ago
(in a report in Sternberg Astron. Inst., Moscow in
1950). Unfortunately they were not published. Later they were proposed
by  \citet{King:1965uy}. 

\section[~~C.  Hydrodynamic equations for spherical stellar systems]{~C.  Hydrodynamic equations for spherical stellar systems}

For a stationary spherical stellar system, only the following equation
remains nontrivial amongst all the hydrodynamic equations up to the
second order 
\be
\frac{\partial\rho\sigma^2_r}{\partial r} + 2\rho \frac{\sigma^2_r -\sigma^2_t }{r} = \rho \frac{\partial\Phi}{\partial r} \label{eq21.A3.1} 
\ee
($r$ is the distance from the centre, $\sigma_r$ and $\sigma_t$ are
the radial and the tangential components of the velocity
dispersions). This equation (we used it in Chapter 15) was proposed by
\citet{Courtez:1950} and later by \citet{Ogorodnikov:1957}.  

For a quasi-stationary spherical stellar system evolving under the
irregular gravitational forces the following equation can be added to
the equations above  
\be
\frac{\rmd\rho}{\rmd t} +\rho\left( \frac{\partial\bar{v}_r }{\partial r} + 2 \frac{\bar{v}_r }{r} \right) = \left( \frac{\rmd\rho}{\rmd t} \right)_d ,
\label{eq21.A3.2} 
\ee
where the right side term accounts for the dissipation of stars from the system; and also the equations
\be
\ba{ll}
\frac{\rmd\sigma^2_r}{\rmd t} + 2\sigma^2_r \frac{\partial\bar{v}_r}{\partial r} + \frac{1}{\rho} \frac{\partial\rho w_{rrr} }{\partial r} + \frac{2w_{rrr}-4w_{rtt} }{ r} &=  \left( \frac{\rmd\sigma^2_r}{\rmd t}\right)_0 \\
\noalign{\medskip}
\frac{\rmd\sigma^2_t}{\rmd t} + 2\sigma^2_t \frac{\bar{v}_r}{r} + \frac{1}{\rho} \frac{\partial\rho w_{rtt} }{\partial r} + \frac{4w_{rtt} }{r} &=  \left( \frac{\rmd\sigma^2_t}{\rmd t} \right)_0 .
\ea
 \label{eq21.A3.3} 
\ee
In these equations the terms containing $w_{rrr}$ and $w_{rtt}$
describe heat conductivity in the usual sense (transfer of the
velocity dispersion tensor). As we can not neglect these terms in
present case, use of these equations causes serious difficulties. An
attempt to overcome this was made by \citet{Agekyan:1963}. 

\section[~~D.  The adiabatic equation and the adiabatic theorem]{~D.  The adiabatic equation and the adiabatic theorem}

If we neglect the third moments of peculiar velocities (heat
conductivity) and the effect of irregular forces then
Eq.~(\ref{eq21.3}) is 
\be
\frac{\rmd w_{mn}}{\rmd t} + w_{lm} \frac{\partial\bar{v}_n}{\partial x_l} + w_{ln} \frac{\partial\bar{v}_m}{\partial x_l} = 0 \label{eq21.A4.1} 
\ee
This equation can be called ``adiabatic''. It is present in the paper by \citet{Courtez:1950}.

Directing the coordinate axis along the axis of the velocity ellipsoid we find
\be
\frac{1}{\sigma^2_i} \frac{\rmd\sigma^2_i}{\rmd t} + 2 \frac{\partial\bar{v}_i }{\partial x_i} = 0 , \label{eq21.A4.2}
\ee
giving us
\be
\frac{\rmd}{\rmd t} \frac{\sigma_1\sigma_2\sigma_3}{\rho} = 0. \label{eq21.A4.3} 
\ee
This is just the Lindblad's adiabatic theorem \citep{Lindblad:1950}.

\section[~~E.  Kinematic oscillations in highly  flattened subsystems]{~E.
  Kinematic oscillations in highly flattened subsystems. Stationarity  in rotating system of coordinates}

In the cylindrical system of coordinates within the assumption of
axial symmetry and stationarity we have the following hydrodynamic
equations of macro-motion 
\be
\ba{ll}
\dot{v}_R - \omega_R kv'_{\theta} - \omega^2_z z w_{Rz}/w_{zz} &= 0,\\
\noalign{\smallskip}
k\dot{v}'_{\theta} + \omega_R v_R - k\omega^2_z z w_{\theta z}/ w_{zz} &=  0, \\ 
\noalign{\smallskip}
\dot{v}_z &=  0 
\ea
\label{eq21.A5.1}
\ee
and micro-motion (adiabatic)
\be
\ba{ll}
\dot{w}_{RR} -2\omega_R k w_{R\theta} + 2 \frac{\partial v_R}{\partial z} w_{Rz} &=  0, \\
\noalign{\smallskip}
k\dot{w}_{R\theta} + \omega_R (w_{RR}-k^2w_{\theta\theta}) + k \frac{\partial v'_{\theta}}{\partial z} w_{Rz} + k \frac{\partial v_R}{\partial z} w_{\theta z} &=  0, \\
\noalign{\smallskip}
k^2\dot{w}_{\theta\theta}+2\omega_R kw_{R\theta} +2k^2 \frac{\partial v'_{\theta} }{\partial z}w_{\theta z} &=  0 \\
\noalign{\smallskip}
\dot{w}_{Rz}-\omega_R kw_{\theta z} + \frac{\partial v_R}{\partial z} w_{zz} &=  0, \\
\noalign{\smallskip}
k\dot{w}_{\theta z} + \omega_R w_{Rz} + k \frac{\partial v'_{\theta}
}{\partial z} w_{zz} & =  0, \\
\noalign{\smallskip}
\dot{w}_{zz} &=  0.
\ea
\label{eq21.A5.2}
\ee
The solutions of these equations are Eqs.~(\ref{eq21.10}) and
(\ref{eq21.12}). They describe the kinematical  oscillations of the
subsystem. Similar results can be easily found with help of integrals
of nearly circular motion (Chapters 19 and 20, Appendices). 

If we assume the stationarity in the system of coordinates rotating
with angular velocity $\omega^* = \omega - \frac{1}{2}\omega_R $
(Chapter 20), the time derivative is $\rmd /\rmd t =
\frac{1}{2}\omega_R \partial /\partial \theta$. Rejecting the terms
containing $w_{Rz}$ and $w_{\theta z}$ we have the expressions 
\be
\ba{ll}
\frac{\partial v_R}{\partial\theta} - 2kv'_{\theta} &=  0,\\
\noalign{\smallskip}
k \frac{\partial v'_{\theta}}{\partial\theta} + 2v_R &=  0 
\ea 
\label{eq21.A5.3}
\ee
and
\be
\ba{ll}
\frac{\partial w_{RR}}{\partial\theta} - 4k w_{R\theta} &=  0, \\
\noalign{\smallskip}
k \frac{\partial w_{R\theta}}{\partial\theta} + 2(w_{RR}-k^2w_{\theta\theta}) &=  0, \\
\noalign{\smallskip}
k^2 \frac{\partial w_{\theta\theta}}{\partial\theta} +4kw_{R\theta} &=  0. 
\ea
\label{eq21.A5.4}
\ee

\section[~~F.  Hydrodynamic equations for highly flattened stellar
system]{~F.  Hydrodynamic equations for quasi-stationary highly
  flattened stellar system. Vertex deviation. Viscosity} 

Let us assume a quasi-stationary highly flattened axially symmetric
stellar system (subsystem). The hydrodynamic equations for this kind
of system are 
\be
\ba{ll}
\dot{\rho} +\rho\left( \frac{\partial v_R}{\partial R} + \frac{v_R}{R} + \frac{\partial v_z}{\partial z} \right) &=  0,  \\
\noalign{\smallskip}
w_{RR} - k^2w_{\theta\theta} &=  0,\\
\noalign{\smallskip}
\dot{w}_{RR} - 2\omega_R kw_{R\theta} + 2w_{RR} \frac{\partial v_R}{\partial R} &=   (\dot{w}_{RR})_0 , \\
\noalign{\smallskip}
k^2\dot{w}_{\theta\theta} + 2\omega_R kw_{R\theta} + 2k^2w_{\theta\theta} \frac{v_R}{R} &=  k^2(\dot w_{\theta\theta})_0 , \\
\noalign{\smallskip}
\dot{w}_{zz}  + 2w_{zz} \frac{\partial v_z}{\partial z} &=  (\dot w_{zz})_0. 
\ea
\label{eq21.A6.1}
\ee

From the second\footnote{taking into account the relation between $k$
  and gradient of $v_R$}, third and fourth equations we find 
\be
4\omega_R kw_{R\theta} = w_{RR} \left( \frac{\partial v_R}{\partial R} - \frac{v_R}{R} \right) - (\dot{w}_{RR} - k^2 \dot{w}_{\theta\theta})_0 .
\label{eq21.A6.2} 
\ee
Now by using also the equation $w_{R\theta} = -\gamma
(w_{RR}-w_{\theta\theta})$ we derive for the vertex deviation the
Eq.~(\ref{eq21.17}) (after eliminating $\omega_R$ and $k$). 

Coefficient of viscosity can be derived from the relation
\be
w_{R\theta} = -\nu \left( \frac{\partial v_{\theta}}{\partial R} - \frac{v_{\theta}}{R} \right) = \nu R \frac{\partial\omega}{\partial R}
\label{eq21.A6.3} 
\ee
(for $w_{R\theta}$ we take a part caused by irregular forces).

%% file: chapter22.tex
\chapter[On the action of irregular forces]{On the action of irregular
  forces.\footnote{\footnotetext ~~Tartu Astron Observatory
    Publications, 34, 26 -- 37, 1963.} }

\section {Basic equations for a quasi-stationary stellar system}

In a stationary stellar system the phase density is a function of
conservative isolating integrals of motion $I_i$. But due to irregular
forces the real systems cannot be precisely stationary. Their
gravitational potential varies slowly,  and as a result $I_i$ are not
precise integrals of motion any more. The phase density $\psi$ is,  in
addition to the integrals $I_i$,  an explicit function of coordinates
and time. 

The Boltzmann equation is in this case 
\be
\frac{\partial\psi}{\partial t} + \sum_i \dot{I}_i \frac{\partial\psi}{\partial I_i} + v \frac{\partial\psi}{\partial s} = \frac{\delta\psi}{\delta t} , \label{eq22.1} 
\ee
where $t$ is time, $s$ and $v$ -- the path length and velocity on the
osculating regular orbit of a star (\ie  orbit corresponding to the
stationary potential), and $\delta\psi /\delta t$ -- direct variation
of $\psi$ per unit of time under the action of irregular forces. 

As the value of $I_i$ remains constant along the osculating orbit, 
\be
\dot{I}_i = \frac{\partial I_i}{\partial t} . \label{eq22.2} 
\ee

Let us divide $\psi$ into two parts
\be
\psi = \Psi + \Delta\Psi , \label{eq22.3} 
\ee
in a way that
\be
\frac{\partial\Psi}{\partial s} = 0, \label{eq22.4} 
\ee
\be
\overline{\Delta\Psi} = 0 , \label{eq22.5} 
\ee
where a bar denotes the ``orbital mean'', \ie  the mean value with the
weight $\rmd s/v = \rmd t$ along the osculating orbit (for
$s\rightarrow\infty$).  

Let us assume that the action of irregular forces is very slow,  and the
system is quasi-stationary. In this case $\Delta\Psi$ is very small, 
when compared with $\Psi$. Rejecting in Eq.~(\ref{eq22.1}) all very
small terms when compared to others we have
\be
\frac{\partial\Psi}{\partial t} + \sum_i \dot{I}_i \frac{\partial\Psi}{\partial I_i} + v \frac{\partial\Delta\Psi}{\partial s} = \frac{\delta\Psi}{\delta t} . \label{eq22.6} 
\ee

Forming the orbital mean in Eq.~(\ref{eq22.6}) we have
\be
\frac{\partial\Psi}{\partial t} + \sum_i \overline{\dot{I}_i} \frac{\partial\Psi}{\partial I_i} = \frac{\overline{\delta\Psi}}{\delta t}, \label{eq22.7}
\ee
and further with help of Eq.~(\ref{eq22.6}) we derive
\be
v \frac{\partial\Delta\Psi}{\partial s} + \sum_i (\dot{I}_i - \overline{\dot{I}_i} ) \frac{\partial\Psi}{\partial I_i} = \frac{\delta\Psi}{\delta t} - \frac{\overline{\delta\Psi}}{\delta t} . \label{eq22.8} 
\ee

Equations (\ref{eq22.7}) and (\ref{eq22.8}) are just the basic
equations for present problem [q.v. Appendix A]. We may add them to
the equations,  resulting from the Poisson's equation. The equations
were proposed by us in 1957 \citep{Kuzmin:1957aa, Kuzmin:1963ad}.  

\section{Very flat subsystem of the Galaxy}

We apply the above equations to a very flat subsystem of the Galaxy.
\vglue 3mm
$\mathrm{1.^o}$ ~If we suppose the stationary axisymmetric and highly
flattened potential, we can use the integrals of nearly circular
motion 
\be
\ba{ll}
I_1 = E_R = & \frac{1}{2} (v_R^2 + \omega_R^2\Delta R^2), \\
\noalign{\medskip}
I_2 = E_z = & \frac{1}{2} (v_z^2 + \omega_z^2 z^2) , 
\ea 
\label{eq22.9} 
\ee
where $v_R$ and $v_z$ are the velocity components along the
cylindrical coordinates $R$ and $z$, $\omega_R$, and $\omega_z$ -- the
angular oscillation frequencies of a star along $R$ and $z$ about the
circular orbit, and $\Delta R$ -- the distance along $R$ counted from
the circular orbit. 

The frequency $\omega_R$ is related to the angular velocity on the
circular orbit $\omega$ via the relation 
\be
\omega_R^2 = 4\omega^2 + R \frac{\partial\omega^2}{\partial R} . \label{eq22.10}
\ee

If $v_{\theta}$ is the velocity along $\theta$ coordinate counted from the circular velocity, then
\be
\Delta R = -k \frac{v_{\theta}}{\omega_R} , \label{eq22.11} 
\ee
where
\be
k = \frac{2\omega}{\omega_R} . \label{eq22.12} 
\ee

Apart from the integrals (\ref{eq22.9}), we have one more integral
\be
I_3 = R_c = R - \Delta R , \label{eq22.13} 
\ee
where $R_c$ is the radius of the circular orbit. For very flattened
subsystems the phase density only weakly depends on $R_c$, when
compared with the dependence on $E_R$ and $E_z$. 
\vglue 3mm

$\mathrm{2.^o}$ ~As $\omega$, $\omega_R$, $\omega_z$ vary continuously
in the neighbourhood of a given circular orbit, the values $R_c$,
$E_R$ and $E_z$ also vary. For the variation of $E_R$ and $E_z$ we
have 
\be
\dot{E}_R = \omega_R \dot{\omega}_R \Delta R^2 - \omega_R^2 \dot{R}_c \Delta R, ~~~~~ \dot{E}_z = \omega_z \dot{\omega}_z z^2 . \label{eq22.14} 
\ee
Now, taking into account that the oscillations about the circular
orbit are harmonic, we find the orbital averages  
\be
\overline{\dot{E}_R} = \frac{\dot{\omega}_R}{\omega_R} E_R , ~~~~~ \overline{\dot{E}_z} = \frac{\dot{\omega}_z}{\omega_z} E_z . \label{eq22.15} 
\ee
Further, from the conservation of angular momentum we have
\be
\dot{R}_c = - \frac{1}{2} \frac{\dot{\omega}}{\omega} R_c . \label{eq22.16} 
\ee
The similar expression is valid also for $\overline{\dot{R}_c}$,
because $\dot{R}_c$ remains constant along the osculating orbit. 
\vglue 3mm

$\mathrm{3.^o}$ ~The effect of irregular forces can be handled as the
diffusion of stars in velocity space with the superposed systematic
shift, so called ``dynamical friction'' \citep{Kuzmin:1957aa}. 

For simplification let us approximate this process by an anisotropic
diffusion, independent of stellar velocity, and oriented along $R$,
$\theta$, $z$. In this case [q.v. Appendix B] 
\be
\frac{\delta\Psi}{\delta t} = A_R \frac{\partial^2\Psi}{\partial v_R^2} + A_{\theta} \frac{\partial^2\Psi}{\partial v_{\theta}^2} + A_z \frac{\partial^2\Psi}{\partial z^2} , \label{eq22.17} 
\ee
where the coefficients $A_R$, $A_{\theta}$, $A_z$ are independent of
velocity. These coefficients are related to the variations of the
velocity dispersions $\sigma_R$, $\sigma_{\theta}$, $\sigma_z$, caused
by the irregular forces according to relations 
\be
A_R = \frac{1}{2} \frac{\delta\sigma_R^2}{\delta t} , ~~~~ A_{\theta} = \frac{1}{2} \frac{\delta\sigma_{\theta}^2}{\delta t} , ~~~~ A_z = \frac{1}{2} \frac{\delta\sigma_z^2}{\delta t} . \label{eq22.18} 
\ee

In Eq.~(\ref{eq22.17}) $\Psi$ is assumed to be a function of $v_R$,
$v_{\theta}$, $v_z$. Now handling $\Psi$ to be a function of $E_R$ and
$E_z$ ($R_c$ can be neglected as an argument) we find 
\be
\frac{\delta\Psi}{\delta t} = A_R \left( \frac{\partial\Psi}{\partial E_R} + \frac{\partial^2\Psi}{\partial E_R^2} v_R^2 \right) + k^2 A_{\theta} \left( \frac{\partial\Psi}{\partial E_R} + \frac{\partial^2\Psi}{\partial E_R^2} k^2v_{\theta}^2 \right) + A_z \left( \frac{\partial\Psi}{\partial E_z} + \frac{\partial^2\Psi}{\partial E_z^2} v_z^2 \right) . \label{eq22.19} 
\ee
For the orbital average we have
\be
\frac{\overline{\delta\Psi}}{\delta t} = (A_R + k^2 A_{\theta} ) \left( \frac{\partial\Psi}{\partial E_R} + \frac{\partial^2\Psi}{\partial E_R^2} E_R \right) + A_z \left( \frac{\partial\Psi}{\partial E_z} + \frac{\partial^2\Psi}{\partial E_z^2} E_z \right) , \label{eq22.20} 
\ee
while we neglected with the dependence of $A_R$ , $A_{\theta}$, $A_z$ on $\Delta R$ and $z$.

\section[The centroid velocity and vertex deviation]{Formation of the
  Schwarzschild velocity distribution. The centroid velocity and
  vertex deviation} 

$\mathrm{1.^o}$ ~Derived Eqs.~(\ref{eq22.15}) and (\ref{eq22.20})
detail Eq.~(\ref{eq22.7}). The solution of the equation in case of
point-like initial velocity distribution  is\footnote{We have in mind
  the velocity distribution concentrated into a point, \ie  in form of
  three-dimensional $\delta$-function. [Later footnote.]}  
\be
\Psi = f (R^2\omega ) \frac{\omega_R\omega_z}{\sigma_R^2\sigma_z^2} e^{-\frac{E_R}{\sigma_R^2} - \frac{E_z}{\sigma_z^2}} , \label{eq22.21} 
\ee
where [q.v. Appendix B]
\be
\sigma_R^2 = \frac{1}{2}\omega_R \int_{t_0}^t \frac{A_R + k^2A_{\theta}}{\omega_R} \rmd t , ~~~~ \sigma_z^2 = \frac{1}{2} \omega_z \int_{t_0}^t \frac{A_z}{ \omega_z} \rmd t , \label{eq22.22} 
\ee
and $f$ is an arbitrary function with $R$ as an argument instead of $R_c$.

Expression (\ref{eq22.21}) corresponds to the Schwarzschild velocity
distribution with the dispersions $\sigma_R$, $\sigma_{\theta} =
\sigma_R/k$, $\sigma_z$ along $R$, $\theta$, $z$. 

We may identify somewhat arbitrarily the moment of time $t_0$ with the
time of subsystem formation. If we have subsystems with different
ages, then velocity distribution is the sum of Schwarzschild
distributions with different dispersions. 
\vglue 3mm

$\mathrm{2.^o}$ ~The results of previous Section detail also
Eq.~(\ref{eq22.8}). Its solution results directly via integration, 
giving us 
$$
\Delta\Psi = \frac{1}{2} \left[ \left( \frac{\dot\omega_R}{\omega_R} \Delta R - 2\dot{R}_c \right) v_R \frac{\partial\Psi}{\partial E_R} + \frac{\dot{\omega}_z }{\omega_z} z v_z \frac{\partial\Psi}{\partial E_z} + \right. $$
\be
\left. + (A_R - k^2 A_{\theta} ) \Delta R v_R \frac{\partial^2\Psi}{\partial E_R^2} + A_z z v_z \frac{\partial^2\Psi}{\partial E_z^2} \right] . \label{eq22.23} 
\ee

Here we may replace $\Delta R$ according to Eq.~(\ref{eq22.11}),  and $R_c$ according to the relation
\be
2\dot{R}_c = - \frac{\dot{\omega}}{\omega} R + \left( \frac{\dot{\omega}_R}{\omega_R} - 3 \frac{\dot{k}}{k} \right) \Delta R ,
\label{eq22.24} 
\ee
(it can be derived from the conservation of the area integral and Eq.~(\ref{eq22.10})). 

If the velocity distribution as a function of $E_R$ and $E_z$ will
have a small shift $\overline{v}_R$ and $\overline{v}_z$ along $R$ and
$z$, and will be slightly inclined by a small angle $\gamma$ about
galactic longitude, then $\Psi$ varies by the quantity 
\be
\Delta\Psi = \left[ - \overline{v}_R v_R + (k^2-1)\gamma v_R v_{\theta} \right]  \frac{\partial\Psi}{\partial E_R} - \overline{v}_z v_z \frac{\partial\Psi }{\partial E_z} . \label{eq22.25} 
\ee
Similar expression results from Eq.~(\ref{eq22.23}), if we express the
second derivatives of $\Psi$ via first, within the assumption of
Schwarzschild velocity distribution. Comparing both expressions we
find 
\be
\overline{v}_R = - \frac{1}{2} \frac{\dot{\omega}}{\omega} R , ~~~~~ \overline{v}_z = - \frac{1}{2} \left( \frac{\dot{\omega}_z}{\omega_z} - \frac{A_z}{\sigma_z^2} \right) z, \label{eq22.26} 
\ee
\be
\gamma = \frac{1}{2} \frac{k}{k^2-1} \frac{1}{\omega_R} \left( -3 \frac{\dot{k}}{k} + \frac{A_R - k^2A_{\theta} }{\sigma_R^2} \right) . \label{eq22.27} 
\ee
These quantities can be interpreted as $R$ and $z$ components of the
centroid velocity and the vertex deviation of the velocity ellipsoid. 

The derived expressions for the velocity dispersion variation, for the
velocity centroid and for the vertex deviation may be derived also in a
more simple way \citep{Kuzmin:1961aa, Kuzmin:1963ae}. 

\section{Variation of the velocity dispersions due to the irregular forces}

In order to detail further the results derived above, one needs to
have the expressions for the variation of velocity dispersions, caused
by the direct action of the irregular forces. In this case by using
Eq.~(\ref{eq22.18}) we may express the coefficients $A_R$,
$A_{\theta}$, $A_z$. 
\vglue 3mm

$\mathrm{1.^o}$ ~On the basis of the theory, presented in our earlier
paper \citep{Kuzmin:1957aa},  the variation of the velocity dispersion
of stars along $j$ coordinate ($R$, $\theta$ or $z$) is 
\be
\frac{\delta\sigma_j^2}{\delta t} = 2\pi G^2 m^2 n \overline{ [w^2-w_j^2+ 2(1+\mu )w_jv_j]L(x)w^{-3} } . \label{eq22.28} 
\ee
Here $G$ is the gravitational constant, $m$ is the mass of disturbing
bodies, $n$ is their spatial number density,  $w$ is the velocity of
perturbing body with respect to a star, $w_j$ -- its $j$ component, and
$\mu$ the stellar mass in units of $m$. Further 
$$x^2 = \frac{wr}{Gm(1+\mu )} , $$
where $r$ means the effective action radius of irregular forces. For
small $x$ the function $L$ increases as $x^4$, thereafter more slowly,
and finally for large $x$ it increases logarithmically. Averaging in
Eq.~(\ref{eq22.28}) is performed over all stars of  a given subsystem
and over all perturbing bodies in the ``element'' of space.  

If we suppose $L$ to be a constant $L_0$, and assume the ellipsoidal
velocity distribution of stars with respect to disturbed bodies, then
the calculation of $\delta\sigma_j^2/\delta t$ is similar to the
calculation of the gravitational virial of the ellipsoidal mass
distribution [discussed in Chapter 12, see also Appendix C]. In case
of Schwarzschild distribution being not too different from the
spherical distribution the result is 
\be
\frac{\delta\sigma_j^2}{\delta t} = \frac{1}{\tau} \{ [ 2\sigma^2+ q(\sigma^2 -\sigma_j^2)]_* - 2\mu\sigma^2 + q (3+2\mu ) (\sigma^2 - \sigma_j^2) \} , \label{eq22.29} 
\ee
where $q = 2/5$, $\sigma^2$ is mean of $\sigma_j^2$,
\be
\tau^{-1} = \frac{2}{3} \frac{\sqrt{2\pi} ~G^2 m^2 n L_0 }{ (\sigma_*^2 + \sigma^2 )^{3/2} } , \label{eq22.30} 
\ee
and a star denotes that given quantity belongs to disturbing bodies.

The case when $L$ is proportional to $x^3$ is even more simple. We
derive again Eq.~(\ref{eq22.29}),  but independently of the velocity
distribution and with $q=1$. If we designate 

$$L/w^3 = \frac{1}{3} \sqrt{\frac{2}{\pi}} (\sigma_*^2+\sigma^2 )^{-3/2} L_0,$$ 
then for $\tau$ we derive Eq.~(\ref{eq22.30}).

It seems that Eq.~(\ref{eq22.29}) may be used within a certain
precision also in more general case by choosing in a suitable way the
effective value of $q$. 
\vglue 3mm

$\mathrm{2.^o}$ ~In the Galaxy irregular forces can be significant
only when caused by quite large masses -- the clouds of stars and of
diffuse matter.  

Thus we may take $\mu =0$.

The values of the first term in Eq.~(\ref{eq22.29}) along three
coordinates are not too different, and there exist arguments in
support of neglecting  this difference \citep{Kuzmin:1961aa}. Further,
in the Galaxy the interval of $x$, where $L$ increases with $x$ quite
rapidly,  is the most important \citep{Kuzmin:1961aa}. Still in this
region the increase of $L$ is not as strong as $x^3$,  and thus $q$ must
taken to be less than one. 

For the Galaxy we may accept the formula
\be
\frac{\delta\sigma_j^2}{\delta t} = \frac{2}{\tau} [ \sigma_*^2 + p
(\sigma^2 -\sigma_j^2) ] , \label{eq22.31}  
\ee
where $p = \frac{3}{2} q\simeq 1$.

The quantity $L_0$ in Eq.~(\ref{eq22.30}) is in case of our Galaxy of the order of unity. 

By using Eq.~(\ref{eq22.31}) and the formula derived for the variation
of the velocity dispersions, we may derive within certain assumptions
the theoretical ratio $\sigma_R/\sigma_z$. Theoretical result agrees
well with observations \citep{Kuzmin:1961aa, Kuzmin:1963ae}.  
\vglue 3mm
{\hfill March 1963}

\vglue 5mm

{\bf\Large Appendices added in 1969}
\vglue 5mm

\section{A.  On the basic equations of the action of irregular forces} 

Replacing the integrals $I_i$ as arguments of the phase density by
adiabatic invariants $P_i$,  and taking into account that 
\be
\overline{\dot{P}_i} = 0 \label{eq22.A1.1}
\ee
Eqs.~(\ref{eq22.7}) and (\ref{eq22.8}) have the  form
\be
\frac{\partial\Psi}{\partial t} = \frac{\overline{\delta\Psi}}{\delta t}, ~~~~ v \frac{\partial\Delta\Psi}{\partial s} + \sum \dot{P}_i  \frac{\partial\Psi}{\partial P_i} = \frac{\delta\Psi}{\delta t} -
\frac{\overline{\delta\Psi}}{\delta t} . \label{eq22.A1.2} 
\ee

In case of slightly variable gravitational field the integrals of motion are
\be
K_i = P_i - \int\dot{P}_i \rmd s/v . \label{eq22.A1.3} 
\ee
Taking them as arguments of the phase density we have
\be
\frac{\partial\Psi}{\partial t} = \frac{\overline{\delta\Psi}}{\delta t} , ~~~~ v \frac{\partial\Delta\Psi}{\partial s} = \frac{\delta\Psi}{\delta t} - \frac{\overline{\delta\Psi}}{\delta t}. \label{eq22.A1.4} 
\ee

The adiabatic invariants and integrals of nearly circular motion were
analysed in Chapter 19. Using them would significantly simplify the
derivations in Section 2 of the paper. 

\section{B.  Detailisation of the effect of irregular forces}

Strictly speaking, the Eq.~(\ref{eq22.7}) can be used only when the
velocity dispersion of perturbing bodies is large compared to the
velocity dispersion of stars, and hence we can neglect dynamical
friction. More precise (although quite rough) is the Chandrasekhar's
formula (Chapter 17) 
\be
\frac{\delta\Psi}{\delta t} = \beta\nabla_v (v\Psi ) + A\nabla^2_v \Psi . \label{eq22.A2.1} 
\ee
Corresponding variation of the dispersion is
\be
\frac{\delta\sigma^2_j}{\delta t} = 2(A - \beta\sigma^2_j ). \label{eq22.A2.2} 
\ee
This equation may be related to Eq.~(\ref{eq22.31}) by supposing
\be
A = \frac{1}{\tau} (\sigma^2_* + p \sigma^2 ), ~~ \beta = \frac{p}{\tau}.
\label{eq22.A2.3} 
\ee

For $\Psi$ we have the same solution in form of Schwarzschild velocity
distribution (\ref{eq22.21}). $\sigma^2_R$ and $\sigma^2_z$ satisfy
the differential equation derived and solved with respect to
$\sigma^2_R/\sigma^2_z$ in Chapter 19. For $\bar v_R$, $\bar v_z$ and
the vertex deviation we have the solutions given in Chapter 19. 

\section[C.  Variation of the velocity dispersions  from the action of
irregular forces]{C.  On the variation of the velocity dispersions
  resulting from the direct action of irregular forces} 

Let us assume that $L=L_0 = \mathrm{const}$ in Eq.~(\ref{eq22.28}). In
this case the calculation of $\delta\sigma^2_j /\delta t$ is similar
to the calculation of the virial of a stellar subsystem caused by
another subsystem (Chapter 12).  

Assuming formally that $GMM^*=-1$, where $M$ and $M^*$ are the
``masses of the subsystems'',  and replacing the spatial coordinates by
the velocity components, we have 
\be
\overline{v_jw_jw^{-3}} = -W_{jj}, ~~~~ \overline{v^*_jw^*_jw^{-3}} = -W^*_{jj} , \label{eq22.A3.1} 
\ee
where $v^*_j$ is the $j$-th component of perturbing bodies and $w^*_j
= -w_j$. ``Virials'' $W$ and $W^*$ are mutual virials of subsystems of
stars and perturbing bodies, but in velocity space, instead of
ordinary space. 

Evidently
\be
\overline{w^2_jw^{-1}} = W_{jj}+W^*_{jj} . \label{eq22.A3.2} 
\ee
Assuming that the velocity distributions of stars and perturbing
bodies are concentric, similar to each other and similar to
Schwarzschild distribution, we have on the basis of the results
derived in Chapter 12 after some calculations 
\be
W_{jj} = \frac{1}{2\sqrt{2\pi}} ~ \frac{\alpha_j s^2_j }{ s_1s_2s_3} ~ \frac{\sigma^2}{ (\sigma^2 +\sigma^2_* )^{3/2}} \label{eq22.A3.3} 
\ee
and similar expression for $W^*_{jj}$, where $\sigma$ and $\sigma^*$
are mutually replaced. In these expressions 
\be
s_j = \frac{\sigma_j}{\sigma} = \frac{\sigma^*_j}{\sigma^*} , \label{eq22.A3.4}
\ee
and $\alpha_j$ are coefficients of gravitational attraction of an homogeneous ellipsoid.

Following relations are valid for the coefficients $\alpha_i$
\be
\sum\alpha_i = 4, ~~~~ \sum\alpha_i s^2_i \simeq 4. \label{eq22.A3.5}
\ee
Taking into account the second of them we derive Eq.~(\ref{eq22.28}) in form
\be
\frac{\delta\sigma^2_j}{\delta t} = \frac{3}{4\tau} \frac{1}{ s_1s_2s_3} \left[ \sigma_*^2 ( 4 -\alpha_j s_j^2) + \sigma^2 ( 4 - (3+2\mu )\alpha_j s^2_j \right] . \label{eq22.A3.6} 
\ee
For coefficients $\alpha_j$ we have the series
\be
\alpha_j = \frac{4}{3} - \frac{4}{5} (s^2_j -1) + ...  . \label{eq22.A3.7} 
\ee
Further
\be
s_1s_2 s_3 = 1+ ... . \label{eq22.A3.8} 
\ee
Limiting the series to the terms linear with respect to $s^2_j -1$, we
derive Eq.~(\ref{eq22.29}) with $q=2/5$.

%% file: ch1973.tex
\chapter[~~On the action of irregular forces in the Galaxy]{~~On the
  action of irregular forces in the Galaxy\footnote{\footnote
    ~~Published in ``Dynamics of galaxies and star clusters'',
    Ed. T.B. Omarov, Nauka, Kaz. SSR, Alma-Ata, p.76 -- 82,
    1973. Report on the meeting of the Commission on Stellar Astronomy
    of the Astron. Council of Acad. Sci. USSR, Oct. 23 -- 26, 1972.}}

To describe the action of irregular gravitational forces, we use the
approximation proposed by \citet{Chandrasekhar:1943}, \ie we assume that the
diffusion in velocity space is isotropic and independent on velocities, and
that the dynamical friction is proportional to velocity and directed in 
the opposite way to the velocity vector. Let us have a highly flattened
galactic subsystem, where the 
stars move in nearly circular orbits. The velocities we count from the
circular velocity and the dynamical friction we assume to vanish in case
of motion in a circular orbit. Let $u$, $v$, $w$ be  the velocity components
along the cylindrical coordinates $R$, $\theta$, $z$. When we take into
account the dynamical friction, the equations, describing the changes of
$u$, $v$, $w$, $z$,  have the following form ($t$ is time)
\be
\frac{\rmd u}{\rmd t} - 2\Omega v = -\kappa u, ~~~~~ \frac{\rmd v}{\rmd t} + \frac{\Omega_e^2
}{2\Omega} u = -\kappa v ; \label{eq-a1.1} 
\ee
\be
\frac{\rmd w}{\rmd t} + \Omega_z^2 z = -\kappa w, ~~~~~ \frac{\rmd z}{\rmd t} - w = 0. 
\label{eq-a1.2} 
\ee
Here $\kappa$ is the dynamical friction coefficient, $\Omega$ is the
circular angular velocity, $\Omega_e$ -- the epicyclic velocity,
$\Omega_z$ -- $z$-oscillation frequency (when the dynamical friction is
absent). We have
\be
\Omega_e^2 = 4\Omega^2 + R \dd\Omega^2 / \dd{R}. \label{eq-a1.3}
\ee

We may assume that the local centroid  moves on a circular orbit, \ie we may
suppose that  mean values $\overline{u} = \overline{v} = \overline{w} =
\overline{z} = 0$. We assume also $\overline{uw} = \overline{vw} =
\overline{uz} = \overline{vz} = 0$.  Taking into account the diffusion
in the velocity space (the fluctuating part in irregular
forces),  we find from Eqs.~(\ref{eq-a1.1}) and (\ref{eq-a1.2}) for the remaining components of the
$u$, $v$, $w$, $z$ the dispersion tensor
\be
\ba{ll}
{\rmd\over \rmd t} \overline{u^2} - 4\Omega\overline{uv} &=  2\kappa (\sigma^2 -
\overline{u^2} ) \\
\noalign{\smallskip}
{\rmd\over \rmd t} \overline{uv} + 2\Omega\overline{u^2} - {\Omega_e^2\over
2\Omega} \overline{v^2} &=  - 2\kappa \overline{uv}  \\
\noalign{\smallskip}
{\rmd\over \rmd t} \overline{v^2} + {\Omega_e^2\over\Omega} \overline{uv} &= 
2\kappa (\sigma^2 - \overline{v^2} ) 
\ea
\label{eq-a1.4}
\ee
and
\be
\ba{ll}
{\rmd\over \rmd t} \overline{w^2} + 2\Omega_z^2 \overline{wz} &=  2\kappa
(\sigma^2 -\overline{w^2} ) \\
\noalign{\smallskip}
{\rmd\over \rmd t}\overline{wz} + \Omega_z^2 \overline{z^2} - \overline{w^2} &=  
-\kappa\overline{wz} \\
\noalign{\smallskip}
{\rmd\over \rmd t}\overline{z^2} - 2\overline{wz} &=  0. 
\ea
\label{eq-a1.5} 
\ee
Here $\sigma^2$ is the equilibrium dispersion. The effect of irregular
forces, described by the right sides of Eqs.~(\ref{eq-a1.4}) and
(\ref{eq-a1.5}), vanishes in the case
of spherical velocity distribution with the dispersion $\sigma^2$.

To simplify the problem, we assume that the characteristics of the regular
gravitational field $\Omega$, $\Omega_e$ and $\Omega_z$, and the
characteristics of the irregular gravitational field $\kappa$ and $\sigma$ do
not vary in time (in reality, surely, it is not so). In this case Eqs.~(\ref{eq-a1.4})
and (\ref{eq-a1.5}) can be easily solved as two independent sets of linear
differential equations with constant coefficients. For the initial
conditions, we accept that the $u$, $v$, $w$, $z$ dispersion tensor is zero
when $t=0$ (conditionally, it may be identified with the moment of star
formation).  We can write the solutions in the form
\be
\left.
\ba{ll}
\overline{u^2} & \\
k^2\overline{v^2} & 
\ea
\right\} 
= {1\over 2} (k^2+1) \sigma^2 \left( 1-
e^{-2\tau} \right) ~\mp ~{1\over 2} {\kappa\over\Omega_e} e^{-2\tau} {\rmd\over \rmd t}
\left( e^{2\tau} k \overline{uv} \right) 
\label{eq-a1.6} 
\ee
\be
k\overline{uv} = {1\over 2} (k^2-1) \sigma^2 {\Omega_e\kappa \over
\Omega_e^2+\kappa^2} \left[ 1 - e^{-2\tau} \left( \cos 2\tau_e +
{\kappa\over\Omega_e} \sin 2\tau_e \right) \right] 
\label{eq-a1.7}
\ee
and
\be
\left. 
\ba{ll}
\overline{w^2} & \\
\Omega_z^2 \overline{z^2} & \\
\ea
\right\} 
= \sigma^2 \left( 1- e^{-\tau} \right) - {1\over 2} \left[ \kappa\overline{wz} ~\mp ~e^{-\tau}
{\rmd\over \rmd\tau} \left( e^{\tau} \kappa \overline{wz} \right) \right] ,
\label{eq-a1.8} 
\ee 
\be
\kappa \overline{wz} = \sigma^2 {\kappa^2\over\omega_z^2} e^{-\tau} \sin^2 \tau_z . 
\label{eq-a1.9} 
\ee
Here we designated
\be
k = 2\Omega /\Omega_e \label{eq-a1.10} 
\ee
($k$ is the ratio of semiaxis of the velocity ellipsoid). In addition we
introduced three dimensionless times
\be
\tau = \kappa t , ~~~~~ \tau_e = \Omega_e t , ~~~~~ \tau_z = \omega_z t .
\label{eq-a1.11} 
\ee
And finally, $\omega_z$ is the frequency of $z$-oscillations in case of
dynamical friction
\be
4\omega_z^2 = 4\Omega_z^2 - \kappa^2 . \label{eq-a1.12} 
\ee

For stellar systems, $\kappa$ is small when compared to $\Omega_e$ and
$\Omega_z$ (the irregular forces are small when compared to the regular
forces),  and in Eqs.~(\ref{eq-a1.6}) and (\ref{eq-a1.8}), in first approximation, we may keep only
the first term. In this case, for small $\tau$
\be
\overline{u^2} = k^2\overline{v^2} = (k^2+1)\sigma^2\tau , ~~~~
\overline{w^2} = \sigma^2\tau , \label{eq-a1.13} 
\ee
\ie there is a valid relation between the velocity dispersions,  proposed
by us earlier \citep{Kuzmin:1961aa, Kuzmin:1963ae, Kuzmin:1965ab} ($\overline{w^2}$ is the smallest
of the dispersions). If $\tau$ is large, then
\be
\overline{u^2} = k^2\overline{v^2} = {1\over 2} (k^2 +1)\sigma^2 , ~~~~~
\overline{w^2} = \sigma^2 . \label{eq-a1.14} 
\ee
This relation between the velocity dispersions (the dispersion
$\overline{w^2}$ is intermediate between $\overline{u^2}$ and
$\overline{v^2}$) corresponds to T.~A. Agekjan's assertion in one of our
earlier discussions about the influence of irregular forces on the
form of velocity ellipsoid. As it is seen, both of us were right in
their  own
region. Our formula corresponds to small $\tau$, the Agekyan's
statement to large $\tau$ (Agekyan's region).

For large $\tau$ (but in mean also for small $\tau$)
\be
\overline{uv} = {(k-k^{-1})\kappa\sigma^2 \over 2\Omega_e} . \label{eq-a1.15}
\ee 
This quantity determines the radial transfer of momenta and thus the
viscosity. For the kinematical  viscosity coefficient we find
\be
\nu = {1\over 2} {\kappa\sigma^2\over\Omega_e^2} . \label{eq-a1.16} 
\ee
Similar result was derived by us earlier \citep{Kuzmin:1965ab}.

\begin{figure}[ht]
\centering
\includegraphics[width=60mm]{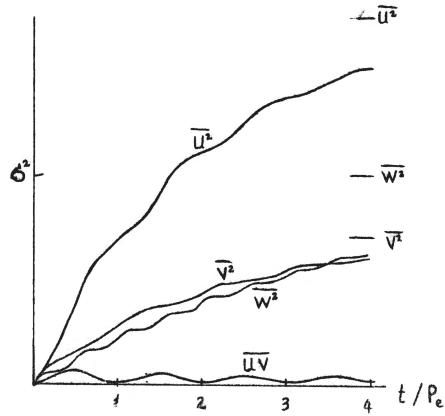}
\caption{Time evolution of the velocity dispersion components.}
\label{ch1973fig1}
\end{figure}

Precise formulae for large $\tau$ are (within the present solution) 
\be
\ba{ll}
\overline{u^2} =  {{1\over 2} (k^2+1)\Omega_e^2 + \kappa^2 \over
\Omega_e^2 + \kappa^2} \sigma^2 ; & 
~~~~~~\overline{v^2} = { {1\over 2} {k^2+1\over k^2}\Omega_e^2 +\kappa^2 \over
\Omega_e^2 + \kappa^2} \sigma^2 ; \\
\overline{uv} =  {1\over 2} (k -k^{-1}) {\Omega_e\kappa\over \Omega_e^2 +
\kappa^2} \sigma^2 ; & 
~~~~~~\overline{w^2} = \sigma^2 . 
\ea
\label{eq-a1.17} 
\ee
From the equation for $\overline{uv}$ we derive
\be
\nu = {1\over 2} {\kappa\sigma^2 \over\Omega_e^2 }. \label{eq-a1.18} 
\ee
If $\kappa$ is large (the irregular forces are significant), then the
velocity distribution becomes naturally nearly spherical. The viscosity
becomes the usual viscosity, proportional to the relaxation time.

The waves, superposed to the smooth curve, increase  
velocity dispersions,  are quite interesting. These waves are seen in
Figure \ref{ch1973fig1}. In calculations, we 
chose $k^2 = 2.5$, corresponding to the solar neighbourhood. The frequency
$\Omega_z$ was taken twice greater than the frequency $\Omega_e$, the
quantity $\kappa$ by the factor $4\pi$ less than $\Omega_e$. The argument in
Figure is $t/P_e$, where $P_e$ is the period of epicyclic motion. The
horizontal bars designate the position of asymptotes.

As a conclusion, we mention  that within the Chandrasekhar approximation, the
velocity distribution, being initially three-dimensional
$\delta$-function, 
becomes Schwarzschild distribution and remains that. In other words, the
Schwarzschild velocity distribution is invariant, if  irregular forces
can be described in Chandrasekhar approximation. But surely the parameters
of the distribution -- the components of the velocity dispersion tensor --
will change.

%% file: ch1986.tex
\chapter[~~Quasi-isothermal models of spherical stellar systems: 
M87 and M105]{~Quasi-isothermal models of
  spherical stellar systems.  
Application to the galaxies M87 and
M105\footnote{\footnotetext ~~Tartu Astron. Obs. Publ. {\bf 51},
232, 1986. Coauthors  \"U.-I. Veltmann and P. Tenjes.}}

It is known that globular clusters are quite well described by
generalised-isochrone models,  comprehensibly discussed by us in our
earlier paper \citep{Kuzmin:1973ab}. But when applied to galaxies
having round form, as for example the galaxies M87 and M105, they fail in
describing the profound central density peak in these galaxies. For
galaxies more useful are the nearly isothermal models. Below we discuss
the generalised-isothermal models. They are similar to the
generalised-isochrone models, and in limiting case approach the
quasi-isothermal models.

First we remind the properties of the generalised-isochrone models. These
models are determined by the gravitational potential
\be
\Phi (r) = \Phi_0 {a\over b +\zeta (r)}, ~~~~~ \zeta (r) = \sqrt{1+
a^2r^2/r_0^2} . \label{eq-a2.1} 
\ee
Here $r$ is the distance from the centre of the model, $\Phi_0$ and $r_0$
are the scale parameters, and $a\ge 1$ and $b\ge 0$ are  structural 
parameters, while
\be
a-b=1 \label{eq-a2.2} 
\ee
indicating that only one structural  parameter is independent. In addition
to $a$ and $b$ we may use also the central concentration parameter 
\be
q=b/a . \label{eq-a2.3} 
\ee
The parameter $q$ may have values from zero to one.

As $\Phi\rightarrow \Phi_0 r_0/r = GM/r$, when $r\rightarrow\infty$
($M$ is the mass of the model and $G$ -- the gravitational constant), 
then 
\be
\Phi_0 = GM/r_0 . \label{eq-a2.4} 
\ee
Taking $r=0$ we have
\be
\Phi (0) = \Phi_0 , \label{eq-a2.5} 
\ee
\ie $\Phi_0$ is the central potential. The constant $r_0$ characterises
the spatial extent of the model. Because $\Phi (0) = GM/r_h$, where $r_h$
is the harmonic mean radius of the model, then
\be
r_0 = r_h . \label{eq-a2.6} 
\ee

It is not difficult to find the mass distribution of the model. Because
$\Phi '(r) = -GM(r)/r^2$, where $M(r)$ is the mass within the sphere of a
radius $r$ (the ``inner mass''), we derive
\be
{M(r)\over M} = \left( {r\over r_0}\right)^3 {a^3\over \zeta (b+\zeta
)^2} = (1-\zeta^{-2})^{3/2} \left( {\zeta\over b+\zeta} \right)^2 .
\label{eq-a2.7} 
\ee
Further, as $M'(r) = 4\pi r^2\rho (r)$, where $\rho (r)$ is the
spatial mass density, we find
\be
\rho (r) = \rho_0 a^3 \left[ {1\over \zeta^3(b+\zeta )^2} + {2\over 3} b
{1-\zeta^{-2} \over \zeta (b+\zeta )^3} \right] . \label{eq-a2.8} 
\ee
The constant $\rho_0$ is determined by the relation
\be
M = {4\pi\over 3} r_0^3 \rho_0 . \label{eq-a2.9} 
\ee
In general, it differs from the central density. For the central density we have
\be
\rho (0) = \rho_0 a . \label{eq-a2.10} 
\ee
If $a=1$ ($q=0$), then $\rho (0) = \rho_0$, but when $a\rightarrow\infty$
($q \rightarrow 1$), then $\rho (0)\rightarrow\infty$.

To compare the model with observations, one needes to know the surface density
$P(R)$, \ie the density projected to the celestial plane ($R$ is the
projection of the radius $r$). It is convenient to use also the surface
inner mass $M_p(R)$, \ie the projected mass within a circle with radius
$R$ (we may compare it with the luminosity integrated within the same circle).
According to Eq.~(2.15) from the paper by \citet{Kuzmin:1973ag} we
have 
\be
M_p(R)/M = (R/R_0)^2 q^{-2} f_2(\zeta /b), \label{eq-a2.11} 
\ee
where $\zeta = \zeta (R) = \sqrt{ 1 + a^2R^2/R_0^2}$ and $R_0 = r_0$, and 
\be
f_n(u) = \int_u^{\infty} {\rmd x\over (1+x)^n \sqrt{x^2-u^2} } =
\int_1^{\infty}  {\rmd y \over (1+uy)^n \sqrt{y^2-1} } . \label{eq-a2.12} 
\ee
Because $M'_p(R) = 2\pi RP(R)$ and $f'_n(u) u = n(f_{n+1} - f_n)$, we
derive
\be
P(R) = P_0 q^{-2} [ \zeta^{-2} f_2(\zeta /b) + (1-\zeta^{-2}) f_3(\zeta /b)]. \label{eq-a2.13} 
\ee
Here the constant $P_0$ is determined by the formula
\be
M = \pi R_0^2 P_0 . \label{eq-a2.14}
\ee
The central surface density is
\be
P(0) = P_0 q^{-2} f_2(b^{-1}) . \label{eq-a2.15} 
\ee

From the formulae for density it results that when $r\rightarrow\infty$
and $R\rightarrow\infty$, then
\be
\rho (r)/rho_0\rightarrow {2/over 3} q(r_0/r)^4, ~~~~
P(R)/P_0 \rightarrow {\pi\over 4} q(R_0/R)^3 , \label{eq-a2.16} 
\ee
\ie $\rho (r)$ and $P(R)$ will be proportional to $r^{-4}$ (the Jeans law)
and $R^{-3}$. An exception is only the case $q=0$,  when there appear
proportionalities to $r^{-5}$ and $R^{-4}$.

When $q=0$ (\ie if $a=1$, $b=0$) in formulae for density, only the first
term in square brackets remains. The model reduces to the known Schuster
model (more correctly to the Schuster-Plummer model), where
\be
\ba{ll}
\Phi (r) = & \Phi_0\left( {r_0^2\over r_0^2+r^2} \right)^{1/2} = {GM\over (r_0^2+r^2)^{1/2}} , \\
\noalign{\smallskip}
{M(r)\over M} = & \left( {r^2\over r_0^2+r^2} \right)^{3/2} , ~~~~
{\rho (r)\over\rho_0} =  \left( {r_0^2\over r_0^2+r^2} \right)^{5/2} ,\\
\noalign{\smallskip}
{M_p(R)\over M} = & {R^2\over R_0^2+R^2} , ~~~~ {P(R)\over P_0} = \left(
{R_0^2\over R_0^2+R^2} \right)^2 . 
\ea
\label{eq-a2.17} 
\ee

If $q=1/2$ (\ie  $a=2$, $b=1$), we have the  usual isochrone model. If
$q\rightarrow 1$ (\ie  $a, b\rightarrow\infty$), we have a limiting model. In
the limiting model only the second term in square brackets in density
expression remains, while the factor $1-\zeta^{-2}$ reduces to one. For
the potential, the densities and the inner masses we derived
\be
\ba{ll}
\Phi (r) = & \Phi_0 {r_0\over r_0+r} = {GM\over r_0+r} , \\
\noalign{\smallskip}
{M(r)\over M} = & {r^2\over (r_0+r)^2} , ~~~~ {\rho (r)\over\rho_0} = 
{2\over 3} {r_0^4\over r(r_0+r)^3}, \\
\noalign{\smallskip}
{M_p(R)\over M} = & \left( {R\over R_0}\right)^2 f_2\left( {R\over R_0}
\right) , ~~~~ {P(R)\over P_0} = f_3\left( {R\over R_0}\right) .
\ea
\label{eq-a2.18}
\ee
\vglue 10mm

Let us turn now to the models,  proposed in the present paper,   which we call
as generalised-isothermal. For these models the potential is 
\be
\Phi (r) = {\Phi_0 \over q}~ \ln \left( 1+ {b\over\zeta}\right) . \label{eq-a2.19}
\ee 
Here again $\zeta = \sqrt{1+a^2r^2/r_0^2}$. The parameters are also the
same. Remains valid also formula (\ref{eq-a2.4}), relating  parameters $\Phi_0$ and
$r_0$ with the mass $M$. But the central potential no more equals to
$\Phi_0$, and the characteristic length $r_0$ no more equals to the
harmonic mean radius. Now
\be
\Phi (0)/\Phi_0 = r_0/r_h = q^{-1}\ln a . \label{eq-a2.20} 
\ee
$\Phi (0) = \Phi_0$ and $r_0=r_h$ only for $q=0$. If $q\rightarrow 1$
then $\Phi (0)\rightarrow\infty$ and $r_h\rightarrow 0$.

When we do the necessary calculations with the new potential, we derive
\be
{M(r)\over M} = \left( {r\over r_0}\right)^3 {a^3\over \zeta^2
(b+\zeta)} = (1-\zeta^{-2})^{3/2} {\zeta\over b+\zeta} \label{eq-a2.21} 
\ee
and further
\be
\rho (r) = \rho_0a^3 \left[ {1\over \zeta^4(b+\zeta )} ~+~ {b\over3}
~{1-\zeta^{-2}\over \zeta^2(b+\zeta )^2} \right] , \label{eq-a2.22} 
\ee
while $\rho_0$ still is determined by Eq.~(\ref{eq-a2.9}). Equations
(\ref{eq-a2.21}) and (\ref{eq-a2.22}) are 
similar to Eqs.~(\ref{eq-a2.7}) and (\ref{eq-a2.8}). But there are differences in powers of
$\zeta$ and $b+\zeta$, suggesting  to more profound central density
concentration in case of large $a$ when compared with the
generalised-isochrone models. For the central density we derive instead of
Eq.~(\ref{eq-a2.10}) the formula
\be
\rho (0) = \rho_0 a^2 , \label{eq-a2.23} 
\ee
giving steeper increase of the central density with the parameter $a$ than
it was the  case in   the generalised-isochrone models.

For $M_p(R)$ the equation (2.15) from \citet{Kuzmin:1973ag} gives
\be
{M_p(R)\over M} = \left( {R\over R_0}\right)^2 q^{-2} ~g_2\left(
{\zeta\over b}\right) . \label{eq-a2.24}
\ee
Here again $\zeta = \zeta (R)$, but instead of the function $f_2(u)$ we
have now the function $g_2(u)$. The functions  $g_n(u)$ are  defined by the
integral 
\be
g_n(u) = \int_u^{\infty} {\rmd x\over x(1+x)^{n-1}\sqrt{x^2-u^2}} = 
{1\over u} \int_1^{\infty} {\rmd y\over y(1+uy)^{n-1}\sqrt{y^2-1}} \label{eq-a2.25}
\ee 
and are related with the functions $f_n(u)$ via relation
\be
f_n(u) = g_n(u)-g_{n+1}(u), \label{eq-a2.26}
\ee
while $g_1(u) = \pi /2 u^{-1}$. 

For the surface density we find
\be
P(R) = {P_0\over q^2} \left[ {1\over\zeta^{-2}} g_2\left( {\zeta\over b}
\right) ~+~ (1-\zeta^{-2}) g_3\left( {\zeta\over b}\right)\right] .
\label{eq-a2.27} 
\ee
The constant $P_0$ is determined as it was before by Eq.~(\ref{eq-a2.14}). But the
central surface density is now
\be
P(0) = {P_0\over q^2} g_2\left( {1\over b}\right) . \label{eq-a2.28} 
\ee

In Table~\ref{tab-a2.1}  we give some values of the functions $g_1(u)$, $g_2(u)$
and $g_3(u)$. 

\begin{table}
\caption{}
\smallskip
\label{tab-a2.1}
\centering
\begin{tabular}{|c | c | c | c |}
\hline\hline
\noalign{\smallskip}
$\log u$  & $\log g_1(u)$ & $\log g_2(u)$ & $\log g_3(u)$ \\
\noalign{\smallskip}
\hline
\noalign{\smallskip}
-2.000& 2.196 & 2.181 & 2.169 \\
-1.699& 1.895 & 1.869 & 1.847 \\
-1.301& 1.497 & 1.443 & 1.398 \\
-1.000& 1.196 & 1.104 & 1.028 \\
-0.699& 0.895 & 0.741 & 0.615 \\
\noalign{\smallskip}
-0.301& 0.497 & 0.210 & -0.033\\
0.000 & 0.196 & -0.244& -0.624\\
0.301 & -0.105& -0.743& -1.310\\
0.699 & -0.503& -1.461& -2.337\\
1.000 & -0.804& -2.032& -3.175\\
\noalign{\smallskip}
1.301 & -1.105& -2.619& -4.044\\
\noalign{\smallskip}
\hline\hline
\noalign{\medskip}
\end{tabular}
\end{table}

The density behaviour at large distances from the centre is the same as it
was for the generalised-isochrone models. When $r\rightarrow\infty$ and $R
\rightarrow\infty$ we derive
\be
\rho (r)/\rho_0 \rightarrow {1\over 3} q \left( {r_0\over r}\right)^4 ,
~~~~~ P(R)/P_0 \rightarrow {\pi\over 8} q \left( {R_0\over R}\right)^3 .
\label{eq-a2.29} 
\ee

If $q=0$,  the model again reduces to the Schuster model. For the limiting
model $q=1$ ($a, b\rightarrow\infty$) we derive
\be
\ba{ll}
{\Phi (r)\over\Phi_0} = & \ln \left( 1+{r_0\over r}\right) , \\
\noalign{\smallskip}
{M(r)\over M} = & {r\over r_0 + r} , ~~~~ {\rho (r)\over\rho_0} = {1\over 3}
{r_0^4\over r^2(r_0+r)^2} , \\
\noalign{\smallskip}
{M_p(R)\over M} = & \left( {R\over R_0}\right)^2 g_2\left( {R\over R_0}
\right) , ~~~~ {P(R)\over P_0} = {1\over 2} g_3\left( {R\over R_0} \right)
\ea
\label{eq-a2.30}
\ee

In the central parts of the limiting model the spatial density is
proportional to $r^{-2}$ (and not $r^{-1}$), \ie  the model approaches in
the central parts to the isothermal model. For this reason it is suitable
to call this model as quasi-isothermal. This name can be used also for
models,  close  to the limiting model, \ie  when $q$ is near to one.

\begin{figure}[h]
\centering
\includegraphics[width=80mm]{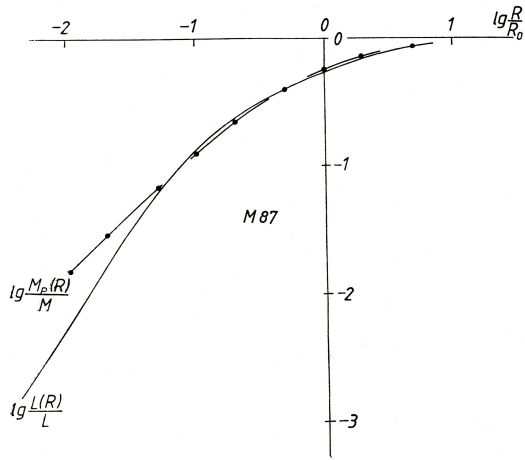}
\caption{The projected mass distribution of the limiting
generalised-isothermal model and the brightness distribution in the
galaxy M87 (NGC 4486).}
\label{fig-a2.1}
\end{figure}

\begin{figure}[h]
\centering
\includegraphics[width=80mm]{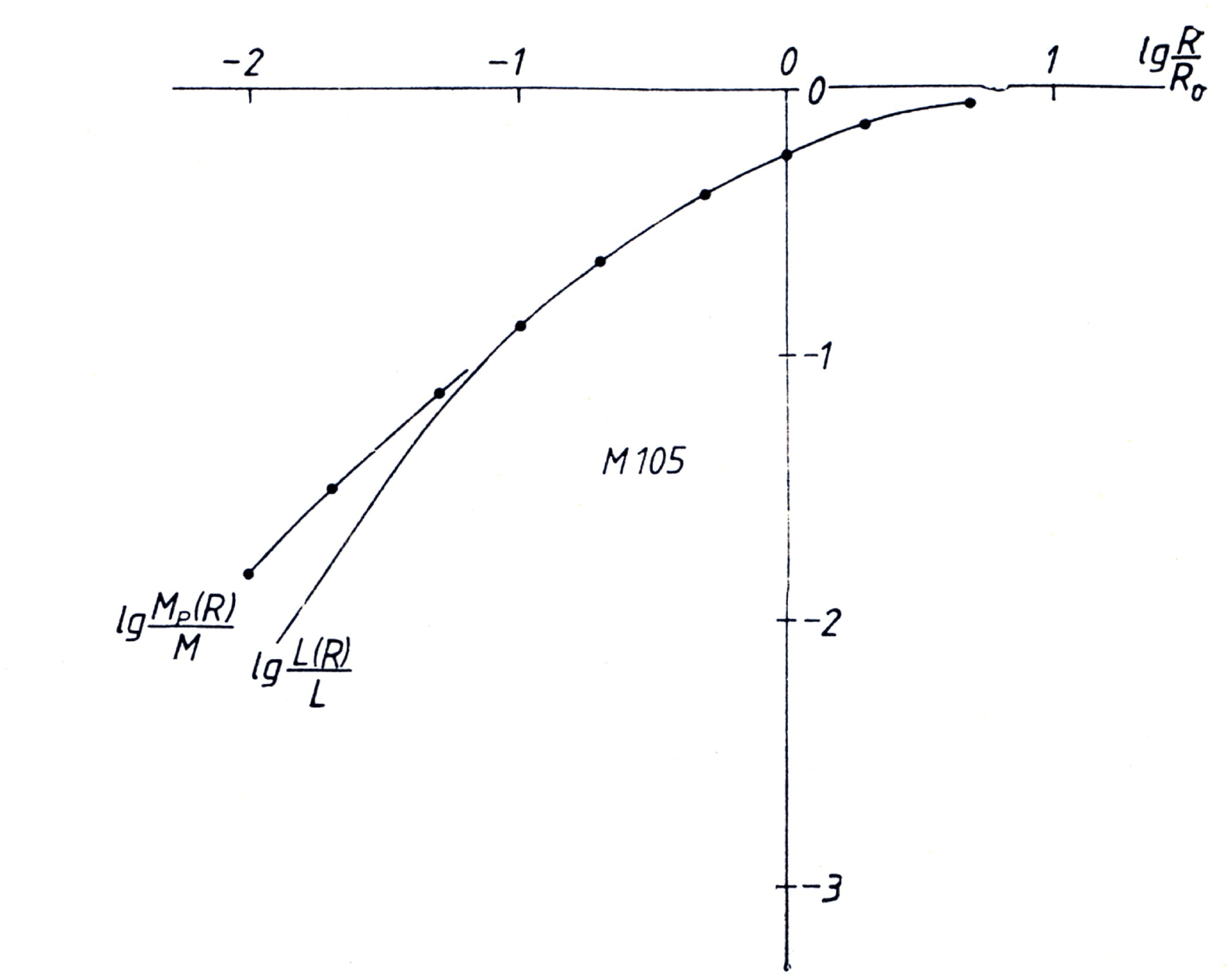}
\caption{The projected mass distribution of the limiting
generalised-isothermal model and the brightness distribution in the
galaxy M105 (NGC 3379).}
\label{fig-a2.2}
\end{figure}

The comparison of the mass distribution of quasi-isothermal models with the
luminosity distribution in galaxies M87 (NGC 4486, E0) and M105 (NGC 3379,
E1) demonstrates quite high similarity. It is seen in Figs.~\ref{fig-a2.1} and \ref{fig-a2.2}, where
$\log M_p(R)/M$ for the limiting quasi-isothermal model (continuous line)
and $\log L(R)/L$ for both galaxies (dashed line) as functions of $\log
r/r_0$ are given ($L(R)$ is the integrated luminosity within the circle
having radius $R$,  and $L$ is the total integrated luminosity). For M87 the
integrated luminosity $L(R)$ was found on the basis of the compilation of
the surface brightness measurements in B-colour by \citet{deVaucouleurs:1978,  deVaucouleurs:1979a}.
For M105 we used the results obtained by \citet{deVaucouleurs:1979b} 
(also in B-colour). The parameters $R_0$ and $L$ we derived

M87: ~~~~~$R_0=$ 240'', $m_B =$ 9.35

M105: ~~~~$R_0=$ ~80'', $m_B =$ 10.18

For both galaxies the behaviour of the integrated luminosity $L(R)$
is similar to the inner surface mass distribution of the model $M_p(R)$ in
very wide distance $R$ interval. Only near to the very centre $L(R)$
decreases with the decreasing $R$ significantly more rapidly. But this
discrepancy can be essentially removed when we choose for both galaxies
$q=0.975$ ($a\simeq 40$) instead of $q=1$.

\begin{figure}[h]
\centering
\includegraphics[width=80mm]{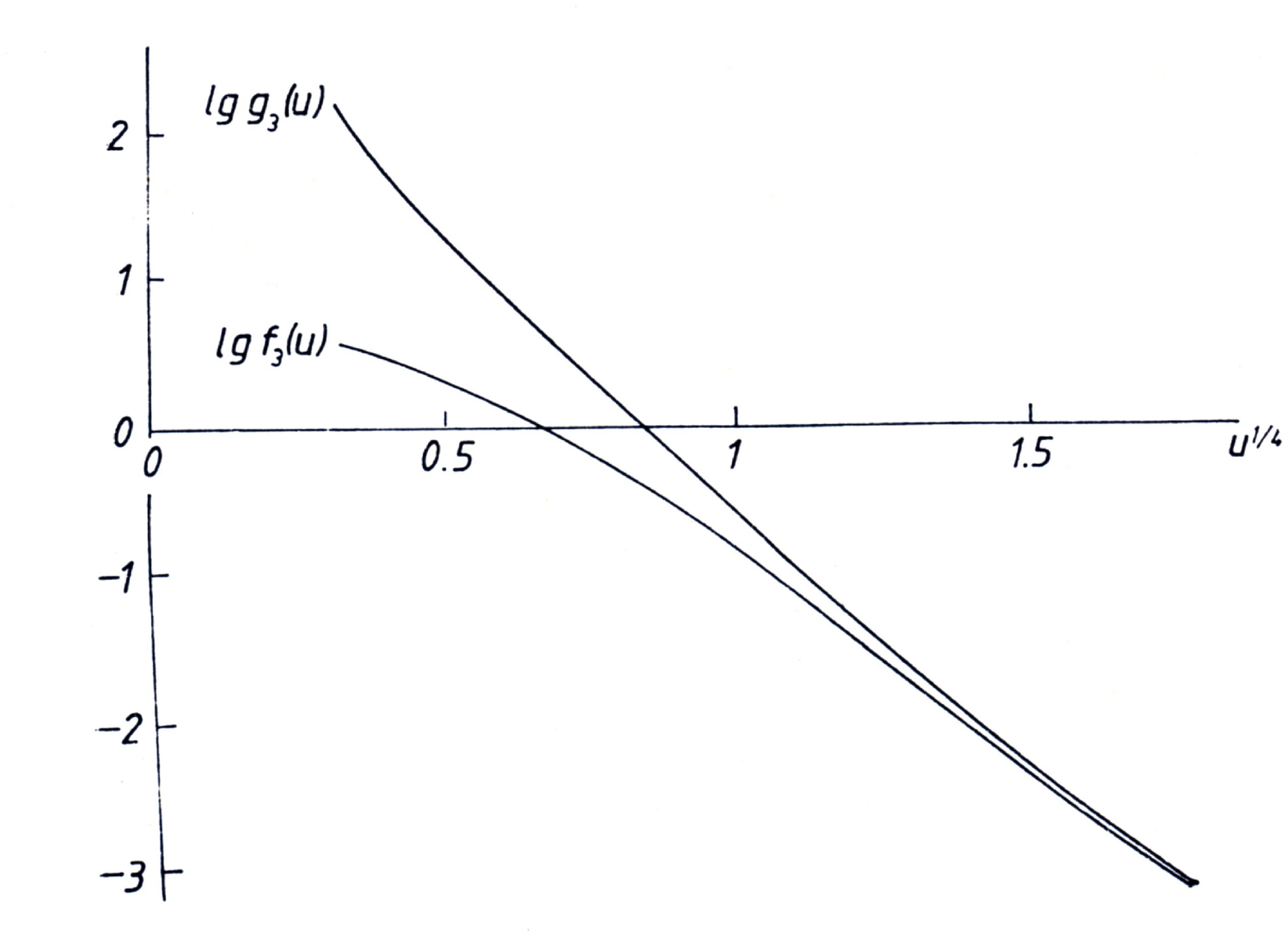}
\caption{The surface density of the limiting generalised-isothermal
and generalised-isochrone laws and the de Vaucouleur's law.}
\label{fig-a2.3}
\end{figure}

Because quite often to describe the light distribution in galaxies, and in
particular in galaxies M87 and M105, it is used the empirical de
Vaucouleur's law, where the logarithm of the brightness varies
proportionally to $R^{1/4}$ we may expect that the quasi-isothermal model
agrees well with this law. The surface density of the limiting
generalised-isothermal model is proportional to $g_3(R/R_0)$. Hence $\log
g_3(u)$ must vary nearly proportionally to $u^{1/4}$. This is seen in
Fig.~\ref{fig-a2.3}, where the curve $\log g_3(u)$ as a function of $u^{1/4}$ has
really long linear part. For comparison we plotted in the same Figure also
the curve $\log f_3(u)$, corresponding to the generalised-isochrone
limiting model. Evidently it agrees much worse with the de Vaucouleur's
law. 

As a conclusion we like to mention that the potentials (\ref{eq-a2.1}) and (\ref{eq-a2.19}) are
the special cases of the potential having more general form
\be
\Phi (r) = {\Phi_0\over mq} \left[ 1-\left( {\zeta\over b+\zeta}
\right)^m \right] , \label{eq-a2.31} 
\ee
where $m$ is an additional independent structural parameter. The
generalised-isochrone models we have when $m=1$, the generalised-isothermal
models when $m\rightarrow 0$. For $M(r)$ and $\rho(r)$ we derive natural
generalisations of Eqs.~(\ref{eq-a2.7}), (\ref{eq-a2.21}) and (\ref{eq-a2.8}), (\ref{eq-a2.22})
\be
{M(r)\over M} = (1-\zeta^{-2})^{3/2} \left( {\zeta\over b+\zeta}
\right)^{m+1}  \label{eq-a2.32}
\ee
and
\be
\rho (r) = \rho_0 a^3 \left[ {1\over\zeta^4(b+\zeta )} ~+~ {m+1\over 3}
b {1-\zeta^{-2}\over \zeta^2(b+\zeta )^2} \right] \left( {\zeta\over b+
\zeta}\right)^m . \label{eq-a2.33} 
\ee

Because the density must be non-negative and smoothly decrease when $r$
increases (for all $q$) it follows the condition
\be
-1\le m\le 2 . \label{eq-a2.34}
\ee
For $m=-1$ the model reduces to the Schuster model (as is was when $q=0$),
but instead of $r_0$ in Eq.~(\ref{eq-a2.17}) there is $r_0/a$.

The parameter $m$ together with the parameter $q$ enables to control the
behaviour of the density in central regions of the model. But in outermost
regions for $q>0$ and $m>-1$ we have the density proportional to $r^{-4}$
independently of $m$ and $q$.

We note that the model  with parameters $m=1/2$ and
$q\simeq 1$ is in better agreement with the de~Vaucouleur's law than
the quasi-isothermal model with parameters $m=0$, $q\simeq 1$.
This model enables to describe the
real behaviour of the density near to the central regions of the galaxies
M87 and M105. But outside of the innermost regions the quasi-isothermal
describes the density distribution in both galaxies still better.
\vglue 10mm

{\bf Note:} 
Basic results of the present paper were derived in 1975 (the models) and
thereafter in 1982 (application for galaxies). In 1983 
the paper by W. Jaffe \citep{Jaffe:1983}  was published,  where he proposed
independently the model,  identical with our limiting generalised-isothermal
model.

\hfill March 1985

%% file: ch1987.tex
\chapter[~~Hydrodynamic models of  flat axisymmetric stationary
stellar system]{~Hydrodynamic models of flat axisymmetric
  stationary stellar system\footnote{\footnotetext ~Report presented
    on the meeting ``Dynamics of gravitational systems and methods of
    analytical celestial mechanics'', Sept. 22 -- 24, Alma-Ata, 1987
    (Report was read by P. Tenjes). The paper remained unpublished. }} 


The construction of self-consistent models of flattened stellar
systems is quite complicated problem when the phase density is a
function of three integrals of motion, corresponding to the triaxial
local velocity distribution ellipsoid. Far more simply the problem can
be solved within the hydrodynamic description. We propose in the
present paper one solution of this kind.

We suppose that the regular gravitational field of a stellar system
allows the third integral of motions, quadratic in respect to
velocities. Some models of the corresponding potential and the density
were constructed by \citet{Kuzmin:1956a}. In another paper
\citep{Kuzmin:1962aa} we found a self-consistent model, but with the
phase density, being the function of only the energy and the angular
momentum integrals.

\begin{figure}
\centering
\includegraphics[width=50mm]{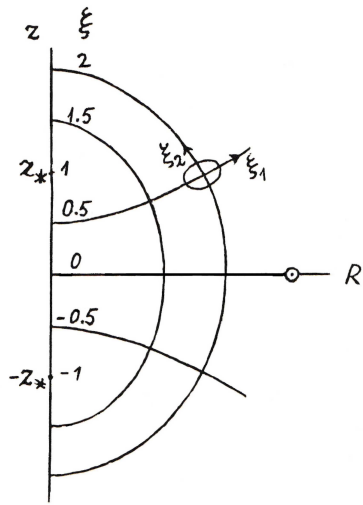}
\caption{Coordinates $(R , z)$ and $(\xi_1 , \xi_2)$ in the galactic meridional plane.}
\label{fig-a3.1}
\end{figure}

If a third integral of motions exist, quadratic about velocities,  then
the velocity distribution at a given point has three orthogonal planes
of symmetry, the intersection of which is along the coordinate lines
of a spheroidal system of coordinates $\xi_1$, $\xi_2$, $\theta$
\citep{Kuzmin:1956a}. In the similar way lies the velocity ellipsoid, 
describing the local velocity distribution. The coordinates $\xi_1$,
$\xi_2$, $\theta$ are related with cylindrical coordinates $R$, $z$,
$\theta$ according to formulae  
\begin{equation}
R = z_0\sqrt{(\xi_1^2-1)(1-\xi_2^2)} , ~~~~ z= z_0\xi_1\xi_2 , ~~~~
z_0 = \mathrm{const}. 	\label{eq-a3.1} 
\end{equation}
In the  meridional plane, where we have $R$ and $z$ as Cartesian
coordinates, the coordinate lines $\xi_1$ and $\xi_2$ are 
intersecting ellipses and hyperbolae with the foci on $z$-axis at
$z=\pm z_0$ (Fig.~\ref{fig-a3.1}). The values of the coordinate
$\xi_2$ are from $-1$ to $+1$, the values of $\xi_1$ are from one to
infinity. The sign of the coordinate $\xi_2$ is the same as of $z$,  and
when $z=0$, then also $\xi_2=0$. On $z$-axis one of the coordinates
$\xi_1$ or $|\xi_2|$ is one, the other is $z/z_0$ or $|z|/z_0$. At
foci both coordinates $\xi_1$ and $|\xi_2|$ are equal to unity.  

Let $\rho$ be  the local density of the system (not necessarily the
mass density). Let us introduce the ``pressures'' along the coordinate
lines $\xi_1$ and $\xi_2$ 
\begin{equation}
p_1 = \rho\overline{v_1^2} , ~~~~~ p_2 = \rho\overline{v_2^2} .	\label{eq-a3.2}
\end{equation}
Here $v_1$ and $v_2$ are the velocity components along $\xi_1$ and
$\xi_2$,  and the bar means averaging. Evidently
\[
\overline{v_1 v_2} = 0,
\]
because the corresponding ``stress'' is zero. The pressure
$p_{\theta}$ along the azimuthal coordinate $\theta$ is not relevant
for us further. The pressure is anisotropic in general, \ie  $p_1 \ne
p_2 \ne p_{\theta}$. The only exceptions are the foci of spheroidal
system of coordinates. Here the velocity distribution is spherical. 

\begin{figure}
\centering
\includegraphics[width=60mm]{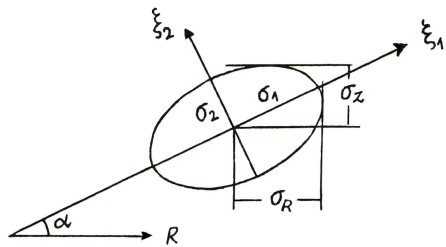}
\caption{Orientation of the velocity ellipsoid in meridional
  plane. $\sigma_1$ and $\sigma_2$ are velocity dispersions along
  coordinate lines $(\xi_1, \xi_2)$.} 
\label{fig-a3.2}
\end{figure}

The mean square velocities in Eq.~(\ref{eq-a3.2}) are the velocity
dispersions, and by designating them as $\sigma_1^2$ and $\sigma_2^2$  
Eq.~(\ref{eq-a3.2}) can be rewritten as 
\begin{equation}
p_1 = \rho\sigma_1^2, ~~~~~ p_2 = \rho\sigma_2^2. \label{eq-a3.2b}	
\end{equation}
The quantities $\sigma_1$ and $\sigma_2$ together with
$\sigma_{\theta}$ are just the semiaxes of  the velocity
ellipsoid. Limiting with two-dimensional case in the meridional plane
we have the velocity ellipse with the semiaxes  $\sigma_1$ and
$\sigma_2$ (Fig.~\ref{fig-a3.2}).  We suppose $\sigma_1$ to be the
longer semiaxis. We designate the inclination angle of $\sigma_1$ about the
$R$-axis as $\alpha$. This is the inclination angle of
the tangent of $\xi_2$-line at a given point. 

In the system of coordinates $R$, $z$, where the corresponding
velocities are $v_R$, $v_z$,  we have the ``pressures'' $p_{RR}$,
$p_{zz}$ and the ``stress'' $p_{Rz}$ 
\begin{equation}
p_{RR} = \rho\overline{v_R^2}, ~~~~ p_{zz} = \rho\overline{v_z^2}, ~~~~ p_{Rz} = \rho\overline{v_Rv_z} , \label{eq-a3.3} 	
\end{equation}
or when introducing the velocity dispersions along $R$ and $z$
\[
p_{RR} = \rho\sigma_R^2 , ~~~~~ p_{zz} = \rho\sigma_z^2 . 
\]
The pressures in both systems of coordinates are related by
\[
p_{RR} = p_1\cos^2\alpha + p_2\sin^2\alpha , 
\]
\begin{equation}
p_{zz} = p_1\sin^2\alpha + p_2\cos^2\alpha , \label{eq-a3.4}
\end{equation}
\[
p_{Rz} = (p_1-p_2)\sin\alpha\cos\alpha . 
\]

From the properties of elliptic coordinates
\begin{equation}
\sin\alpha = \sqrt{g_1}\xi_2 , ~~~~ \cos\alpha = \sqrt{g_2}\xi_1 ,	\label{eq-a3.5}
\end{equation}
where
\begin{equation}
g_1 = \frac{\xi_1^2 - 1}{\xi_1^2 - \xi_2^2} , ~~~~ g_2 = \frac{1 - \xi_2^2}{\xi_1^2 - \xi_2^2}	\label{eq-a3.6}
\end{equation}
are the coefficients,  introduced earlier \citep{Kuzmin:1956a}. In
addition to Eqs.~(\ref{eq-a3.5})--(\ref{eq-a3.6}) we add the equation, 
relating the differential of the coordinate $\xi$ with the
corresponding differential of the path length  of coordinate lines 
\begin{equation}
	z_0 ~\dd\xi = \sqrt{g} ~\dd{s}. \label{eq-a3.7}
\end{equation}
The quantity $z_0 / \sqrt{g}$ is the Lam\'e coefficient for the present case. 

In hydrodynamic description the equilibrium of a stationary stellar
system with respect to $R$ and $z$ coordinates is determined by two
hydrodynamic equations of stellar systems \citep{Kuzmin:1965ab}, which
is suitable to call as the hydrostatical equations. In the present paper we
use only the equilibrium equation about $z$. It has the form 
\begin{equation}
	\frac{\partial p_{zz}}{\partial z} + \frac{\partial
          p_{Rz}}{\partial R} + \frac{p_{Rz}}{R} = \rho
        \frac{\partial\Phi}{\partial z} . \label{eq-a3.8}  
\end{equation}
Here $\Phi$ is the gravitational potential of the system. Moving to
the pressures $p_1$ and $p_2$ and to elliptic coordinates, we derive
the equation  
\begin{equation}
g_1 \frac{\partial p_1}{\xi_1\partial\xi_1} + g_2 \frac{\partial p_2}{\xi_2\partial\xi_2} + 2 \frac{p_1-p_2}{\xi_1^2-\xi_2^2} = \rho z_0 \frac{\partial \Phi}{\partial z} . \label{eq-a3.9}	
\end{equation}

Known functions in Eq.~(\ref{eq-a3.9}) are the density $\rho$ and the
potential $\Phi$, and unknown functions are the pressures $p_1$ and
$p_2$. Thus we have two unknown functions and one equation, relating
them. It is needed one more relation between them. Using also the
equilibrium equation about $R$ does not help us, because there appear
two additional unknown functions -- the azimuthal pressure
$p_{\theta}$ and the rotation velocity of the system
$\overline{v_{\theta}}$. To solve the problem with respect to $p_1$
and $p_2$ most simply  is to assume some form of the ratio of pressures
as a function of coordinates. In this case the partial differential
equation (\ref{eq-a3.9}) reduces to the usual linear differential
equation of the first order. The integration of this equation must be
done along the curves  determined by the following differential
equation 
\begin{equation}
\frac{\xi_2 d\xi_2 }{\xi_1 d\xi_1} = \frac{(1-\xi_2^2) }{ (\xi_1^2 -1)} ~ \frac{p_1}{p_2} . \label{eq-a3.10} 	
\end{equation}
To simplify the problem we propose the separation of variables, \ie
to suppose that the ratio $p_2/p_1 = f(\xi_2)/g(\xi_1)$, where $f, g$
are some functions. Here we assumed 
\begin{equation}
\frac{p_2}{p_1} = \frac{\sigma_2^2}{\sigma_1^2} = \frac{c^2+\xi_2^2}{c^2+\xi_1^2} ; ~~~ c^2 = \mathrm{const.} \label{eq-a3.11}
\end{equation}
At foci of elliptical coordinates $\sigma_1 =\sigma_2$ just as it must
be. Outside of the foci $\sigma_2$ is smaller than $\sigma_1$, while
the ratio $\sigma_2/\sigma_1$ tends to zero.   When the coordinate
$\xi_1$ increases, \ie  when moving away from the system centre, the
velocity ellipse becomes more and more elongated,  and the motion in $R,z$ plane
approaches to a pure radial motion. But if the parameter $c^2$ increases
up to infinity,  then $\sigma_1$ and $\sigma_2$ are equal
everywhere. This particular case corresponds to the case, when the
phase density is a function of the energy and angular momentum
integrals. These models is suitable to call Jeans models, because
they were studied first by \citet{Jeans:1915, Jeans:1922}. 

The solution of Eq.~(\ref{eq-a3.9}) we derive in form
\begin{equation}
\frac{p_1}{c^2+\xi_1^2} = \frac{p_2}{c^2+\xi_2^2} = \frac{1}{(c^2 +
  z^2/ z_0^2)^2 } \int_z^{\infty}\left[ (c^2 + z^2/z_0^2) -
\frac{\partial\Phi}{\partial z} \rho\right] \dd\,z, \label{eq-a3.12}  	
\end{equation}
where the integration is along the hyperbolae
\begin{equation}
\frac{R^2}{a^2} - \frac{z^2}{c^2} = z_0^2 . \label{eq-a3.13} 
\end{equation}
The parameter $a$ determines the point, where the hyperbola intersects
the $R$-axis. As it must be, for the Jeans case the hyperbola
degenerates into straight  lines, parallel to the $z$-axis,  and the
pressure $p_1=p_2=p$ results via simple integration of
$-\rho{\partial\Phi\over\partial z} \dd{z}$ along these lines. 

The solution (\ref{eq-a3.13}) we used for our model $n=3$
\citep{Kuzmin:1956a}. This is the same model, we used in a subsequent
paper \citep{Kuzmin:1962aa} to construct the phase model, where the
phase density was a function of the energy and the angular momentum
integrals. In studying this model it is suitable to introduce instead
of $\xi_1$, $\xi_2$ new variables $\zeta_1$, $\zeta_2$, related with
the old ones as 
\begin{equation}
\zeta^2 = \zeta_0^2 + (1-\zeta_0^2)\xi^2 , ~~~~~ \zeta_0^2 = \mathrm{const.} \label{eq-a3.14}
\end{equation}
The variables $\zeta_1, \zeta_2$ are also elliptic coordinates,  but
defined in a different from $\xi_1, \xi_2$ way. With help of variables
$\zeta$ we introduce new variables 
\begin{equation}
Y = \zeta_1 + \zeta_2, ~~~~~ Z = \zeta_1 \zeta_2 . \label{eq-a3.15}	
\end{equation}
The formula for $Z$ is similar to the equation (\ref{eq-a3.1}) for
$z$. For $\zeta_0=0$ we have simply $Z = z/z_0$. Via the variables $Y$
and $Z$ the potential $\Phi$ and the mass density $\rho$ can be
expressed in the following way 
\begin{equation}
\Phi = \frac{\Phi_0}{Y} , \label{eq-a3.16}	
\end{equation}
\begin{equation}
\rho = \rho_0\zeta_0^2 \frac{Y^2+Z^2+Z }{Y^3Z^3} , \label{eq-a3.17} 	
\end{equation}
while $4\pi G\rho_0 = \frac{1-\zeta_0^2}{z_0^2} \Phi_0$. When
$\zeta_0=0$, the parameter $\Phi_0$ has the meaning of the central
potential (in general the central potential is
$\Phi_0/(1-\zeta_0)$). The parameter $\zeta_0$, taking values from zero
to one, has the meaning of the model flattness. If $\zeta_0\rightarrow
0$ we have ``nearly flat disk'' embedded into ``slight
atmosphere''. If $\zeta_0\rightarrow 1$, the model approaches to
spherically symmetrical one (the isochrone model by H\'enon). The
ratio $z_0/(1-\zeta_0^2)^{1/2}$ is a characteristic length. When we
exclude the nearby to the centre regions and the outermost regions,  the
model quite well resembles the Galaxy, while $\zeta_0\sim
0.1$. Advantages of the model $n=3$ are,  first, a simple expression for
the potential and not too complicate expression for the density, and 
second, that the potential and the density depend on $Y$ and $Z$ in
the same way for all flattnesses (only the region of possible values
of $Y$, $Z$ varies). 

Solution for the pressures $p_1$ and $p_2$ results
\begin{equation}
\frac{p_1}{c_0^2+\zeta_1^2} = \frac{p_2}{c_0^2+\zeta_2^2} = \frac{\Phi_0}{(c_0^2 +Z^2)^2} \int_Z^{\infty} (c_0^2 + Z^2) (1+Z) Y^{-3} \rho \dd{Z} , \label{eq-a3.18} 	
\end{equation}
while the integration is along the curves
\begin{equation}
X = \frac{Y^2 - (1+Z)^2}{c_0^2 + Z^2} = \mathrm{const}, \label{eq-a3.19}
\end{equation}
where
\begin{equation}
c_0^2 = (1-\zeta_0^2)c^2 - \zeta_0^2 . \label{eq-a3.20} 	
\end{equation}
In Eq.~(\ref{eq-a3.17}) $\rho$ is not necessarily the mass density of
the whole system. It may be the density of a subsystem. But if we want
to obtain a self-gravitating model of the system as a whole, then we
must use for the density Eq.~(\ref{eq-a3.17}). Eliminating $Y^2$ we
have 
\begin{equation}
Y^{-3} \rho = \rho_0\zeta_0^2 \frac{(1+Z)(1+2Z) + X^2(c_0^2+Z^2) }{
  [(1+Z)^2 + X^2(c_0^2+Z^2)]^3 Z^3} . \label{eq-a3.21}	 
\end{equation}
Evidently the integral in Eq.~(\ref{eq-a3.18}) can be taken in
elementary functions. But the result is very long. The solution
results in a form 
\begin{equation}	
\frac{p_1}{c_0^2+\zeta_1^2} = \frac{p_2}{c_0^2+\zeta_2^2} = \Phi_0
\rho_0 \zeta_0^2 F(X,Z;c_0^2), \label{eq-a3.22} 
\end{equation}
where $F$ is a known function. Thus  the problem is solved. By
giving the coordinates $R,z$ we move to coordinates $\xi$, thereafter
to coordinates $\zeta$, and find from Eq.~\ref{eq-a3.22} the
pressures. Dividing the pressures by density,  we have the velocity
dispersions $\sigma_1^2$ and $\sigma_2^2$, and thereafter  $\sigma_R^2$
and $\sigma_z^2$.

\begin{figure}[h]
\centering
\includegraphics[width=80mm]{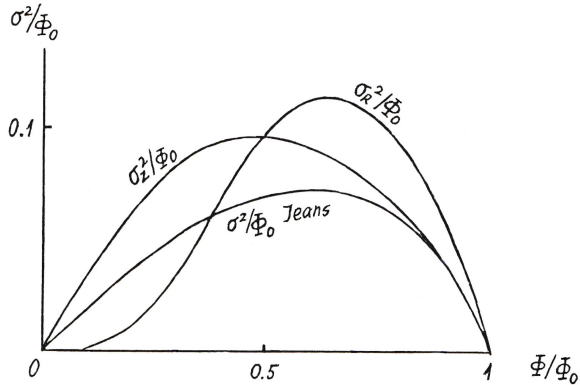}
\caption{Calculated velocity dispersions $\sigma_R$ and $\sigma_z$ as
  functions of the potential. The dispersions and the potential are
  given in units of the central potential $\Phi_0$.} 
\label{fig-a3.3}
\end{figure}

Solution is valid for all flatnesses. If we take $\zeta_0=0$, we have
the solution for ``flat disk''. It is similar to the one, resulting
for a highly flattened stellar system like our Galaxy. To have the
solution, we must choose the parameter $c_0^2$. In the solar
neighbourhood $\sigma_z^2/\sigma_R^2\simeq 1/2.5$, and here $\xi_2 =0$,
$\xi_1^2\simeq 5$. It gives $c^2\simeq 2$. Because $\zeta_0 \sim 0.1$, 
then $c_0^2$ is approximately the same. Thus we chose $c_0^2 =2$. In
Fig.~\ref{fig-a3.3} the velocity dispersion curves $\sigma_R^2$,
$\sigma_z^2$ on $z$-axis for the ``flat disk'' model are plotted. The
dispersions are plotted as functions of the potential, while the unit
of dispersions as well as of potential is the central potential
$\Phi_0$. The curves intersect at $\Phi /\Phi_0=1/2$,  corresponding to
the foci of the elliptic system of coordinates. For comparison, in the
same figure the dispersion curve for the Jeans case is presented. It
is seen that the difference is very essential.

Similar curves we may construct for the velocity dispersions along the
lines $R = \mathrm{const}$, parallel to the $z$-axis. In this case for the unit
of the dispersions and the potential is suitable to take the potential
on $R$-axis at given $R$. The resulting picture is in general
similar. But the intersection of the curves $\sigma_z^2$ and
$\sigma_R^2$ shifts to smaller potential when $R$ increases. The curve
$\sigma_R^2$ becomes higher, the curve $\sigma_z^2$ becomes lower,  and
finally even lower than the Jeans curve. 

The proposed model is only one of the most simple models. It is clear
that the construction of hydrodynamic self-consistent models of
flattened stellar systems has no principal difficulties.

%% file: conclusions.tex
\chapter{Conclusions}

This thesis concludes with a summary. The most important results of
the work taken as a basis for this dissertation can be summarised
(without going into too much detail) 
in the form of the following points.

\begin{enumerate}
  \item{}
The solution to the one-dimensional problem of stellar dynamics
and its applicability to the Galaxy  is developed   (Chapters 1, 3, 6).

  \item{}
The dynamical constant $C$ is introduced and  its value is derived,  and the Galactic
density in the vicinity of the Sun is found (Chapters 1 and 5). 
  \item{}
The stellar dynamical method for 
determining the Galactic flattening and equatorial half-thickness is
specified and applied in practice  
(Chapters 1 and 5). 
  \item{}
A method for determining the radial mass distribution of stellar
systems is developed and applied in practice (Chapters 2, 3, 7, 9.
10). 
  \item{}
The theory of empirical superposition models of the mass
distribution consisting of inhomogeneous spheroids of different
flatness is suggested  (Chapter 9). 
  \item{}
The system of hydrodynamic equations of stellar dynamics is studied
(Chapter 21) and the second Jeans equation is refined (Chapters 3 and 8). 

  \item{}
The question which integrals of motion can serve as
arguments for the phase density at stationarity is clarified (Chapter 4). 
  \item{}
A third integral of stellar motion quadratic with respect to
velocities has been proposed as an argument of phase density together
with integrals of energy and areas (Chapters 3 and 4). 
  \item{}
It is shown that the restrictions imposed on the gravitational
potential by the third quadratic integral of motion are compatible
with the requirement of non-negativity of the mass density
 (Chapter  6). 
   \item{}
On the basis of the third quadratic integral theory  models of
 mass distribution in the Galaxy are constructed (Chapters 3 and 6). 
   \item{}
 On the basis of the same theory, a method of calculating
 three-dimensional galactic orbits of stars  is developed (Chapter 6). 
   \item{}
 A theory of the third quadratic integral as an approximate
 quasi-integral is developed (Chapters 4 and 8). 
  \item{} 
 Theoretical relations are found  between the gradient of the slope of the
 velocity ellipsoid and the radial gradient of the mass density
 (Chapter 8). 
   \item{}
 The first hydrodynamic and phase description model of the
 spatial-kinematical structure of the Galaxy as a self-gravitating stellar
 system is found (Chapter 11). 
   \item{}
  The hydrodynamic and phase description models of the
 spatial-kinematical  structure of spherical stellar systems are found,  satisfying
 the theory of irregular gravitational forces (Chapters 15 and 16).
   \item{}
  It is proposed to use the virial theorem in the tensor form. The
 virial of an inhomogeneous ellipsoid is found (Chapter 12). 
   \item{}
 Some theorems for the presence of symmetry planes in stationary
 stellar systems are proved (Chapter 13). 
   \item{}
 The classification of integrals of motion depending on the
 restrictions imposed on the potential is given (Chapter 14). 
   \item{}
 The foundations of the theory of irregular forces in stellar
 systems have been developed (Chapters 17, 18, 22). 
   \item{}
     It is pointed out that irregular forces should cause in spherical
 stellar systems a radial stretching of velocity distribution (Chapter
 17). 
   \item{}
     Taking into account irregular forces, the theory of very
 flattened subsystems of the Galaxy in the hydrodynamical description
 is developed  (Chapters 19 to 21). 
   \item{}
     The theoretical expression for the ratio of the lengths of the
 semi-axes of the velocity ellipsoid is found (Chapter  19). 
   \item{}
     The theory of very flattened subsystems in the phase description
is developed; a theoretical distribution of velocities is derived (Chapter 22).
  \item{}
    The oscillatory variations in stellar kinematics, which can
explain the ellipsoidal velocity vertex deviation, are considered
(Chapters 20 and 21). 
\end{enumerate}

\chapter{Acknowledgements}

In conclusion I consider it my pleasant duty to express my deep
gratitude to the Academician of the Academy of Sciences of the
Estonian SSR A. Kipper who insisted on the realization of this work
and rendered his full assistance as the Director of the Institute of
Physics and Astronomy of the Academy of Sciences of the Estonian SSR.
\\

I am also very grateful to Prof. K. F. Ogorodnikov, Prof. T. A. Agekyan
and Prof. G. M. Idlis for their friendly advice and constant interest
in my work.
\\

I express my gratitude to J. Einasto, A. Kivila, U. R\"ummel, 
 J. Silvet, H. Albo, V. D. Malyuto,
J.-I. Weltmann, S. A. Kutuzov, H. Eelsalu, M. J\~oeveer, M. R\"ummel, E.
Kurvits, M. Kull and A. Linnas for their great assistance in the
preparation of my thesis.